\documentclass{article}
\usepackage{epubtk}    
\usepackage{natbib}
\usepackage{booktabs}  
\usepackage{graphicx}  
\usepackage{epsfig}
\usepackage{eqnarray,amsmath}

\usepackage{longtable}
\usepackage{rotating}

\usepackage[usenames, dvipsnames]{color}   

\newcommand{\arcsec}{\hbox{$^{\prime\prime}$}}
\newcommand{\arcmin}{\hbox{$^{\prime}$}}
\newcommand{\lo}[1]{\mbox{$\rm _{#1}$}}
\newcommand{\up}[1]{\mbox{$\rm ^{#1}$}}

\newcommand{\x}{$\times$}

\makeatletter 
\def\eqalign#1{\null\,\vcenter{\openup\jot\m@th 
  \ialign{\strut\hfil$\displaystyle{##}$&$\displaystyle{{}##}$\hfil 
      \crcr#1\crcr}}\,} 
\makeatother

\def\skf{\vskip 0.2truecm\noindent}

\def\orb[#1 #2]{{$#1^{#2}$}}
\def\tm[#1 #2 #3]{{$^{#1}{#2}_{#3}$}}

\def\red[#1]{{ \color{red} #1}}
\def\eul{{\rm e}}

\def\rsun{$R_\odot$}
\newcommand{\beq}{\begin{equation}}
\newcommand{\eeq}{\end{equation}}

\DeclareMathAlphabet{\mathsc}{OT1}{cmr}{m}{sc}
\def\testbx{bx}%
\DeclareRobustCommand{\ion}[2]{%
\relax\ifmmode
\ifx\testbx\f@series
{\mathbf{#1\,\mathsc{#2}}}\else
{\mathrm{#1\,\mathsc{#2}}}\fi
\else\textup{#1\,{\mdseries\textsc{#2}}}%
\fi}

\begin{document}

\title{Solar UV and X-Ray Spectral Diagnostics}

\author{\epubtkAuthorData{Giulio Del Zanna}{%
DAMTP, Centre for Mathematical Sciences,  University of Cambridge \\
Wilberforce Road Cambridge CB3 0WA, UK
}{%
gd232@cam.ac.uk}{%
}%
\and
\epubtkAuthorData{Helen E.\ Mason}{%
DAMTP, Centre for Mathematical Sciences,  University of Cambridge \\
Wilberforce Road Cambridge CB3 0WA, UK
}{%
hm11@damtp.cam.ac.uk}{%
}
}

\date{}
\maketitle

\begin{abstract}
 X-Ray and Ultraviolet (UV) observations of the outer solar atmosphere have
been used for
many decades to measure the fundamental parameters of the solar plasma.
This review focuses on the optically thin emission from the solar atmosphere,
mostly found at UV and X-ray (XUV) wavelengths, and discusses some of
the diagnostic methods
that have been used to measure electron densities, electron temperatures,
differential emission measure (DEM), and
relative chemical abundances. We mainly focus on methods and results
obtained from high-resolution spectroscopy, rather than broad-band
imaging.
However, we note that the best results are often obtained by combining
imaging and spectroscopic observations. We also mainly focus the review on
 measurements of  electron densities and temperatures obtained from single
ion diagnostics, to avoid issues related to the ionisation state of the plasma.
We start the  review with a short historical
introduction on  the main XUV high-resolution spectrometers,
then review the basics of optically thin emission and the main processes
that affect the formation of a spectral line.
We mainly discuss plasma in equilibrium, but briefly mention non-equilibrium
ionisation and non-thermal electron distributions.
We also summarise the status of atomic data, which are an essential
part of the diagnostic process.
We then review the methods used to measure
electron densities, electron temperatures, the DEM,  and
relative chemical abundances, and the results obtained for the lower
solar atmosphere (within a fraction of the solar radii),
for coronal holes, the quiet Sun, active regions and flares.
\end{abstract}

\epubtkKeywords{atomic processes; Sun: corona; atomic data - Line: formation - Techniques: spectroscopic - Sun: abundances }

\newpage

\tableofcontents






\newpage

 \section{Introduction}

The solar corona is the tenuous outer atmosphere of the Sun, revealed in its full glory  during a total solar eclipse. The visible spectrum of the solar corona has two major components: the continuum (the K-corona) due to Thomson scattering of photospheric light by the free electrons in the corona; and weak absorption lines (corresponding to the Fraunhofer lines -- the F-corona) superimposed on the continuum emission. The latter is due to scattering by interplanetary dust particles
in the immediate vicinity of the Sun.
From white light coronagraph observations, and using a model for the distribution of electrons in the corona \citep{vandehulst50}, it is possible to estimate the electron number density, which has a value of the order of 10$^8$ cm$^{-3}$ in the inner corona.

In addition, strong \emph{forbidden} emission lines  of highly-ionised atoms formed around 1\,--\,2~MK (e.g., the green and red coronal lines  Fe XIV 5303~\AA\ and  Fe X 6374~\AA) are also observed during eclipses. The forbidden lines allow measurements of electron densities and also of chemical abundances.

The solar corona is a very hot plasma (1 MK or more) that is mostly optically thin. The emission is due to  highly-ionised atoms, which emit principally in the X-rays (5\,--\,50~\AA), soft X-rays (50\,--\,150~\AA), Extreme Ultra-Violet (EUV, 150\,--\,900 \AA) or far Ultra-Violet (UV, 900\,--\,2000~\AA) region  of the spectrum. 
Since radiation at these wavelengths cannot penetrate to the Earth's surface,
most of the observations and spectral diagnostics have  been obtained from XUV (5\,--\,2000~\AA) observations from space. These observations and associated spectroscopic diagnostics are the main focus for this review.

When imaged in the EUV at 1~MK,  the solar corona shows a wide range of different structures that are magnetically linked to the underlying and cooler regions of the solar atmosphere, the chromosphere and the photosphere. Between the chromosphere and the corona there is a thin but highly complex region, the `transition region' \citep[TR, see e.g.,][]{gabriel:76}, where the temperature dramatically increases.
In addition we know, from both a theoretical and observational perspective, that 
there is a multitude of cooler  loops at transition region temperatures that
are not connected  with the corona, as discussed e.g. by 
\cite{feldman:1983,antiochos_noci:1986,landi_etal:2000,hansteen_etal:2014,sasso_etal:2015}.
The transition region emission  is highly dynamic and very complex to interpret, with the likelihood of
non-equilibrium and high-density effects that are normally not considered when studying the solar corona.

\begin{figure}[htbp]
 \centerline{\includegraphics[width=0.6\textwidth]{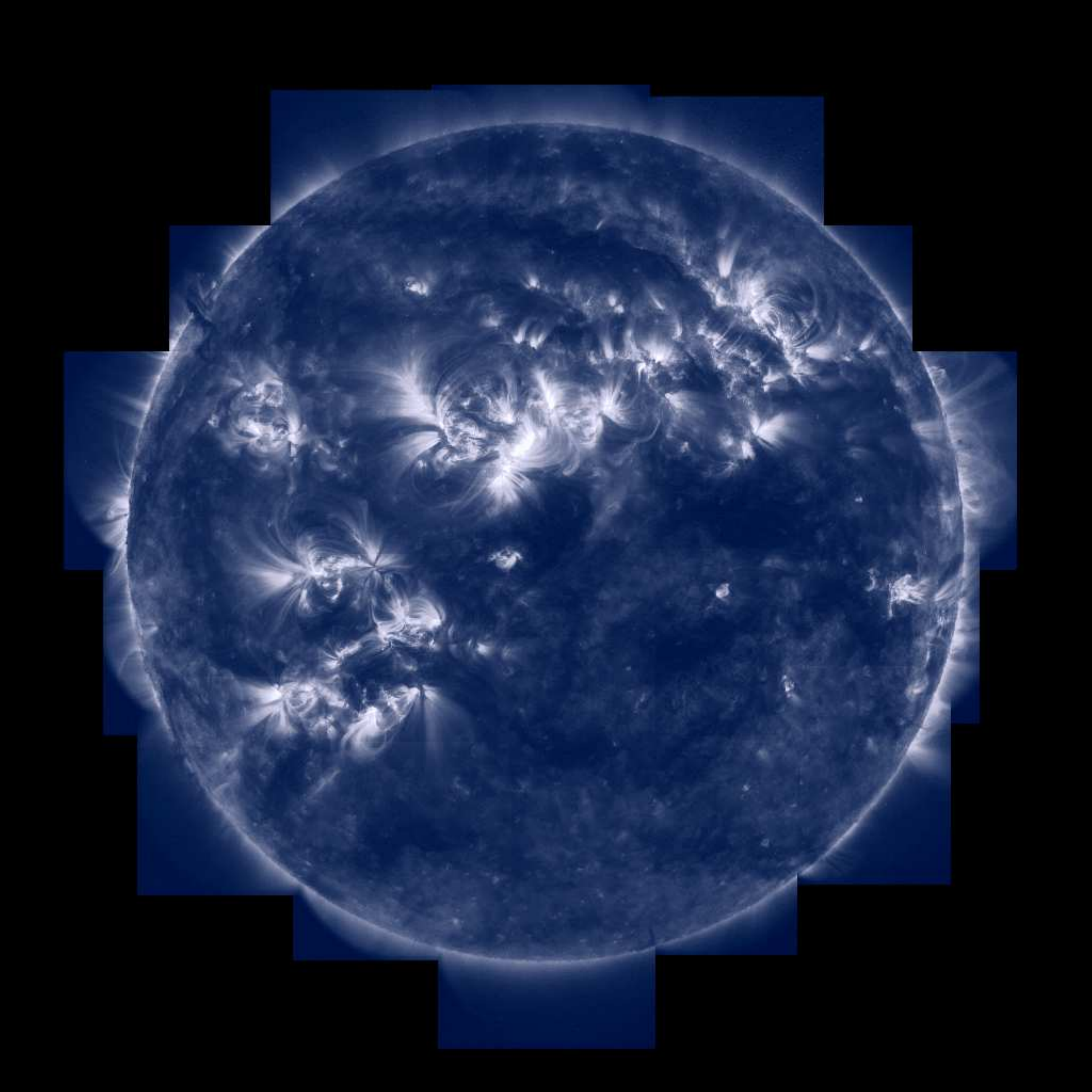}}
  \caption{A TRACE composite image of the solar corona at 171~\AA, formed at around 1 MK
(courtesy of the TRACE consortium, NASA).}
 \label{fig:corona}
\end{figure}

 Much of the corona appears to have a diffuse nature (at modest spatial resolutions) and is referred to as `quiet Sun (QS)'. This quiet corona, which corresponds to a mixed-polarity magnetic field,  is spattered with  small bipolar regions which give rise to `bright points (BP)' in the EUV and X-ray wavelength ranges. Then, in regions with enhanced magnetic field (which in the corresponding visible photosphere appear as sunspots), bright `active regions (AR)' form, with a multitude of extended loop structures (see  Fig.~\ref{fig:corona}).

The other large-scale features of the solar  outer atmosphere are the coronal holes (CH), which appear as  dark areas in EUV and soft X-ray images. At the photospheric level, they correspond to a prevalence of unipolar magnetic fields, corresponding to open magnetic field lines extending out into space. Inside polar coronal holes,  large-scale ray-like extended features are usually observed, at various wavelengths \citep[see e.g.,][]{deforest01a}. Due to their appearance these are named coronal hole plumes.

Remote-sensing XUV spectroscopy allows detailed measurements of plasma parameters such as electron temperatures and densities, the differential emission measure (DEM), the chemical abundances, Doppler and  non-thermal motions, etc.

These topics have a vast literature associated with them. 
The present review aims to provide a synthetic 
up-to-date summary of some of the spectral  diagnostics that have been used with data from recent 
missions or are currently routinely used, focusing on measurements of 
electron temperatures, electron number  densities and chemical abundances in the lower solar corona. 
The diagnostic techniques used to study the plasma thermal emission measure (EM) distribution 
are only briefly described, as more emphasis is given to  direct measurements of 
electron temperatures from  ratios of lines from the same ion.

Although there are no current or planned  instruments which will 
observe the X-rays with high-resolution spectrometers, we briefly discuss the 
rich set of diagnostics that 
in the past were available using satellite lines, i.e. lines formed 
by inner-shell excitation or dielectronic recombination.

We review the main processes underlying the formation of spectral line emission 
but we do not intend to replace in-depth presentations
of basic material that can be found in books such as that one on 
solar UV and X-ray spectroscopy by \cite{phillips_etal:08} or specialised ones, such as the 
older but very good review on the transition region by \cite{mariska:92}.

We provide a short review on atomic data, with emphasis on the most recent results. We also briefly 
describe some of the commonly-used  atomic codes to calculate atomic data, 
and mention some of the issues related to uncertainties and line identifications,
again not providing in-depth details on each of these topics, which were developed
over more than forty years. 
This review does not replace standard textbooks of atomic spectroscopy such as 
\cite{condon_shortley:1935, grant_book:2006, landi_book:2014}, nor standard 
books and articles on atomic calculations in general. 

The material presented here builds on and updates the useful 
review articles  on atomic processes and spectroscopic diagnostics for the solar transition region and corona 
that have been previously written by 
\cite{dere_mason:1981, gabriel_mason:82,doschek:1985_review,mason_monsignori:94,delzanna_mason:2013_book,bradshaw_raymond:2014}.

We briefly discuss non-equilibrium effects such as  time-dependent ionization
and non-thermal distributions, two areas that have recently received more attention.
Processes relating to hard X-ray emission from e.g. solar flares, as observed
by e.g. RHESSI are not discussed in this review. Recent reviews on this topic
have been provided by \cite{benz:2008_review, krucker_etal:2008_review}.

Plasma processes such as radiative transfer, relevant to the lower solar atmosphere
(e.g. chromosphere), are not covered in this review. Such information can be found in 
standard textbooks such as 
\cite{athay:1976, mihalas_book:1978}.
In the future, this Living review will be extended to cover the diagnostics of the outer 
solar corona, where densities become so low that photo-excitation and resonance excitation from the 
disk radiation need to be included in the modelling.

\section{The Solar XUV Spectrum}
\label{sec:instruments}

We first briefly review some of the main spectrometers which
have been used to observe the Sun from the X-rays to the UV.
In this review, we do not discuss hard X-ray spectrometers,
nor spacecrafts  which carried spectrometers, but ultimately
did not produce spectra.
The emphasis is on high-resolution spectra.
For most diagnostic applications, having an accurate
radiometric calibration is a fundamental requirement, but particularly
difficult to achieve, especially in the EUV and UV.
Significant degradation typically occurs in space, due to
various effects (see, e.g. the  recent review of  \cite{benmoussa_etal:2013}).
We also mention some  EUV imaging instruments which have been used 
extensively.

\subsection{Historic perspective}

The solar corona has been studied in detail since the early 1960s
 using data from a number of rocket flights.
Some of them produced  the best XUV spectra of the solar corona
and transition region to date.
An overview of these early days can be found in
\cite{doschek:1985_review,mason_monsignori:94,  wilhelm_etal:2004}.

\subsubsection{Rocket flights}

The X-rays are mostly dominated by  L-shell  ($n=2,3,4 \to 2$)
emission from highly ionised atoms.
Early (but excellent) X-ray spectra of the Sun were obtained
by a large number of rocket flights, see
for example \cite{evans_pounds:1968}, \cite{davis_etal:1975},
and the reviews of  \cite{neupert:1971,walker:1972}.

The best X-ray spectrum of a quiescent active region
was obtained with an
instrument, built by the University of Leicester (UK),
 which consisted of Bragg crystal spectrometers
with a collimator having a  FOV (FWHM) of 3\arcmin,
and flown on a British Skylark sounding rocket  on 1971  Nov 30
\citep{parkinson:75}.
The instrument  had an excellent spectral resolution,
 was radiometrically calibrated and
the whole spectral region was observed simultaneously, unlike
many other X-ray instruments which  scanned  the spectral regions.

The soft X-ray (50--170~\AA) spectrum of the quiet and active
 Sun is rich in $n=4 \to n=3$ transitions from  highly ionised iron ions,
from \ion{Fe}{vii} to \ion{Fe}{xvi}
(see, e.g.  \citealt{fawcett_etal:68}).
 \cite{manson:72} provided an excellent list of
calibrated soft X-ray irradiances
 observed in quiet and active conditions in the 30--130~\AA\ range
by two rocket flights, on 1965 November 3 and 1967 August 8.
The spectral resolution was moderate, about 0.23~\AA\ (FWHM)
for the quiet Sun, and 0.16~\AA\ for the active Sun observation.

\cite{behring_etal:72}
published a line list from a
high resolution (0.06 \AA) spectrum in the 60-385 \AA\ region
of a moderately active  Sun. The instrument
was  built at the Goddard Space Flight Center (GSFC) and flown on
an Aerobee 150 rocket flight on 1969 May 16.
A similar line list for the EUV was produced by  \cite{behring_etal:76}.
\cite{behring_etal:72} and \cite{behring_etal:76} represent the 
best solar EUV line lists in terms of accuracy of wavelength measurements 
and spectral resolution. 
Unfortunately, these EUV spectra were not radiometrically calibrated.

\cite{malinovsky_heroux:73}
 presented  an integrated-Sun spectrum
covering the 50-300 \AA\ range with a medium resolution (0.25 \AA),
taken with  a grazing-incidence spectrometer  flown on a
rocket on 1969 April 4.
The photometric calibration of the EUV part of the spectrum
was exceptionally good (about 10--20\%), but the soft X-ray part was 
recently shown to be incorrect by a large factor
\citep{delzanna:12_sxr1}.

\cite{acton_etal:85}  published a high-quality solar spectrum
recorded on photographic film during the  rocket flight on 1982 July
13, two minutes after the GOES X-ray peak emission of an M1-class flare.
The instrument was an X-Ray Spectrometer/Spectrograph Telescope (XSST).

The spectrum was radiometrically calibrated, and it provided accurate line intensities
from 10~\AA\ up to about  77~\AA.
The spectral resolution was excellent, clearly resolving
lines only 0.04~\AA\ apart.
Excellent agreement between predicted and measured line intensities
has been found (see, e.g. the recent study of \citealt{delzanna:12_sxr1}).

At longer wavelengths, the best EUV  spectra have been obtained
by the series of  GSFC  Solar Extreme Ultraviolet
Rocket Telescope and Spectrograph (SERTS) flights.
The  one  flown in 1989 (SERTS-89)
(\citealt{thomas_neupert:94}; hereafter TN94) observed the
235-450 \AA\ range in first order.
The SERTS-95 covered shorter wavelengths \citep{brosius_etal:98b}.
The SERTS-97 \citep{brosius_etal:00}  covered the 300--353~\AA\
spectral region.
Both SERTS-89 and SERTS-97 were radiometrically calibrated, although 
the calibration of the SERTS-89 spectra has been questioned \citep{young_etal:98}.
Other  SERTS and EUNIS  (see, e.g. \citealt{wang_etal:10,wang_etal:2011}) 
sounding rockets built at GSFC  have served for the
calibration of in-flight EUV spectrometers of several satellites, but have also
returned many scientific results.

\begin{sidewaystable}
\centering
\caption[Overview of XUV spectroscopic instruments]{Some of the older space-borne spectroscopic 
instruments that observed   the  solar corona in the XUV.}
\begin{center}
\begin{tabular}{|l|c|c|c|c|c|c|l|}
\toprule
Instrument 	&Dates	&$\lambda$ [\AA]&$\delta \lambda [\AA]$	&min $\delta S$ 	&$\delta t $	&FOV	& Calibrated \\ 
\midrule

\hline
OSO-4		&1968	&300-1400	&3.2		&60 		&900 full scan 	&1x1		& \\ 

\hline
Rocket     	&1969	&50-300		&0.25		&-		&-		&integrated Sun	& \\ 
\hline

GSFC Rocket	&1969	&60-385		&		&		&		&		& \\ 

\hline

OSO-5		&1969	&25-400		& $\sim$ 0.4	&		&900 full scan	&integrated Sun	& \\ 

\hline

OSO-5		&1969	&280-370	&integrated	&none		&2		&integrated Sun	& \\ 
		&	&465-630	&		&		&		& 		&\\
		&	&760-1030	&		&		&		& 		&\\
\hline
OSO-6		&1969	&280-1390	&3.2		&35		&900 full scan 	&1x1 		&  \\ 
\hline
OSO-7		&1972	&150-400	&0.85		&20		&120 	 	&5x5 & \\ 

\hline

OSO-8  UV spectrometer 	& 1975--1978 	& 1200-2000	&	0.02	&  2.5\arcsec	&  20-50s 	&	variable  & Yes \\ 

OSO-8  UV/vis. & 1975--1978 & Lyman $\alpha,\beta$ & 2-10 pm &  $\simeq$2	&  variable	& variable  & Yes \\ 
 \quad\quad polychromator  &  &  \ion{Mg}{ii},\ion{Ca}{ii} &  &  	&  	&   &  \\ 

OSO-8 graphite crystals & 1975--1978  & 1.5--6.7      &   0.03--0.06        &    -  &  10s  & full-Sun & Yes (10\%)  \\ 

OSO-8 PET crystals  & 1975--1978   & 5.13--7.18        &           &    -  &  10s  & full-Sun & Lab (30\%) \\ 

\hline
\hline


Skylab SO55 (HCO)&1973-4	&296-1340	&1.6		& 5$\times$ 5\arcsec	&$\sim$ 330		& variable 	& Yes (35\%) \\

Skylab SO82A (NRL)&1973-4	&171-630	& $\ge$0.03		&$\ge 2$\arcsec slit-less & 		& full-Sun      & \\
Skylab SO82B (NRL)&1973-4	&970-3940	&0.04-0.08	& 2$\times$60\arcsec &		&	 &    Yes \\

\hline
P78-1 SOLEX A &   1979--  &  7.8--25 & 0.02  at 16~\AA &  20\arcsec   &  56s & variable &  \\ 
P78-1 SOLEX B &   1979-- &  3--10 & 0.001 at 8~\AA &  60\arcsec  &  56s &  variable &  \\ 
P78-1 SOLFLEX &   1979-- &  1.82--8.53 & 0.00024-0.001 &   -  &  56s & full-Sun &  \\  

\hline

HRTS (8 rockets) & 1975-1992 & 1170-1710  & 0.05	&1\arcsec  &		&		&   Yes (some) \\
\hline

CHASE		& 1985	& 160-1344 & 0.25-0.4	& 15\arcsec		&few sec.	&3x1\arcmin max	&  No \\

\hline


SMM XRP FCS & 1980--1989 &  1.8--25    &    &  15\arcsec$\times$14\arcsec  & $\simeq$ 10 m  &  variable & Yes  \\ 
SMM XRP BCS & 1980--1989 &   1.7--3.2 & $\simeq $ 0.0005  & 6'$\times$6' & $\le$ 1s & & Yes \\

SMM UVSP & 1980--1981 & 1750--3600 & 0.04 (0.02 IIo) & 3 & minutes &  variable &  Yes  \\
\hline

Hinotori SOX1,2 & 1981-1982 &   1.7--1.95    & 0.00015        &  -  &  $\simeq$ 1   &  full-Sun  & Yes  \\

\hline
 
XSST  & 1982 Jul 12 &  10--77 & 0.04  &  -  &  145s & 625 arcsec$^2$  & Yes  \\

\hline
Yohkoh BCS & 1991--2001 & 1.8--5.0  & 0.0004--0.002  &  -  &  $\simeq$ 1   &  full-Sun  & Yes  \\

\hline

SERTS		&1989-& 235-450 &0.06		&6		&none		&5x8		& \\

\noalign{\smallskip}
\hline


\bottomrule
\end{tabular}
\end{center}
\label{tab:instruments}
\end{sidewaystable}

\subsection{OSO}

After the early rocket flights, the first series of  small satellites
were the  Orbiting Solar Observatories (OSO).
The first observations of $n=3 \to 2$  lines in solar flares
were made with the OSO-3 satellite in the 1.3--20~\AA\ region
 \citep{neupert_etal:67}.
OSO-5 produced spectra
of solar flares in the 6--25~\AA\ region \citep{neupert_etal:73},
and the first solar-flare spectra containing the $n=2 \to 2$ L-shell
iron emission, in the  66--171~\AA\ range \citep{kastner_etal:1974_flare}.
OSO-6 also provided solar flare spectra
(see the  \citealt{doschek:1972,doschek_etal:1973} line lists). 
OSO-7 produced EUV spectra in the 190--300~\AA\ range
of the coronal lines \citep{kastner_etal:1974_limb}, later  studied in detail 
by  \cite{kastner_mason:1978}.

OSO-8 (1975--1978)
obtained the first high spatial (2\arcsec) and spectral observations
of the chromosphere and the transition region 
with an UV spectrometer which operated in the 1200--2000~\AA\ range
\citep{bruner:1977}.
In its scanning mode, a line profile would typically be scanned
across 1\AA\ with very high spectral resolution (0.02~\AA) but 
low cadence (30--50s).
OSO-8 also carried a UV/visible polychromator which observed the
\ion{H}{i} Lyman $\alpha,\beta$,  the \ion{Mg}{ii} h,k, and
\ion{Ca}{ii} H,K lines.
Results from these  instruments are reviewed by \cite{bonnet:1981}. 
Unfortunately, the instruments suffered a 
 drop in sensitivity very early on during the mission. The degradation
was so dramatic that measures were put in place by Bonnet and others for a strict 
cleanliness program for the SOHO spacecraft.
This cleanliness program was an overall success,
as most instruments suffered little degradation, compared to other 
missions (see below).

OSO-8 also carried two X-ray spectrometers,  with co-aligned graphite and PET
crystals \citep[see][and references therein]{parkinson_etal:1978}.
The graphite had a large geometrical area (100 cm$^2$) but lower
spectral resolution than the PET system.
The spectrometers were uncollimated so viewed the whole Sun. 
The overlapping of spectra from active regions located in different parts of the solar surface 
was therefore a problem.
The great advantages of the OSO-8 spectrometers over previous ones
 were the high sensitivity and the fact that the crystals were
fixed, i.e. the entire wavelength ranges were observed with a high cadence,
about 10s.
The graphite crystal spectrometer was well calibrated (10\%) in the laboratory
and in flight \citep{kestenbaum_etal:1976}.

\subsection{Skylab}

More detailed studies of the solar corona from space started
in May 1973, when  Skylab, the first NASA space station, was launched.
The Apollo Telescope Mount (ATM) on
Skylab carried several  solar instruments, observing from the UV to the X-rays
between June 1973 to February 1974.
Three successful series of Skylab workshops held in Boulder,
Colorado,  summarised the main results:
Coronal Hole and High Speed Wind Streams \citep{zirker:77}, Solar Flares
\citep{sturrock:80} and Solar Active Regions \citep{orrall:81}.

The  Harvard College Observatory  (HCO)
 EUV spectrometer  SO55 \citep{reeves_etal:77a}  on the ATM
had  a good spatial resolution in the 296~\AA\ -- 1340~\AA\ range,
but  low spectral resolution ($\simeq$ 1.6 \AA\ or more, depending on the
spectral range).
In the standard grating position, the instrument scanned
with a 5\arcsec\  $\times$ 5\arcsec\  slit covering  six wavelengths
typical of chromospheric to coronal temperatures:
Ly$\alpha$ (1216 \AA, $T \simeq 2  \times  10^4$ K);
C II (1336 \AA, $T \simeq 3.5  \times  10^4$ K);
C III (977 \AA, $T \simeq 7  \times  10^4$ K);
O IV (554 \AA, $T \simeq 1.5  \times  10^5$ K);
O VI (1032 \AA, $T \simeq 3  \times  10^5$ K) and
Mg X (625 \AA, $T \simeq 1.1  \times  10^6$ K).
With other grating positions, spectroheliograms of the lines
Ne VII (465 \AA, $T \simeq 5  \times  10^5$ K) and
Si XII (521 \AA, $T \simeq 1.8  \times  10^6$ K) were also recorded.
The good spatial resolution of the HCO spectrometer
enabled \cite{vernazza_reeves:1978}  to
produce a list of line intensities for different solar regions,
which was  a standard reference for many years.
The radiometric calibration of the instrument (about 35\% uncertainty)
is described by \cite{reeves_etal:77b}. This instrument
suffered severe in-flight  degradation.

\begin{figure}[htbp]
 \centerline{\includegraphics[width=\textwidth]{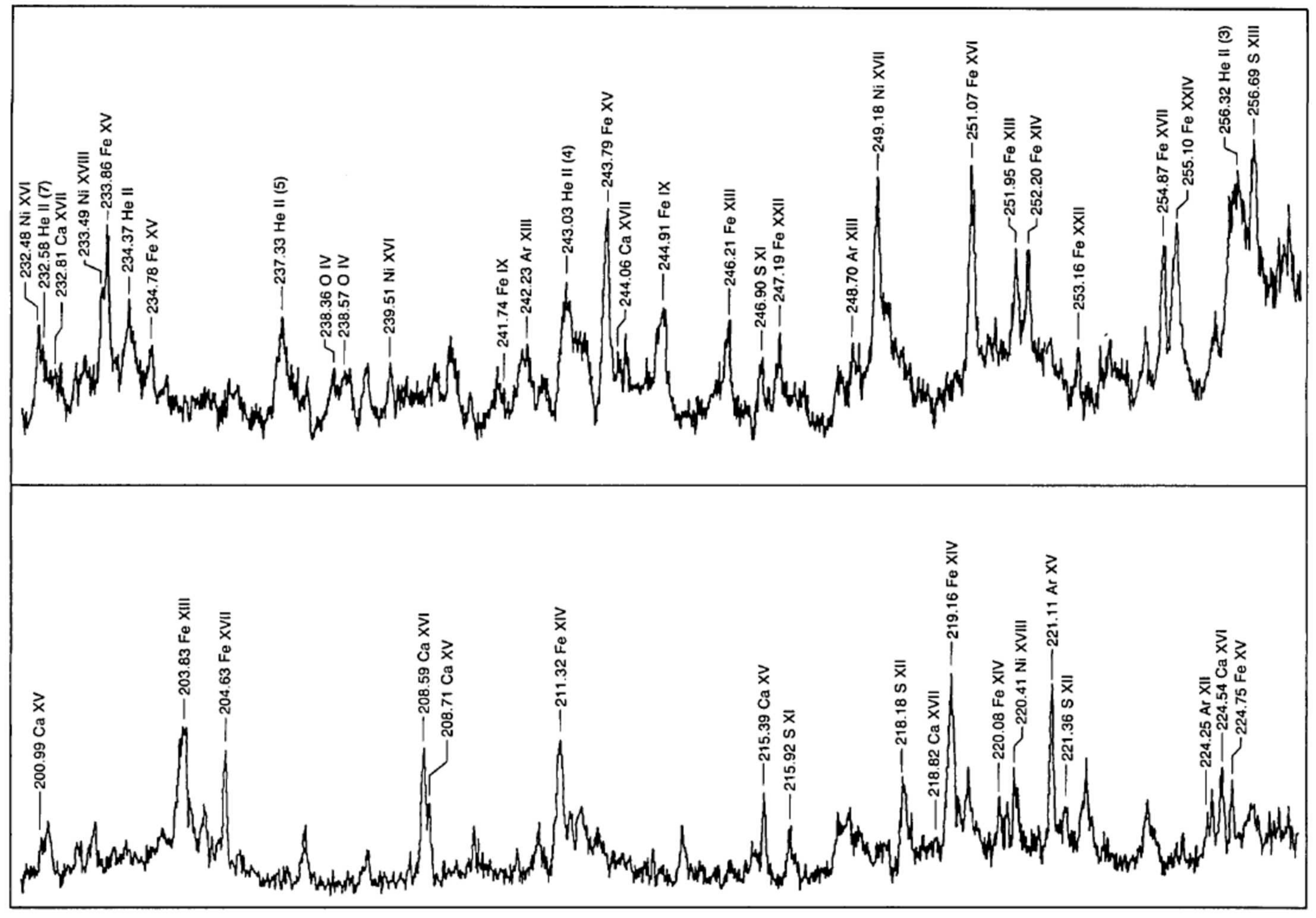}} 
  \caption{A portion of the NRL S082A slitless spectrum of a solar flare.
Image reproduced with permission from \citep{feldman_etal:1988}, copyright by OSA.
}
 \label{fig:skylab_slitless}
\end{figure}

The  Naval Research
Laboratory (NRL) S082A  slitless spectroheliograph on the ATM
\citep{tousey_etal:77} had  a very good  wavelength
coverage (170--630~\AA), with a spatial resolution reaching 2" for
small well defined features.
The instrument obtained 1023 spectroheliograms of
the whole Sun. However, the dispersion direction
coincided with one spatial dimension (normally oriented E-W), so the
images of the solar disk in the nearby spectral lines
 were overlapped (the instrument was fondly called the `overlappograph').
For this reason, most of
scientific results have been  obtained from spectra emitted by small
well defined regions,
such as
compact flares, active region loop legs, the limb brightening, etc.
For the smallest features, the instrument achieved an excellent spectral
resolution of about 0.03~\AA. As an example of the 
excellent quality of the instrument, a portion of a flare spectrum is 
shown in Fig.~\ref{fig:skylab_slitless}.
A complete list of lines observed during flares was
produced by \cite{dere:78}.

\begin{figure}[!htbp]
 \centerline{\includegraphics[width=.8\textwidth]{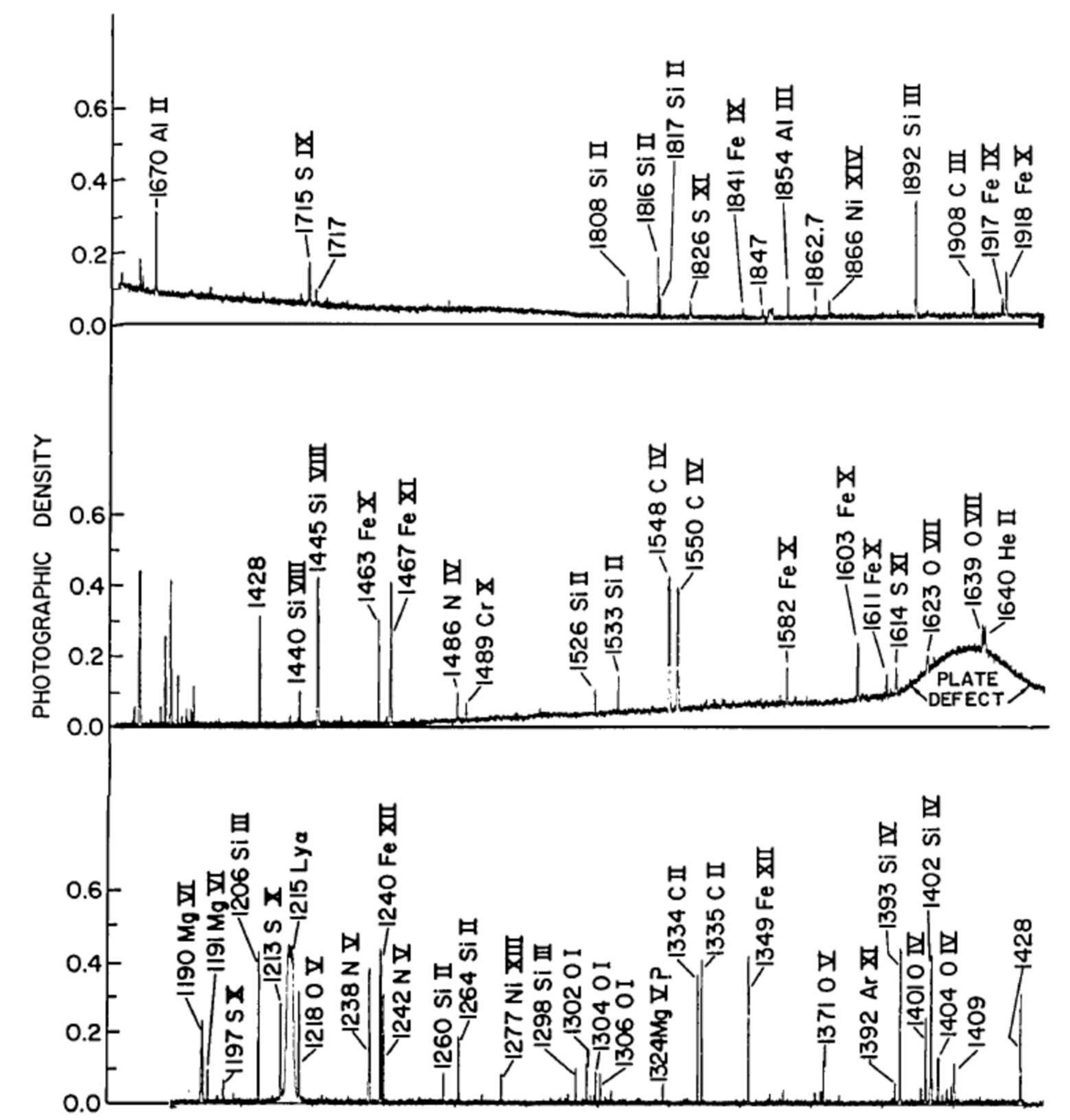}} 
  \caption{A portion of the NRL S082B   spectrum taken about 30\arcsec\ off the 
solar limb, showing several coronal forbidden lines. 
Image reproduced with permission from \cite{feldman_etal:1988}, copyright by OSA.
}
 \label{fig:skylab_slit}
\end{figure}

The NRL Skylab SO82B \citep{bartoe_etal:1977} had a
2\arcsec $\times$60\arcsec\ non-stigmatic slit and an excellent
spectral resolution (0.04-0.08~\AA) over the 970-3940~\AA\ spectral range.
\cite{sandlin_etal:77, sandlin_tousey:79} produced lists of coronal
forbidden lines.  Fig.~\ref{fig:skylab_slit} shows  a spectrum taken about 30\arcsec\ off the 
solar limb, showing several coronal forbidden lines
\citep{feldman_etal:1988}. 
Line lists of chromospheric lines were provided by
\cite{doschek_etal:77} and \cite{cohen:1981}. Many excellent papers using the
S082A and S082B instruments were produced by the groups at NRL and collaborators.

\subsection{ P78-1}

Significant improvements in terms of  spectral resolution
 were achieved in the X-rays with the
 SOLEX  and SOLFLEX crystal spectrometers aboard the US P78-1 satellite,
launched in 1979
(for a description of the solar instruments on the P78-1 spacecraft
see \citealt{doschek:1983}).

The SOLEX crystals scanned the 3--25~\AA\ spectral region 
with a resolution of 10$^{-3}$~\AA\ at 8.2~\AA\ and
two multi-grid collimators  of 20 or 60\arcsec. For more details, see 
\cite{landecker_etal:1979}.
The SOLFLEX crystals observed the full Sun and 
covered  four spectral bands in the   1.82--8.53~\AA\ range 
(1.82--1.97; 2.98--3.07; 3.14--3.24; 8.26--8.53~\AA)
with a resolution varying from 
2.4 $\times$ 10$^{-4}$~\AA\ at 1.9~\AA\ to 10$^{-3}$~\AA\ at 8.2~\AA.
The bands were chosen to observe the strong resonance lines from 
Fe XXV, Ca XX,  and Ca XIX with their associated satellite lines, 
as well as  several other lines from high-temperature ions.
The spectral ranges were  scanned by rotating the crystals, with a 
typical cadence of 56s.

These spectrometers allowed seminal discoveries related to solar flares.
During the rise (impulsive) phase of solar flares, strong 
blue asymmetries in the resonance lines were observed, interpreted as 
upflows during the  chromospheric evaporation. The crystals also 
enabled the observation of  non-thermal broadening and the measurement of  
temperature and emission measure variations during flares.
For a review of the SOLEX 
results see, e.g.  \cite{mckenzie_etal:80a,mckenzie_etal:80b,doschek_etal:1981,mckenzie_etal:85}.
For a review of SOLFLEX spectra and their interpretation
 see e.g.  \cite{doschek_etal:1979,doschek_etal:1981_solflex}.

\subsection{HRTS}

Superb  UV solar spectra were obtained by the series of
High Resolution Telescope and Spectrometer
(HRTS) instruments, developed at  the Naval Research Laboratory (NRL)
\citep{brueckner83}. 
HRTS was flown eight times on sounding rockets 
 between 1975 and 1992 which enabled  many results.
It was also flown on Spacelab 2, together with the CHASE instrument (see below)
in 1985.
HRTS covered the 1170--1710~\AA\ spectral region with very high
spectral (0.05~\AA) and spatial (1\arcsec) resolution.
HRTS had a stigmatic slit and also provided slit-jaw images.
Each flight was unique, in that different slits, wavelength ranges or pointing
were chosen. A good review of the flights can be found in 
\url{http://wwwsolar.nrl.navy.mil/hrts_hist.html}, written by K. Dere.

\cite{sandlin_etal:1986} published a well-known list of  HRTS
observations of different regions on the Sun, with accurate wavelengths
and line intensities, in the 1175--1710~\AA\ spectral range.
This list represents the most complete coverage in this wavelength range.
A modification of the HRTS was  flown on Spacelab 2
\citep{brueckner_etal:86}.

\begin{figure}[htbp]
 \centerline{\includegraphics[width=1.\textwidth]{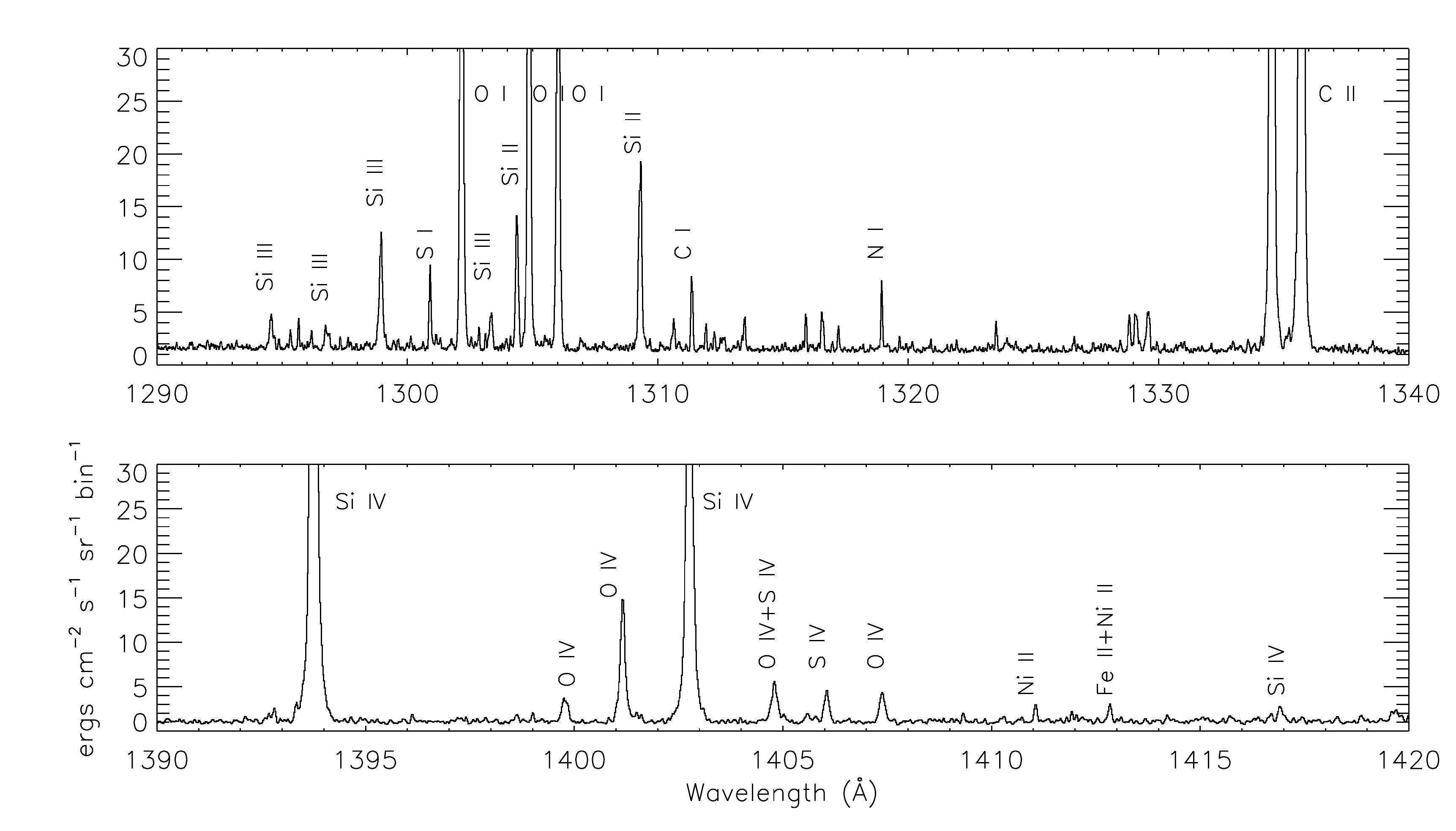}}
  \caption{An  averaged spectrum of an active region at the 
limb  from HRTS, in two spectral regions, dominated by  the 
important UV lines from \ion{Si}{iii}, \ion{C}{ii}, \ion{Si}{iv}, 
\ion{O}{iv} and \ion{S}{iv}. Data from  \citep{brekke_etal:1991}.
}
 \label{fig:hrts}
\end{figure}

\cite{brekke_etal:1991} published  excellent HRTS spectra, obtained during the
second rocket flight in February 1978. The spectrum was radiometrically
calibrated by matching the quiet Sun intensities with those measured by the
Skylab S082B calibration rocket flight CALROC.
The long stigmatic slit of the HRTS instrument covered  many solar regions.
 Fig.~\ref{fig:hrts} shows an averaged spectrum of an active region at the 
limb from HRTS.

\subsection{CHASE}

The Coronal Helium Abundance Spacelab Experiment (CHASE, see \citealt{breeveld_etal:1988}) on
the Spacelab 2 Mission (1985) was  specifically designed to determine  the helium
abundance from the ratio of He~II 304 \AA\ to Lyman-$\alpha$ 1218 \AA,
on the disc and off the limb.
A value of 0.079$\pm$0.011 for the quiet corona was obtained
 by \cite{gabriel_etal:95}.
However, the instrument also recorded several spectral lines, used by 
\cite{lang_etal:1990} to describe the temperature structure of the 
corona. One limiting aspect of  CHASE  was the lack of a specific 
radiometric calibration for the flight instrument.

\subsection{SMM}

The Solar Maximum Mission (SMM) was dedicated to the study of
active regions and solar flares.
SMM was launched in February 1980 but encountered  some problems.
It was repaired  in-orbit  by the NASA Space Shuttle in
April 1984, and then resumed full  operations until December 1989.
It carried several instruments.

\begin{figure}[htb]
 \centerline{\includegraphics[width=0.7\textwidth,angle=90]{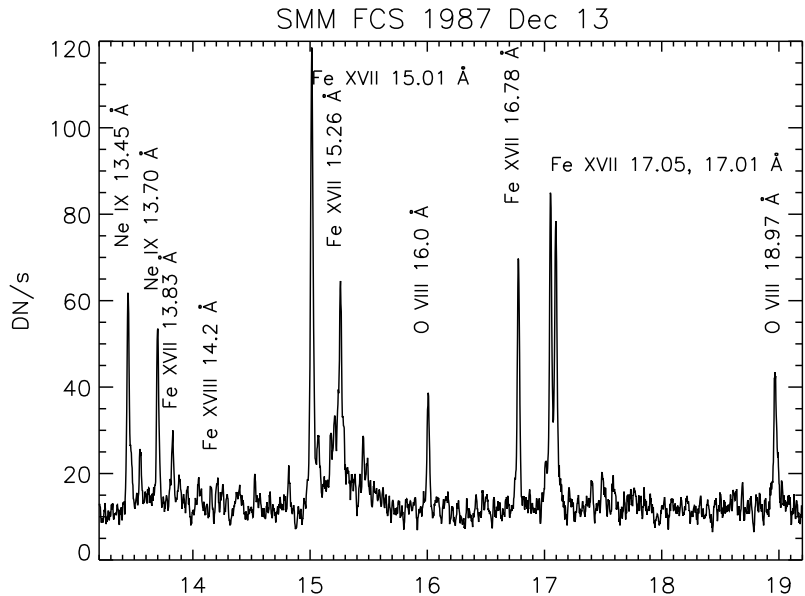}}
  \caption{The SMM/FCS AR spectrum of 1987 Dec 13.
Image reproduced with permission from \cite{delzanna_mason:2014}, copyright by ESO.
 }
 \label{fig:fcs_sp}
\end{figure}

The SMM  X-ray polychromator (XRP)
Flat and Bent Crystal Spectrometers  (FCS and BCS)
instruments \citep{acton_etal:1980} produced excellent X-ray spectra of the
solar corona, active regions and solar flares.
The XRP/FCS had a collimator of about 15\arcsec\ $\times$14\arcsec, so had a limited
spatial resolution, although it could raster  an area 
as large as 7'$\times$7' with 5\arcsec\ steps.
The FCS crystals could be rotated to 
provide the  seven detectors  access to the spectral range 1.40-22.43~\AA.
The FCS  5--25~\AA\ spectral region is 
 dominated by \ion{Fe}{xvii}, \ion{O}{viii},  \ion{Ne}{ix} lines.
A sample spectrum of a quiescent active region is shown in
Fig.~\ref{fig:fcs_sp}.
The sensitivity of the FCS instrument 
decreased significantly early on in the mission, in particular at  the longer wavelengths,  
so the \ion{O}{vii} lines around 22~\AA\  became very weak.
The in-flight  radiometric calibration was based on the assumption of carbon
deposition on the front filter (see, e.g. \citealt{saba_etal:99}).
The FCS  data allowed measurements of the temperature distribution and
electron density of solar active regions and flares together with
measurements of relative coronal abundances
of various elements.
A complete list of lines observed with the SMM FCS during flares
was published by \cite{phillips_etal:1982}.
Another excellent  SMM/FCS solar flare spectrum,
this time with $n=4 \to 2$ calibrated  line intensities,
was published by \cite{fawcett_etal:87}.
 A significant limitation of these solar observations
was that the spectral range of each detector was scanned,
 hence  lines within the same channel were not
observed simultaneously.  This considerably complicated
the analysis (cf. \citealt{landi_phillips:2005}), particularly for solar flares.

The XRP/BCS, with a collimator field of view of about 6'$\times$6' (the size of a large 
active region), 
was able to obtain spectra with eight position-sensitive proportional counters 
in the range 1.7-3.2~\AA\ simultaneously, at a resolving power of about 10$^4$.
The XRP/BCS  observed X-ray line complexes of high
temperature (in excess of 10~MK) coronal lines:
\ion{Fe}{xxvi}, \ion{Fe}{xxv}, \ion{Ca}{xix} and \ion{S}{xv}.
This  spectral range allowed a wide range of plasma
diagnostics to be applied. The He-like ions allowed
measurements of electron densities and temperatures \citep{gabriel_jordan:1969a}.
The  satellite lines also allowed some important diagnostic
measurements
(see, e.g.
\citealt{gabriel:1972,gabriel:1972b,gabriel_phillips:1979,doschek:1985_review,phillips_etal:08}).

The Ultraviolet Spectrometer and Polarimeter (UVSP)
 on board SMM was used to observe many features, including solar active regions and flares in
the  1750--3600~\AA\ range in first order of diffraction
and 1150--1800~\AA\ in the second
\citep{woodgate_etal:1980}.  The spatial resolution was very good,
about 3\arcsec, and the spectral resolution was excellent
(0.04 \AA\ in first order and 0.02 in second order).
Various slits were available, from 1 to 15\arcsec\ wide.
As with previous UV instruments, the UVSP also suffered severe degradation
in orbit \citep{miller_etal:1981} during its ten months of operations.

\subsection{Hinotori}

Hinotori was a Japanese  spacecraft in orbit  
between 1981 and 1982 which was used to observe high-energy
X-ray emission produced by solar flares.
The most important results were obtained with the Bragg
spectrometers. 
The SOX2 scanning spectrometer  had an excellent  spectral resolution 
(0.15 m\AA) and produced excellent spectra of \ion{Fe}{xxvi} and \ion{Fe}{xxv}
during large flares. 
Very high temperatures were measured. For a summary of the results  from these 
spectrometers  see \cite{tanaka_etal:1982,tanaka:1986}.

\subsection{Yohkoh}

Yohkoh (Japanese for {\it sunbeam}) was used to observe the Sun in X-ray emission from
1991 to 2001. The Bragg Crystal Spectrometer (BCS, see 
\citealt{culhane_etal:1991})
produced similar measurements to the SMM/BCS of the
 X-ray H- and He-like  line complexes.
It used 4 bent crystals which observed the 
 X-ray lines of highly ionized S, Ca, and Fe 
produced by flares and active regions in the 1.76--5.1~\AA\
wavelength range
(1.76--1.80: \ion{Fe}{xxvi}; 1.83--1.89: \ion{Fe}{xxv}; 
3.16--3.19: \ion{Ca}{xix}; 5.02--5.11: \ion{S}{xv}).
Many scientific results have been obtained, in particular 
those regarding the temperatures during flares  
\citep{phillips_feldman:1995,feldman_etal:1996}.

One potential problem with instruments such as  XRP/BCS and the 
Yohkoh BCS  is that 
sources in different spatial locations  produce superimposed
spectra at different wavelengths. For example, extended sources produced a 
broadening of the lines, and complex spectra 
could arise  in the case of multiple flares occurring at the same time
within  the field of view. 
This was not normally an issue for XRP/BCS, given its field of view (6$\times$6 arc minutes),
but was occasionally more of a problem for Yohkoh/BCS which observed the full Sun.

\subsection{SoHO}

The Solar and Heliospheric Observatory  (SoHO), 
a joint NASA and ESA mission which was launched in December 1995
to the L1 position,  is still operational, although the attitude loss in
1998
caused significant degradation to  some of its instruments, many of which 
have now been switched off.
First results from SoHO were published in a special issue of
Solar Physics in 1997 (volume 170).
With 24 hour monitoring, SoHO has produced a wealth of data and 
has changed our view of the Sun.
SoHO carried a suite of several instruments, performing in-situ and 
remote-sensing observations.
The radiometric calibration of various instruments on-board SoHO
during the first few years of the mission
was discussed during two workshops held at the
International Space Science Institute (ISSI), in Bern.
The results were summarised in the 2002 ISSI Scientific Report SR-002
\citep{SOHO-calib:02}. 

 Here we summarise the spectroscopic instruments used to study the 
solar corona.
The Coronal Diagnostic Spectrometer (CDS),
 a  UK-led instrument \citep{Harrison-etal:95} was routinely operated from
1996 to 2014.
It comprised  of a Wolter-Schwarzschild
 type II  grazing incidence telescope, a scan mirror, a set of
different slits (2, 4, 90\arcsec),   and two spectrometers, a Normal
Incidence Spectrometer (NIS) and a Grazing Incidence Spectrometer
(GIS). The wavelength range covered by the two detectors (150 - 800~\AA)
contains many emission lines emitted from the chromosphere,
the transition region and  the corona.
The NIS had  two wavelength bands, NIS-1, 308 - 381~\AA\ and NIS-2, 513 -
633~\AA.
To construct monochromatic images (rasters), a scan mirror was moved
across a solar region to project onto the detectors
the image of a stigmatic slit (2" or 4"). For the NIS instrument,
the spectral resolution was about 0.3~\AA\ before the SoHO loss
of contact in 1998, then degraded to about 0.5~\AA\ afterwards.
The effective spatial resolution was about 4\arcsec.

The in-flight radiometric calibration of the CDS instrument
 was found to be very different from that which had been measured on the
ground (actually better than expected,
for the NIS see, e.g. \citealt{landi_etal:97,delzanna01_cdscal,lang_etal:07},
while for the GIS see \citealt{delzanna01_cdscal}).

The CDS team believed that 
the variation in the long-term radiometric calibration of the NIS instrument was 
mainly caused by the degradation of the microchannel plate  detectors
following the use of the wide slit. A standard correction was 
implemented in the calibration software. 
However, \cite{delzanna_etal:10_cdscal} showed that this standard correction
was quite different  from expectation, and 
overestimated by large factors (2--3) for the stronger lines.
The NIS instrument only degraded by about a factor of two in 
13 years, which is quite remarkable. 
The new calibration by \cite{delzanna_etal:10_cdscal}
was confirmed by sounding rocket flights
(e.g. EUNIS-2007, see \citealt{wang_etal:2011}) and was adopted for the 
final calibration of the instrument.

\begin{figure}[htbp]
\centerline{\includegraphics[width=12cm,angle=90]{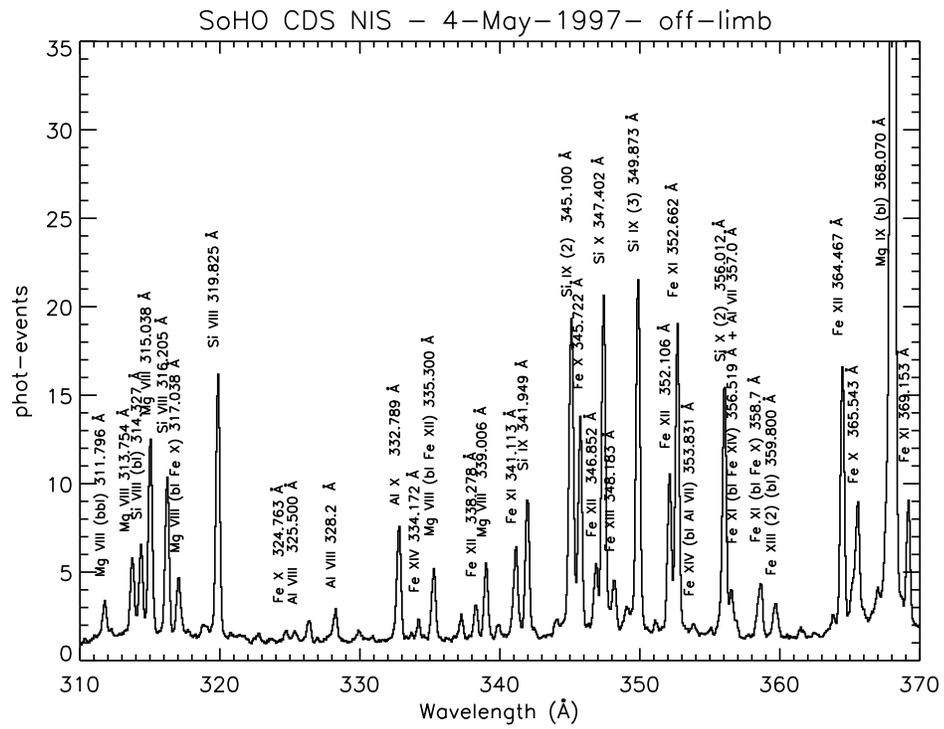}} 
  \caption{CDS  NIS 1 averaged spectrum of the quiet Sun (off-limb)
 \citep[adapted from ][]{delzanna_thesis99}.}
  \label{fig:nis1}
\end{figure}

\begin{figure}[htbp]
\centerline{\includegraphics[width=12cm,angle=90]{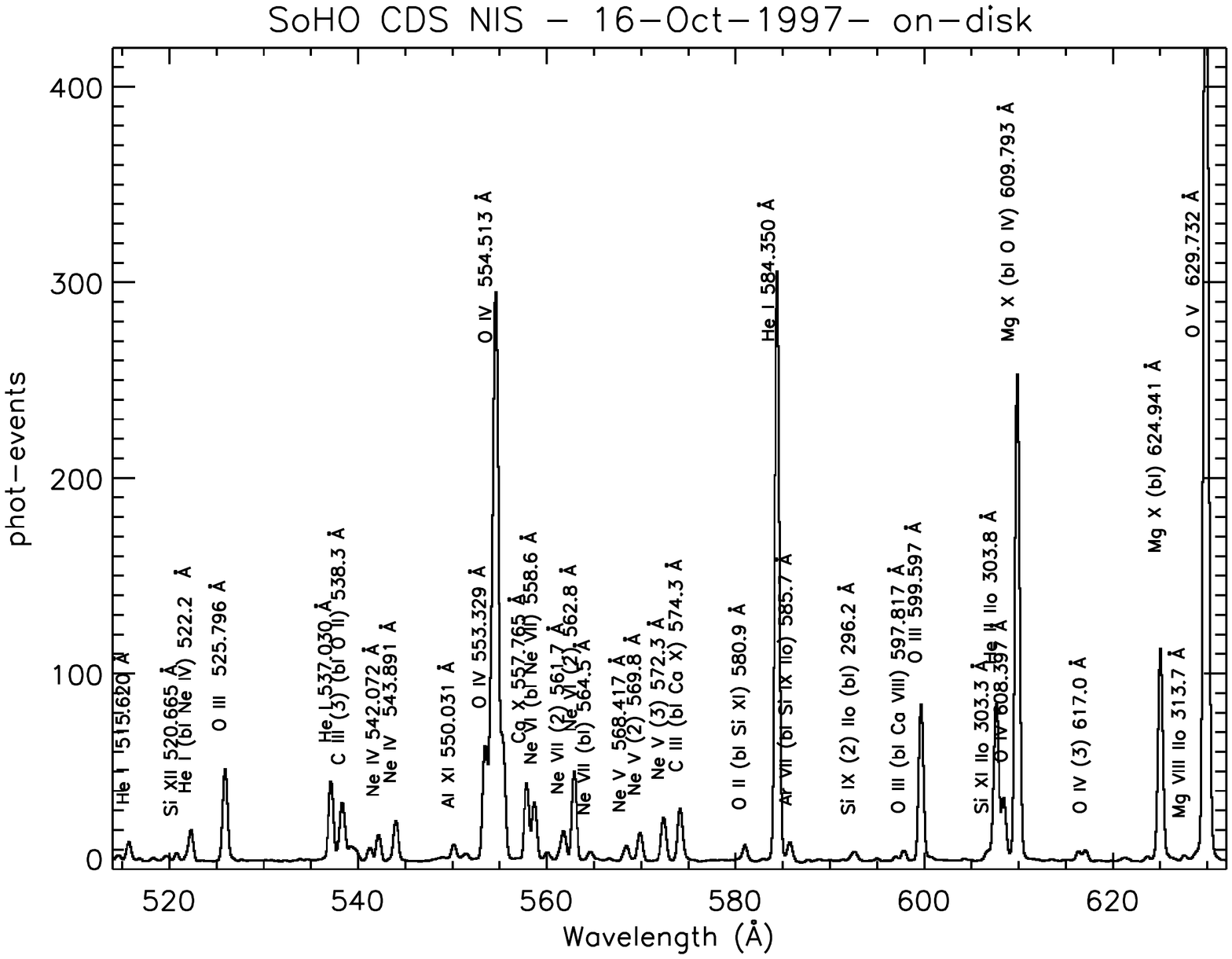}} 
  \caption{SoHO CDS  NIS 2 averaged spectrum of the quiet Sun (on-disk)
\citep[adapted from ][]{delzanna_thesis99}.}
  \label{fig:nis2}
\end{figure}

The diagnostic potential of CDS was discussed by  
\cite{mason_etal:1997}.
A NIS spectral line list for the quiet Sun
can be found in \cite{brooks_etal:99}, while  more
extended lists for different regions,
with line identifications based on CHIANTI  are  given in
\cite{delzanna_thesis99}.
Sample NIS spectra are given in Figures~\ref{fig:nis1},\ref{fig:nis2}.

The  grazing incidence spectrometer used a
grazing incidence  spherical grating  that disperses the incident light
 to four  microchannel plate  detectors placed along the Rowland Circle
(GIS 1: 151 \,--\,221 \AA,
GIS 2: 256 \,--\,341 \AA, GIS 3: 393 -- 492 \AA\ and GIS 4: 659 \,--\,785 \AA).
The spectral resolution of the GIS detectors was about 0.5~\AA.
The GIS  was astigmatic, focusing the image of the slit along the direction of
dispersion but not perpendicular to it.
The in-flight radiometric calibration of the GIS channels
(only the  pinhole 2\arcsec$\times$2\arcsec\ and 4\arcsec$\times$4\arcsec\  slits) is described in
\cite{delzanna01_cdscal} and \cite{kuin_delzanna:07}.
No significant degradation of the GIS sensitivity was found.
The GIS suffered badly from `ghost lines' due to the spiral nature of
the detector. This made data analysis somewhat complicated.
A full list of GIS spectral lines can be found in
\cite{delzanna_thesis99}.

\begin{figure}[htbp]
\centerline{\includegraphics[width=1.1\textwidth]{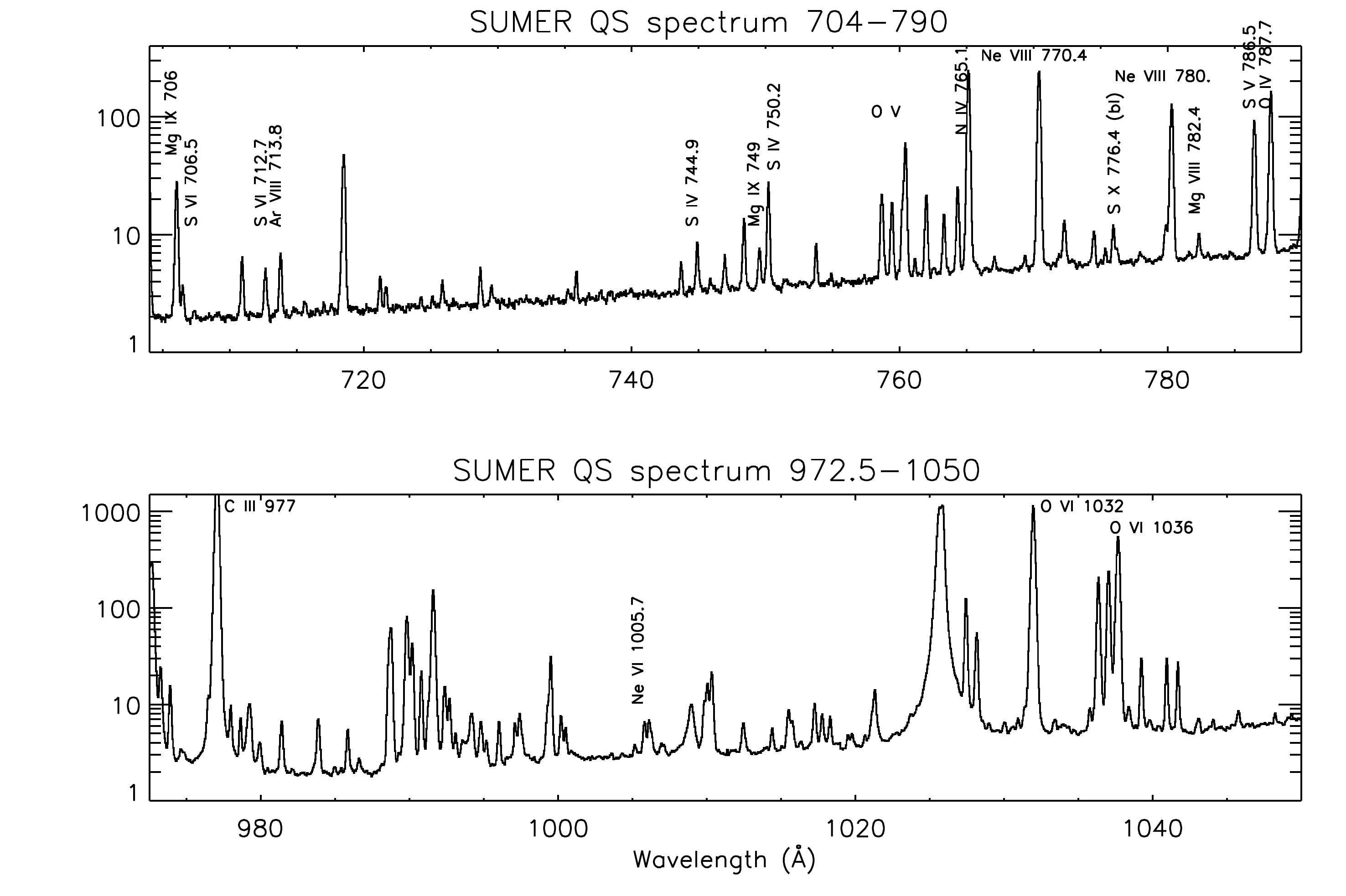}}
 \caption{An on-disc quiet Sun SUMER spectrum in the two spectral
ranges which will be covered by the Solar Orbiter SPICE spectrometer. }
  \label{fig:sumer}
\end{figure}

The Solar Ultraviolet Measurement of Emitted Radiation
(SUMER) was a joint German and French-led instrument \citep{wilhelm_etal:95}.
It was a high-resolution (1\arcsec\ spatial) spectrometer
covering the wavelength range 450-1600~\AA,  dominated by lines
emitted by the chromosphere and transition region.
Detector A observed  the 780--1610~\AA\ range, while detector
B covered  the 660--1500\AA\ range. 
Second-order lines were superimposed on the first order spectra, however
the second order sensitivity was such that only the strongest lines 
in the 450--600~\AA\ range were observable.

SUMER had an excellent  spectral resolution
($\lambda \over d \lambda$ = 19 000 to 40 000), and was able
to measure Doppler motions (flows) with an accuracy better than 2~km/s,
using photospheric lines as a reference.
As in the case of the CDS, SUMER was able to scan solar
regions to obtain monochromatic images in selected spectral lines.
One main difference was that only lines within  a wavelength range 
(typically  40~\AA) could be
recorded simultaneously by SUMER. 

One disadvantage with SUMER was the amount of 
time it took to scan a  spatial area. 
Some difficulties were encountered with  the scanning mechanism, so it was
used sparingly during the latter part of the mission. 
The SUMER  radiometric calibration for the first few years of the mission
was discussed at the ISSI workshops, see 
ISSI Scientific Report SR-002 \citep{SOHO-calib:02}.

A spectral atlas of SUMER on-disk lines was published by
\cite{curdt_etal:2001}.
Figure~\ref{fig:sumer} shows the spectrum in the two spectral
ranges that will be observed with the Solar Orbiter SPICE instrument.
Several strong transition region lines are present.
A list of SUMER on-disk  quiet Sun radiances in the 800-1250~\AA\ range
 was published by \cite{parenti_etal:2005a}, where radiances of 
a prominence were also provided.
Within the SUMER spectral range,
several coronal forbidden lines become  visible off the 
solar limb, see e.g.  the spectral atlases by  \cite{feldman_etal:97, curdt_etal:2004}.
Some of the high-temperature  forbidden lines become visible even on-disk
in active regions and during  flares (see, e.g. the lists of
\cite{feldman_etal:98a,feldman_etal:00}).
The last observations were carried out with SUMER in 2014,
due to the significant degradation of the detectors.

The Ultraviolet Coronagraph Spectrometer (UVCS) was an
 instrument that was built and operated by a 
USA-Italy collaboration \citep{kohl_etal:1995}.
It observed the solar corona from its base out to 10~\rsun.
Its heritage was the Spartan Ultraviolet Coronagraph Spectrometer,
which flew several times between 1993 and 1998 \citep{kohl_etal:2006}.
Most of the UVCS scientific results are based on the
measurements of the strong  H~I Lyman-$\alpha$
and  O~VI (1032 and 1037~\AA) lines, which are
partly collisionally excited and partly
 resonantly scattered \citep{raymond_etal:97}.
Several coronal lines such as Si~XII (499 and 521~\AA)
and Fe~XII (1242~\AA) were also observed.
UVCS produced measurements of chemical abundances,
proton velocity distribution, proton outflow velocity, electron temperature,
and ion outflow velocities and densities.
Some of scientific results from UVCS have been reviewed by
\cite{kohl_etal:2006,antonucci_etal:2012}.
The instrument suffered a significant degradation
(factor of ten) at first light, but continued to operate
for a long time.

\subsection{CORONAS}

CORONAS-F, launched in 2001, provided XUV spectroscopy with the SPIRIT
(Russian-led, see  \citealt{zhitnik_etal:2005})
and  RESIK (REntgenovsky Spektrometr
s Izognutymi Kristalami, Polish-led, see \citealt{sylwester_etal:2005})
instruments, especially of flares during solar maximum.
The SPIRIT spectroheliograph had two wavelength ranges, 
176--207 and 280--330~\AA, and a relatively high 
spectral resolution of about 0.1~\AA. 
The instrument was  slitless. The solar light was deflected at a 
grazing angle of about 1.5$^o$ by a grating, and then 
focused by a mirror coated with a multilayer with a high
reflectivity in the EUV. This resulted in  
`overlappogram' images of the spectral lines, highly
compressed in the solar E-W direction, but with a good 
spatial resolution along the N-S direction. 
Most solar flares have a typical small spatial extension during the 
impulsive phase, so it was relatively straightforward to 
obtain flare spectra, only slightly contaminated by nearby
emission at similar latitudes.
The radiometric calibration was only approximate, obtained 
with the use of line ratios.
SPIRIT observed several flares. For details and 
a line list  see \cite{shestov_etal:2014}.

RESIK was a full-Sun spectrometer employing bent crystals 
to observe simultaneously four spectral bands within  the
range 3.4--6.1~\AA, with a spectral resolution of 
about  0.05~\AA. 
 The instrumental fluorescence was a major limiting effect
in the spectra, creating a complex background emission,
which needed to be subtracted for continuum analysis,
and to measure the signal from weaker lines.

\begin{figure}[!htb]
\centerline{\includegraphics[width=16cm]{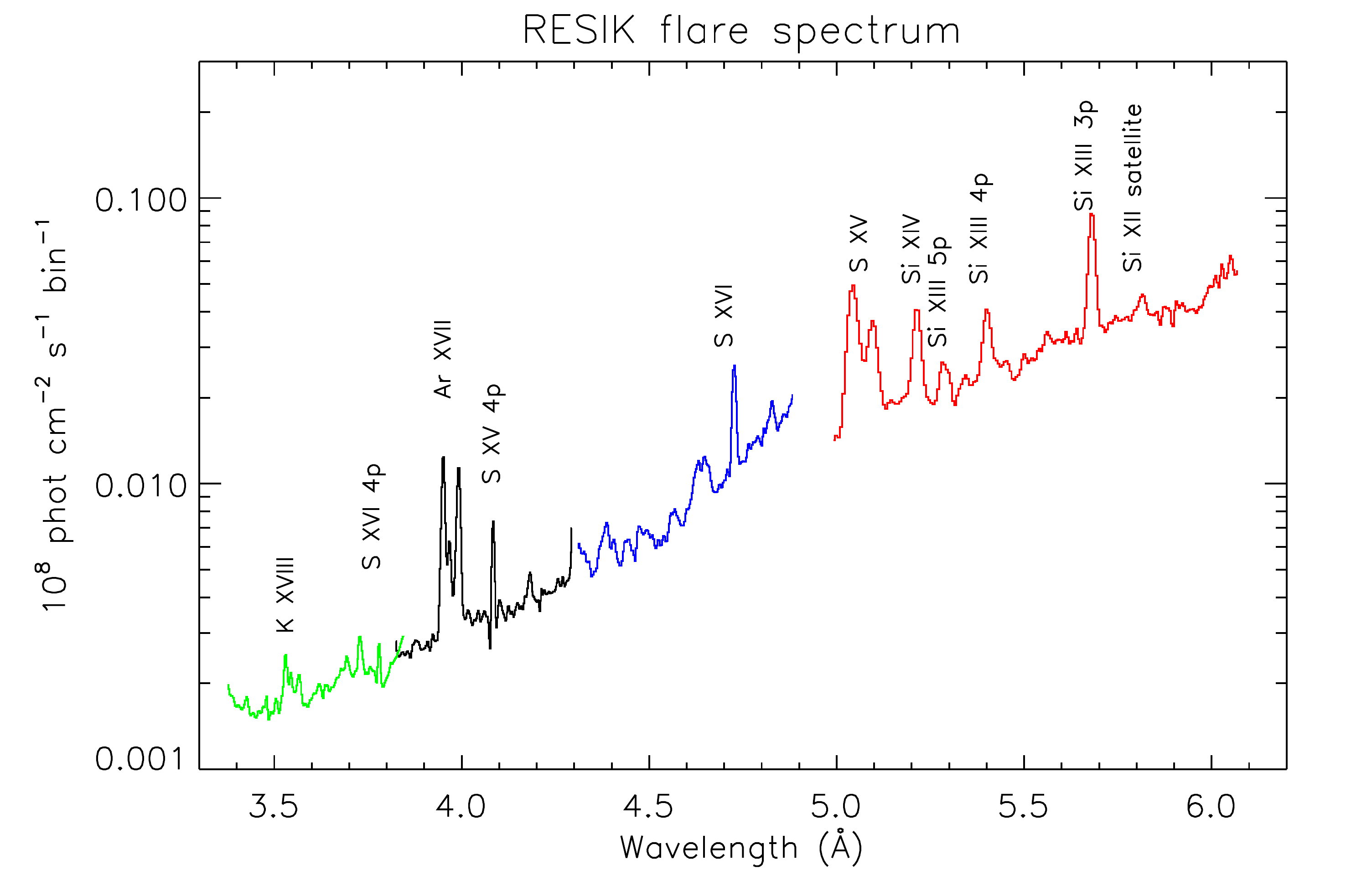}}
 \caption{RESIK spectrum of a flare (adapted from \citealt{chifor_etal:2007}). 
Different colours indicate the four channels. }
 \label{fig:resik}
\end{figure}

The RESIK wavelength range (see  Figure~\ref{fig:resik})
is of particular interest because line-to-continuum
measurements  can be used for absolute abundance determinations
(see, e.g. \citealt{chifor_etal:2007}) and because of the presence of 
dielectronic satellite lines (see, e.g. \citealt{dzifcakova_etal:08}). 
RESIK observed a large number of flares during 
2001-2003.


Some XUV spectroscopy was provided by
CORONAS-Photon  with the SPHINX (Polish-led) instrument
\citep{sylwester_etal:08},
although the satellite, launched in 2009, unfortunately had a failure in 2010.

\subsection{RHESSI}

The NASA \textit{Reuven Ramaty High-Energy Solar Spectroscopic Imager} 
\citep[RHESSI][]{Lin02}, despite not being a purely high-resolution spectrometer,
provided nevertheless very important X-ray spectral observations since 2002
at high energies, above 3\,keV. 
The instrument achieved spatial and spectral resolutions significantly 
 higher than those of earlier missions.
Depending on the signal, it was possible to obtain imaging
in selected energy bands at about 2.5'' resolution. 
The spectra have about 1 keV resolution, 
just allowing the Fe line complex  at 6.7 keV to be resolved
 from the continuum emission.

\subsection{Hinode}

Hinode (Japanese for {\it sunrise}) 
is a Japanese mission developed and launched in
September 2006 by ISAS/JAXA, collaborating with NAOJ as a domestic partner, 
NASA and STFC (UK) as international partners. 
Scientific operation of the Hinode mission is conducted by the
 Hinode science team organized at ISAS/JAXA. 
This team mainly consists of scientists from institutes in the partner countries.
 Support for the post-launch operation is provided by JAXA and NAOJ (Japan), 
STFC (UK), NASA, ESA, and NSC (Norway).

Initial results from the Hinode satellite have been  published in
 special issues of the PASJ, Science and A\&A journals in 2007 and 2008.
Hinode \citep{kosugi_etal:2007} carried 3 instruments, 
the Solar Optical telescope (SOT, see \citealt{tsuneta_etal:2008}), 
the X-ray imaging telescope (XRT, see \citealt{golub_etal:2007}), and the 
Extreme-ultraviolet Imaging Spectrometer (EIS, see \citealt{culhane_etal:07}).
We focus here on the latter instrument.

\begin{figure}[htbp]
\centerline{\includegraphics[width=\textwidth]{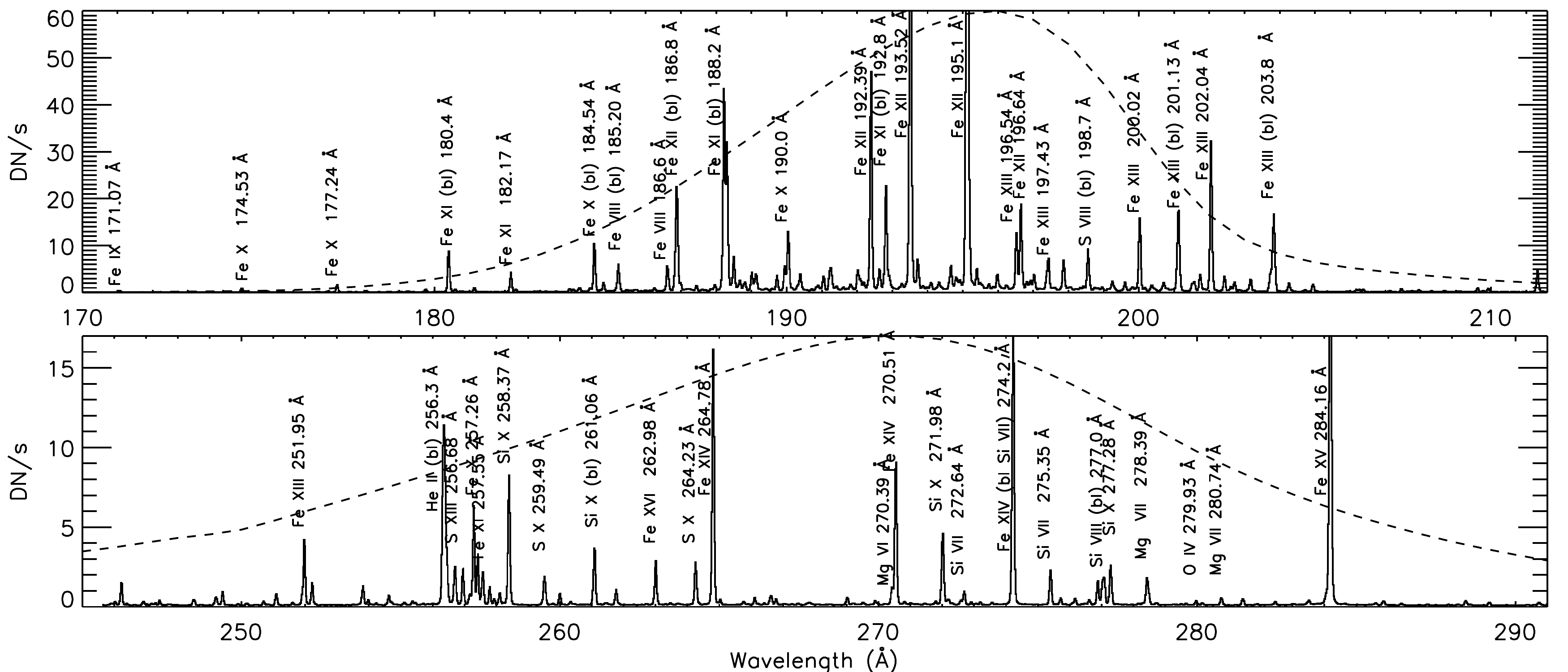}}
\caption{Hinode EIS spectrum of an active region. The dashed lines are
the (scaled) effective areas of the two channels (adapted from \cite{young_etal:07a}).}
 \label{fig:eis_ar_spectrum}
\end{figure}

EIS  has two  wavelength bands,
170--211~\AA\ and 246--292~\AA\ (see Fig.~\ref{fig:eis_ar_spectrum}),
 which include spectral lines
formed over a  wide range of temperatures, from
chromospheric to flare temperatures (log~T (MK) = 4.7 - 7.3).
The instrument has  an effective  spatial resolution of about 3--4".
The high spectral resolution (0.06~\AA) allows velocity
measurements of a few km/s. 
 However with no chromospheric lines, these velocity 
measurements are difficult to calibrate.
Rastering is normally obtained with the narrow
slits (1" or 2").

The ground  radiometric calibration \citep{lang_etal:06} was revised
by \cite{delzanna:13_eis_calib}.
A significant degradation e of about a factor of 
two within the first two years was
found  in the longer-wavelength channel. 
 This degradation  was  confirmed  by \cite{warren_etal:2014}.

Earlier spectroscopic diagnostic applications were
described in \cite{del_zanna_mason_eis:05,young_etal:07a,young_etal:07b}.
A tabulation of spectral lines observed by Hinode/EIS
was provided by \cite{brown_etal:08}. A more complete list of coronal lines
with their identification based on CHIANTI was provided by  \cite{delzanna:12_atlas}.
A comprehensive list with identifications of cool ($T \le $1 MK) emission lines can
be found in \cite{delzanna:09_fe_7} and  \cite{landi_young:2009}.
A discussion of the high-temperature flare lines and their blends
can be found in
\cite{delzanna:08_bflare,delzanna_etal:2011_flare}.

%

\subsection{SDO}

The NASA Solar Dynamics Observatory (SDO, see \citealt{pesnell_etal:2012}) 
was launched in February 2010,
carrying a suite of instruments.
The Helioseismic and Magnetic Imager (HMI, see \citealt{schou_etal:2012}), 
led from Stanford University in Stanford, California, 
measures the photospheric magnetic field.
The Atmospheric Imaging Assembly (AIA, see \citealt{lemen_etal:12}), led from the Lockheed Martin Solar 
and Astrophysics Laboratory (LMSAL), 
provides continuous full-disk observations of the solar chromosphere and corona in 
seven extreme ultraviolet  channels.

The only coronal spectrometer on  SDO is the 
Extreme ultraviolet Variability Experiment
(EVE) instrument \citep{woods_etal:12}. It provided  solar EUV irradiance with an
unprecedented
wavelength range (1 to 1220~\AA) and temporal resolution
(10 seconds).
The EVE spectra are from the Multiple EUV Grating Spectrographs (MEGS) and have
about 1~\AA\  spectral resolution.
 The MEGS~A channel is a grazing incidence spectrograph for the
50--380~\AA\ range. It ceased operations in May 2014.
The MEGS~B channel is a double pass normal incidence spectrograph for the
350--1050~\AA\  range.
A list of MEGS flare lines and their identifications is presented in
\citep{delzanna_woods:2013}.
Additionally, EVE has an EUV
Spectrophotometer (ESP),  a transmission grating and photodiode instrument
similar to the SOHO Solar EUV Monitor (SEM).
ESP has four first-order channels centered on 182, 257, 304, and
366~\AA\ with  approximately 40~\AA\ spectral width, 
and  a zero order channel covering the region 1--70~\AA. 

 MEGS~B suffered a significant degradation of its sensitivity 
from the beginning of the mission (factor of 10), while the degradation 
of MEGS~A was more contained (see, e.g. \citealt{benmoussa_etal:2013}).
Various procedures such as the line ratio technique, 
previously applied to other instruments
(as SoHO CDS, \citealt{delzanna01_cdscal}), were used to obtain 
in-flight corrections. A recent evaluation of the EVE version 5 calibration
showed relatively good agreement  (to within 20\%) with the 
SoHO CDS irradiance measurements for most lines \citep{delzanna_andretta:2015}.

\begin{figure}[htb]
\centerline{\includegraphics[width=\textwidth]{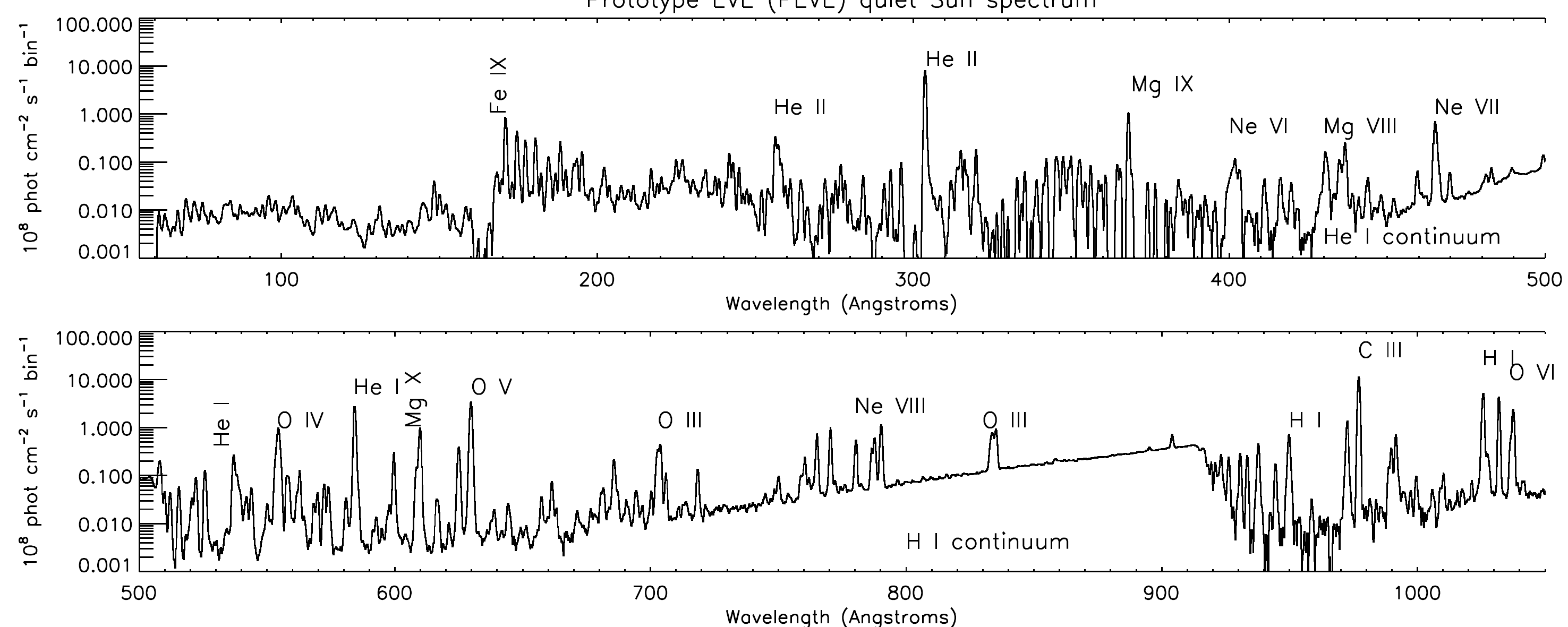}}
\caption{The EUV  spectrum of the whole Sun, as measured by the
prototype SDO/EVE instrument flown aboard a rocket in 2008 April 14. 
Data from \cite{woods_etal:2009,chamberlin_etal:2009}.}
\label{fig:peve_spectrum}
\end{figure}

The combined spectral range of the two channels was observed
with a sounding rocket in 2008 April 14 carrying a prototype EVE
MEGS instrument \citep{woods_etal:2009,chamberlin_etal:2009,hock_etal:2010,hock_etal:2012}.
The spectrum was radiometrically calibrated on the ground,
and is shown in  Fig.~\ref{fig:peve_spectrum}.
On that day, and during the whole long deep solar minimum around 2008,
the Sun was extremely quiet, so in principle the prototype EVE 
observation should represent the best EUV solar spectrum at minimum.
However, significant discrepancies 
were found for several of the strongest lines when comparisons of 
the SoHO CDS and EVE observations were made (from May 2010, when the Sun started to be more active),
 as discussed in  \cite{delzanna_andretta:2015}.
The irradiances of the strongest lines appear to have been overestimated 
by 30--50\%.

\begin{figure}[htb]
\centerline{
\includegraphics[width=8cm]{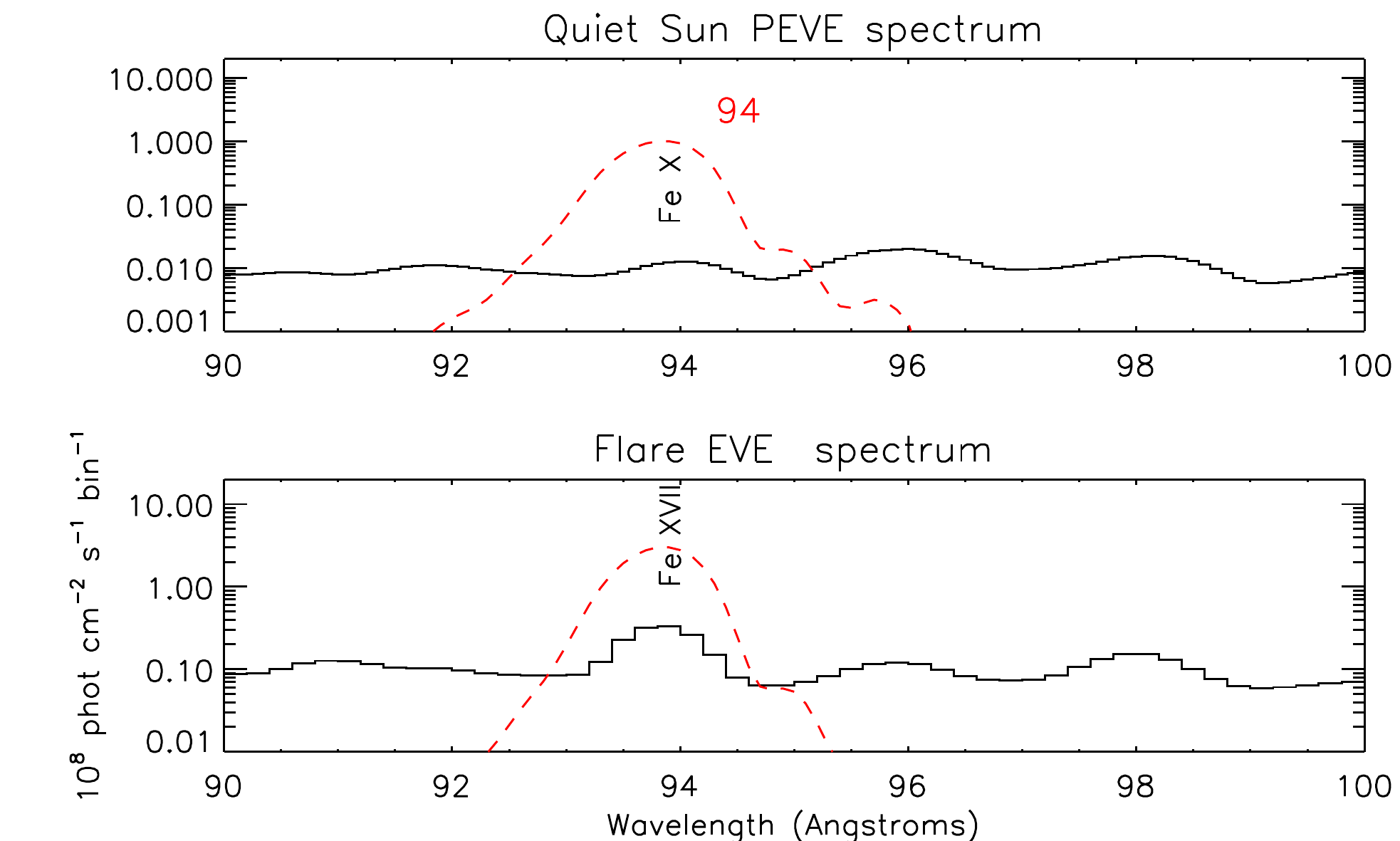}
\includegraphics[width=8cm]{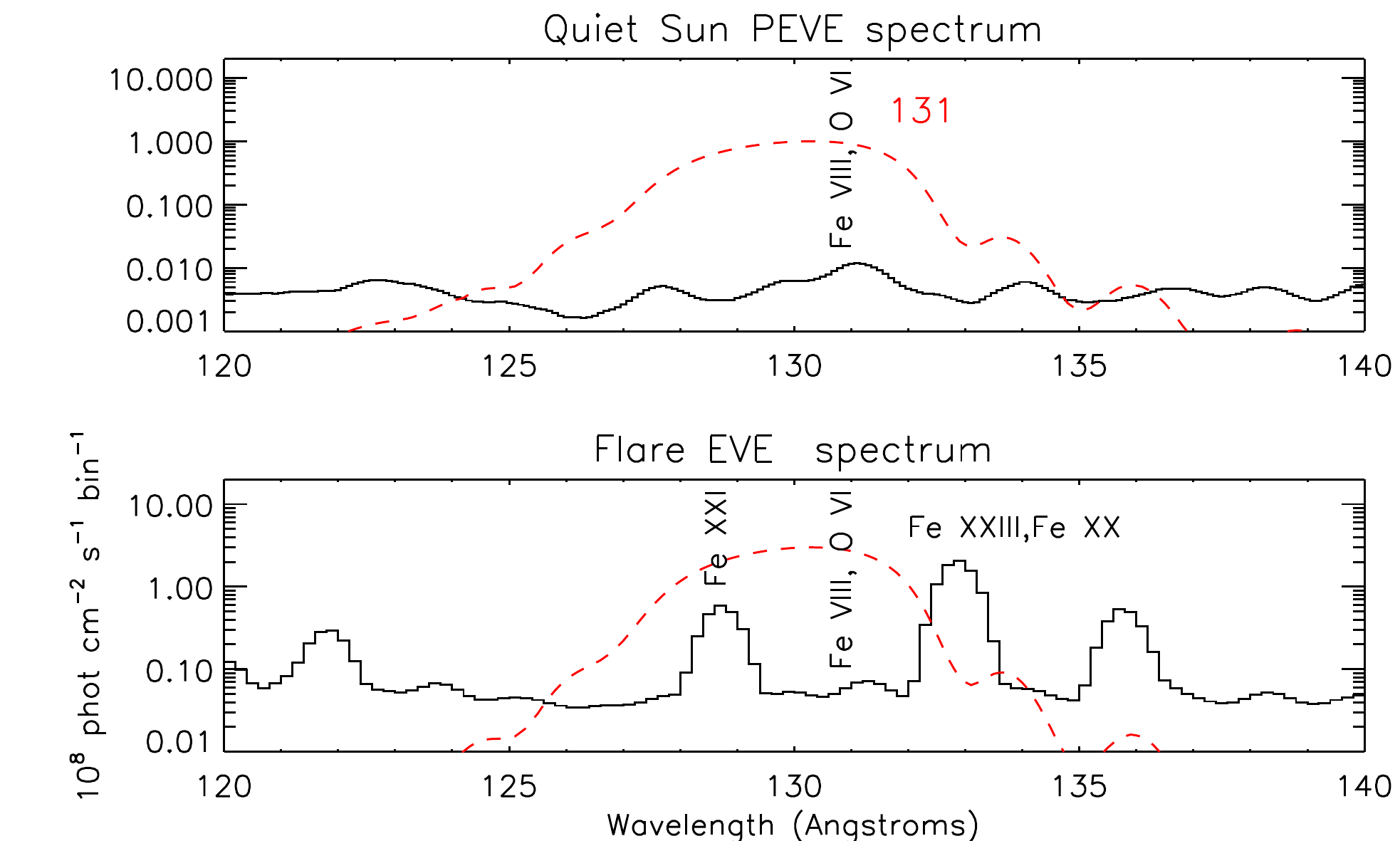}
}
\centerline{\includegraphics[width=9cm]{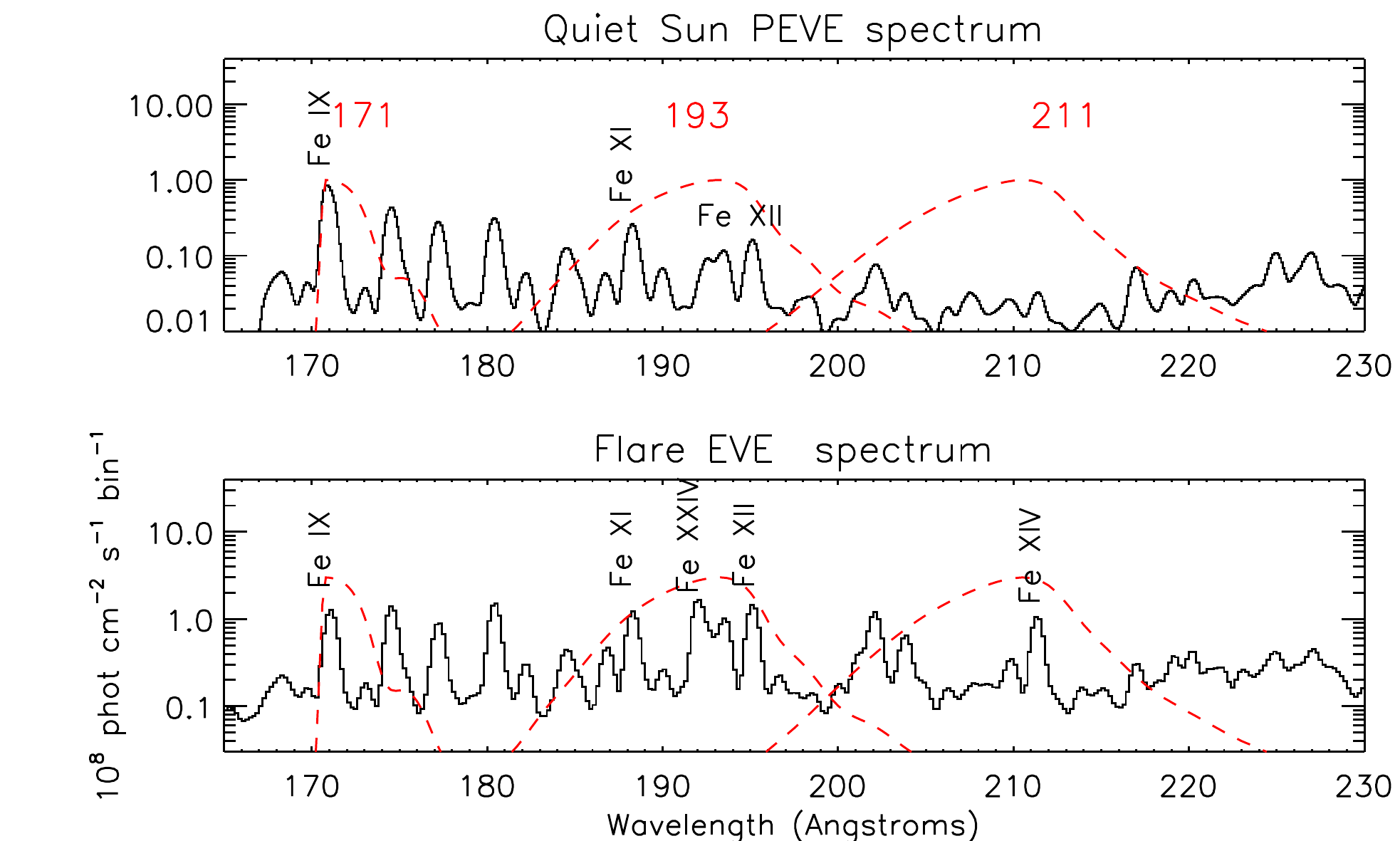}}
\centerline{\includegraphics[width=9cm]{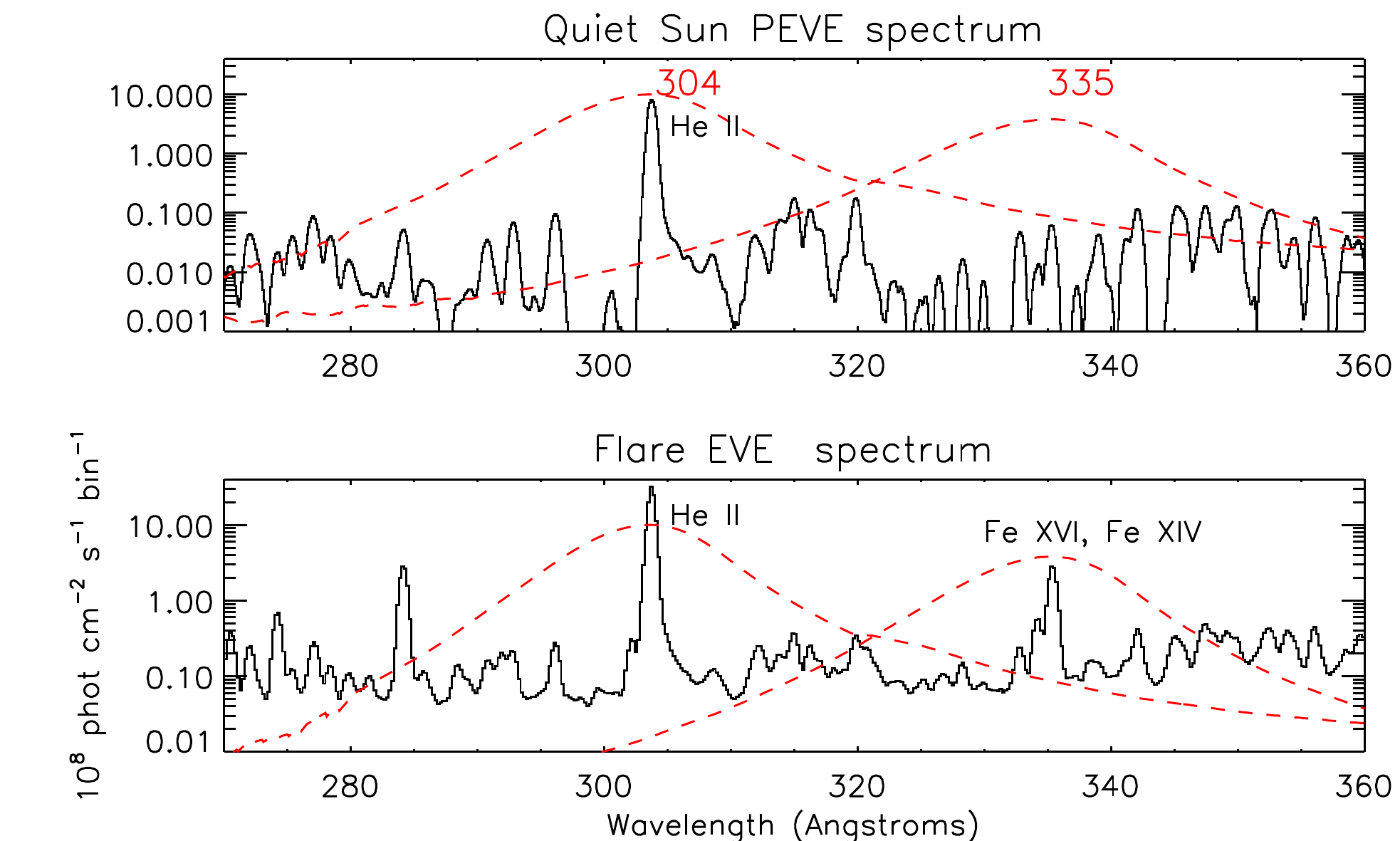}}
\caption{The EVE  spectrum of the quiet Sun (above)
and of an X-class flare (below, see \citealt{delzanna_woods:2013}),
in the spectral ranges covered by
the six SDO/AIA EUV bands (their effective areas are
shown, rescaled, with dashed lines).}
\label{fig:peve_aia}
\end{figure}

It is worth noting that the EVE spectra vary a lot
 in some spectral ranges, depending on the level of activity on the Sun. An example
is shown in Fig.~\ref{fig:peve_aia}, where the quiet Sun
spectrum is  shown together with an X-class flare spectrum
 in the spectral ranges covered by
the six SDO AIA  EUV bands.
Even  the coarse EVE spectral resolution clearly
shows that each AIA band has contributions from several spectral lines,
and that different spectral lines become dominant under different
solar conditions.

\subsection{IRIS}

The Interface Region Imaging Spectrograph (IRIS, 
\citealt{depontieu_etal:2014}) was launched in 
July 2013, and has been producing excellent 
spectra and images of the solar atmosphere at
very high temporal (2s) and spatial (0.33-0.4\arcsec) resolution,
but with a limited field of view. 
The high resoloution has enabled many new scientific results. 
A special section of \textit{Science} in October 2014  (volume 346) 
was dedicated to the first results from  IRIS.

As with earlier  UV instruments, IRIS suffered 
significant in-flight degradation during the first few years 
of the mission.

\begin{figure}[htbp]
\centerline{\includegraphics[width=\textwidth]{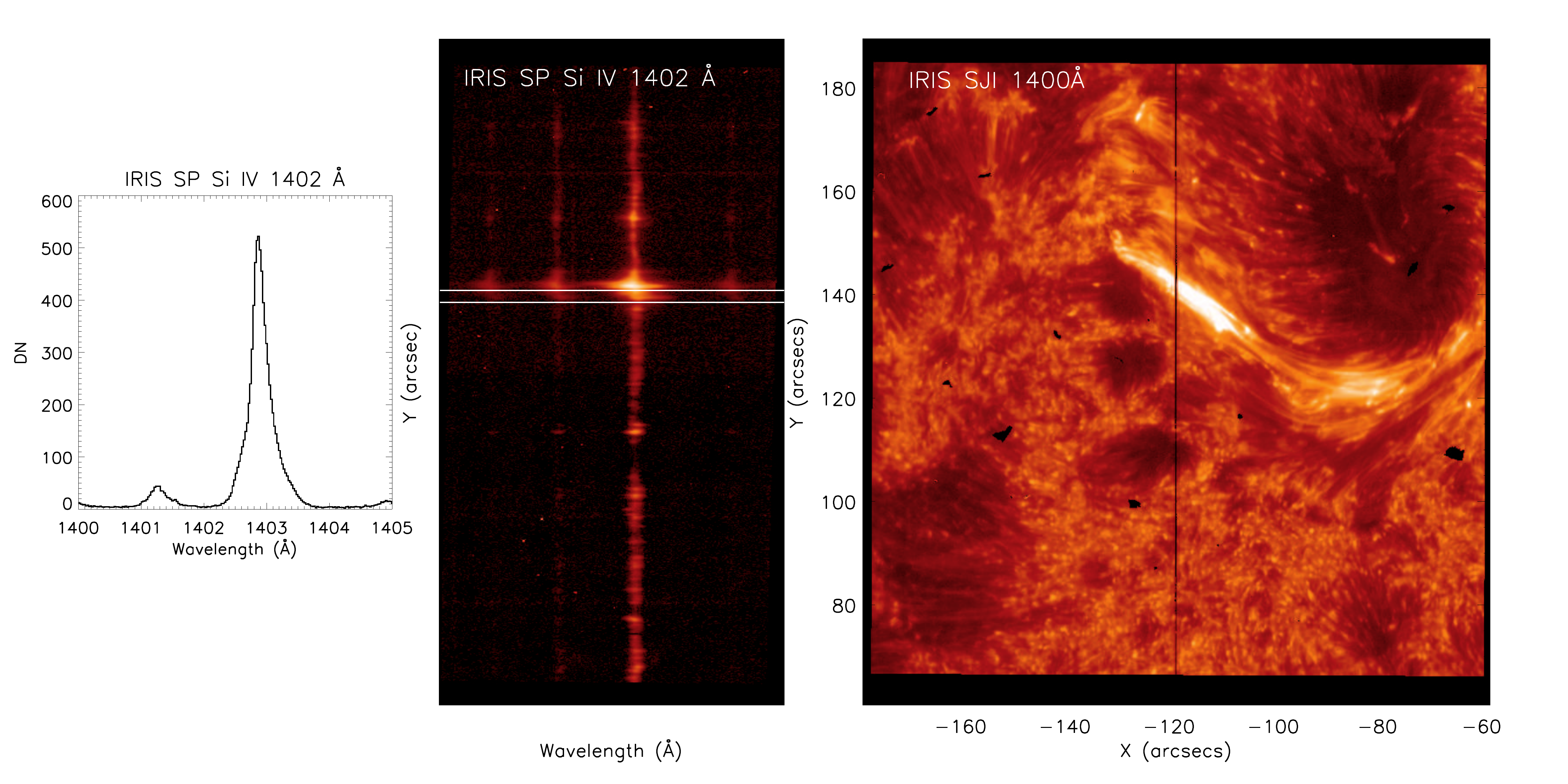}}
\centerline{\includegraphics[width=\textwidth]{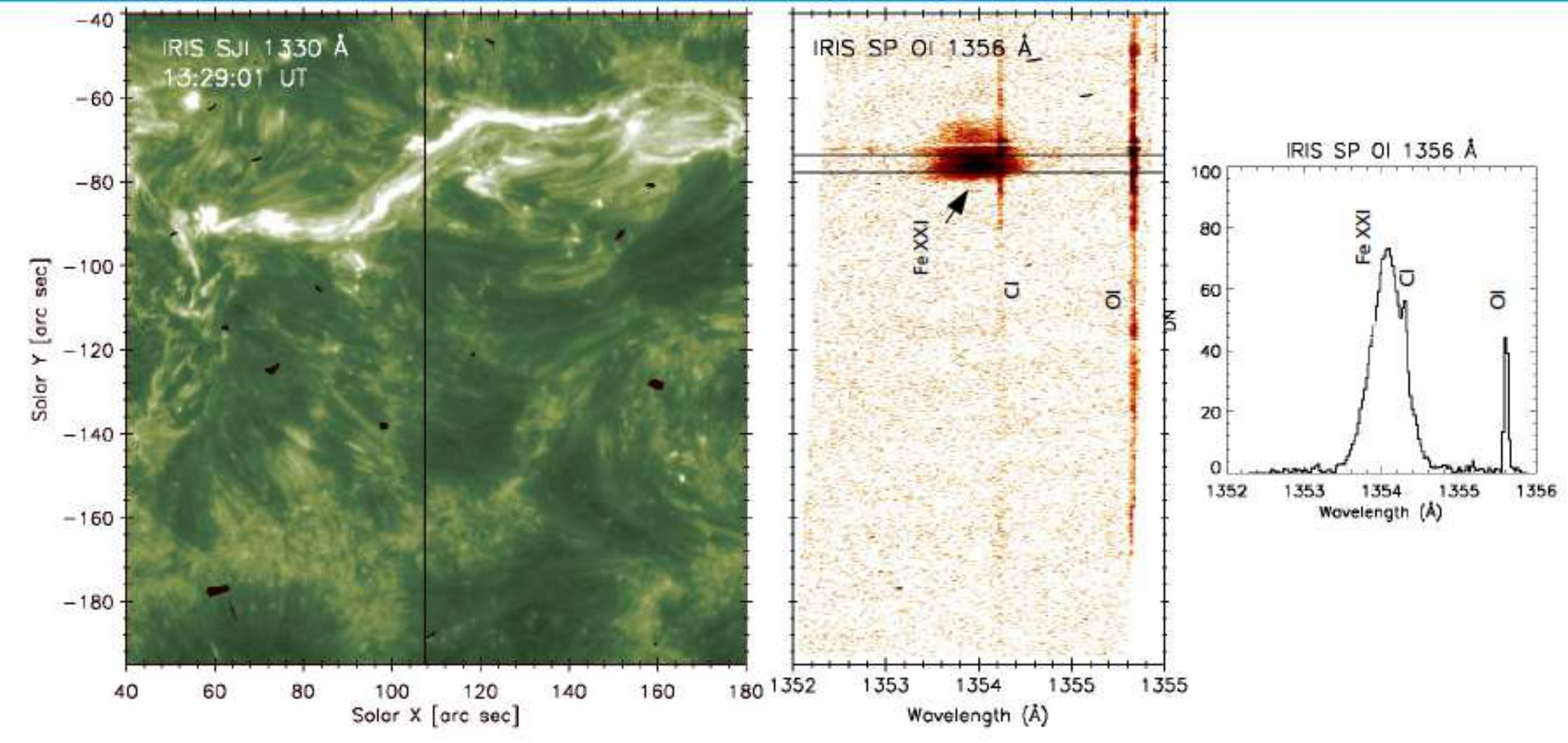}}
\caption{Top: IRIS spectra in the Si IV spectral band (courtesy of V. Polito).
The right panel shows the Si IV  slit-jaw image and the location of the slit.
The middle panel shows the detector image in the Si IV spectral window,
and the left panel shows the averaged spectrum from the region indicated 
by the two parallel lines in the middle panel.
Bottom: IRIS spectra in the Fe XXI band during a flare. The left image shows the 
C II slit-jaw image and the location of the IRIS slit across two ribbons
during a flare. The middle image shows a detector image with the 
Fe XXI 1354.08~\AA\ line and  photospheric lines from C I and O I. The rightmost panel
shows the spectrum averaged along the slit in between the two lines shown in 
the middle panel.}
\label{fig:iris}
\end{figure}

The IRIS Slit Jaw Imager (SJI) provides high-resolution
images in four different passbands (C II 1330~\AA,
Si IV 1440~\AA,  Mg II k 2796 and Mg II wing at 2830~\AA).
The IRIS spectrograph (SP) observes 
 spectra in the 1332--1358, 1389--1407, and 2783--2834~\AA\
spectral regions, where there are several emission
lines formed in the photosphere, chromosphere, as well as 
in the transition region (Si IV, O IV, S IV).
The highest temperature line observed by IRIS is the 
Fe XXI  1354.08~\AA\ line, formed at high
temperatures (12 MK) typical of flares
\citep[][, see Fig.~\ref{fig:iris}]{young_etal:2015,polito_etal:2015}.
  This interesting flare line was previously observed with Skylab SO82B 
(see e.g., \citealt{doschek_etal:1975}) and SMM UVSP 
(see, e.g. \citealt{mason_etal:86}).

The use of a slit together with a slit-jaw image enables the precise location
of the spectra to be established, and in addition high-cadence observations
can be obtained.
HRTS also had a slit-jaw camera but lacked the 
high-cadence which IRIS has, used photographic plates, and was generally flown on a 
rocket (with relatively short duration).
 Nonetheless, HRTS provided some interesting observations which can now
be explored in more detail with IRIS.

\clearpage
\section{The formation of the XUV spectrum}


\subsection{Spectral line intensity in the optically thin case}

In the majority of cases, the XUV emission from the solar corona
and transition region is optically thin, i.e. all the radiation 
that we observe remotely has freely escaped the solar atmosphere.
In this case, the observed intensity of a spectral line is directly related 
to its bound--bound emissivity. 
The radiance  (or more simply intensity)  $I(\lambda_{ji})$ of a
 spectral line of wavelength $\lambda_{ji}$ (frequency $\nu_{ji} = c / \lambda_{ji}$) is therefore:
\beq
{I(\lambda_{ji})} = {{{h \nu_{ji}\over{4\pi }}\;{\int\limits \; N_j(Z^{+r}) \;A_{ji}\;ds}}}
\quad [\mathrm{erg\ cm^{-2}\ s^{-1}\ sr^{-1}}] 
\label{eq:radiance}
\eeq
where $Z^{+r}$  indicates the element of atomic number  $Z$ 
which is  $r$ times ionized, 
$i, j$ are the lower and upper levels of the ion $Z^{+r}$, $h$ is Planck's constant,
$A_{ji}$\footnote{Note: the indexing of a rate or cross-section is indicated 
 either with a $ji$  or $j,i$  in this manuscript} 
is the transition probability for the spontaneous  emission (Einstein's A value), 
$N_j(Z^{+r})$ is the number density (i.e. the number of particles per unit volume in $\mathrm{cm}^{-3}$) 
of the upper level $j$ of the emitting ion, and $s$ is the  line-of-sight coordinate.

The term
\beq
P_{ji} = h \nu_{ji} \;  N_j(Z^{+r})  \; A_{ji} \quad.
\eeq 
is the power (or \textit{emissivity}) per unit volume ($\mathrm{erg\ cm^{-3}\ s^{-1}}$) 
emitted in the  spectral line.

When the Sun is  observed as a star, the flux (i.e. total irradiance in a line),
$F(\lambda_{ji})$, for 
an optically thin line of wavelength $\lambda_{ij}$  is defined as:
\beq
F(\lambda_{ji}) = {1 \over 4 \pi d^2 }~ \int_V P_{ji}~ dV \quad
[\mathrm{erg\ cm^{-2}\ s^{-1}}]
\eeq
where $d$ is the Sun/Earth distance.

There are many processes that affect the population of an upper level of an ion.
Those processes occurring between  the levels of an ion are, under normal
conditions, much faster  than those  affecting the charged
state of the ions, so the two groups of processes (described below) 
can usually be considered separately. 
 For example, for  \ion{C}{iv}  at $N_{\rm e}=10^{10}$ cm$^{-3}$ and $T=10^5 $K we have 
\citep{mariska:92}  for the
allowed  transition at 1548~\AA\ the following time scales:
collisional excitations and de-excitations: $2 \times 10^{-3} s$;
spontaneous radiative decay:  $4 \times 10^{-9} s$;
radiative recombinations + dielectronic recombinations: $ 88 s$;
collisional ionization + autoionization: $107 s$.

The population of a level is therefore normally calculated
by separately calculating the excited level populations and the 
ion population.
The radiance (intensity) of a spectral line (see Eq.\ref{eq:radiance})
 is therefore usually rewritten using the identity:
\beq
  N_j(Z^{+r}) \equiv {N_j(Z^{+r})\over N(Z^{+r})} \;  { N(Z^{+r}) \over N(Z)} \; 
{N(Z) \over  N\lo{H}} \;  {N\lo{H} \over N_{\rm e}}  \;     N_{\rm e} 
\eeq

where the various terms are defined as follows:

\begin{itemize}
\item
   ${N_j(Z^{+r}) / N(Z^{+r})}$
is the population of level $j$ relative to the total $N(Z^{+r})$
number density of the ion $Z^{+r}$. 
As we shall  see below, the  population of level $j$ is calculated by solving
the statistical equilibrium equations for the ion $Z^{+r}$.
 It is a function of the electron  temperature and density.

\item
  ${N(Z^{+r}) / N(Z)}$ 
is the ion abundance, and is predominantly a function of
temperature, but also has some electron density dependence.

\item
  ${N(Z) / N\lo{H}} \equiv Ab(Z)$ 
is the element abundance relative to hydrogen.

\item
 ${N\lo{H} / N_{\rm e}}$ 
is the hydrogen abundance relative to the electron number density.
This ratio is usually in the range $\sim 0.8 - 0.9$,
since H and He,  are fully ionized at  coronal temperatures. 
If we neglect the contribution of the heavier elements to the electron density
($\le$ 1\%),
and assume fully ionized H and He, the only parameter that changes
this ratio is the relative
abundance of He, which is variable. If we assume e.g.
 the \cite{meyer:85} 
abundances, N\lo{H}/N(He)=10, then  ${N\lo{H} / N_{\rm e}}=0.83$

\end{itemize}

The intensity  of a spectral line can then be written in the form:
\beq
I(\lambda_{ji})= \int_s  G(N_{\rm e},T,\lambda_{ji}) \, N_{\rm e} \, N_H \, {\rm d}s
\eeq

with the \textit{contribution function} $G$ given by:
\beq
G(N_{\rm e},T,\lambda_{ij}) =  Ab(Z)~ A_{ji} ~{h \nu_{ij} \over{4\pi } } {N_j(Z^{+r})\over N_{\rm e} N(Z^{+r})}
{ N(Z^{+r}) \over N(Z)}   \quad~~[erg~cm^{3} ~sr^{-1} ~s^{-1}]
\eeq
The  contribution function contains all of the relevant atomic physics parameters and 
for most of the  transitions  has a narrow peak in temperature, 
and therefore effectively confines the emission to a limited temperature range. 
In the literature there are various definitions of  contribution function,
depending on which terms are included.
Aside from constant terms, sometimes the elemental abundance and/or 
 $N\lo{H}/N_{\rm e}=0.8$ and/or the term ${1 / 4 \pi  }$
are included in the definition.

Sometimes, as in the CHIANTI package, the contribution function 
$ C(N_{\rm e},T,\lambda_{ij})$ is 
calculated without the abundance factor:
\beq
G(N_{\rm e},T,\lambda_{ij}) =  Ab(Z)~ C(N_{\rm e},T,\lambda_{ij})
\eeq
Also note that  sometimes (as in the CHIANTI {\it emiss\_calc} program)
the emissivity is defined as 
$Emiss = h \nu_{ij} \;  N_j(Z^{+r}) / N(Z^{+r})  \;A_{ji}$, 
i.e. with the fractional population of the upper level, in which case
 the emissivity is basically the contribution function without the elemental abundance 
$Ab(Z)$ and the ion abundance $N(Z^{+r}) / N(Z)$, and without dividing for the electron density.

\subsection{Collisional rates and  Maxwellian distributions}

The dominant  mechanisms for the level populations in the low solar corona are 
 collisional excitation and ionization of the ions by the free electrons. 
Considering excitation, the number of transitions in an ion from a 
state  $i$ to a state $j$ due to electron collisions, per 
unit volume and  time, is $N_i \sigma_{ij} N_{\rm e} v f(v) {\rm d}v$, 
where $N_i$ is the number density of the ion in the initial state,
 $\sigma_{ij}$ is the cross section for the process,
 $v$ is the velocity (in absolute value)
of the  electron, and $f(v)$ the distribution function of the  electrons.

In general, the collisional  excitation rates are proportional to the 
total  number of transitions integrated over the free electron distribution,
the so-called rate coefficients
\beq
C^{\rm e}_{ij}  = \int_{v_0}^\infty v\,  \sigma_{ij}  (v) \, f(v)\, {\rm d}v \;\; ~ [{\rm cm}^3~{\rm s}^{-1}] \; ,
\eeq
where the limit of integration $v_0$ is the threshold velocity, i.e. the minimum
velocity for the electron to be able to excite the atom from level
 $i$ to  $j$:
\beq
 {1 \over 2}\, m\, {v_0}^2 = E_j - E_i
\eeq 
where $m$ is the mass of the electron.

It is normally assumed that the 
electrons have enough time to thermalise, i.e. follow a Maxwell-Boltzmann (thermal)
distribution (but see Section~\ref{sec:non-maxwell}) in the lower  solar corona.
 In this case, the  probability $f(v) \, {\rm d}v$ that the electron has a velocity 
(in the 3-D space) between $v$ and $v+{\rm d}v$ is: 
\beq
 f(v) = 4\, \pi \, v^2  \left( {m \over 2\, \pi \, kT_{\rm e}} \right)^{3/2} \, \eul^{-{m v^2 / 2kT_{\rm e}}}
  \;\; .
\eeq
where $k$ indicates Boltzmann's constant, and $T_{\rm e}$ the electron temperature.
On a side note,  the most probable speed $v_p$, i.e. the 
maximum value of the distribution is found by imposing that 
$ { {\rm d} f(v) \over {\rm d}v }$=0:
\beq
v_p = \left({2\, k T_{\rm e} \over m}\right)^{1/2} \;\; ,
\eeq
while the average speed $<v>$ is:
\beq
<v> = \int v\, f(v)\, {\rm d}v =  \left({8\, k T_{\rm e} \over \pi m}\right)^{1/2} \;\;
\eeq

The collisional  excitation rates can then be written using the 
Maxwell-Boltzmann distribution. As a function of the kinetic energy of the incident electron 
$E = {1 \over 2}\, m\, {v}^2$, the rate coefficient can be written as:
\beq
\eqalign{
C^{\rm e}_{ij} & = \left( {m \over 2 \pi\, k T_{\rm e}}  \right)^{3/2} \, 4 \pi \,
  \int_{v_0}^\infty v^3\,  \sigma_{ij} (v) \,  \eul^{-{m v^2 / 2kT_{\rm e}}}\, {\rm d}v \cr 
& = \left({8 \over  \pi\, m } \right)^{1/2} \, (k T_{\rm e})^{-3/2} \int_{E_0}^\infty  E\, \sigma_{ij} (E) \, \eul^{-{E / kT_{\rm e}}} \, {\rm d}E  \cr
 & = 
 5.287 \, 10^{13} \, (k T_{\rm e})^{1/2} 
\int_{E_0}^\infty  {E \over k T_{\rm e}} \,  \sigma_{ij} (E) \, \eul^{-{E / kT_{\rm e}}} \,  {\rm d}\left({E \over k T_{\rm e}}\right) \;\; , \cr } 
\label{eq:coll_coeff}
\eeq
where $E_0$ is the threshold energy of the electron, i.e.
$E_0 = E_j-E_i$, the energy difference between the ion states $i$ and $j$.

In the case of collisional ionization of an atom or ion by a free electron, the 
expressions for the number of transitions are similar, as described below. 


\subsection{Excitation and de-excitation of ion levels}

Inspection of Eq.\ref{eq:coll_coeff} indicates a way to simplify the expression,
by introducing a dimensionless  quantity, the collision strength for electron excitation:
\begin{equation}
\sigma_{ij} =  \pi a_0^2 \, \Omega_{ij}(E) \,  { I_{\rm H} \over  g_i  E }
	\label{Eq:Omega}
\end{equation}
where where $g_i$ is the statistical weight of the initial level,
$I_{\rm H}$ is the ionization energy of hydrogen, and $a_0$ the Bohr radius.
The collision strength is a symmetrical quantity, such that
$\Omega_{ij}(E)=\Omega_{ji}(E^\prime)$, 
where $E^\prime=E-E_{ij}$ is the kinetic energy of the electron after the scattering.

The electron collisional excitation rate
coefficient for a Maxwellian electron velocity distribution
with a temperature $T_{\rm e}$(K) is then obtained by integrating:

\beq
\label{eq:sum}
\eqalign{
	C^{\rm e}_{ij}  & =  a_0^2 \left( { 8 \pi I_{\rm H} \over m } \right)^{1/2} 
 \left( { I_{\rm H} \over k T_{\rm e} } \right)^{1/2} 
 {\Upsilon_{ij} \over g_i} \exp \left(- {E_{ij} \over k T_{\rm e}} \right)  \cr
	& ={8.63\times10^{-6} \over T_{\rm e}^{1/2}}
{\Upsilon_{i,j}(T_{\rm e}) \over g_i} \exp \left( {- \Delta E_{i,j} \over kT_{\rm e}} \right) 
  ~ {\rm cm}^3~{\rm s}^{-1} \cr }
\eeq

where $k$ is the Boltzmann constant and $\Upsilon_{i,j}$ is the 
thermally-averaged collision strength:

\begin{equation}
\Upsilon_{i,j}(T_{\rm e}) = \int_0^\infty
\Omega_{i,j} \exp\left(-{E_{j} \over kT_{\rm e}}\right) d
\left({E_j \over kT_{\rm e}}\right)  \quad ,
\label{u-ij}
\end{equation}
where $E_j$ is the energy of the scattered electron
relative to the final energy state of the ion.
Some details on  electron-ion scattering calculations are 
provided in Section~\ref{sec:atomic}.

The electron de-excitation rates $C_{j,i}^{\rm d}$ 
from the upper level $j$ to the lower level $i$ are obtained by applying  the
principle of detailed balance, assuming thermodynamic equilibrium, following
Milne, who repeated Eintein's reasoning on the radiative transition 
probabilities. In thermodynamic equilibrium, the processes of 
excitation and de-excitation must equal:
\beq
 N_i N_{\rm e} C^{\rm e}_{i,j} =  N_j N_{\rm e} C^{\rm d}_{j,i} \;\;,
\eeq
and the populations of the two levels are in Boltzmann equilibrium:
\beq
{N_i \over N_j} = { g_i \over g_j}  \exp \left( {
\Delta E_{i,j} \over kT_{\rm e}} \right) \quad ,
\eeq
where $g_i, g_j$ are the statistical weights of the two levels.
So we obtain
\beq
C_{j,i}^{\rm d} = { g_i \over g_j}  C_{i,j}^{\rm e}  \exp \left( {
\Delta E_{i,j} \over kT_{\rm e}} \right) \quad .
\eeq
This relation holds also outside of  thermodynamic equilibrium, as long
as the plasma is thermal.
If the electron distribution is not Maxwellian, the definitions of the 
excitation and de-excitation rates are somewhat different,
as described below in Section~\ref{sec:non-maxwell}.

\subsection{The ion level population and the metastable levels}

The variation in  $N_j$, the population of level $j$ of the ion $Z^{+r}$,  
 is calculated by solving the statistical equilibrium equations for the ion 
including all the important excitation  and de-excitation mechanisms:

\beq
\label{eq:sum}
\eqalign{
{dN_j\over dt} & =   \sum_{k>j} N_k A_{k,j} + 
\sum_{k>j} N_k N_{\rm e} C^{\rm d}_{k,j} +  \sum_{i<j} N_i N_{\rm e} C^{\rm e}_{i,j} + 
 \sum_{i<j} N_i B_{i,j}\; J_{\nu_{i,j}} + \sum_{k>j} N_k B_{k,j}\; J_{\nu_{k,j}}  \cr 
 & - N_j (\sum_{i<j} A_{j,i} +  N_{\rm e} \sum_{i<j} C^{\rm d}_{j,i} +  N_{\rm e} \sum_{k>j} C^{\rm e}_{j,k} + 
 \sum_{i<j} B_{j,i}\; J_{\nu_{j,i}}  +  \sum_{k>j} B_{j,k} \; J_{\nu_{k,j}}  ) \cr }
\eeq
where the first five terms are processes which populate the level $j$:
the first is decay from higher levels, the second is de-excitation from 
higher levels, the third is excitation from lower levels, and the other two 
are photo-excitation and de-excitation. The other five terms are the corresponding 
depopulating processes. Note: 

*  $N_{\rm e}$ (cm$^{-3}$) is the electron number density.

* $C^{\rm e}, C^{\rm d}$ (cm$^{-3}$ s$^{-1}$) are the electron 
collisional excitation and de-excitation rate  coefficients 
defined above. 

* $A_{j,i}$ (s$^{-1}$) are Einstein's coefficients for spontaneous emission,
also called  transition probabilities or A-values.

* $B_{i,j}$ ($i<j$) are Einstein's coefficients for absorption.

* $B_{k,j}$ ($k>j$) are Einstein's coefficients for stimulated emission.

* $J_{\nu} = { 1 \over 4 \pi} \int I_\nu(\vec  \Omega) \, {\rm d}\Omega  \;\;,  $ 
i.e. is the average of the intensity of the radiation field over the solid angle.

Note that the terms associated with stimulated emission by radiation are 
 normally negligible for the solar corona, while  
the terms associated with absorption of radiation are only important
in the outer corona, where electron densities become sufficiently 
small. 
Also note that Einstein's coefficient for stimulated emission
is related to the   A-value by:
\beq
  B_{ul} = {c^2 \over 2 \, h \, \nu_{ul}^3} \, A_{ul} \;\;, 
\eeq
while Einstein's coefficient for absorption is related to the other two
coefficients by:
\beq
B_{lu} = {c^2 \over 2 \, h \, \nu_{ul}^3} \, {g_u \over g_l} \,
   A_{ul} = {g_u \over g_l} \, B_{ul} \;\;,
\eeq
where we have indicated the lower level with $l$ and the upper level 
with $u$ for simplicity, and $g$ indicates the statistical weight of 
the level.
Typical $A_{j,i}$ values for transitions that are dipole-allowed are of the order
of 10$^{10}$ s$^{-1}$, while those of forbidden transitions can be as low as 
100 s$^{-1}$.

For most  solar (and astrophysical) applications, 
the time scales of the relevant processes are so short that 
the plasma is normally assumed to be in a steady state (${dN_j\over dt}=0$).
The set of Eqs.~\ref{eq:sum} is then solved for a number of low lying levels,
 with the additional requirement that the total population of the levels
equals the population of the ion: $N(Z^r) =\sum_j N_j$.

In the simplified case of a two-level ion model
(a ground state $g$ and an excited state $j$) and 
neglecting other processes such as photoexcitation, we have:

\beq
N_{\rm g} N_{\rm e} C^e_{g,j} = N_j (N_{\rm e} C^e_{j,g} + A_{j,g})
\eeq
so the relative population of the level $j$ is 

\begin{equation} 
{N_j \over N_{\rm g}} = { N_{\rm e} C^e_{g,j} \over {N_{\rm e} C^e_{j,g} + A_{j,g}}} 
\end{equation}
i.e. depends strongly on the relative values between the radiative rate 
$A_{j,g}$ and the collisional de-excitation term $N_{\rm e} C^e_{j,g}$.
Levels that are connected to the ground state by a dipole-allowed transition
have  $A_{j,i}$ values typically several orders of magnitude larger than 
the de-excitation term (at typical coronal densities, $N_{\rm e}=10^8$--$10^{12}$ cm$^{-3}$),
so their population is negligible, compared to the population of the ground state.
When  all the upper levels of the ion are of this kind, the statistical equilibrium 
equations are simplified, so that only direct excitations from the ground state 
need to be included.  This is the so called \emph{coronal-model approximation}, where
 only the electron \emph{collisional excitation} from the ground state
 of an ion and the \emph{spontaneous radiative decay}  are competing.

However, ions often have  so called 
{\it metastable levels}, $m$,  which have 
 a small radiative decay rate (e.g. corresponding to intersystem or 
forbidden transitions), so that collisional de-excitation
starts to compete with radiative decay as a depopulating process
 at sufficiently high electron densities
($A_{m,g}~\simeq~N_{\rm e}~ C^e_{m,g}$).  
In such cases, the population of the metastable levels 
becomes comparable to that of the ground state.
Whenever ions have metastable levels, it is necessary to include 
all collisional excitation and de-excitation rates to/from the 
metastable levels when solving the  statistical equilibrium 
equations. 
All the ions that have more than one level in the ground configuration 
have metastable levels (i.e. all the excited levels within the ground configuration), because the 
transitions within a configuration are forbidden, i.e. they have small
 radiative decay rates.
The ion population is shifted from the ground level 
into the metastable(s) as the electron density of the plasma increases.

\subsubsection{Proton excitation}

Proton collisions become non-negligible  when excitation energies are small,
$\Delta E_{i,j} \ll kT_{\rm e}$.
This occurs, for example, for transitions
between fine structure levels, as in the   Fe~XIV transition in the ground configuration
(3s$^2$3p $^2$P$_{1/2}$ - $^2$P$_{3/2}$), as discussed in \cite{seaton:64_protons}. 
Normally, only the 
 fine structure levels within the ground configuration of an ion 
have a significant population, so only the proton  collisions 
among such levels are important. 
The inclusion of proton excitation has some effects
on the relative population of the levels. 
Proton collisional excitation and de-excitation  are easily included 
  as  additional terms  $C^{\rm p}$ (cm$^{3}$ s$^{-1}$)
 in the level balance equations.

\subsubsection{Photoexcitation}

Photoexcitation is an  important process which needs to be 
included in the level balance equations  when electron densities are
sufficiently low.
Photoexcitation is the process by which 
the excitation of an ion from a level i to a level j is caused 
by absorption of a photon. For this to occur, the photon has to have the 
same energy of the transition from i to j.
From the statistical equilibrium 
equations (Eq.~\ref{eq:sum}) and considering the relations between the 
Einstein coefficients, it is obvious that photoexcitation and 
de-excitation
can easily be included as  additional terms which modify the A-value.
These terms are proportional to $J_{\nu}$. It is common to assume that 
the intensity of the radiation field originating from the solar photosphere
$I_\nu$ does not vary with the solid angle 
(no limb brightening/darkening), in which case we have:

\beq
J_{\nu} = { \Delta\;  \Omega  \over 4 \pi} \;  \overline{I_\nu} = 
W(r) \; \overline{I_\nu} 
\eeq
where $W(r)$ is the {\it dilution factor} of the radiation, i.e. the 
geometrical factor  which accounts for the weakening of the radiation
field at a distance $r$ from the  Sun, and $\overline{I_\nu}$ is the 
averaged disk radiance at the frequency $\nu$.

Assuming spherical symmetry (i.e. the solar photosphere a 
perfect sphere),  and indicating  with  $r$ the distance from  Sun centre 
$R_{\odot}$ the solar radius,  we have:
\beq
W(r) = { 1 \over 4 \pi} \int_0^{2 \pi} \int_0^{\theta_0} sin\;\theta \; d\theta \; d\phi =
 { 1 \over 2} \; (1- cos\;\theta_0) = { 1 \over 2} \left(1- \left[1- \left({R_{\odot} \over r}\right)^2\right]^{1/2} \right) 
\eeq
where  $\theta_0$ is the angle sub-tending $R_{\odot}$ at the distance $r$,
i.e. $sin \theta_0 = R_{\odot} / r $.

In  terms of the energy density per unit wavelength, $U_\lambda$,
the photoexcitation rate for a transition $i\rightarrow j$ is:
\begin{equation}\label{photoexc2}
P_{ij}=A_{ji} \; W(r) \; {g_j \over g_i} \; {\lambda^5 \over 8\pi hc}
       U_\lambda
\end{equation}
where $A_{ji}$ is the Einstein coefficient for spontaneous emission from
$j$ to $i$, $g_j$ and $g_i$ are the statistical weights of
levels $j$ and $i$, and $W(r)$ is the radiation dilution factor.

Metastable levels affect the population of an ion,
in particular those of the ground configuration. The 
transitions between these levels are normally in the visible and 
infrared parts of the spectrum, where almost all the photons 
are emitted by the solar photosphere. Therefore, 
the main contributions of photoexcitation to the level population
is due to visible/infrared  photospheric emission.
A reasonable approximation for the photospheric radiation field at 
visible/infrared wavelengths is a black-body of temperature 
$T_*$, for which the  photoexcitation rate becomes:
\begin{equation}
P^{\rm bb}_{ij}=A_{ji} W(r) {g_j \over g_i} {1 \over \exp (E/kT_*) -1}
\label{photoexc}
\end{equation}

The inclusion of photoexcitation 
can simply be carried out  by replacing the ${A}_{ji}$ value in the 
statistical equilibrium equations with a generalized 
radiative decay rate (as coded in the CHIANTI atomic package), 
which in the black-body  case is:

\begin{equation}
{\cal A}_{ij} = \left\{ \begin{array}{l@{\quad}l}
                W(r) A_{ji} {g_j\over g_i} 
                {1 \over \exp({\Delta\kern-1ptE}/kT_*) -1} & i<j \\
                \\
                A_{ji} \left[ 1 + W(r)
                {1 \over \exp({\Delta\kern-1ptE}/kT_*) -1} \right] & i > j
                \end{array}
                \right. 
\end{equation}

Clearly, the photospheric radiation field might depart significantly 
from  black-body radiation, for example with absorption lines, so 
an observed solar spectrum should be used for more accurate calculations. 
Photoexcitation typically becomes a significant 
process  at densities of about 10$^8$ cm$^{-3}$ and below. 
It therefore  becomes a non-negligible effect
for off-limb observations of the  corona above  a fraction of the solar radius,
where electron densities (hence electron excitations) decrease
quasi-exponentially. 

\begin{figure}[htb]
\centerline{\includegraphics[width=9cm,angle=90]{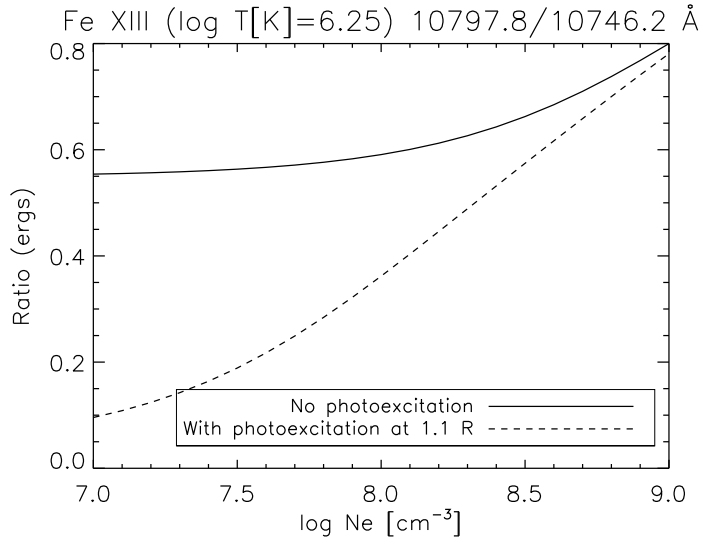}}
\caption{Ratios of the two infrared forbidden \ion{Fe}{xiii} lines, 
as a function of density, without and with photoexcitation 
(assuming a blackbody spectrum seen at 1.1~\rsun). CHIANTI version 8 was used.
The wavelengths indicated are the  wavelengths in air.
}
  \label{fig:fe_13_ratios_infrared}
\end{figure}

By affecting the populations of the levels of the ground configuration,
photoexcitation  has a direct effect on the intensities of the 
(visible and infrared) forbidden lines which are emitted by these levels.
One typical example is for the \ion{Fe}{xiii} infrared forbidden lines
(see, e.g. \cite{chevalier_lambert:69,flower_etal:73, young_etal:03} and 
 Figure~\ref{fig:fe_13_ratios_infrared}).
These lines  are used to measure electron densities (see, e.g. \citealt{fisher_pope:1971}),
the orientation and strength of the magnetic field 
and small Doppler shifts in the solar corona. Such measurements are currently being made
with  the Coronal Multi-Channel Polarimeter (CoMP) instrument, 
now located at Mauna Loa (see, e.g. \citealt{tomczyk_etal:2007}).

By changing the population of the metastable levels of the ground configuration,
photoexcitation also indirectly  affects some EUV/UV spectral lines,
in particular those connected with the ground configuration metastable levels.

\subsection{ Atomic processes affecting the ion charge state} 
\label{sec:charge_state}

Various processes can affect the ionisation state of an element. 
If we consider  two ionization stages, we have the following processes,
denoting for simplicity  the population of the ion $r$-times ionised
with $N_{r}=N(Z^r)$:
\begin{enumerate}
\item
 radiative recombination, induced by the radiation field  $\sim N_{r+1} N_{\rm e} \alpha^I_{r+1}$;
\item
 radiative recombination, spontaneous $\sim N_{r+1} N_{\rm e} \alpha^R_{r+1}$;
\item
photoionisation,     induced by the radiation field  $\sim N_{r}  \alpha^{\rm PI}_{r}$;
\item
collisional ionization by direct impact of free electrons $\sim N_{r} N_{\rm e} C^{\rm I}_{r}$;
\item
 dielectronic recombination $\sim N_{r+1} N_{\rm e} \alpha^{\rm D}_{r+1}$;
\end{enumerate}

where $C^{\rm I}_{r}$ ($cm^{3} s^{-1}$) are the  rate coefficients for collisional ionization
by electrons,  and 
 $\alpha$  ($cm^{3} s^{-1}$) are the various  recombination coefficients.

\subsubsection{Collisional ionisation by electron impact and three-body recombination}

The cross section for collisional ionization of an atom or ion from an 
 initial state $i$ to a final state $j$  by a free electron, 
differential in the energy $E_1$ of the incident and that 
of the ejected electron $E_2$ can be expressed, as in the 
excitation process, in terms of a collision strength $\Omega$:
\begin{equation}
\label{eq_cross}
\sigma(E_1, E_2) = \frac{1}{k_1^2 g_i} \Omega_{ij},
\end{equation}
where $g_i$ is the statistical weight of the initial state $i$,  $k_1$ is the  kinetic momentum of the
incident electron, while the collision strength $\Omega_{ij}$ can be calculated
 by replacing the bound orbital in
the final state with the free orbital of the ejected electron and summing 
over its angular momentum \citep[see, e.g.][]{gu:2008}.
The total ionization cross section is obtained by integrating
over the energy $E_2$ of the ejected electron:
\begin{equation}
\label{eq:eq_total}
\sigma_{ij}(E_1) = \int\nolimits_0^{\frac{E_1-I}{2}}  \sigma_{ij}(E_1, E_2) {\rm d}E_2  \;,
\end{equation}
where $I$ is the ionization energy. Note that 
$E_1 =I + E_s + E_2$, where $E_s$ is the energy of the
scattered electron. By energy conservation, indicating with $E_i$ the energy
of the $i$ state of the ion, we have: $E_j - E_i = E_1 - E_2 - E_s$.
Also note that the energy of the incident electron must be above threshold:
$E_1 > E_j - E_i $, and  for the ejected electron to exist it must 
have an energy $E_2 > E_1 - (E_j - E_i)$.

The total number of ionisations is found by 
integrating over the  distribution of the incident (free) electron.
The total ionisation rate between two ions can be obtained by summing 
the  rate coefficients for each initial state $i$ over the final states $j$,
and then over all the initial states, although the main contribution to the total
is typically the ionisation between the two ground states of the ions.

Note that ionization by direct impact (DI) is the main process, although 
for some isoelectronic sequences additional non-negligible 
ionisation can occur via  inner-shell excitation into a state 
above the ionization threshold which then auto-ionizes.
This is referred as  excitation--autoionization (EA). 
\cite{goldberg_etal:1965} were among the first to point out the importance of 
this process, which was later confirmed with experiments 
\citep[see, e.g.][]{crandall_etal:1979}.
The EA provides additional contributions at higher energies of the 
incident electron, and increases with charge. 
This contribution is calculated by multiplying the inner-shell 
excitation cross section with the branching ratios associated with the 
doubly-excited level.

The main DI process occurs via neighboring ionisation stages, however
double ionisation processes can sometimes be non-negligible, 
 as e.g.  shown by \cite{hahn_etal:2017}. 
More details and references about these 
processes can be found below in the atomic data section.

The study by \cite{bell_etal:1983} presented a review of 
calculated and measured  cross sections 
 between  ground states of the main ions relevant for astrophysics.
This was a landmark paper which formed a reference for a long time.
A significant revision of the collisional ionization by direct impact
was produced by \cite{dere:07}, where most of the DI and EA cross sections 
 between  ground states were recalculated 
and compared to experimental data, whenever available.
\cite{urdampilleta_etal:2017} recently also provided a review of 
ionisation rates, but without providing new calculations.

Assuming a Maxwellian distribution, and indicating for simplicity 
with $E$ the energy of the incident electron,  
we have,  as in  Eq.\ref{eq:coll_coeff},  that the rate coefficient 
for collisional ionization is:

\beq
\label{eq:coll_ionization_coeff2}
\eqalign{
C^{\rm I} & = \left( {8 \over  \pi\, m } \right)^{1/2} \, (k T)^{-3/2} \int_{I}^\infty  E\, \sigma_{ij} (E) \, \eul^{-{E / kT}} {\rm d}E \cr
         & = \left ({8 \over  \pi\, m } \right)^{1/2} \, (k T)^{1/2} \, \eul^{-{I / kT}} \int_0^\infty (k T x + I)\,  \sigma_{ij} (k T x + I) \, \eul^{-x} \; {\rm d}x \;\; , \cr}
\eeq
where  we have applied the substitution $E = k T x + I$.

The function in the integrand is very steep. For typical temperatures 
where the ions are formed in the solar corona in equilibrium, the 
dominant values the cross section to the integral are those 
from threshold until the peak. The integral in 
Eq.\ref{eq:coll_ionization_coeff2} is of the type 
\beq
\int_0^\infty f(x) \, \eul^{-x} \,  {\rm d}x \;\; ,
\eeq
 and is therefore often evaluated using a Gauss-Laguerre quadrature: $\sum_i w_i f(x_i)$,
where $x_i$ is the root of a Laguerre polynomial, and $w_i$ is the weight:

\beq
C^{\rm I} = \left({8 \over  \pi\, m }\right)^{1/2} \, (k T)^{1/2} \, \eul^{-{I / kT}}
\sum_i w_i \, \left(x_i + {I\over k T} \right)\,  \sigma_{ij} (k T x_i + I)  \;\; .
\label{eq:coll_ionization_coeff}
\eeq
The numerical factor $\sqrt{8  / (\pi\, m)} = 5.287 \times 10^{13}$ as before in the case of 
excitation.
Also note that the analogous expression in the landmark paper by \cite{bell_etal:1983} 
(their equation 8), is incorrect (surprisingly).

Three body recombination is the inverse process of collisional ionization. 
If we indicate with $C^{\rm I}_{ij}$ the collisional rate coefficient for ionisation 
by direct impact of the ion $Z^{+r}$ in its state $i$ to the 
ion  $Z^{+r+1}$ in its state $j$, 
the rate coefficient for the three body  recombination $C^{\rm 3B}_{ji}$
can be obtained by applying the principle of detailed balance:
\beq
N_{\rm e} \,  N_i(Z^{+r}) \, C^{\rm I}_{ij} = N_{\rm e} \,  N_j(Z^{+r+1}) \, C^{\rm 3B}_{ji} \quad ,
\eeq
which leads to 
\beq
C^{\rm 3B}_{ji} =  C^{\rm I}_{ij} = {g_i \over g_j} {N_{\rm e} \over 2}
\left( {h^2 \over 2\, \pi\, m\, k T_{\rm e}  } \right)^{3/2} \; \eul^{-{(E_i -E_f) / kT_{\rm e}}} \,
\eeq
in the case of a Maxwellian electron distribution. 
In the more general case of non-Maxwellian distributions,
 a simple relation between the rates does not hold.
The detail balance applied to the two processes leads to the Fowler relation  
between the differential cross-sections and the 
three  body recombination rate involves an integral over the 
energies of the electrons involved in the process.

\subsubsection{Photoionisation and radiative recombination}

Photoionisation is the process by which a photon of  energy 
higher than the ionisation threshold is absorbed by an ion, leaving 
the ion in the next ionisation stage. 
For the solar corona, photoionisation is normally a negligible process.
For this reason, it is not discussed in detail here.

We note, however, that photoionisation 
can become important in a number of cases, 
for example  for cool prominence material in the corona, and
for low-temperature lines formed in the chromosphere / transition-region, 
especially during flares, when the photoionising coronal 
radiation can become significant.
This particularly affects lines from  H I  and He I, He II, as  
photoionisation is  followed by recombination into 
excited levels, which can then  affect the population of lower
levels via cascading. For a  discussion of this 
photoionisation-recombination process for He see e.g. \cite{zirin:75,andretta_etal:03}.

The photoionisation rate coefficient from a bound level 
$i$ to the continuum $c$ is:
\beq
 \alpha^{\rm PI}_{ic} = 4 \pi \; \int\limits_{\nu_0}^\infty {\sigma^{\rm (bf)}_{ic}(\nu) \over h\nu} \; J_\nu \; {\rm d}\nu \;\; ,
\eeq
\noindent where $\nu_0$
is the threshold frequency below which the bound-free cross section $\sigma^{\rm (bf)}_{ic}(\nu)$  
is zero.

The photoionisation cross-sections increase roughly
 as the cube of the wavelength, until threshold.
For H-like ions, the modified Kramers' semi-classical expression  is often used:
\begin{equation}
\sigma^{\rm (bf)}_\nu = { 64 \pi^4 m e^{10} \over 3 \sqrt{3} c h^6 } \, {Z^4 \, g^{\rm (bf)} \over n^5 \, \nu^3} =  
2.815 \times 10^{-29} {Z^4 \, g^{\rm (bf)} \over
n^5 \, \nu^3} \qquad   \nu \ge \nu_0 \; ,
\end{equation}
where $n$ is the principal quantum number of the level 
from which the ion of charge $Z$ is ionised.
 $g^{\rm (bf)}$ is the dimensionless bound-free Gaunt factor 
(which is close to 1.), introduced as a correction. 
Values of the bound-free Gaunt factor for 
H-like ions are tabulated by \cite{karzas_latter:61}.

Quantum-mechanical calculations of photoionisation cross-sections 
for ions in general are quite complex. 
There are two main approaches 
required to generate opacities: the R-matrix method 
\citep{berrington_etal:1987} and the perturbative,  or ``distorted wave'' (DW) method 
\citep[see, e.g.][]{badnell_seaton:2003}.  
The R-matrix method is accurate but
computationally expensive. The perturbative approach is much faster but
approximates resonances with symmetric line profiles and neglects their
interaction with the direct background photoionization.
There is generally good agreement between the DW cross sections
and the background $R$-matrix photoionization cross sections.

The Opacity Project \citep[OP, see][]{seaton_etal:1994}
involved many researchers under the coordination of M.J. Seaton (UCL)
for the calculations of the cross-sections with the R-matrix method. 
A significant improvement  was the inclusion of inner-shell data calculated 
with the DW method, which formed the Updated OP data \citep[UOP][]{badnell_etal:2005}.
Further updates are made available via the web site: 
\url{http://opacity-cs.obspm.fr/opacity/index.html} maintained by F.  Delahaye
at the Paris Meudon Observatory. 
On a side note, there is an extended recent literature where other groups
used similar approaches but found discordant results. 
Work within several groups is ongoing, to try and resolve the various issues, 
as they are of fundamental importance to calculate the opacities
for stellar interiors.

Photoionisation cross-sections for transitions from the ground level
are often  available in the literature as analytic fits to the 
slow-varying component of the cross-sections 
\citep[see, e.g.][]{verner_yakovlev:1995}.

Radiative recombination is the inverse process of photoionisation,
i.e. when a free electron recombines with the ion and a photon is emitted.
The cross-sections for recombination from a level $f$ to a level $i$ are  normally
calculated from the photoionisation cross-sections using  the principle
of detailed balance in  thermodynamic equilibrium, which leads to the 
Einstein-Milne relation:
\beq
\sigma^{\rm (RR)}_{fi} (E) = {g_i \over g_f} { (h \nu)^2 \over 2 m E c^2}  \sigma^{\rm (bf)}_{if}(E) \; .
\eeq

Radiative recombination rates for all  ions of astrophysical interest
 have recently been obtained  by \cite{badnell:06} using the 
 photoionisation cross-sections calculated with the DW method 
\citep[][]{badnell_seaton:2003} and the principle of detail balance 
(i.e. for Maxwellian electron distributions).
Finally, we note that the process of 
radiative recombination, induced by the radiation field, should also be included,
although it is normally a negligible process for the solar corona.

\subsubsection{Dielectronic recombination }

 Dielectronic recombination occurs when a free electron is 
captured into an autoionization state of the recombining ion. The ion can 
then autoionize (releasing a free electron) or produce a radiative 
transition into a bound state of the recombined ion. The transition can 
only occur at specific wavelengths.
The process of dielectronic recombination was shown by 
\cite{burgess:64,burgess:65} and \cite{seaton:64} to 
be  very important  for the solar corona.

By applying the principle of detailed balance, 
the rate for dielectronic capture should equal the 
spontaneous ejection of the captured electron, i.e, autoionization.
If we indicate with $C^{\rm dc}_{nj}$ the rate coefficient for the capture of the 
free electron by the ion $Z^{r+1}$ in  the state $n$ 
 into a doubly-excited state $j$ of the recombined ion $Z^{+r}$, we have

\begin{equation} 
 N_n(Z^{r+1}) N_{\rm e} C^{\rm dc}_{nj} = N_j(Z^{+r}) A^{\rm auto}_{jn} 
\end{equation}
where $A^{\rm auto}_{jn}$ is the autoionization probability for the transition 
from the doubly-excited state $j$ to the state $n$.

By applying the Saha equation for thermodynamic equilibrium,
we obtain 

\begin{equation}
	C^{\rm dc}_{nj} = { h^3 \over (2\pi m kT)^{3/2} } {g_j \over 2 g_n} 
	A^{\rm auto}_{jn} \exp \left( - { E_j - E_n \over kT } \right) 
    \label{Eq:Cap_auto}
\end{equation}
which is a general formula that also holds  outside of thermodynamic equilibrium
(as long as the electrons have a Maxwellian distribution),
and relates the rate for dielectronic capture to the autoionisation rate
(the two inverse processes).
The dielectronic capture can then be followed by 
a radiative stabilization  into a bound state of the recombined ion. 
For  coronal plasmas, this normally occurs with a 
decay of the excited state within the recombining ion.

To calculate the dielectronic recombination coefficient $C^{\rm d}_{p,u,l}$ 
of an ion in a state $p$ that captures an electron to form a state $u$ 
of the recombined ion, which then decays to a stable state $l$ we have 
\begin{equation}
	C^{\rm d}_{p,l,u} (T) = C^{\rm dc}_{pu}(T) \left[ { A_{ul} \over \sum_k A_{uk} + \sum_q A^{\rm a}_{uq} } \right]
	\label{Eq:DR_partial}
\end{equation}
where the sum over $k$ is over all states in the recombined ion $Z^{r}$  that are below $u$,
and the sum over $q$ is over all possible states of the ion  $Z^{r+1}$ and 
the free electron associated with the autoionization of the doubly-excited state $u$.

\cite{burgess:64,burgess:65} showed that for coronal ions the main DR process is 
capture into high-lying levels ($n\approx 10$ to 100), and obtained a 
general formula for the DR rates that provides a good approximation at low 
densities and has been used extensively in the literature.
More recent  DR rates have been computed from 
the autoionization rate for the reverse process, as described in the atomic data
section.

Dielectronic recombination for ions of  several isoelectronic sequences have been 
calculated in a series of papers by N.R. Badnell and colleagues 
(see e.g. \citealt{badnell_etal:03}).
More details on these atomic data can be found in Section~\ref{sec:atomic}.

\subsubsection{Charge transfer}

Charge transfer is also an efficient ionisation/recombination process, but
only at very low temperatures, and is normally 
expected to be negligible in the solar corona. 
In principle, in the low transition region, charge transfer could be 
an important process for some ions. 
For example, \cite{baliunas_butler:1980} discuss the effect of charge transfer 
between H, He and low charge states of Si, finding that, for example,
the \ion{Si}{iii} ion population becomes broader in temperature, 
with a peak around 30\,000 K instead of 50\,000 K. 
Such an effect could be significant  on a range of spectroscopic
diagnostic applications.
Such charge transfer effects were included in 
\cite{arnaud_rothenflug:85} but not in subsequent  tabulations 
of ion abundances. 
In principle, it would be possible to estimate these effects. 
For example, \cite{yu_etal:1986} used \ion{Si}{iii} line ratios to 
obtain a temperature of about 70\,000 K in a fusion device.
The temperature of low-Z ions in fusion devices is normally close to that
of ionization equilibrium, although particle transport effects 
could shift the ion populations towards higher temperatures.

\subsubsection{Charge state distributions}

In the case of local thermodynamic equilibrium ( LTE), if we write the
  detailed balance  of processes 1., 2., 3., we obtain the  Saha equation.
At low densities, plasma becomes optically thin and most of radiation escape,
therefore processes 1. and  3. are attenuated and the plasma is no longer  
in  LTE. 
In this condition, the degree of ionization of an element 
is obtained by equating the total  ionization
and recombination rates that relate successive ionization stages:
\beq
 {1 \over N_{\rm e}} {dN_{r}\over dt} =N_{r-1} S_{r-1} - N_{r} (S_{r}
+\alpha_{r}) + N_{r+1}\alpha_{r+1} 
\label{eq:time_dep_ionisation} 
\eeq
for transitions of ion $Z^r$ from and to higher and lower stages, obtaining
a set of coupled equations with the additional condition $N(Z)=\sum_r N_{r}$.
Here, $S_{r}$ and $\alpha_{r}$ are the total ionization and recombination 
rates, i.e. those that include all the relevant processes.

\begin{figure}[htb]
\centerline{\includegraphics[width=0.7\textwidth, angle=90]{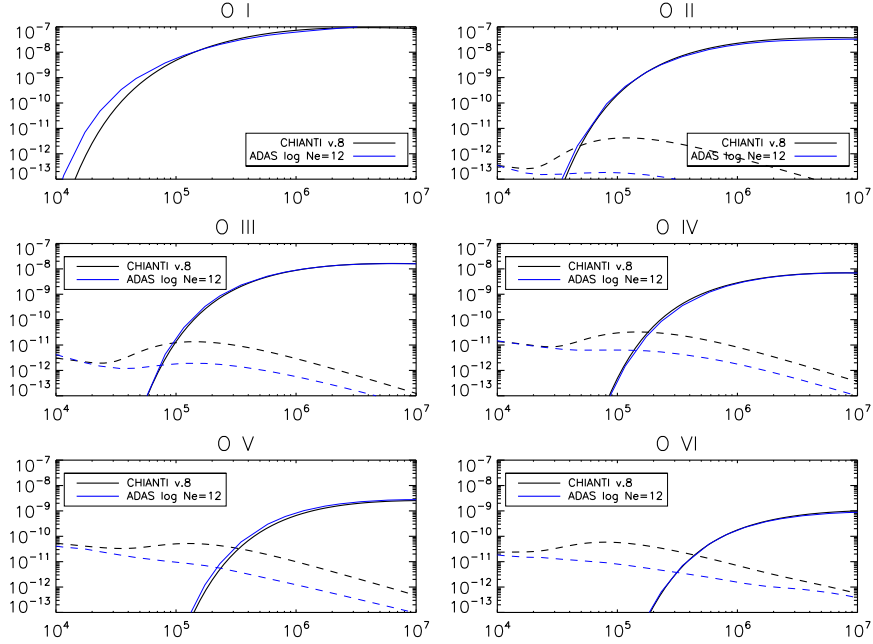}}
\caption{Ionization and recombination (total) rates for a few oxygen
ions, as a function of 
temperature. The first plot in the top left (\ion{O}{i}) shows the total ionization rate 
for neutral oxygen. The second plot  (\ion{O}{ii})  shows the total ionization rate
for \ion{O}{ii}, and the total recombination rate from \ion{O}{ii} to \ion{O}{i}, 
and so on. 
The full curves are the  ionization rates, while the dashed lines 
are the  recombination rates. The black curves are the CHIANTI v.8 rates (zero density),
while the blue ones are the effective rates 
obtained from OPEN-ADAS, and  calculated at a density  
$N_{\rm e}=10^{12}$ cm$^{-3}$.
}
  \label{fig:ionrec_rates_O_log_ne=12}
\end{figure}

Figure~\ref{fig:ionrec_rates_O_log_ne=12} shows 
as an example the total ionization and recombination rates for a few
 oxygen ions, as available with CHIANTI  version 8 (black).
Note that in addition the plots also show the ionization and recombination 
effective rates calculated at a density  $N_{\rm e}=10^{12}$ cm$^{-3}$, 
 as available in OPEN-ADAS (see next Section).

Whenever the time scales of the observed phenomena are less than those for 
ionization and recombination, we can assume that 
the population of ions lying in a 
given state is constant (${dN_{r}\over dt} =0$)
 and so the number of ions leaving this state
per unit time must exactly balance the number arriving into that state.
This is the so called collisional ionization equilibrium (CIE),
which is normally assumed for the solar corona
(for very low densities, as in the case of 
planetary nebulae, photoionisation becomes important and 
dominates the ion charge state distribution).
 In the case of two successive stages:

\beq
{ N_{r+1}\over N_{r} } = { S_{r} \over \alpha_{r+1} } \; .
\eeq

Many ionization equilibrium calculations have been published,
perhaps the most significant for the iron ions was that one from 
\cite{burgess_seaton:1964}, where the newly discovered dielectronic
recombination was included.  
Later ones include:  \cite{jordan:69},
\cite{arnaud_rothenflug:85}, \cite{landini_monsignori-fossi:1991}, 
\cite{arnaud_raymond:92}, \cite{mazzotta_etal:98}.
Other calculations which also included density effects are those from 
\cite{summers:1972,summers:1974}.

\begin{figure}[htb]
\centerline{\includegraphics[width=9cm,angle=90]{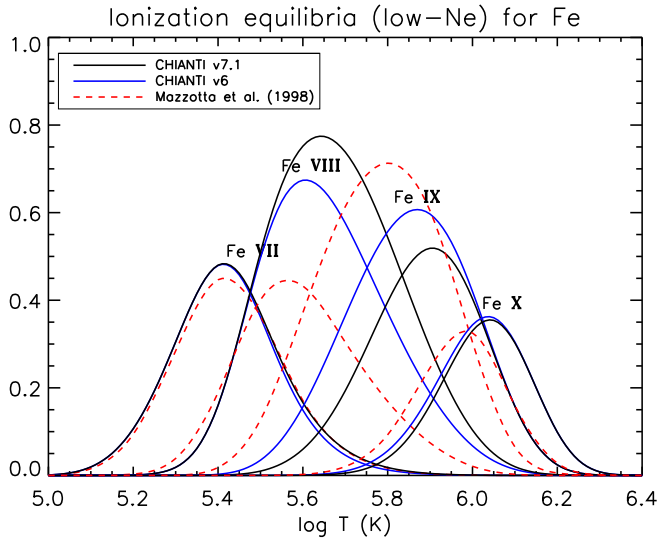}}
\caption{Ion populations of some important coronal ions
 in equilibrium in the low-density limit. Results from the 
tabulations of CHIANTI v.6, v.7.1 and \cite{mazzotta_etal:98} are shown.}
  \label{fig:comp_bal3}
\end{figure}

The improvements in the rates as recalculated in recent years have led to a revision of the 
ion populations in equilibrium, published by \cite{dere_etal:09_chianti_v6}
for  CHIANTI version~6.
 In some cases, significant differences with 
previous ionisation tables were present. An example is given in Figure~\ref{fig:comp_bal3}.
Clearly, such large differences affect any measurements that 
rely on the ion abundances, such as temperatures from DEM analyses, or 
relative elemental abundances. 
The \cite{bryans_etal:2009} ion populations are based on 
almost the same atomic data and rates as used in CHIANTI v.6, so are very similar.

Finally, a few remarks about the formation temperature of a spectral line.
This can be quite different from the temperature $T_{\rm max}$, corresponding to the peak 
charge state of an ion, a common  assumption in the past.  
This is often found especially in transition region lines,
where the temperature gradient of the solar atmosphere is very steep.
As we shall see below, there are indications that several 
structures in the solar corona are nearly isothermal, with temperatures
typically very different from $T_{\rm max}$. This issue is particularly relevant
when lines from different ions are used for diagnostic applications such as
measuring relative elemental abundances (see Section~\ref{sec:abund}).

As a first approximation, an effective temperature $T_{\rm eff}$  
(cf. Eq.~\ref{eq:t_eff}), where most of the line
is formed, can be defined when a continuous distribution of temperatures
in the plasma volume is present, as described below in Section~\ref{sec:dem_diagn}.

Diffusion processes within the transition
region, where strong flows and strong temperature gradients exist can also
affect the temperature of formation of a line, see e.g. 
\cite{tworkowski:1976}.

\subsubsection{Density-dependent effects on the ion balances}

Most ionization equilibrium calculations are in the so-called \textit{low-density limit
(the coronal approximation)}, where 
 all the population in an ion is assumed to be in the ground state and all the rates
are calculated at low densities.

However, as shown by  \cite{burgess_summers:1969}, 
the dielectronic recombination rates decrease significantly at high densities.
This is caused by the fact that the intermediate excited states below the ionization limit
can easily be re-ionized via electron impact, since this ionization
increases linearly with the electron density. When this re-ionization occurs, 
recombination does not happen. 
 The effect is particularly strong for 
lower charge states, i.e. transition-region ions, since 
at typical TR densities the  dielectronic recombination becomes 
suppressed significantly. 
As a consequence, these ions  become populated at progressively lower temperatures. 
To estimate this suppression,  collisional-radiative (CR) modelling was carried out
by Summers and Burgess, as described in \cite{summers:1972,summers:1974}, 
and further refined in following studies.

Recently,  \cite{Nikolic13} suggested an empirical formula to reproduce the 
suppression factors as calculated by \cite{summers:1974}
as a function of the ion charge, isoelectronic sequence, 
electron density, and temperature. The principle idea was that one would apply these
suppression factors to the most recent dielectronic recombination 
 results from the DR  project \citep{badnell_etal:03},
to assess for the importance of this effect for a particular application. 
If the effect is found to be important, appropriate CR modelling should be carried out.
We note that such suppression should only be applied to the total DR rates.
One problem with the approach is that the recombination rates calculated by  \cite{summers:1974}
were actually effective rates, i.e. included many other density effects in addition to
the DR suppression, such as the three body recombination, and the changes
related to the presence of  metastable levels. 
Another problem is that  the \cite{Nikolic13} formulae were trying to reproduce
the  \cite{summers:1974} tables which neglected secondary autoionization, 
with the result that the suppression would be over-estimated, if those
factors were applied to the  DR  project rates, as they included secondary autoionization. 
Indeed for the boron, carbon, aluminium and silicon sequences
\cite{summers:1974} produced
tables with secondary autoionization which show less suppression with density.
A revision of the \cite{Nikolic13} formulae is underway.

The effect of including the  metastable levels in the ion balance
calculations for transition region ions can be as important as the 
suppression of dielectronic recombination.
An approximate calculation for \ion{C}{iv} was carried out by 
\cite{vernazza_raymond:79} using a rough estimate of the 
 DR suppression based on \cite{summers:1974} (assuming that it would be the same
for all fine-structure levels), and adding 
collisional ionisation from the metastable levels in  \ion{C}{iii}.
The two effects appeared equally important.

One way to estimate  these  effects is the 
Generalised Collisional-Radiative (GCR) modelling \citep{McWhirter84,Summers06}
which has been implemented within the Atomic Data and Analysis Structure (ADAS). 
Once the level-resolved ion balance equations are solved, 
 {\it effective} ionisation and recombination rates can be obtained.
These rates are currently  available via OPEN-ADAS\footnote{open.adas.ac.uk}. 

Figure~\ref{fig:ionrec_rates_O_log_ne=12} shows as an example the total effective
ionization and recombination rates (blue curves) for a few oxygen ions,
calculated at a density  $N_{\rm e}=10^{12}$ cm$^{-3}$, as available in 
 OPEN-ADAS (1996 version), compared to the CHIANTI v.8 rates (black curves, at 
zero density).
The effect of the suppression in the DR rates is clearly present in the 
ADAS  effective recombination rates.

The effect of these processes on diagnostic ratios and predicted radiances
 has been discussed in the literature  
(cf. \citealt{vernazza_raymond:79,delzanna_etal:02_aumic,doyle_etal:2005}).
Generally, the  predicted radiances increase by factors of 2--3, thus reducing the 
discrepancies with observation
 for the so-called anomalous ions, i.e. those that have  predicted radiances typically 
a factor of 5--10 lower than observed, as discussed below in Section~\ref{sec:anomalous_ions}.

\begin{figure}[!htbp]
\centerline{\includegraphics[width=0.6\textwidth, angle=90]{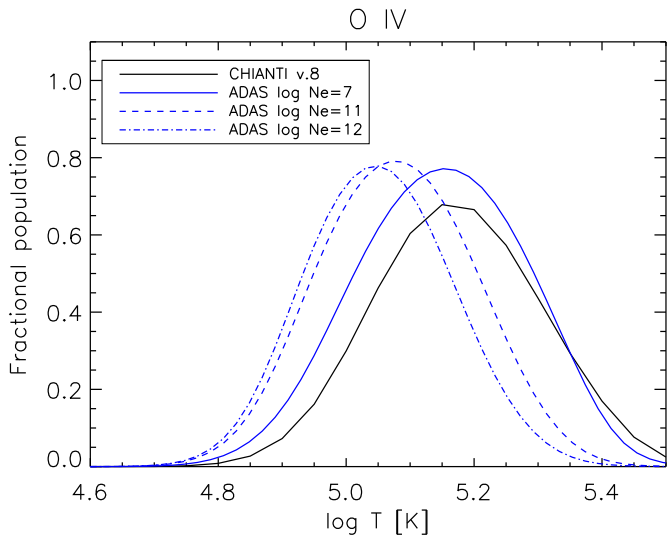}}
\caption{Fractional \ion{O}{iv} abundances in equilibrium at 
different electron densities (blue, from OPEN-ADAS),
 and the low-density values as in CHIANTI version 8 (black). 
}
\label{fig:O+3_ioniz}
\end{figure}

The density effect on the ion charge state distribution  are particularly important for the IRIS 
\ion{O}{iv}  and \ion{S}{iv}  lines, as shown by \cite{polito_etal:2016b}.
 Fig.~\ref{fig:O+3_ioniz} shows as an example the fractional \ion{O}{iv} 
abundance in equilibrium at 
different electron densities (blue) as calculated with the 
OPEN-ADAS rates and the low-density value as in CHIANTI version 8.

\subsection{Optical depth effects}

In most cases, spectral lines in the XUV originating from the 
solar corona and transition region are optically thin, i.e. 
the emitted photons freely escape.
There are however several cases, especially for strong transition-region
 lines, where lines could  be optically thick. 
One question naturally arises: how can we check if a spectral line is 
optically thin?
There are various ways. One obvious effect of optical depth effects 
is a flattening, widening (or even self-reversal of the core) of the line profile.
Estimates of the optical depth of lines from their width and 
their centre-to-limb variation have been obtained by e.g. 
\cite{roussel-dupre_etal:1979} and \cite{doschek_feldman:2004}.
 However, a shape that is for example non-Gaussian does not necessarily 
mean that a line is not optically thin. In fact, a non-Gaussian 
shape can be produced by a superposition of Doppler motions.
We further discuss such effects together with other broadening mechanisms
below in Section~\ref{sec:widths}.

\begin{table}[htbp]
\caption[Some UV line ratios that can be used to investigate  optical depth effects.]
{Some UV line ratios that can be used to investigate  optical depth effects.
The last column indicates  the theoretical value in the optically thin case.}
\centering
\begin{tabular} [c]{lclcc}
\toprule
Ion & Terms & Wavelengths (\AA) & Ratio &  \\
\midrule
C~II &  $^2$P$_{3/2}$ -  $^2$S$_{1/2}$ / & &  &  \\
	& $^2$P$_{1/2}$ -  $^2$S$_{1/2}$   & 1037.02/1036.33  & 2 & \\

Si~II   & $\prime\prime$   & 1533.43/1526.71  & 2 &  \\

C~III	& $^3$P$_{0}$ - $^3$P$_{1}$ / & &  &  \\
	& $^3$P$_{2}$ - $^3$P$_{1}$   & 1175.26/1176.37  & 0.8 &  \\

Si III  & $\prime\prime$   & 1296.73/1303.32   & 0.8 &     \\ 

C~III	& $^3$P$_{2}$ - $^3$P$_{2}$ / & &  &  \\
	& $^3$P$_{1}$ - $^3$P$_{2}$ &1175.71/1174.93   & 3 &    \\

Si III  & $\prime\prime$   & 1298.94/1294.54  & 3 &     \\

Si~IV  & $^2$S$_{1/2}$ - $^2$P$_{3/2}$ / & &  &  \\
	& $^2$S$_{1/2}$ - $^2$P$_{1/2}$  & 1393.75/1402.77 & 2 &   \\ 

C~IV	& $\prime\prime$   &1548.20/1550.77 & 2 &    \\ 
N~V	& $\prime\prime$   &1238.82/1242.80 & 2 &     \\ 
O~VI	& $\prime\prime$   &1031.91/1037.61 & 2 &     \\ 

Ne~VIII & $\prime\prime$  & 770.41/780.32& 2 &     \\ 
\bottomrule
\end{tabular}
\label{tab:opacity}
\end{table}

A direct way to estimate if lines are optically thin is to consider 
the ratios of lines that  originate from 
a common upper level (a branching ratio). 
In the optically thin case, the ratios are equal to the 
ratios of the A-values (see, e.g.  \citealt{jordan67}), 
which are normally known to a   good accuracy. 
If opacity effects are present, the ratio would be different.
Table~\ref{tab:opacity} lists a few commonly used ratios
from C~II, Si~II, Si~III, and C~III
(see, e.g.  \citealt{doyle_mcwhirter:80,keenan86_si3,brooks00,delzanna_etal:02_aumic}).

Another  direct method  applies to the doublets of the 
Li-like ions. 
In the optically thin case, the ratios of the two lines of the doublet
should be equal to the ratio of the oscillator strengths, i.e. two.
Due to opacity effects, the intensity of the 
brightest component, the $^2$S$_{1/2}$ - $^2$P$_{3/2}$ typically decreases
(relative to the weaker one), because it has a larger oscillator strength. 

From the observed departure of a ratio from the optically thin case,
the optical depths of the lines and the path lengths of the 
emitting layer can be obtained. The process is not straightforward though.
Various approximations and approaches exist in the literature, 
 involving the probability that a photon emitted from
 a layer of certain optical depth will escape along the line of sight. 
For details, see, e.g. 
\cite{holstein:1947,jordan67,irons:1979,doyle_mcwhirter:80,kastner_kastner:1990,brooks00}
and references therein.

We now introduce a simplified discussion of optical depth.
The one-dimensional radiative transfer equation for the 
specific intensity of the radiation field that propagates, at frequency 
$\nu$ along the direction $\vec \Omega$  inside a plasma is: 
\beq
{{\rm d} \over {\rm d}s} I_\nu(\vec \Omega) =  
   -k_\nu^{({\rm a})} \, I_\nu(\vec \Omega) \, +k_\nu^{({\rm s})} 
    I_\nu(\vec \Omega) \, +\epsilon_\nu \;\; , 
\eeq
where $s$ is the spatial coordinate measured along the 
direction of propagation, and 
 the three quantities $k_\nu^{({\rm a})}$, $k_\nu^{({\rm s})}$, and $\epsilon_\nu$, 
are the absorption coefficient, the coefficient of stimulated emission, and
the emission coefficient,  respectively.
The equation is often written in this form:
\beq
{{\rm d} \over {\rm d}s} I_\nu(\vec \Omega) =  
   -k_\nu \, [I_\nu(\vec \Omega) - S_\nu] \;\; , 
\eeq
where the source function $S_\nu = \epsilon_\nu / k_\nu$ and 
$k_\nu =  k_\nu^{({\rm a})} - k_\nu^{({\rm s})}$. 

The specific optical depth
\beq
{\rm d} \tau_\nu = - k_\nu \, {\rm d}s \;\; 
\eeq
is defined in the direction opposite to that of the propagation of the radiation, 
which reflects the point of view of an observer  receiving the radiation.
The optical thickness at frequency 
$\nu$ of a  plasma slab of geometrical thickness $D$ is
defined as:
\beq
 \tau_\nu = \int_0^D  k_\nu(s) \,  {\rm d}s \;\; . 
\eeq
A plasma is  optically thick when $\tau_\nu \gg 1$, i.e. when 
the  photon has a probability  practically equal to unity to 
be absorbed within the slab.

Considering a simple two-level atom with lower and upper level 
populations $N_l, N_u$, it is possible to show that these
monochromatic coefficients are related to Einstein's coefficients:

\beq
k_\nu^{({\rm a})} ={h \, \nu \over 4 \pi} \, N_l \, B_{lu} \, \phi(\nu - \nu_0) \; ; \quad
k_\nu^{({\rm s})} = - {h \, \nu \over 4 \pi} \, N_u  B_{ul} \, \chi(\nu - \nu_0) \; ; \quad 
\epsilon_\nu ={h \, \nu   \over 4 \pi} \, N_u \, A_{ul}   \, \psi(\nu - \nu_0)
\eeq
where $ \phi(\nu - \nu_0), \chi(\nu - \nu_0), \psi(\nu - \nu_0)$ 
are the area-normalised line profiles for the 
extinction,  induced emission, and spontaneous emission, respectively.
 $\nu_0$ indicates the line centre frequency.

When the incident (exciting)
and emitted photons are not correlated, we have 
{\it complete redistribution} and the three local line profiles are equal, in 
which case the source function $S_\nu$ becomes frequency independent:
\beq
S_{\nu_0} = { N_u \, A_{ul}  \over N_l \, B_{lu} - N_u  B_{ul}} \;\; ,
\eeq
and the bound-bound opacity of a line can be expressed in terms of the A-value as: 
\beq
 \tau_\nu = \int_0^D {c^2 \over 8\,\pi \, \nu^2} \, {g_u \over g_l} \, N_l \, A_{ul}  \,
 \left\{ 1- {N_u g_l \over N_l g_u}   \right\} \, \phi(\nu - \nu_0)  {\rm d}s \;\; , 
\eeq
which could also be written in terms of the oscillator strength  $f_{lu}$ 
of the transition, using the relation:
\beq
 g_u \, A_{ul} = {8 \pi^2 \, e^2 \, \nu_{ul}^2 
   \over m \, c^3} \, g_l \, f_{lu} \;\; .
\eeq

The line centre optical thickness (or opacity) is an useful quantity to assess
if a spectral line is optically thin:
\beq
\tau_{\nu_0} = \int  k_{\nu_0}  N_l  {\rm d}s \;\; ,
\eeq
where $k_{\nu_0}$ is the absorption coefficient at line center frequency $\nu_0$ 
which can be written, neglecting the induced emission and assuming a Doppler line profile:
\beq
 k_{\nu_0} = {h \, \nu_0 \over 4 \pi} \, B_{lu} {1 \over  \pi^{1/2} \, \Delta \nu_D} \;\; ,
\eeq
where $ \Delta \nu_D$ is the Doppler width of the line, in frequency 
(for more details on Doppler widths see Section~\ref{sec:widths}).

Considering the relations between Einstein's coefficients
and the oscillator strength, and the expression for the population
of the lower level  (cf previous notation):

$$N_l = {N_l \over N(Z^{+r}) } \, {N(Z^{+r}) \over N(Z)} \,  Ab(X)  \,
\frac{N_\mathrm{H}}{N_\mathrm{e}}  \, N_\mathrm{e} \;\; , $$
the opacity at line centre can be written as:
\begin{equation}
\tau_{0} = {\pi \, e^2 \, \over m \, c} {1 \over  \pi^{1/2} \, \Delta \nu_D} f_{lu} \; D = 
 {\pi^{1/2} \, e^2 \, \over m \, c \, \Delta \nu_D} f_{lu} \; D \,
{N_l \over N(Z^{+r}) } \, {N(Z^{+r}) \over N(Z)} \,  Ab(X)  \,
\frac{N_\mathrm{H}}{N_\mathrm{e}}  \, N_\mathrm{e} \;\; ,
	\label{Eq:tau}
\end{equation}
where  $N_\mathrm{e}$ is the average electron density of
 the plasma emitting the line, and $D$ is the path length 
along the line of sight through the source.
This is a very useful definition commonly used, and easy to estimate,
once the various factors are known. 
If $\tau_{0}$ is of the order of one  or less, opacity effects can be neglected.
Even if $\tau_{0}$ is much larger, it does not mean that photons do not 
escape. 
For example, values of  $\tau_{0} \simeq 10^4$ are needed for no photons 
 to escape \citep{athay:1971}.

With the above definition, we can show that even the strongest 
lines in the transition region are not affected by opacity in typical quiet Sun 
regions. Following \cite{mariska:92}, we consider one of the strongest 
lines, the \ion{C}{iv} 1548~\AA\ resonance transition, which has 
$f=0.2$. Assuming typical quiet Sun  values of  
$N_{\rm e}=10^{10}$ cm$^{-3}$ and $T=10^5 $K and observed values
of the line width we have $\tau_{0} \simeq 10^{-8}\, D$ as an 
order of magnitude (this value can differ depending on the chosen 
element and ion abundances). Therefore, a slab thickness larger than 
$10^{8}$ cm would be needed before opacity effects come into play.
This is more than the typical size of the region where  \ion{C}{iv} is emitted.
Indeed opacity effects are typically seen in lower temperature lines 
formed in the chromosphere or occasionally in active regions.

Another effect that often decreases the intensity of a spectral line
is absorption by cool plasma that is along the line of sight.
The main absorption is by neutral hydrogen 
(with an edge at 912~\AA), neutral helium (with an edge at 504~\AA) and 
ionised helium (with an edge at 228~\AA).
Such effects were noticed long ago from Skylab observations
(see e.g. \citealt{orrall_schmahl:1976}).
They are particularly evident when surges or filaments are present.

From the observed absorption at different wavelengths and some 
assumptions about  the underlying emission, it is possible to estimate
the column density of the hydrogen and helium absorbing plasma.
 \cite{kucera_etal:1998} used SOHO CDS while 
\cite{delzanna_etal:2004_filaments} combined
 SOHO CDS with SUMER. Others have used the absorption seen in 
EUV images, such as SDO AIA (see, e.g. \citealt{williams_etal:2013}).

\subsection{Continuum radiation}
\label{sec:continuum}

\textit{Free-free} emission is produced when an electron interacts with a 
charged particle $Z$ and looses its kinetic  energy 
$E= m v^2/2$  releasing a photon of energy $h\nu$: $ Z + e(E) 
\Longrightarrow Z + e(E') + h\nu $. 
The process is called \textit{bremsstrahlung} (`braking radiation').
The emission is a continuum.
The calculation of the free-free emission in the classical way can be found in 
textbooks, and is based on the radiation emitted by a charged particle
in the Coulomb field of the ion, with an impact parameter $b$.
Classically, the minimum impact parameter is set so the 
kinetic  energy of the free electron is greater than the 
binding energy: $E > {e^2 Z \over b }$, otherwise we would 
have recombination. With a quantum-mechanical treatment,
a correction factor needs to be introduced.  This need to take into account
that  $m v^2/2 \ge h\nu$, otherwise a photon could not be created. 
The correction  is the 
free-free Gaunt factor $g_{\rm ff}(\nu\ v)$ which is
 close to unity and has a weak dependence on 
the frequency $\nu$ and the electron velocity/energy. 
The energy emitted per unit time, volume and frequency is  

\begin{equation}
\frac{dW}{dt dV d\nu}= \frac{16 \pi}{3^{3/2}}  \frac{Z^{2}e^{6}}{m^{2} v c^{3}} N_{\rm e} N_{\rm i} g_{\rm ff}
\end{equation}
where $N_{\rm e} N_{\rm i}$ are the electron and ion number densities.
We note that slight different definitions are found in the literature,
depending on how the Gaunt factor is defined. We follow \cite{rybicki_lightman:1979}.

For a Maxwellian velocity distribution of electrons the process is called 
\textit{thermal bremsstrahlung}. The calculation of the emitted energy per unit of time and 
volume requires an integration over the electron velocity 
distribution. The integral with the Gaunt factor produces the mean Gaunt factor
 $<g_{\rm ff}>$  (see, e.g. \cite{karzas_latter:61}).  
We obtain:

\begin{equation}
\frac{dW}{dt dV d\nu}=  \frac{32 \pi e^6}{3 m c^3} \Big(\frac{2 \pi}{3 k m}\Big)^{1/2}~
\frac{Z^{2} N_{\rm e} N_{\rm i}}{T_{\rm e}^{1/2}}~\eul^{-{h\nu / kT_{\rm e}}}~<g_{\rm ff}>
\end{equation}

\begin{equation}
\frac{dW}{dt dV d\nu}=
6.8\times 10^{-38} \frac{Z^{2} N_{\rm e} N_{\rm i}}{T_{\rm e}^{1/2}}~\eul^{-{h\nu/kT_{\rm e}}}~<g_{\rm ff}>
 \rm erg\,cm^{3}\,s^{-1}\,Hz^{-1}
\end{equation}
\noindent
where $k$ is  Boltzmann constant and $h$ is the Planck constant. 

Free-free emission is the main radiative loss mechanism for low 
density plasmas at $T>10^7 $K.

\textit{Free-bound} emission
is produced when a free electron of energy $E$ is captured by
an ion ($Z^{r+1}$) into a bound state of $Z^{r}$:

$$ Z^{r+1} + e(E) \Rightarrow Z_{n}^{r} + h \nu $$
\noindent
a photon of energy $h \nu = E + I_{n}$ is emitted and
$I_n$ is the ionization energy of the bound state $n$.
For a Maxwellian electron velocity distribution, the  continuum emission is
characterized by discontinuities at the ionization thresholds.

The free-bound continuum emissivity produced from recombination onto
an ion of charge $Z$ can be written as

\begin{equation}\label{fb-eqn}
P_{{\rm fb},\lambda} =  3.0992 \times 10^{-52}
    N_{\rm e} N_{Z+1} {E_\lambda^5 \over T^{3/2}} \sum_i
    {\omega_i \over \omega_0} \sigma^{\rm bf}_i
    \exp  \left( - {E_\lambda - I_i \over kT} \right) \qquad
    [{\rm erg}~{\rm cm}^{-3}~{\rm s}^{-1}~{\rm \AA}^{-1}]
\end{equation}
where $ N_{\rm e}$ and $N_{Z+1}$ are the number densities of electrons and
recombining ions, respectively, in units of cm$^{-3}$; $E_\lambda$ is
the energy in cm$^{-1}$ of the emitted radiation; $T$ is the plasma
temperature in K; $\omega_i$ is the statistical weight of the level
$i$ in the recombined ion; $\omega_0$ is the statistical weight of the
ground level of the recombining ion; $\sigma^{\rm bf}_i$ is the
photoionization cross-section from the level $i$ in the recombined ion
to the ground level of the recombining ion, in units of Mb
($=10^{-18}$~cm$^2$); $I_i$ is the ionization energy in units of
cm$^{-1}$ from the level $i$ in the recombined ion, 
and the sum is over all levels $i$ below the
recombined ion's ionization limit.

Finally, there is another process which produces continuum radiation, the so-called
\emph{Two-photon} continuum. It is caused by two-photon decay processes in 
H-like and He-like ions. 
Compared to the free-free and the bound-free, this continuum is nearly negligible,
except at low temperatures  T$\le$ 3 $\times$10$^4$ K. 
However, it is important in the population modelling of such ions. 
The transition from the metastable 1s 2s $^1$S$_0$ state of  Helium-like ions 
to their ground state 1s$^2$ $^1$S$_0$ is strictly forbidden, hence the 
two-photon process becomes an important  depopulating process. 
The same occurs in H-like ions, where the transition from the metastable 
 2s $^2$S$_{1/2}$ state to the ground state 1s $^2$S$_{1/2}$ is also strictly 
forbidden. 
Calculations of the rates for these processes have been carried out 
by several authors, see e.g. \cite{drake:1986} for He-like and 
\cite{parpia_johnson:1982} for the H-like ions.

\begin{figure}[htb]
\centerline{\includegraphics[width=0.7\textwidth, angle=90]{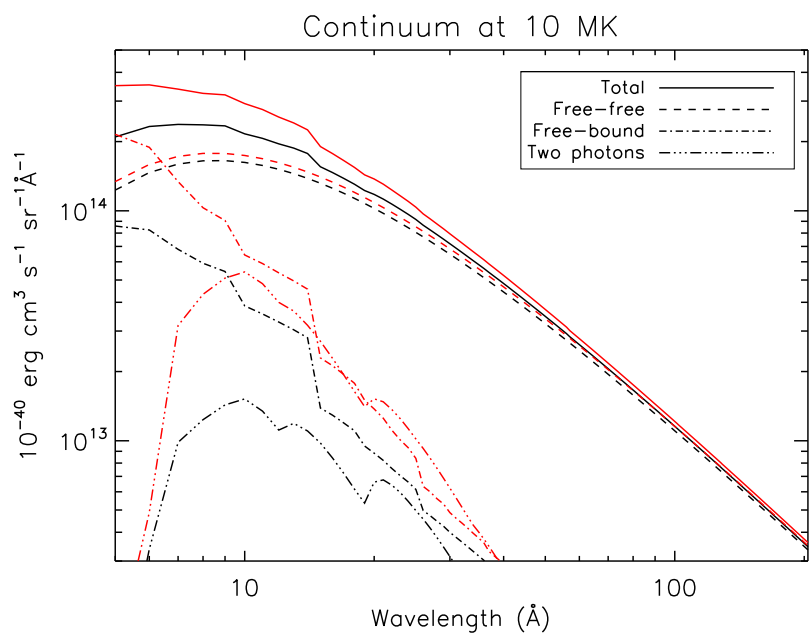}}
  \caption{The continuum calculated with CHIANTI version 8 
at a temperature of 10 MK, from 5 to 200~\AA. 
The black curves are calculated with the photospheric abundances recommended by
\cite{asplund_etal:09}, while the red ones with the `coronal' abundances of \cite{feldman:92}.}
\label{fig:continuum}
\end{figure}

The available atomic data (including relativistic effects) 
for the continuum were assessed for CHIANTI version 3 by \cite{young_etal:03}.
 Figure~\ref{fig:continuum} shows the continuum calculated with CHIANTI version 8 
at a temperature of 10 MK, from 5 to 200~\AA. 
The black curves are calculated with the photospheric abundances recommended by
\cite{asplund_etal:09}, while the blue ones with the `coronal' abundances
of \cite{feldman:92}. Note that at the EUV wavelengths, most of the continuum is
free-free radiation. Also note the significant variation with elemental 
abundances in the X-rays. This means that when chemical abundances are 
estimated from line-to-continuum measurements (see below in Section~\ref{sec:abund}),
a self-consistent approach needs to be adopted.

\section{The satellite lines}
\label{sec:satellites}

Satellite lines were discoverd by \cite{edlen_tyren:1939}
in laboratory vacuum spark spectra of the He-like  carbon. 
They were called satellites because they were
close  to the resonance line, mostly at longer wavelengths,
although some were also present at shorter wavelengths. 
 They  were correctly interpreted as 
1s$^2$ $nl$ -- 1s 2p $nl$ transitions, satellites of the 
He-like resonance transition 1s$^2$--1s 2p (the so-called {\emph parent line}). 
They also observed the 1s$^2$ $nl$ -- 1s 3p $nl$ transitions, satellites of the 
He-like  1s$^2$--1s 3p transition, and the 1s $nl$--2p $nl$
satellites of the H-like 1s--2p resonance line.

Satellite lines have later been observed in solar spectra, and 
soon it was recognised that they have many important and  unique 
diagnostic applications.

\begin{figure}[!htb]
 \centerline{\includegraphics[width=0.8\textwidth,angle=0]{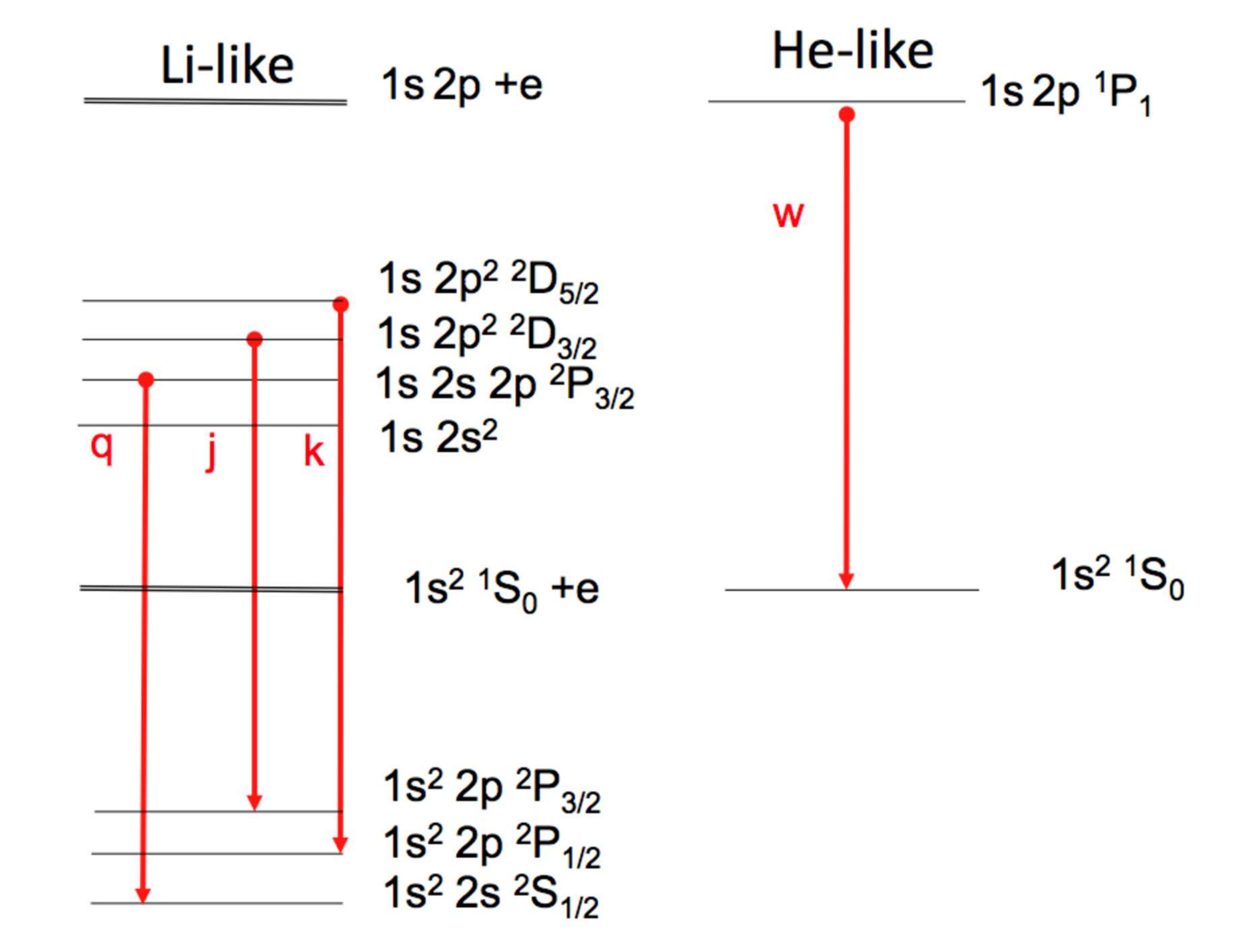}} 
 \caption{A sketch of a few of the main satellite lines of the He-like resonance line $w$.}
 \label{fig:he-like_satellite}
\end{figure}

The satellite lines of the He-like ions involve the presence of doubly-excited 
states of the Li-like ions, which can then auto-ionize or produce the satellite line
(see Fig.~\ref{fig:he-like_satellite}).
The doubly excited states exist because of the interaction with 
at least one continuum, in this case, the continuum of the ground state
of the  He-like ions. 
The formation of the doubly-excited state follows the usual selection rules
in that the autoionising state and the final state plus free electron 
have to have the same parity and total angular momentum $J$
(and $L$ and $S$ in case of LS coupling).

In principle the doubly excited states can be formed by 
inner-shell electron impact ionisation
of the Be-like ion or by two-electron impact excitation, but these processes
are normally negligible, compared to the 
inner-shell excitation of one electron in the Li-like ion (which can 
subsequently auto-ionize or produce the satellite line), or 
dielectronic capture, the inverse process of autoionisation.
Both processes need to be calculated. The relative importance of the two 
then depends on the relative values of the relevant rates, 
as well as the status of the ion (i.e. if it is out of equilibrium, ionising 
or recombining).

 Dielectronic capture  occurs when a free electron is 
captured into an autoionization state of the recombining Li-like ion. The ion can 
then autoionize (releasing  a free electron) or produce a satellite 
line, i.e. a transition into a bound state of the recombined ion. The transitions  
occur at specific wavelengths, close to the wavelength of the He-like resonance line.
As we have mentioned in Sect.~\ref{sec:charge_state},
 A. Burgess with M.J Seaton  developed the theory of dielectronic recombination,
however those  calculations were focused on the states where the spectator electron
is in a highly excited state with $n$ typically around 100, as these 
give the highest  contribution to the total dielectronic recombination of an ion.

On the other hand, for spectral  diagnostic purposes, the transitions that 
are important and are observed are those arising from lower autoionising 
levels.

Early theories and computations for these satellite lines  
were initially carried out by A. Gabriel and colleagues in the late 1960's and 
early 1970's, and later by a number of researchers, in Europe, Russia and 
USA.  Notable papers, where the main diagnostics and notation were developed
for the He-like lines are \cite{gabriel_paget:1972} and  \cite{gabriel:1972}. 
Several excellent laboratory observations of the satellite lines exist,
taken by Doschek, Feldman, Presnyakov, Boiko, Aglitskii and several other colleagues
\citep[see, e.g.][]{aglitskii_etal:1974,feldman_etal:1974,boiko_etal:1977_h-like,boiko_etal:1978_h-like,boiko_etal:1978}.
There are several excellent in-depth reviews on the satellite lines, such as 
\cite{dubau_volonte:1980,doschek:1985_review,mewe:1988,doschek:1990_review,phillips_etal:08}.

Below, we summarise the main processes and highlight some of the main 
diagnostic applications for the solar plasma. Many more diagnostic applications 
are available for other types of plasma.

\subsection{Inner-shell of Li-like ions}

We consider first the inner-shell excitation. The simultaneous 
excitation of two electrons has a low probability and 
therefore the 1s 2p$^2$ population comes mainly from excitation
of the  excited state 1s$^2$ 2p. On the other hand, the 1s 2s 2p can be excited
from either 1s$^2$ 2s or 1s$^2$ 2p, although the first one usually dominates.
Following \cite{gabriel:1972}, the intensity of the 
satellite line from level $s$ to the final level $f$ produced by  inner-shell excitation is 
(in photon units):

\beq
I^{inner}_{sf} = \beta\, N_{\rm Li-like}\, N_{\rm e}\, C^e_{is} \, {A_{sf} \over 
 \left( \sum_k A^{auto}_{sk} + \sum_{f<s} A_{sf} \right)}  \sim \beta  N_{\rm Li-like} \, N_{\rm e} \, C^e_{is} \,  {A_{sf} \over A^{tot}_s}
\eeq
where: $ N_{\rm Li-like}$ is the abundance of the Li-like ion;
 $ C^e_{is}$ is the electron impact excitation rate coefficient by inner-shell; 
$A_{sf}$ the transition probability by spontaneous emission from level $s$ to the final level $f$;  
 $\sum_k A^{auto}_{sk}$ is 
the total decay  rate via autoionisation from level $s$ to into all
available continua $k$ in the He-like ion;  
and $\sum_{f<s} A_{sf} $ is the total decay rate by spontaneous radiative 
transitions to all possible final states $f$.
For He-like ions, the main autoionisation is to the ground configuration 
of the recombining ion (which is a single level), so the  $\sum_k A^{auto}_{sk}$ is normally 
replaced by $A^{auto}_s$, the total autoionisation rate into the 
He-like ground-level continua.
The ratio of A values $A_{sf} / A^{tot}_s$ on the right of the equation is basically 
a branching ratio. This is normally small, meaning that the majority of
inner-shell excitations decay by autoionisation, rather than by emission
of the satellite line.

$\beta \, N_{\rm Li-like}$ is the population of the lower level which 
populates the  level $s$ by collisional excitation. The value of 
$\beta$ is obtained by solving the level population for the Li-like ion.
At coronal densities, the population of the Li-like ions is all in the ground state,
so $\beta=1$ for the 1s 2s 2p, and $\beta=0$ for the 1s 2p$^2$. 
At increasing densities, $\beta$ varies but can easily be calculated.

The ratio of the satellite line produced by  inner-shell excitation
to the resonance line $w$ in the He-like ion is an excellent diagnostic 
for measuring  the relative population of the Li-like vs. the He-like ion, i.e. 
to assess if departures from ionisation equilibrium are present. 
In fact, since most of the population of the He-like ion $ N_{\rm He-like}$ is in the ground 
state $g$, the intensity of the resonance line $w$ can be approximated with 

\beq
I_w = N_u A_{ug} \simeq  N_{\rm He-like} N_{\rm e} C^e_{gu}
\eeq
where $ C^e_{gu}$ is the excitation rate from the ground state.
Therefore, the  ratio of the satellite line with the resonance is approximately

\beq
{I^{inner}_{sf} \over I_w} = {\beta N_{\rm Li-like} \over  N_{\rm He-like} } {C^e_{is} \over  C^e_{gu}} {A_{sf} \over A^{tot}_s}
\eeq
which  is independent of electron density.
We recall that the electron collisional excitation rate 
$C^e_{ij} \sim  T_{\rm e}^{-1/2} {\Upsilon_{i,j}(T_{\rm e}) \over \omega_i} \exp \left( {- \Delta E_{i,j} \over kT_{\rm e}} \right)$,
so the ratio does not  depend much on 
electron temperature, because  the excitation energies $\Delta E_{i,j}$ of the two transitions are very similar,
so also the exponential factors  are very similar. 
In summary, the ratio of the satellite line with the resonance line depends directly on the 
relative population of the Li-like vs. the He-like ion, aside from some factors 
which depend on the atomic data. 
In the original study by \cite{gabriel:1972}, the actual collision strengths were approximated with 
effective oscillator strengths.

The above ratio is usually small, because the branching ratio $A_{sf} / A^{tot}_s$ is small,
and because  the $N_{\rm Li-like}$ abundance is normally  lower than 
 $N_{\rm He-like}$. 
Finally, we note that a correction factor to the resonance line intensity
needs to be included. This factor should take into account the increase in the 
line intensity due to $n>3$ satellite lines formed by dielectronic recombination
(see below). 
Once the  ratio of the satellite line with the resonance is measured, 
the relative population of the Li-like vs. the He-like ion can be obtained,
and compared to the values predicted by assuming ionisation equilibrium.
In this way, one can assess if the plasma is ionising or recombining.

\subsection{ He-like satellite lines formed by dielectronic capture}

As  mentioned earlier, the dielectronic capture of a free electron 
colliding with the He-like ion can produce an autoionising state $s$ of the 
Li-like ion, which can then produce a satellite line. 
The intensity of the satellite line, decay to the final level $f$ is  
\beq
I^{\rm dr}_{sf} = N_s A_{sf}
\eeq
where as before $A_{sf}$  the transition probability by spontaneous emission from level $s$ to the final level $f$,
and  $N_s$ is the population of the autoionising state, which is 
 determined by the 
balance between the dielectronic capture (with rate $C^{\rm dc}$),
autoionisation and radiative decay to all possible lower levels:

\beq
N_{\rm He-like}  N_{\rm e} C^{\rm dc} = N_s  \left( \sum_k A^{auto}_{sk} + \sum_{f<s} A_{sf} \right)
\eeq
where we have seen that the sum of autoionising rates in this case reduces to a single
term $A^{auto}_s$, the total autoionisation rate into the He-like ground-level continua.
The intensity of the satellite line can therefore be written as 
\beq
I^{\rm dr}_{sf} =  N_{\rm He-like}  N_{\rm e} C^{\rm dc}  {A_{sf} \over A^{auto}_s + \sum_{f<s} A_{sf} } \,.
\eeq
We have seen in Sect.~\ref{sec:charge_state} that, by applying the Saha equation for thermodynamic equilibrium
we obtain a direct relation between the rates of the two inverse processes
of  dielectronic capture and autoionisation (Eq.~\ref{Eq:Cap_auto}), which in our case is

\begin{equation}
	C^{\rm dc} =   {h^3 \over (2\pi \, m \, kT_{\rm e})^{3/2} } {\omega_s \over  \omega_1} 
 \exp \left( - { E_s - E_1 \over k T_{\rm e} } \right)  	A^{\rm auto}_{s}
\end{equation}
where $\omega_s, \omega_1$ are the statistical weights of the level $s$ and the ground 
state of the He-like ion, and $E_s - E_1$ is the energy difference between the two 
states.

The intensity of the  satellite line formed by dielectronic recombination can therefore
be written as
\beq
I^{\rm dr}_{sf} =  3.3\times 10^{-24} N_{\rm He-like}  N_{\rm e}  {I_H \over (kT_{\rm e})^{3/2} } {\omega_s \over  \omega_1} 
 \exp \left( - { E_s - E_1 \over k T_{\rm e} } \right)   {A_{sf} A^{auto}_s \over A^{auto}_s + \sum_{f<s} A_{sf} } 
\eeq
where $I_H$ is the ionisation energy of neutral hydrogen (13.6 eV).

The ratio of the intensities of the  satellite line formed by dielectronic recombination with the 
parent line is therefore 
\beq
{I^{\rm dr}_{sf} \over I_w}  \simeq  3.3\times 10^{-24} {I_H \over (kT_{\rm e})^{3/2} C^e_{gu}} {\omega_s \over  \omega_1} 
 \exp \left( - { E_s - E_1 \over kT_{\rm e} } \right)   {A_{sf} A^{auto}_s \over A^{auto}_s + \sum_{f<s} A_{sf} }
\eeq
i.e. is independent of the population of the ions and the electron density. 
We recall the the  electron collisional excitation rate $C^e_{gu}$ depends on the electron temperature,
so in effect the above ratio depends only on $T_{\rm e}$, aside from the 
transition  probabilities.
 In other words, the ratio is an excellent temperature diagnostic.
It depends on  $T_{\rm e}$ as 

\beq
{I^{\rm dr}_{sf} \over I_w} \sim T_{\rm e}^{-1/2}  \exp \left( {E_0 - E_s + E_1 \over kT_{\rm e} } \right)
\eeq
where $E_0$ is the energy of the resonance transition.
In normal coronal conditions, $E_0 - E_s + E_1 <  kT_{\rm e}$ so the exponential factor is 
a slowly varying function of $T_{\rm e}$, so effectively the intensities of the 
satellite lines increase, with respect to the resonance line, when the electron 
temperature is lower. Indeed, observations show that during the gradual phase
of solar flares, when the temperature decreases, the satellite lines become prominent.

When considering ions along the isoelectronic sequence,  since $A^{auto}_s $ does not depend much on 
$Z$, the main dependence with $Z$ comes from the transition probability $A_{sf}$, which 
scales as $Z^4$. Therefore, while the intensities of the satellite lines of say oxygen 
are weak, they become significant in heavier elements like iron. 
Indeed the best measurements of the satellite lines are from iron.
The literature usually follows the notation from  \cite{gabriel:1972}, which is noted here in 
Table~\ref{tab:he-like_satellites}.

\begin{table}[!htb]
\caption[List of the main He-like satellite lines]{List of the main He-like satellite lines,
following the notation of \cite{gabriel:1972}.}
\centering
\begin{tabular}{llllllll}
\toprule
Key & Transition & \\

a & 1s$^2$ 2p $^2$P$_{3/2}$ -- 1s 2p$^2$  $^2$P$_{3/2}$ & \\
b & 1s$^2$ 2p $^2$P$_{1/2}$ -- 1s 2p$^2$  $^2$P$_{3/2}$ & \\
c & 1s$^2$ 2p $^2$P$_{3/2}$ -- 1s 2p$^2$  $^2$P$_{1/2}$ & \\
d & 1s$^2$ 2p $^2$P$_{1/2}$ -- 1s 2p$^2$  $^2$P$_{1/2}$ & \\
\\
e & 1s$^2$ 2p $^2$P$_{3/2}$ -- 1s 2p$^2$  $^4$P$_{5/2}$ & \\
f & 1s$^2$ 2p $^2$P$_{3/2}$ -- 1s 2p$^2$  $^4$P$_{3/2}$ & \\
g & 1s$^2$ 2p $^2$P$_{1/2}$ -- 1s 2p$^2$  $^4$P$_{3/2}$ & \\
h & 1s$^2$ 2p $^2$P$_{3/2}$ -- 1s 2p$^2$  $^4$P$_{1/2}$ & \\
i & 1s$^2$ 2p $^2$P$_{1/2}$ -- 1s 2p$^2$  $^4$P$_{1/2}$ & \\
\\
j & 1s$^2$ 2p $^2$P$_{3/2}$ -- 1s 2p$^2$  $^2$D$_{5/2}$ & \\
k & 1s$^2$ 2p $^2$P$_{1/2}$ -- 1s 2p$^2$  $^2$D$_{3/2}$ & \\
l & 1s$^2$ 2p $^2$P$_{3/2}$ -- 1s 2p$^2$  $^2$D$_{3/2}$ & \\
\\
m & 1s$^2$ 2p $^2$P$_{3/2}$ -- 1s 2p$^2$  $^2$S$_{1/2}$ & \\
n & 1s$^2$ 2p $^2$P$_{1/2}$ -- 1s 2p$^2$  $^2$S$_{1/2}$ & \\
\\
o & 1s$^2$ 2p $^2$P$_{3/2}$ -- 1s 2s$^2$  $^2$S$_{1/2}$ & \\
p & 1s$^2$ 2p $^2$P$_{1/2}$ -- 1s 2s$^2$  $^2$S$_{1/2}$ & \\
\\
q & 1s$^2$ 2s $^2$S$_{1/2}$ -- 1s 2p 2s ($^1$P) $^2$P$_{3/2}$ & inner-shell \\ 
r & 1s$^2$ 2s $^2$S$_{1/2}$ -- 1s 2p 2s ($^1$P) $^2$P$_{1/2}$ & inner-shell \\ 

s & 1s$^2$ 2s $^2$S$_{1/2}$ -- 1s 2p 2s ($^3$P) $^2$P$_{3/2}$ & inner-shell \\ 
t & 1s$^2$ 2s $^2$S$_{1/2}$ -- 1s 2p 2s ($^3$P) $^2$P$_{1/2}$ & inner-shell \\ 

u & 1s$^2$ 2s $^2$S$_{1/2}$ -- 1s 2p 2s $^4$P$_{3/2}$ & inner-shell \\ 
v & 1s$^2$ 2s $^2$S$_{1/2}$ -- 1s 2p 2s $^4$P$_{1/2}$ & inner-shell \\ 
\\
\bottomrule 
\end{tabular}
\label{tab:he-like_satellites}
\end{table}

\subsection{Satellite lines of other sequences}

The process of formation of satellite lines is a general one,
and  satellite lines are present also for ions of 
other sequences. We refer to the above-cited reviews for more details,
but would like to mention two important sets of satellites. 
 The most important ones  are those of H-like ions.

\begin{figure}[!htb]
 \centerline{\includegraphics[width=0.8\textwidth,angle=0]{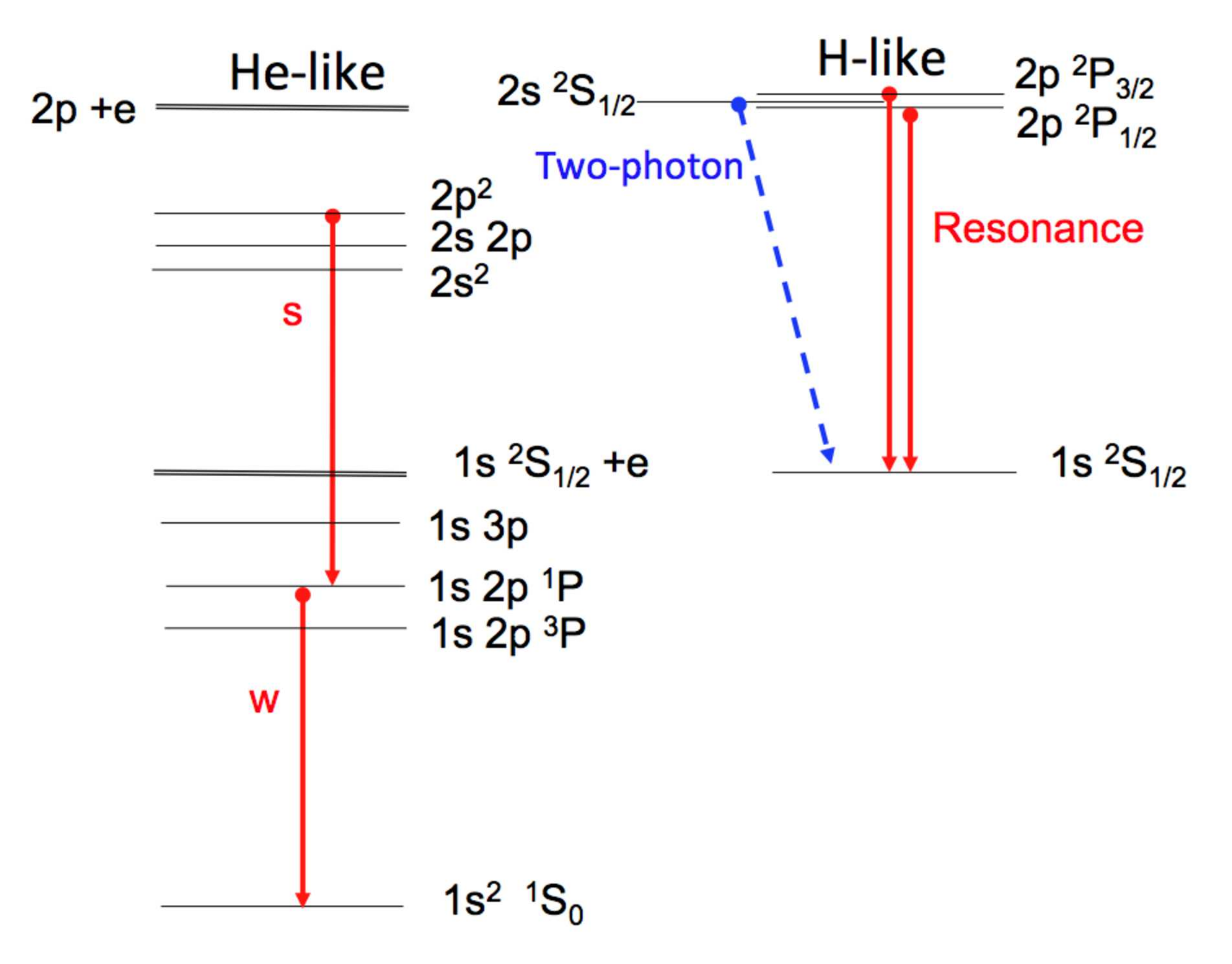}} 
 \caption{A sketch of  the process of formation of
 satellite lines of the H-like $n=2$ resonance line.}
 \label{fig:h-like_satellite}
\end{figure}

H-like satellite lines have some similarities and differences with the He-like ones.
The main difference is that the only process of formation is dielectronic capture. 
The main satellites have, as in the He-like case, the $n=2$ 
parent line, the L-$\alpha$ doublet, the decay of the 2p $^2$P$_{3/2,1/2}$
to the ground state (see Fig.~\ref{fig:h-like_satellite}).

The 2p $^2$P$_{3/2,1/2}$ and 2s $^2$S$_{1/2}$ are mainly collisionally excited from 
the ground state. The 2s $^2$S$_{1/2}$ decays mainly via a two-photon emission
at low densities (even for solar flare conditions), and collisional transfer between 
the  2s $^2$S$_{1/2}$ and 2p $^2$P is negligible. 
For a list of the main transitions and 
description of the calculations see e.g. \cite{boiko_etal:1977_h-like,boiko_etal:1978_h-like}.

Satellite lines in ions of 
other sequences are also important at some wavelengths. Most notably, the region 
between 1.85 and 1.9~\AA\ (cf. Figure~\ref{fig:smm_bcs_fe_25})
 is reach in satellite lines from  \ion{Fe}{xx},  \ion{Fe}{xxi},
 \ion{Fe}{xxii}, and  \ion{Fe}{xxiii}, 
 as discussed e.g. in \cite{doschek_etal:1981_solflex}, where a detailed list
of transitions is provided.
The most notable transition is the  \ion{Fe}{xxiii} inner-shell $\beta$ line,
which can be used, in conjunction with the the He-like \ion{Fe}{xxv},
to assess for departures from ionisation equilibrium
 \citep{gabriel:1972b}.

\subsection{Measuring departures from Maxwellian distributions}

\begin{figure}[!htb]
 \centerline{\includegraphics[width=0.6\textwidth,angle=90]{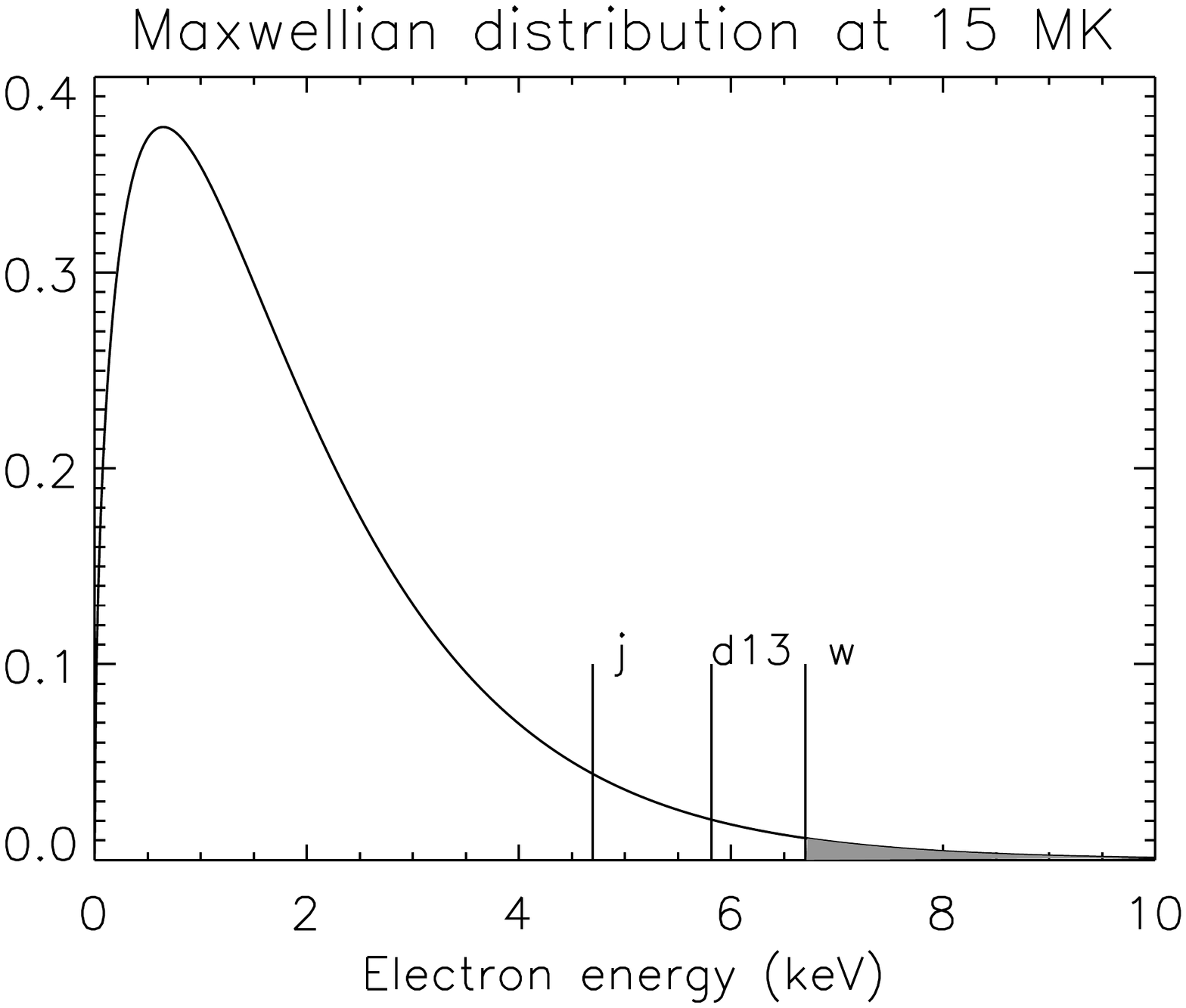}}
 \caption{A Maxwellian distribution of electrons at 15 MK, with the 
energies of the satellite lines $j$ and $d13$, and 
the threshold energy to excite the resonance line $w$. 
}
 \label{fig:gabriel_phillips_1979}
\end{figure}

The fact that the satellite lines can only be excited by free electrons 
having specific energies means that, if one could measure the 
intensities of several satellites, information on the energy 
distribution of the free electrons could be obtained. 
A method to  assess if  
 non-Maxwellian distributions exist during solar flares was suggested by 
\cite{gabriel_phillips:1979}. It relies on the measurement of 
the $j$ and $n=3$ complex of satellites in the He-like Fe, of which the transition 
$d13$ is the main one.

The  $j$ and $n=3$ complex are produced by electrons having two specific energies, 
shown in  Fig.~\ref{fig:gabriel_phillips_1979} with two vertical lines, superimposed
on a Maxwellian distribution of a plasma with 15 MK, a temperature typical 
of moderately-large solar flares.

On the other hand, any electron with energy above the resonance $w$ threshold 
can excite the line. Those energies are indicated with the grey area under the 
Maxwellian curve. 
If the electron distribution is Maxwellian, then the temperatures 
obtained from the observed $j/w$ and $d13/w$ and the above prescriptions 
should be consistent, assuming that 
all the atomic data are accurate. On the other hand, if a tail 
of high-energy electrons is present 
(as it is known to be present for large flares from hard X-ray 
bremsstrahlung observations), then the resonance line would be greatly 
enhanced, as all the electrons above the threshold would increase the 
collisional population of the upper level of the resonance line.
The temperatures obtained (with the Maxwellian assumption) 
from the observed $j/w$ and $d13/w$ should therefore be different. 

The method does not allow one  to infer the complete electron distribution, but just to 
confirm if the distribution is Maxwellian or not. 
Several measurements of satellite 
lines that can only be excited at progressively higher energies would 
be needed to describe the electron distribution.

The d13 line complex has been resolved in  laboratory plasmas, and  the effects of 
non-thermal electrons on the spectra have been detected 
(see, e.g., \citealt{bartiromo_etal:1983,lee_etal:1985,bartiromo_etal:1985}).
Satellite lines have also been observed in solar spectra (see next Section), 
and some applications of the present diagnostics have been published. 
They are summarised later in Section~\ref{sec:non-eq} where we discuss non-equilibrium
effects.

\subsection{Observations}

Among the earliest observations, there are those reported by 
\cite{neupert_swartz:1970, neupert:1971_satellites} obtained  with the OSO-5, where 
 the satellite lines 
from the 1s 2s 2p configuration in the He-like Si, S, Ar, and Fe
ions were observed.
\cite{walker_rugge:1971} observed several satellites of 
the H- and He-like Mg, Al, Si and S ions, form the OV1-17 satellite.
\cite{acton_etal:1972} observed the satellites of the He-like O and Ne using 
a rocket,
while \cite{parkinson:1971,parkinson:1972_mg,parkinson_etal:1978} 
provided excellent spectra of He-like satellites of Ne, Mg and Si
obtained with  Skylark rockets and with OSO-8 observations of active regions 
and flares.
\cite{parkinson_etal:1978} also reported observations of the 
H-like \ion{Si}{xiv} with OSO-8.

\cite{doschek_etal:1971_ca,doschek_etal:1971_fe} presented 
an analysis of the observations of the He-like Ca and Fe
obtained with the NRL Bragg crystal spectrometer on-board OSO-6.
Further earlier observations were those of the He-like Fe obtained by the
Intercosmos-4 Satellite and the `Vertical-2' Rocket
\citep{grineva_etal:1973}, and of the H-like lines by the same
rocket and Intercosmos-4 and Intercosmos-7
\citep{aglitskii_etal:1978}.

Later, several  observations were obtained with the 
excellent NRL SOLFLEX instrument \citep[cf.][]{doschek_etal:1980,feldman_etal:1980,doschek_etal:1981_solflex}
and then from the crystals on-board the  Hinotori, Yohkoh and SMM satellites.

\begin{figure}[!htb]
 \centerline{\includegraphics[width=0.6\textwidth,angle=90]{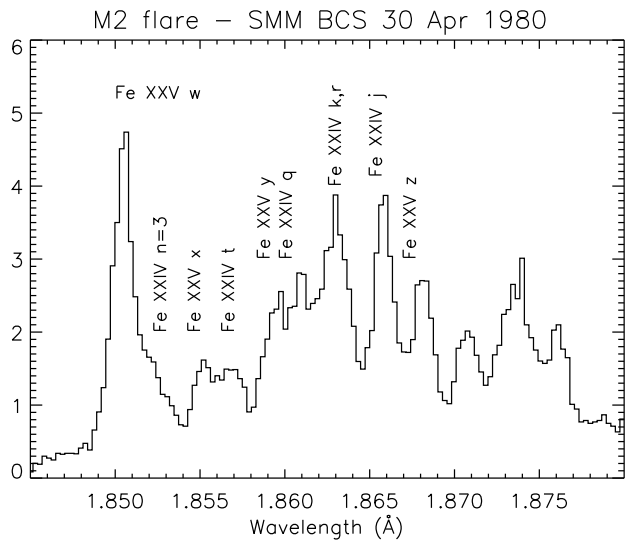}}
 \caption{An SMM BCS spectrum  during the peak phase of a flare, containing 
 the He-like lines from iron and many satellites, some of 
which are labelled.}
 \label{fig:smm_bcs_fe_25}
\end{figure}

\begin{figure}[!htb]
 \centerline{\includegraphics[width=0.6\textwidth,angle=90]{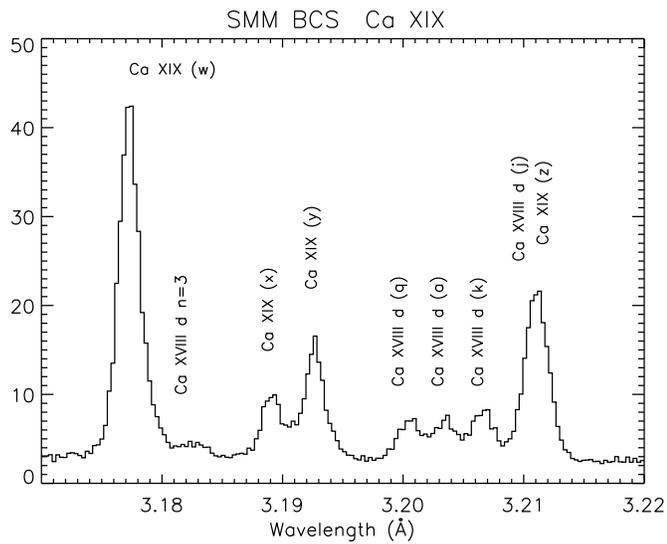}}
 \caption{ An SMM BCS spectrum  during the peak phase of a flare, 
containing  the Ca He-like lines and  associated satellites.}
 \label{fig:smm_bcs_ca_19}
\end{figure}

\begin{figure}[!htb]
 \centerline{\includegraphics[width=0.8\textwidth,angle=0]{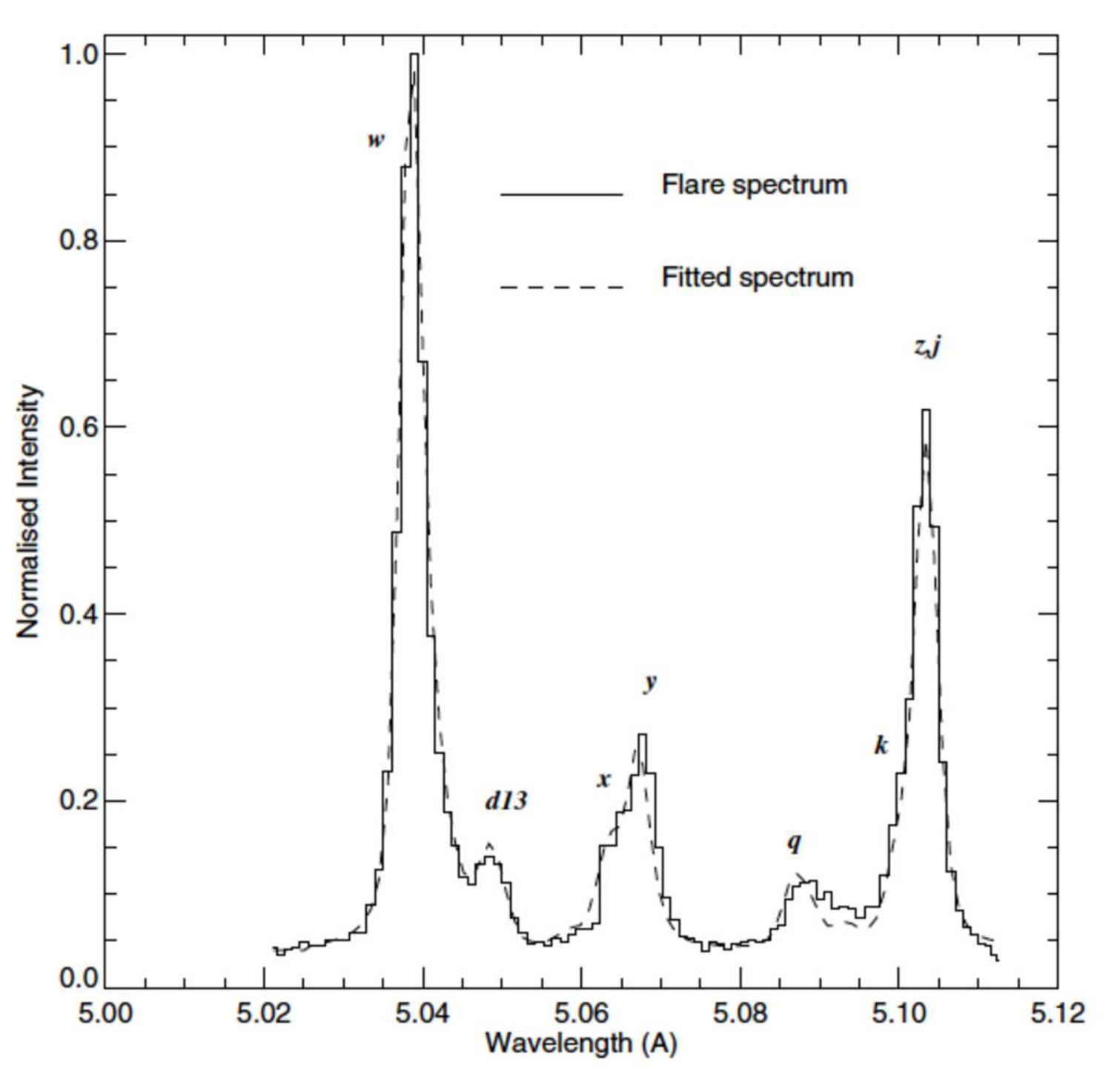}} 
 \caption{ A Yohkoh BCS spectrum of the He-like lines from S and associated satellites.
The dashed line is a theoretical spectrum. Figure from \cite{phillips_etal:2004}.}
 \label{fig:s_15_yohkoh_bcs}
\end{figure}

 Fig.~\ref{fig:smm_bcs_fe_25} shows as an example a spectrum obtained by SMM BCS,
 the He-like lines from iron,
while Fig.~\ref{fig:smm_bcs_ca_19} shows the corresponding spectrum of the 
He-like Ca lines.
In addition, Fig.~\ref{fig:s_15_yohkoh_bcs} shows  a 
Yohkoh BCS spectrum of the sulphur He-like lines  and associated satellites
during a flare.

\clearpage

\section{Atomic calculations and  data}
\label{sec:atomic}

This review is primarily concerned with diagnostics of 
electron densities and temperatures from spectral lines of the same ion.
In this case, the primary atomic data are the cross-sections for 
electron-ion collisions and the radiative data (A-values), 
with their uncertainties. 
We therefore mainly review the latest work on these atomic data,
and where references and data can be obtained, with emphasis on the 
CHIANTI database. 
We also  introduce to the unfamiliar reader a few basic aspects about the 
relevant atomic calculations for such data. 
Finally, we  also briefly review  
ionization and recombination data, 
line identification and benchmarking  studies.

\subsection{The CHIANTI atomic package}

The CHIANTI package consists of a critically evaluated set of atomic
data for a large number of ions of astrophysical interest.
It also includes a quantity  of ancillary data and a 
suite of Interactive Data Language (IDL) programs to calculate
optically thin synthetic spectra (including continuum emission) 
and to perform general spectral analysis and plasma diagnostics,
from the X-rays to the infrared region of the spectrum.

The CHIANTI database started in 1995 as a collaboration  between  
K.\ P.\ Dere (NRL, USA), H.\ E.\ Mason (University of Cambridge UK), 
and B.\ Monsignori Fossi (Arcetri Observatory, Italy). 
Sadly, the latter died prematurely in 1996. M.\ Landini (University of Florence, Italy) 
then  worked with the CHIANTI team for several years.  
Currently, the project involves 
K.\ P.\ Dere and P.\ R.\ Young (George Mason University, USA), E.\ Landi 
(University of Michigan, USA), and G.\ Del Zanna and H.\ E.\ Mason 
(University of Cambridge, UK).

The database relies almost entirely on the availability of 
published atomic data. Each version release is accompanied by
a publication that discusses the new atomic data and any other 
new features of CHIANTI.
References to the original calculations, line identifications,
wavelength measurements  and 
benchmark work are provided in the published papers and 
throughout the database. It is very important to acknowledge the hard
work of the atomic data providers.

The  CHIANTI package is freely available  at 
the  web page:
 \url{http://www.chiantidatabase.org/}
and within SolarSoft,  a programming and data analysis
    environment written in IDL  for the solar physics community.
There is also a Python interface to some of the CHIANTI routines
(CHIANTIPy: \url{http://chiantipy.sourceforge.net/}).

The first version  of the CHIANTI database was released in 1996
and is described in \cite{dere_etal:97}.
v.2 is described in \cite{landi_etal:99}. In v.3 \citep{dere_etal:01}
the database was extended to wavelengths shorter than
50~\AA, while v.4 \citep{young_etal:03} included  
proton excitation data, photoexcitation, and new continuum routines.
v.5 \citep{landi_etal_v5:06} included new datasets, while v.6 
\citep{dere_etal:09_chianti_v6}
included a complete database of ionization and recombination coefficients.
v.7 \citep{landi_etal:12_chianti_v7}  improved ions  important for the 
EUV and the UV, while
v.7.1 \citep{landi_etal:12_chianti_v7.1} 
included new DW atomic data for the soft X-ray lines
 and new identifications from \cite{delzanna:12_sxr1}.
V.8 \citep{delzanna_chianti_v8} included  new atomic data
for several isoelectronic sequences and for a few coronal 
ions from iron and nickel.

As with any atomic data package, CHIANTI has been 
developed to suit  some specific applications in solar physics and
astrophysics, although with time it has become more generally used and 
is often used as a reference atomic database rather than a modelling
code. CHIANTI basic data are in fact included in many 
 other atomic databases  and modelling packages, for example:
\begin{itemize}
\item
VAMDC (\url{http://portal.vamdc.eu}),
\item
CLOUDY (\url{http://www.nublado.org}),
\item
MOCASSIN (\url{http://mocassin.nebulousresearch.org}),
\item
XSTAR (\url{http://heasarc.gsfc.nasa.gov/docs/software/xstar/xstar.html}), 
\item
ATOMDB (\url{http://www.atomdb.org}), 
\item
PINTofALE (\url{http://hea-www.harvard.edu/PINTofALE/}), 
\item
XSPEC (\url{http://heasarc.gsfc.nasa.gov/docs/xanadu/xspec/index.html}).
\end{itemize}

The main assumptions within the CHIANTI codes are that the 
plasma is optically thin and  in thermal equilibrium. 
The plasma ionization is dominated by collisions
(i.e. no photo-ionization is included).
Tabulations of ion populations in equilibrium are provided, 
in the low-density limit.

Line emissivities are reliable only within given 
 temperature and density ranges. 
The maximum densities for  an ion model 
depend on which excitations/de-excitations are included.
Normally, all those from the ground configuration and the main 
metastable levels are included, 
although for most ions 
the complete set of excitation rates has been included 
in v.8 \citep{delzanna_chianti_v8}. This allows the use of the 
CHIANTI data for higher densities.
The  electron collision strengths were, until v.8, 
interpolated over a fixed temperature grid in the scaled
domain as formulated by \cite{burgess_tully:92}.
However, this occasionally reduced the accuracy at lower temperatures,
typical of photoionised plasma. 
The rates for the new ions in v.8 were normally released as 
published, without interpolation.

Electrons  and protons have  Maxwellian distribution 
functions. Indeed CHIANTI data include 
 Maxwellian-averaged electron and proton collision strengths.
However, it is possible to study the effects of 
 particle distributions that are linear combinations 
of Maxwellians of different temperatures.
An implementation of the CHIANTI programs that 
calculates line emissivities assuming {\em kappa ($k$)} distributions
has recently been made available \citep{dzifcakova_etal:2015},
based on v.7.1.

Finally, various other databases of atomic data exist, e.g.
 ADAS (\url{http://open.adas.ac.uk}, primarily for fusion work) and 
ATOMDB (\url{http://www.atomdb.org}, primarily for high energies).


\subsection{Atomic structure calculations }

We refer the reader to standard textbooks of atomic spectroscopy such as 
\cite{condon_shortley:1935, grant_book:2006, landi_book:2014}, 
the various articles cited below, and the  Springer handbook of atomic and
molecular physics \citep{springer_handbook:2006}, in particular 
Chapter 21 on atomic structure and 
multi-configuration Hartree-Fock theories by Charlotte Froese Fischer,
and  Chapter 22 by Ian Grant on  relativistic atomic structure.
For  recent reviews of multiconfiguration methods for complex atoms
see \cite{bieron_etal:2015,froese_fischer:2016_review}.

The atomic structure of an ion is obtained by solving the 
time-independent  Schr\"odinger equation: 
\begin{equation}
H\Psi_i = E_i \Psi_i
\end{equation} 
where  $\Psi_i$ are the wavefunctions of the system, $E_i$ are the 
eigenvalues, and $H$ is the Hamiltonian.

For lighter elements of astrophysical importance, the approach that is 
often  adopted is to describe the system with a 
non-relativistic  Hamiltonian, and then add relativistic 
corrections using nonrelativistic wavefunctions and
the Breit-Pauli approximation. 
For heavier elements,  a fully relativistic approach is 
usually needed. This involves solving the Dirac equation and 
taking into account the 
Breit interaction  \citep[see, e.g.][]{grant_book:2006}.

The non-relativistic  Hamiltonian describing 
an ion  with $N$ electrons 
and a central nucleus having charge number $Z$ can be written in the form

\beq
 {H} = \sum_{i=1}^N \left({p_i^2 \over 2 \, m} - {Z e^2 \over r_i}
              \right) + \sum^{N-1}_{i < j} {e^2 \over r_{ij}} \;,
\eeq             
where $\vec r_i$ is the position vector of the $i$-th electron
(relative to the nucleus), $\vec p_i$ is its momentum, and 
$r_{ij}$ is the  distance between the $i$ and $j$ electrons:
$ r_{ij} = r_{ji} = \left| \vec r_i - \vec r_j \right| $.
The  terms in the Hamiltonian are the total 
kinetic and potential energy of the 
electrons in the  field of the nucleus, and the 
repulsive Coulomb energy  between  the electrons.
This equation neglects the spin of the electron.

It can be shown that this Hamiltonian commutes with the total
angular momentum $L$ and the total spin $S$ of the electrons.
Additionally, it commutes with the parity operator. 
This allows the usual description of the wavefunction in terms of
the associated good quantum numbers LSp.

For complex highly-ionised atoms, it is common to assume that 
the electron  repulsions 
are  small perturbations relative to the much stronger nuclear 
central potential, and to simplify the equation by assuming that
each electron experiences a central potential, caused by 
 the electrostatic interaction  with the nucleus, screened by  the other electrons.
In this way, the zero-order Hamiltonian is first solved:

\beq
{ H}_0 = \sum_{i=1}^N \left( {p_i^2 \over 2 \, m} + V_{\rm c}(r_i)
               \right) 
\eeq
\noindent
where $V_{\rm c}(r_i)$ is the potential energy of each electron in 
the central field.  
The first-order Hamiltonian

\beq
 { H}_1 = \sum_{i=1}^N \left( -V_{\rm c}(r_i) -{Z e^2 \over r_i}
               \right) + \sum_{i<j} {e^2 \over r_{ij}} \;\; .  
\eeq
\noindent
is then considered as a perturbation. 
There are several approximate numerical methods to find a solution
for the central potential and to evaluate 
 $V_{\rm c}(r)$. For example,  the 
Thomas-Fermi statistical method  and the  Hartree-Fock 
self-consistent method using the variational principle.

 When configuration interaction (CI) is considered, one atomic state is 
described by a linear combination of eigenfunctions of 
different   configurations (but with the same parity, as ${H}_1$ 
commutes with the parity operator).
 Configuration interaction becomes increasingly important for 
excited states whenever the difference in energy between the states 
becomes small.
The Hamiltonian ${H}_1$ commutes with the operators $L^2$ and 
$S^2$, so the total $L$ and $S$ values are still good quantum numbers
representing the state.

There are several relativistic corrections
\citep[see, e.g.][]{bethe_salpeter:1957}, some of which arise directly from the 
solution of the Dirac equation. They are normally grouped into 
two classes, one-body and two-body operators \citep[see,e.g.][]{badnell:1997}.
The non-fine-structure operators commute 
with the operators associated with
the total angular momentum $L$ and the total spin $S$ of the electrons
(plus their azimuthal components) 
so they do not break the $LS$ coupling
and  are often included as an additional term in the Hamiltonian $H_0$.
On the other hand, the fine-structure operators only commute 
with the operators associated with
the total angular momentum $J$ and its azimuthal component. 
The nuclear spin-orbit interaction, i.e. 
the interaction between the orbital angular momentum in the field of the nucleus
 and the intrinsic spin of the electron, is the operator
causing the main splitting of the $J$-resolved levels.
Normally, atomic structure calculations have to be carried out in 
intermediate coupling, where the atomic states, 
characterized  by the quantum numbers $J$ and $M$ 
are expressed as  linear combinations of the form (in Dirac notation)

\beq
|\alpha J M > = \sum_{LS} {\cal C}_{LS} \,
   |\alpha L S J M > \;\; , 
\eeq
where the sum is extended to all the $L$ and $S$ values that are compatible
with the configuration $\alpha$ and with the value of  $J$.
The ${\cal C}_{LS}$ are the coefficients of the expansion,
obtained by diagonalising the Hamiltonian of the perturbations.

Once the  Schr\"odinger equation is solved and the eigenfunctions
and eigenvalues are calculated, it is relatively straightforward 
to calculate the radiative data such as line strengths and A values
for the different types of transitions.

The first requirement in any structure calculation is a good representation
of the target, i.e. accurate wavefunctions for the target ion. 
Such representation  requires the
inclusion of Configuration Interaction (CI) with a large 
set of configurations
\citep[see, e.g.][]{layzer:1954}.  

Often, significant discrepancies between the experimental energies
and those calculated ab-initio have been found. 
Various semi-empirical corrections based on the
observed energies have therefore  often been applied to the calculations.
This  improves  the oscillator strengths.
One example is the term energy correction (TEC) to 
the Hamiltonian matrix, introduced by \cite{zeippen_etal:77}
and \cite{nussbaumer_storey:78} within the 
{\sc superstructure} program to improve the 
description of the spin-orbit mixing between two levels,
which requires their initial term separation to be accurate.
Semi-empirical corrections are often applied 
within  the other atomic structure  programs, see e.g. 
B.\ C.\ Fawcett work using the Cowan's Hartree--Fock code
(see the \citealt{fawcett:1990_rev} review).

Traditionally, there have been  two sets of atomic structure
calculations that have been carried out in the literature.
The first one is to provide the wavefunctions to be used for 
the scattering calculation.
 For such calculations, the emphasis is 
to include in the CC expansion all the levels that are deemed
important (e.g. by cascading or by adding resonances in the 
excitation cross sections) for the spectroscopically-relevant 
levels.

The second set of calculations normally include a large set
of CI and are focused on obtaining the most accurate energies and
A-values for the lower levels in an ion, especially for the 
ground configuration and in general for the metastable levels.
In fact, such sets of A-values establish the population 
of the lower levels, so it is important to obtain accurate values.

\subsubsection{Codes}

Over the years, many codes have been developed 
 to calculate the atomic structure  and associated radiative data.
Among the most  widely used earlier codes  are Cowan's Hartree--Fock 
\citep{cowan:1967,cowan:1981},  the {\sc superstructure} program
\citep{eissner_etal:74}, 
the {\sc CIV3} program \citep{hibbert:1975}, and 
the HULLAC code \citep{bar-shalom_etal:1988}.

 {\sc Autostructure} \citep{badnell:11}, originally a development of
 the {\sc superstructure} program, has now become a  multi-purpose program
for the calculations of a  wide range of atomic processes. 
The code and relative information
are available at \url{http://www.apap-network.org/codes.html}. 
Ian Grant's general-purpose relativistic atomic structure program (GRASP)
 \citep{grant_etal:1980,dyall_etal:1989}, based on multiconfiguration
Dirac-Hartree-Fock theory, was  significantly modified
and extended \citep[GRASP2K, see ][]{grasp2k:2007,grasp2k:2013}. Among other features, 
the new version implements a bi-orthogonal transformation method that allows initial and
final states in a transition array to be optimized separately.
This often leads to an improvement in the accuracy of the rates.
The multi-configuration Hartree--Fock (MCHF) \citep{froese-fisher:1991}
code has also been improved and modified over the years. 
The most up-to-date version is called ATSP2K. 
As in GRASP2K, to calculate transition probabilities the orbitals of the 
initial and final state do not need to be orthogonal. 
The international collaboration on computational atomic structure
(COMPAS) has a useful description of various codes (especially GRASP2K and  ATSP2K) and
their applications. The codes can be downloaded from their website: 
 \url{ddwap.mah.se/tsjoek/compas/software.php}.
The Flexible Atomic Code (FAC)  \citep{gu:04_FAC} is  available 
at GitHub: \url{https://github.com/fnevgeny/fac/}.

\subsection{Uncertainties in atomic structure calculations}

The accuracy of a particular atomic structure calculation depends on several factors. 
One important factor  is the representation which is used 
for the target wavefunctions.
The target must take into account 
configuration interaction and allow for intermediate coupling for the 
higher stages of ionization \citep[cf.][]{mason:1975a}.

It is customary to assess the accuracy of the target wavefunctions 
by comparing the level energies with the experimental ones. 
This is a good zero-order approach, although there are cases 
when, despite relatively good agreement between theoretical and
experimental energies, significant problems are still present.

In the literature, the studies which focus on the atomic structure of an atom 
normally perform a series of calculations, where the size of the CI 
(that is the number of configurations) is increased
until some convergence is achieved for the (lower) levels of spectroscopic 
importance. It is also common to study the effects of which electrons 
are allowed to be promoted/excited. One recent example is given by 
\cite{gustafsson_etal:2017_atom}, where the 
 MCDHF method, as implemented in the GRASP2K program, was used to 
study correlation effects for  Mg-like iron. 

\begin{figure}[htbp]
\centerline{\includegraphics[width=0.6\textwidth, angle=90]{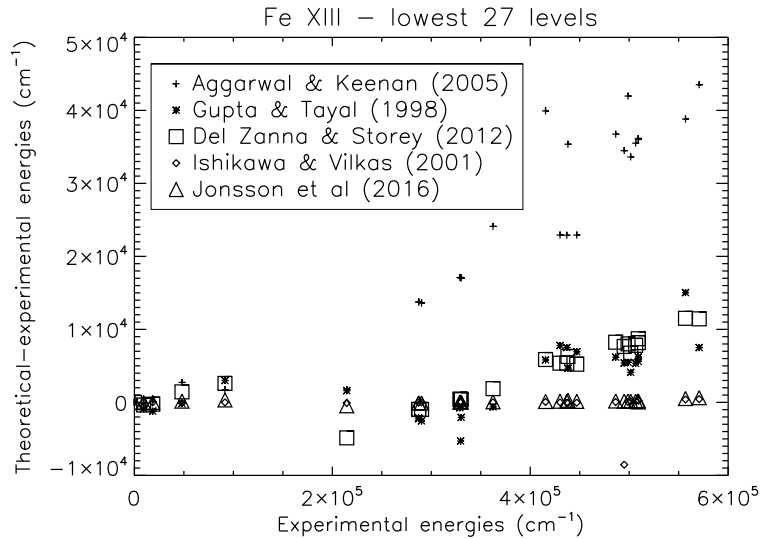}}
\caption{A comparison between experimental  and
 ab-initio energies for the lowest 27 levels
(3s$^2\,$3p$^2$, 3s\,3p$^3$, 3s$^2$\,3p\,3d)  in Si-like \ion{Fe}{xiii}
with different target wavefunctions (see text).
}
\label{fig:fe_13_comp_e}
\end{figure}

As an example of how different calculated energies can be,
 we consider here the ab-initio energies for the lowest 27 levels
(3s$^2\,$3p$^2$, 3s\,3p$^3$, 3s$^2$\,3p\,3d)
 as calculated by different authors  with different codes in recent years 
for Si-like iron (\ion{Fe}{xiii}). These are plotted in Fig.~\ref{fig:fe_13_comp_e},
relative to the experimental energies (all known) as available in CHIANTI v.8.
\cite{gupta_tayal:1998} used the CIV3 computer code and considered for the
CI expansions one- and two-electron excitations. 
\cite{aggarwal_keenan:2005} used the GRASP code to calculate the ab-initio
energies for the lowest 97 levels. 
\cite{delzanna_storey:12_fe_13} used the AUTOSTRUCTURE code and the 
Thomas-Fermi-Amaldi central potential, with scaling parameters, including 
in the CI expansion all the configurations up to $n$=4, giving rise to 
2186 fine-structure levels.
Vilkas and Ishikawa developed a 
multiconfiguration Dirac–Fock–Breit self-consistent field 
 plus state-specific multireference M{\o}ller–Plesset (MR-MP) perturbation procedure to
obtain very accurate level energies for open-shell systems.
They applied this procedure to Si-like ions as described in \cite{vilkas_ishikawa:2004}. 
Recently, \cite{jonsson_etal:2016_si-like} applied the GRASP2K code to 
carry out large-scale calculations for Si-like ions, including correlations 
up to $n$=8.
It is clear that the  GRASP2K and MR-MP calculations are outperforming the 
other codes, achieving spectroscopic accuracy, i.e. deviations from 
the experimental energies of a few hundred Kaysers. 
However we should note that the other calculations were mostly concerned in 
defining the wavefunctions for the scattering calculations. 
For this relatively simple ion, the AUTOSTRUCTURE energies
 are close to those from  CIV3, while larger deviations are clear for the 
 GRASP calculations. In turn, such deviations often affect the oscillator
strengths to the levels, and as a consequence the cross-sections 
for electron excitations, as pointed out by \cite{delzanna:11_fe_13} for this 
particular ion, and also as discussed below.

\begin{figure}[htbp]
\centerline{\includegraphics[width=0.7\textwidth, angle=0]{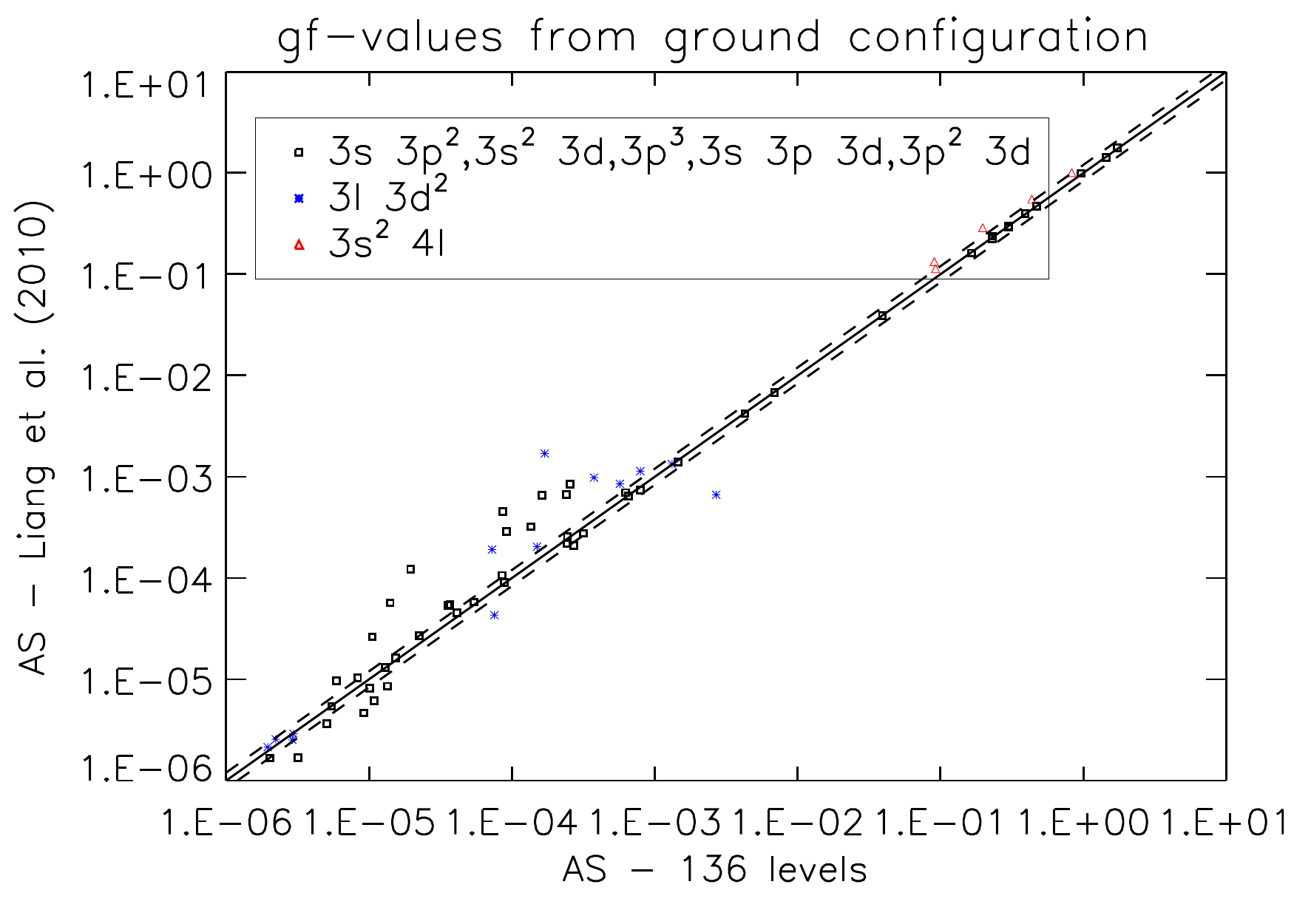}}
 \caption{Oscillator strengths ($gf$ values) for all transitions from the  
 ground configuration of \ion{Fe}{xiv} as calculated with a smaller CI 
 (136 levels) vs. those calculated  with a much larger CI (2985 levels) by \cite{liang_etal:10_fe_14}. 
The dashed lines indicate $\pm$20\%. Figure adapted from \cite{delzanna_etal:2015_fe_14}. }
\label{fig:fe_14_comp_gf}
\end{figure}

Another way to assess the accuracy of the target wavefunctions  
is to compare oscillator strengths ($gf$ values) for the dipole-allowed
transitions as computed with different sets of configurations 
or parameters. Similarly, A-values for the forbidden transitions are usually
compared. 
Typically, the $gf$ values for transitions to lower levels do not 
vary significantly, while it is always the transitions to the highest
levels in the calculations that are  more uncertain. They are 
usually also the weakest.
Fig.~\ref{fig:fe_14_comp_gf} shows as an  example the 
$gf$ values for transitions from the  ground configuration
of \ion{Fe}{xiv} as calculated with a set of configurations giving rise to 
136 fine-structure levels \citep{delzanna_etal:2015_fe_14}, and a much larger CI  
(2985 levels) as calculated by \cite{liang_etal:10_fe_14}. 
There are some clear discrepancies for transitions to the higher levels,
despite the fact that the same code ({\sc autostructure}) and approximations were used.

Generally, larger  CI calculations lead to more accurate 
oscillator strengths. However, for complex ions (e.g. the coronal iron ions)
a large calculation does not necessarily provide accurate values. 
For example, 
the  spin-orbit mixing among fine-structure levels of the same J and parity is 
very sensitive to the difference between energy level values, 
and small variations can lead to large differences in oscillator strengths. 
This can even occur for some of the  strongest transitions to low-lying levels,
 as  shown e.g. for  \ion{Fe}{xi} \citep{delzanna_etal:10_fe_11}
and \ion{Fe}{viii} \citep{delzanna_badnell:2014_fe_8}.

Any difference in the $gf$ values is directly reflected in
differences in the A values, which directly affect the line intensity.
One  way to assess the accuracy in the oscillator strengths (or A values)
is to compare the values as 
computed in the Babushkin (length) and  the Coulomb (velocity) gauges.
A recent example is from the \ion{Fe}{xiv}  GRASP2K calculations 
by \cite{ekman_etal:2017}, where the accuracy was estimated from 
\begin{equation}
dT = \frac{|A_{B}-A_{C}|}{\mbox{max}(A_{B},A_{C})},
\end{equation}
where $A_{B}$ and $A_{C}$ are the transition rates obtained 
in the Babushkin and Coulomb gauge, respectively.

\begin{figure}[htbp]
\centerline{\includegraphics*[width=0.5\textwidth, angle=0]{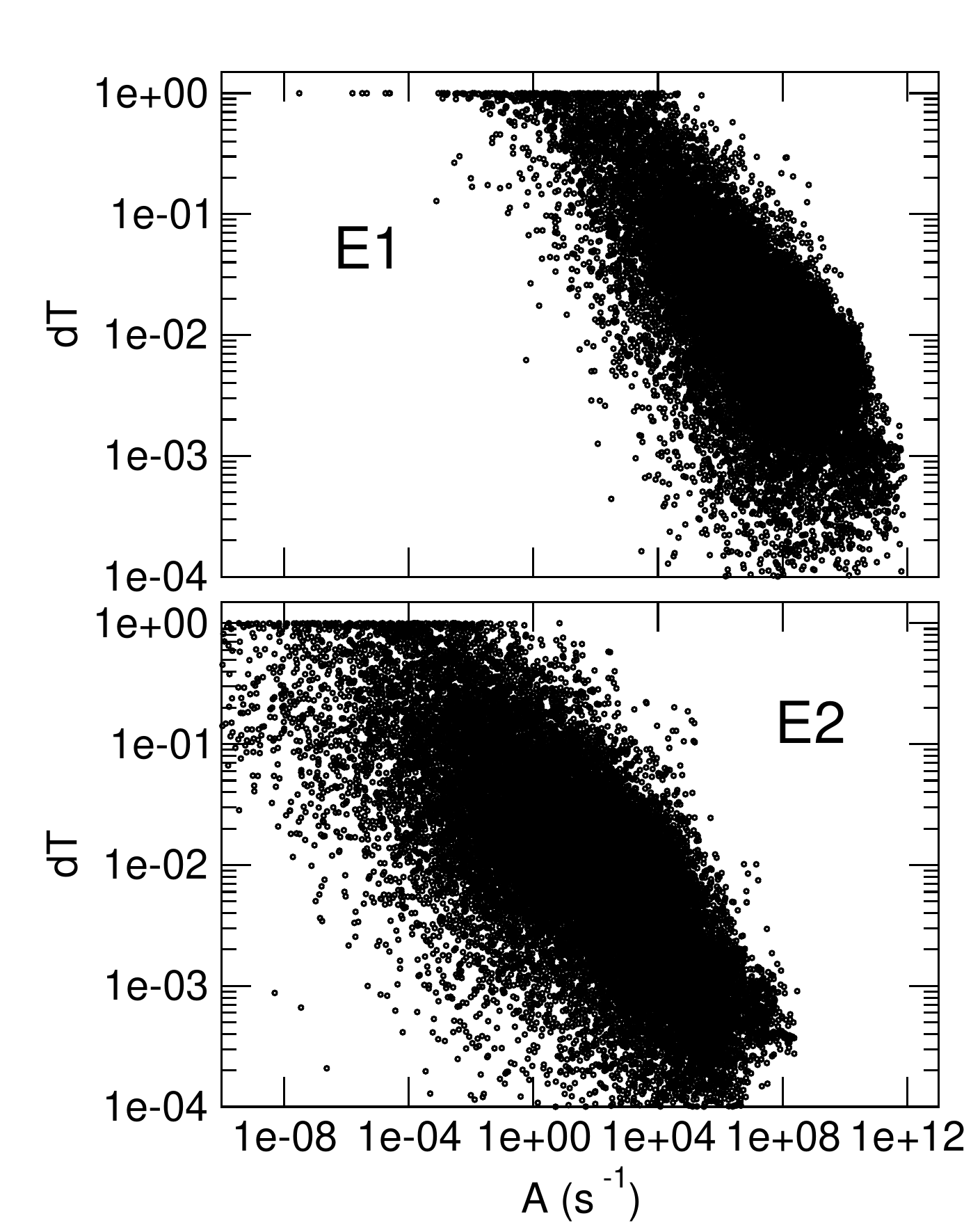}}
 \caption{Scatter plots of accuracy estimates $dT$ vs. A-value
  for  E1 and E2 transitions in \ion{Fe}{xiv}.
Image reproduced with permission from  \citep{ekman_etal:2017}, copyright by Elsevier.
}
\label{fig:fe_14_comp_gf2}
\end{figure}

\begin{figure}[!htb]
 \centerline{\includegraphics[width=0.7\textwidth,angle=0]{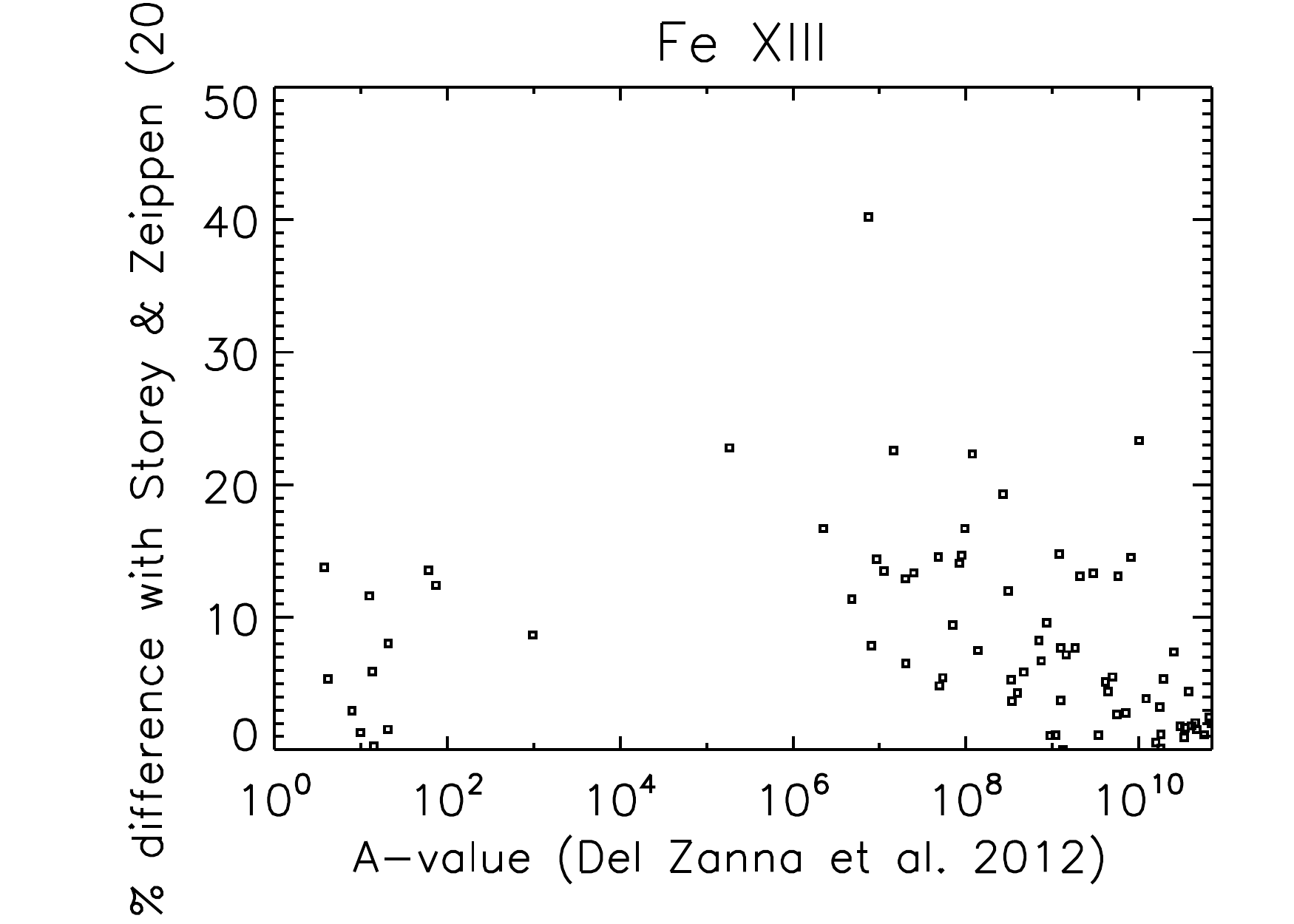}}
  \caption{Percentage difference in the A values of 
 \ion{Fe}{xiii} transitions within the energetically lowest 27 levels, 
as obtained with two different calculations.
}
 \label{fig:fe_13_a}
\end{figure}

One other way to assess the accuracy of the A-values is to compare 
the results of different calculations, where e.g. the size of the CI 
is different, but the same methods and codes are used. 
 Fig.~\ref{fig:fe_13_a} shows as an example the percentage difference in the A values of 
 \ion{Fe}{xiii} transitions within the energetically lowest 27 levels, 
as obtained by \cite{storey_zeippen:10}  and \cite{delzanna_storey:12_fe_13}.
In both cases, the AUTOSTRUCTURE code was used. 
In \cite{storey_zeippen:10}, 114 fine-structure levels within the $n=3$ complex were included
in the CI expansion. In \cite{delzanna_storey:12_fe_13}, 
2985 $n=3,4$ levels were included.
Clearly, very good agreement within 5--10\%  is present for the 
strongest transitions. A similar agreement is found with the A-values 
calculated by \cite{young:2004} with SUPERSTRUCTURE.

\subsection{Measurements of level lifetimes}
\label{sec:beam-foil}

Laboratory measurements of radiative lifetimes of atomic levels 
have been very important in order  to assess  the accuracy
of the theoretical radiative rates. Even  with large computations,
it is often found that theoretical values disagree significantly
with the observed ones, especially for forbidden and intersystem transitions.

Many results have been obtained using   beam-foil spectroscopy. 
Indeed there is a large body of literature  on beam-foil measurements 
of level lifetimes.  
For recent reviews  see e.g. \cite{traebert:2005,traebert:2008}.
 Beam-foil spectroscopy is based on a production of 
beams of fast ions that target a thin foil,
where they are excited by the collisions with the electrons that 
are present in the  foil (at the solid density).
As a result, the ions reach a status where high-lying levels
(with large  values of the quantum numbers $n,l, J$) 
are multiply excited (i.e. several electrons are excited). 
The experimental setup is such that the 
ions travel through the thin foil and continue to travel in 
a low-density environment. 
During this travel, the various excited levels decay to the ground state
in a complex way. The allowed transitions quickly depopulate most
of the levels, while the metastable levels last much longer.
In the mean time, the high-lying levels tend to decay via 
$\Delta n$=1 transitions, and maintain the populations of the 
low-lying levels for a while via cascading.
The beam-foil technique consists in measuring the transitions
at varying distance from the foil. The decay curves provide
measurements of the lifetimes of the levels, while the 
spectra at larger distances have been extremely useful 
to identify the transitions from metastable levels.

\begin{figure}[htbp]
\centerline{\includegraphics[width=0.65\textwidth, angle=0]{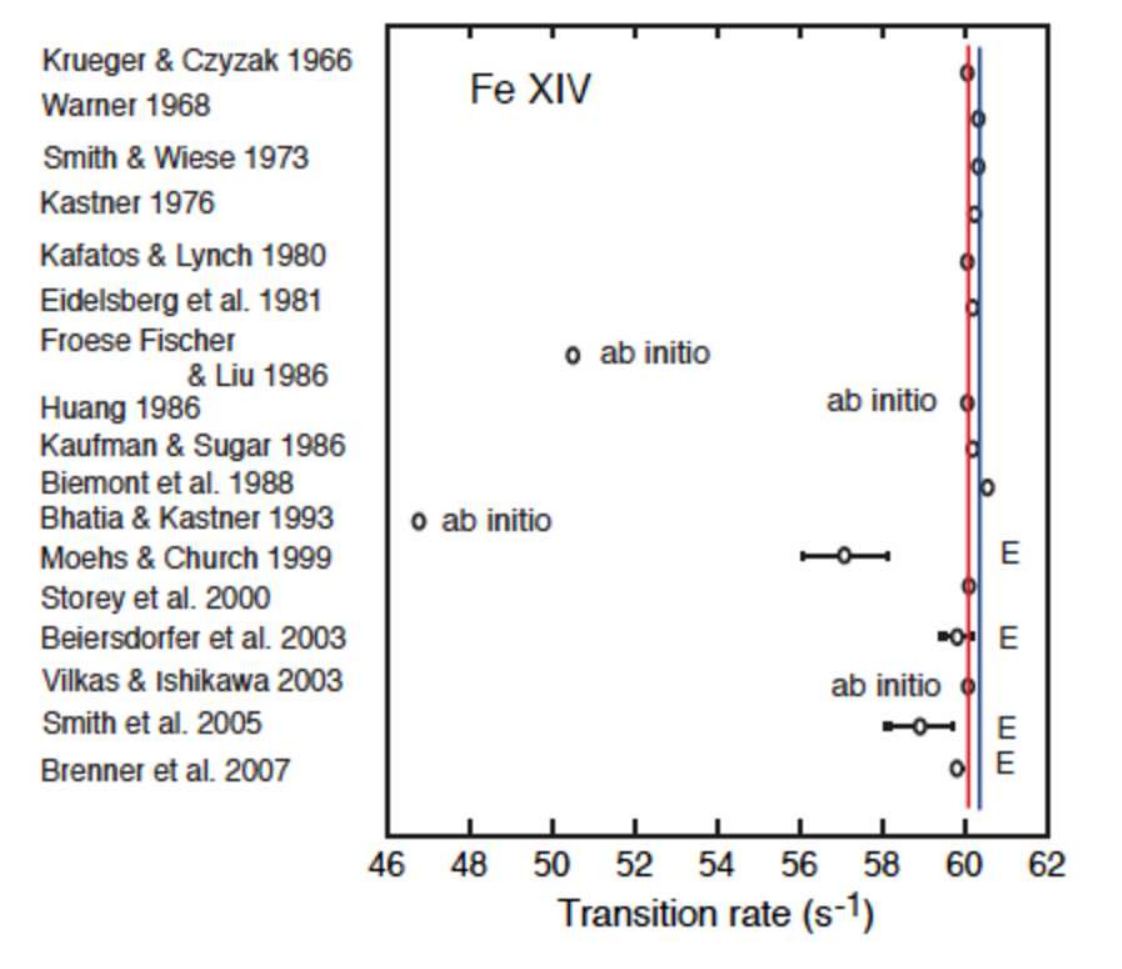}}
\caption{A comparison of theoretical and experimental 
 A-value  measurements for the green coronal line in \ion{Fe}{xiv} 
 \citep{traebert:2014}. The red line indicates the value obtained from 
ab-initio calculations, adjusted to the experimental transition energy.
The blue line indicates the value obtained by additionally taking
into account  the anomalous magnetic moment of the electron. }
\label{fig:lifetime_fe_14}
\end{figure}

From the lifetimes of the levels, direct measurements of the 
A-values are sometimes possible. There is an extensive literature
on theoretical and experimental 
 lifetime measurements for important metastable levels, mostly in 
the ground configurations of abundant ions.
Such levels typically have large quantum numbers $J$.
  Fig.~\ref{fig:lifetime_fe_14} shows as an example the results
for the A-value measurements for the famous {\it green coronal line}
 in  \ion{Fe}{xiv}, obtained directly from the 
lifetime of the first excited level within 
the ground configuration of this ion. Theoretical values 
are also plotted in the figure.
We can see  good agreement between most calculations and experimental 
data, although  a scatter in the values is present.
 Note that  \cite{brenner_etal:2007} 
used the  Heidelberg EBIT to obtain a lifetime of 16.73 ms
(equivalent to an A-value of 59.8 s$^{-1}$) with a 
very small uncertainty, which would indicate disagreement with theory
(blue line, taking into account the  experimental transition energy
and the anomalous magnetic moment of the electron). 
However,  this small uncertainty has been questioned \citep{traebert:2014_asos11}.

Quite often, however, levels decay
via multiple branching pathways, so in the literature it is common 
to compare level lifetimes as calculated by theory (taking into
account the branching ratios) and by experiment. This offers 
an excellent way to benchmark/validate the atomic structure 
calculations. 
For  general reviews of how atomic structure calculations
can be assessed by e.g. lifetime measurements see e.g. \cite{traebert:2010,traebert:2014}.
\cite{traebert:2005} reviewed lifetime measurements for the important 
iron ions.
Table~\ref{tab:traebert_etal:2008_fe_12} shows as an example
the   lifetime measurements and predictions
 for \ion{Fe}{xii}, as described in \cite{traebert_etal:2008_fe_12}.

\begin{table}[htbp]
\label{tab:traebert_etal:2008_fe_12}
\caption[\ion{Fe}{xii} level lifetimes]{\ion{Fe}{xii} level lifetimes 
in the millisecond range and known or predicted principal decay
channels (M1, E2, or M2)  \citep[adapted from][]{traebert_etal:2008_fe_12}. 
Legend: $^a$ \cite{Garstang_P-like}, 
$^b$ \cite{SmithWiese}, 
$^c$ \cite{Smitt_76}, 
$^d$ \cite{MendozaZ},
$^e$ \cite{Huang_P}, 
$^f$ \cite{KaufmanS}, 
$^g$ \cite{BiemontHansen}, 
$^h$ \cite{Moehs_2001},
$^i$ \cite{TSR_Fe_A}, 
$^j$ \cite{Biemont_2002} (two approximations),
$^k$ \cite{delzanna_mason:05_fe_12}, 
$^l$ \cite{CFF_FeXII}, 
$^m$ \cite{traebert_etal:2008_fe_12}.
}
\begin{tabular} [c]{lclccc}
\toprule
 Upper level & $\lambda $(nm)$^k$ &     \multicolumn{2}{l}{ Lifetime $\tau $
(ms)} \\
 &       &       Experiment &  Theory \\
\midrule
$3s^{2}3p^{3}$ $^{2}$D$^{\rm o}_{3/2}$ & 240.641 &   20.35 $\pm $
1.24$^h$ & 18.9$^a$, 18.9$^b$, 5.0$^c$, 16.8$^d$, 16.0$^e$, \\
 & &  18.0 $\pm $ 0.1$^i$ & 20.8$^f$, 18.4$^g$, 22.57$^h$ 18.0/18.0$^j$,
17.7$^k$ \\
$3s^{2}3p^{3}$ $^{2}$D$^{\rm o}_{5/2}$ & 216.976 &   306
$\pm $ 10$^i$ & 324$^a$, 115$^b$, 326$^c$, 294$^d$, 313$^e$,\\
 & & &  544$^f$, 314$^g$ 323/323$^j$, 311$^k$ \\
$3s^{2}3p^{3}$ $^{2}$P$^{\rm o}_{1/2}$ & 307.206, 356.6 &  4.38
$\pm $ 0.42$^h$ & 3.84$^a$, 3.84$^b$, 3.84$^c$, 3.64$^d$, 3.58$^e$, \\
 & &  4.10 $\pm $ 0.12$^i$ & 4.05$^f$, 3.81$^g$ 3.59/3.79$^j$, 3.8$^k$  \\
$3s^{2}3p^{3}$ $^{2}$P$^{\rm o}_{3/2}$ & 256.677, 290.470 &  1.85
$\pm $ 0.24$^h$ & 1.61$^a$, 1.61$^b$, 2.39$^c$, 1.55$^d$, 1.53$^e$,\\
 & &  1.70 $\pm $ 0.08$^i$ & 1.67$^f$, 1.59$^g$ 1.59/1.59$^j$, 1.6$^k$ \\
$3s^{2}3p^{2}$($^{3}$P)$3d$ $^{4}$F$_{9/2}$ & 25.187, 59.2600,1421.868 & 11
$\pm $ 1$^m$ &  11.6/9.2$^j$, 9.7$^k$, 9.56$^l$   \\
$3s^{2}3p^{2}$($^{1}$D)$3d$ $^{2}$G$_{9/2}$ & 184.723, 163.484, 64.292 & 4 $\pm
$ 1$^m$ &  4.00/4.27$^j$, 4.03$^k$, 4.43$^l$ \\
 &  22.167 & &  \\
\bottomrule
\end{tabular}
\end{table}

\begin{figure}[htbp]
\centerline{\includegraphics[width=0.65\textwidth, angle=0]{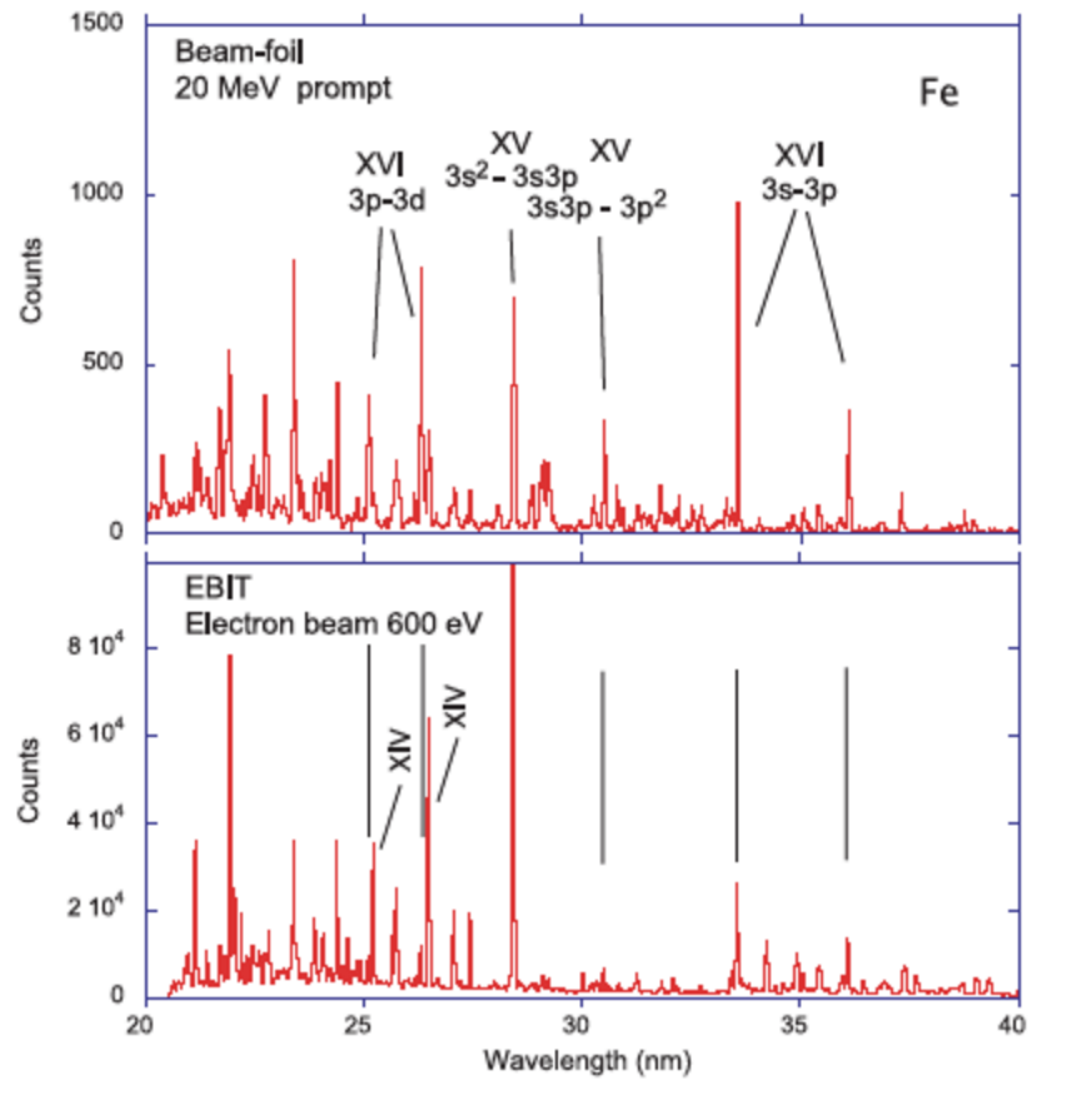}}
\caption{A beam-foil spectrum of iron ions (top), compared to 
an EBIT spectrum  \citep{traebert:2017}.}
\label{fig:beam-foil-spectrum}
\end{figure}

The ion beam energy determines the average charge state reached. 
Therefore, by tuning the energy,
it is possible to obtain spectra where  particular charge state ions 
are  enhanced.
Beam-foil spectroscopy has therefore been 
fundamental for the identification of spectral lines, in particular those arising
from levels with long lifetimes.
 However, the spectra always contain a mixture of
charge states. A sample spectrum is shown in Figure~\ref{fig:beam-foil-spectrum}
(above).

Many other methods and devices have been used to measure atomic lifetimes,
mostly with heavy-ion storage rings and  Electron Beam Ion Trap (EBIT) devices. 
EBIT devices are extremely useful for assessing  solar spectra
 because the plasma density is normally in the range 10$^{11}$--10$^{13}$ cm$^{-3}$,
i.e. not far from typical densities found in active regions and flares.
The beam energy can be adjusted so only a range of charge states from an element
(all those below the threshold for ionization of the highest charge) are 
observed. This aids the identification of the spectral lines.
Figure~\ref{fig:beam-foil-spectrum} (below) shows 
 a spectrum obtained with an Electron Beam Ion Trap (EBIT), as an example.

For a review of 
measured  lifetimes for iron ions using  heavy-ion storage rings see
\cite{traebert_etal:2003}, while \cite{traebert:2002} also discusses 
EBIT measurements.

\subsection{Magnetically-induced transitions}

We note here  a relatively new
and potentially interesting diagnostic to measure solar magnetic 
fields using XUV lines. 
A few cases have been discovered where transitions which are 
strictly forbidden are actually observed in laboratory plasma,
being induced by very strong magnetic fields. They are called 
magnetically-induced transitions (MIT).
For sufficiently strong magnetic fields and  atomic states
which are nearly degenerate and have the same magnetic quantum
number and parity,  a strong external field could break 
the atomic symmetry (by  mixing the atomic states),
and induce new magnetically-induced transitions. 

\begin{figure}[!htbp]
\centerline{\includegraphics[width=0.7\textwidth, angle=0]{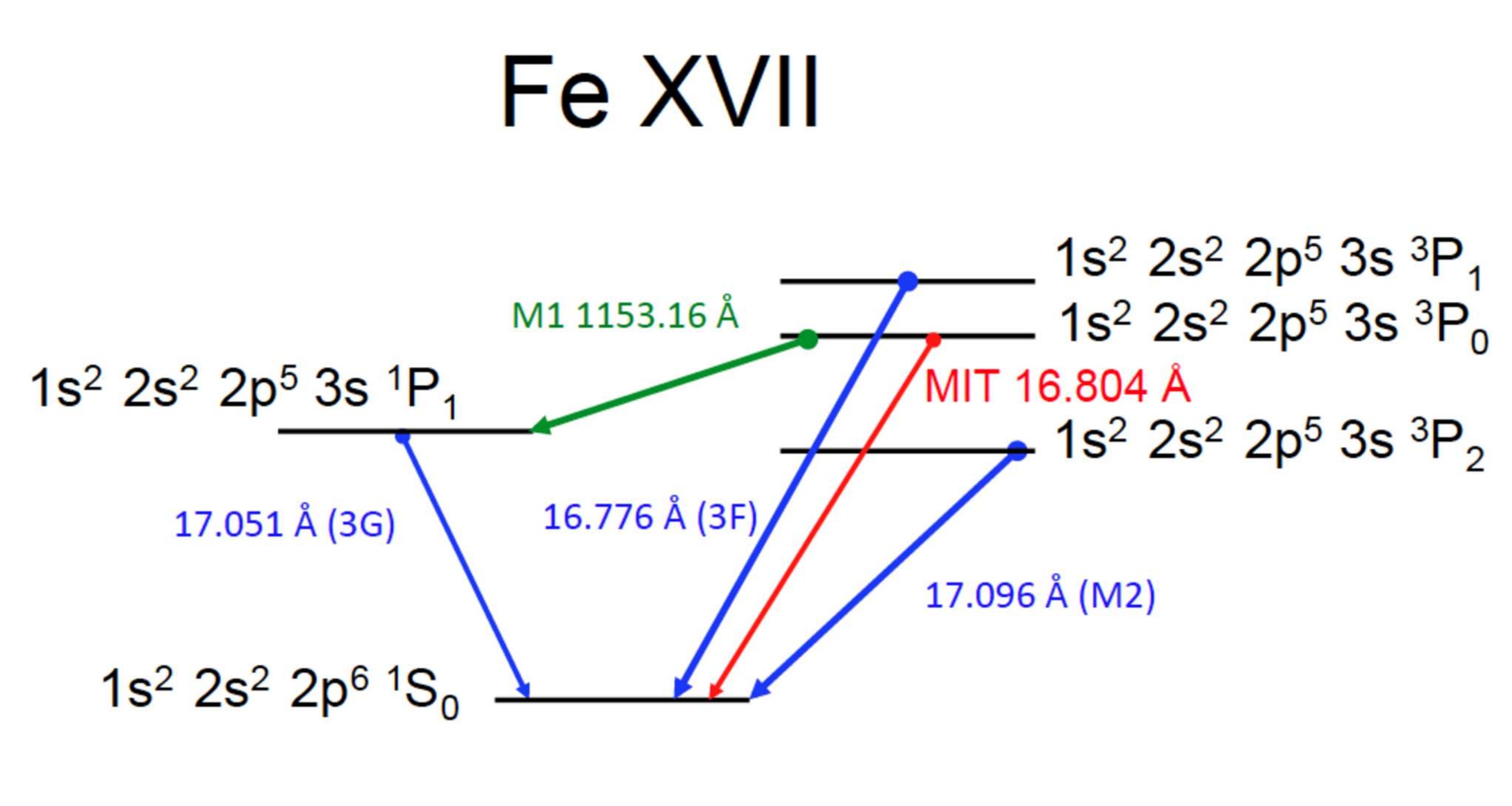}}
\centerline{\includegraphics[width=0.7\textwidth, angle=0]{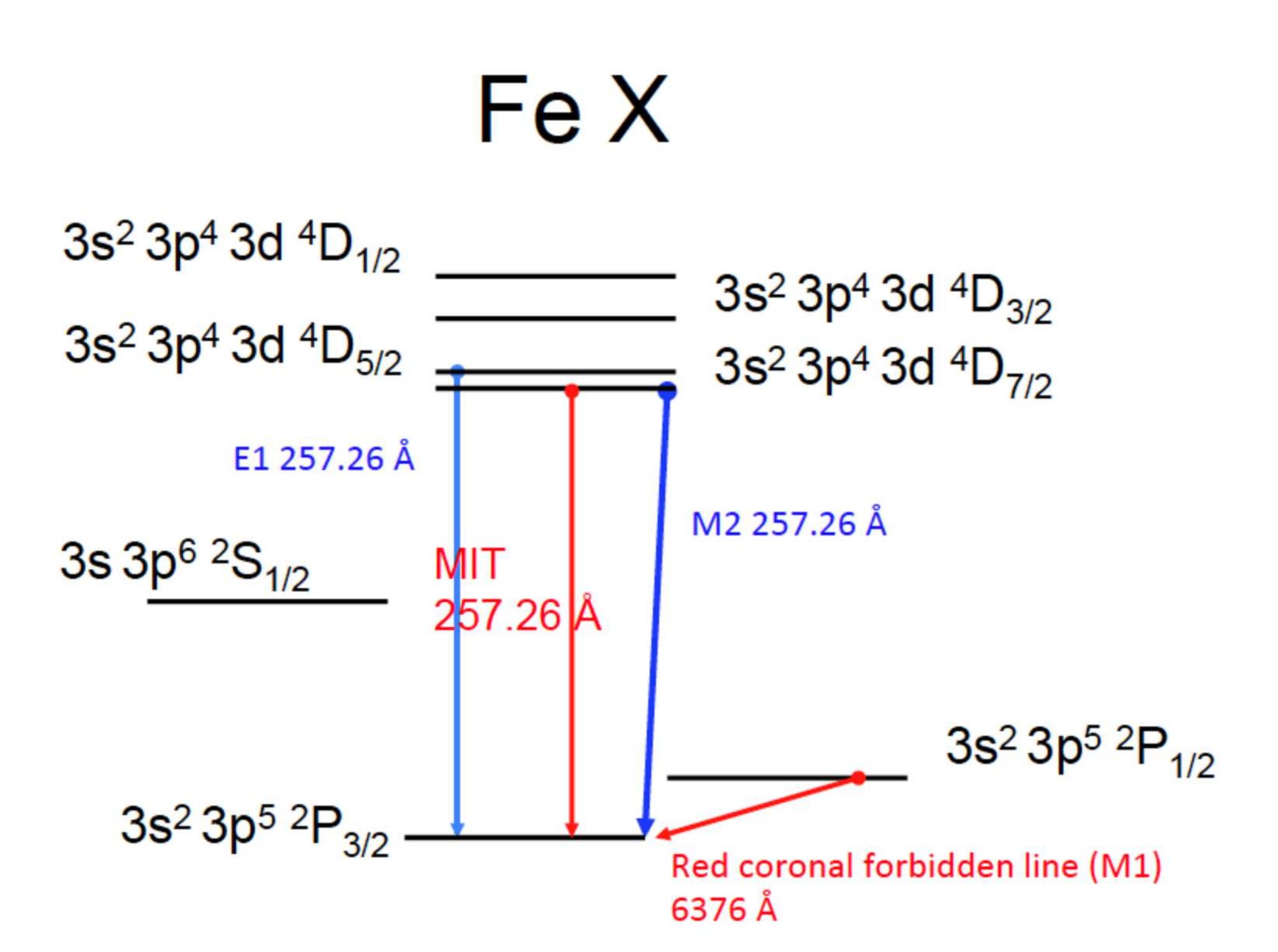}}
\caption{Grotrian diagrams (not to scale) of the 
lowest  levels in  \ion{Fe}{xvii} and  \ion{Fe}{x}, indicating the levels
involved in the magnetically-induced transitions. 
}
 \label{fig:mit}
\end{figure}

\cite{beiersdorfer_etal:2003} reported  an observation of 
an X-ray MIT transition at 47~\AA\  in Ne-like Ar by an EBIT with
strong magnetic fields  of the order of 30 kG. 
\cite{li_etal:2013} reported calculations of the strength of  MIT
along the Ne-like sequence.
The same transition in Ne-like Iron (\ion{Fe}{xvii}) 
originates from  the metastable 
  2s$^{2}$2p$^{5}$ $^3$P$_{0}$ 
level (hereafter metastable $J=0$ level), which normally decays with a well-known 
M1 forbidden transition (at 1153.16~\AA)  to the lower $^1$P$_{1}$ level, 
see  Fig.~\ref{fig:mit} (top).
The decay to the $J=0$ ground state is strictly forbidden, 
however the MIT line at 16.804~\AA\ was recently 
 measured by \cite{beiersdorfer_etal:2016_fe_17} using the 
Livermore EBIT in the presence of a strong magnetic field, allowing them to provide an  estimate 
of the lifetime of the metastable $J=0$ level in \ion{Fe}{xvii}.
This MIT transition is very close to one of the strong X-ray lines from 
this ion, at 16.776~\AA, usually labelled with (3F). 
Fig.~\ref{fig:mit} (top) also shows two other main X-ray lines
(see \cite{delzanna_ishikawa:09} for a discussion on line identifications and
wavelength measurements for this ion).

Another example of a  MIT  occurs in Cl-like Iron.
The  \ion{Fe}{x} spectrum was discussed  in detail in 
\cite{delzanna_etal:04_fe_10}.
This ion gives rise to  famous red coronal line, within the 
3s$^2$ 3p$^5$ $^2$P$_{3/2, 1/2}$ ground configuration states, 
and several important 
forbidden transitions in the EUV and visible which 
ultimately decay into the 3p$^{4}$3d $^4$D$_{5/2, 7/2}$ levels
(see  Fig.~\ref{fig:mit}, bottom).
The  3p$^{4}$3d $^4$D$_{7/2}$ level is a metastable, and can 
only decay to the ground state via a M2 transition. 
This line is the fourth strongest line in the EUV spectrum of \ion{Fe}{x}.
The $^4$D$_{5/2}$ level decays via an  E1 transition to the ground 
state. This transition is relatively strong, but much weaker than the M2. 
\cite{smitt:1977} was the first to tentatively  
assign an energy (of 388708 cm$^{-1}$) to both  3p$^{4}$3d $^4$D$_{5/2, 7/2}$ 
levels, which were then confirmed 
in \cite{delzanna_etal:04_fe_10}. The Hinode EIS measurements
however showed a discrepancy of about a factor of 2 between predicted and observed
intensities of these lines \citep{delzanna:12_atlas}, and it was only
with the latest large-scale scattering calculation \citep{delzanna_etal:12_fe_10} 
that agreement has been found.

The MIT in \ion{Fe}{x} was discussed in detail by 
\cite{li_etal:2015_fe_10}. 
In the presence of a strong  magnetic field, the two 3p$^{4}$3d $^4$D$_{5/2, 7/2}$ 
states will  mix, which will produce a new MIT (E1) from the 
$^4$D$_{7/2}$ level to the ground state 3s$^2$ 3p$^5$ $^2$P$_{3/2}$, 
see  Fig.~\ref{fig:mit} (bottom).
\cite{li_etal:2015_fe_10} carried out increasingly large atomic structure 
calculations, which  were used to estimate the splitting of the 
3p$^{4}$3d $^{4}$D$_{5/2}$ and 3p$^{4}$3d $^{4}$D$_{7/2}$ levels, 
which turns out to be a major uncertainty in the diagnostic. 
The best calculations provided a  splitting of 20 cm$^{-1}$.
The authors did not take into account the  review by 
\cite{delzanna_etal:04_fe_10}
of previous experimental data, where  a much smaller 
energy difference of 5 cm$^{-1}$ was suggested. The assessment was
 partly based on  previous work, which included Skylab observations. 
\cite{judge_etal_2016} carried out a similar analysis of the 
same Skylab observations to obtain 3.6 $\pm$2.7  cm$^{-1}$.
A similar energy difference was estimated from recent 
EBIT measurements \citep{li_etal:2016}, although a more accurate 
measurement would be useful.
It remains to be seen if  magnetic fields can be measured 
with Hinode EIS, given the various uncertainties:
aside from the uncertainty in the  energy splitting, 
and the difficulty in obtaining accurate atomic data for these levels,
the transitions from the  $^4$D are significantly affected by 
variations in the electron temperature and density. 
They are also sensitive to non-Maxwellian electron distributions
\citep{dudik_etal:2014_fe}. Finally, a significant uncertainty is associated with 
 the Hinode EIS calibration.

Finally, we point out that in principle the MIT are a general 
tool available for  other states/sequences. 
For example, \cite{grumer_etal:2013} presented  calculations for the 
2s 2p $^3$P$_{0} \to$ 2s$^2$ $^1$S$_{0}$  MIT transition in Be-like ions.
while \cite{grumer_etal:2014} and \cite{johnson:2011} also discuss the effects
that are present in ions/isotopes with non-zero nuclear spin,
which cause hyperfine structure and  hyperfine-induced E1 transitions.

\subsection{Atomic structure data}

Excellent sets of A-values have been obtained with 
semi-empirical corrections and the {\sc superstructure}
and {\sc CIV3} programs. The MCHF have also been very accurate 
for the ground configurations, and there are several studies by
C. Froese Fischer and collaborators. An on-line database of  
MCHF calculations is available at \url{http://nlte.nist.gov/MCHF/}.
Very accurate ab initio multi-reference M{\o}ller-Plesset calculations
have been carried out for a limited set of ions and configurations
by Y. Ishikawa and collaborators. For example, see 
\cite{ishikawa_vilkas:2001} for the Si-like ions,  and 
\cite{ishikawa_vilkas:2008} for S-like ions.

The new development of the GRASP program,
GRASP2K \citep{grasp2k:2007},  allows large-scale 
calculations to be performed even for the most complex ions.
Such calculations reach spectroscopic accuracy, in the sense that 
theoretical wavelengths in the EUV can be within a fraction
of an \AA\ of the observed values. Previous ab-initio calculations
typically had an accuracy of a few \AA, which is often not sufficient 
to enable the line identification process (see below).
A recent review of  GRASP2K calculations can be found in 
\cite{jonsson_etal:2017}. As a measure of the accuracy, one could look at the 
differences between the length and velocity forms of the oscillator strengths,
or at the differences in the A-values as calculated by different codes.
Typical uncertainties of the order of a few percent are now achievable
for the main transitions.

The literature on radiative data on each single ion is too 
extensive to be provided here. We refer the reader to the  on-line
references 
(\url{http://www.chiantidatabase.org/chianti_direct_data.html}) in the CHIANTI database,
 where details about the radiative data for each ion are provided.
We also note that 
the NIST database at \url{http://physics.nist.gov/PhysRefData/ASD/lines_form.html} 
provides an extensive and very useful bibliography of all 
the published radiative  data for each ion. 
NIST also provides a series of assessment reports of theoretical and 
experimental data.

Here, we just mention the most recent GRASP2K calculations not just because they are very accurate,
but also because they  cover entire isoelectronic sequences, and references to previous 
studies on each ion in the sequence can be found. 

\cite{jonsson_etal:2011_c-like}  carried out GRASP2K calculations 
for states of the 2s $^{ 2 }$ 2p $^{ 2 }$ , 2s 2p $^{ 3 }$ , and 2p $^{ 4 }$ configurations in 
C-like ions between F IV and Ni XXIII. These calculations have been supplemented with those
from \cite{ekman_etal:2014} 
for ions in the C-like sequence, from Ar XIII to Zn XXV,
by including levels up to $n$=4. 

\cite{rynkun_etal:2012} carried out calculations  for  transitions among the lower 15 levels
(2s$^{2}$2p, 2s 2p$^2$, 2p$^3$) in ions of the B-like sequence, from 
 N III to Zn XXVI.
\cite{rynkun_etal:2013} calculated radiative data for transition originating from 
the 2s$^{2}$2p$^{4}$, 2s 2p$^{5}$, and 2p$^{6}$ configurations in all O-like ions 
between F II and Kr XXIX. These calculations have recently been extended 
to include many $n$=3 levels, for ions from  \ion{Cr}{xvii} to \ion{Zn}{xxiii}
by \cite{wang_etal:2017_o-like}.
\cite{rynkun_etal:2014} calculated  transition rates for states of the 2s$^{2}$2p$^{3}$, 2s2p$^{4}$, and 2p$^{5}$ 
configurations in all N-like ions between F III and Kr XXX.
These calculations have recently been extended 
to include levels up to  $n$=4, for ions from  Ar XII to Zn XXIV, by 
\cite{wang_etal:2016_n-like}.
\cite{jonsson_etal:2013_f-like} carried out GRASP2K calculations for the
 2s$^{2}$2p$^{5}$ and 2s2p$^{6}$ configurations in F-like ions, from  Si VI to W LXVI.
The calculations on F-like ions has recently been extended by 
\cite{si_etal:2016_f-like} to include higher levels up to $n$=4 for ions from 
Cr to Zn.
\cite{jonsson_etal:2013_b-like} calculated radiative data for levels up to $n$=4
 for the B-like  Si X and all the B-like ions between  Ti XVIII and Cu XXV.
\cite{jonsson_etal:2014} carried out GRASP2K calculations 
for the 2p$^{6}$ and 2p$^{5}$3l configurations in Ne-like ions,  between Mg III and Kr XXVII.
These calculations have recently been extended 
to include levels up to  $n$=6, for all Ne-like ions between \ion{Cr}{xv} and \ion{Kr}{xxvii}
by \cite{wang_etal:2016_ne-like}.

\cite{wang_etal:2015_be-like} calculated radiative data for ions along the 
Be-like sequence. 

\cite{jonsson_etal:2016_si-like} calculated radiative data for the 
3s$^2$3p$^2$, 3s 3p$^3$ and 3s$^2$ 3p 3d configurations of  all the Si-like ions
from  Ti IX to Ge XIX, plus  Sr XXV, Zr XXVII, Mo XXIX.
\cite{gustafsson_etal:2017} performed GRASP2K calculations
for 3l 3l', 3l 4l', and 3s 5l states in Mg-like ions from 
Ca IX to  As XXII,  and Kr XXV. 
\cite{ekman_etal:2017} performed GRASP2K calculations
for ions in the Al-like sequence, from Ti~X through Kr~XXIV,
plus  Xe~XLII, and W~LXII. The radiative data are for a large set
of 30 configurations:
  $3s^2\{3l,4l,5l\}$, $3p^2\{3d,4l\}$, $3s\{3p^2,3d^2\}$, $3s\{3p3d,3p4l,3p5s,3d4l'\}$, $3p3d^2$, $3p^3$ and $3d^3$ with
  $l=0,1,\ldots,n-1$ and $l'=0,1,2$.

\subsection{Electron-ion scattering calculations }

As the electron-ion collisions are the dominant populating 
process for ions in the low corona, significant effort
has been devoted to calculate 
 electron-ion scattering collisions over the past decades.
The collision strengths usually have a slowly varying 
part and spikes, that are due to resonances (dielectronic capture). 
The theory of electron-ion collisions has received significant contributions
by M.J. Seaton  (University College London), 
 P. Burke (Queens University of Belfast), and A. Burgess (University of Cambridge),
among many other members of a large community of atomic physicists.
Standard textbooks and reviews on electron-ion collisions 
 are \cite{mott_massey:1949, burgess_etal:1970, seaton:1976,  henry:1981, burke:2011}.

Using Rydberg units, the nonrelativistic Hamiltonian 
describing an ion  with $N$ electrons 
and a central nucleus having charge number $Z$ and a free electron
(i.e. the  $(N+1)$-electron system) is

\beq
 {H (Z,N+1)} = - \sum_{i=1}^{N+1} \left(\nabla_i^2 + {2Z  \over r_i}
              \right) + \sum^{N}_{i < j} {2 \over r_{ij}} \;.
\eeq             

We search for solutions of the time-independent  Schr\"odinger equation 
for the  $(N+1)$-electron system 
\begin{equation}
H(Z,N+1) \Psi_c = E \Psi_c
\end{equation} 
where $E$ is the total energy of the $(N+1)$-electron system,
and $\Psi_c$ is the wavefunction of the $(N+1)$-electron system.
The $\Psi_c$  are usually  expanded in terms
of products of wavefunctions of the $N$-electron target  and those of the free electron.

Within the so called  \emph{Close-Coupling} (CC) approximation,
 the scattering electron sees individual target electrons, 
 and a set of integro-differential equations need to be solved.
An efficient way to solve the scattering problem is to use the 
$R$-matrix method, described in \cite{burke_etal:1971,burke_robb:1976,burke:2011}.
Such method was further developed over many years 
by the Iron Project, an international collaboration of atomic physicists, 
which was set up to calculate electron excitation rates for all the iron ions.
A significant contribution to the development of the Iron Project programs
 was made  by K. Berrington, see e.g. 
\cite{berrington_etal:1987}.

The analysis of the scattering interaction 
\beq
 < \Psi_i | H(Z,N+1) - E |  \Psi_{i'} >  
\eeq
involves the 
calculation of  the elements of the so-called reactance matrix,  $ K_{ii'}$.
The transmission matrix $T_{ii'}$  is related to the 
reactance matrix:
\begin{equation}
   T = \frac{-2{\rm i} K}{1-{\rm i} K} \;\;,
\label{eq:unitary}
\end{equation}
and the resulting scattering matrix $ \cal S$
\beq
 {\cal S} = 1-T = { 1+ {\rm i} K \over 1- {\rm i} K }
\eeq
is unitary.
The dimensionless collision strength ($\Omega_{if}$) for any transition $i-f$ 
from an initial $i$ to a final state $f$ is then obtained from the transmission matrix
\begin{equation}
  \Omega_{if} \propto |T_{if}|^2 \,.
\end{equation}

The collision strengths are usually calculated over 
a finite range of the energy of the incoming electron.
Within the Iron Project codes, the collision strengths are extended to high
 energies by interpolation using  the 
appropriate high-energy limits in the 
 \cite{burgess_tully:92} scaled domain.
The high-energy limits are calculated 
following \cite{burgess_etal:1997} and \cite{chidichimo_etal:03}.

\subsubsection{DW  UCL codes}

In the \emph{Distorted Wave} (DW) approximation,  the coupling between different 
target states is assumed to be negligible and 
the system of coupled integro-differential equations is significantly reduced.
A general method which includes the effects of 
exchange between the free and a bound electron, was developed
by \cite{eissner_seaton:1972} at UCL.
The method requires that the free-electron wavefunction $\theta_i$ 
are  orthogonal to those of the one-electron bound orbitals 
with the same angular momentum.
This is achieved by adopting a general expression for the 
total wavefunction of the $N+1$ system of the form

\beq
\Psi={\cal A} \sum_i  \chi_i(\vec x_1,..., \vec x_{N}) \theta_i(\vec x_{N+1})  + \sum_j \phi_j c_j
\eeq
where $\phi_j $ is a function of bound state type for the whole system.
The functions $\chi_i$ and  $\phi_j $ are constructed from  one-electron orbitals,
and $\cal A$ is an operator that guarantees antisymmetry.
The functions $\phi_j$ are called correlation functions, which are mainly
introduced to obtain orthogonality.
Note that the above general expression with the correlation functions 
was adopted by the UCL DW code, mainly developed by W. Eissner and M. Seaton
\citep{eissner_seaton:1972, eissner:1998}. This code has been
widely used for a very long time and has produced  a very large 
amount of atomic data.

The first accurate scattering calculations for the coronal iron ions
were carried out with the UCL DW codes 
(see, e.g.  \citealt{mason:1975a}). These  were prompted by ground-based 
eclipse observations of the forbidden lines. These were also needed for 
the   early EUV Skylab and OSO-7 observations in the 1970's.
There was a series of papers from A.K. Bhatia and H.E. Mason. 
Such calculations were further revised and improved by several groups.
These calculations were generally carried out in intermediate coupling 
that is transforming the LS coupling calculation using term coupling 
coefficients.
This approximation was found adequate for most coronal ions.

\begin{figure}[!htbp]
\centerline{\includegraphics[width=0.5\textwidth, angle=90]{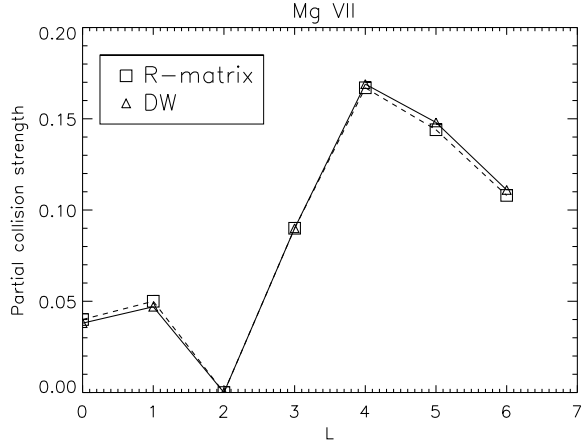}}
\caption{A comparison of partial collision strengths calculated 
with the  DW and  $R$-matrix codes, 
for the \ion{Mg}{vii} 2s$^2$ 2p$^2$ $^1$S--2s  2p$^3$ $^1$P transition
at 10 Ry \citep{burgess_etal:1991}.
}
 \label{fig:burgess_etal:1991}
\end{figure}

Although the effects of resonance enhancement can be included 
in DW calculations, usually it is not.  DW calculations typically 
agree with the background collision strength values at lower energies,
lower than the maximum threshold. At energies higher than the 
 maximum threshold, DW and $R$-matrix calculations should agree, if 
 the same wavefunctions are used. 
For a comparison between DW and $R$-matrix calculations 
see e.g. \cite{burgess_etal:1991}.
Excellent agreement between the collision strengths calculated with the
UCL DW and the $R$-matrix codes was found, to within a few percent.
Fig.~\ref{fig:burgess_etal:1991} 
shows as an example the partial collision strengths calculated at 10 Ry 
for the \ion{Mg}{vii} 2s$^2$ 2p$^2$ $^1$S--2s  2p$^3$ $^1$P transition.

\subsubsection{Other codes - excitation by electron impact}

Several other  codes have been developed over the years
to calculate cross-sections for excitation by electron impact, and have produced a
large amount of atomic data.  
For example,  D. Sampson and H.L. Zhang  developed  several codes based on the 
Coulomb-Born-exchange method \citep[cf.][]{sampson_etal:1979}.
Widely used DW codes are the 
Flexible Atomic Code (FAC)~\citep{gu:2003, gu:2008}, and 
HULLAC \citep{bar-shalom_etal:1988}. 
Recently, the {\sc autostructure} DW code \citep{badnell:11} has also been implemented.
FAC and  {\sc autostructure}  are publicly available, as the UCL DW.

We note that different assumptions are made in the various DW codes.
For example, the {\sc autostructure} DW code does not impose the 
orthogonality condition and the second term in the expansion is not present,
but it does calculate all the  appropriate exchange overlaps.
In the DW approximation, the scattering equations are uncoupled and 
not all the elements of the reactance matrix $ K_{ii'}$  need to 
be calculated.
Another important difference is whether the DW method is unitarized. 
 Some of the DW codes, such as the FAC 
 and  HULLAC,  use the relation 

\begin{equation}
   T\ =\ \frac{-2{\rm i}K}{(1-{\rm i} K)} \times 
         \frac{(1+{\rm i} K)}{(1+{\rm i} K)}\ 
   =\ \frac{-2{\rm i} K + 2K^2}{1 + K^2} \approx -2{\rm i} K\,,
\label{eq:nonunitary}
\end{equation}
in which case the DW method is called non-unitarized.
As pointed out by  \cite{fernandez-menchero_etal:2015_unitary},
the  non-unitarized  DW method can lead to large errors (factors of 10) for a 
few weak transitions, where the coupling in  the scattering equations
becomes important. The original and default version of the 
{\sc autostructure} DW was also non-unitarized, but a unitarized
option, which provides good agreement with the fully close-coupling
calculations, has recently  been implemented \citep{badnell_etal:2016}.

The $R$-matrix codes,  used within the  Iron Project 
for the scattering calculations,  are described
in \cite{hummer_etal:93,berrington_etal:95, burgess:74,badnell_griffin:01}, and 
have been applied to the calculations of many  ions.
The main repository for the codes is the UK APAP web page 
maintained by N.R. Badnell at \url{http://www.apap-network.org/codes.html}. 

Many of the calculations for the astrophysically important (not heavy)
ions are carried out in $LS$-coupling, and  relativistic corrections 
are applied later. The  $R$-matrix calculations are carried out in an 
inner and outer region.  The 
$R$-matrix calculations in the  inner region normally
 include  the  mass and Darwin relativistic energy corrections. 
Many of the outer region calculations are carried out 
with  the intermediate-coupling frame transformation
method (ICFT),  described by \citet{griffin_etal:98}. 
This method is computationally much faster than the 
 Breit-Pauli $R$-matrix method (BPRM). 

A fully relativistic Dirac $R$-matrix code, called  DARC, was 
developed by  P. H. Norrington and I. P. Grant 
\citep[see, e.g.][]{norrington_grant:1981}, and is also 
available at the UK APAP web page.
Several comparisons between the results obtained by the 
various codes have been carried out, and are normally satisfactory.
For example, \cite{badnell_ballance:2014}  compared the $R$-matrix results of 
ICFT, BPRM and DARC on \ion{Fe}{iii}.
However, discrepancies for weaker transitions have also 
been reported \citep[see, e.g., the discussion in ][]{badnell_etal:2016}.

A different approach, the 
B-spline R-matrix (BSR) method was developed by other authors
\citep[see, e.g. ][ for details]{zatsarinny:2006,zatsarinny_bartschat:2013}.
The method uses term-dependent
non-orthogonal orbital sets for the description of the target
states. The wave functions for different states are then optimized 
independently, which result in a much better target representation.
This method is computationally demanding and has mostly been used 
for the calculations of cross sections for neutrals and low charge
states, where an accurate representation of the  target states is typically  difficult
\citep[see, e.g.][]{wang_etal:2013}.
In a recent work, a detailed comparison of similar-size calculations
for the same ion with the BSR, ICFT, and DARC was carried out by 
\cite{fernandez-menchero_etal:2017_n_4}. As in the previous comparisons,
good agreement was found among the lower and stronger transitions. 
 Significant differences  were however  found for the weaker transitions
and for transitions to the higher states.
The differences were mainly due to  the structure description
and  correlation effects, rather than due to 
the  different treatment of the relativistic effects in the three codes.

Other approaches and codes exist. For example, the 
Dirac R-matrix with pseudo-states method (DRMPS), described in 
\cite{badnell:2008_drmps}. For H-like systems, the 
the convergent close-coupling (CCC) method of \cite{bray_stelbovic:1992} and
the relativistic  convergent close-coupling (RCCC) method \citep{bostock:2011}
provide extremely accurate cross sections.

\subsection{Uncertainties in electron-ion excitation rates and in the level population}

In this Section we provide some examples on various effects
which can reduce the accuracy of the  collision strengths and the derived 
rates. 
Clearly, if the atomic structure is not accurate, this can have 
a direct effect on the rates. For example, the high-energy limits of 
the collision strengths for  dipole-allowed 
transitions are directly related to the $gf$ values
\citep{burgess_etal:1997}. Therefore, any differences in $gf$ values 
obtained with different structure calculations are directly reflected as
 differences in the collision strengths.
The comparisons of the $gf$ values in Fig.~\ref{fig:fe_14_comp_gf} were used by 
\cite{delzanna_etal:2015_fe_14} to explain the discrepancies in the 
rates for  \ion{Fe}{xiv} calculated by  \cite{aggarwal_keenan:2014}.

\begin{figure}[htbp]
\centerline{\includegraphics[width=0.7\textwidth, angle=0]{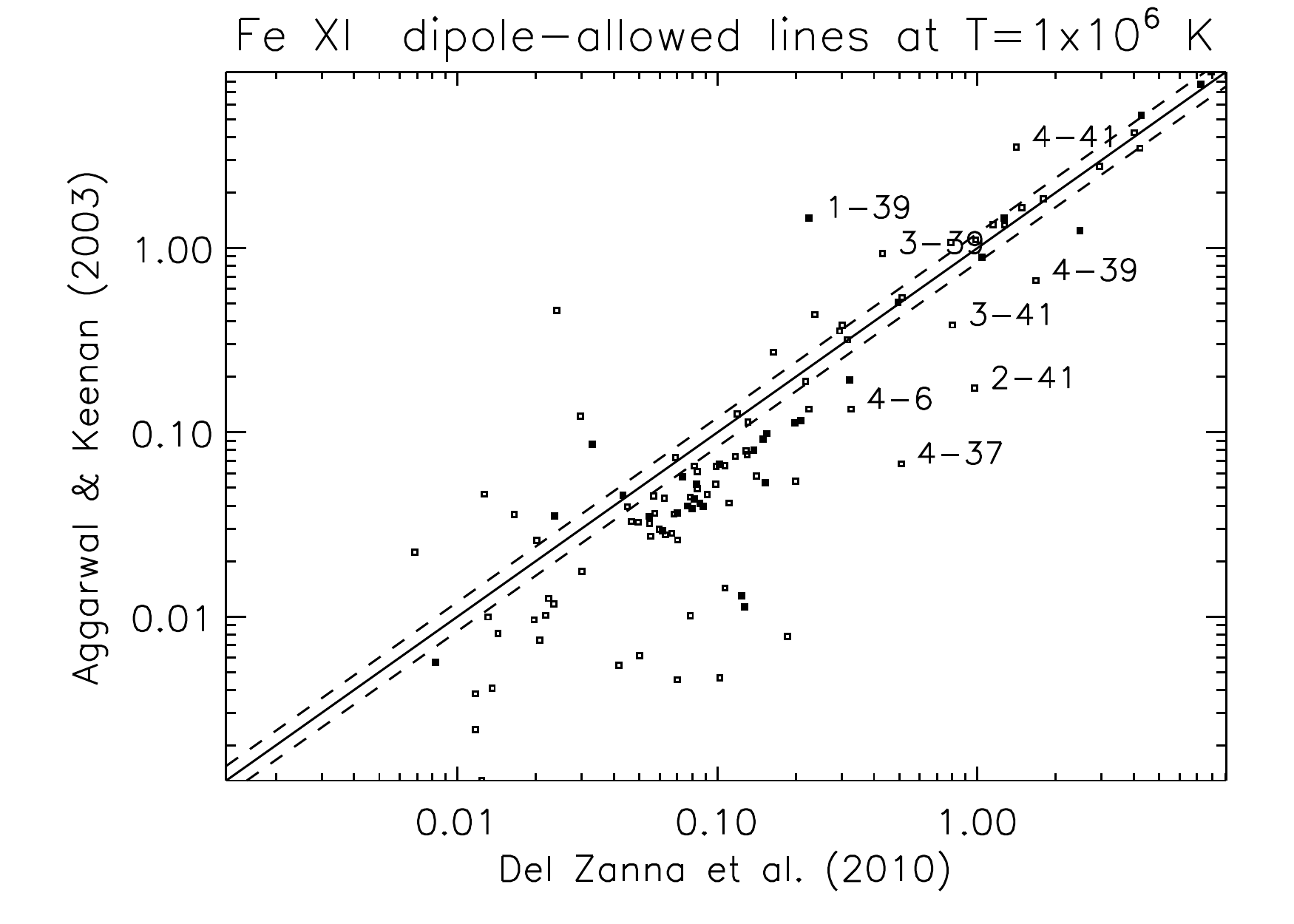}}
 \caption{ Effective collision strengths for the strongest \ion{Fe}{xi} transitions
as calculated by \cite{aggarwal_keenan:2003b} and \cite{delzanna_etal:10_fe_11}.
The dashed lines indicate $\pm$20\%. The largest differences for the strongest transitions
are related to decays from  $J=1$ levels 
(e.g. 4--37, 1--39, 3--39, 4--39, 2--41, 3--41, 4--41 indicated in the figure).
Figure adapted from \cite{delzanna_etal:10_fe_11}.
}
\label{fig:fe_11_comp_ups}
\end{figure}

The calculations of dipole-allowed collision strengths for  simpler ions 
obtained with different targets are normally in agreement,  within 10--20\%,
while results for  weaker transitions can differ by significant factors.
However, even strong transitions in complex ions can sometimes be difficult to 
calculate accurately. For example, as we have previously mentioned, 
transitions to levels with a  strong spin-orbit interaction are very sensitive
to the target wavefunctions.
Fig.~\ref{fig:fe_11_comp_ups} shows as  an example a comparison between 
rates for \ion{Fe}{xi} lines, as calculated in two different ways.
 The largest differences are for a few of  the brightest transitions, and 
are related to decays from three $J=1$ levels which have a strong spin-orbit interaction.
Indeed, as shown in \citep{delzanna_etal:10_fe_11}, large differences in the 
$gf$ values for these transitions are present.

Semi-empirical corrections are normally not applied to the scattering 
calculations. However,  they can provide significant improvements to the rates
 (see, e.g. \citealt{fawcett_mason:1989,delzanna_badnell:2014_fe_8}).

\begin{figure}[!htbp]
\centerline{\includegraphics[width=7.5cm,angle=0]{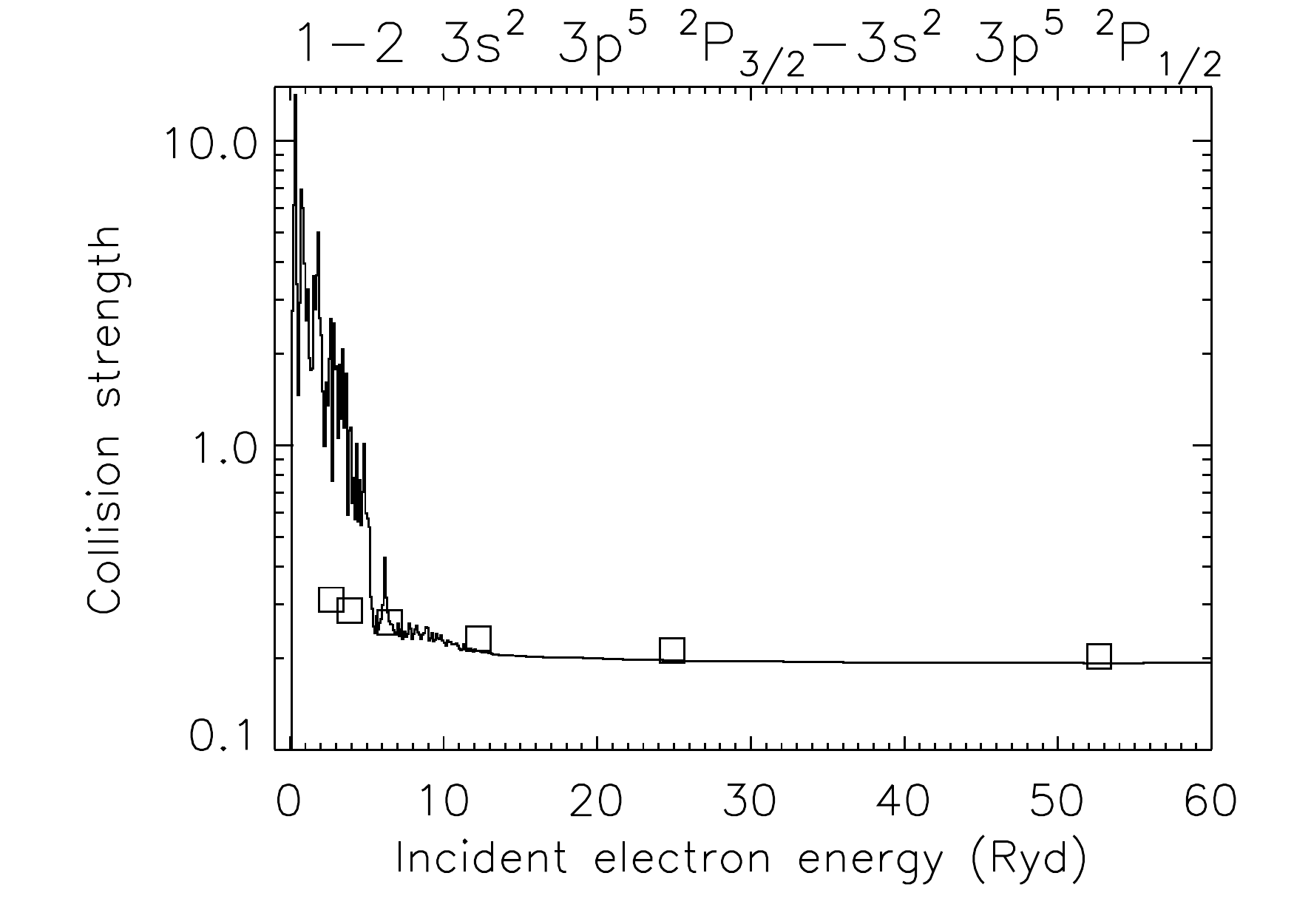}
\includegraphics[width=7.5cm,angle=0]{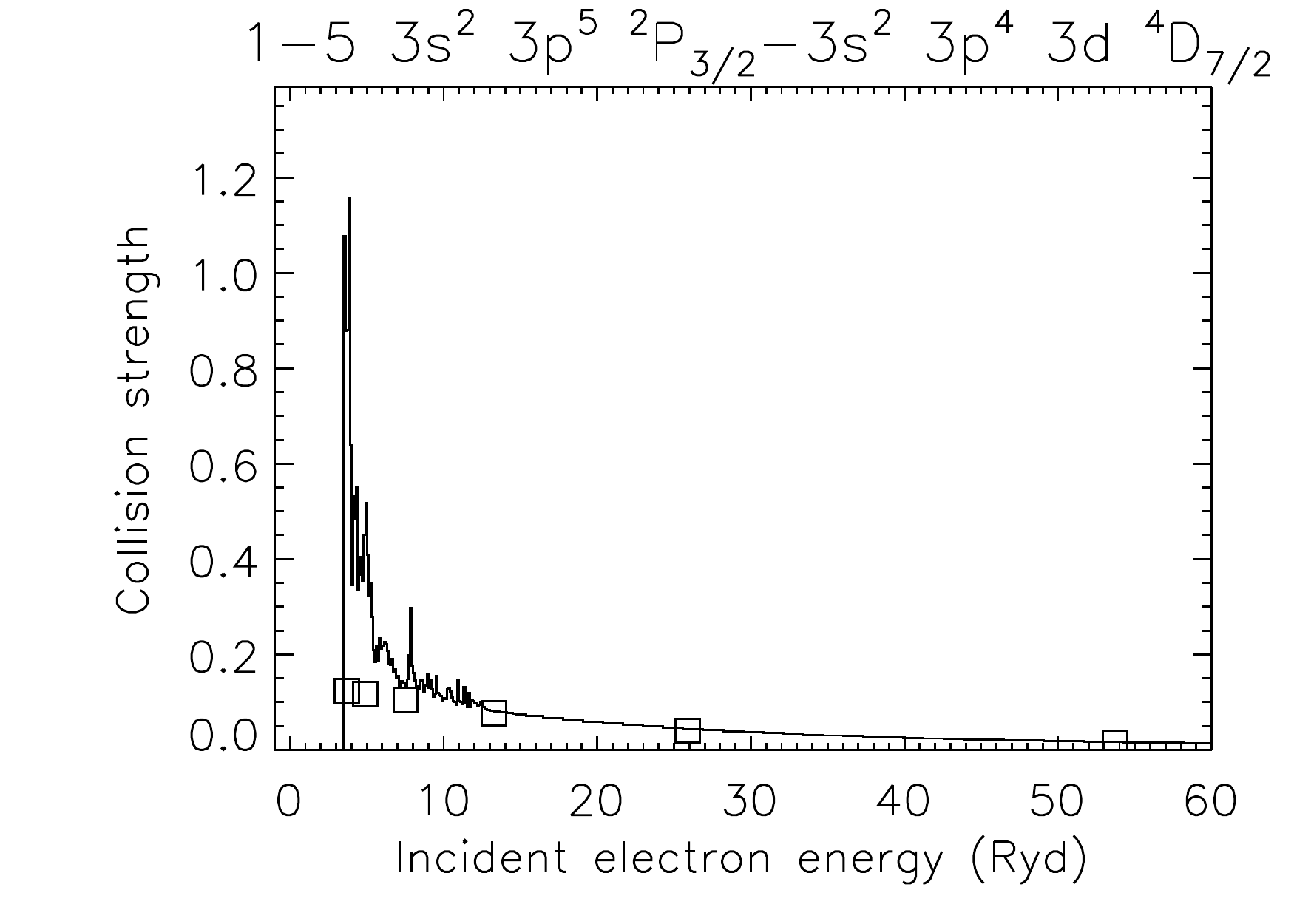}
}
\centerline{\includegraphics[width=7.5cm,angle=0]{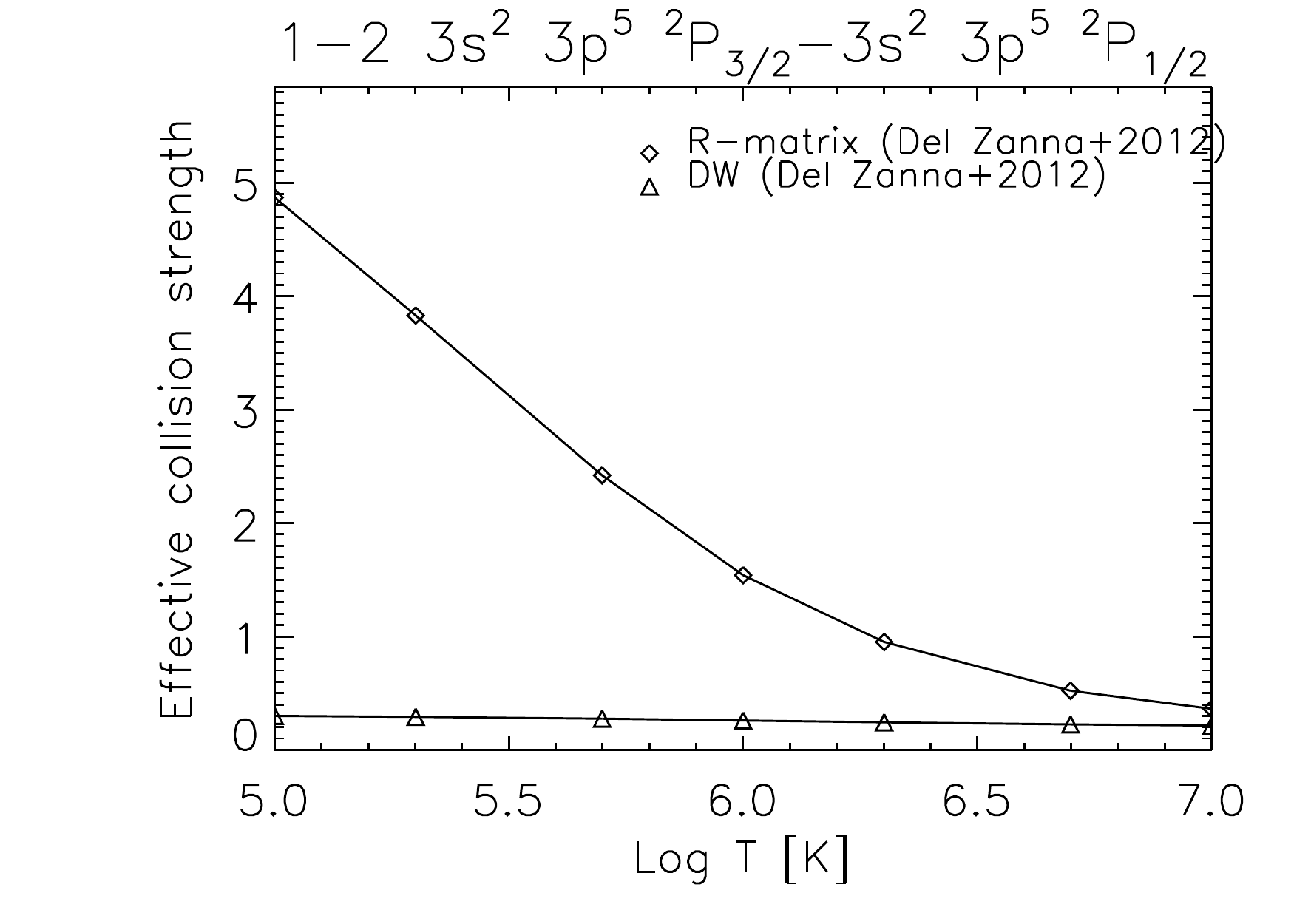}
\includegraphics[width=7.5cm,angle=0]{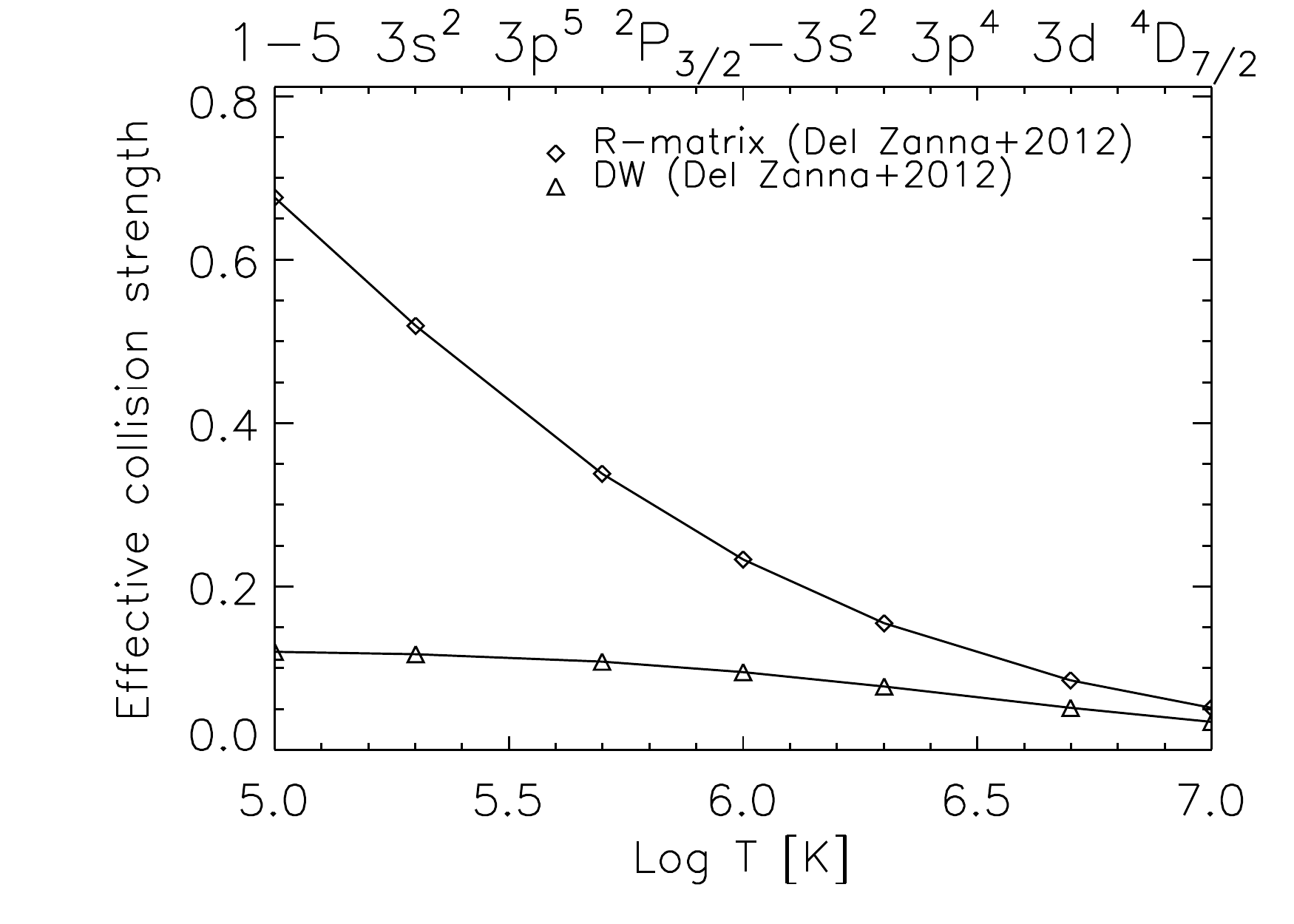}
}
 \caption{Above: collision strengths,  averaged over 0.1 Ryd in the resonance region. 
The data points are displayed in histogram mode.
Boxes indicate the DW values.
Below: thermally-averaged collision strengths, with other 
calculations. The plots on the left are for  
the \ion{Fe}{x} forbidden red coronal line within the ground 
configuration, those on the right
for the \ion{Fe}{x}  forbidden 257.26~\AA\ transition 
(adapted from \citealt{delzanna_etal:12_fe_10}).
}
\label{fig:fe_10_omega_ups1}
\end{figure}

\begin{figure}[!htbp]
\centerline{\includegraphics[width=7.5cm,angle=0]{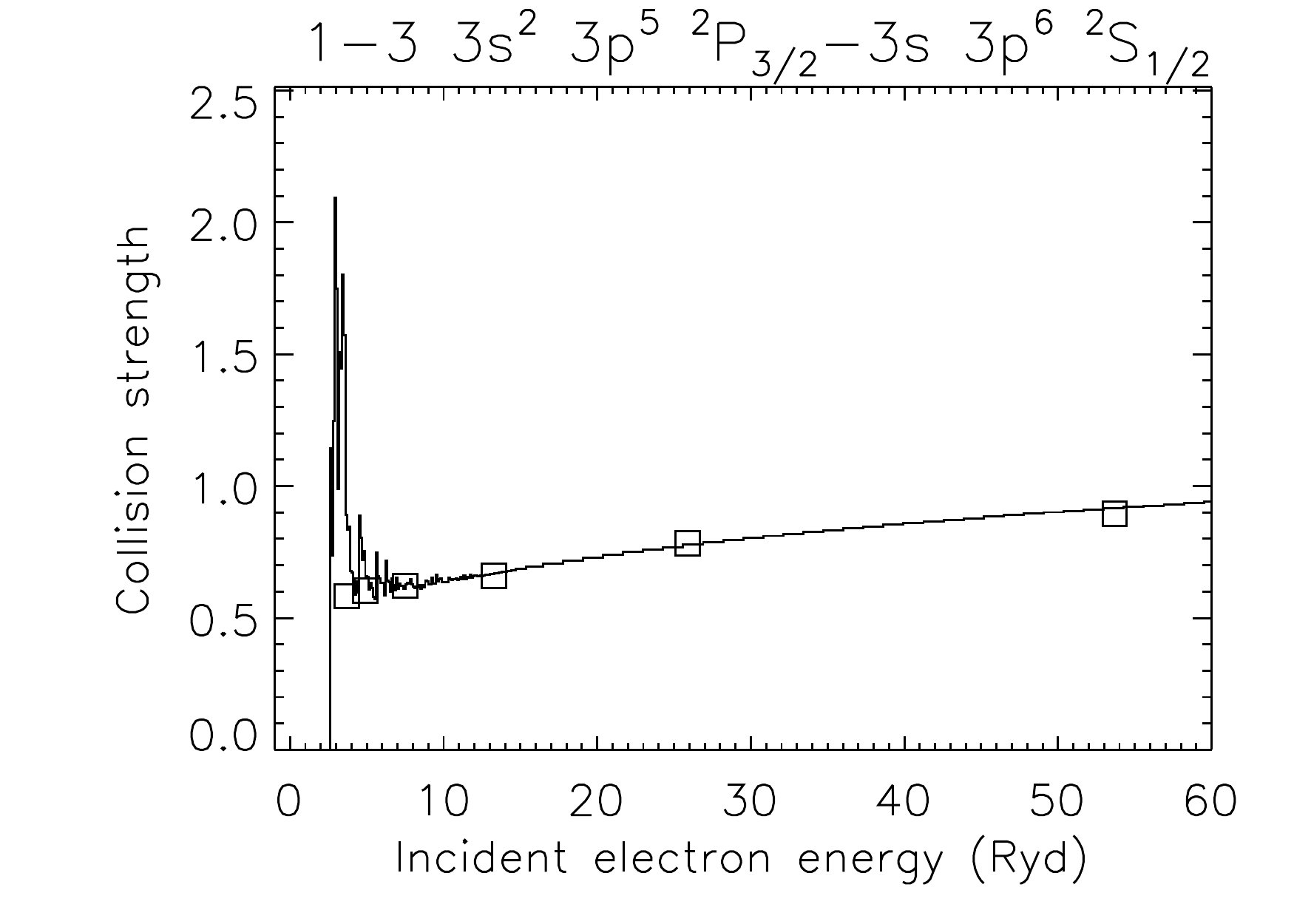}
\includegraphics[width=7.5cm,angle=0]{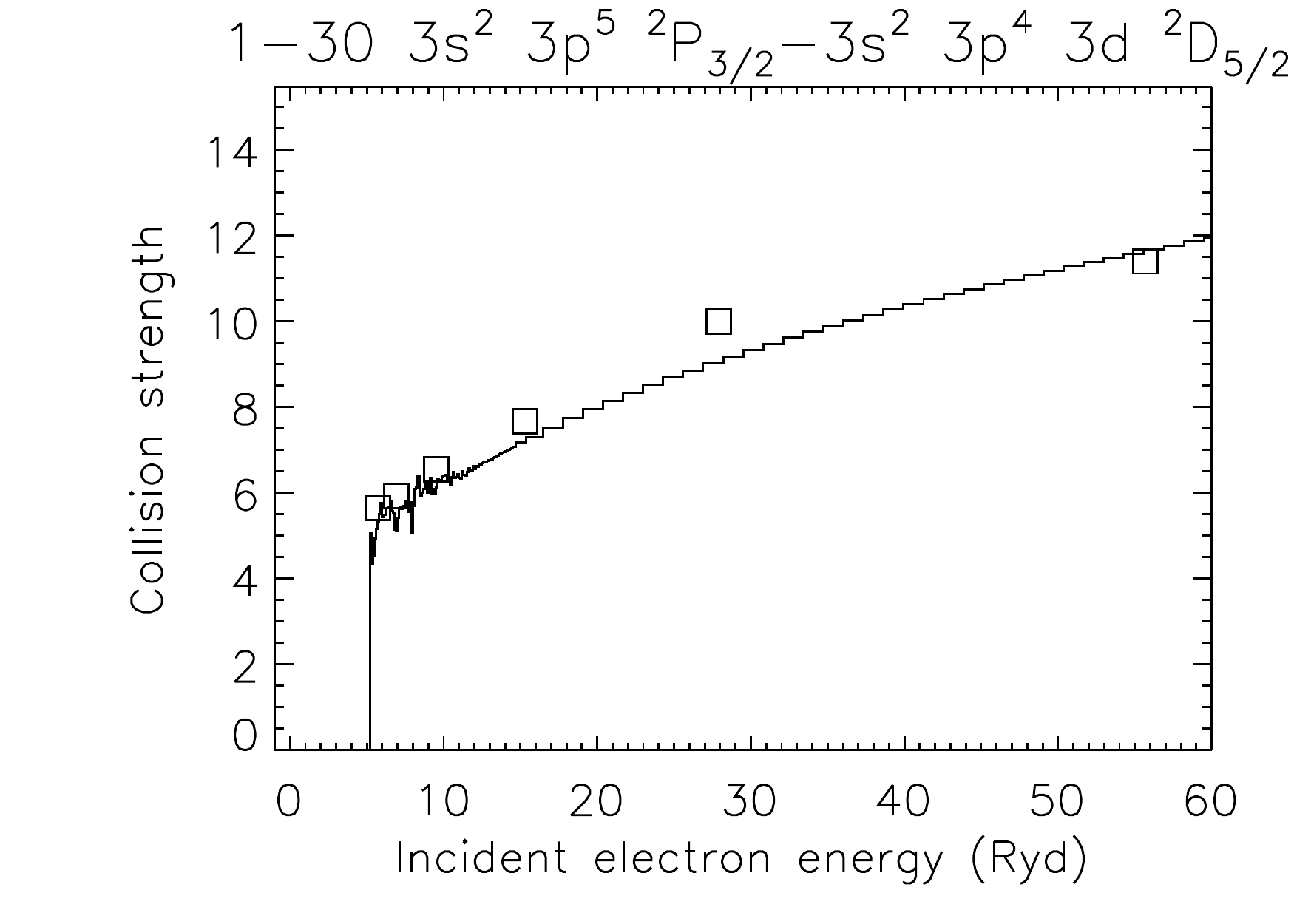}
}
\centerline{
\includegraphics[width=7.5cm,angle=0]{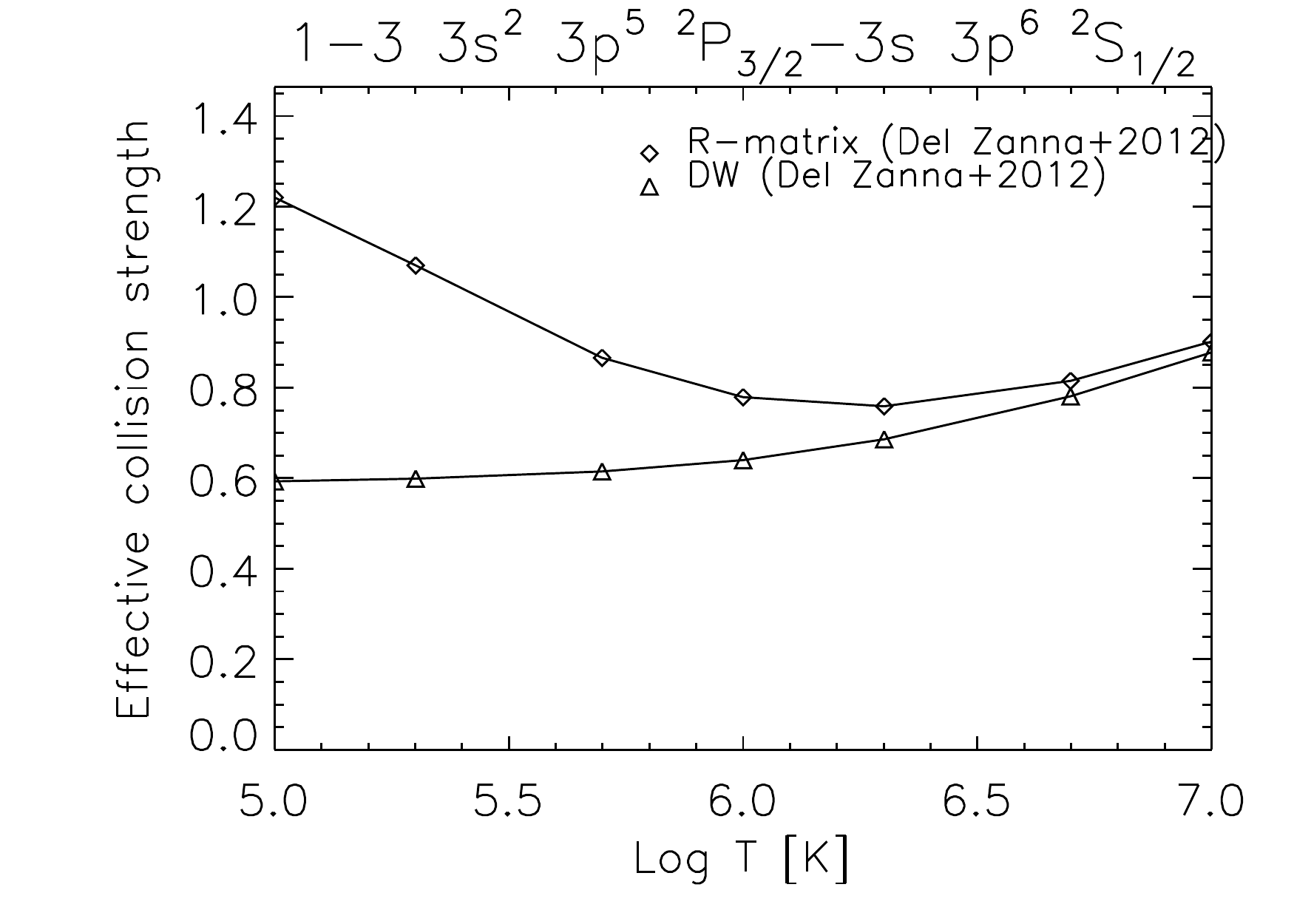}
\includegraphics[width=7.5cm,angle=0]{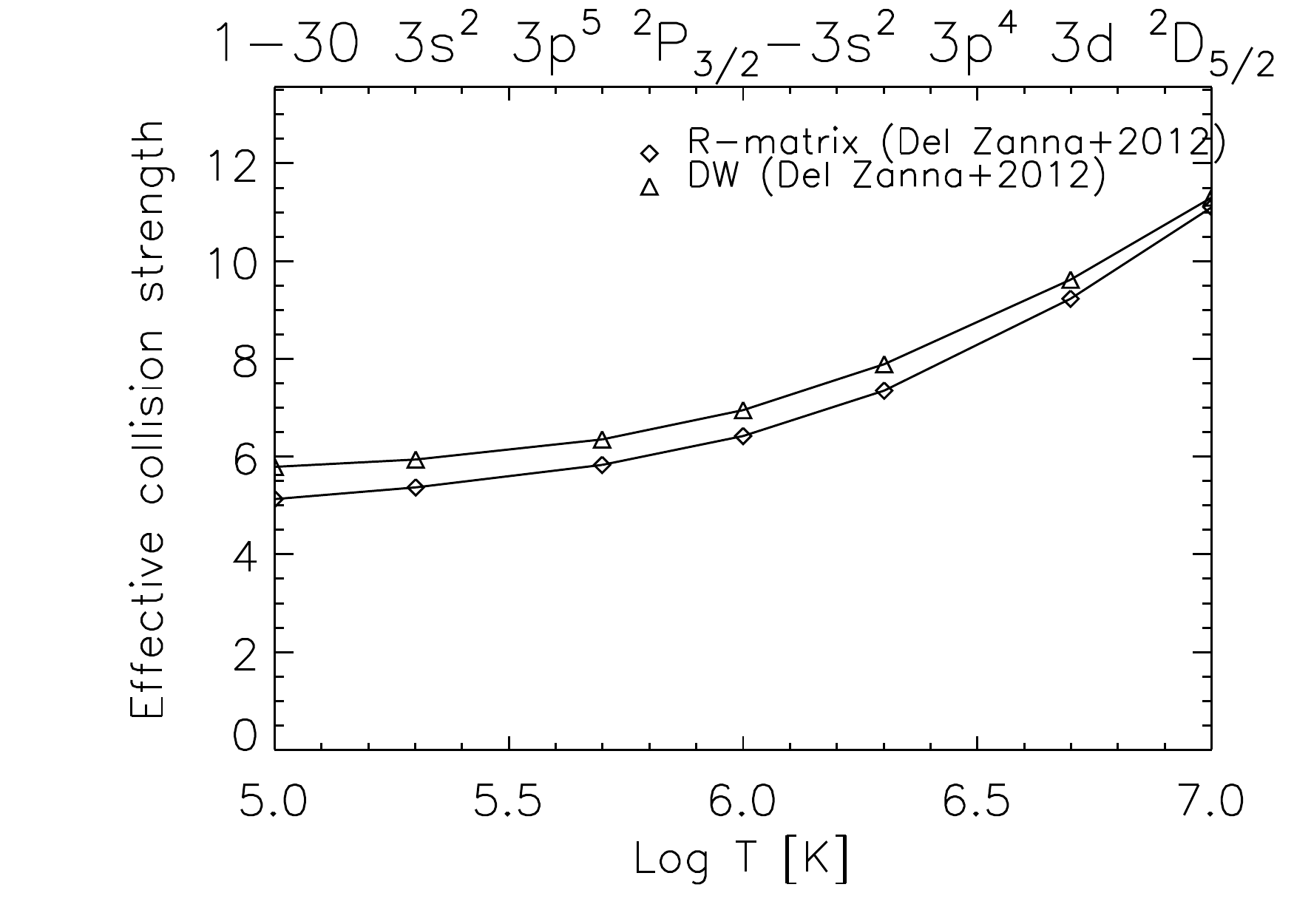}
}
 \caption{Above: collision strengths,  averaged over 0.1 Ryd in the resonance region. 
The data points are displayed in histogram mode.
Boxes indicate the DW values.
Below: thermally-averaged collision strengths, with other 
calculations. The plots on the left are for  
the \ion{Fe}{x}  allowed 
345.7~\AA\ transition, those on the right
for the allowed  1--30 174.5~\AA\ transition, the strongest in the EUV.
(adapted from \citealt{delzanna_etal:12_fe_10}).
}
\label{fig:fe_10_omega_ups2}
\end{figure}

Another factor in the accuracy of a calculation is the type of scattering 
approximation used. DW calculations are known to considerably underestimate
collision strengths for some types of transitions.
The resonances at lower energies typically affect the 
rates at lower temperatures, where the DW and $R$-matrix calculations
can provide very different values for the rates. 
The contribution of the resonances varies with the type of transition.
For dipole-allowed transitions, resonance effects can be very small.
On the other hand, they are usually significant for forbidden
and intersystem  lines,
because the background collision strength values are much lower. 
Fig.~\ref{fig:fe_10_omega_ups1} shows as an example the 
collision strengths and rates for forbidden transitions in \ion{Fe}{x},
one within the ground configuration, and one not.
In contrast, Fig.~\ref{fig:fe_10_omega_ups2} shows  
collision strengths and rates for two allowed transitions in \ion{Fe}{x}.
We can see that  resonances can occasionally  have a significant contribution to 
the collision strengths  even for strong dipole-allowed transitions.

\begin{figure}[!htbp]
\centerline{\scalebox{0.5}{\includegraphics{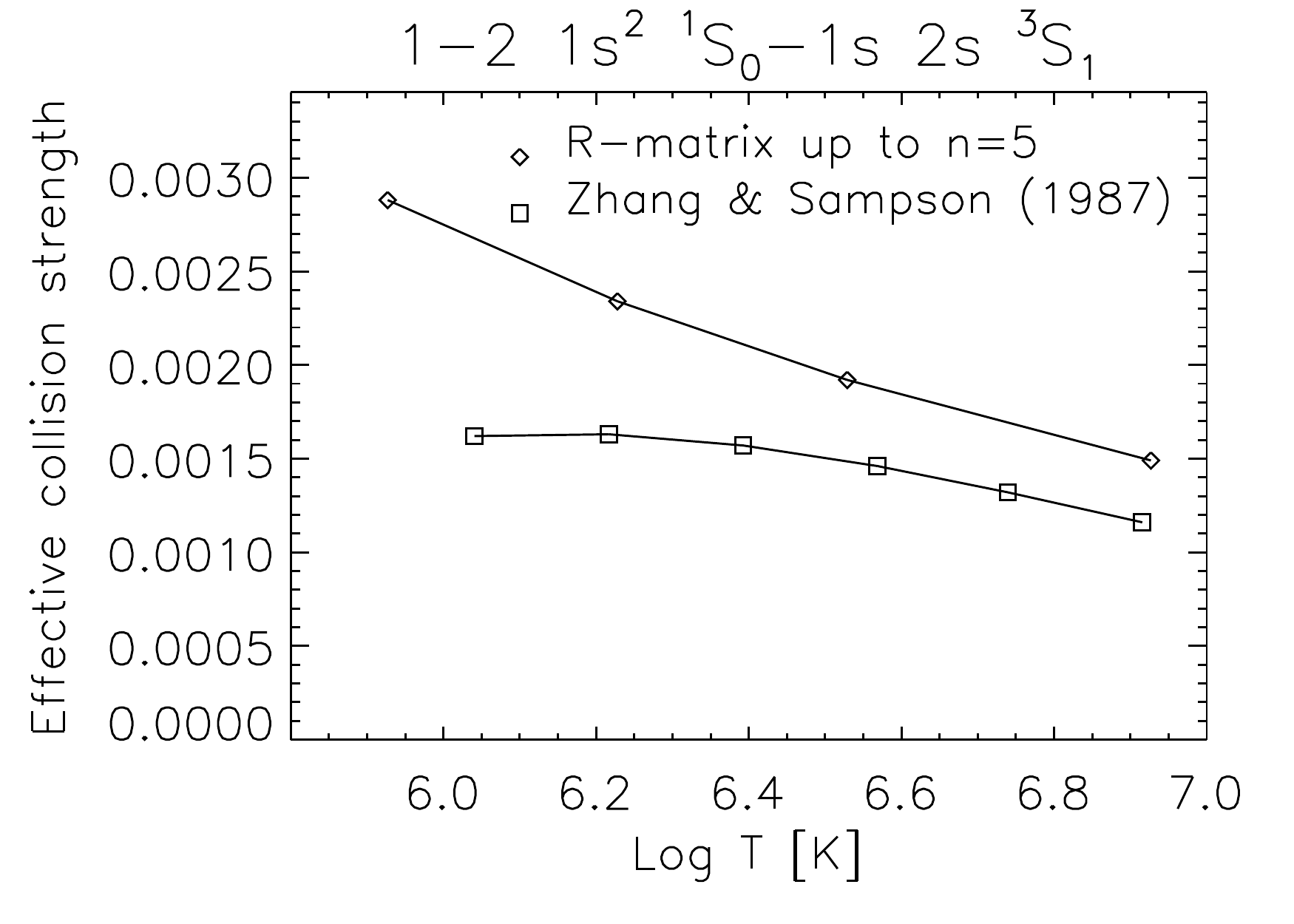}}}
\caption{The rate for the forbidden line in \ion{Si}{xiii} as calculated
with the  $R$-matrix  codes by  \cite{whiteford_etal:01}  and
by \cite{zhang_sampson:1987} with the DW method, including resonance enhancements.}
\label{fig:si_13_forbidden}
\end{figure}

We note that resonances can be included within DW calculations
with a perturbative treatment (DW + resonances), as 
e.g. in H.L. Zhang \& D. Sampson codes and FAC (although they are not normally included). 
We also note that generally the resulting collision strengths,
even with the resonances included, are lower than those 
calculated with the close-coupling $R$-matrix calculations.
An example is shown in  Figure~\ref{fig:si_13_forbidden},
 where the collision strengths for a  He-like
forbidden transition as calculated by \cite{zhang_sampson:1987} 
is compared to the result of an $R$-matrix 
calculation by  \cite{whiteford_etal:01} (see 
further details in \cite{badnell_etal:2016}).
This issue has been studied by \cite{badnell_etal:1993},
where significant differences between the DW + resonances
treatment and  close-coupling  $R$-matrix calculations for Mg-like ions
were found. 
Recently, \cite{menchero_etal:2016_fe21} compared two similar
large-scale calculations for \ion{Fe}{xxi}, an 
$R$-matrix one  and an earlier DW + resonances calculation of \cite{landi_gu:2006}. 
Again, the comparison showed a systematic underestimate of the 
cross sections by the perturbative results. 

\begin{figure}[!htbp]
\centerline{\includegraphics[width=0.7\textwidth, angle=0]{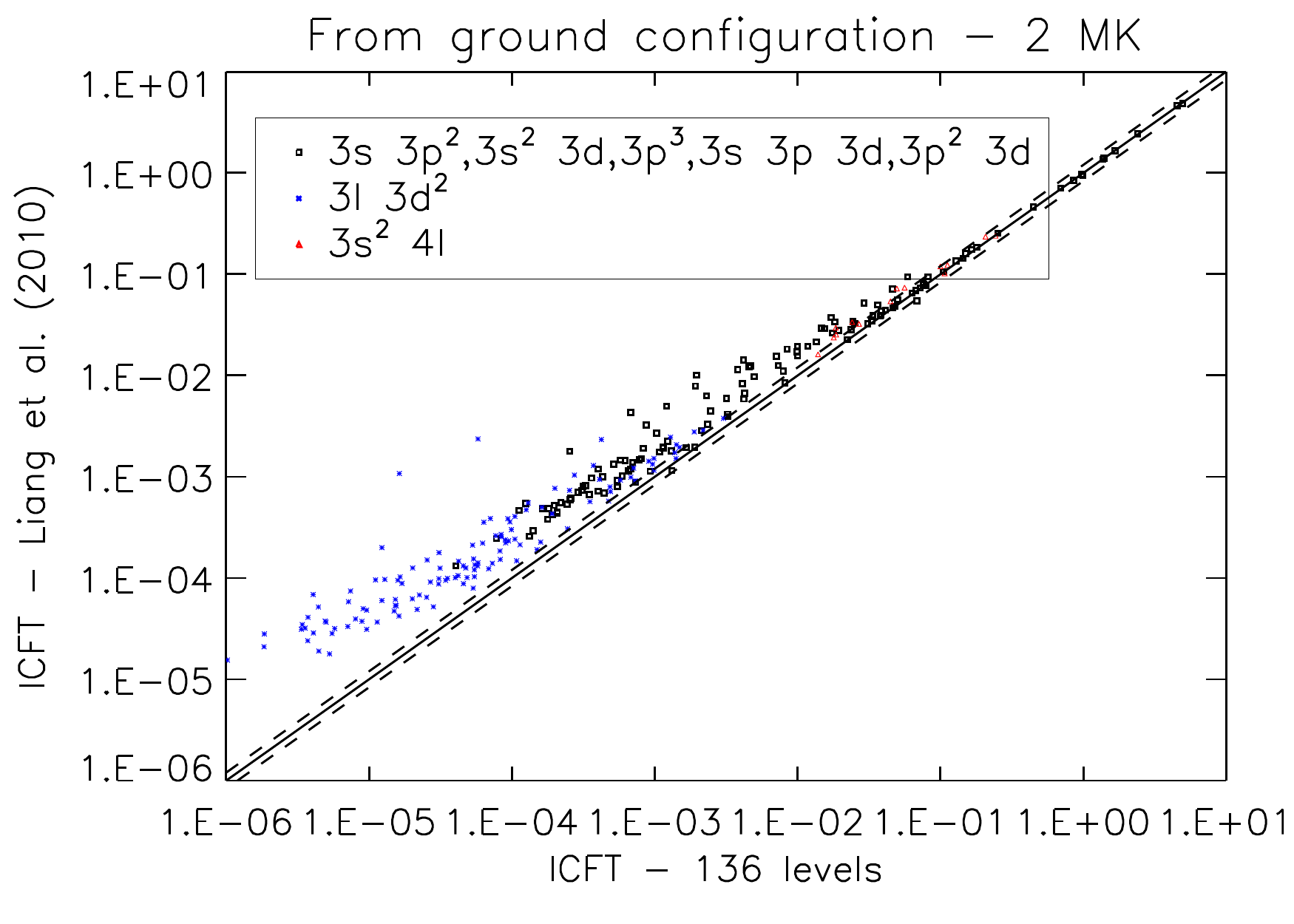}}
\caption{Thermally-averaged collision strengths $\Upsilon$
for  \ion{Fe}{xiv} transitions  from the ground  configuration 3s$^2$ 3p 
levels at 2 MK, as calculated with a smaller  target of 136-levels CI/CC, compared to 
those obtained by \cite{liang_etal:10_fe_14} with a  larger  CC (197).
 The dashed lines indicate $\pm$20\%.
Figure adapted from \cite{delzanna_etal:2015_fe_14}. }
\label{fig:fe_14_comp_ups}
\end{figure}

The resonances included in e.g. the $R$-matrix calculations 
are never complete, but depend on the number of levels included in the 
CC calculations. In other words,  the number of configurations/levels included in the 
close-coupling calculations has a relevance to the accuracy of the 
effective collision strengths, especially towards lower temperatures.
There are countless examples in the literature. 
Fig.~\ref{fig:fe_14_comp_ups} shows the values for  \ion{Fe}{xiv}
as calculated (with the same codes and approximations) with a 
more limited CC (136-levels) and a larger CC
(197 lowest-lying levels). It is clear that there is an overall 
enhancement in the values obtained from the larger calculation.
Some of the scatter is due to the difference in CI, related to the 
differences in $gf$ values as we have shown in  Fig.~\ref{fig:fe_14_comp_gf}.

Resonance effects are more important for transitions to 
low-lying levels, as the number and strength of the resonances decreases 
with the energy of the levels. 
Until recently, it was thought that resonances would have a negligible
effect on transitions to $n$=4 levels. However, non-negligible 
contributions are present, as discussed e.g. for the \ion{Fe}{x}
case in \cite{delzanna_etal:12_fe_10}.

Another factor in the accuracy of a calculation is the 
energy resolution in the resonance region.
Calculations with progressively finer energy grids and 
more levels can be used to provide some estimates of the 
accuracy and convergence of the calculations. 
One explicit example for \ion{Fe}{xi} was
provided in  \cite{delzanna_etal:10_fe_11}.

So far, we have discussed some of the uncertainties  in the basic 
atomic rates. Ultimately, the best way to provide an estimate of the 
uncertainty of an atomic dataset is to compare calculated
line intensities  with observed ones, so long as the instrumental
calibration is well known. 
However, two main questions still need to be addressed:

\noindent
(1) how do the uncertainties on the rates affect the line intensities?

\noindent
(2) how can one provide an uncertainty on the rates?

The question on how the uncertainties  on the rates
affect the spectral line intensities is not a trivial one to 
answer.  One actually needs to identify the main populating processes 
for the levels of interest and for the densities of interest, and
then assess which rates are the most important ones.

\begin{figure}[!htb]
 \centerline{\includegraphics[width=0.7\textwidth,angle=0]{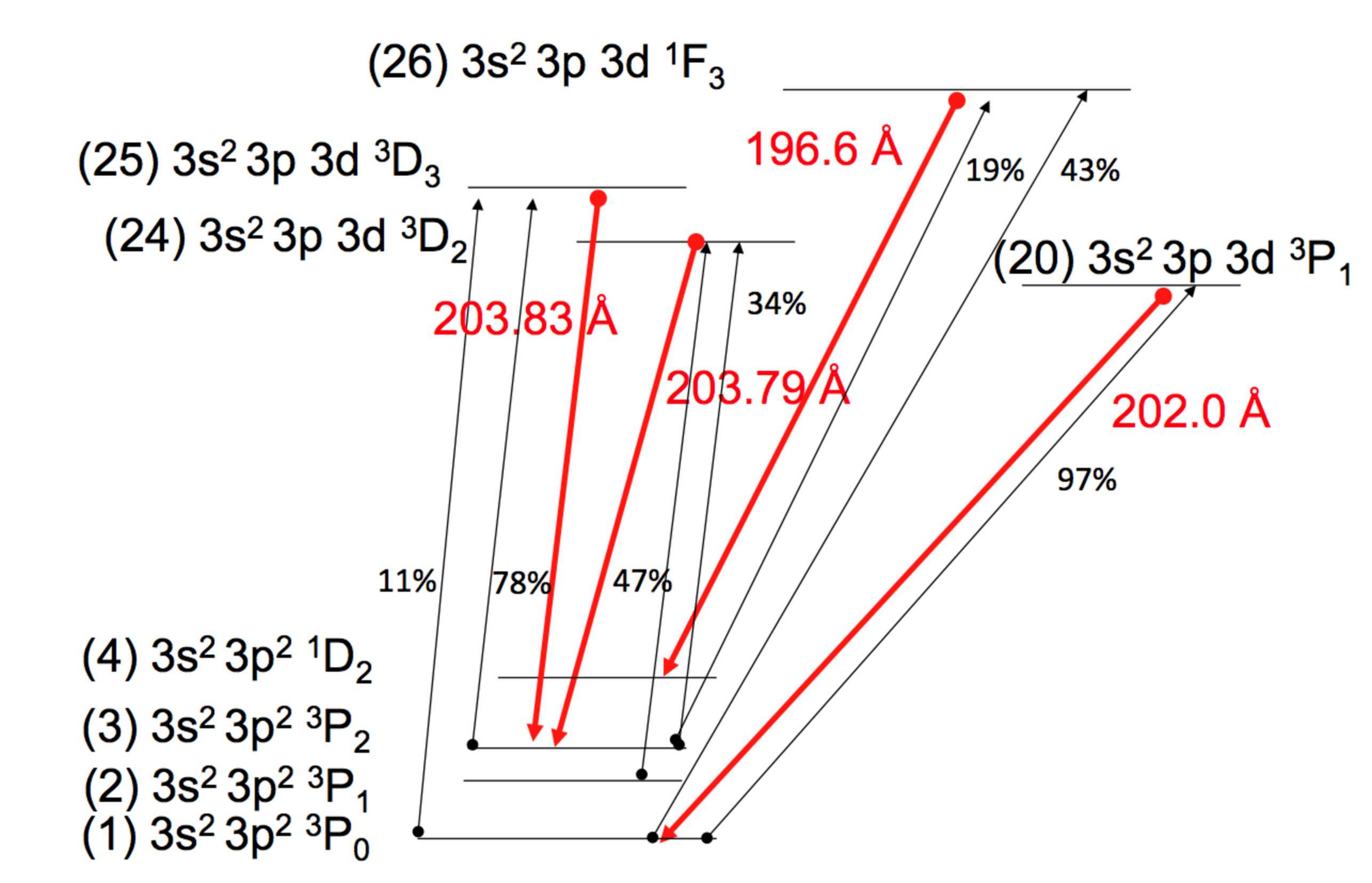}}
\centerline{\includegraphics[width=0.7\textwidth,angle=0]{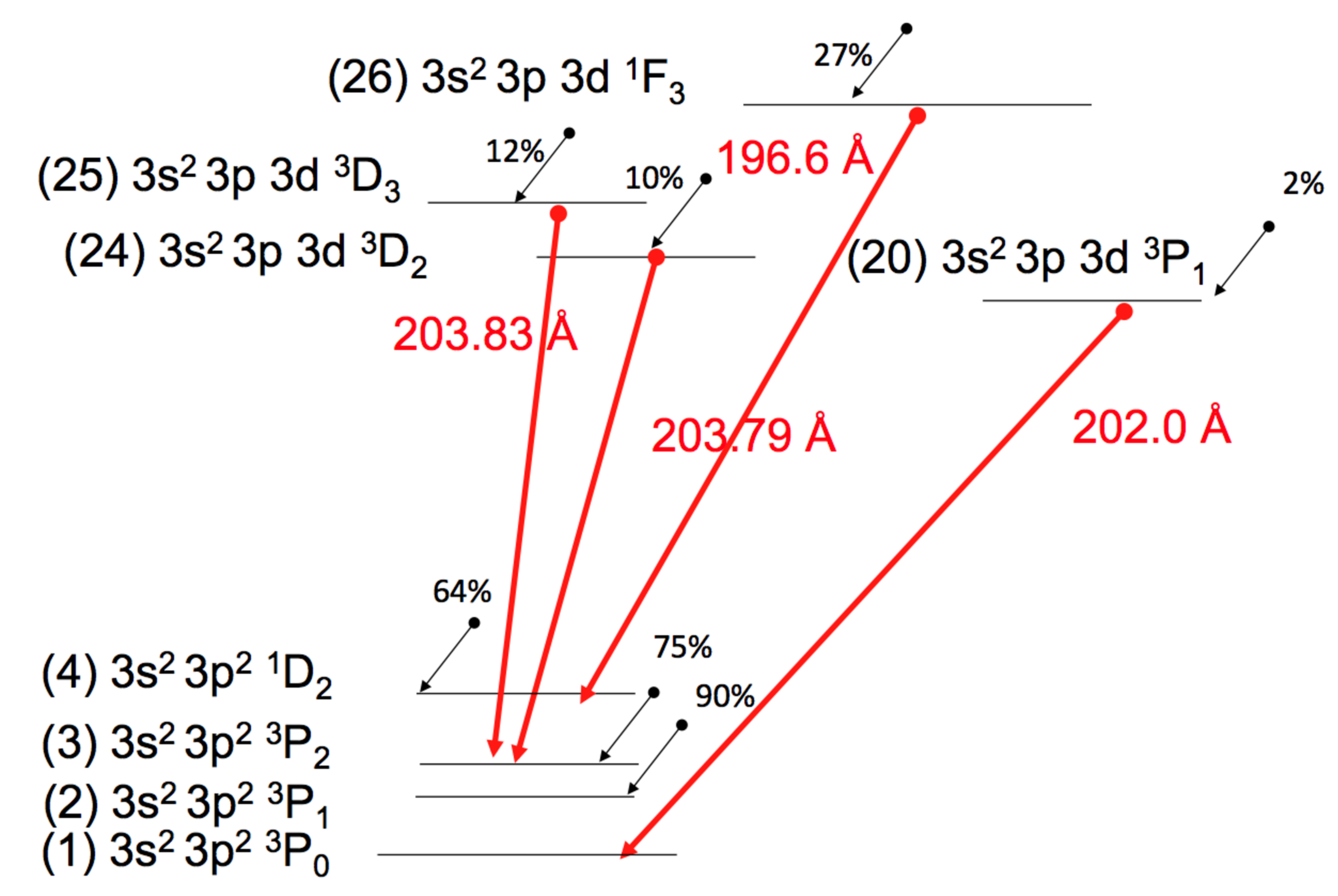}} 
  \caption{Main populating processes for \ion{Fe}{xiii} 
at a relatively low density of 10$^{8}$ cm$^{-3}$.
The red downward arrows indicate the main density diagnostic lines in the EUV.
The upward black arrows indicate the main populating transitions;
the percentage contribution to the population of the upper level is indicated.
The downward arrows and percentages on the right plot indicate the contributions
to the level populations due to cascading from all the higher levels. 
}
 \label{fig:fe_13_dens}
\end{figure}

\begin{figure}[!htb]
 \centerline{
\includegraphics[width=0.55\textwidth,angle=0]{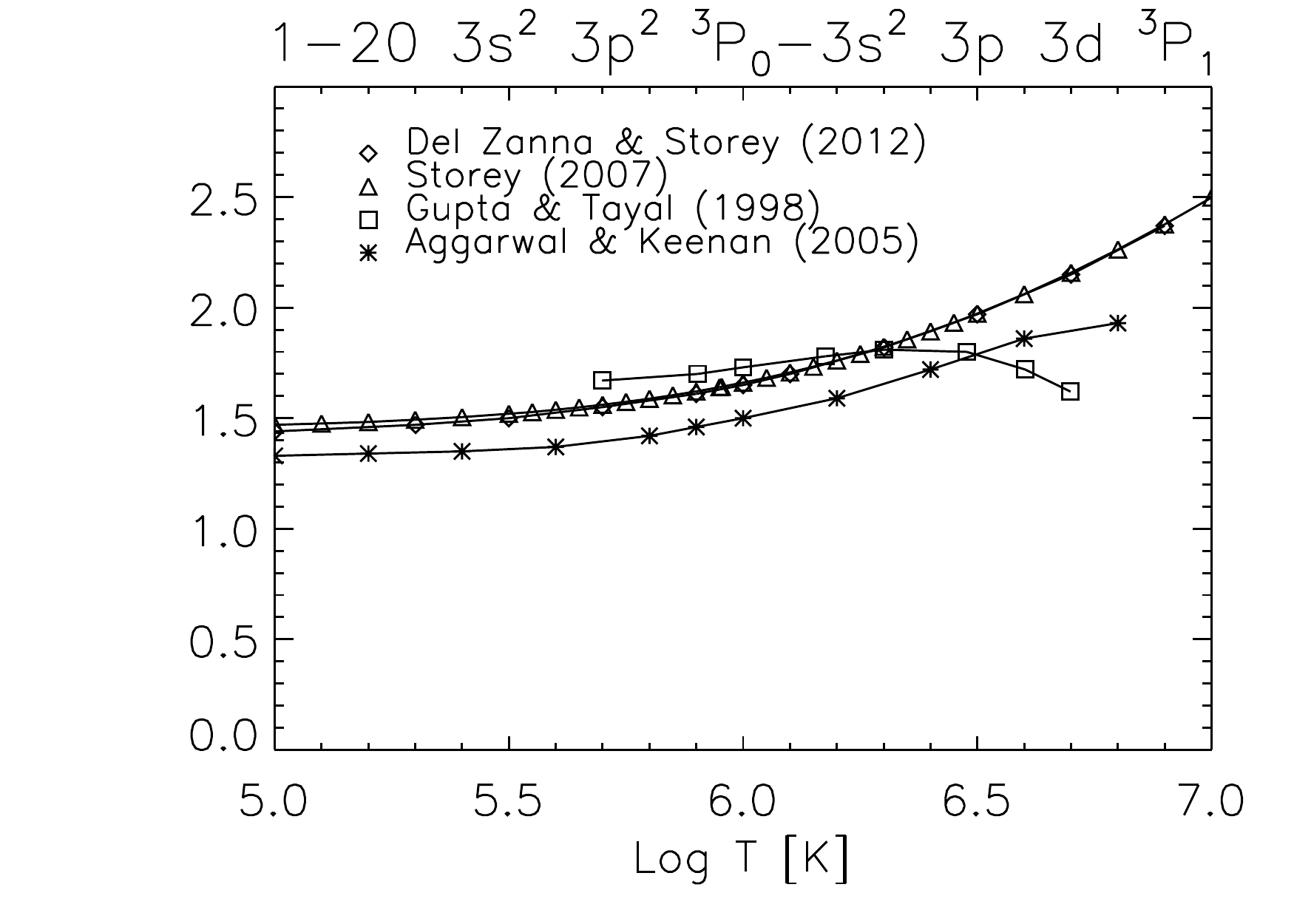}
\includegraphics[width=0.55\textwidth,angle=0]{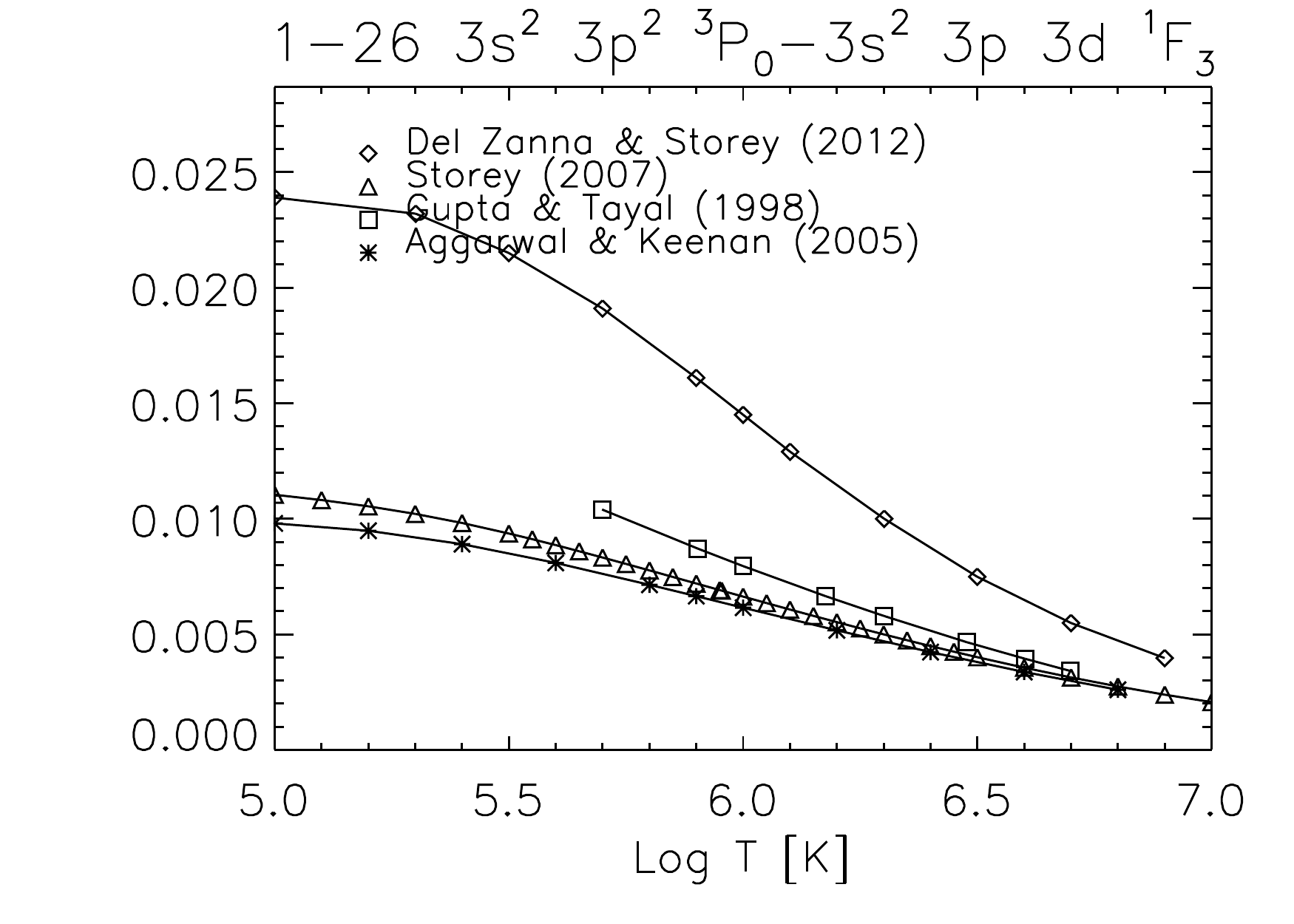}}
  \caption{Rates for electron excitation for two important allowed transitions in 
 \ion{Fe}{xiii}, as obtained from different calculations.}
 \label{fig:fe_13_ups}
\end{figure}

 Fig.~\ref{fig:fe_13_dens} shows the main 
populating processes for \ion{Fe}{xiii} at a relatively low density of 10$^{8}$~cm$^{-3}$,
using the atomic data calculated by \cite{delzanna_storey:12_fe_13}, 
which included cascading from $n=5,6$ levels.
The 202.044~\AA\ line is mainly (97\%) populated by 
direct excitation from the 3s$^2$ 3p$^2$ $^3$P$_{0}$ ground state via a strong dipole-allowed
transition. This means that the uncertainty for the  intensity of the  202.044~\AA\ line is solely
related to the uncertainty in the rate for direct excitation and the A value. 
The strength of this transition means that it is relatively easy to calculate the 
rate accurately.
Fig.~\ref{fig:fe_13_ups} (left) shows the effective collision strength 
for this transitions, as calculated by different authors.
With the exception of Gupta and Tayal's calculation, there is excellent 
agreement within a few percent between all the calculations.

On the other hand, the populations of the levels which provide the density 
sensitivity for this ion are due to excitation from several levels, 
and cascading from higher levels. For example, the 
3s$^2$ 3p 3d $^1$F$_{3}$ level, which produces a strong allowed transition at 196.6~\AA,
has a contribution to its population of about 43\% from direct excitation from  the ground state,
and about 24\% from cascading from higher levels (at a density of 10$^{8}$ cm$^{-3}$). 
The transition from the ground state is weak and forbidden, and as shown in 
Fig.~\ref{fig:fe_13_ups}, different calculations have provided quite 
different rates. For these reasons, the intensity of the 196.6~\AA\ line is intrinsically
more uncertain than the  202.044~\AA\ line.

Similarly, the populations of the 3s$^2$ 3p 3d $^3$D$_{3,2}$ levels which produce the 
strong allowed transitions at 203.8~\AA\ are mainly due to 
excitation from the 
 3s$^2$ 3p$^2$ $^3$P$_{1,2}$ levels, although non-negligible
contributions  (10\%) comes also from cascading from higher levels.
In turn, the populations of the 3s$^2$ 3p$^2$ $^3$P$_{1,2}$ levels 
are mainly due to cascading from higher levels, 
as indicated in Fig.~\ref{fig:fe_13_dens} (right).
Therefore, many rates are actually responsible for the intensities of the lines
at 203.8~\AA.

The fact that cascading has a considerable effect on the populations of the lower levels
means that  the size of the target can also in principle indirectly 
affect the intensities of the lines.
One recent example where the size of the 
target was found to have a  significant effect on the 
populations of the levels within the ground configuration
regards the important coronal iron ion \ion{Fe}{xii}  \citep{delzanna_etal:12_fe_12}. 
The increased population of the ground configuration
levels  in turn affected all the density diagnostics that 
were associated with these levels, and the forbidden lines within the ground configuration. 

Another issue is that the main populating processes change with density.
Overall, obtaining the correct theoretical behaviour for most lines requires accurate 
scattering and radiative data  for a large set of levels. 

The question of  how  one can provide an uncertainty on the rates is also non trivial.
One possibility is to compare the rates as obtained from the same set of codes and 
approximation but by changing some parameters, such as the CI/CC expansions in the 
calculations. 

\begin{figure}[!htb]
 \centerline{
\includegraphics[width=0.5\textwidth,angle=0]{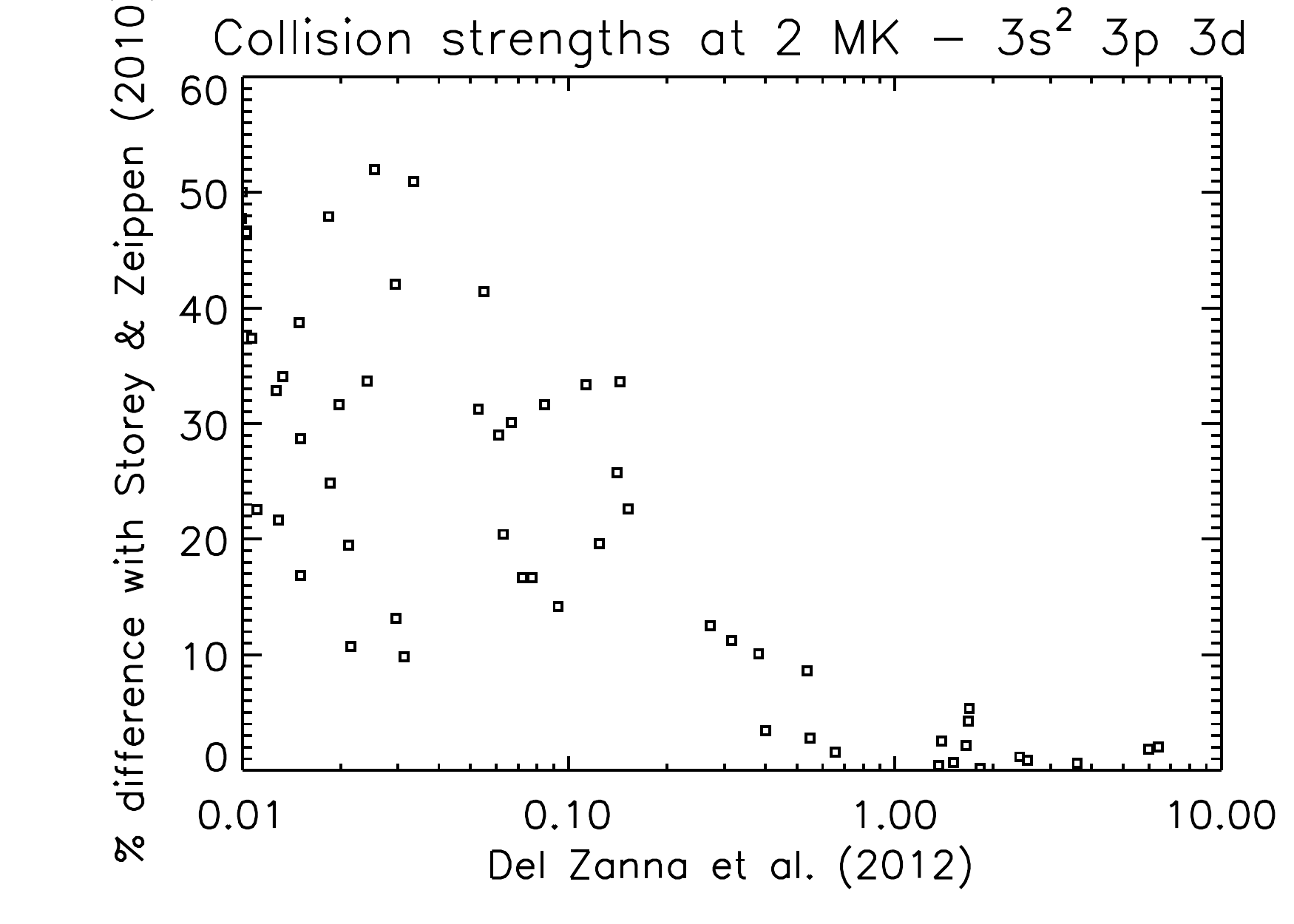}}
  \caption{Percentage difference in the A values of 
 \ion{Fe}{xiii} transitions within the energetically lowest 27 levels, 
as obtained with two different calculations.
}
 \label{fig:fe_13_ups_comp}
\end{figure}

For the excitation rates, as we have seen there are many factors that can 
affect the end result, so sometimes comparisons can show very large differences.
A more instructive comparison is when two calculations obtained with the same
codes and approximations are considered, and only the CI/CC expansions are 
different. Such a comparison is shown in  Fig.~\ref{fig:fe_13_ups_comp} (right),
where the rates at peak \ion{Fe}{xiii} abundance as calculated by 
\cite{delzanna_storey:12_fe_13} which included a total of 749 levels up to $n=4$ 
in the CC expansion, and by \cite{storey_zeippen:10}, where 
114 fine-structure levels within the $n=3$ complex were included.
It is clear that very good agreement is present for the strongest transitions,
while significant variations are present for the weaker ones.

\subsection{Electron excitation data }

A review of effective collision strengths for coronal ions 
 was published in a single volume (No. 57) of 
Atomic Data and Nuclear Data Tables in preparation
for the SoHO mission. This is based on a meeting that was held in 1992
\citep{lang:1994}. The volume contains  review articles, and includes 
discussions of the atomic data for  the main isoelectronic sequences.
Since then, a 
 significant number of new calculations  have been published  by various groups. 
Many are detailed in each of the CHIANTI database papers,  
where the most up-to-date calculations are assessed.
As in the case of the radiative data,  
the references for the data included in the CHIANTI database
can be found on-line at  
\url{http://www.chiantidatabase.org/chianti_direct_data.html}.

Atomic data for several ions have also recently  been calculated at Queens 
University of Belfast. They 
 are available at \url{http://web.am.qub.ac.uk/apa/data/}.
A large set of excitation data have  been 
produced by the Iron project and the earlier Opacity Project.
A  series of Iron Project papers have been published in 
Astronomy and Astrophysics. 
These data are available from the TIPTOPbase (\url{http://cdsweb.u-strasbg.fr/OP.htx}).

An even larger set of excitation data has been produced 
recently by the UK Rmax and UK APAP network (data are available at \url{http://www.apap-network.org}).
using mostly  the Iron Project R-matrix codes 
\citep[see the review in][]{badnell_etal:2016} 
but also the new DW code within the {\sc Autostructure} program \citep{badnell:11}.

Below we summarise the latest calculations for each isoelectronic sequence,
focusing on the  coronal and transition-region ions of diagnostic importance. 
Further details on previous calculations can be found in the  literature cited.

\subsubsection{H-like ions}

$R$-matrix 
calculations for several ions in this 
sequence (from He to Ne) were carried out by \cite{ballance_etal:2003}.
$R$-matrix  calculations for other ions also exist, and they are referenced in the 
CHIANTI database.
Recently, \cite{fernandez-menchero_etal:2016_h_he} showed that excitation rates to 
 levels higher than $n=5$ can  be estimated quite accurately with extrapolation procedures.

\subsubsection{He-like ions}

Many calculations for the He-like ions exist in the literature.
Earlier DW  calculations such as those of \cite{zhang_sampson:1987}
for the $n=2$ transitions have been supplemented by several $R$-matrix 
calculations. For the ions from  \ion{C}{v} to \ion{Zn}{xxix},
the most accurate $R$-matrix 
calculations are those of \cite{whiteford_etal:01}, mainly 
because  radiation damping  is included, unlike other calculations. 
In fact, radiation damping  is an important effect for H- and He-like ions, 
as discussed e.g. in \cite{gorczyca_badnell:1996,griffin_ballance:2009}.
\cite{whiteford_etal:01} calculated atomic data for 
all the transitions among the 49 levels up to  1s~5$l$.
These $R$-matrix rates  show a significant enhancement for the forbidden transition
in several ions of the sequence, compared to the  \cite{zhang_sampson:1987}
data, despite the fact that resonance 
enhancement associated with the 
1$s$ 3$l^{\prime\prime}$ $n^{\prime}l^{\prime\prime\prime}$ autoionizing levels
of the Li-like ions was added to the DW calculations.
Recently, \cite{fernandez-menchero_etal:2016_h_he} showed that excitation rates to 
 levels higher than $n=5$ can  be estimated quite accurately with extrapolation procedures.

\subsubsection{Li-like ions}

The most complete scattering 
calculations for electron-impact excitation of all Li-like ions from Be$^+$ to Kr$^{33+}$
is the APAP one from \cite{liang_badnell:2011}, where the radiation- and Auger-damped 
ICFT $R$-matrix approach was used. 
The model ions included 204 close-coupling (CC) levels, with valence electrons up to $n=5$ 
and core-electron excitations up to $n=4$.

\subsubsection{Be-like ions}

The early excitation rates  for many ions in this sequence were not very accurate
 since they  were interpolated  (see e.g.  \citealt{keenan_etal:86}).
This was pointed out  by \cite{delzanna_etal:08_mg_9} who carried out  an ICFT $R$-matrix 
calculation for  \ion{Mg}{ix}, an  ion of particular importance for 
its temperature diagnostic applications. 
Significant (up to 50\%) differences were found in some of the main 
lines and diagnostics, when using the $R$-matrix data rather than  the 
interpolated ones.

The largest calculations for many ions in the sequence 
(from  $\mathrm{B}^{+}$ to $\mathrm{Zn}^{26+}$)
 are from the APAP team \citep{fernandez-menchero_etal:2015_be-like}, where the 
ICFT $R$-matrix method was used. 
The CI and CC expansions included atomic states up to $nl=7{\rm d}$, for  a total
of $238$ fine-structure levels.
Good agreement with the 
previous  $R$-matrix calculations for \ion{Mg}{ix} 
\citep{delzanna_etal:08_mg_9} and \ion{Fe}{xxiii} \citep{chidichimo_etal:05}
were found.

However, these results were recently questioned by 
\cite{aggarwal_keenan:2015}, because of large differences with 
the results of their DARC calculation.
As shown by  \cite{fernandez-menchero_etal:2015_be-like}, 
 the DARC calculation of \cite{aggarwal_keenan:2015} 
were less accurate, because of their more limited CI and CC expansions.
As a follow-up, \cite{aggarwal_keenan:2016} carried out a 
larger DARC calculation for \ion{N}{iv}, using the same set of configurations
as that one adopted by  \cite{fernandez-menchero_etal:2015_be-like}. 
Large differences were still found, for transitions to highly excited levels.
We should point out that good agreement (within 10--20\%) is present for the strong 
transitions, and that transitions to high levels are typically more uncertain.  
This was  shown by a recent calculation for \ion{N}{iv} using a completely 
different approach and set of codes (the B-spline $R$-matrix) by 
\cite{fernandez-menchero_etal:2017_n_4}. The  B-spline $R$-matrix is 
computationally intensive but provides much more accurate energies
than the other methods. Significant differences in the rates  to highly excited levels 
as calculated by the ICFT, DARC and B-spline $R$-matrix were found.

\subsubsection{B-like ions}

The most up-do-date calculation for all boron-like ions from C$^+$ to Kr$^{31+}$
is the APAP one by \cite{liang_etal:2012}, where the ICFT  $R$-matrix method was used.
204 close-coupling levels were included in the target.
These data are a significant improvement for many ions where 
 only $n=2,3$ DW data were previously available.

\subsubsection{C-like ions}


 Collisional data for 49 fine structure levels of
\ion{Ne}{v} have been calculated by \citet{griffin_badnell:2000} using the ICFT
 $R$-matrix approximation.
For the important coronal ion 
\ion{S}{xi},  DW calculations from \cite{landi_bhatia:2003_s_11} are available.

For the important \ion{Fe}{xxi}, 
\cite{badnell_etal:2001} provided  ICFT
 $R$-matrix calculations with a large-scale model ion including 
 564 levels  in the CI expansion and 200 levels, up to $n=4$
in the CC expansion.
Recently, \cite{menchero_etal:2016_fe21} recalculated the cross sections
with the same codes, but including in the CC expansion all the 
 564 levels. A larger calculation, including some $n=5$ levels
was also carried out. 
Significant differences between this larger calculation and the earlier 
one were found in the collision strengths,  
 due to the lack of convergence of the earlier data, 
particularly for the  $n=3$ levels.

\subsubsection{N-like ions}


\citet{zhang_sampson:1999} provided collision strengths
 for all transitions between the 15 levels of the $2s^22p^3$,
$2s2p^4$ and $2p^5$ configurations of N-like ions, using the DW approximation.
Several other DW calculations for these ions exist.


\noindent
A few  $R$-matrix calculations for some important ions also exist.
For example, for  \ion{S}{x} \cite{bell_ramsbottom:2000} 
included the 22 levels of the
$2s^22p^3$, $2s2p^4$, $2p^5$ and $2s^22p^23s$ configurations.

\noindent
\cite{ramsbottom_bell:1997} produced  $R$-matrix calculations for Mg VI.

\noindent
The most complete  scattering calculations for the important \ion{Fe}{xx} ion
are the APAP ICFT $R$-matrix by \cite{witthoeft_etal:2007}, which included all
the main levels up to $n=4$.

\subsubsection{O-like ions}



For ions of this sequence, several  DW calculations exist,
with  some additional $R$-matrix calculations for the lowest levels.

\noindent
 \ion{S}{ix} is an important coronal ion. 
Several DW calculations exist, the most recent is from 
\cite{bhatia_landi:2003_s_9} which included 
\orb[2s 2]\orb[2p 4], \orb[2s2p 5], \orb[2p 6] and
\orb[2s 2]\orb[2p 3]\orb[3l ] ($l=s,p,d$), corresponding to 86 fine
structure levels.

\noindent
For \ion{Fe}{xix}, the most up-do-date calculations are the  
$R$-matrix ones by \cite{butler_badnell:2008}. 
The target included 342 close-coupling levels up to $n=4$.

\subsubsection{F-like ions}

The most important coronal ion of this sequence is
\ion{Fe}{xviii}. Large discrepancies between theory and 
observations in the strongest X-ray lines were finally resolved, as shown in 
\cite{delzanna:2006_fe_18}, with the  APAP 
ICFT $R$-matrix calculations of \cite{witthoeft_etal:06}. 
The most complete calculations for ions along this sequence are the 
APAP ones by \cite{witthoeft_etal:2007}, where the 
ICFT $R$-matrix approach was adopted. 
The model ions include  195 close-coupling levels up to $n=3$.

\subsubsection{Ne-like ions}

The most important coronal ion of this sequence is
\ion{Fe}{xvii}. As in the \ion{Fe}{xviii} case,
 large discrepancies between theory and 
observations in the strongest X-ray lines were present, until large-scale $R$-matrix
calculations were performed. 

\noindent
The most up-do-date calculations for ions from from Na$^+$ to Kr$^{26+}$
of this sequence are the APAP ones by \cite{liang_badnell:10_ne-like} data,
calculated with the ICFT $R$-matrix method.
The target included  209 levels, up to outer-shell promotions to $n=7$.

\noindent
As shown by \cite{delzanna:2011_fe_17}, the discrepancies in 
astrophysical observations were finally 
resolved with the  \cite{liang_badnell:10_ne-like} data, although the
Breit-Pauli $R$-matrix calculations of  \cite{loch_etal:06}
also showed good agreement with observation.

\subsubsection{Na-like ions}

The latest calculations for  Na-like ions from Mg$^+$ to Kr$^{25+}$ 
are the APAP ones by \cite{liang_etal:2009} using the ICFT $R$-matrix approach.
The close-coupling expansion included  configurations
up to $n=6$. 
Inner-shell  excitation data with the ICFT $R$-matrix  method 
with both Auger and radiation damping included were 
produced by \cite{liang_etal:2009b}.

\subsubsection{Mg-like ions}

The most complete set of scattering data for ions in this sequence have
been produced by the APAP network \citep{fernandez-menchero_etal:2014_mg-like}
with  ICFT $R$-matrix calculations for all the ions from $\mathrm{Al}^{+}$ 
to $\mathrm{Zn}^{18+}$. The target includes
 a total of 283 fine-structure levels in both the 
CI target and CC collision expansions,  from the configurations  
$1\mathrm{s}^2\,2\mathrm{s^2p^6}\,3\{\mathrm{s,p,d}\}\,nl$ with
$n=4,5$, and $l=0 - 4$.
 \ion{Fe}{xv} is one of the most important coronal ions in the sequence.

\noindent
 \ion{Si}{iii} is also an important ion for the 
transition region, with lines useful for a wide range of diagnostics.
The latest atomic data and 
diagnostics are discussed in \cite{delzanna_etal:2015_si_3}.

\subsubsection{Al-like ions}

The latest $R$-matrix scattering calculation for  \ion{S}{iv} was carried
out by \cite{delzanna_badnell:2016} within the APAP network.
A few problems with the previous calculations by 
\cite{tayal:2000_s_4} were found, but good agreement was obtained for the 
intersystem lines around 1400~\AA, which are the main diagnostic lines.

\noindent
{\bf \ion{Fe}{xiv} }

\noindent
Significant discrepancies between observed and 
predicted intensities of the 
very strong  \ion{Fe}{xiv} EUV coronal lines existed until the 
APAP scattering calculation of \cite{storey_etal:00}.
These calculations were further improved by the  APAP ICFT $R$-matrix calculations
by  \cite{liang_etal:10_fe_14}.
\cite{aggarwal_keenan:2014} carried out a DARC $R$-matrix calculation for this ion and 
showed large differences with the ICFT ones, suggesting that 
there might be a problem in the ICFT data. 
However,  \cite{delzanna_etal:2015_fe_14} showed that a smaller 
ICFT calculation with the same smaller target used in the DARC 
calculations (136 levels) for
 the  CI and CC expansions agreed quite well with the DARC one.
The \cite{liang_etal:10_fe_14} calculations were further improved 
in \cite{delzanna_etal:2015_fe_14} 
by retaining the full set of 228 levels for the CC expansion.

\subsubsection{Si-like ions}

The latest data for \ion{S}{iii} have been calculated with the 
$R$-matrix suite of codes  by \cite{hudson_etal:2012}. 

\noindent
{\bf \ion{Fe}{xiii}}

\noindent
This ion is important for the solar corona. 
For example, lines observed by Hinode EIS are used to measure electron densities.
The largest calculation for this ion 
is the  APAP ICFT $R$-matrix calculation
which included a total of 749 levels up to $n=4$ 
\citep{delzanna_storey:12_fe_13}.

\noindent
{\bf \ion{Ni}{xv}}

\noindent
 \ion{Ni}{xv} also produces several EUV 
lines  observed by Hinode EIS  that are useful for density and 
chemical abundance diagnostics of solar active regions.
The largest calculation for this ion 
is the  APAP ICFT $R$-matrix calculation
which included  levels up to $n=4$ \citep{delzanna_etal_2014_ni_15}.

\subsubsection{P-like ions (\ion{Fe}{xii})}



\ion{Fe}{xii} lines are prominent in the EUV region of the spectrum.
Several of them are routinely observed by Hinode EIS and are used to 
measure electron densities.
The density values obtained from these lines were significantly higher 
than those from other ions such as Fe XIII \citep[see, e.g.][]{young_etal:2009}.

\ion{Fe}{xii} is a notoriously difficult ion to carry out accurate calculations for.
It was only with the
APAP  $R$-matrix \cite{storey_etal:04} calculations that theory came close to experiment. 
The largest calculation for this ion is the APAP ICFT $R$-matrix
one by  \cite{delzanna_etal:12_fe_12}, with 912 levels up to  $n=4$.
This new data  provides densities from the  Hinode EIS lines 
about a factor of three lower
than the previous model by  \cite{storey_etal:04}.
This is due to the  combined effect of extra cascading 
and increased excitation, which changed  
the populations of the levels of the ground configuration significantly.
This affected the intensities of the forbidden lines and
the density diagnostics of the EUV lines.

\subsubsection{S-like ions (\ion{Fe}{xi})}


\ion{Fe}{xi} lines are prominent in the EUV and are 
observed routinely by instruments such as Hinode EIS.
\ion{Fe}{xi} is another  notoriously difficult ion to deal with. 
Several DW and $R$-matrix calculations have been published, but 
large discrepancies between observation and theory were present
for some of  the strongest lines of this ion.
The main problem was  with the decays from  three $n$=3, $J$=1  levels
which have a  strong spin-orbit interaction. 
The collision strengths to these levels are very sensitive to the target.
Good agreement with observations was found with an ad-hoc 
target and APAP ICFT $R$-matrix calculations by \cite{delzanna_etal:10_fe_11}.
A larger calculation,  including  996 levels up to $n=4$ 
by \cite{delzanna_storey:2013_fe_11}, showed similar effects as found in 
\ion{Fe}{xii}, i.e.  significantly increased (30--50\%) intensities of strong EUV lines 
from lower levels due to the  combined effect of extra cascading 
and increased excitation which affected the 
 populations of the (energetically) lowest levels in this ion.

\subsubsection{Cl-like ions }

\noindent
{\bf \ion{Fe}{x}}

\noindent
 \ion{Fe}{x} is another important coronal ion. 
The largest scattering calculation for this ion is the 
 APAP ICFT $R$-matrix one by \cite{delzanna_etal:12_fe_10}
which included levels up to $n=4$.
As in the \ion{Fe}{xii} and \ion{Fe}{xi} cases, the larger
calculation resulted in significant increases (30--50\%) in the intensities of several
EUV lines, in particular the strong  decays from the 
3s$^2$ 3p$^4$  3d  configuration observed by Hinode EIS 
(the 257.26~\AA\ self-blend), 
and from  the  3s 3p$^6$ $^2$S$_{1/2}^{\rm e}$ level.
The intensities of the   visible forbidden lines were also significantly 
affected,  as shown in \cite{delzanna_etal:2014_fe_9}.

\noindent
{\bf \ion{Ni}{xii}}

\noindent
The largest scattering calculation for this ion is the APAP
ICFT $R$-matrix one by \cite{delzanna_badnell:2016_ni_12}, which included levels 
up to $n=4$.
 With the exception of the three 
strongest soft X-ray transitions, large differences 
with previous calculations are found, mainly affecting the 
the decays from the lowest 3s$^2$ 3p$^4$ 3d levels and 
the forbidden UV lines.

\subsubsection{Ar-like ions }

\noindent
{\bf \ion{Fe}{ix}}

\noindent
\ion{Fe}{ix} is another important  ion for the solar corona,
as it produces strong EUV lines  from the 3s$^2$ 3p$^4$ 3d$^2$ and
3s$^2$ 3p$^5$ 4p configurations, some of which are useful for measuring electron 
temperatures.
As in the \ion{Fe}{xii} case, significant problems in the 
scattering calculations for this complex ion were present, until the 
\cite{storey_etal:02} APAP calculation.
The largest scattering calculation for this ion is the 
 APAP ICFT $R$-matrix one by \cite{delzanna_etal:2014_fe_9},
which included the main levels up to $n=5$.

\noindent
{\bf \ion{Ni}{xi}}

\noindent
\ion{Ni}{xi} lines are observed in the soft X-ray and EUV
by e.g. Hinode EIS and are useful density diagnostics.
The largest calculation for this ion is the 
 APAP ICFT $R$-matrix one by \cite{delzanna_etal_2014_ni_11},
which included the main configurations up to $n=4$.
Significant  differences with the  smaller $R$-matrix results of 
\cite{aggarwal_keenan:2007} and the DW calculations 
by \cite{bhatia_landi:2011} were found.

\subsubsection{K-like \ion{Fe}{viii} }

\ion{Fe}{viii} is another coronal ion that has been notoriously 
difficult to calculate, mostly because of the difficulty in 
obtaining an accurate atomic structure, as described by
\cite{delzanna:09_fe_8}, where Hinode EIS observations were
used. Several calculations have been carried out 
\citep[see, e.g.][]{griffin_etal:00,tayal:11_fe_8}, but it was only recently 
that agreement with the  Hinode EIS observations was achieved,
with the APAP ICFT $R$-matrix calculations by \cite{delzanna_badnell:2014_fe_8}.
The agreement was achieved by adopting 
a new method which employs semi-empirical corrections within the 
scattering calculation.

\subsubsection{Ca-like \ion{Fe}{vii}}

The most important ion of this sequence is 
 \ion{Fe}{vii}, which produces several EUV lines in the Hinode EIS 
spectral range. 
The most complete scattering calculation is the ICFT 
$R$-matrix one by the APAP team \citep{witthoeft_badnell:08}.
A benchmark of these atomic data against Hinode EIS 
observations has however indicated several problems 
\citep{delzanna:09_fe_7}, some of which are caused by incorrect identifications,
as also found by \cite{young_landi:09}. However, it is still 
unclear if the scattering calculations should  be revised.
The \ion{Fe}{vii} atomic structure is as complex as the 
\ion{Fe}{viii} one, with several  strong lines arising from levels 
which have a strong spin-orbit mixing.

\subsection{Ionisation/recombination rates}


 Direct ionization (DI) via electron impact 
is a smoothly-varying function of energy. 
Earlier approximate studies were those of \cite{lotz:1968}
and \cite{seaton:64} who introduced  semi-empirical formulae. The former 
was more appropriate for non-equilibrium plasma, while Seaton's approach provided
good cross sections near threshold.
\cite{burgess:1964acp} introduced 
the exchange classical impact parameter (ECIP) method which was used extensively.
\cite{burgess_etal:1977} later found out that the ECIP method produced 
good agreement with laboratory data, as good as Coulomb Born calculations,
especially for low charge states and simple ions such as those of the H- and He-like.
These authors also suggested a way to include the 
excitation of auto-ionizing states in the calculations, an approach 
that was later refined in 
\cite{burgess_chidichimo:1983}, where good agreement with laboratory data
was found.

DI  has often been provided
for astrophysical applications, 
as the excitation--autoionization (EA),   in the form of parametric fits. 
Examples can be found in \citet{younger:1981,arnaud_rothenflug:85}.

Various close-coupling 
calculations of DI of atomic ions have been carried out by 
Pindzola, Griffin  and collaborators for a number of years
\citep[see, e.g. the review by][]{pindzola_etal:2014}.

\begin{figure}[htbp]
\centerline{\includegraphics[width=0.4\textwidth, angle=90]{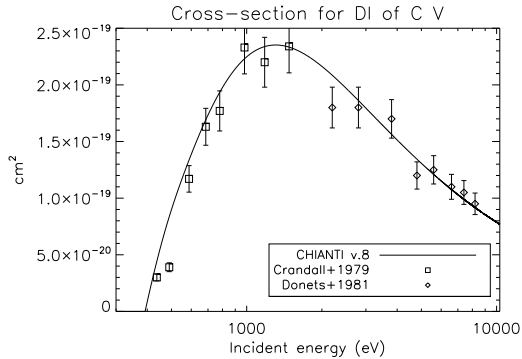}}
\caption{ Cross-section for direct ionization of \ion{C}{v}, as calculated and
measured in the laboratory (see  \citealt{dere:07} for details). }
 \label{fig:c_5_ioniz}
\end{figure}

A significant revision of the ionization rates was produced by \cite{dere:07},
where total cross-sections for all elements and ions of isoelectronic sequences up to zinc
were calculated with the DW code FAC. 
The calculations only considered the ground states, but  both the direct 
an the EA processes were included.
The cross-sections were fitted with splines in a  scaled energy domain,
also to check for the correct limit behaviour at high energies, as
in the case of the cross-sections for collisional excitation.  
Whenever available, experimental cross-sections were compared to the 
calculated ones, and sometimes used to apply empirical corrections to the 
 theoretical values.
The results, in terms of cross-sections and rates, are stored in the CHIANTI database.
An example is shown in Fig.~\ref{fig:c_5_ioniz}.

Further comparisons between these cross-sections
and  laboratory measurements can be found in a series of papers,
based on measurements of electron impact ionisation
carried out using the TSR storage ring  at
the Max-Planck-Institut fur Kernphysik in Heidelberg, Germany.
Details can be found in the review by \cite{hahn:2014} and in the 
cited references. A range of ions from many 
isoelectronic sequences (Li-like to K-like) have been studied,
with the exclusion of those ions with 
long-lived metastable levels which make the measurements of 
the total cross sections difficult.
Good overall agreement between theory and experiment  was found. 
It should be said that in most cases the recent calculations  by \cite{dere:07}
are in good agreement with previous ones, although in some cases 
differences of the order of 10--20\% have been found.

Radiative recombination (RR) and dielectronic recombination (DR) are normally
calculated separately, as the two processes are quite 
independent \citep{pindzola_etal:1992}.
RR rates are usually calculated from the  photoionization cross-sections using
the principle of detailed balance.
Earlier calculations only considered transitions between the ground states of the ions,
plus hydrogenic cross-sections for photoionisations from excited levels of the recombined ion to the ground level of the recombining ion \citep[see, e.g.][]{aldrovandi_pequignot:1973}.
More complete calculations along entire isoelectronic sequences were 
carried out by \citet{gu:2003} using the FAC code  \citep{gu:2008} and \citet{badnell:06} using 
AUTOSTRUCTURE.
\citet{badnell:06} calculated  rates 
 for all isoelectronic sequences up to the Na-like sequence.
Further AUTOSTRUCTURE calculations  for the Mg, Al and Ar sequences 
were published in  \cite{altun_etal:2007,abdel-naby_etal:2012} and \cite{nikolic_etal:2010}, 
respectively.
Additional rates for specific  iron ions were published  by \citet{badnell:2006b}.
The above rates plus additional ones for the minor ions, originating from a 
variety of sources, are available within the CHIANTI database.

\begin{figure}[htbp]
\centerline{\includegraphics[width=0.6\textwidth, angle=90]{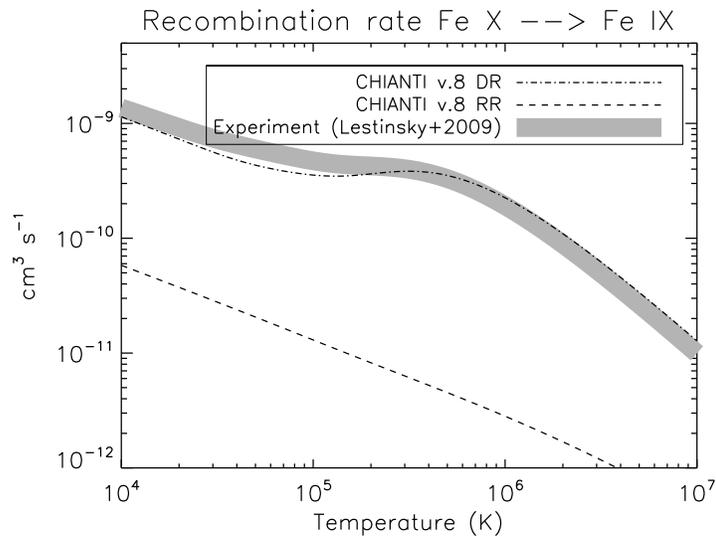}}
\caption{ Recombination rates (radiative, RR, and dielectronic, DR)
 for \ion{Fe}{x} (i.e. to recombine  \ion{Fe}{x} into  \ion{Fe}{ix}), as stored in 
the CHIANTI database version 8. The grey area indicates the experimental 
estimate for the DR rate from \cite{lestinsky_etal:2009}. 
The thickness of the grey area represents the approximate experimental uncertainty. 
}
 \label{fig:fe_10_recomb}
\end{figure}

Most of the 
DR rates currently used in atomic physics codes have been  obtained within the 
DR project \citep{badnell_etal:03} with AUTOSTRUCTURE calculations.
One of the most recent papers in the series is 
\cite{abdel-naby_etal:2012} (Al-like). 
Comparisons between DR rates and laboratory measurements 
show overall good agreement, although discrepancies are present in 
some cases.
The most significant series of studies employed the 
electron-ion merged-beams method applied to the observations 
obtained at the
heavy-ion storage ring TSR of the Max-Planck-Institute
for Nuclear Physics in Heidelberg, Germany. The DR rates 
have been  obtained by estimating the contribution of the RR rates.
A series of papers, a collaboration between N.R. Badnell 
and the Heidelberg group have been published. 
\cite{schippers_etal:2010} reviewed such studies for the 
important iron ions. 
 Fig.~\ref{fig:fe_10_recomb} shows as an example a comparisons 
between the experimental DR rate for recombination of   \ion{Fe}{x}
as estimated by \cite{lestinsky_etal:2009} (grey area) and the theoretical 
DR rate as available from the CHIANTI database version 8 (dot-dash line).
Note the good agreement between experiment and theory, and also 
how the RR contribution is nearly negligible for this ion.

\subsection{Proton excitation data}

A significant number of cross-section for ion  excitation by proton impact
has been calculated following the semi-classical
impact-parameter approach introduced by  \cite{seaton:64_protons},
which is accurate only  for low energy collisions. 
A number of methods have been developed to improve the calculations.
Details can be found in the reviews by  
 \cite{reid:1989} and especially \cite{burgess_tully:2005},
where some problems in earlier calculations were highlighted.

For example, \cite{bely_faucher:1970} 
used a symmetrised first
order semi-classical approximation to calculate the proton
cross-sections  for a large number of
ions with configurations $2p$, $2p^5$, $3p$ and $3p^5$.

\cite{kastner:1977}
used the first order semi-classical
approximation for low energies while for higher energies 
different approximations were adopted.

Landman, in a series of papers, computed proton rates 
retaining the classical treatment for the proton trajectory, but 
using  a symmetrised, semi-classical close-coupling method 
to calculate  transition probabilities 
\citep[see, e.g.][]{landman:1973}.

Faucher used  a fully  close-coupling method to 
compute proton cross-sections for a number of ions \citep[see, e.g.][]{faucher_etal:1980}.

\cite{reid_schwarz:1969} developed a symmetrised, semi-classical
close-coupling  approximation including  polarization effects, which was used 
by V.J.~Foster, R.S.I.~Ryans and others  (see, e.g. \citealt{ryans_etal:1999})
 to calculate proton rates for a large number of ions.

Reviews of  proton excitation data can be found in 
\cite{copeland_etal:1997} and in \cite{young_etal:03}, where the data included in 
the CHIANTI database are listed.

\subsection{Benchmarking  atomic data }

Our current  knowledge of the solar spectrum is not only 
based on a long history of observations (as we summarised 
in the first Section) and of atomic calculations 
(as we briefly summarised), but also on a continued effort to
establish line identifications, blends, and benchmark the 
reliability of the atomic calculations against experiment.
In this section, we provide a few examples and references to 
recent studies. As the subject is vast, we mainly focus on 
Iron lines, as they  are abundant in the XUV spectra of the 
solar corona, and provide many diagnostic applications.

\subsubsection{Line identifications}

For simple ions, atomic structure calculations have been 
accurate enough to easily identify all the main spectral lines,
hence establish  all the experimental energies.
For more complex ions, and until recently, ab-initio
calculations of wavelengths were typically off  by several \AA,
hence it has not been easy to establish experimental energies,
and occasionally identifications have been revised several times.
Surprisingly, there are many spectral ranges in the XUV where a large fraction
of the observed lines are still unidentified.

Most of the identifications in the past have been based on 
laboratory spectra, where typically the sources have very different
conditions to solar plasma  (e.g. where densities are very low in 
comparison). The identifications were therefore biased in some sense.

A  review of EUV $n=3 \to n=3$ transitions 
was published in a Culham  report \citep{fawcett:1971_rev}, while 
 a comprehensive bibliographical review of all the identifications in the XUV 
($n=2 \to n=2$, $n=3 \to n=3$, $n=2 \to n=3$, $n=4 \to n=4$)
 was  published in a later Culham  report by \cite{fawcett:1990_rev}.

General compilations of line identifications and wavelengths in the XUV
were produced by  \cite{kelly:87,shirai_etal:00}. 
These compilations, with more recent updates, have been included in 
the  National Institute of Standards and Technology
(NIST) database\footnote{http://physics.nist.gov}. 
However, for several important ions the NIST 
energies are still in need of revision and are not up-to-date with the
most recent literature.
In what follows, we mostly  mention recent literature for the 
important Iron ions, 
 because this is not included in the NIST compilations.

The identifications of the strongest XUV lines of the most important  
Iron ions have been assessed in a series of benchmark papers,
where a significant number of new identifications have  been proposed,
and where detailed references to previous original identifications can be found: 
\ion{Fe}{vii} \citep{delzanna:09_fe_7},
\ion{Fe}{viii} \citep{delzanna:09_fe_8},
\ion{Fe}{ix} \citep{delzanna:09_fe_7},
\ion{Fe}{x} \citep{delzanna_etal:04_fe_10},
\ion{Fe}{xi} \citep{delzanna:10_fe_11},
\ion{Fe}{xii} \citep{delzanna_mason:05_fe_12},
\ion{Fe}{xiii} \citep{delzanna:11_fe_13},
\ion{Fe}{xvii} \citep{delzanna_ishikawa:09,delzanna:2011_fe_17},
\ion{Fe}{xviii} \citep{delzanna:2006_fe_18},
\ion{Fe}{xxiii} \citep{delzanna_etal:2005_fe_23},
\ion{Fe}{xxiv} \citep{delzanna:2006_fe_24}. 
This work was mostly focused on the Hinode EIS and SDO AIA spectral ranges,
where we knew atomic data and line identification were lacking.

\begin{figure}[!htbp]
\centerline{\includegraphics[width=0.6\textwidth]{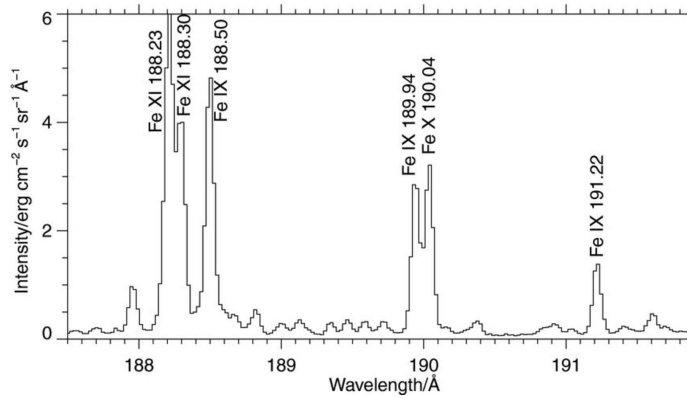}}
\vspace{1cm}
\centerline{\includegraphics[width=0.63\textwidth]{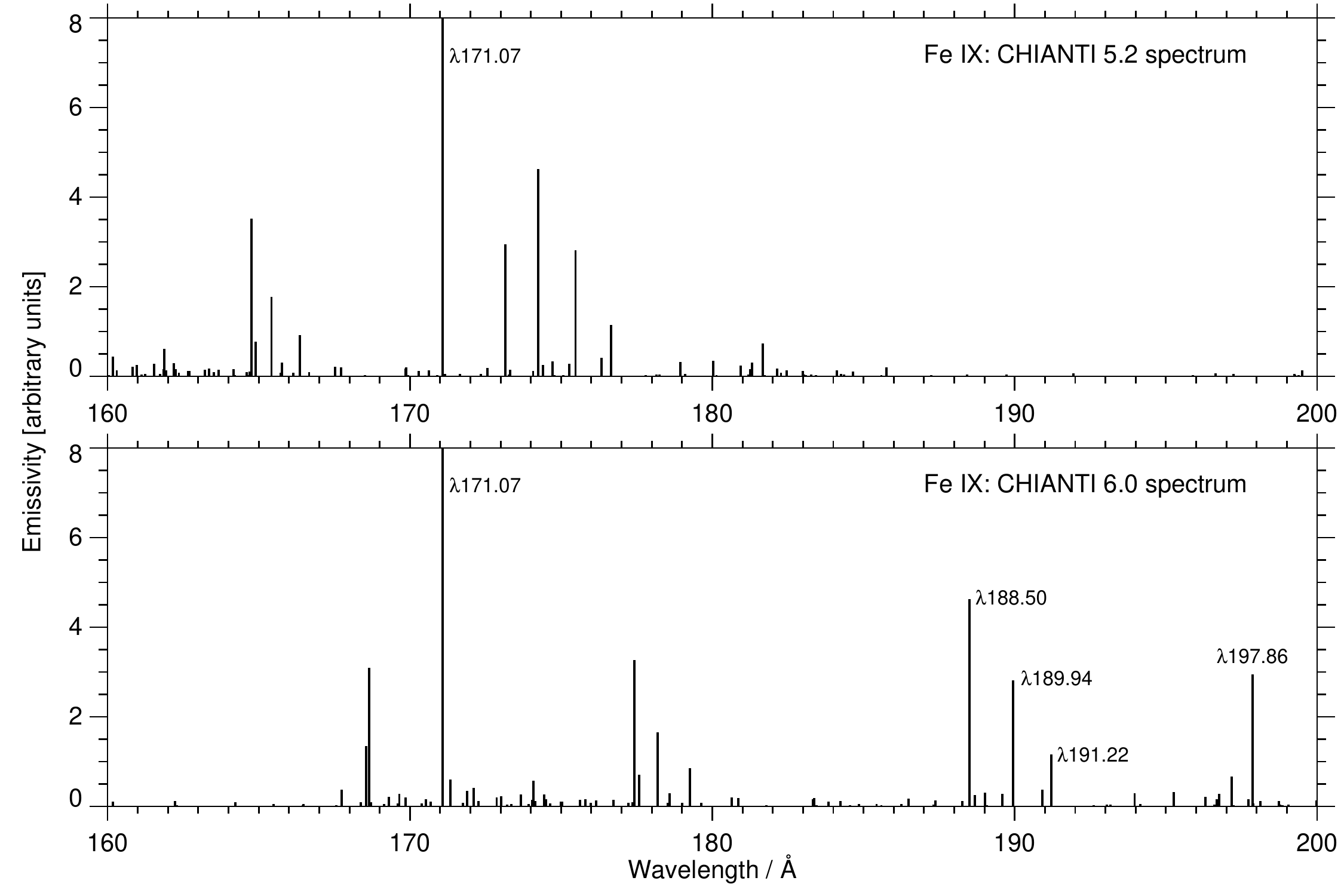}}
\caption{Top: Hinode EIS spectra of the newly identified \ion{Fe}{ix} lines 
by \cite{young:09}. Bottom:  simulated spectra \citep{young_landi:09}.}
\label{fig:young_2009_fe_9}
\end{figure}

Similar work on  \ion{Fe}{vii}, \ion{Fe}{viii} and \ion{Fe}{ix}
has been carried out by \cite{young:09,landi_young:2009,young_landi:09}. 
Most of these identifications have been included in the CHIANTI
version 7.1 and 8 database, and later confirmed by several analyses of 
solar and laboratory spectra. 
Many of the new identifications in the EUV have been possible thanks to the 
accuracy and spectral resolution of the Hinode EIS spectrometer. 
They have also been greatly aided by the ability to spatially resolve
spectra of very different sources. By simply comparing monochromatic images
of unidentified lines with those of well-known lines, it is straightforward 
to know the formation temperature of a line.
Fig.~\ref{fig:young_2009_fe_9} shows the  
\ion{Fe}{ix} lines  identified by \cite{young:09}.

We now briefly mention some of the most relevant studies,
broadly dividing them by wavelengths.

\smallskip\smallskip\noindent
{\bf Soft X-rays}
\smallskip

Perhaps the most famous identifications in the XUV are those
based on Edl{\'e}n's pioneering (and to date best) soft X-ray
spectra in the 1930s, dominated by $n=4 \to n=3$ transitions in highly 
ionised  ions from Iron and Iron-peak elements.
Among these, the most famous ones are those of 
\ion{Fe}{x}, where  Edl{\' e}n identified several transitions
\citep{edlen:37_cl-like}. These identifications established the 
splitting of the $^2$P within the ground configuration.
This allowed  Edl{\' e}n, in a seminal paper for 
solar physics \citep{edlen:42}, following a suggestion from 
\cite{grotrian:39}, to identify  the famous 
bright red coronal forbidden line at 6374.6~\AA\
as the \ion{Fe}{x} transition
$^2$P$_{3/2}$--$^2$P$_{1/2}$  within the ground configuration \citep[see][]{swings:43}.
The forbidden line had 
been observed for seventy years during total solar eclipses,
and it was with this identification that it was realised that 
the solar corona is a million degree plasma.

{Edl{\' e}n} had also identified  several lines from other ions in the soft X-rays,
from Iron  and several other elements, so he  
later extended similar identifications for a large number of 
ions in the visible. Most of his 
identifications in the soft X-rays were correct, but some
identifications of forbidden lines had to be revised several times,
sometimes by {Edl{\' e}n} himself. Various other researchers at Lund
and other institutes continued {Edl{\' e}n}'s work. He developed 
simple methods to check the identifications and suggest new ones
by studying ions along isoelectronic sequences. Even today,
many energies remain unknown experimentally but are relatively
 well known thanks to such analyses at Lund.

\begin{figure}[!htbp]
\centerline{\includegraphics[width=0.6\textwidth]{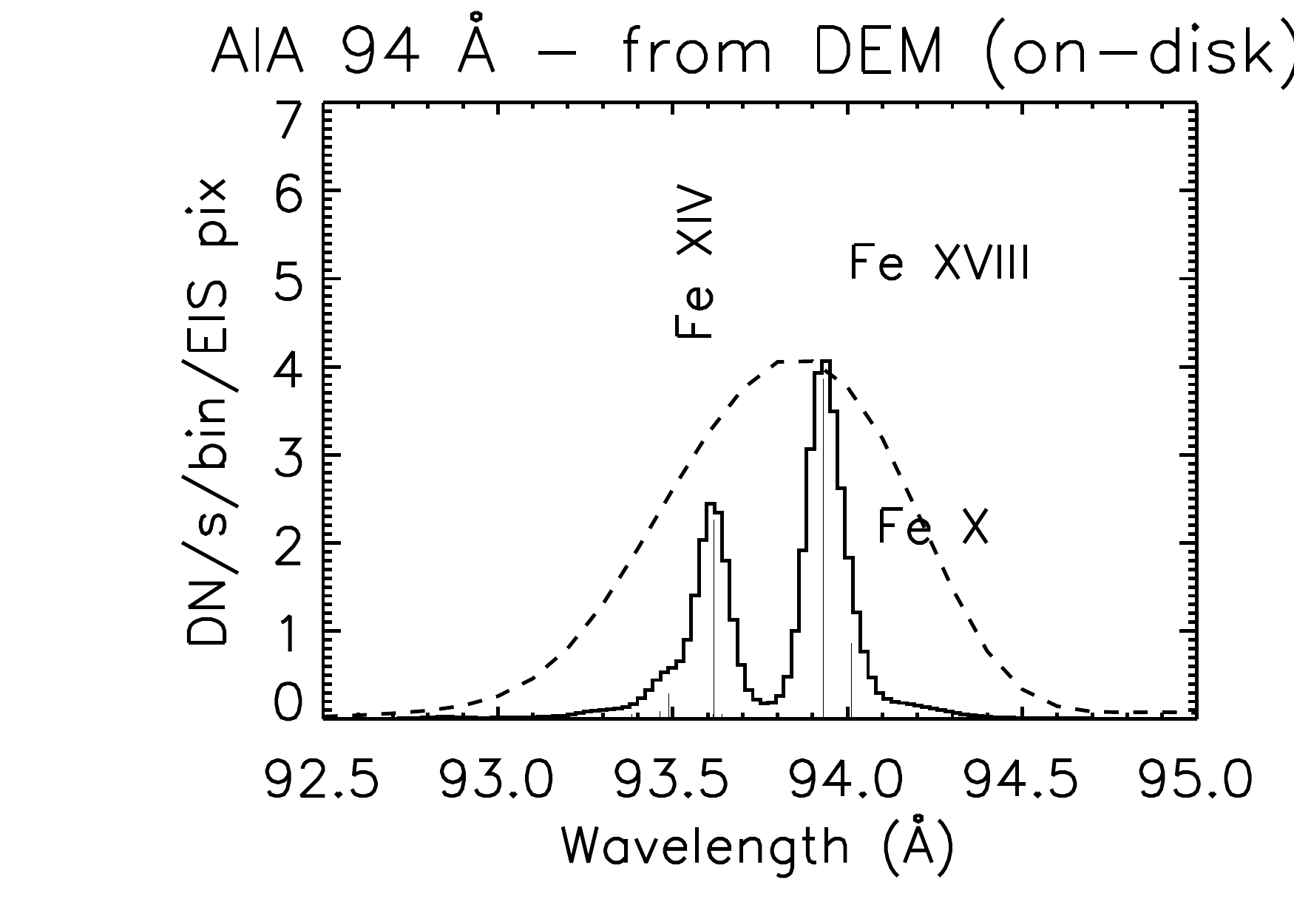}} 
\caption{AIA 94~\AA\ count rates for the core of an active region,
as obtained with a DEM analysis using Hinode EIS observation.
The dashed curve is the normalised AIA  effective area 
\citep{delzanna:2013_multithermal}.}
\label{fig:aia_94}
\end{figure}

Edl{\' e}n's work in the soft X-rays was extended to the coronal 
Iron ions (3s$^2$3p$^2$~4$l$, $l=$s,p,d,f levels) by
 the fundamental  laboratory work of \cite{fawcett_etal:72}.
However, in both cases, many but not all of the observed lines
with strong  oscillator strengths were identified. 
It turns out that some of the strongest $n=4 \to n=3$ transitions in solar 
spectra produced by the coronal Iron ions were actually not 
identified by Edl{\' e}n or Fawcett, partly because they were not very strong
in the higher density  laboratory spectra. 
Their identification was provided by  \cite{delzanna:12_sxr1}.
This  was  possible thanks to large-scale
scattering calculations within the APAP network
 and significant benchmark work using solar spectra and 
the original laboratory glass photographic plates from Fawcett. 
Some of the \cite{fawcett_etal:72} identifications have been
revised, in particular that of the 
 \ion{Fe}{xiv} 3s$^2$ 3d $^2$D$_{3/2}$--3s$^2$ 4p $^2$P$_{1/2}$ 
transition with the 93.61~\AA\ line, which turns out 
to be the strongest contribution to the SDO AIA 94~\AA\ 
band in the cores of active regions (whenever  \ion{Fe}{xviii}
is not present), as discussed  in \cite{delzanna:2013_multithermal}
and shown in  Fig.~\ref{fig:aia_94}.

Further studies on the lower charge states of Iron, from 
 \ion{Fe}{vii} to  \ion{Fe}{ix} were obtained with laboratory spectra 
from an EBIT by \cite{lepson_etal:2002}.

The soft X-ray are also rich in Iron lines produced during 
solar flares.
Such lines, from highly-charged Fe ions of the type 2s$^2$ 2p$^k$--2s 2p$^{k-1}$, 
fall around 94--135~\AA\ and have been  identified mostly using laboratory
spectra of laser plasma
(e.g. \ion{Fe}{xviii}: \citealt{boiko_etal:1970}; 
 \ion{Fe}{xix}: \citealt{feldman_etal:1973}).
\cite{kastner_etal:1974_flare} reported the first 
solar-flare spectra containing the $n=2 \to 2$ L-shell
Iron emission, in the  66--171~\AA\ range from OSO-V.
\cite{lawson_peacock:80,lawson_peacock:84}
reviewed observations and diagnostics of  $n=2 \to n=2$ transitions.

\smallskip\smallskip\noindent
{\bf  X-rays}
\smallskip

Within the X-rays, the strongest lines emitted by the Sun
(actually active regions) during quiet conditions are lines
from \ion{Fe}{xvii} around 15~\AA.
The identification of \ion{Fe}{xvii}  lines started with the 
excellent work of \cite{tyren:38}.

A significant effort in the identification of flare lines, from the 
X-rays to the soft X-rays was carried out by several groups, 
using both solar observations and high-power lasers.
Most notably, 
G. Doschek and U. Feldman at NRL, USA, B. Fawcett 
at RAL,  and various researchers
in the USSR such as  E. Kononov at the Institute for Spectroscopy,
and others at the Lebedev Institute.

Transitions from highly-charged Fe ions of the type
2p$^k$-- 2p$^{k-1}$ 3l are as strong as the lines from 
 H- and He-like ions, and fall around 10--20~\AA.

The first observations of $n=3 \to 2$ Fe lines in solar flares 
were made with the OSO-III satellite in the 1.3--20~\AA\ region,
OSO-V in the 6--25~\AA\ region
 and were reported by \cite{neupert_etal:67,neupert_etal:73}.
Soon, similar spectra were obtained by other groups 
(see the review of these early observations by \cite{doschek:1972}).
 Several identifications were suggested by Neupert and colleagues.  
Further firm identifications came  from the NRL
group, see e.g. \cite{doschek_etal:1972,doschek_etal:1973,feldman_etal:1973}

Laser spectra  were also published 
in a series of papers by Boiko and colleagues (see \citealt{boiko_etal:1978}
and references therein).
The \cite{boiko_etal:1978} spectral accuracy 
 and resolution ($\simeq$0.002~\AA\ on average)
 were excellent. 
Very good laser spectra were also obtained at RAL 
\citep{bromage_etal:77_RAL_rep, bromage_etal:77a} and were used to identify lines.
The  literature is too extensive  to be summarised here.
A good review is provided by \cite{shirai_etal:00}, one of the 
NIST compilations. 

\smallskip\smallskip\noindent
{\bf  EUV}
\smallskip

In the early 1960's, 
B.C. Fawcett and collaborators established  the identifications
of the majority of  the strongest EUV lines (from Iron), with a large number of publications.
The breakthrough came when laboratory spectra showed that 
highly-charged Iron lines were the strongest lines
in the EUV spectra of the Sun in the 150--300~\AA\ spectral 
region, \cite{fawcett_etal:1963,fawcett_gabriel:65},
which led to the first identifications 
\citep{gabriel_fawcett:1965,gabriel_etal:1966,fawcett_gabriel:1966,fawcett_etal:67}.

Significant studies in the EUV are those of 
\cite{svensson_etal:1974} on \ion{Fe}{ix}; those of 
\cite{smitt:1977, bromage_etal:77} on \ion{Fe}{x} and \ion{Fe}{xi}; 
\cite{fawcett:1971} on \ion{Fe}{xi}, \ion{Fe}{xii}, \ion{Fe}{xiii},  \ion{Fe}{xiv}, and   \ion{Fe}{xv};
\cite{bromage_etal:78_fe_12} on \ion{Fe}{xii} and \ion{Fe}{xiii}; 
\cite{churilov_etal:1985} and \cite{cowan_widing:1973} on  \ion{Fe}{xv}.
Many \ion{Fe}{xvii} EUV lines were identified by 
\cite{jupen:1984}, with further significant revisions by 
\cite{feldman_etal:1985} and \cite{delzanna_ishikawa:09}.
Many high-temperature Iron lines in the EUV have been identified
using flare spectra obtained with the NRL slitless spectrometer
\citep[see, e.g.][]{sandlin_etal:76,dere:78}.

In terms of laboratory spectra, the best measurements in the EUV
have recently been carried out with Electron Beam Ion Traps (EBIT),
see for example \cite{beiersdorfer_lepson:2012,beiersdorfer_etal:2014}. 
 These studies have confirmed the accuracy of the atomic data 
in the CHIANTI database, which are now more  up-to-date
than those in the NIST database. 
The above-mentioned identification papers also 
revised previous assessments in a few cases. All the new 
or revised identifications have so far been confirmed, with one 
 one notable exception of an \ion{Fe}{xii} line 
\citep{beiersdorfer_etal:2014}. The CHIANTI v.8  atomic data for the 
\ion{Fe}{xiii}, \ion{Fe}{xiv}, and \ion{Fe}{xv} in terms of measurements of 
electron densities have recently been checked with 
high-density EBIT spectra around 10$^{13}$ cm$^{-3}$ by \cite{weller_etal:2018}.
Very good agreement (to within about 30\%) between 
observed and predicted line intensities was found.

In a series of studies,  the Livermore EBIT was used to 
study the emission of the most abundant elements in the SDO AIA 
EUV wavelength channels, which are, with one exception, dominated 
by Iron lines:  131~\AA\ \citep{traebert_etal:2014_131}, 193~\AA\ \citep{traebert_etal:2014_193}, 
and 304~\AA\ \citep{traebert_etal:2016_304}.

Beam-foil spectroscopy has also been 
fundamental  to identify spectral lines (mostly in the EUV) from levels with long
lifetimes, see e.g. the reviews by \cite{traebert:1998,traebert:2005}.

\smallskip\smallskip\noindent
{\bf  UV}
\smallskip

In the UV,  several forbidden lines from the coronal Iron ions
are present. Notable identifications are those by  
\cite{mason_nussbaumer:77} on \ion{Fe}{x}, 
\cite{sandlin_etal:77}  on \ion{Fe}{ix},  \ion{Fe}{xi}, \ion{Fe}{xii};
\cite{sandlin_tousey:79} on   \ion{Fe}{x}, \ion{Fe}{xi},  and 
\cite{burton_etal:1967} on \ion{Fe}{xii}.

\subsubsection{Benchmarking atomic data using line intensities}

Comparisons between observed and predicted line intensities 
can be found throughout the literature. 
However, the comparisons have  normally been limited to a few line ratios
within an ion, or to a small wavelength range. 
Until recently, either the instrument calibrations or the 
atomic data were uncertain, so comparisons between observed and predicted line intensities
were not very satisfactory, especially for the complex Iron ions.

A comprehensive approach, whereby the intensities of all the 
strongest lines within an ion are  compared  to all available astrophysical 
and laboratory spectra has been carried out by one of us (GDZ).
The work 
started with the coronal Iron ions, in a series of benchmark 
papers following \ion{Fe}{x}  \citep{delzanna_etal:04_fe_10}. 
The intensities of all the lines within an ion 
were compared using the emissivity ratio method, which often pointed to 
problems in either the radiometric calibration, the atomic data
(either in the scattering calculations or in the line identification) or 
sometimes indicated that an observed line was blended. 

Another approach is to compare predicted and observed line 
intensities within a spectral range. A few of such studies have been carried
out. Below we mention a few examples of benchmarking studies,
with particular emphasis on the Iron ions, because of their 
diagnostic importance in the X-rays and EUV.

\smallskip
\noindent
{\bf X-rays}
\smallskip

\begin{figure}[!htbp]
 \centerline{ \includegraphics[width=7cm,angle=90]{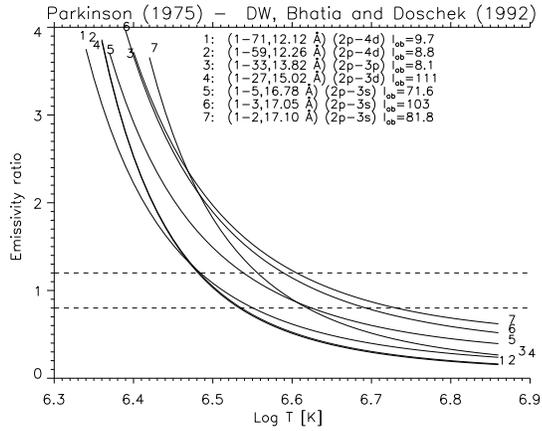}}
 \centerline{ \includegraphics[width=7cm,angle=90]{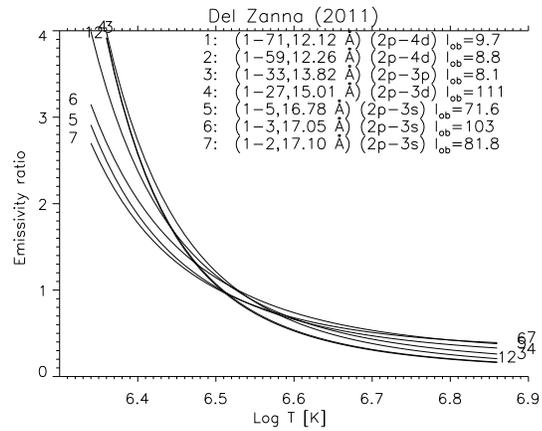}}
 \caption{Emissivity ratio curves for the strongest \ion{Fe}{xvii} lines, using the 
active region core observations of  \citep{parkinson:75}. 
Top: with the DW scattering calculations of \cite{bhatia_doschek:1992}.
Bottom: with the $R$matrix calculations of \cite{liang_badnell:10_ne-like}.
Figures adapted from  \cite{delzanna:2011_fe_17}.
}
 \label{fig:delzanna:2011_fe_17a}
\end{figure}

\noindent
Within the X-rays, perhaps the most important benchmark studies 
have been those on  \ion{Fe}{xvii}  \citep{delzanna:2011_fe_17}, 
which produces the strongest X-ray lines in active regions, and of 
\ion{Fe}{xviii} \citep{delzanna:2006_fe_18}, which produces 
strong lines in more active conditions. 
For both these complex ions, resonance effects are very important.
Therefore,  earlier DW scattering calculations provided poor agreement 
between theory and observations, with discrepancies of factors of 2--3.
Such discrepancies have led to a large number of papers where 
possible explanations for such deviations were sought, especially
for the \ion{Fe}{xvii} case. We do not review those studies here as 
 it turned out that the  
more recent $R$-matrix calculations resolved all the main 
discrepancies. 
 Fig.~\ref{fig:delzanna:2011_fe_17a} shows as an example 
the differences for \ion{Fe}{xvii} between one set of 
DW calculations and one set of   $R$-matrix calculations. 
Similar results were obtained for \ion{Fe}{xviii}.

Other benchmarking studies in the X-rays were carried out on 
\ion{Fe}{xxiii} \citep{delzanna_etal:2005_fe_23} and 
\ion{Fe}{xxiv} \citep{delzanna:2006_fe_24}. 
A more complete benchmark of the X-rays was carried out by 
\cite{landi_phillips:2005}. They  reviewed older SMM FCS observations
of two solar flares, using CHIANTI  atomic data. A few
line identifications and blends were revised. The spectra
had excellent resolution, but were limited by the fact that 
spectral lines at different wavelengths were not observed simultaneously.

\smallskip
\noindent
{\bf Soft X-rays}
\smallskip

\begin{figure}[!htbp]
 \centerline{
 \includegraphics[width=\textwidth]{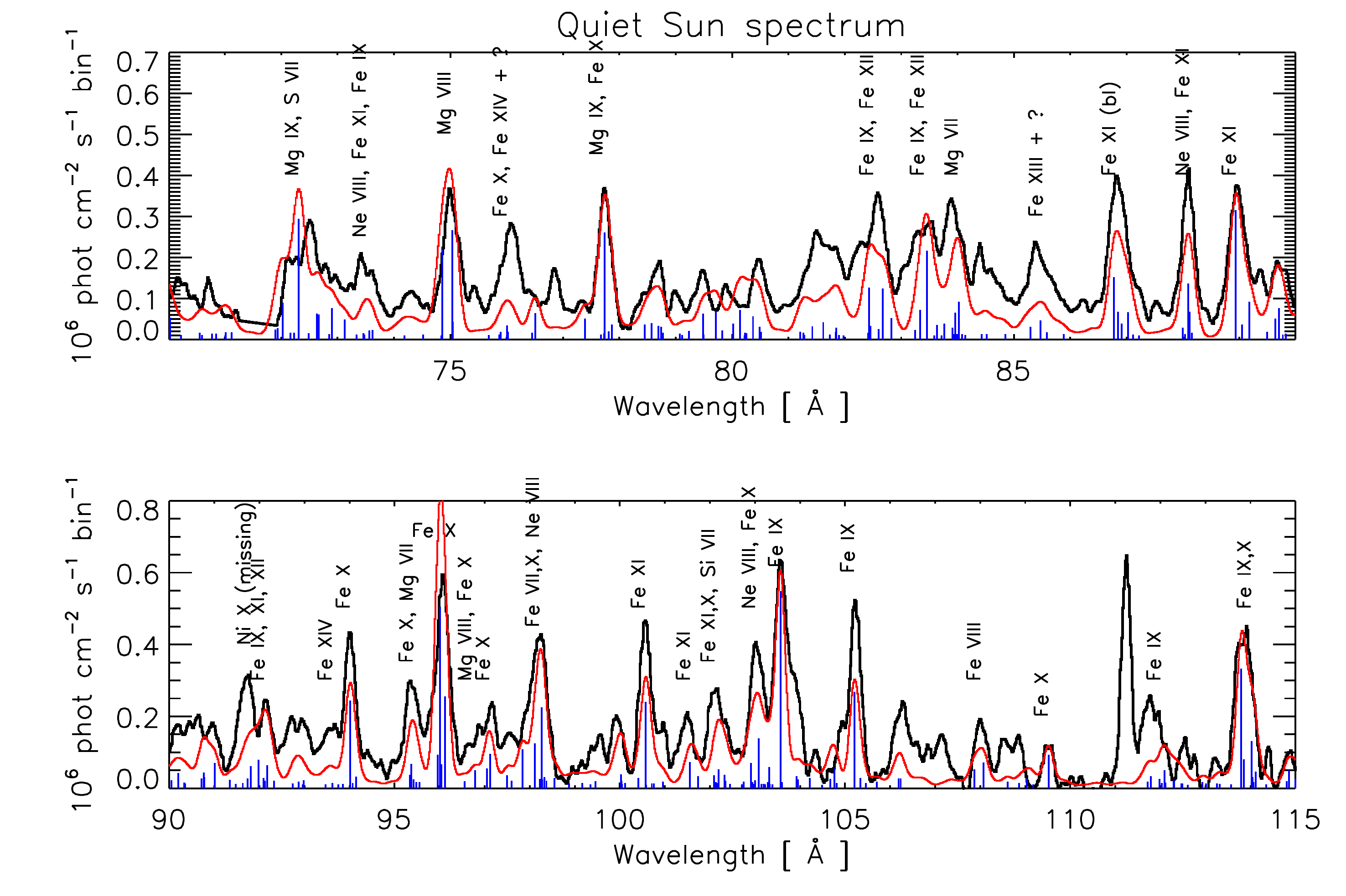}}
\caption{The quiet Sun soft X-ray spectrum (black) obtained from the 
\cite{manson:72} paper, with superimposed a predicted one,
based on the atomic data which were subsequently made
available within version 8 of the CHIANTI database
 \citep[adapted from ][]{delzanna:12_sxr1}.}
 \label{fig:delzanna_sxr1}
\end{figure}

\noindent
As already mentioned, little experimental and theoretical data 
has been available to study the soft X-rays. 
The high-temperature 2--2 transitions from \ion{Fe}{xix},
\ion{Fe}{xx}, \ion{Fe}{xxi}, and \ion{Fe}{xxii} were benchmarked against 
SDO EVE spectra of solar flares by \cite{delzanna_woods:2013}. 
\cite{delzanna:12_sxr1} used the recent calculations of the 
coronal Iron ions and  benchmarked them,
using the few high-resolution solar  soft X-ray spectra available.
Relative good agreement was found, although it is clear that 
a significant number of lines still needs to be identified,
and atomic data are still missing, as shown in Fig.~\ref{fig:delzanna_sxr1}.

\smallskip
\noindent
{\bf EUV}
\smallskip

\noindent
Within the EUV, a significant benchmark study  based on the first 
SERTS rocket flight \citep[][]{thomas_neupert:94} was provided by 
\cite{young_etal:98}.
Relatively good agreement between theory and observations was found.
Other benchmark studies carried out using a wide range of
SoHO CDS NIS and GIS observations also showed relatively good agreement 
(within 30-50\%) between observation and theory \citep{delzanna_thesis99},
especially above 300~\AA.

\begin{figure}[!htbp]
\centerline{\includegraphics[width=12cm,angle=90]{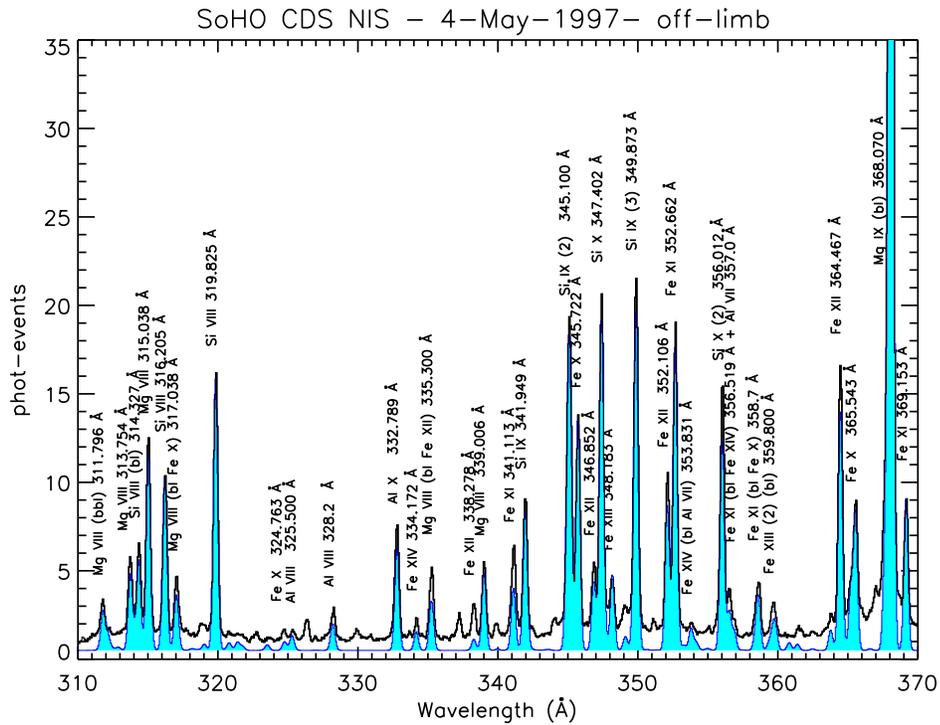}}
  \caption{SoHO CDS  NIS 1 averaged spectra of the quiet Sun, with the simulated spectra
obtained with CHIANTI v.1 shown with shaded blue areas  \citep[adapted from ][]{delzanna_thesis99}.}
  \label{fig:nis_obs_sint}
\end{figure}

 Fig.~\ref{fig:nis_obs_sint} shows an example of CDS NIS 1 spectra
with superimposed simulated spectra obtained with CHIANTI version 1. 
The coronal Iron lines, as well as the others from Si, Mg are 
quite well reproduced, although the recent large-scale calculations
for the Iron ions have further improved agreement.

\cite{landi_etal:2002_cds} also performed a benchmark of CHIANTI data 
for the coronal lines using off-limb observations from SOHO CDS NIS.
Some uncertainties associated with the instrument calibration 
were present.

With the launch of Hinode in 2006, the superior resolution and sensitivity
of the EIS spectrometer indicated several problems  in the 170--300~\AA\ range,
with about half of the lines not firmly identified, and discrepancies 
of large factors (up to 2) for several lines, including 
some of the strongest ones, from \ion{Fe}{xi}. All the main lines are 
from the complex coronal Iron ions.
A long-term benchmarking  programme was initiated by one of us (GDZ).
It involved recalculating atomic data and reassessing line identifications 
using both laboratory and astrophysical spectra. It took several years to 
complete the work on the most complex ions.
 Most of the unknown levels
(which were about half for the lower configurations of 
\ion{Fe}{x}, \ion{Fe}{xi}, and \ion{Fe}{xii})
have been identified in a series of papers:
\ion{Fe}{vii} \citep{delzanna:09_fe_7},
\ion{Fe}{viii} \citep{delzanna:09_fe_8},
\ion{Fe}{ix} \citep{delzanna:09_fe_7},
\ion{Fe}{x} \citep{delzanna_etal:04_fe_10},
\ion{Fe}{xi} \citep{delzanna:10_fe_11},
\ion{Fe}{xii} \citep{delzanna_mason:05_fe_12},
\ion{Fe}{xiii} \citep{delzanna:11_fe_13},
\ion{Fe}{xvii} \citep{delzanna_ishikawa:09}.
These benchmark studies identified a number of problems, some of which 
were in the atomic data and 
have been resolved with the latest calculations on these ions,
where further benchmarking was also carried out.

The  identifications  of lines in the Hinode/EIS wavelengths 
 proposed by \cite{delzanna_etal:04_fe_10} have been confirmed in \cite{delzanna:12_atlas}.
The \cite{delzanna_etal:10_fe_11} scattering 
 calculation, obtained with an optimal target, 
finally produced  improved agreement and 
allowed the identifications of all the strongest transitions  \citep{delzanna:10_fe_11}.

\begin{figure}[!htbp]
 \centerline{ \includegraphics[width=0.6\textwidth,angle=90]{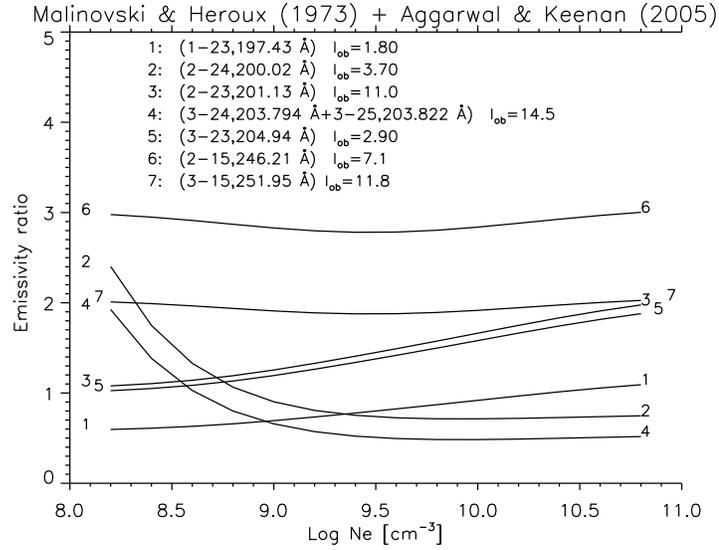}}
  \centerline{\includegraphics[width=0.6\textwidth,angle=90]{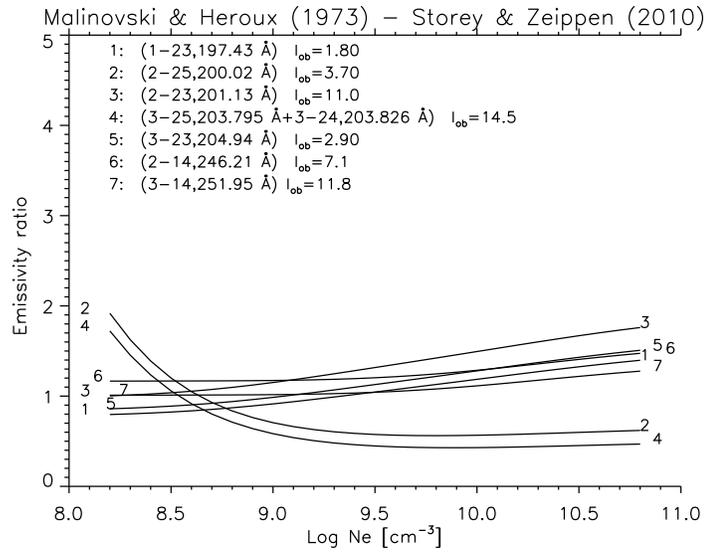} }
\caption{Emissivity ratio plots of the \ion{Fe}{xiii} EUV lines 
reported  by  \citep{malinovsky_heroux:73}. Top: with the 
\cite{aggarwal_keenan:2005} atomic data. Bottom:: with the 
\cite{storey_zeippen:10} atomic data. 
Note that the curves should cross if the plasma is isothermal.
Figure adapted from \cite{delzanna:11_fe_13}.}
 \label{fig:fe_13_emplot}
\end{figure}

One example is Fig.~\ref{fig:fe_13_emplot}, where the well-calibrated 
EUV observations of the \ion{Fe}{xiii} reported by 
 \cite{malinovsky_heroux:73} were benchmarked against two $R$matrix
calculations. It is clear that the  \cite{storey_zeippen:10} calculations
are in good agreement with observations, while those of 
\cite{aggarwal_keenan:2005} have several problems, with discrepancies of 
factors of 2--3. Similar problems have been found in other ions.

Further benchmark studies not on single ions but on 
full Hinode EIS spectra were those of 
\cite{delzanna:09_fe_7,landi_young:2009} where CHIANTI atomic data were
benchmarked for cooler transition-region lines.
\cite{delzanna:12_atlas} benchmarked the available atomic data 
for coronal lines observed by Hinode EIS off-limb
spectra of  the quiet Sun and an active region.

\smallskip
\noindent
{\bf EUV/UV}
\smallskip

\noindent
At wavelengths longer than 600~\AA, excellent spectral data 
have been obtained by SoHO SUMER. As previously mentioned, 
there are several line lists, but comparatively few benchmark 
studies of the atomic data. 
A comprehensive study of the 
coronal lines observed off-limb by SOHO SUMER 
was carried out by \cite{landi_etal:2002_sumer}, using CHIANTI 
atomic data and the EM loci method. 
Relatively good agreement was generally found. 

\cite{doschek_etal:1999} carried out a systematic study of the 
cooler (transition-region) lines observed on-disk by SUMER. Observed line ratios
were compared to those predicted with CHIANTI and with other 
atomic data. Significant discrepancies were found for many ions. 
We recall however that only lines within a 40~\AA\ band were 
observed simultaneously, so  despite averaging some of the 
discrepancies could be due to real temporal variability. 
Transition-region lines are in fact well-known to have a strong variability.

\subsubsection{Rest wavelengths}

The issue of rest wavelengths is a non-trivial one.
In the X-rays (1-50~\AA), the best measurements have
normally been obtained from laboratory plasma
(cf. \citealt{boiko_etal:1978}), with a typical accuracy 
around 10~\AA\ of 0.002~\AA. High-resolution crystal spectrometers
(cf. SMM/FCS, \citealt{phillips_etal:1982}) have achieved similar accuracies.

However, laboratory wavelengths are often not accurate enough in the 
EUV/UV, or are at odds with solar measurements.
Databases such as NIST mostly rely on laboratory measurements
and are therefore  sometimes inaccurate.

The most effective way to 
obtain  rest wavelengths is to measure them directly with high-resolution
spectroscopy of the Sun.
This brings in a series of problems, given that, depending on the 
temperature of formation of the lines and the source, 
significant Doppler motions are normally present. 

It is now well established that lines formed around 1 MK 
show small Doppler motions (see, e.g. \citealt{peter_judge:99,delzanna:08_flows}).
These lines mostly emit at  soft X-ray and EUV wavelengths.
The most accurate wavelengths for these lines come 
from full-Sun spectra  obtained by two  rocket flights with a spectrograph built at 
the Goddard Space Flight Center. 
The first  rocket flight was on  1969 May 16. On that day, the Sun was moderately active,
(the F10.7 radio flux was 159.4).
The instrument observed the  60--385~\AA\ region with 
high resolution (0.06~\AA) \citep{behring_etal:72}.
 A second flight, on 1973 Sept 21, observed the 
160--770~\AA\ region  \citep{behring_etal:76}.
A few of these coronal measurements have been revised with 
high-resolution EUV spectra from  Hinode EIS \citep[cf.][]{delzanna:12_atlas} and 
the SERTS rocket flights \citep[cf.][]{thomas_neupert:94}. 

The issue of the EUV lines formed in the 1--3 MK range is more
complex. Most of these lines are typically emitted by active regions.
The hot core loops typically show small  Doppler shifts, even 
at their footpoints (near sunspots and plage regions, where the 
`moss' is, see e.g. \citealt{winebarger_etal_2013}).
However, there are large spatial areas where significant 
blue-shifts, increasing with the formation temperature of the lines,
are present \citep{delzanna:08_flows,doschek_etal:08}.
 
Lines formed at higher temperatures (above 3 MK) are normally associated with 
very dynamic events such as microflares and
solar flares, and typically show strong 
Doppler motions. It is only during the gradual phase in the post-flare loops
that the wavelengths of the flare lines  normally become at rest.

On the other hand, it is well established that 
lower-temperature EUV  lines
are red-shifted, following various trends depending on the source
\citep{peter_judge:99,delzanna:08_flows}.
Accurate rest wavelengths can still be established, 
see e.g.  \cite{hassler_etal:99} for a revision of the 
Ne VIII  wavelengths using SUMER, and  \cite{delzanna:09_fe_7} for  wavelengths of 
cool lines observed by Hinode EIS.

In the UV, the best wavelength measurements 
have been obtained by Sandlin and colleagues 
\citep{sandlin_etal:1986}  with a careful analysis and averaging of 
Skylab limb observations (where it is assumed that lines should be at rest), 
as well as HRTS data. These measurements
have an accuracy of about 0.005~\AA, and agree, within uncertainties, with the best 
laboratory measurements, for the few ions we have checked.

\clearpage

\section{Non-equilibrium effects}
\label{sec:non-eq}

Several non-equilibrium  effects can significantly modify XUV spectra.
In this section, we provide a brief overview of some of these effects.
 For a fuller recent review on this topic see \cite{dudik_etal:2017_review}.

\subsection{Time-dependent ionisation}

We recall (Eq.~\ref{eq:time_dep_ionisation}) 
 that the charge state distribution of an element can be 
calculated:
$$ {1 \over N_e} {dN_{r}\over dt} =N_{r-1} S^e_{r-1} - N_{r} (S^e_{r}
+\alpha_{r}) + N_{r+1}\alpha_{r+1}  $$
once the total ionisation and recombination rates are known,
and a model for the plasma flow is assumed.

In the low solar corona, the timescales for $N_{r}$ to ionise ($1/(N_e S^e_{r})$)
or to recombine ($1/(N_e \alpha_{r})$) are of the order of 100 seconds, 
 and  ionisation equilibrium is a reasonable approximation.
However, in regions where strong flows and strong temperature gradients are 
present, significant departures from equilibrium could  occur.
In this Section we briefly summarize a few cases, to provide 
the reader with an idea of where departures from ionization equilibrium 
are expected and what kind of effects they could  have on the plasma 
diagnostics. 

A classical example is the  solar transition region, where  downflows 
are ubiquitous, so the ions are moving into regions with much reduced
temperatures (there is a steep gradient).
A simple semi-empirical model for the quiet Sun transition region 
(e.g. \citealt{gabriel:76}) shows gradients of about 2000 K/km.
Electron number densities in the transition region are of the order 
of 10$^{10}$  cm$^{-3}$, and the timescales
for ionisation and recombination are of the order of tens of seconds,
so any flow of a few km/s would have a significant effect. 
 The main effect  would be to shift the formation temperature of a line 
to  much lower temperatures, compared to those expected if ionization
equilibrium conditions. 
Observationally, variations of line intensities on timescales faster than
ionisation / recombination times are common, so one would indeed 
expect that time-dependent ionisation has a major effect in the 
transition region.

In the literature, there are several early  studies of the 
effects that departures from ionization equilibrium could have on 
the various diagnostics.
For example, \cite{raymond_dupree:78} showed that a 5 km/s downflow 
would have a significant effect on the line intensities of \ion{C}{iii},
in particular on the electron densities estimated from  line ratios of this ion.
Significant effects were later confirmed with 
subsequent 1-D hydrodynamical modelling of loop structures by e.g. 
\cite{noci_etal:89,raymond:1990, hansteen:93}, 
although a complete self-consistent treatment is still lacking.
An improvement in the modelling was introduced within the 
HYDRAD code \citep{bradshaw_mason:03}, where the  radiative losses 
are calculated self-consistently using the time-dependent ion populations.
HYDRAD was used in \cite{bradshaw_etal:04}
to model  a small compact flare.
 An 80\,Mm loop was heated by a 300\,s heating pulse  at the apex, 
raising its temperature from 1.6\,MK to about 7\,MK. 
 Strong enhancements of \ion{He}{I}, \ion{He}{II}, and \ion{C}{IV} 
line intensities  were found during the impulsive phase,
while the hot lines such as  \ion{Fe}{XVIII}, \ion{Fe}{XIX} we significantly
suppressed.  \cite{bradshaw_etal:04} also noted that the  enhancements 
could perhaps indicate a viable way to explain 
the anomalously high intensities of  lines from these ions, which have long
recombination times (see Section~\ref{sec:anomalous_ions}). 
 
Several further studies of non-equilibrium ionization 
in  transition-region lines, particularly important for the IRIS mission,
have reached similar conclusions. 
For example, \citet{Doyle13} applied the GCR technique  
to study the response of \ion{Si}{iv} and \ion{O}{iv} lines 
to short heating bursts. They found that the \ion{Si}{iv} intensity 
is significantly enhanced (factor of three or more)
 shortly after the burst, while the \ion{O}{iv} lines showed little
variation. 

 \citet{Martinez16} ran  MHD simulations using the Bifrost code \citep{Gudiksen11}
with the inclusion of non-equilibrium ionization as described by 
 \citet{Olluri13a}, using atomic data included in the  DIPER package \citep{Judge94}.
They also found significant increases of \ion{Si}{iv} relative to \ion{O}{iv}
in active regions, and although were not able to fully reproduce 
the IRIS observations, confirming  that non-equilibrium ionization is an important 
effect which brings closer agreement between observations and theory.

 \citet{Olluri13b} used the same codes to study the effects that 
 non-equilibrium ionization have on the \ion{O}{IV} density diagnostics.
They found that significant differences in the derived densities occur, because the 
the \ion{O}{IV} abundance is shifted to lower temperatures.

Flows in the transition region can not only affect the measurements of 
electron densities, but also those of relative elemental abundances. 
For example, \cite{edgar_esser:00} showed that a significant enhancement of 
\ion{Mg}{vi} emission compared to \ion{Ne}{vi} is present, if 
non-equilibrium effects in a recombining downflow are considered.
This would  significantly affect the FIP effect measurements in the 
transition region (see Section~\ref{sec:abund}).
We should note, however, that \cite{edgar_esser:00} considered 
cooling at constant pressure or at constant density. 
The formation mechanism producing the transition region emission is 
still not clear, as with the general coronal heating.
However, there is some evidence that the plasma in the legs of coronal 
loops (where \ion{Mg}{vi}  and  \ion{Ne}{vi} are emitted) 
is radiatively cooling, i.e. where the heating 
has been switched off and the dominant cooling process is by radiation losses.
The footpoints of active region 1~MK loops show ubiquitous redshifts 
in transition-region lines of 10--30 km/s (see, e.g. \citealt{delzanna:08_flows}).
The observed downflows and densities are consistent with the 
radiative losses sustained by an enthalpy flux while the loops
radiatively cool \citep{bradshaw:08}.
In the case of radiative cooling, $T \sim N_{\rm e}^2$ \citep{bradshaw_cargill:05},
which is quite different to the cases considered by \cite{edgar_esser:00}.

Time-dependent ionisation can also increase  the cooling times of 
coronal loops, as discussed e.g. in \cite{bradshaw_mason:03}.
Also, it was found to have a significant effect in the low corona
after interchange reconnection occurred in the high corona:
\cite{bradshaw_etal:11} studied the hydrodynamic response of the 
plasma undergoing reconnection  between a high-pressure 
(high temperature and density) hot loop in the core of an active region
and a surrounding open field with much lower pressure.
Once the reconnection has taken place, a 
 rarefaction wave would start travelling towards the chromosphere,
causing a rapid expansion and cooling of the plasma.
The higher-temperature lines formed  around 3 MK were found to be 
enhanced by a factor of about two when time-dependent ionization was considered.

Stationary  plasma that is heating or cooling 
quickly, on timescales shorter than the local 
ionisation and recombination timescales, will also be out of equilibrium. 
In the case of heating, the plasma is under-ionised since the 
ions do not have enough time to ionise, despite the high temperatures. 
In the case of cooling, the plasma is over-ionised since the ions do not
have enough time to recombine, whilst in a low-temperature environment.

Plasma is most likely to be out of ionisation equilibrium during the 
impulsive phase of solar flares. Indeed non-equilibrium may 
be a more common process in the solar corona, depending on 
the frequency and type of  heating, and the local density. 
 For example, \cite{bradshaw_cargill:06} 
showed that fast heating to high temperatures (10 MK) of a low-density
plasma would be considerably out of equilibrium, so there would be 
very little emission (if any)  from the ions that are normally formed 
(in equilibrium) at 10 MK. 
The same conclusions were reached by \cite{reale_orlando:08},
and could explain the fact that very little emission is observed at 
high temperatures in active regions, despite the fact that 
some nanoflare  modeling predicts that high temperature plasma should be present.

 At the end of the cooling phase of a flare 
the plasma undergoes a thermal instability, which also involves
strong non-equilibrium effects, as decribed e.g. by \cite{reale_orlando:08}.

Signatures of non-equilibrium in X-ray spectra 
during solar flares have long been sought by several authors.
As we have discussed in  Section~\ref{sec:satellites},
the best diagnostic to assess for  departures from ionisation equilibrium
is the $q/w$ ratio of the inner-shell vs. the resonance in He-like ions,
which 
depends  on the  relative ion abundance between the  Li-like and the He-like ion.
 SOLFLEX observations of this
ratio for the He-like Fe and Ca are shown in 
 Fig.~\ref{fig:solflex_satellite} as an example. 
If one uses the temperatures obtained from the $j/w$ and $k/w$ ratios
(see Section~\ref{sec:satellites})
to calculate the $q/w$ ratio, one often finds that the ratio of the 
Li-like vs. the He-like ion abundance is much higher, compared to what 
it should be in ionisation equilibrium. This is because of the significant
intensity of the $q$ line. 
The plasma therefore appears to be in transient ionisation. 
This result was obtained by \cite{doschek_etal:1979,doschek_etal:1980,doschek_feldman:1981,feldman_etal:1980}
 using SOLFLEX observations. However, given the high densities observed during the 
peak phase of a flare, typical equilibration times should be of the order of 
a second or less, so it is hard to accept that 
the plasma is out of equilibrium. Indeed the authors suggest that there might 
be problems in the atomic data. 

Later observations were obtained with other instruments, most notably
with the SMM \citep[see, e.g.][]{gabriel_etal:1983,antonucci_etal:1984,antonucci_etal:1987} 
and Hinotori  spectrometers,
but considering the various uncertainties, it was not very clear if
departures from ionisation equilibrium were present or not.
See the  discussion  by \cite{doschek_tanaka:1987} in their
review of  SMM and Hinotori X-ray observations.

In the high corona, we have the opposite situation,
with  the plasma  flowing out into the heliosphere
at progressively faster speeds, also with  a significant geometrical expansion. 
At some point, local electron densities become so low that 
the ionisation/recombination timescales become much longer than the 
expansion timescales, so the ion populations do not change anymore 
(`are frozen-in').  Typical solar wind models show that 
the region where this occurs is within a few 
solar radii from the Sun (see, e.g. \citealt{esser_etal:1998}
and references therein).

\subsection{Non-Maxwellian electron distributions}
\label{sec:non-maxwell}

A central assumption for most of the spectroscopic diagnostics 
applied to the solar corona is that the electrons 
have  a Maxwellian distribution. 
The classical textbook plasma physics shows that  
the relaxation time for 
e-e collisions in the low corona is  $\tau_{ee}\simeq 0.01~ T^{3/2}~ N_e^{-1}~ s$,
i.e.   $\simeq$0.01s for an active region loop ($T=10^6$ K and $N_e=10^9$ cm$^{-3}$), 
but $\simeq$300s for a plasma suddenly heated at $T=10^7$ K (with $N_e=10^8$ cm$^{-3}$).
After a time longer by a factor $(m_i/m_e)^{1/2} \simeq 43 $,  the ions also become thermal,
and after the same time factor, both species are thermalised, and the electron and ion 
temperatures are the same:  $T\equiv T_{e}=T_{i}$.
Therefore, as in the non-ionization equilibrium case, there could 
be cases in the solar corona where the electrons have non-Maxwellian distributions.

This topic has been studied by a number of authors.
Theoretical calculations based on the  Fokker-Planck (sometimes also called Vlasov-Landau) equation
have shown that a high-velocity tail of the electron distribution can form
in some circumstances. 
\cite{shoub:1983} solved the  Landau Fokker-Planck equation to model the 
 transition region, and found that significant deviations from the 
local Maxwellian approximation can form.
This was later confirmed by \cite{ljepojevic_burgess:1990} where solutions to the 
equation in cases where large temperature and density gradients are present
(as in the TR) were discussed in detail. 
Physically, the bulk of the electrons is found to follow the  Spitzer-Harm 
solution, i.e.  Maxwellian, while high-velocity tails of the distribution
form and remain present  where strong  temperature gradients are present.
This is because the mean free path of an electron $\sim v^4$ so the high-velocity
electrons from the hotter region stream down towards the chromosphere almost freely.
Towards the chromosphere, the gradients decrease substantially, and the 
high-velocity tail should disappear.

\cite{ljepojevic_mac_niece:1988} simulated a flaring coronal loop
and also found for this case  an enhancement of the tail population 
 for electrons moving down the temperature gradient.
\cite{ljepojevic_mac_niece:1989} used the Landau Fokker-Planck equation to show that the 
classical Spitzer-Harm approximation for the heat flux completely breaks down in the 
presence of non-Maxwellian electrons, with fundamental consequences for the 
energy balance. 

Below we focus on  few diagnostic applications and observational aspects.

\subsubsection{Modelling the XUV spectra}

The presence of a non-Maxwellian distribution  affects the 
excitation and de-excitation rates within an ion and 
the ionisation and recombination rates. 
A complete self-consistent treatment of all the rates is still not
available, although significant advances have been made in recent years. 

Regarding the excitation and de-excitation collision rates, a simple way to 
include the effects of a non-Maxwellian distribution is to model the 
distribution as a superposition of Maxwellian distributions with 
different temperatures. The non-Maxwellian collision rates can then be calculated 
as a linear combination of the Maxwellian ones.
 Such an approach was included in 
the CHIANTI database version 5 \citep{landi_etal_v5:06}, and proposed 
again by \cite{hahn_savin:2015} in terms of combinations that would reproduce a 
$\kappa$-distribution.

In order to obtain excitation rates integrated over non-Maxwellian 
velocities, one requires the collisional cross sections. 
Since these are rarely available,  a way to 
 recover the behaviour of the collision 
cross sections as a function of energy from the rates via a parametrisation method
was developed and implemented by \citep{Dzifcakova15} using CHIANTI version 7.1 data,
and released as a separate package, KAPPA\footnote{http://kappa.asu.cas.cz}.
This method typically provides cross sections 
sufficiently accurate, within a few percent \citep[see, e.g.,][]{DzifcakovaMason08,Dzifcakova15}.

Of course, whenever the original cross sections are available, it 
is better to use them directly for the calculations of the rates.
The UK APAP team  is building a database of such cross-sections.

It is worth noting that, for a distribution that is not Maxwellian,
 the effective collision strengths for the  excitation and de-excitation
are not the same. In the case of  $\kappa$-distributions, the
$\Upsilon_{ij}(T,\kappa)$ for the excitation and 
\rotatebox[origin=c]{180}{$\Upsilon$}$_{ji}(T,\kappa)$ are 
\citep[see ][]{Bryans06, dudik_etal:2014_fe}:

\begin{eqnarray}
	\Upsilon_{ij}(T,\kappa) &= A_\kappa \mathrm{exp}\left(\frac{E_{ij}}{k_\mathrm{B}T}\right) \int\limits_0^{+\infty} \frac{\Omega_{ji}(E^\prime)}{ \left(1+ \frac{E^\prime +E_{ij}}{(\kappa-3/2)k_\mathrm{B}T}\right)^{\kappa+1}} \,\mathrm{d}\left(\frac{E^\prime}{k_\mathrm{B}T}\right)\,,
 \\
	\rotatebox[origin=c]{180}{$\Upsilon$}_{ji}(T,\kappa) &= A_\kappa \int\limits_0^{+\infty} \frac{\Omega_{ji}(E^\prime)}{\left(1+ \frac{E^\prime}{(\kappa-3/2)k_\mathrm{B}T}\right)^{\kappa+1}} \,\mathrm{d}\left(\frac{E^\prime}{k_\mathrm{B}T}\right)\,.
	\label{Eq:Downsilon_kappa}
\end{eqnarray}
%
Once these rates are calculated, they can be used directly instead of the 
thermally-averaged cross-sections.

Regarding the ionization and recombination rates, similar issues arise.
The calculations are again straightforward if the actual cross-sections 
are available. In the case of ionization by direct impact by electrons, 
parametrised versions of the ionization cross-sections 
have been stored in the CHIANTI database version 6 \citep{dere_etal:09_chianti_v6}.
They have been used by \cite{dzifcakova_dudik:2013} to calculate ionization
rates for  $\kappa$ distributions.

For the radiative recombination, \citet{Dzifcakova92} provided a method
to estimate the rates from the usual parametrised formulas
which describe the Maxwellian rates.

Regarding the DR rates, a similar approach was developed 
for the $\kappa$-distributions by \cite{Dzifcakova92,Dzifcakova13a}. 
 The total DR rate for an ion is  fitted with the formula:
\begin{equation}
	\alpha^{\rm d}(T) = T^{-3/2} \sum_i c_i \exp( -E_i / T ) ~~ {\rm cm}^{3}~{\rm s}^{-1}
	\label{Eq:Diel_approx_Maxwellian}
\end{equation}
where the $c_i$ and $E_i$ are the fit parameters.
The dielectronic recombination rate can then be calculated with 
\begin{equation}
	\alpha^{\rm d}(T) = \frac{A_\kappa}{T^{3/2}} \sum_i \frac{c_i}{\left( 1 +\frac{E_i}{(\kappa -3/2)T}\right)^{\kappa+1} } ~~ {\rm cm}^{3}~{\rm s}^{-1}
	\label{Eq:Diel_approx_kappa}
\end{equation}

\begin{figure}[htb]
\centerline{\includegraphics[width=9cm,angle=90]{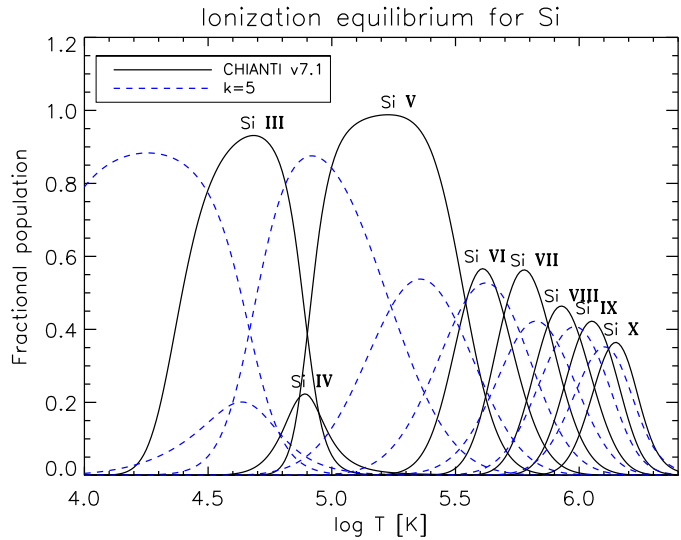}}
 \caption{Fractional population of Si ions in the Maxwellian case (full line) 
and a strong  non-thermal distribution with  $\kappa$=5 (dashed line).
The figure was obtained from the KAPPA package. }
 \label{fig:si_ion_kappa5}
\end{figure}

Once all the rates are calculated, it is straightforward to calculate ion charge state 
distributions in equilibrium.  A tabulation of values
calculated with a  $\kappa$-distribution can be found in 
\cite{dzifcakova_dudik:2013}.
An example is shown in Figure~\ref{fig:si_ion_kappa5}, where 
the distribution of silicon ions are shown, for the Maxwellian case and for a 
strongly non-thermal distribution with  $\kappa$=5. 
It is worth noting the shift of the low charge states towards 
lower temperatures, and the increase in the range of temperatures
where each ion is formed.

\subsubsection{Observational constraints}

\skf
{\bf Solar  flares}
\skf

Departures from a Maxwellian distribution,
especially the presence of high-energy tails, are
expected to arise due to magnetic reconnection
or wave-particle interactions, in particular during the impulsive phase of flares.
Indeed, hard X-ray  (HXR)  observations  above a few keV
routinely show non-thermal bremsstrahlung emission which (at least partly)
is caused by non-thermal electrons that have traveled from the flare heating
site. 
 Strong HXR emission is
 found at the footpoints of flare loops,
but  {\it Yohkoh}  \citep[see, e.g.][]{masuda_etal:1994} and 
  RHESSI  \citep[see, e.g.][]{krucker_etal:2010} observations have also shown that much weaker emission 
in coronal sources is often present.
This indicates that  non-equilibrium processes are also occurring in the solar corona
during flares.

The HXR emission is caused by bremsstrahlung of the free electrons,
and has traditionally been modelled with two components, a thermal one
at low energies (usually with the isothermal 
approximation), and a non-thermal one, usually with a power law 
for the electron distributions, with a cutoff at lower energies.
However, it has been shown that the non-thermal part of the spectrum 
can also be modeled with  $\kappa$-distributions
\citep[see, e.g.][]{kasparova09,oka_etal:2013}.

One issue that hard X-ray observations at lower energies cannot easily resolve is related to 
the  multi-thermality of flare plasma. The lower energy emission from RHESSI is  often 
approximated with a single temperature distribution, but the emission 
is likely to be  multi-thermal, or thermal plus non-thermal, as suggested by 
\cite{oka_etal:2013}.

A more direct spectroscopic method to assess if  
 non-Maxwellian distributions exist during solar flares was suggested by 
\cite{gabriel_phillips:1979}, as described previously in  Section~\ref{sec:satellites}.
 It relies on the measurement of the 
dielectronic satellite lines that are observed in the wings of the 
\ion{Fe}{xxv} resonance line at 1.85~\AA. 
These  satellite lines are formed by electrons dielectronically captured by \ion{Fe}{xxv} ions,
which have a very narrow energy distribution (each transition can only 
occur  at a single energy). 
SOLFLEX observations of solar flares were analysed by 
\cite{seely_etal:1987}. The authors deblended 
the d13 line complex, and found some evidence, during the impulsive phase,
of departures from equilibrium.

One problem with this  method for solar flare diagnostics
is that it requires a very high spectral resolution.
In fact, the  $n=3$  satellites lines are very close to the resonance line. 
In solar flare spectra during the impulsive phase, when  one might expect departures from a
Maxwellian, the satellite lines  have not been fully resolved.
This is because the resonance  lines are always broadened. 

\begin{figure}[htb]
\centerline{\includegraphics[width=9cm]{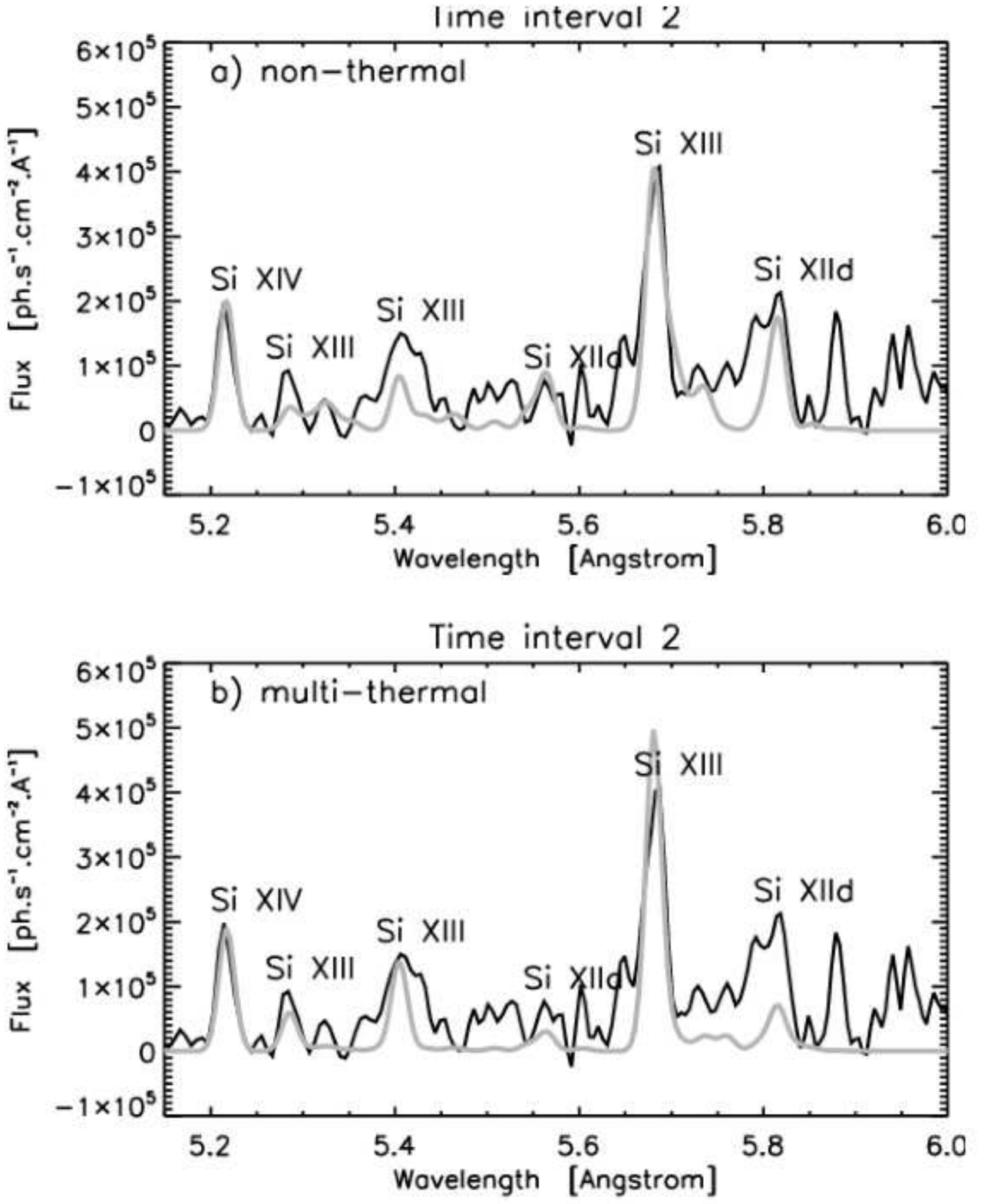}}
 \caption{RESIK spectra of a flare, with modelled lines (grey curves) assuming a non-thermal
electron distribution (top)  or a multi-thermal plasma (bottom, adapted from 
\citealt{dzifcakova_etal:08}). }
 \label{fig:dzifcakova_etal:08}
\end{figure}

More success has been obtained in the X-ray spectral range around 5~\AA,
with the satellite lines observed by the RESIK instrument during 
the impulsive phase of solar flares \citep{dzifcakova_etal:08,kulinova_etal:2011},
although even in this case it is not trivial to assess  how much
a non-thermal electron distribution is present and how much the 
plasma  is multi-thermal (see  Figure~\ref{fig:dzifcakova_etal:08}).

\begin{table}\caption{ Main  RESIK spectral lines.
diel.:  lines formed by dielectronic recombination.
} \label{tab_resik}
\begin{tabular}{llclc}
\hline \hline  \\
 RESIK&Ion & $\lambda$ ({\AA})&Transition& Log $T_{max}$ \\
Cha. No.&      &  & & (K)\\
 \hline
1\\
\hline
   & Ar~{\sc xvii}  &3.365 & 1s$^2$ $^1$S$_{0}$ - 1s 3p $^1$P$_{1}$ &  7.2 \\
   & Ar {\sc xvi} diel.   &3.428 & 1s$^2$ 2p $^2$P$_{3/2}$ -  & 7.1 \\
   &                     &      &1s 2p ($^3$P) 3p$^2$D$_{5/2}$ & \\
   & K {\sc xviii}   & 3.532 & 1s$^2$ $^1$S$_{0}$ - 1s 2p $^1$P$_{1}$ &  7.2\\
   & K {\sc xviii}  &3.571 & 1s$^2$ $^1$S$_{0}$ - 1s 2s $^3$S$_{1}$ &  7.2 \\
   & S {\sc xvi} &   3.696, 3.696 & 1s $^2$S$_{1/2}$ - 5p $^2$P$_{3/2, 1/2}$ & 7.2 \\
   & S {\sc xvi} &  3.784, 3.785 & 1s $^2$S$_{1/2}$ - 4p $^2$P$_{3/2, 1/2}$ &   7.2 \\

2\\
\hline
   & Ar {\sc xvii}  &  3.949 & 1s$^2$ $^1$S$_{0}$ - 1s 2p $^1$P$_{1}$ &   7.2 \\
   &                &  3.966 & 1s$^2$ $^1$S$_{0}$ - 1s 2p $^3$P$_{2}$ &   7.2 \\
   &                &  3.969 & 1s$^2$ $^1$S$_{0}$ - 1s 2p $^3$P$_{1}$ &   7.2 \\
   & S {\sc xv}     &  4.089 & 1s$^2$ $^1$S$_{0}$ - 1s 4p $^1$P$_{1}$ &  7.2 \\

3\\
\hline
   & S {\sc xv} &   4.299 & 1s$^2$ $^1$S$_{0}$ - 1s 3p $^1$P$_{1}$    & 7.2 \\
   & S {\sc xvi} &  4.727, 4.733 & 1s $^2$S$_{1/2}$ - 2p $^2$P$_{3/2, 1/2}$  & 7.2 \\
   & Si {\sc xiv} & 4.831, 4.831 & 1s $^2$S$_{1/2}$ - 5p $^2$P$_{3/2, 1/2}$ &   7.2 \\

4\\
\hline
   & S {\sc xv} &   5.039 & 1s$^2$ $^1$S$_{0}$ - 1s 2p $^1$P$_{1}$ &  7.1 \\
   &             &   5.102 & 1s$^2$ $^1$S$_{0}$ - 1s 2p $^3$S$_{1}$ &  7.1 \\

   & Si {\sc xiv} &   5.217, 5.218 & 1s $^2$S$_{1/2}$ - 3p $^2$P$_{3/2, 1/2}$ &   7.2\\
   & Si {\sc xiii} &  5.286 & 1s$^2$ $^1$S$_{0}$ - 1s 5p $^1$P$_{1}$   & 7.1 \\
   & Si {\sc xiii} &   5.405 & 1s$^2$ $^1$S$_{0}$ - 1s 4p $^1$P$_{1}$  &   7.1 \\
   &               &   5.408 & 1s$^2$ $^1$S$_{0}$ - 1s 4p $^3$P$_{1}$  &   7.1 \\

   & Si {\sc xii} diel. &  5.56  &  1s$^2$ 2p $^2$P$_{1/2, 3/2}$ -   &  7.0  \\
   &                 &      &1s 2p  4p $^2$D$_{3/2,5/2}$ & \\

   & Si {\sc xiii} &    5.681 & 1s$^2$ $^1$S$_{0}$ - 1s 3p $^1$P$_{1}$  &   7.1 \\
   & Si {\sc xii} diel. &  5.816  &  1s$^2$ 2p $^2$P$_{1/2, 3/2}$ -   &  7.0  \\
   &                 &      &1s 2p ($^3$P)  3p $^2$D$_{3/2,5/2}$ & \\

\hline
\end{tabular}
\end{table}

Another possible diagnostic method could be high-resolution observations of the 
free-bound edges in the X-rays, as suggested by \cite{dudik_etal:2012}.
However,  observations to date have not had enough sensitivity and 
spectral resolution to be able to detect  the presence of 
 non-thermal electron distributions.

\skf
{\bf Solar wind}
\skf

Another place where we know from in-situ measurements that the electron distributions
are very anisotropic and non-Maxwellian is the solar wind.
 They can  typically be modelled with  $\kappa > 2.5$ distributions.
For a list of references, 
see the recent review by \cite{dudik_etal:2017_review}. 
Two question  naturally arise: ``why are there non-Maxwellian distributions?'' 
and ``do they  originate  in the solar corona?'' 
If there are correlations between the particles in the system, 
induced by any long-range interactions in the emitting plasma 
such as wave-particle interactions, shocks, or particle acceleration
then the distribution function may depart from the Maxwellian one.
It is therefore possible that non-Maxwellian distributions
originate in the corona.

We do not yet have direct in-situ measurements of the distributions functions 
close to the Sun to reach a conclusion. The closest observations were those based on the 
\textit{Helios} spacecrafts, at 0.3 AU, although { \color{blue} Parker  Solar Probe
 will provide such observations}. Therefore,  studies of the particle distributions 
have so far been based on modelling. For example, 
\cite{esser_edgar:00} allowed for the presence of non-Maxwellian electrons 
to try and resolve a  discrepancy between the 
freezing-in temperatures estimated from the in-situ measurements of the 
fast solar wind, about 1.5 MK at 1.4~\rsun, and remote-sensing measurements
(in particular those from SUMER), which indicate electron temperatures
below 1 MK. 
The freezing-in temperatures are normally obtained from in-situ measurements 
of the charge state distributions of several ions, and with various assumptions,
such as the ion flow speed and Maxwellian electrons (see, e.g. \citealt{geiss_etal:95}).
\cite{esser_edgar:00} were able to reconcile the two methods with a 
 Maxwellian  distribution close to the coronal base, but that 
becomes rapidly non-Maxwellian as a function of height.
We should note, however, that  remote-sensing measurements are 
uncertain and have recently been revised (see Section~\ref{sec:ch_te}), so it is not clear if such 
discrepancies really exists.

\skf
{\bf Low corona}
\skf

Various studies based on remote-sensing measurements and modelling have attempted
to assess for the presence of non-Maxwellian electrons in the low corona
 (outside of flares), but results are still inconclusive.

 \cite{Feldman07} used SOHO SUMER active region observations 
of He-like lines to  search for signatures of a secondary Maxwellian
component at high temperatures (10 MK), but nothing was found.
\cite{Hannah10} used  RHESSI observations of the quiet Sun to place 
upper limits  on the high-energy non-thermal emission.
Recent direct HXR imaging  (FOXI, NuSTAR) of  quiescent active regions 
typically does not show any non-thermal emission 
\citep[see, e.g.][]{ishikawa_etal:2014,hannah_etal:2016}.

One promising aspect for some time was associated with the anomalous 
 \ion{Si}{iii} line intensities. \cite{dufton_etal:1984} and subsequent
authors (e.g. \citealt{keenan_etal:1989, pinfield_etal:1999,dzifcakova_kulinova:2011}) 
suggested that these discrepancies
could have been caused by the presence of  non-Maxwellian electron
distributions in the low transition region.
For example, \cite{dzifcakova_kulinova:2011}
used CHIANTI atomic data, with the excitation cross sections for $\kappa$-distributions
 being recovered using an approximate parametric method \citep{DzifcakovaMason08},
to show that the \ion{Si}{iii} intensities observed  with SUMER by 
\cite{pinfield_etal:1999} could  be explained by $\kappa$-distributions.
However, \cite{delzanna_etal:2015_si_3} showed, using  new atomic data,
that the   \ion{Si}{iii} line intensities are generally consistent with 
Maxwellian electrons, if one takes into account the  temperature
sensitivity of the different \ion{Si}{iii} lines.

\begin{figure}[htbp]
\centerline{
\includegraphics[width=8cm]{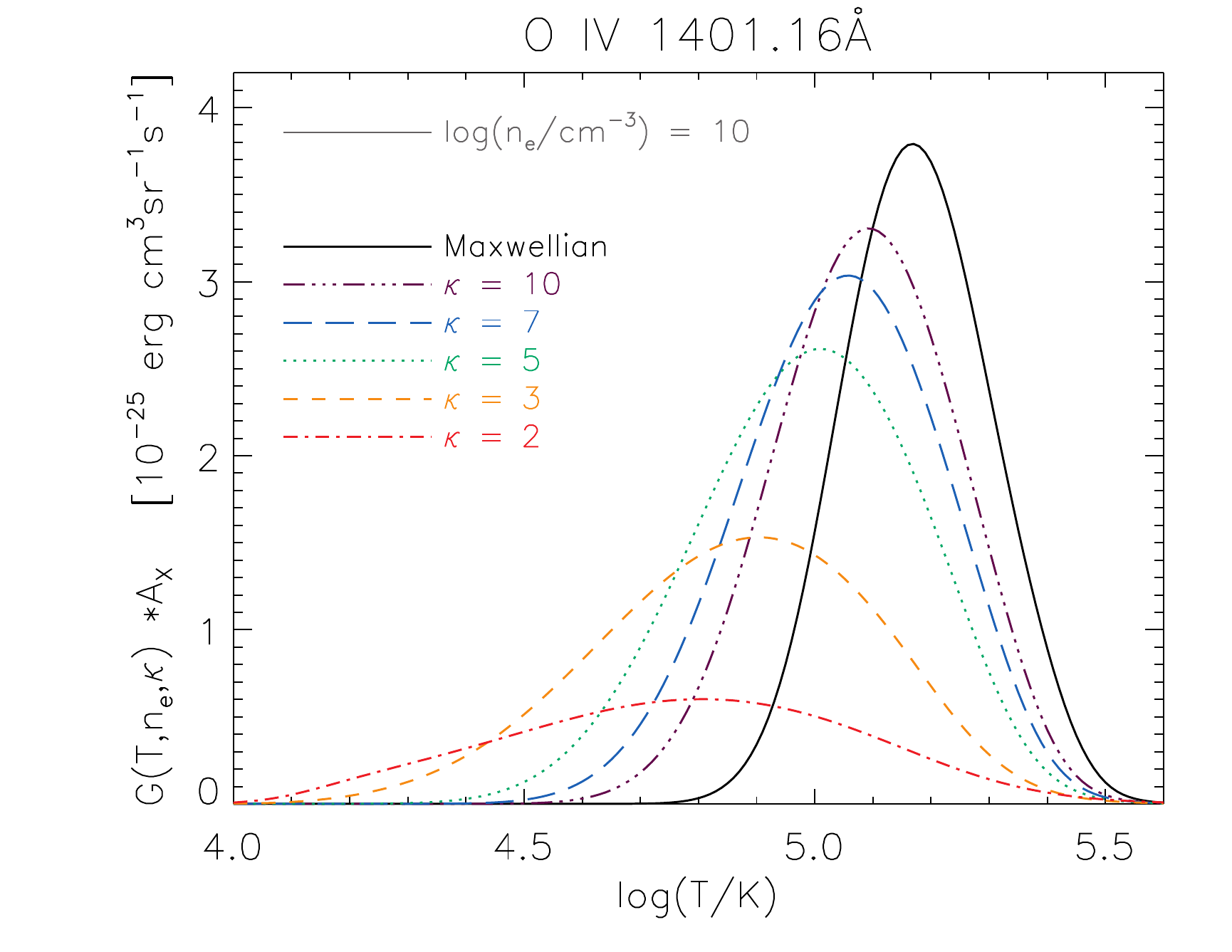}
\includegraphics[width=8cm]{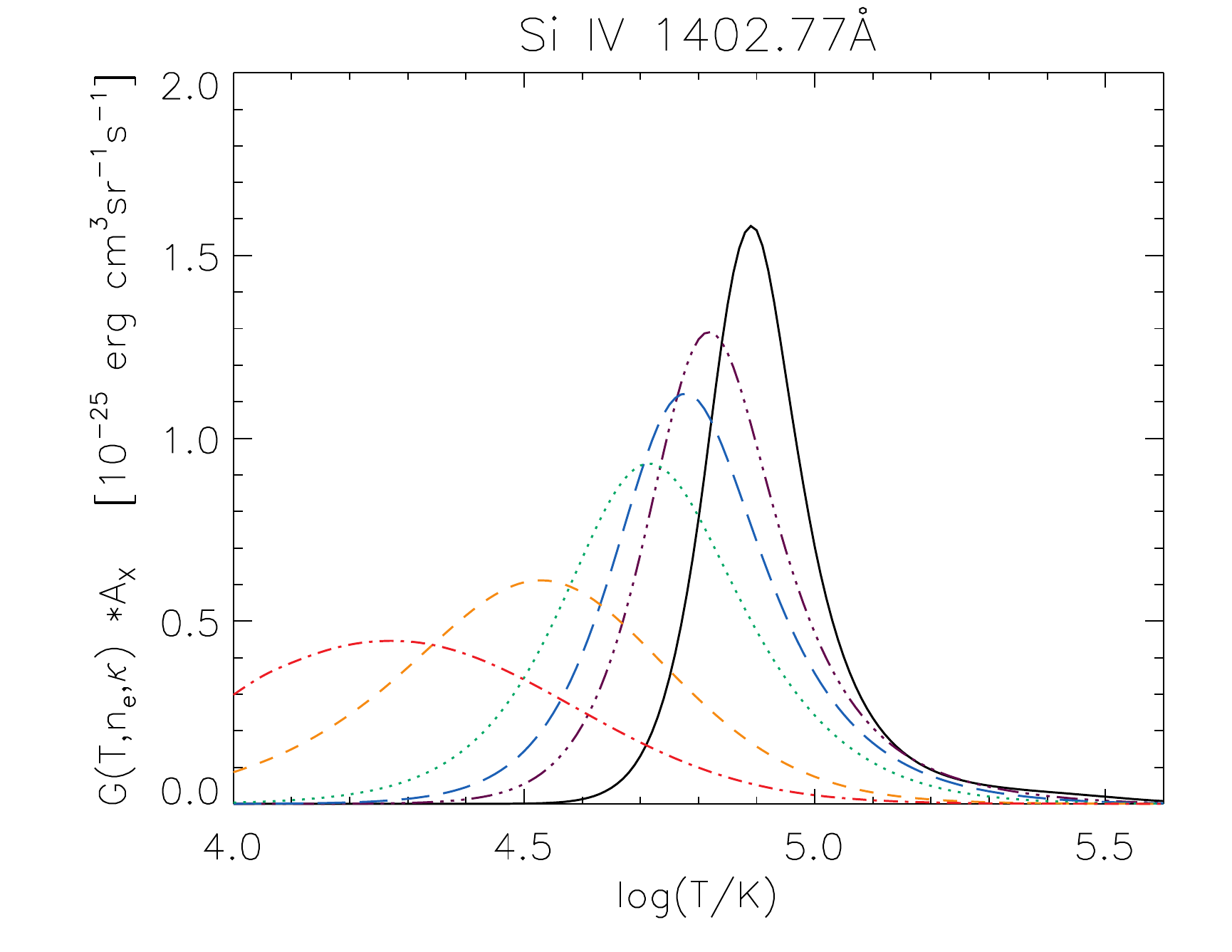}
}
\centerline{\includegraphics[width=9cm]{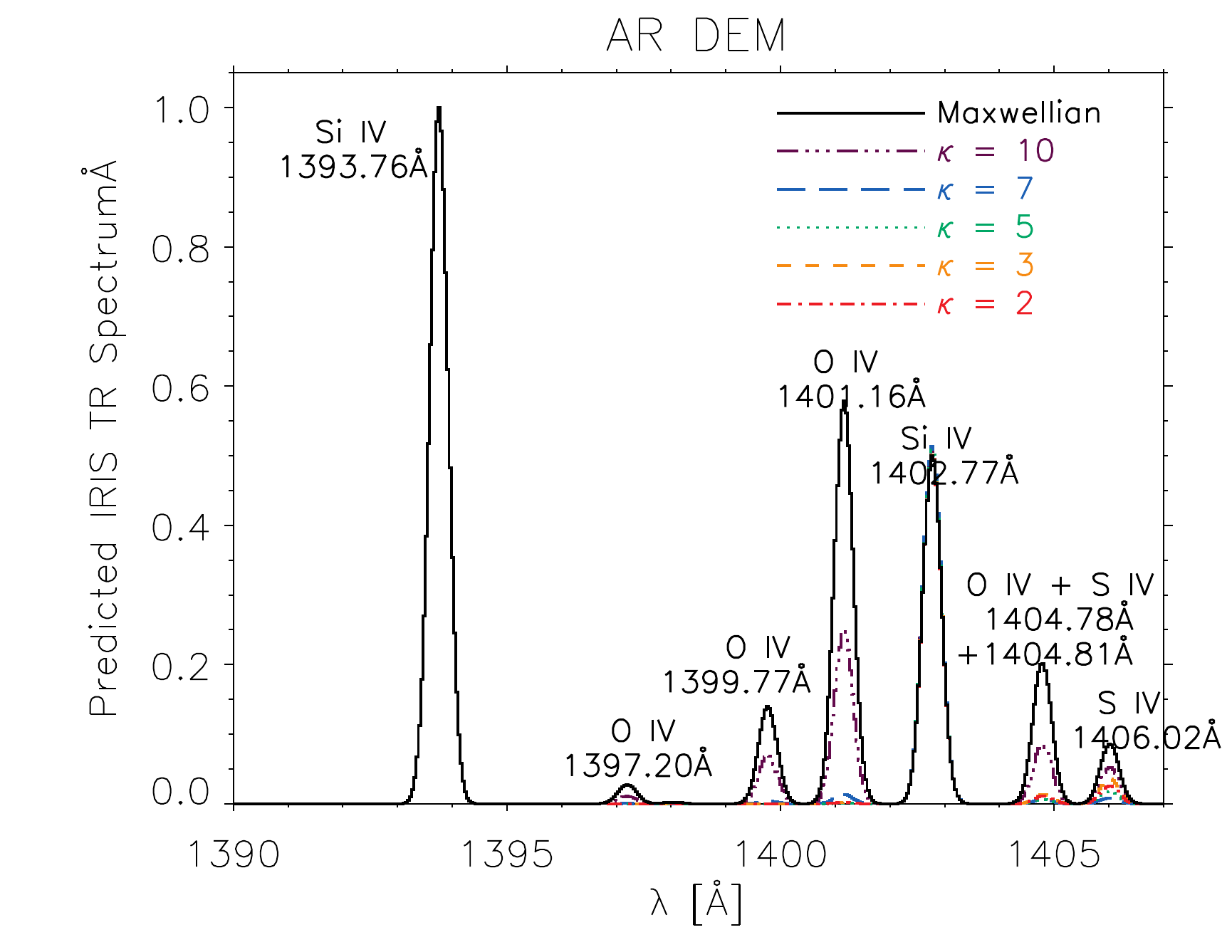}}
 \caption{ Top: contribution functions $G(T)$ for the  O IV 1401.16~\AA\ (left) and 
Si IV 1402.77~\AA\  (right) lines, calculated at fixed density and for different 
electron distributions. Note that the  $\kappa$-distributions  progressively
shift the formation temperature of the lines  (especially Si IV) towards lower
values.  
Bottom: IRIS simulated  O IV and Si IV line profiles in the FUV channel, normalised to the 
Si IV intensities. The simulations  show 
how the O IV lines become strongly depressed when  $\kappa$-distributions 
are considered.  These figures have been adapted from \cite{dudik_etal:2014_o_4}.}
 \label{fig:dudik_etal:2014}
\end{figure}

One general problem with these types of diagnostics is the
fact that line ratios that are good diagnostics of non-Maxwellian distributions 
are normally also sensitive to changes in the electron temperature. Hence, a 
good independent knowledge of the temperature distribution is often needed.
A few recent studies have used the original atomic cross sections to 
search for diagnostics of non-thermal electron distributions.
 \cite{dudik_etal:2014_o_4}
studied the effects of non-Maxwellian distributions on the 
the \ion{O}{iv} and  \ion{Si}{iv} lines observed by IRIS, showing that strong 
enhancements of the \ion{Si}{iv} / \ion{O}{iv} ratios  are expected in the case of a $\kappa$-distribution
(with high-energy tails), see Figure~\ref{fig:dudik_etal:2014}. 
Such enhancements are indeed
very common, although it is still not clear if time-dependent ionisation, 
non-Maxwellian distributions or other effects such as 
the DR suppression could be  the main factors \citep{polito_etal:2016b}.

\cite{dudik_etal:2014_fe}  discussed the excitation of \ion{Fe}{ix}\,--\,\ion{Fe}{xiii}
 under the assumption of a $\kappa$-distribution of electron energies,
reviewing the main line ratios for these important coronal ions,
which can be used to measure densities, temperatures and departures
from  Maxwellian distributions. 
\cite{dudik_etal:2014_fe} pointed out that a good diagnostic is given by the ratios 
of the EUV allowed lines with the visible forbidden lines. 
However, such a comparison requires a good knowledge of the relative calibrations.

\cite{dudik_etal:2015_loop} used Hinode EIS observations of a transient loop
to study the ratios of the observed iron coronal lines that are
sensitive to non-Maxwellian distributions. 
The observations appear to be inconsistent with a Maxwellian distribution.

\begin{figure}[htbp]
\centerline{\includegraphics[width=11cm]{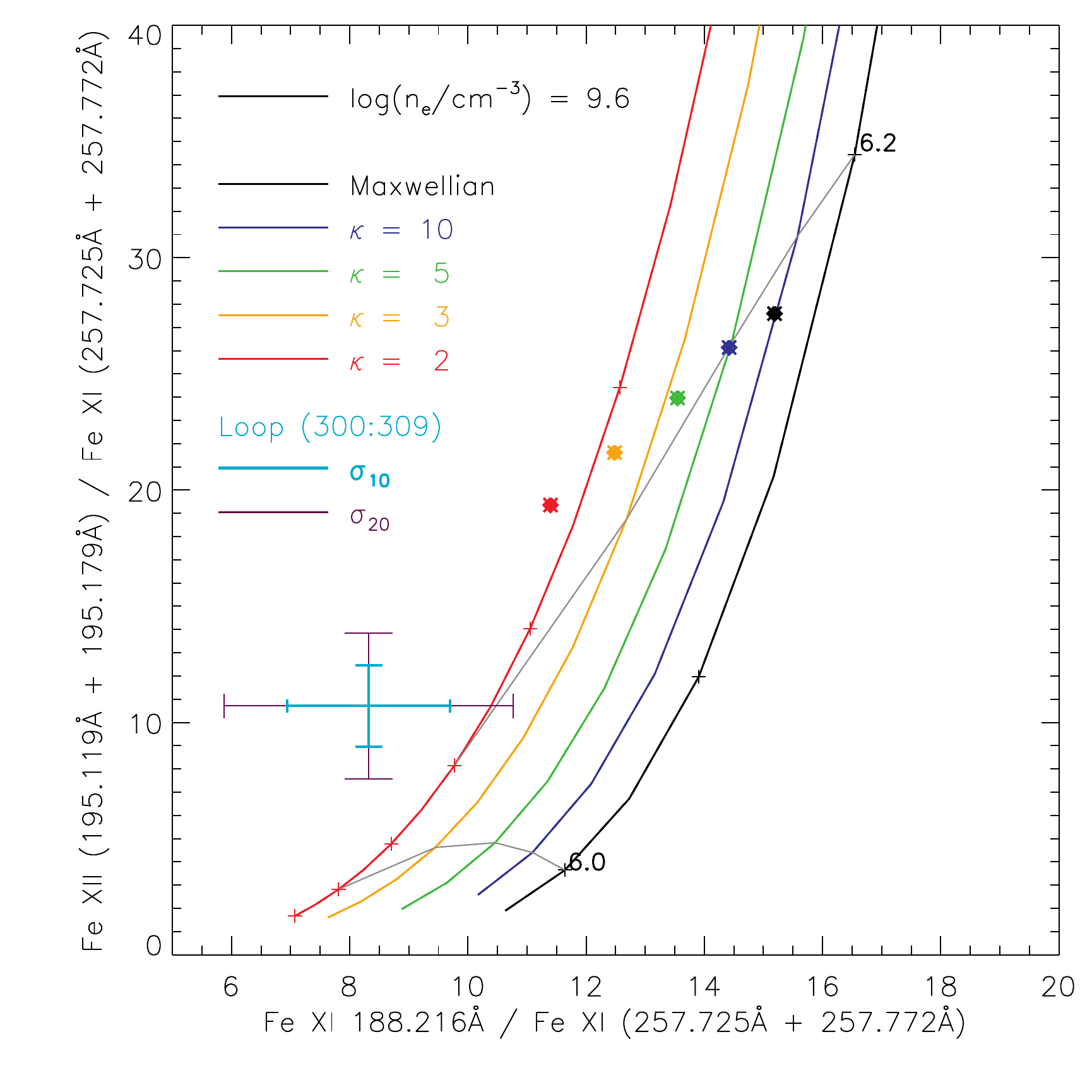}}
 \caption{Ratio-ratio diagram of Hinode EIS line intensities for a transient loop.
Lines of different colours indicate the theoretical curves for a 
Maxwellian and different $\kappa$-distributions. The cross, with a 
10\% and 20\% uncertainty, indicates the measured value
(adapted from \citealt{dudik_etal:2015_loop}).}
 \label{fig:dudik_etal:2015}
\end{figure}

\clearpage

\section{Emission measure - diagnostics}
\label{sec:dem_diagn}

If a unique relationship exists between $N\lo{e}$ and $T$,
a column differential emission measure   $DEM(T)$, a  function of only the plasma temperature, can 
be defined (\cite{withbroe78}):
\beq
\int\limits_T DEM (T) dT = \int\limits_h N_e ~N_H dh 
\eeq
i.e.
\beq 
DEM (T) = N_e N_H {dh \over dT} \quad [\mathrm{cm^{-5}\ K^{-1}}]
\eeq
For example, 
if we assume that the plasma pressure along the line of sight is constant,
then from the perfect-gas law 
$N_e^2\sim ({P^2 / T^2})$,  and  the electron density is only a 
function of temperature, and the $DEM(T)$ can be defined as a single-value 
function of the temperature.
The DEM gives an indication of the amount
of plasma along the line of sight that is emitting the radiation observed and 
has a temperature between $T$ and $T+dT$.

With this definition, the  intensity integral becomes:
\begin{equation}
{I(\lambda_{ij})}= {Ab(Z)} {\int\limits_T ~{C(T,\lambda_{ij},N_e)} ~DEM (T) ~ dT},
\label{eq:ab_int_dem_gt}
\end{equation}
assuming that the abundance of the element  $Ab(Z)$ is constant
along the line of sight.

We therefore have a system of Fredholm integrals of the first kind to 
be inverted, in order to deduce the DEM from a set of observed intensities.
 More details on various approximations and
 inversion techniques  can be found below.
This inversion procedure is notoriously 
difficult (see, e.g. \citealt{craig_brown:76, judge_etal:97}), but has many 
advantages over other diagnostics based e.g. on single line ratios.
For example, it has the advantage that at the same time 
the DEM distribution and the elemental abundances can be obtained, 
if a sufficient number of lines over a broad temperature 
range is observed. This minimizes the effects of errors 
in the observation or theory associated with individual lines.

Once the  DEM is known, it is possible to define a 
 column  emission measure $EM_T$ over some temperature
interval  $\Delta T$:
\beq
EM_T (T_i) \equiv  \int_{T_i- {\Delta T \over 2}}^{T_i+{\Delta T \over 2}}  DEM(T) dT 
\eeq
The  total column  emission measure $EM$  can also be calculated, integrating the 
DEM over the whole  temperature range:
\beq
 EM  \equiv \int_h N_e N_H dh = \int_T DEM(T)~ dT  \quad~~~ [{\rm cm}^{-5}]   
\eeq

Sometimes the total column  emission measure is defined as 
$EM  = \int N_e^2 dh = \int  DEM(T) dT $, 
with the column differential emission measure defined as 
$DEM(T)=  N_e^2  ({dT / dh})^{-1} $.

If the DEM is well defined and known, it is then useful to define 
an  \emph{effective temperature}.
This  is a weighted average of the temperature, where the 
weight is given by the product of the $G(T,N)$ with the $DEM$:
\beq
\log T_{\rm eff} = \int G{\left({T, N}\right)}~DEM{\left({T}\right)} \log T~dT / 
{\int G{\left({T}\right)}~DEM{\left({T}\right)}~dT}
\label{eq:t_eff}
\eeq
The effective temperature $T_{\rm eff}$ gives an indication of where 
most of the line is formed. This is often  quite different to   
 $T_{\max}$, the temperature where the  $G(T)$ of a line has a maximum,
or where the maximum ion abundance is.

\begin{figure}[htb]
\centerline{\includegraphics[width=9cm,angle=90]{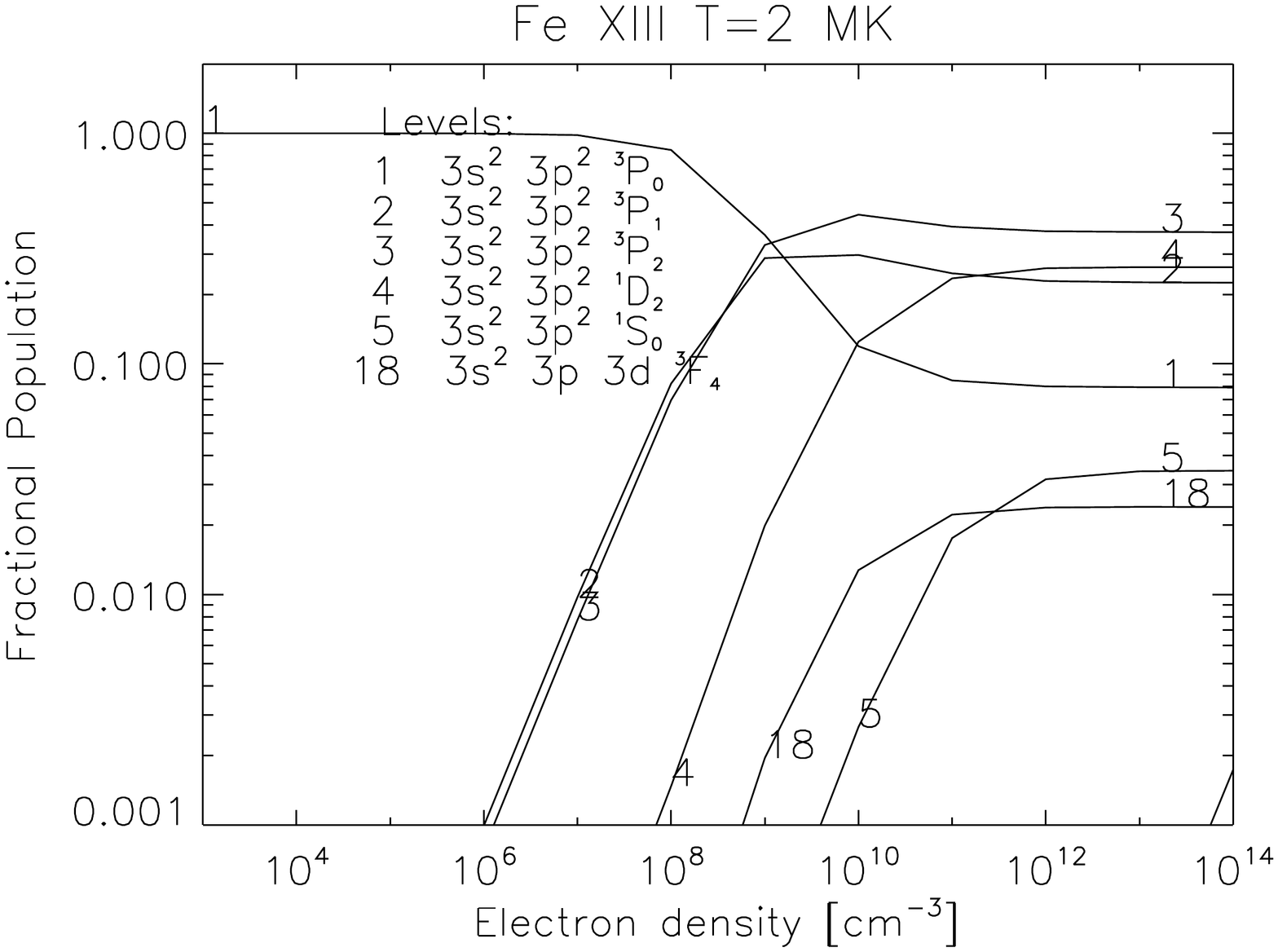}}
\centerline{\includegraphics[width=9cm,angle=90]{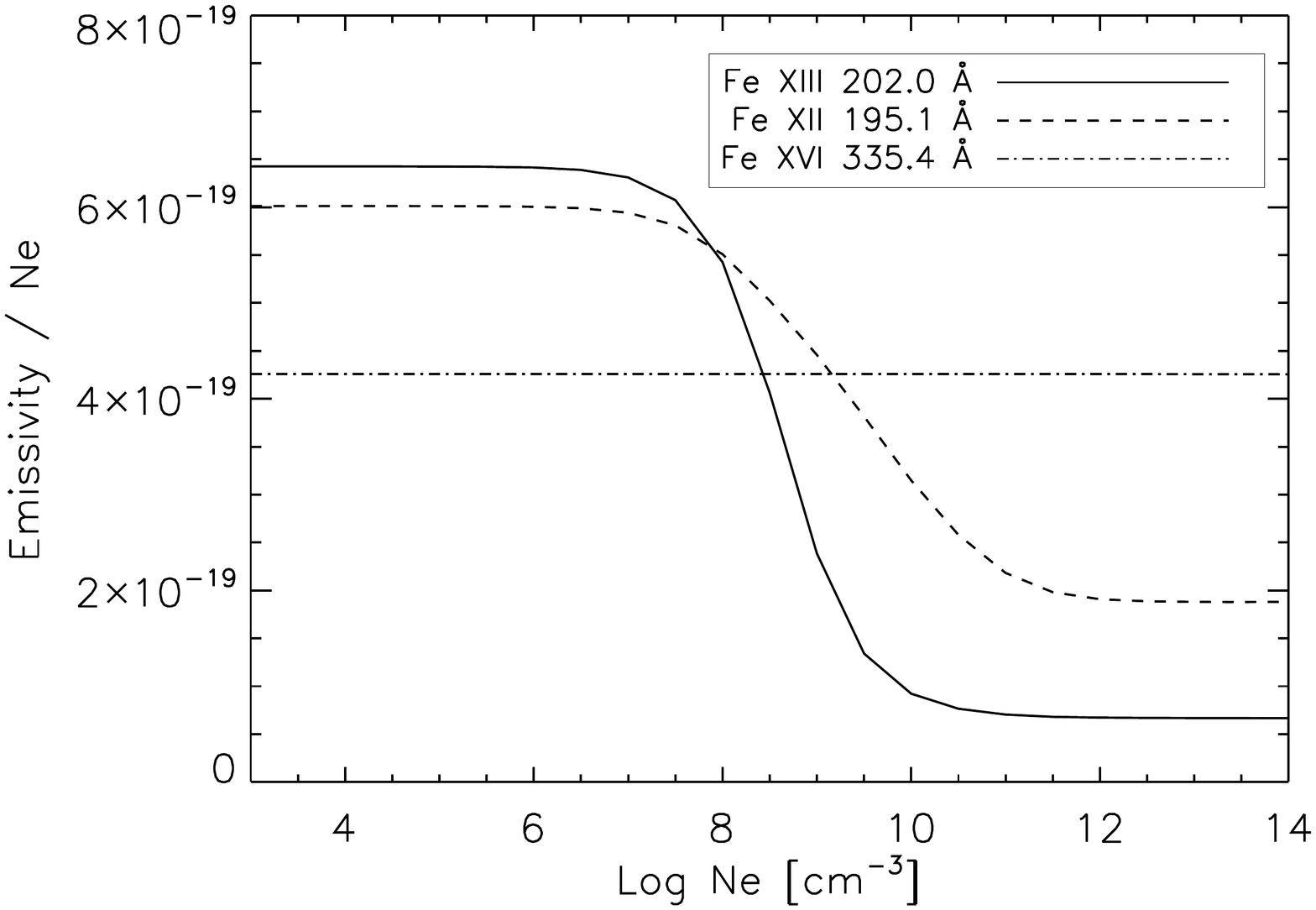}}
  \caption{Top: the fractional population of the \ion{Fe}{xiii} levels; 
note the significant decrease in the population of the ground state.
Bottom: the emissivity (divided by the density) of the strongest
dipole-allowed decays to the ground state of \ion{Fe}{xiii}
at 202.0~\AA,  \ion{Fe}{xii} at 195.1~\AA, and  \ion{Fe}{xvi}
at 335.4~\AA. CHIANTI v.8 was used for the emissivities.
 }
  \label{fig:fe_13_pop}
\end{figure}

Further refinements and considerations of the integral inversion problem
and extensions to  an emission measure differential in 
both temperature and density can be found e.g. in 
 \cite{almleaky_etal:1989,hubeny_judge:95}.
However, observations are often limited by practical considerations,
for example the number of spectral lines which can be observed with a given instrument. 
 The simplest approach,
used in most literature, is to 
 assume that the $DEM(T)$ is a  function of only the plasma temperature,
and for the inversion use only spectral lines that have contribution functions  $G(T)$  that 
do not vary with the density. Normally, these are dipole-allowed lines
that decay to the ground state. 
However, even in these cases some 
dependence on density is present. This occurs when  metastable levels 
 reach a non-negligible population at sufficiently high densities.
An example is shown in  Figure~\ref{fig:fe_13_pop}: the population of the 
\ion{Fe}{xiii} metastable levels becomes so high that the population of the ground state
decreases  significantly above $N_{\rm e}=10^{8}$ cm$^{-3}$. Therefore, lines such as the 202.0~\AA\
that are mainly populated directly from the ground state have $G(T)$ 
that are affected by density changes. 
Another ion strongly affected is \ion{Fe}{ix}.
Similar effects are  present 
in many ions, e.g. \ion{Fe}{xii} and  \ion{Fe}{xxi}. 
 On the other hand, lines of ions such as  \ion{Fe}{xvi}
which do not have metastable levels,  have $G(T)$ which are independent of density
and are therefore excellent diagnostics for $DEM$ analyses.

\subsection{Volume vs. column EM}

As we have previously discussed, most solar measurements are of 
radiances over a  volume along the line of sight, hence the inversion methods produce 
a \emph{column emission measure}, i.e. the amount of plasma along the line of 
sight $dh$ within the observed area ($A$). Sometimes in the 
literature the  \emph{volume emission 
measure} $ EM_{\rm V} = \int_V N_{\rm e}^2 dV ~ (\mathrm{cm^{-3}})$ and its differential 
in temperature $ DEM_{\rm V}(T) = ~N_{\rm e}^2 {dV \over dT}$ are used
instead of the column  emission measures, with $dV= A \times dh$.

There are, however, measurements of the Sun as a star (i.e. irradiances), 
where the relation between column and volume emission measures is non-trivial.
One could in principle assume that the distribution of the emitting 
plasma is homogeneous and has spherical symmetry, i.e. 
the emitting region is a layer $dh$ distributed 
over the entire solar disk, i.e. $dV=4\pi R^2_* dh$
($R_*$ is the solar radius).
Then the  \emph{volume} $DEM$
can be scaled with the factor $4\pi R^2_* \over d^2$ to obtain 
a \emph{column} $DEM$. 
However, the layers over which lines emitted at different temperatures
would be different, so a simple relation is not straightfoward. 

The same issue occurs when  relating the 
radiances observed over the solar disk  to the  irradiances. 
Again if the Sun were homogeneous
and only emitting from its disk, the measured irradiance 
at the Earth in a line $F(\lambda_{ji})$ ($\mathrm{erg\ cm^{-2}\ s^{-1}}$)
would be related to the radiance on the solar disk $ I(\lambda_{ji})$
($\mathrm{erg\ cm^{-2}\ s^{-1}\ sr^{-1}}$) by 
\beq
F(\lambda_{ji}) = \pi {R^2_* \over d^2} ~ I(\lambda_{ji})  ~~.
\eeq

However, a significant fraction of the emission in most spectral lines
comes from the limb brightening and from the off-limb corona. 
A correction factor for the limb-brightening for chromospheric and 
transition-region lines
would need to be introduced, to relate the on-disk Sun-centre radiance in a 
line to its irradiance.
A recent review of limb-brightening estimates
for EUV/UV lines is given by \cite{andretta_delzanna:2014}.
A further off-limb contribution would also need to be estimated
(see, e.g. \cite{delzanna_andretta:10}).

\subsection{EM, DEM approximations }
\label{sec:em_methods}

Over the years, various approximations 
and methods have been introduced 
in order to approximate the DEM and to deduce elemental abundances.
We describe one method below, while the other ones are described later
when we discuss abundance measurements, since those methods were mainly 
used for this purpose.

\subsubsection{The emission measure loci approach}
\label{sec:em_loci}

One direct  approach to assess how the plama is 
distributed in temperature is to plot the ratio 
of the observed intensity $I\lo{ob}$ of a line with its 
contribution function,
$I\lo{ob} / G(T)$,  as a function of temperature.  
 The loci of these curves are an upper limit to the  emission measure distribution.
In fact, for each line and temperature $T_i$ the value 
$I\lo{ob} / G(T_i)$ represents an upper limit to the  value of the 
\emph{line emission measure} $EM_L$ (Eq.~\ref{eq:em_l}, below)
 at that temperature, assuming that all the observed emission $I\lo{ob}$ 
is produced by plasma at temperature $T_i$.

The method was first introduced 
 by \cite{strong:78} and later applied by \cite{veck_etal:84} 
and following authors to the analysis of solar X-ray  spectra, mainly to 
obtain relative abundances. 
It was also used to study stellar emission measures 
\citep{jordan_etal:87}.

\begin{figure}[htb]
\centerline{\includegraphics[width=8cm,angle=90]{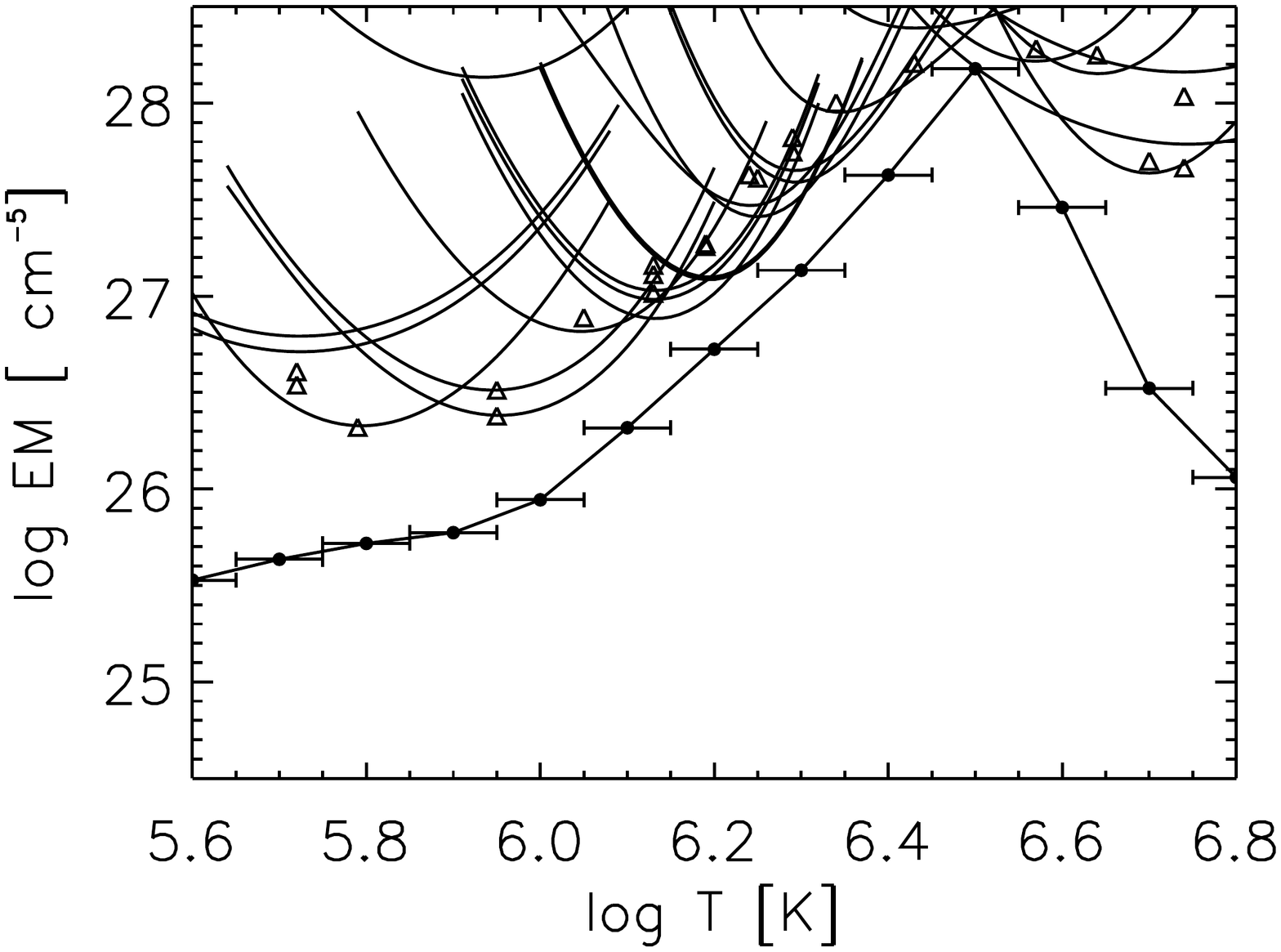}}
  \caption{The EM loci ($I\lo{ob} / G(T)$) curves for a selection of lines 
observed by Hinode EIS in an active region core \citep{delzanna_mason:2014}.
The filled circles are   the $EM_T$ values calculated from the  DEM
with  $\Delta$~log~$T=0.1$ K.
The triangles, plotted at the temperature T\lo{max}, are 
the  \cite{jordan_wilson:71} values.
 }
  \label{fig:em_pott}
\end{figure}

The  $I\lo{ob} / G(T)$ curves for a selection of lines 
observed by Hinode EIS in an active region core \citep{delzanna_mason:2014}
are  displayed as an example  in Figure~\ref{fig:em_pott}.
The filled circles are   the $EM_T$ values calculated from the  DEM
with  $\Delta$ log $T=0.1$ K.
The triangles, plotted at the temperature T\lo{max}, are 
the  \cite{jordan_wilson:71} values (see below in Sect.~\ref{sec:abund}).

\subsection{The anomalous EM problem}
\label{sec:anomalous_ions}

\begin{figure}[htb]
\centerline{\includegraphics[width=12cm]{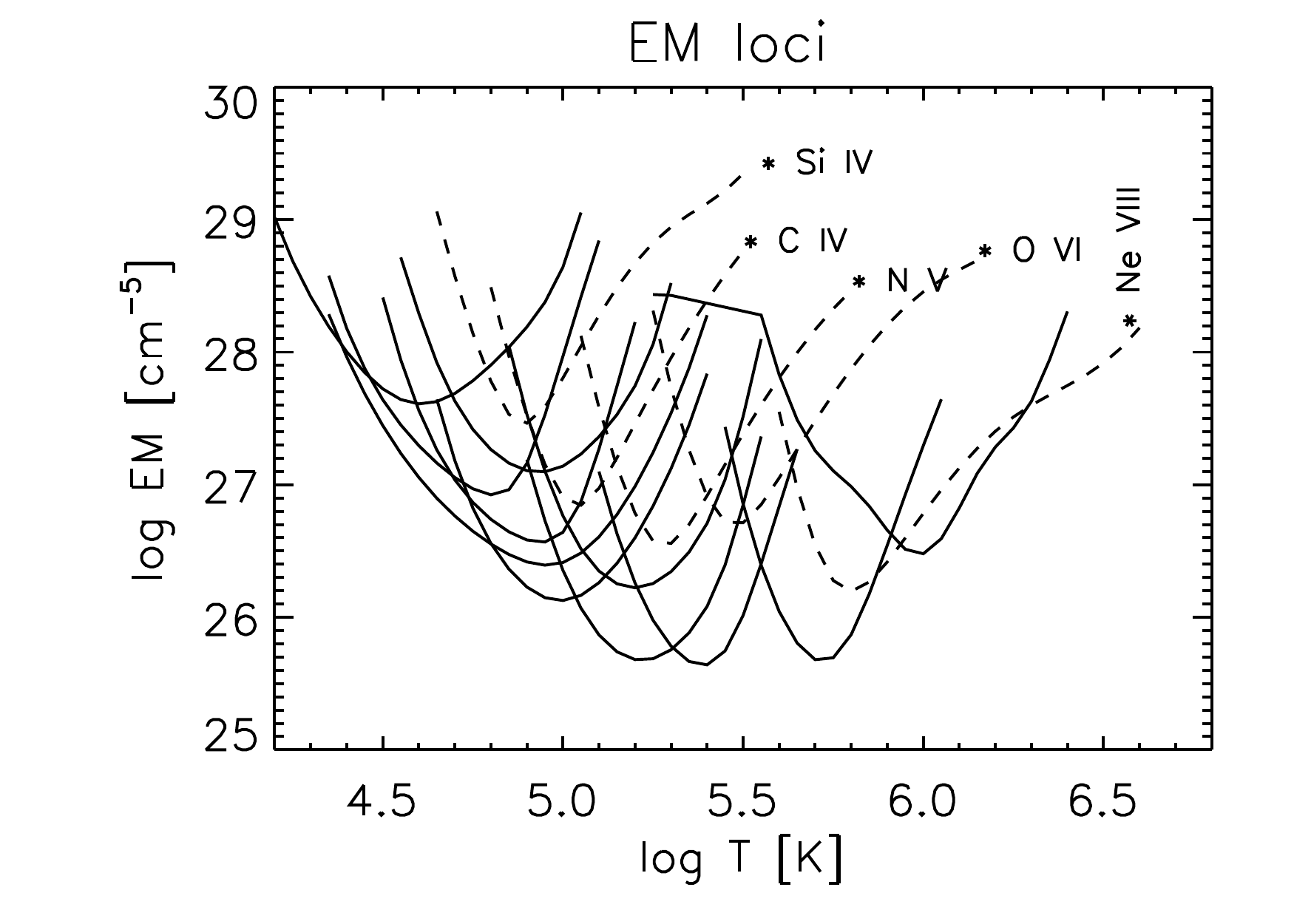}}
 \caption{EM loci curves from the solar irradiance measurements presented in 
\cite{judge95} using  CHIANTI v.8. The EM loci curves 
of the spectral lines of the Li- and Na-like sequences 
are displayed as dashed lines. The other curves are from C II, 
O III, Si III, N III, O IV, O V, Ne VII, Mg IX.}
 \label{fig:judge_etal_95}
\end{figure}

The Li and Na-like ions give rise to some of the strongest lines in the UV. 
However,  the emission measures for 
lines emitted by  ions of the Li- and Na-like isoelectronic sequences are at odds
with those of the other ions.
The problem is clearly present even in the early \cite{pottasch:63}
irradiance measurements, although it was not noted by that  author
(see Figure~\ref{fig:pott}).  
The issue was mentioned by \cite{burton_etal:71,dupree72}, and later 
discussed by several authors, e.g. \cite{judge95}. 
For a detailed discussion see \cite{delzanna_etal:02_aumic} and references therein. 
This anomaly is illustrated in Figure~\ref{fig:judge_etal_95}, where the 
the  solar irradiance measurements presented in 
\cite{judge95} are combined with the atomic data in  CHIANTI v.8.
 Note that the irradiances 
were approximately converted to radiances by \cite{judge95}.
It can be seen that the lines from the  Li- and Na-like isoelectronic sequences are clearly at odds
with the other isoelectronic sequences.

\cite{delzanna_etal:02_aumic} showed that the same problem occurs when 
stellar emission measures are considered. This was shown by  
combining XUV observations from different instruments (FUSE, HST/STIS, EUVE). All 
previous literature considered stellar observations at wavelengths where 
by far the brightest lines are those from Li- and Na-like ions (e.g. IUE and 
HST/STIS), so all the published results (in terms of DEM, abundances, 
densities) were incorrect.
 In fact, using ions from the Li- and Na-like sequences together with 
those from other sequences leads to incorrect estimates of elemental abundances, 
 densities,  and temperatures.

There are clearly some processes by which the lines of the anomalous ions are 
enhanced. The enhancements vary depending on the region observed, but are
typically about a factor of 5. 
They affect lines formed in the lower and upper transition region 
(0.2--0.9 MK), but occasionally also coronal lines (1-3 MK). 
One possible explanation that has recurrently been suggested is the 
possible effect of high densities in the charge state distribution.
However, this effect is not very large (see, e.g. \cite{bradshaw_thesis:2003} and 
\citealt{doyle_etal:2005}),
and would require high densities ($N_{\rm e}=10^{11}$ cm$^{-3}$ or more), 
so it  would presumably only affect lines formed in the low transition region.

The contribution functions of the Li-like and Na-like ions 
have a significant  high-temperature tail, i.e. are more sensitive to high-energy
electrons. Strong departures from Maxwellian distributions could in principle 
explain the discrepancies.

It is interesting to note that even larger enhancements are observed in 
all the EUV helium lines  in the quiet Sun.  
There is an extensive literature on this subject, where many 
possible explanations have been considered. 
One issue that is not clear, for the helium lines, is their formation
process. Some authors believe (see, e.g. \citealt{zirin:75}) 
that the dominant process is 
 photoionisation followed by recombination to the
upper level of the line (either directly or in a cascade through intermediate
levels).
If collisional excitation is the dominant process 
(see e.g. \citealt{Athay:88}), then the large 
excitation energy compared to their ionisation 
equilibrium temperature 
would make those lines very sensitive to non-thermal effects.
Also, the large non-thermal
velocities observed imply that many of the ions would be subject 
to higher-temperature thermal
electrons in the regions where the temperature gradient is large, as in the solar 
transition region  \citep{Jordan:75,Jordan:80}.
Ambipolar diffusion is also an important process (see, e.g. 
\citealt{fontenla_etal:1993,fontenla_etal:02}). 
Non-equilibrium ionization effects can also increase 
helium and TR line intensities \citep{raymond:1990,bradshaw_etal:04}.
Further details on the helium EUV lines can be found in 
\cite{andretta_etal:03,pietarila_judge:2004,jordan_etal:05,andretta_etal:2008} 
and references therein.

\subsection{Methods to estimate the Differential Emission Measure}

There are many methods and codes available in the literature 
to obtain estimates of the  DEM, using either spectral lines 
or broad-band imaging. 
Each method has advantages and limitations and  results can often be
misleading.  Only a brief summary of some of the most widely used methods
is given here.

The inversion problem itself is not simple and requires some
assumptions about the nature of the solution.
A series of workshops was sponsored  
in 1990/91 to study differential emission measure 
techniques \citep{harrison92}. It was found that 
most codes eventually gave consistent results, but that the DEM derived depends
rather critically on the methods used to constrain the solution
and the errors in the observed intensities and atomic data.

Perhaps one of the earliest methodd developed to estimate the DEM was the iterative 
approach of \cite{withbroe:1975}, later implemented by several other authors, mostly for the 
analysis of solar X-ray observations.
A statistical method based on the Bayesian formalism and using the 
Markov-Chain Monte Carlo (MCMC) approach  was developed
by \cite{kashyap_drake:98}.
This MCMC method is part of the PINTofALE%
\footnote{\url{http://hea-www.harvard.edu/PINTofALE/}}
modelling package and is widely used.
The advantages of this method are that it provides positive solutions,
a measure of the associated uncertainty, and does not require a
prescribed functional form, although the program forces
the solution to be smooth. The program starts with a guess solution
and applies  random adjustments in a Markov-Chain, producing a family
of DEM solutions that are a  representation of
the probability distribution function of the actual DEM. The program is
robust but very slow, compared to other methods.

Most codes adopt a simple $\chi^2$ approach. However, since the solution
is normally undetermined (has more degrees of freedom than the input values),
the DEM solutions have to be `regularised' in some way, to obtain a smooth 
result.
One way to enforce a smooth and positive solution is to impose  a functional 
form to the DEM. One of simplest functional forms is a single Gaussian,
and such an approach was developed by e.g. 
\cite{aschwanden_boerner:2011} and \cite{guennou_etal:2012}.
Another  common example is to assume that 
the DEM can be modelled with a spline function. 
Such an approach, combined with a simple $\chi^2$ minimization (using MPFIT), is adopted by 
the XRT\_DEM iterative  method (available within SolarSoft), 
coded by M.\ Weber, and widely used. 
The minimization provides  DEM values at the spline nodes 
which best describe the observations. 
It is possible to evaluate uncertainties by running  the code many times,
by varying the input intensities to within their uncertainties
(a sort of Monte Carlo forward modelling).

Additional constraints are often added using Lagrange multipliers, as 
with the `maximum entropy method', as discussed by A. Burgess in one of the 
DEM workshops \citep{harrison92}. 
A common  implementation of this method has the additional assumption that 
the DEM is  a cubic spline. 
This procedure has the advantage of  giving only positive $DEM$
values  and converging  quickly. 
This method was  adopted by several  authors, 
 e.g. \cite{monsignori91,landi_landini:1997,delzanna_thesis99}.
Note that the smoothness of the DEM distribution depends on the number and position 
of the selected mesh
points, which should be chosen at those temperatures where there are 
constraints from the observations.

Another approach is the so-called {\it zero-order regularization},
which was implemented by e.g. \cite{hannah_kontar:2012} and 
\cite{plowman_etal:2013}.
The DATA2DEM\_REG%
\footnote{\url{http://www.astro.gla.ac.uk/\~iain/demreg/}}  routine implemented by 
\cite{hannah_kontar:2012}
recovers the DEM using a Tikhonov regularization and a Generalized
Singular Value Decomposition (GSVD) approach.
 The solution is not always positive which is a major disadvantage,
but the code is faster than most other methods.

\begin{figure}[htbp]
\centerline{\includegraphics[width=8cm,angle=90]{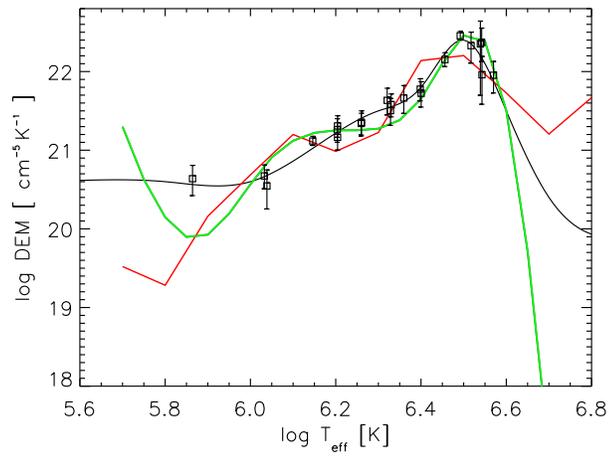}}
  \caption{DEM of an  AR core observed with Hinode EIS, obtained
with the spline (MEM\_DEM) method (black smooth curve), with the MCMC\_DEM
program (red curve) and XRT\_DEM method (green curve). The points
are plotted at their effective temperature and the corresponding 
DEM value, multiplied by the ratio of the observed vs.\ predicted intensity 
(figure adapted from \citealt{delzanna_mason:2014}).}
  \label{fig:dem_comp} 
\end{figure}

In the temperature regions where the DEM is well constrained, the results
of the various methods normally agree quite well. 
The best constraints are given by a range of individual  emission lines
formed over a wide range of temperatures. 
Fig.~\ref{fig:dem_comp} shows a comparison of different DEM methods using 
Hinode EIS observations in an active region core \citep{delzanna_mason:2014}.

One major limitation which can occurr is when the plasma is nearly isothermal.
In this case, most codes generally fail, as they enforce a  continuous  DEM solution.
A possible way to recover both a nearly isothermal and a continuous solution
was presented by  \cite{delzanna:2013_multithermal}, where the DEM 
was modelled by a superposition of Gaussian distributions. A 
simple $\chi^2$ minimization  iterative  method was used to obtain DEM 
solutions from SDO AIA data. The solution could be nearly isothermal, with a 
suitable choice of the width of the Gaussians, or continuous across
the entire temperature range.
A similar approach (in terms of basis functions) was introduced by 
\cite{cheung_etal:2015}  to obtain DEM solutions from SDO AIA data,
although the actual inversion method is  different and its implementation is fast.

Another significant limitation of all the available codes is 
when the  kernels (i.e. the line contribution function or a temperature
response for a broad-band imaging instrument) have a double peak 
feature. For example, several of the SDO AIA channels have this problem.
 Naturally, the codes tend to find  continuous DEM curves 
which cover both peaks,  but these are not necessarily the correct solutions.
A way to improve the solution is to have different kernels (and input) 
for different cases, as suggested for SDO AIA by 
\cite{delzanna:2013_multithermal}.

It has been shown by \cite{dere:78b}  that if the DEM is approximated 
with a  non-negative smooth 
function, then the solutions have errors that are not significantly 
greater than the errors in the spectral data, in the temperature
ranges  where there are observational constraints.
In the regions with limited observational constraints, the 
DEM values can have large errors, as shown by e.g. \cite{jakimiec_etal:1984}.

\clearpage

\section{Emission measure - observations}
\label{sec:dem_obs}

The above methods have been applied in  a wide range of cases.
As we have mentioned, the main purpose is to define the temperature
distribution in the plasma, which is a fundamental parameter
for any coronal heating study. 

Once the DEM/EM is known, estimates of densities can be obtained
(as described below), as well as the optically thin radiative losses, which are needed 
to study the energy budget in any coronal structure.
Earlier studies used the DEM slope below 1 MK to infer the temperature 
gradient in the transition-region and build semi-empirical hydrostatic models
\citep[see e.g.][]{mariska:92}. 
Later studies, briefly mentioned below, focused more on the forward-modelling,
i.e. built theoretical models and then predicted  the DEM distribution.

A second and important purpose of the DEM/EM methods is to obtain the 
relative elemental abundance, as described below in Section~\ref{sec:abund}.

The literature on EM/DEM measurements is too extensive to even be summarised here.
Measurements across the XUV spectrum on all solar features, using both spectral 
lines and broad-band instruments have been carried out for over fifty years.
We focus in this review on more direct measurements of temperatures. 
A selection of  results on some of the main solar features is 
however presented here,  to provide an overview to the unfamiliar reader.

\subsection{Quiet Sun and Coronal holes}

\begin{figure}[!htbp]
\centerline{\includegraphics[width=7cm,angle=0]{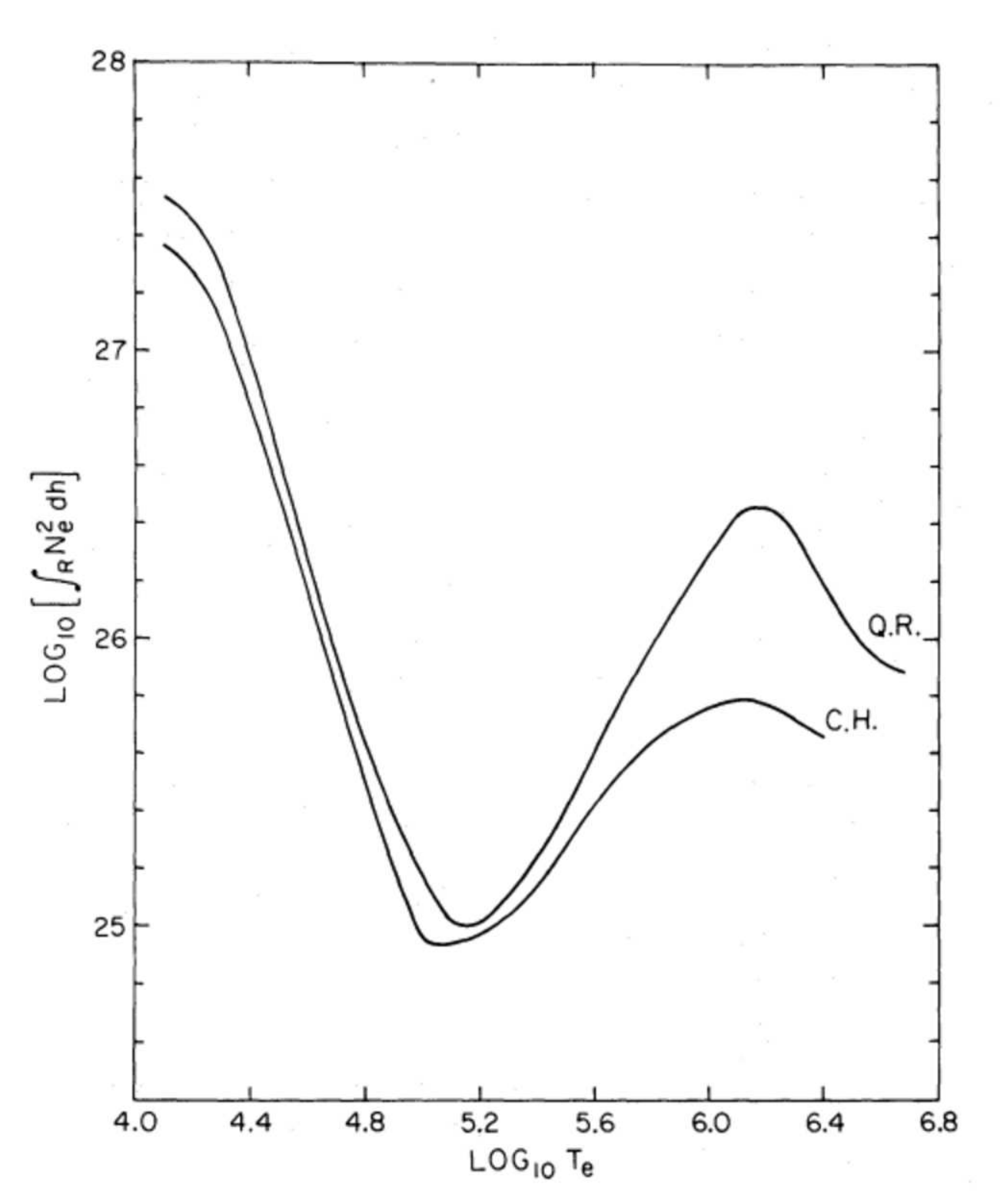}} 
  \caption{ The DEM distributions of the average quiet Sun
and coronal hole obtained by \citealt{raymond_doyle:1981b}.}
\label{fig:raymond_doyle:1981}
\end{figure}

Following \cite{pottasch:63}, all subsequent measurements based on 
solar irradiances clearly showed that the quiet Sun has a peak 
EM/DEM around 1 MK, a minimum around log $T$[K]=5.2, and a steep 
increase towards the chromosphere. Radiance measurements of the quiet Sun 
on-disk produced the same picture. Notable examples are the results 
obtained by \cite{withbroe:1975}, based on OSO-IV and OSO-VI,
and those based on  Skylab observations (see, e.g. \citealt{raymond_doyle:1981b}
and Fig.~\ref{fig:raymond_doyle:1981}).
The slope above 2 MK is difficult to establish, as there is very little
emission from the quiet Sun.
All the observations from later instruments gave similar results.

In off-limb observations of the quiet Sun, the EM distribution
tends to be very narrow around 1 MK, as e.g. clearly shown with the EM loci method
applied to SOHO SUMER observations 
(see, e.g. \citealt{feldman_etal:97,landi_feldman:2003}) or Hinode EIS
\citep[see, e.g.][]{warren_brooks:2009, delzanna:12_atlas}. 
The peak temperature of the EM distribution typically increases
with the height above the limb.

Radiance measurements of coronal holes typically produce EM distributions
that are  similar to the quiet Sun ones, up to  1 MK, where a 
sharp drop occurs (i.e. the coronal distribution is nearly isothermal).
In fact, transition region lines have almost the same intensity distribution inside and outside
coronal holes, with the main difference that the brightest supergranular cell boundaries
(network) are depressed in coronal holes (by about 50 \%), and the height of the 
TR layer is much higher.
Notable results are those
based on OSO-IV and  Skylab observations, such as those of 
\cite{munro_withbroe:1972,raymond_doyle:1981b}.
All later results based on SOHO CDS 
\citep[see, e.g.][]{mason_etal:1997,delzanna_jgr99a,young_etal:1999}, 
SOHO SUMER \citep[see, e.g.][]{feldman_etal:98b,landi:2008}, Hinode EIS \citep[see, e.g.][]{hahn_etal:2011},
 and other missions were similar.

Within coronal holes, plumes are long-lasting features (see the recent 
review from \citealt{poletto_lrsp}). 
These were 
spatially resolved with SOHO CDS, and EM/DEM analyses clearly showed that near their 
bases, the cross-sections of plumes are almost  isothermal
\citep{delzanna_bromage:99,delzanna_etal:03}. This  confirmed an earlier suggestion of 
\cite{ahmad_withbroe:1977} based on off-limb Skylab HCO observations of 
\ion{Mg}{x} and \ion{O}{vi}.
The isothermality of the plumes has important implications in terms of the 
derived elemental abundances, as discussed below in Section~\ref{sec:abund}.

\subsection{Active region cores}

A popular theory for the coronal heating assumes that 
nanoflare storms  occurring in the corona  heat the plasma
to $T_{\rm e} >$3 MK, with subsequent cooling to form the $\simeq$3 MK loops
typical of AR quiescent cores 
(see e.g. in particular \citealt{parker:1988,klimchuk:2006,patsourakos_klimchuk:09, klimchuk:2010,bradshaw_klimchuk:2011,cargill:2014}
and in general the Living Review by \citealt{reale:2012_lr}).  
Other theories, such as the dissipation of Alfven waves
\citep{van_ballegooijen_etal:2011} also predict  high-temperature 
plasma heated by short-lived and frequent heating events.
Measuring the slopes  $a$  [EM(T)~$\propto$~T$^{-a}$] at  temperatures above 3 MK 
is therefore  fundamental for constraining the heating of the core loops.
The frequency of the heating events also leaves an imprint in the 
plasma emission. 
In the high-frequency heating scenario, the duration between successive 
heating events is shorter than the cooling time, which 
produces  a narrow EM distribution with a 
steep slope  $b$  [EM(T)~$\propto$~T$^b$]
in the 1--3~MK range. However, in the low-frequency heating scenario, 
the time duration is longer than the cooling time, and the plasma 
has sufficient time to cool down before being  heated again. 
Hence, there is a substantial amount of cooler material,
producing  shallower EM slopes (see e.g. 
\citealt{mulu-moore_etal:2011,tripathi_etal:2011,bradshaw_etal:2012,cargill:2014} 
and references therein).
Generally, the models predict  that
low-frequency nanoflares can only account for slopes $b$  that are below 3.

There is a vast literature with contradictory results in terms of 
the slopes $a, b$. Our view has always been that the AR cores are nearly
isothermal around 3 MK because it was clear from CDS images that they were 
only visible in \ion{Fe}{xvi} (see, e.g. \citealt{delzanna_mason:03}).
This view is not new. For example, \cite{rosner_etal:78} found that the 
excellent 
Skylab X-ray imaging was consistent with nearly  isothermal plasma at 3 MK.
 \cite{evans_pounds:1968} found similar results using  X-ray spectroscopy.
However, the limitation of X-ray spectroscopy is that only lines 
formed above 2--3 MK are observed. Many analyses of X-ray spectra
such as those of  SMM FCS indeed assumed that the plasma was 
isothermal (see, e.g. \citealt{schmelz_etal:1996}).
Results from other instruments have also confirmed that the emission
in the cores is always around 3 MK. For example, this was obtained from 
Yohkoh BCS \citep[see, e.g. ][]{sterling_etal:1997}.

EUV spectroscopy with SOHO CDS  \citep[see, e.g.][]{mason_etal:99,delzanna_mason:03}
and Hinode EIS \citep[see, e.g.][]{odwyer_etal:11,tripathi_etal:2011,winebarger_etal_2011,warren_etal:2012,schmelz_pathak:2012}
have shown a relatively steep  plasma distribution in the 1--3~MK range, 
but with a wide range of values of $b$,  between 2 and 5. 
The revised Hinode EIS calibration of \cite{delzanna:13_eis_calib}
substantially increased the slopes $b$ to values around 5, 
which increase even further if 
 foreground/background emission is taken into account
\citep{delzanna:2013_multithermal,delzanna_etal:2015_emslope}. 
These results suggest that low-frequency nanoflare modeling should be 
 ruled out. However, evidence of significant lower-temperature emission in AR cores,
interpreted as signatures of cooling,  
 has also been presented \citep[see, e.g.][]{viall_klimchuk:2012}.

Clearly, it is not easy to disentangle line of sight issues without
high spatial resolution  stereoscopy. 
Our view \citep[see, e.g.][]{delzanna_mason:03, odwyer_etal:11,delzanna:2013_multithermal},
based on the spatial distribution of structures seen with SOHO CDS and 
Hinode EIS in lines formed over 
different temperatures, is that AR cores present a near isothermal 
distribution near 3 MK in the core, a less isothermal (around 2 MK) `unresolved' and 
more extended emission, plus a near isothermal emission around 1 MK in warm loops. 
SOHO SUMER observations of an active region above the limb also indicate
the presence of three near-isothermal structures \citep{landi_feldman:2008}.

Regarding the high-temperature slopes $a$, a much larger scatter of values is found 
in the literature. One main problem is the limitation of EUV 
spectroscopy to observe lines above 3 MK. 
For example,  Hinode EIS lines from e.g. 
 \ion{Fe}{xvii},  \ion{Fe}{xxiii},  \ion{Fe}{xxiv}, 
 \ion{Ca}{xvi}, \ion{Ca}{xvii} are all blended 
\citep[see, e.g.][]{young_etal:07a,delzanna:08_bflare,ko_etal:2008,delzanna_etal:2011_flare}. 
Several studies complemented these EUV observations with imaging
data from the X-rays (e.g. Hinode XRT) or soft-Xrays (e.g. SDO AIA),
however the results are always uncertain, given the multi-thermal 
nature of imaging data. For example, the 
SDO AIA 94~\AA\ band was used by \cite{warren_etal:2012}
to find  slopes $a$ ranging between 6 and 10. 
This band was thought to be dominated by \ion{Fe}{xviii}.
However, \cite{delzanna:2013_multithermal} showed that a significant fraction of the 
 AIA 94~\AA\ counts could  be due to a newly identified  \ion{Fe}{xiv} line, 
and that most of the \ion{Fe}{xviii}
weak emission is formed at 3 MK and not at higher temperatures.

A few other studies measured the hot emission at different 
wavelengths. For example, \cite{teriaca_etal:2012} 
measured \ion{Fe}{xviii} 974~\AA\ emission in SOHO SUMER, while
\cite{brosius_etal:2014} measured the \ion{Fe}{xix}  592.2~\AA\ emission 
with the EUNIS-13 rocket flight. 

Another problem is the fact that AR cores often exhibit microflaring
activity, so single measurements are not necessarily representative of the 
quiescent emission. 

The above limitations were overcome by \cite{delzanna_mason:2014}
with the use of SMM observations. 
Only observations of quiescent cores of several active regions were chosen, based on the lightcurves
of the BCS instrument. The FCS X-ray spectra were then used to find 
steep slopes ($a \simeq 14$ in one case) from a peak around 3 MK.
An upper limit to the EM at 7 MK was obtained from the SMM spectra. 
It was about three orders  of magnitude lower than the value of the EM at 3 MK.  
It was also found that the \ion{Fe}{xvii} and  \ion{Fe}{xviii} lines 
were mainly formed  around 3 MK,
and not at higher $T_{\rm e}$  as one would expect. 
Similar results have recently been obtained by combining 
EIS and SUMER observations of an active region observed off-limb 
\citep{parenti_etal:2017}, with the difference that in several 
places faint \ion{Fe}{xix} emission was observed, providing 
the need for the presence of an EM three orders of magnitude below the peak
in those regions. 

Steep slopes ($a$ about 10) have recently also been found
with direct focusing X-ray optics by FOXSI \citep{ishikawa_etal:2014,glesener_etal:2016}.
Several results from the hard X-ray astrophysical mission 
Nuclear Spectroscopic Telescope ARray (NuSTAR) 
have recently been presented, confirming steep slopes 
\citep[see, e.g.][]{grefenstette_etal:2016,hannah_etal:2016}.

Finally, we recall that high temperatures could actually exist,
if the plasma was out of ionization equilibrium (see Section~\ref{sec:non-eq}).

%

\subsection{Moss}

The term `moss' was  introduced 
\citep[cf.][]{berger_etal:1999a} because 
of the textured appearance of this  low-lying  emission,
best seen in the TRACE EUV 173~\AA\ passband.
However, the first description of this emission is due 
to \cite{peres_etal:1994}, that reported high-resolution
 Normal Incidence X-ray Telescope (NIXT) observations.
The NIXT passband had a peak response at $T \sim$ 1 MK,
similar to the  TRACE 173 \AA\ one.
\cite{berger_etal:1999a} presented TRACE and Yohkoh 
data, together with ground-based observations.
They found that moss was mostly  associated with 
magnetic plage as seen in the Ca~II~K line.
Yohkoh/SXT images showed emission overlying the 
regions where the moss was located, and therefore
led the authors to conclude that 
moss regions occur only above the magnetic plage that
underlie soft X-ray coronal loops.
\cite{martens_etal:2000} proposed that moss is the transition-region 
emission  of the 3 MK core loops seen with 
Yohkoh/SXT, a view that is now commonly accepted.

\cite{fletcher_depontieu:1999} analysed 
SoHO CDS observations of the same active region discussed by 
  \cite{berger_etal:1999a}.
They found, with a $DEM$ analysis, that a patch of moss 
emission mostly emits at  1-3 MK temperatures.
Hinode EIS observations have confirmed that 
moss emission  peaks around 2 MK
\citep[see, e.g.][]{tripathi_etal:2010}.

\subsection{Active region unresolved  emission}

The cores of active regions are normally surrounded by a `halo'
of diffuse, unresolved emission between 1 and 3 MK. This emission is 
more extended than the core 3 MK emission and is typically multi-thermal,
as shown e.g. by \cite{delzanna_mason:03} with SoHO CDS and e.g. 
\cite{odwyer_etal:11,delzanna:12_atlas,subramanian_etal:2014} with Hinode EIS.

\subsection{Coronal warm (1 MK) and fan loops }

\begin{figure}[htbp]
\centerline{\includegraphics[width=7cm,angle=0]{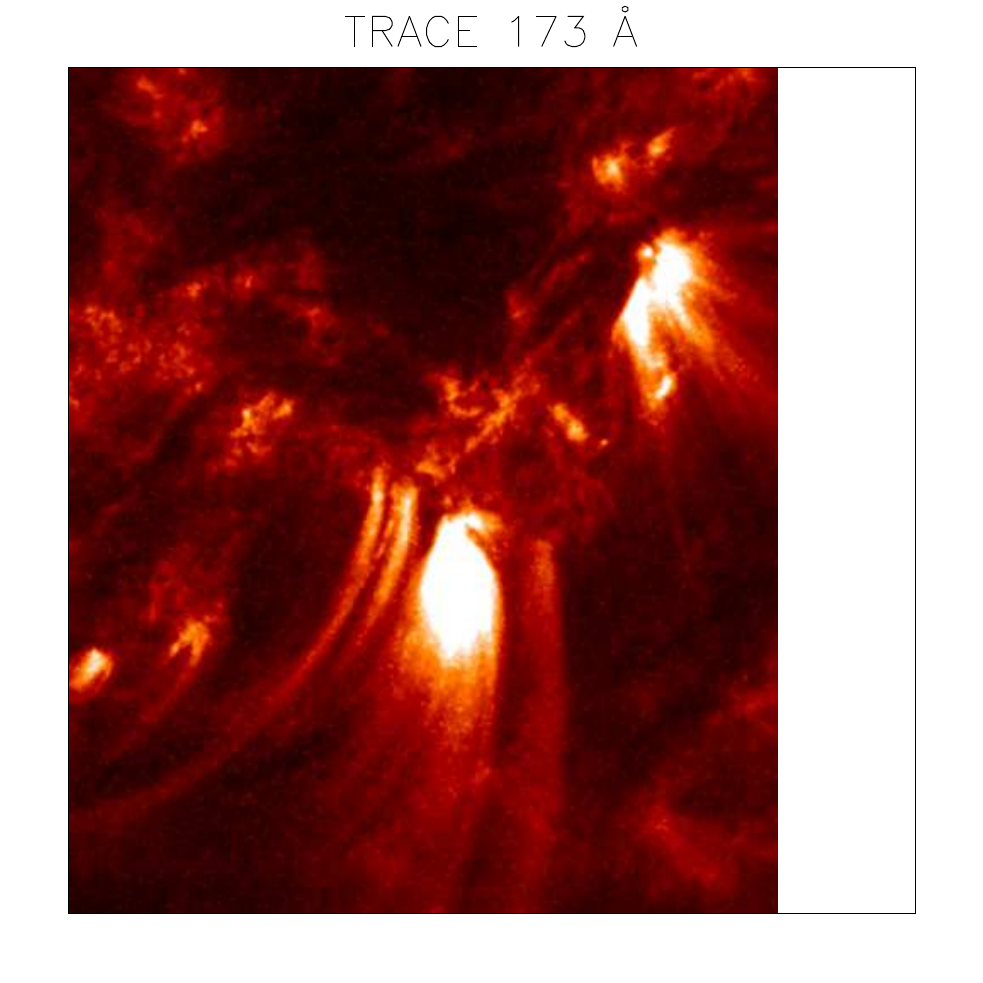}
\includegraphics[width=8cm,angle=0]{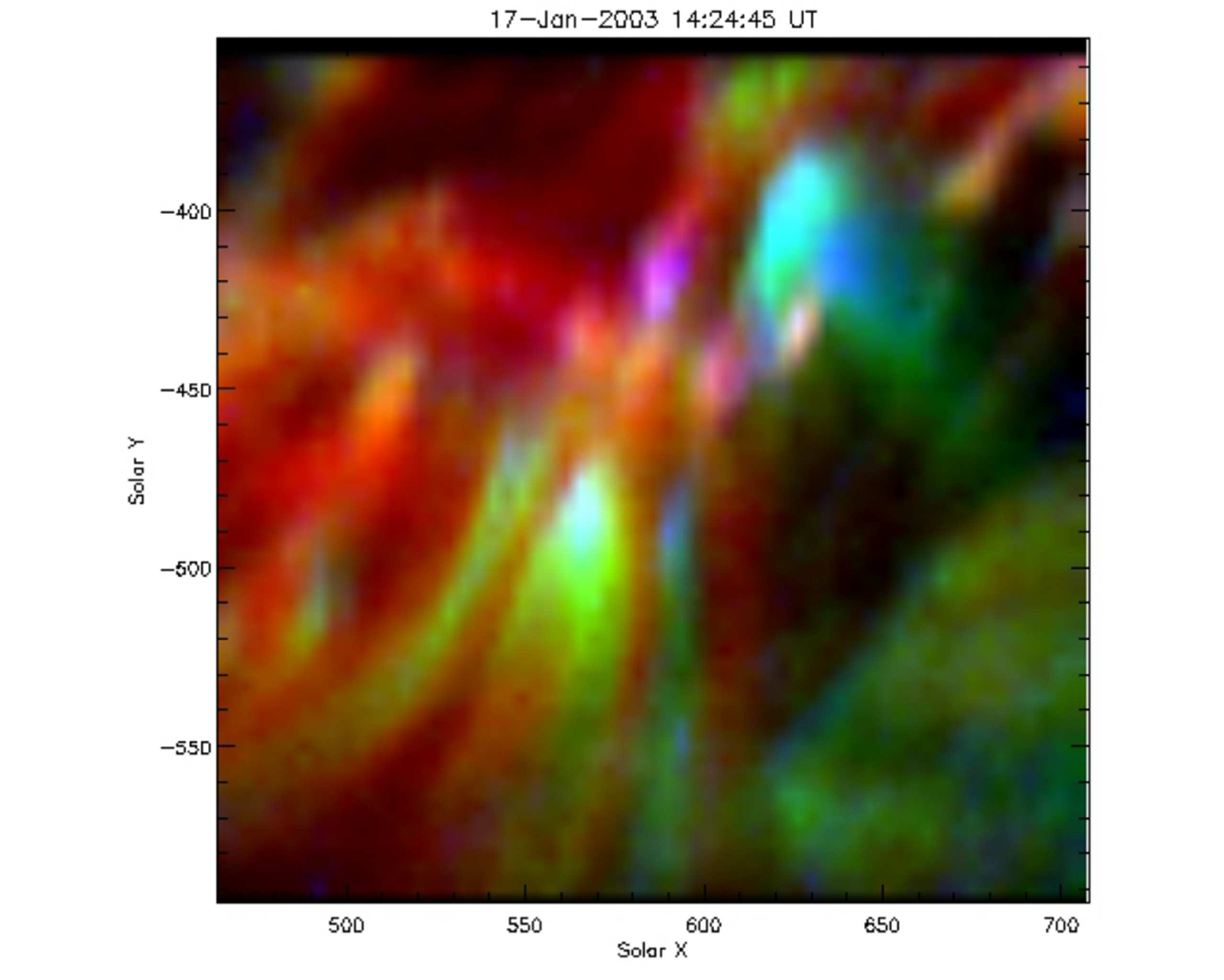}}
  \caption{Left: TRACE 173~\AA\ (1 MK) images of the legs of warm 1 MK loops 
(bottom) and fan loops (on the right). Right: false-colour image
obtained from monochromatic CDS images (Ne VII, 0.7 MK, in blue; Ca X, 1 MK, in green; and Si XII, 3 MK, in
red). The warm loops (green) are clearly seen in TRACE, however 
hotter loops  (red) are not visible in this band. CDS also shows that 
fan loops are cooler (adapted from \cite{delzanna_etal:2006_ESASP}).
}
\label{fig:warm_loops}
\end{figure}

Active regions also contain the so-called warm loops, i.e.
structures which reach about 1 MK at their top and appear to be nearly
resolved. These loops have been observed in their full glory 
in the TRACE 173~\AA\ band, and more recently with the 
SDO AIA 171~\AA\ images.

\cite{delzanna:03,cirtain_etal:2007} used the EM loci method applied to 
SOHO CDS observations (at about 4\arcsec\ resolution) 
of  legs of `warm' (1 MK) loops, which have strong emission in 
upper TR lines. They found 
that these  loops appear to be  nearly isothermal in their cross-section.
In fact, if the $I\lo{ob} / G(T)$  curves all cross at one temperature,
this means that the observations are consistent with the plasma 
being isothermal (although this is not necessarily always the case).
Cooler and hotter loops are intermingled and fill the entire AR volume
\citep{delzanna_etal:2006_ESASP}.

Hinode EIS, with its effective 3\arcsec\ resolution, allowed 
measurements of some parts of the coronal warm loops, which overall
also show near isothermal distributions  
\citep[see, e.g. ][]{warren_etal:2008,tripathi_etal:2009}.

 The so-called fan loops are typically anchored in sunspots and their penumbra,
 are cooler than the warm loops, and are condensed in the form of a fan. 
This was obvious from 
 SoHO CDS observations (see, e.g. \citealt{cirtain_etal:2007} 
and  Fig~\ref{fig:warm_loops}), and later with Hinode EIS 
\citep[see,e.g.][]{delzanna:09_fe_7,brooks_etal:2011}.

\clearpage

\section{Diagnostics of Electron Densities} 
\label{sec:ne_diagn}

\subsection{$N_{\rm e}$ from line ratios}

The main principle for the spectroscopic diagnostic of the 
electron density involves a ratio of two lines,
where at  least one line is connected to a 
metastable level.
We recall that  metastable levels have very  small radiative decay rates, 
so that collisional de-excitation
competes with radiative decay as a depopulating process.

The ratio of two spectral lines, where one is 
populated from the ground state and one is either a decay from a 
 metastable level (a forbidden or intersystem transition) or is an allowed transition that is 
populated from a  metastable level is therefore
strongly dependent on the electron density. 
A ratio of two forbidden or intersystem lines, decays from two different metastable levels,
is also a density diagnostic. 
 This is the basic principle of 
most of the electron density diagnostics (see \cite{gabriel_mason:82}).

For illustration purposes we consider  a simplified model ion.
For forbidden transitions
the radiative decay rate  is  very small
($A_{m,g} \simeq  10^0 - 10^2$ sec$^{-1}$), so the 
collisional de-excitation term ($N_{\rm e}~ C^e_{m,g}$)
becomes an important depopulating mechanism at sufficiently high densities.
We recall that 

\begin{equation} N_m~=~ { N_{\rm g} N_{\rm e} C^e_{g,m} \over {N_{\rm e} C^e_{m,g} + 
A_{m,g}}} 
\end{equation}

For small electron densities, $N_{\rm e} \rightarrow 0$,
$ A_{m,g} \gg N_{\rm e} C^e_{m,g}$, then the intensity
has the same dependence on the
density as an allowed line ($I^A$):

\begin{equation}
I_{m,g}~  \simeq N_{\rm e}^2  
\end{equation}

For very large values of electron density,
$N_{\rm e}\rightarrow \infty$ ,
the collisional de-population dominates,
$ N_{\rm e} C^e_{m,g} >> A_{m,g}$ ;
the metastable level is in Boltzmann equilibrium with the ground level:

\begin{equation}
{N_{\rm m} \over N_{\rm g} }~=   ~{ C^e_{\rm g,m} \over C^e_{\rm m,g}}~
= ~{ \omega_m \over \omega_{\rm g}}~exp\left({-\Delta E_{\rm g,m} \over
kT}\right)  \label{boltz}
\end{equation}

The line intensity has the form:

\begin{equation}
I_{\rm m,g} \simeq  N_{\rm e} 
\end{equation}

For intermediate values of electron density 
($A_{m,g} \simeq N_{\rm e} C^e_{m,g}$), 
 the population of the metastable
level is significant and the intensity varies as:

\begin{equation}
I_{m,g} \simeq N_{\rm e}^\beta  \quad\quad   1< \beta <2 
\end{equation}

The intensity ratio of a forbidden to an allowed transition
($I^F/I^A$) for different spectral lines from the
same ion can be used to determine an average electron density for the
emitting volume. This value is independent of the elemental abundance,
ionisation ratio and any assumptions about the size of that volume.

If the population of metastable level ($m$) is comparable
with the ground level ($g$), then other excited levels ($k$) can
be populated from this metastable level as well as from the ground level
and the dependence of the intensity on electron density becomes:

\begin{equation}
I_{k,m} \simeq N_{\rm e}^\beta  \quad\quad 2< \beta <3 
\end{equation}

The classical simple example is the \ion{Fe}{xiv} case.
The ground
configuration 3s$^2$3p has two levels - $^2$P$_{1/2}$ and 
$^2$P$_{3/2}$ - the transition between these two levels gives rise to
the coronal green line at 5303~\AA, with a 
probability to the ground level of 
only 60~sec$^{-1}$. For low electron densities (less than 10$^8$~cm$^{-3}$), 
almost all the population for Fe~XIV is in
the ground level, but as the electron density  increases
(greater than 10$^{10}$~cm$^{-3}$), the upper level becomes more populated.

\begin{figure} 
\centerline{\includegraphics[width=7cm]{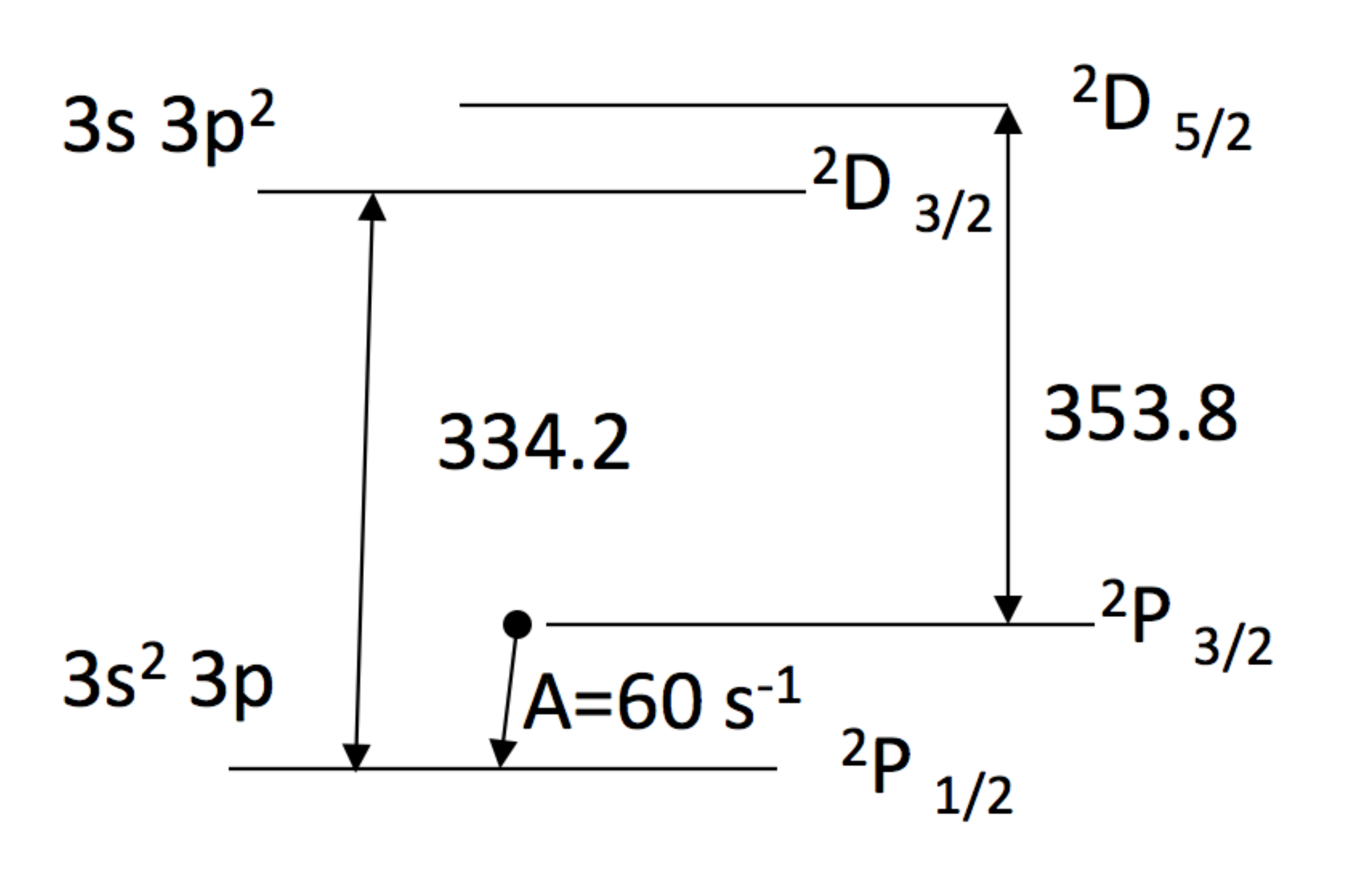}} 
\centerline{\includegraphics[width=12cm]{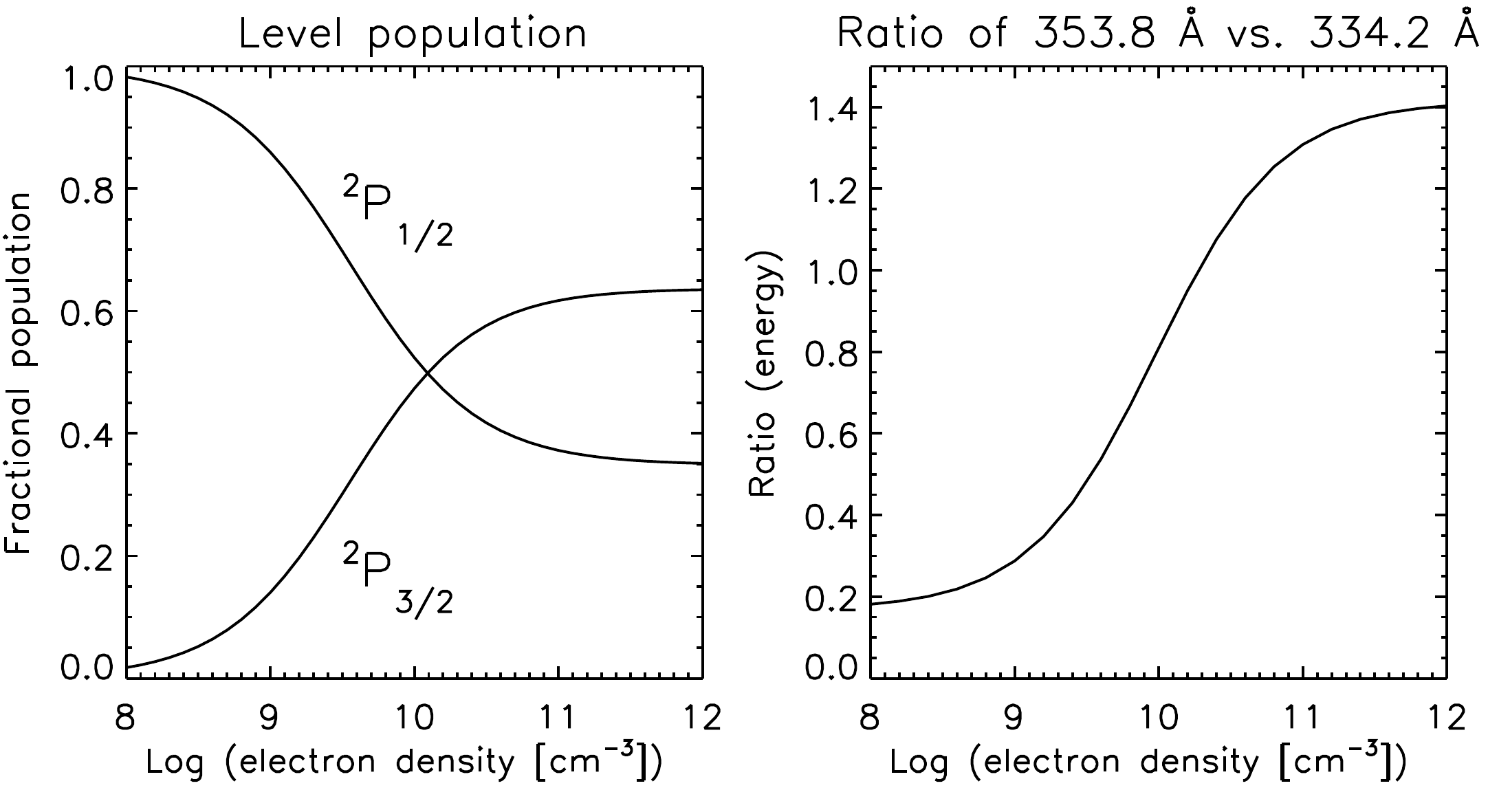}} 
\caption{Electron density diagnostics  from \ion{Fe}{xiv}. }
\label{fig:fe_14_ne}
\end{figure}

The upper level of the spectral line at 334.2~\AA\  is mainly populated by 
direct excitation from the
ground level(3s$^2$3p~$^2$P$_{1/2}$), whereas the upper level of the 
353.8~\AA\ line  is mainly excited from the upper level of the 
ground configuration (3s$^2$3p~$^2$P$_{3/2}$).
The intensity ratio of these two lines 
therefore varies with the electron density, following the 
relative change in the populations of the two levels in the 
ground configuration, as shown in Figure~\ref{fig:fe_14_ne}.
This ratio was observed e.g. by SOHO CDS (see  \citealt{mason_etal:99}).

To avoid additional uncertainties, lines from the same ion 
should be used to measure the density. 
Also, lines that have a similar temperature dependence
should be used.
The temperature dependence comes in when the lines have 
very different excitation energies, which means that 
lines far away in wavelength should in general not be used.

Each ion and line ratio has a range of density sensitivity.
The number of diagnostics depends on the number of metastable levels,
the most important ones being those in the ground configuration 
(the most populated).
As the density increases, it reaches a point where the levels
reach Boltzmann equilibrium, so the level population does not 
change with density.
For more details, see \cite{gabriel_mason:82}.

In the visible range of the spectrum, density-sensitive line ratios 
 have been studied for a long time \citep{aller_etal:1949}.
We note that lines in  the visible
wavelength range are strongly affected by photoexcitation, a process
that needs to be included in the modeling.
Ratios of a forbidden or intersystem to an allowed line 
are very common diagnostics, but they can also be sensitive 
to the temperature. Examples are the Be-like
C III 1909/977 and O V 1218/629~\AA\ (see below).
Ratios of two intersystem lines such as 
the \ion{O}{iv} 1407.3/1404.8~\AA\ are not sensitive to the temperature,
hence are better in this respect (but see below).

Ratios of two allowed lines are very common, for example the 
\ion{Fe}{xii} 186.8/195.1~\AA\ (see below). Both lines are usually populated from the 
ground configuration so have a similar temperature dependence.
However, if one of the lines is populated by a metastable level
that is in an excited configuration, then there will be 
some temperature dependence. Examples are 
the strong TR lines from the Be-like ions C III 1176/977~\AA\ 
and O V 760/629~\AA\ (see below), which are ratios of 
multiplets of lines with the resonance lines.

In any case, the density that can be obtained is always going to be
an averaged value along the line of sight over the formation
temperature for that ion, weighted by the 
(unknown) density distribution. If the plasma distribution is
homogeneous, then the measured value is the real value.
However, if the plasma is inhomogeneous,
different line ratios can produce different density values,
 as e.g. \cite{doschek:1984}  pointed out.
%
%
The dominant contribution obviously comes from the regions with the
highest densities. 
In principle, if several line ratios are observed, one could
try to estimate the density distribution. However, 
to date  uncertainties on the radiometric calibration and on the atomic data
have been significant and made such measurements difficult.
We note that the situation in recent years has dramatically improved
with regard to the accuracy of the 
atomic data, but not significantly improved with regard to the
reliability of the instrument calibration.

\subsection{ $N_{\rm e}$ from multiple lines within  the same ion}

Plots of theoretical ratios of lines 
as a function of the electron density are common in the literature. Quite often, 
some line ratios  show discrepancies between theory and observations,
and it is not obvious which line is most affected. 
When several lines from the ion are observed, it is much more 
instructive to plot all the information about the lines at once.
This was originally suggested by  Brunella Monsignori Fossi, and has been implemented 
in two slightly different ways, the  $L$-function method
\citep{landi_landini:1997} and the emissivity ratio method
\citep{delzanna_etal:04_fe_10}.

\begin{figure}[htbp]
 \centerline{\includegraphics[width=8.0cm,angle=0]{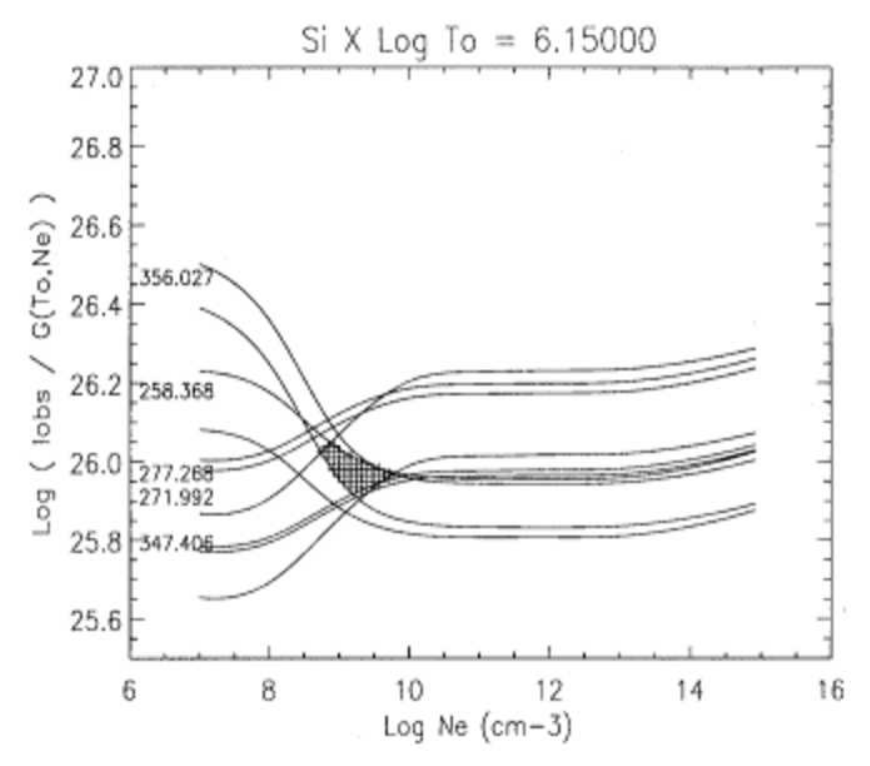}}
  \caption{The $L$-function  curves  of a few among the strongest 
EUV lines of  \ion{Si}{x} \citep{landi_landini:1997}.
}
 \label{fig:l-function}
\end{figure}

The  $L$-function method consists in basically plotting 
the ratio of the observed intensity (or flux) in a line 
with its contribution function $G(T,N)$, calculated at an 
 effective temperature $T_{\rm eff}$:

\begin{equation}
L = { I_{\rm ob}  \over G(N_{\rm e}, T_{\rm eff}) } 
\end{equation}
\noindent
An example is shown in Fig.~\ref{fig:l-function}.
We recall that the effective temperature gives an indication of where 
most of the line is formed. This is often  quite different to  
 $T_{\rm max}$, the temperature where the  $G(T)$ of a line has a maximum,
or where the maximum ion abundance is.
If the plasma distribution along the line of sight has constant density,
all the  $L$-function curves  should be either overlapping 
(if the lines have the same density sensitivity) or intersect.
One non-trivial issue is obtaining the 
 effective temperature of the lines as it requires the calculation
of the DEM,  with all its associated uncertainties. However, in principle the
method can also be used when the plasma is isothermal.

\begin{figure}[htbp]
 \centerline{\includegraphics[width=8.5cm,angle=90]{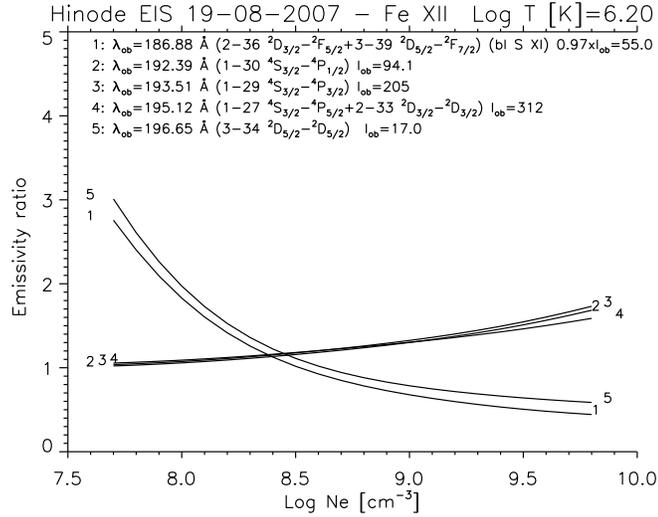}}
  \caption{The emissivity ratio curves  of a few among the strongest 
lines of  \ion{Fe}{xii}, as observed with Hinode EIS and using the 
new atomic data (adapted from \citealt{delzanna_etal:12_fe_12}).
}
 \label{fig:emr}
\end{figure}

The other similar approach is the emissivity ratio method, 
where the ratios of the observed ($I_{\rm ob}$, energy units) 
and the calculated line emissivities:
\begin{equation}
R_{ji}= { I_{\rm ob}  N_{\rm e} \lambda_{ji} \over N_j(N_{\rm e}, T_{\rm e})  \;A_{ji}} \; Const
\end{equation}
\noindent
are plotted as a function of the electron density $N_{\rm e}$.
Here, $N_j (N_{\rm e}, T_{\rm e})$ is the
population of the upper level $j$ relative to the total
number density of the ion,  calculated at a fixed temperature  $T_{\rm e}$ 
(the same for all the lines),  $\lambda_{ji}$ is the wavelength of the transition, 
$A_{ji}$ is the spontaneous radiative transition probability,
and $Const$ is a scaling constant that is the same for all the lines within one observation.
 The value of $Const$ is chosen so that the emissivity ratios $R_{ji}$ are  near unity
where they intersect, so  the scatter of the curves around unity provides
a direct measure on the relative agreement between predicted and 
observed intensities.
One example is shown in  Fig.~\ref{fig:emr}. There is agreement within 
about 10\% for all the lines at log $N_{\rm e}$ [cm$^{-3}$] =10$^{8.5}$ 

The advantage of the emissivity ratio method is that it does not depend on the 
ion population, and does not require
knowledge of the temperature distribution. The disadvantage is that it cannot
be used if the lines have a very different temperature dependence.

\subsection{$N_{\rm e}$ from other methods}
\label{sec:filling_factor}

The previous approaches can also be followed whenever lines from different 
elements are observed, as long as the relative chemical abundances are known,
and the temperature of formation of the lines is also known. 
For example, an extension of the L-function method for an isothermal 
plasma at temperature $T_{0}$ is to plot the emissivity ratios 
$ER$ of a series of lines, to measure the density:

\begin{equation}
ER_{ji}= { I_{\rm ob}   \lambda_{ji} \over G(N_{\rm e}, T_{0}) }  \; Const  \;.
\end{equation}

\noindent
These curves are effectively EM loci curves, scaled by a constant 
(see Fig.~\ref{fig:ne_loop_s_4} below for an example).

If the volume of the emitting plasma can be estimated, then 
an average value of the  
electron density $\langle N_{\rm e} \rangle$ can be deduced from the total emission measure.
In the case of radiance observations: 

\beq
 \langle N_e^2 \rangle= EM  / dh 
\eeq
where $dh$ is the path length of the emitting plasma 
and EM is the column emission measure.
Alternatively, the path length $dh$ 
can be estimated from the column EM
once the average electron density $\langle N\lo{e} \rangle$ is measured from
a line ratio.
In most cases, when the values of $\langle N_e^2 \rangle$ are compared to the 
 $\langle N_e \rangle^2$ values which are calculated from direct 
measurements of the average (along the line of sight) density $\langle N_e \rangle$ 
in the transition region, 
large discrepancies are often found.

\cite{dere_etal:1987} used HRTS observations 
to obtain EM estimates from \ion{C}{iv} lines and 
 densities from  \ion{O}{iv} (both formed at 
around 10$^5$~K). The resulting path lengths were in the range
 0.1-10 km, which is much 
smaller than the observed sizes of the spicular structures 
at the HRTS resolution (2400 km).
These measurements  suggest that the  transition region has a filamentary 
structure, with most of the plasma occupying only a small fraction 
(0.01-0.00001 -- the filling factor) of the observed volume. %
 However, alternative interpretations have also been proposed (see, e.g. \citealt{judge:00}).

In principle, the EM could also be obtained from the 
free-free and the free-bound continuum. From these EM values, 
$N_{\rm e}$ can then be estimated.
Added uncertainties are that the continua depend on the 
ion and elemental abundances, which are normally unknown.

The broadening of the neutral hydrogen lines 
depends on thermal motions, but also 
on the Stark effect, and on the collisions with free electrons.
It therefore depends, among other things, on $N_{\rm e}$,
so in principle the broadening can be used to estimate $N_{\rm e}$
(see, e.g. \citealt{kurochka_maslennikova:1970}).

The method has been used with Skylab spectra to get densities via Stark
broadening in \cite{rosenberg_etal:1977}  (quiet Sun and polar hole) and by
\cite{feldman_doschek:77_stark} (active regions).

\subsection{$N_{\rm e}$ from specific ions}

There are a  large number of papers where  density diagnostics are discussed, from both a 
theoretical and/or observational point of view.
In what follows we review the main diagnostics in the EUV/UV,
providing lists of the strongest unblended line ratios that have a strong 
density sensitivity. We flag  a line when there is a known blend.

The lists are not comprehensive, i.e. there are always other 
options/combinations of lines. The choice mostly depends on the 
spectral range covered by an instrument, and its resolving power.
Indeed most of the lines turn out to be blended in medium 
(and some even in high) resolution spectra.

The improvement in atomic calculations  has gone hand in hand with better
observations. For many ions, electron densities 
obtained with the earlier atomic data were inaccurate. 
The  atomic data  for the simpler ions reached sufficient accuracy in 
the 1980's. However, the atomic data for the most complex isoelectronic 
sequences and 
the iron coronal ions 
(which are in many ways the most important ones)
have only recently been available to a high accuracy.

Earlier reviews on spectroscopic diagnostics are given in 
\cite{feldman_doschek:77,feldman_etal:1978_review,dere_etal:1979, dere_mason:1981, feldman:81,doschek:1985_review, feldman_etal:92a,mason_monsignori:94}. They cover many  spectral ranges of 
earlier instruments. 

A comprehensive review of the density diagnostics available across the wide SERTS-89 
spectral range (235--450~\AA) is given by \cite{young_etal:98}. 
A review of the density diagnostics available to SoHO CDS is given in 
\cite{delzanna_thesis99}.  The density diagnostics of coronal lines 
available to SoHO SUMER have been discussed in several papers,
see e.g. \cite{laming_etal:1997,mohan_etal:2003}.
The density diagnostics for the coronal lines observed by Hinode EIS 
were reviewed by \cite{delzanna:12_atlas}. 

\cite{keenan:1996} provided an extended bibliographical list of previous work.
There are also several papers and reviews that discuss atomic data and diagnostics
within an isoelectronic sequence. Some of these are cited below.

\subsubsection{$N_{\rm e}$ from He-like ions}

\begin{figure}[!htb]
\centerline{\includegraphics[width=0.8\textwidth]{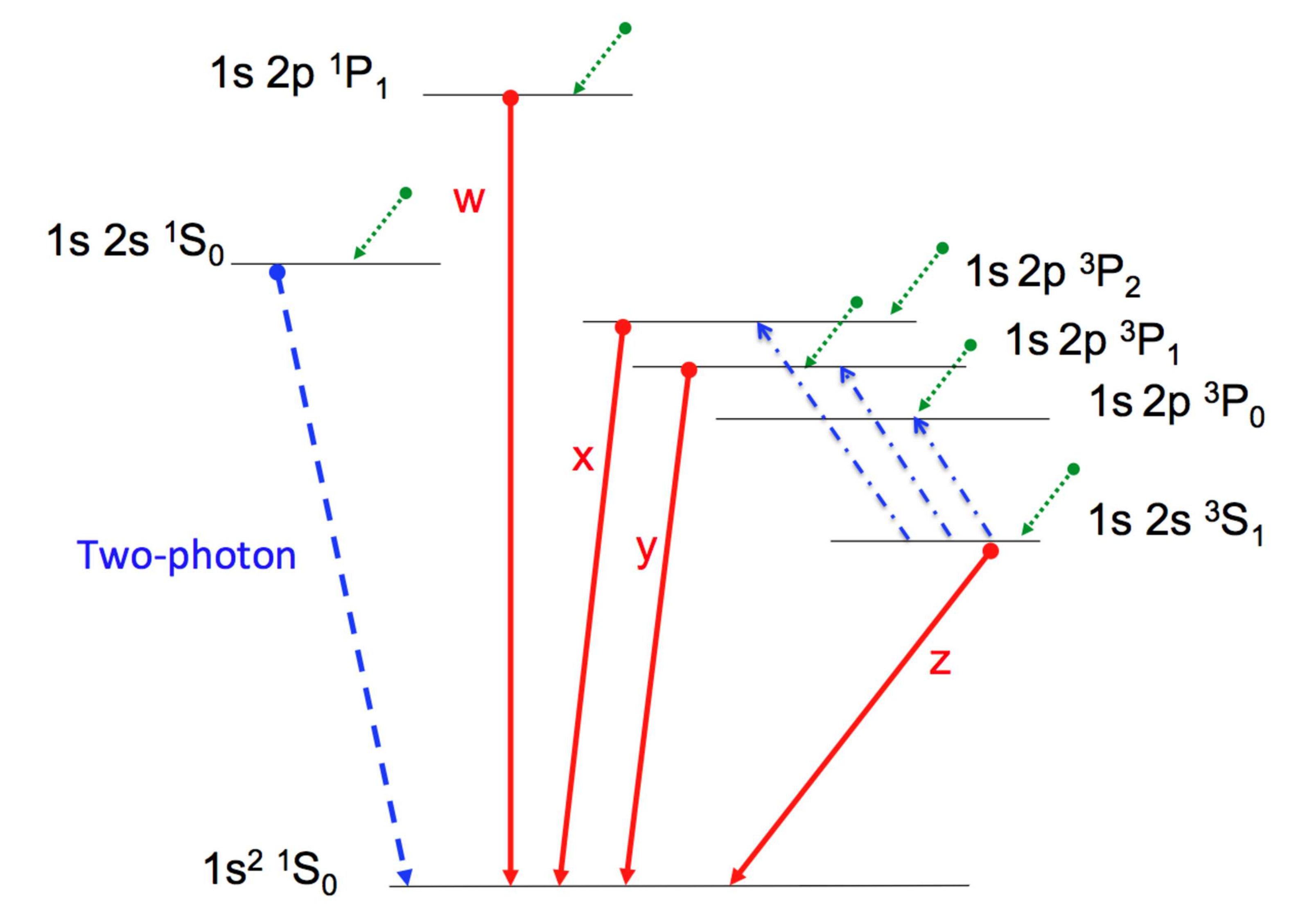}} 
\caption{A diagram (not to scale) of the main transitions in He-like ions.
The red downward arrows indicate the main spectral lines, while the 
blue dashed line indicates the two-photon decay. The downward green dotted
lines indicate contributions from cascading/recombination. 
The blue dot-dash lines show where photoexcitation is an important 
excitation mechanism. 
}
\label{fig:he-like}
\end{figure}

\cite{gabriel_jordan:1969a} showed that the ratios 
$R$ of the forbidden $z$ with the intercombination lines $x+y$
(see Fig.~\ref{fig:he-like}) in 
the He-like C, O, Ne, Mg  are mostly 
density-sensitive, with little temperature sensitivity.
Lines from these ions are observed in the X-rays, as shown in 
 Table~\ref{tab:he-like}.
The $x$ line is a magnetic quadrupole (M2) transition, while the $y$ is an electric dipole
one (E1). The cross-section for excitation to their upper levels is relatively 
easy to calculate accurately.  Photoexcitation and 
cascading from higher levels (and recombination) need to be taken into account as they
can significantly affect the upper level populations.  
The  excitation rate to the forbidden $z$ magneitc dipole (M1) transition is more difficult to 
calculate accurately, as it is very sensitive the effects of the 
resonances. 
We note that in previous literature, resonance effects were sometimes added 
to the DW calculations. However, when results have been compared to 
more recent full CC $R$ matrix calculations, significant differences
were found \citep[see ][]{badnell_etal:2016}.

\begin{table}[!htb]
\caption[He-like main lines.]{He-like main lines. 
The log $T$ [K] values 
indicate the temperatures of peak ion abundance in equilibrium,
although we note that the temperature of formation  can be quite different.
Wavelengths are in \AA\ and are from CHIANTI v.8.
}
\centering
\begin{tabular}{llllllll}
\toprule
Ion            & $w$ & $x$ & $y$ & $z$ & log $T$ \\
        & 1s$^2$ $^1$S$_{0}$ - 1s 2p $^1$P$_{1}$ &  1s$^2$ $^1$S$_{0}$ - 1s 2p $^3$P$_{2}$ &
       1s$^2$ $^1$S$_{0}$ - 1s 2p $^3$P$_{1}$ &  1s$^2$ $^1$S$_{0}$ - 1s 2s $^3$S$_{1}$ &   \\
\midrule

\ion{C}{v}     & 40.268  & 40.728 & 40.731 & 41.472 & 5.5 \\
\ion{N}{vi}    & 28.787  & 29.082 & 29.084 & 29.534 & 5.75  \\   
\ion{O}{vii}   & 21.602  & 21.804 & 21.807  & 22.101 & 6.0 \\  %
\ion{Ne}{ix}   & 13.447 (bl Fe XIX) & 13.550  & 13.553 (bl Fe XIX) & 13.699  & 6.2 \\
\ion{Mg}{xi}   & 9.169   & 9.228 (bl Fe XXIII) & 9.231  & 9.314 & 6.45 \\
\ion{Si}{xiii} & 6.648   & 6.685  & 6.688  & 6.740 & 6.65 \\
\ion{S}{xv}    & 5.039   & 5.063  & 5.066  & 5.101 &  6.65 \\
\ion{Ca}{xix}  & 3.177   & 3.189  & 3.193  & 3.211   & 7.15 \\
\ion{Fe}{xxv}  & 1.8504  & 1.8554 & 1.8595 & 1.8682 &  7.5 \\
\bottomrule 
\end{tabular}
\label{tab:he-like}
\end{table}

\begin{figure}[htb]
 \centerline{\includegraphics[width=0.6\textwidth,angle=90]{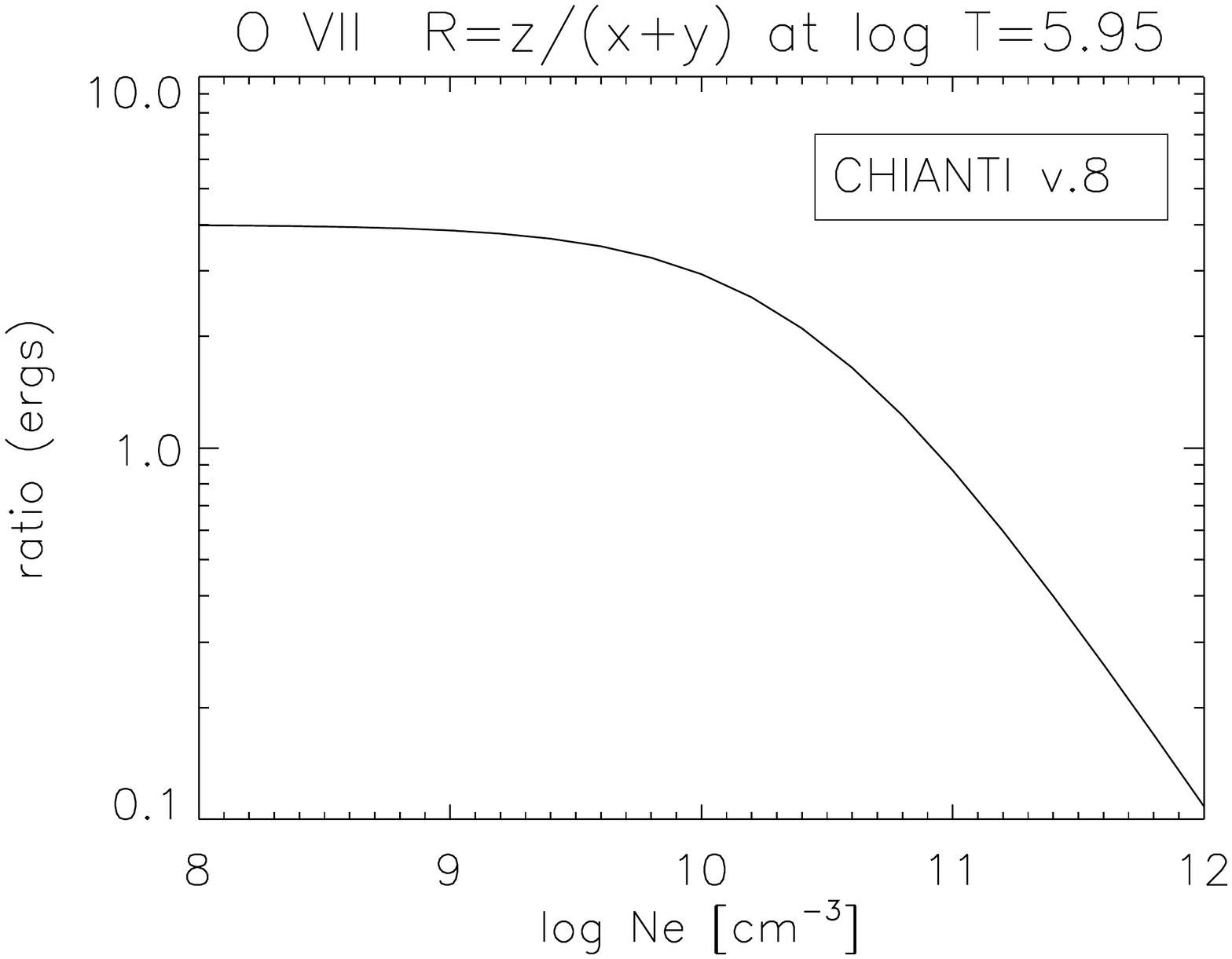}}
  \caption{ $R$ ratio for the He-like \ion{O}{vii}.}
  \label{fig:he-like_r-ratio}
\end{figure}

One problem is that the range of densities 
for which the line ratios are sensitive is higher  than
usual  solar flare densities for all ions, with the exception of \ion{O}{vii}.
 Figure~\ref{fig:he-like_r-ratio} shows the $R$ ratio for this ion.
Another non-trival issue is the fact that satellite lines can 
contribute to the observed intensities of the lines.

Another problem is that the lines need to be observed with 
high-resolution spectrometers.
One example are the  flare observations obtained by a sounding rocket, used to obtain 
densities by \cite{brown_etal:1986}.
Another one are the SOLEX observations reported by \cite{mckenzie_etal:80a,doschek_etal:1981}.
For more details on the He-like ions, see the review by 
\cite{porquet_etal:2010}.

\subsubsection{$N_{\rm e}$ from Be-like ions}

\begin{figure}[!htb]
\centerline{\includegraphics[width=0.8\textwidth]{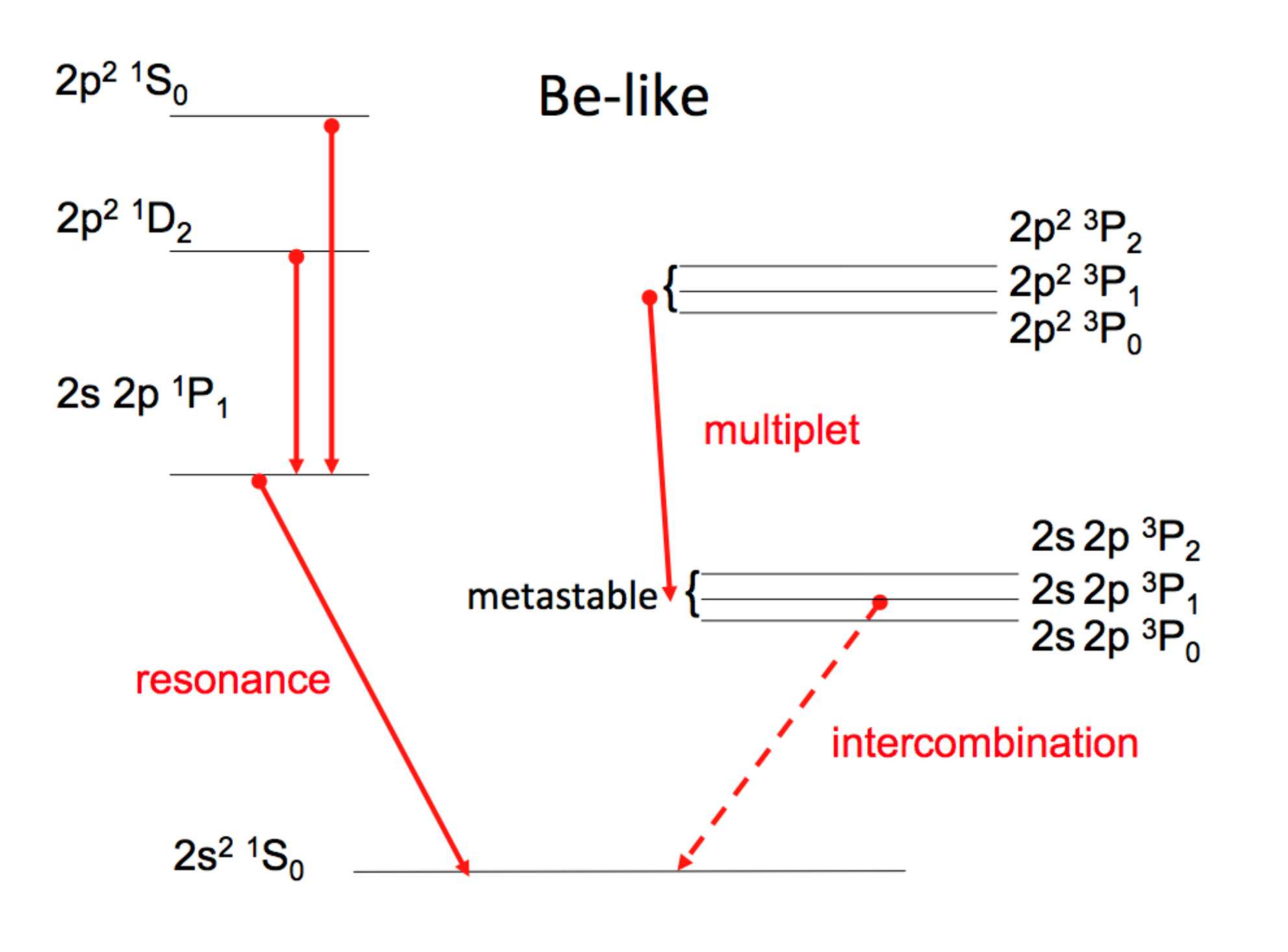}} 
\caption{A diagram (not to scale) of the main levels  in Be-like ions.
The red downward arrows indicate the main spectral lines.
}
\label{fig:be-like}
\end{figure}

\begin{table}[!htb]
\caption[Be-like density diagnostics.]{ Be-like density diagnostics. 
The log $T_{\rm e}$ [K] values (in parenthesis)
indicate the  temperatures of peak ion abundance in equilibrium.
Wavelengths are in \AA. The density can be measured from e.g. the ratios of 
the resonance (a) line with any of the multiplet (b,c), or from within the 
multiplet between line (c) and any of the (b) lines.
}
\centering
\begin{tabular}{llllllll}
\toprule
Transition &  \ion{C}{iii} & \ion{O}{v} & \ion{Ne}{vii} & \ion{Mg}{ix} & \ion{Al}{x} & \ion{Si}{xi} \\ 
           &     (4.8)       &      (5.4) & (5.7)     & (6.0) &  (6.1) & (6.2) \\
\midrule


2s 2p $^3$P$_{2}$ - 2s 3d $^3$D$_{3}$ & & 192.904  & & & & \\
2s 2p $^3$P$_{2}$ - 2s 3d $^3$D$_{2}$ & & 192.911  & & & & \\

2s 2p $^3$P$_{1}$ - 2s 3s $^3$S$_{1}$ & & 215.103  & & & & \\
2s 2p $^3$P$_{2}$ - 2s 3s $^3$S$_{1}$ & & 215.245  & &  & & \\
2s 2p $^1$P$_{1}$ - 2s 3d $^1$D$_{2}$ & & 220.353  & & & & \\
2s 2p $^1$P$_{1}$ - 2s 3s $^1$S$_{0}$ & & 248.460  & & & & \\
 & & & & \\
2s$^2$ $^1$S$_{0}$ - 2s 2p $^1$P$_{1}$ (a) & 977.020 &   629.732  & & & & \\
\\
2s 2p $^3$P$_{1}$ - 2p$^2$ $^3$P$_{2}$ (b) & 1174.933 &  758.677  & & 439.176 & & \\
2s 2p $^3$P$_{0}$ - 2p$^2$ $^3$P$_{1}$ (b) & 1175.263 &  759.442  & & 441.199 & & \\
2s 2p $^3$P$_{1}$ - 2p$^2$ $^3$P$_{1}$ (b) & 1175.590 &  760.227  & & 443.404 & & \\
2s 2p $^3$P$_{2}$ - 2p$^2$ $^3$P$_{2}$ (b) & 1175.711 &  760.446  & & 443.973 & & \\
2s 2p $^3$P$_{1}$ - 2p$^2$ $^3$P$_{0}$ (c) & 1175.987 &  761.128  & & 445.981 & & \\ 
2s 2p $^3$P$_{2}$ - 2p$^2$ $^3$P$_{1}$ (b) & 1176.370 &  762.004  & & 448.294 & & \\
\\


2s$^2$ $^1$S$_{0}$ - 2s 2p $^3$P$_{1}$ & 1908.734 & 1218.344 & 895.18 & 706.06 & 637.76 & 580.91 \\
2s 2p $^1$P$_{1}$ - 2p$^2$ $^1$D$_{2}$ & 2296.870 & 1371.296 \\

2s$^2$ $^1$S$_{0}$ - 2s 2p $^3$P$_{2}$ &  &  & 887.28 & 693.98 & 623.32 & 564.02 \\

\\

\bottomrule 
\end{tabular}
\label{tab:be-like}
\end{table}

Table~\ref{tab:be-like} lists the lines most useful for density 
diagnostics in Be-like ions.
Lines from the Be-like  \ion{C}{iii} 
can be used to measure densities 
\citep[see e.g.,][]{munro_etal:1971,jordan:1974,dupree_etal:1976,dufton_etal:1978}.

The first excited configuration for the Be-like ions is 
2s 2p, which has  $^3$P and $^1$P$_{1}$ levels (see Fig.~\ref{fig:be-like}).
The $^3$P are metastable, so ratios of lines populated from these levels
such as  the  2s 2p $^3$P  - 2p$^2$ $^3$P 
multiplet at 1175~\AA\ with the resonance 2s$^2$ $^1$S$_{0}$ - 2s 2p $^1$P$_{1}$ 
line at 977.022~\AA\ are density-sensitive.
However, this diagnostic is only sensitive up to  $N_{\rm e}$=10$^{10}$ cm$^{-3}$,
does not have a large variation with density, and is also temperature sensitive. 
Moreover,  opacity effects can affect  the  resonance line.
Another possibility is to use the ratio of the 1175.987~\AA\ line
with any other line of the multiplet, if high-resolution spectra are available.
In this case, the ratios do not depend on the  temperature.

The same ratio for  \ion{O}{v} involves the resonance 
 629.73~\AA\ line and the multiplet at 760~\AA. This 
is a good density diagnostic in the range N$_{\rm e}$ = 10$^{10}$ -- 10$^{12}$ cm$^{-3}$,
although it is not ideal because of the strong  temperature dependence.
This ratio was  studied by several 
authors, see e.g. \cite{munro_etal:1971,jordan:1974,dufton_etal:1978} for 
observations pre-SOHO.
SOHO CDS and SUMER observed the  \ion{O}{v} 
 2s 2p $^3$P -- 2p$^2$ $^3$P transitions around 760 \AA\
(760.444, 758.675,  762.002, 759.439, 760.225, 761.126 \AA) 
and the  629.73~\AA\ resonance line.
The 761.126~\AA\ line is a good density-diagnostic in conjunction with 
any of the other lines of the multiplet, because there is no 
dependence on temperature.
SUMER was able to clearly separate the lines in the multiplet, 
allowing this ratio to be used to derive densities 
(see, e.g. \citealt{doschek_etal:1998}).
The  \ion{O}{v} ratio of the 1218.390 and 1371.292~\AA\ lines 
is an excellent diagnostic (because it has a small 
temperature dependence) at high densities, 
above 10$^{11}$ cm$^{-3}$, although the lines are weak.
Other lines from \ion{O}{v}  
suffer from various problems (continuum, blends)
as summarised in \cite{mariska:92}.
The same transitions in \ion{N}{iv} are in principle useful, but 
the lines are significantly blended \citep[see, e.g.][]{dufton_etal:1979}.

The other possibility is to use the forbidden lines which fall in the UV,
as suggested by \cite{munro_etal:1971}.
However, even there the diagnostics are not straightforward, and line
ratios have to be carefully selected. 
For example, \cite{doschek_feldman:1977} pointed out that 
the ratio of the C III multiplet at  1176~\AA\
with the intercombination line at 1909~\AA\ cannot be used without a knowledge of the 
temperature structure of the atmosphere.

At densities well above 10$^{10}$ cm$^{-3}$, 
$n$=3$\to$$n$=2 transitions in 
 \ion{O}{v} in the 190--250~\AA\ range are  useful, as suggested by \cite{widing_etal:1982}.
They have been observed with the Skylab NRL spectrograph,
and some of them more recently with Hinode EIS.

The 192.750, 192.797, 192.801~\AA\ lines of  \ion{O}{v} are self-blended
and also blended with other transitions from \ion{Fe}{xi} and
\ion{Ca}{xvii}, so are not listed in  Table~\ref{tab:be-like}.  
The 192.904 and  192.911 are self-blended,
and  can be blended with  \ion{Ca}{xvii} during flares,
when high temperatures are present. 
However, the ratio of the 192.9~\AA\ self-blend with the 
248.460~\AA\ line is in principle a good density diagnostic
for Hinode EIS, although the 248.460~\AA\ line is very 
weak, especially after the instrument degradation, and there is some 
temperature dependence on this ratio. 
Other ratios of the lines listed in  Table~\ref{tab:be-like} are possible, but
were only available to Skylab.

Another  density ratio is the one between the 
intercombination 2s$^2$ $^1$S$_{0}$ - 2s 2p $^3$P$_{1}$ 
line and the 2s$^2$ $^1$S$_{0}$ - 2s 2p $^3$P$_{2}$.
 Such ratios have been observed by SoHO SUMER,
see e.g. \cite{laming_etal:1997}.
One of the most used ratios is the \ion{Mg}{ix} 694 vs. 706.06~\AA.
One problem with such a ratio is that both lines 
are  relatively weak. The \ion{Mg}{ix} were often observable by SUMER (off-limb), 
but  the \ion{Al}{x} and \ion{Si}{xi} were only visible in more active spectra.
Another problem is that the ratio is significantly temperature-sensitive.

For flare densities, two other line ratios are potentially
useful. The first is the \ion{S}{xiii} 308.95 vs. 256.68~\AA, sensitive 
 up to  10$^{11}$ cm$^{-3}$,
although the first line is normally very weak (at most 2.5\% of the stronger 
line in ergs). The second is the \ion{Ar}{xv} 266.23 vs. 221.13~\AA,
sensitive in the 10$^{10}$ -- 10$^{12}$ cm$^{-3}$ range,
although again the weaker line is at most 
 2\% of the stronger line (ratio value in ergs). 
The above \ion{S}{xiii} and  \ion{Ar}{xv} lines have been observed in 
Skylab flare spectra.

\subsubsection{$N_{\rm e}$ from B-like ions}

\begin{figure}[!htb]
\centerline{\includegraphics[width=0.8\textwidth]{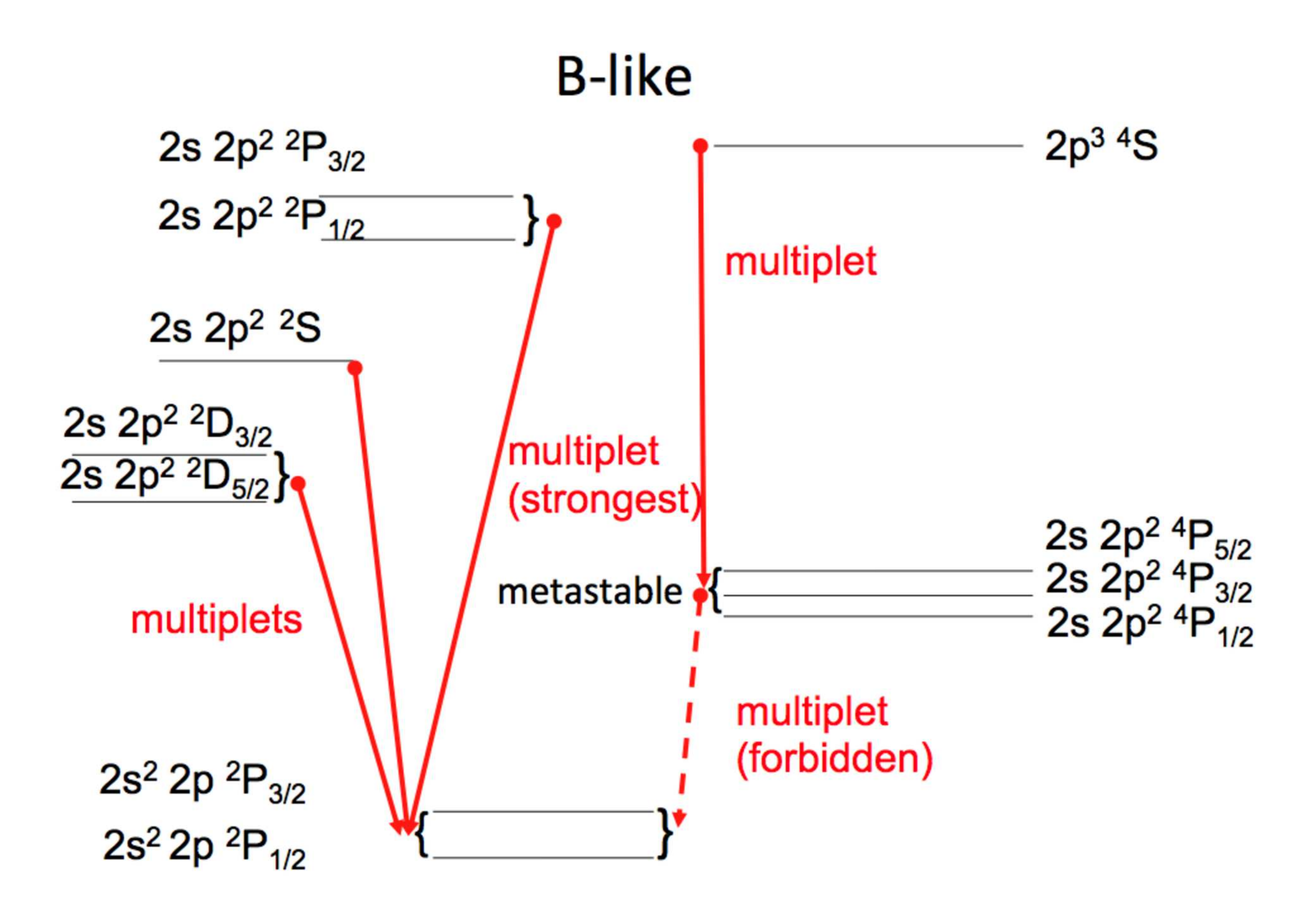}} 
\caption{ A diagram (not to scale) of the main levels  in B-like ions.
The red downward arrows indicate the main spectral lines.
}
\label{fig:b-like}
\end{figure}

The boron sequence contains some particularly useful electron density sensitive
line ratios. 
Na~VII, Si X, S XII  were studied by \cite{flower_nussbaumer:1975_b-like}.
N III,  Ne VI, Mg VIII and Si X   were studied by \cite{vernazza_mason:1978}.
\cite{nussbaumer_storey:1979} studied N III.
The ground levels are 2s$^2$ 2p $^2$P$_{1/2,3/2}$ and the first excited 
levels are 2s 2p$^2$ $^4$P$_{1/2,3/2,5/2}$ which are metastable 
(see Fig.~\ref{fig:b-like}).
The second excited term is the 2s 2p$^2$ $^2$D which produces allowed 
decays to the ground term.
As in the previous cases, the density sensitivity involves 
levels that are collisionally excited from the metastable levels, for example the  
decays from 2p$^3$ $^4$S. Ratios of this type are listed in the 
first line of Table~\ref{tab:ne_b-like}. 
The \ion{N}{iii} 772/989~\AA\ is a good diagnostic. The 
\ion{O}{iv} 625.9~\AA\ line  is also an excellent diagnostic  \citep{doyle_etal:1985}, 
e.g. in combination with the 790.1~\AA\ line,  as the 
ratio varies by a factor of 35.
However, this line was blended in Skylab spectra with the 
Mg X 624.9 \AA\ line, which is normally much stronger.
This  \ion{O}{iv} 625.9 \AA\ line, at the SoHO CDS NIS resolution, 
becomes measurable (even though  it is in the wing of  the Mg X 624.9 \AA\ line)
especially  when the Mg X intensity is reduced, as in coronal holes 
\citep{delzanna_thesis99}  or in 
 transition region brightenings  \citep{young_mason:97}. 
The ratio of the 625.9 \AA\ line with any other \ion{O}{iv} line
 in the SOHO CDS wavelength range 
(553.329, 554.513, 555.263,   608.4 \AA) is also density-sensitive.
The analogous \ion{Ne}{vi} 454/562~\AA\ ratio is not very 
useful as the 454~\AA\ line is very weak.

\begin{figure}[!htb]
 \centerline{\includegraphics[width=0.8\textwidth,angle=90]{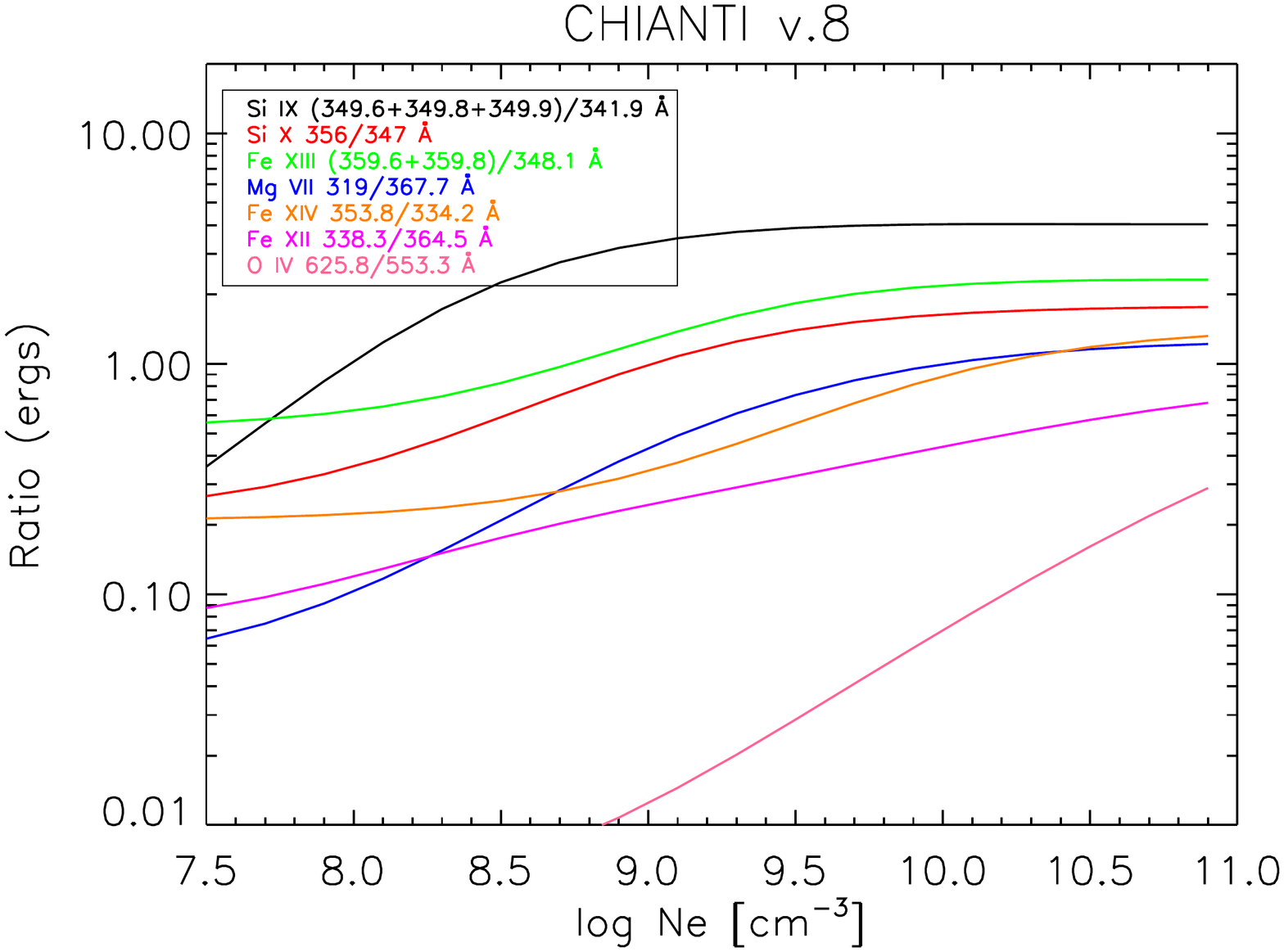}}
  \caption{The main diagnostic ratios available to the SoHO CDS NIS.
The atomic data are from CHIANTI version 8.
}
 \label{fig:ratios_ne}
\end{figure}

Other types of diagnostic ratios are available. For example, 
ratios of allowed lines, where one decays to the ground state 
and one to the metastable level 2s$^2$ 2p $^2$P$_{3/2}$. 
Table~\ref{tab:ne_b-like} shows two examples. 
The first one is excellent, since the lines are close in wavelength.
Very good diagnostics are offered by \ion{Mg}{viii}, \ion{Si}{x}, and \ion{S}{xii}.
The Mg VIII 436.7 (436.735+436.672) \AA / 430.465~\AA\ and the 
Si X 356.0 (self-blend) / 347.40 ratios were among the best 
diagnostic ratios available for the  SOHO GIS and NIS.
They are shown, together with a few others, in 
Fig.~\ref{fig:ratios_ne}.

\begin{table}[!htb]
\caption[B-like density diagnostics.]{ B-like density diagnostics.  $T_{\rm e}$ [K]
indicates the  temperature of peak ion abundance in equilibrium.
The N$_{\rm e}$ values [cm$^{-3}$] indicate the approximate range of densities 
where the ratios are usable.}
\centering
\begin{tabular}{llllllll}
\toprule
Transitions  & Ion & $\lambda_1$  (\AA)  & $\lambda_2$ (\AA) & log $T_{\rm e}$ & log N$_{\rm e}$  \\
\midrule

\\[-1.0ex]
&  \multicolumn{4}{l}{
 2s 2p$^2$ $^4$P$_{3/2,5/2}$--2p$^3$ $^4$S$_{3/2}$ \quad\quad
2s$^2$ 2p $^2$P$_{3/2}$--2s 2p$^2$ $^2$D$_{5/2,3/2}$ } \\
 \\[-1.0ex]

 & \ion{N}{iii} & 772  & 991.5 & 4.9 &  9--11 \\
 & \ion{O}{iv}  & 625  &  790.1   & 5.1 & 9--12 \\


\\[-1.0ex]
&  \multicolumn{4}{l}{
2s$^2$ 2p $^2$P$_{3/2}$--2s 2p$^2$ $^2$D$_{5/2,3/2}$ \quad\quad
2s$^2$ 2p $^2$P$_{1/2}$--2s 2p$^2$ $^2$D$_{3/2}$ } \\
 \\[-1.0ex]

 & \ion{Mg}{viii} & 436.73 & 430.45  & 5.9 & 7--9 \\ 


 & \ion{Si}{x}  &  356.03 &  347.40  & 6.0 & 8--10 \\
 & \ion{S}{xii} & 299.54   & 288.42  & 6.4 & 8--11 \\

 & Ar XIV &  257.44 (bl) &  243.8 (bl) &  &  9--12 \\

 &  Ca XVI & 224.54 & 208.57 (bl Ca XV) & 6.7 & 10--12 \\

 & Fe XXII & 156.0  & 135.79 &  7.2 &  11.5--15 \\


\\[-1.0ex]
  &  \multicolumn{4}{l}{
2s$^2$ 2p $^2$P$_{3/2}$--2s 2p$^2$ $^2$P$_{3/2}$ \quad\quad
2s$^2$ 2p $^2$P$_{1/2}$--2s 2p$^2$ $^2$S$_{1/2}$} \\
\\[-1.0ex]

 &  \ion{Mg}{viii} & 315.01 & 335.23 (bl) & 5.9 & 7--8.5 \\

 & \ion{Si}{x} & 258.37 & 271.99 & 6.0 &  8--10 \\

 & \ion{S}{xii} &  218.20 &  227.50 & 6.4 & 9--11 \\

 & Ar XIV 187.97 (bl) & 194.41 (w) &  9--12 \\

 & Ca XVI &  164.17 & 168.85 & 6.7 & 10--13 \\

 & Fe XXII & 114.41 & 117.15 (bl Fe XXI) & 7.2 &  11.5--15 \\

\\[-1.0ex]
 &  \multicolumn{4}{l}{
2s$^2$ 2p $^2$P$_{1/2}$ -- 2s$^2$  4d  $^2$D$_{3/2}$ \quad\quad
2s$^2$ 2p $^2$P$_{3/2}$ -- 2s$^2$  4d  $^2$D$_{5/2,3/2}$ } \\
\\[-1.0ex]

&  Fe XXII & 8.976  & 9.075 & 7.1 & 12.5--15 \\

\\[-1.0ex]
 &  \multicolumn{4}{l}{
2s$^2$ 2p $^2$P$_{3/2}$ -- 2s 2p$^2$ $^4$P$_{5/2}$ \quad\quad
2s$^2$ 2p $^2$P$_{3/2}$ -- 2s 2p$^2$ $^4$P$_{3/2}$} \\
\\[-1.0ex]

 &  \ion{Na}{vii} & 872.12 & 880.33  & 5.8  & 6--8 \\
 & \ion{Mg}{viii} & 772.28 & 782.36  & 5.9 & 6.5--8.5 \\ 
 & \ion{Al}{ix}  & 691.54  & 703.66 & 5.9 & 7--9 \\
 & \ion{Si}{x}  & 624.70  &  638.94  & 6.0 & 8--10 \\

\\[-1.0ex]
 &  \multicolumn{4}{l}{
2s$^2$ 2p $^2$P$_{3/2}$ -- 2s 2p$^2$ $^4$P$_{5/2}$ \quad\quad
2s$^2$ 2p $^2$P$_{3/2}$ -- 2s 2p$^2$ $^4$P$_{1/2}$} \\
\\[-1.0ex]

 &  \ion{Na}{vii} & 872.12 &  885.98  & 5.8  & 6--8 \\
 & \ion{Mg}{viii} & 772.28 & 789.44  & 5.9 & 6.5--8.5 \\ 
 & \ion{Al}{ix}  & 691.54  & 712.23 & 5.9 & 7--9 \\
 & \ion{Si}{x}  & 624.70  &  649.19  & 6.0 & 8--10 \\
 
\bottomrule 
\end{tabular}
\label{tab:ne_b-like}
\end{table}

One of the  Ca XVI lines (208.57~\AA)  is blended with a Ca XV 208.70~\AA\
even at the Hinode EIS spectral resolution.
At flare temperatures, Fe XXII soft X-ray lines within the 
2s$^2$ 2p--2s 2p$^2$ transition array are  useful 
density diagnostics \citep{doschek_etal:1973,mason_storey:1980,mason_etal:84,delzanna_woods:2013},
but at rather high densities (above 10$^{11.5}$ cm$^{-3}$), i.e.
 higher than those normally found in solar flares.
Other ions produce either  weak or blended lines, as is the case for 
the  Ar XIV lines.

\cite{laming_etal:1997} suggested the use 
of two other diagnostic ratios, 
one is the ratio of the decays from the 2s 2p$^2$ $^4$P$_{3/2,5/2}$
levels to the 2s$^2$ 2p $^2$P$_{3/2}$, while the other is 
the ratio of the decays from the  2s 2p$^2$ $^4$P$_{1/2,5/2}$
levels to the 2s$^2$ 2p $^2$P$_{3/2}$.  The lines are relatively close in
wavelength but are quite weak, although lines from 
a few ions have been observed with SoHO SUMER.

A few diagnostics in the X-rays (around 10~\AA) are available, from 
 \ion{Fe}{xxii} 2s$^2$ 2p-- 2s$^2$ 4d transitions 
 (cf.  Table~\ref{tab:ne_b-like} and  \citealt{fawcett_etal:87,phillips_etal:1996}). 
However, these lines are weak and only sensitive to high densities.

Ratios of forbidden transitions  between the metastable 
2s 2p$^2$ $^4$P and the ground configuration 2s$^2$ 2p $^2$P 
in \ion{N}{iii}  are also 
density-sensitive in the 10$^{9}$ -- 10$^{11}$ cm$^{-3}$ range
\citep{nussbaumer_storey:1979}. It should be noted that these ratios
are virtually independent of temperature. However, the variations are not large
(a factor of 2.5) 
and some lines are affected by blending \citep[see e.g.][]{mariska:92}.
The same ratios for \ion{O}{iv} fall around 1400~\AA\ and 
have been used extensively. They are discussed below.

\smallskip\smallskip\noindent
{\bf \ion{O}{iv} 1400~\AA\ multiplet }
\smallskip

The \ion{O}{iv}  multiplet around 1400~\AA\  has been studied 
by \cite{flower_nussbaumer:1975_o_4} and many other authors.
\cite{feldman_doschek:1979} also discussed the potential of 
N III and \ion{O}{iv} intersystem multiplets as density diagnostics.

Several solar instruments have observed the \ion{O}{iv} lines: 
Skylab S082B, HRTS,  SMM UVSP (see, e.g. \citealt{hayes_shine:1987}), 
SOHO SUMER (see, e.g. \citealt{teriaca_etal:2001}), and more recently 
IRIS (see, e.g. \citealt{dudik_etal:2014_o_4}).

Table~\ref{tab:ne_o4} lists the main diagnostic lines and the 
densities that can be measured. We note that some discrepancies 
in the wavelengths of these lines can be found in the literature. 
This table lists the wavelength values as reviewed in \cite{polito_etal:2016b},
which are mainly based on Skylab  observations at the limb, 
as reported by \cite{sandlin_etal:1986}, with an  accuracy of about 0.005~\AA,
and laboratory measurements from \cite{bromander:1969}. 

The densities obtained from \ion{O}{iv} have been particularly important
because they have been used to obtain a very small filling factor 
(less than 1\%)  for the transition region emission, corresponding to a path length of around 10 km
\citep[see, e.g.][ using HRTS observations]{dere_etal:1987}.

\begin{table}[!htbp]
\caption[\ion{O}{iv} density diagnostics]{\ion{O}{iv}  transitions  commonly used to measure densities
in the log N$_{\rm e}$ [cm$^{-3}$] = 9.5--11.5 range (wavelengths from \cite{polito_etal:2016b}).}
\centering
\begin{tabular}{llllll}
\toprule
  Transitions & $\lambda$  (\AA) &    \\
\midrule

2s$^2$ 2p $^2$P$_{1/2}$ -  2s 2p$^2$ $^4$P$_{3/2}$ & 1397.226 &  \\

2s$^2$ 2p $^2$P$_{1/2}$ -  2s 2p$^2$ $^4$P$_{1/2}$ & 1399.776 &     \\
 & &  &  \\
2s$^2$ 2p $^2$P$_{3/2}$ - 2s 2p$^2$ $^4$P$_{5/2}$ & 1401.163 &  \\

2s$^2$ 2p $^2$P$_{3/2}$ - 2s 2p$^2$ $^4$P$_{3/2}$ & 1404.806 (bl \ion{S}{iv}) &    \\

2s$^2$ 2p $^2$P$_{3/2}$ - 2s 2p$^2$ $^4$P$_{1/2}$ & 1407.384  &    \\

\bottomrule 
\end{tabular}
\label{tab:ne_o4}
\end{table}

\begin{figure}[!htbp]
\centerline{\includegraphics[width=0.6\textwidth, angle=90]{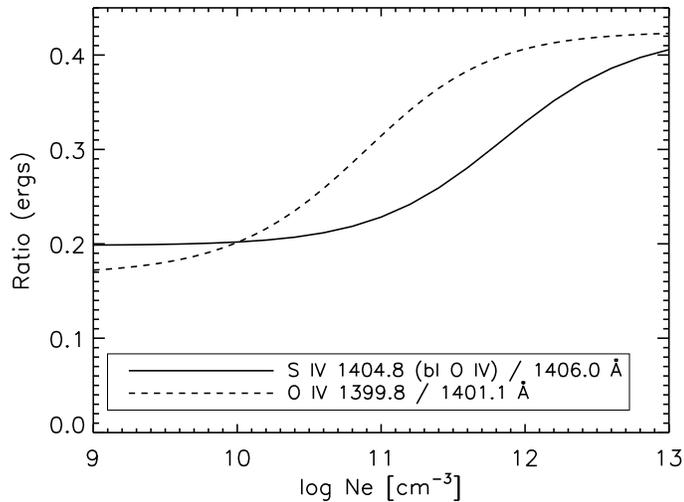}}
\caption{Electron density sensitivity of the  \ion{O}{iv},
  \ion{S}{iv} lines, observed with IRIS. 
}
\label{fig:ne_so_4_ratios}
\end{figure}



One problem is that the \ion{O}{iv} lines are often weak and blended. The 
 1397.22 and 1399.78~\AA\ lines are particularly  weak.
The ratios that include the 1401.16~\AA\ line 
have a  small density sensitivity, as shown in 
Fig.~\ref{fig:ne_so_4_ratios} and already noted by \cite{feldman_doschek:1979}.  
Ratios involving the 1404.78 \AA\ line are better; however,  
 the 1404.78~\AA\ line is a well-known  blend with  
\ion{S}{iv} (1404.8~\AA). 
It is common practice to use the \ion{S}{iv}  theoretical ratio of the 
1404.8~\AA\ and 1406.01 \AA\ lines  to infer 
the contribution of the 1404.81~\AA\ line to the 
observed blend with \ion{O}{iv}, however the \ion{S}{iv} ratio is density sensitive as well,
so deblending the line is not straightforward.
Another issue is that the \ion{O}{iv} lines within the multiplet 
have a density sensitivity only up to about 10$^{11}$ cm$^{-3}$, 
as Fig.~\ref{fig:ne_so_4_ratios} shows. On the other hand, the 
\ion{S}{iv} lines are much better since flare densities up to about
10$^{13}$ cm$^{-3}$ can be measured.

The \ion{O}{iv} and \ion{S}{iv} density
sensitive ratios have been analysed  by several authors in both solar 
and stellar spectra
(cf. \citealt{keenan_etal:1994,cook_etal:1995,delzanna_etal:02_aumic,polito_etal:2016b}).
Inconsistencies between densities of \ion{O}{iv}  and \ion{S}{iv} 
have been noted \citep{cook_etal:1995}.
Some of the discrepancies were subsequently resolved 
by \cite{keenan_etal:2002} by taking into account line blending and 
the then new atomic data. Nevertheless, some inconsistencies still remain, in particular 
regarding the 1404.8~\AA\ blend, which is about 30\% stronger than predicted,
as  discussed in \cite{delzanna_etal:02_aumic}.

Recently, several studies have been carried out 
with IRIS.
\cite{peter_etal:2014} used the \ion{O}{iv} lines in combination with 
the \ion{Si}{iv} lines to find very high densities
(about 10$^{13}$ cm$^{-3}$ or higher) in the so called IRIS plasma `bombs'. 
Several problems with the use of such diagnostic were however pointed out by 
\cite{judge:2015}. In fact, the \ion{O}{iv}  and \ion{Si}{iv} 
ratios are sensitive to changes in the temperature and the 
relative abundances of O and Si. In addition,  high density effects
such as the suppression of the  dielectronic recombination rate
can shift the formation temperature of these ions \citep{polito_etal:2016b}.
\cite{judge:2015} also mentioned the fact that the intensity of the 
\ion{Si}{iv} lines is  known to be anomalous.

The  important piece of work carried out by 
\cite{hayes_shine:1987} using SMM/UVSP, where the densities obtained from the 
two methods (\ion{O}{iv}  vs \ion{Si}{iv} and \ion{O}{iv} line ratios)  
were compared is often overlooked in the literature. 
Indeed,  they found  a large discrepancy of about an 
order of magnitude,  with a large scatter. 
Large discrepancies with the use of these two methods ( \ion{O}{iv} line ratios
vs. the intensities of the \ion{O}{iv}  vs \ion{Si}{iv} lines)  were also
found using IRIS data by  \cite{polito_etal:2016b}.
The differences vary substantially depending on the source region
(e.g. AR plage or solar flare). The direct \ion{O}{iv} line ratio
method clearly provides more accurate measurements, although 
the  variations in the densities can also be roughly 
estimated with the other method (using the ratio of lines
from\ion{O}{iv} versus \ion{Si}{iv}  \citep[see][]{doschek_etal:2016}.

\begin{figure}[!htbp]
\centerline{\includegraphics[width=0.7\textwidth]{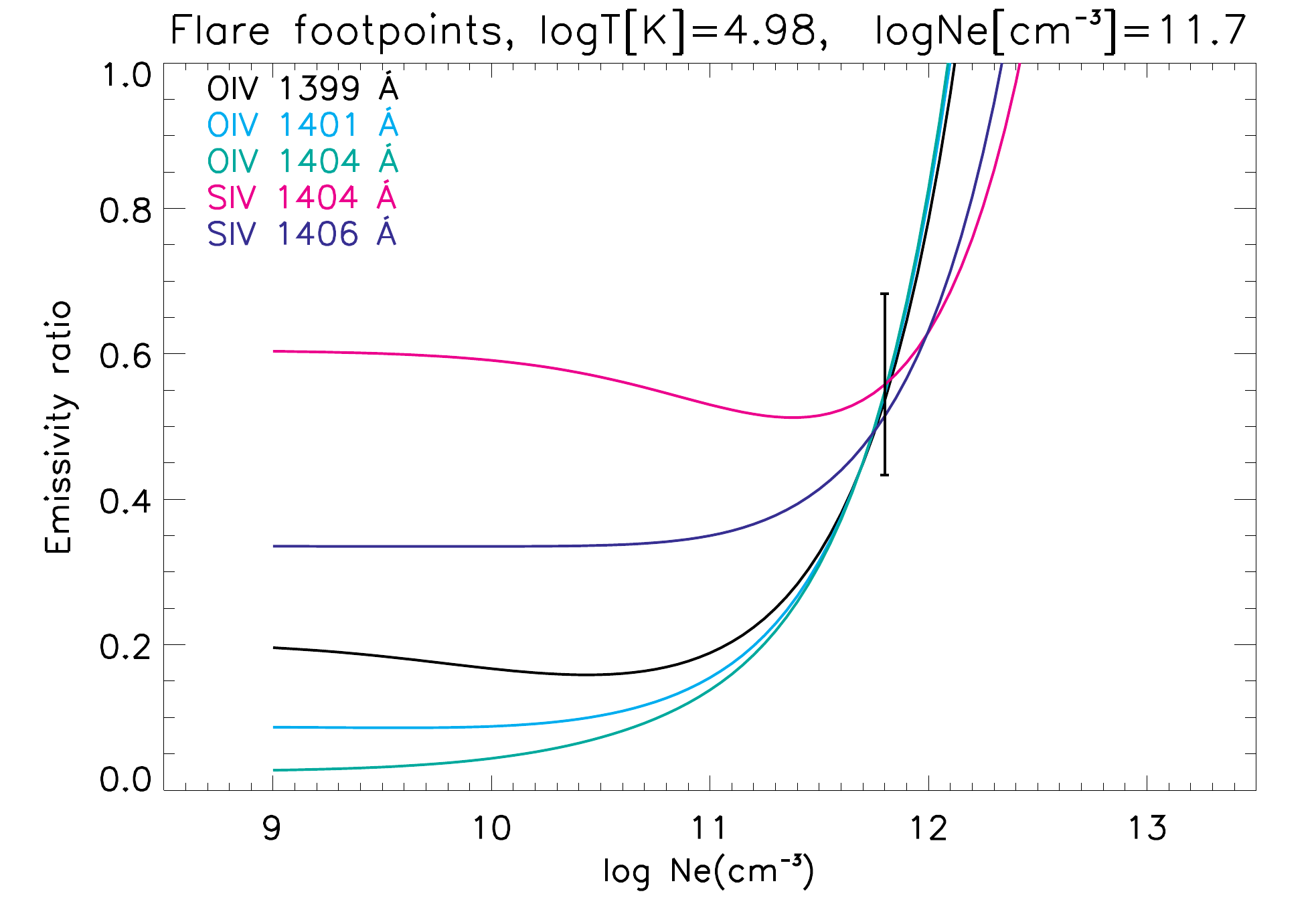}}
\caption{Electron densities obtained from  
the emissivity ratio ($ER$) curves of \ion{S}{iv} and \ion{O}{iv} IRIS lines 
observed in an active region  (adapted from \citealt{polito_etal:2016b}).
The  curves are obtained assuming that both  \ion{O}{iv}  and \ion{S}{iv} 
are formed  at the same isothermal temperature of log $T$[K]=4.98.
}
\label{fig:ne_loop_s_4}
\end{figure}

Regarding the discrepancies between densities found using the
 \ion{O}{iv}  and \ion{S}{iv} lines,
\cite{polito_etal:2016b} has shown that agreement is 
found if  the plasma is assumed to be isothermal.
Fig.~\ref{fig:ne_loop_s_4} shows the emissivity ratio ($ER$) curves
obtained assuming that both  \ion{O}{iv}  and \ion{S}{iv} 
are formed at the same isothermal temperature.

Ideally one would need to have an independent way to measure 
the plasma temperature, i.e. the ion charge state distribution.
It is well known that various effects can shift the temperature
of formation of the \ion{O}{iv} lines (as well as  other 
TR lines), as discussed in \cite{polito_etal:2016b}. For example, 
taking into account high density effects on the ion balance shifts
the ion formation towards lower temperatures (cf. Fig.~\ref{fig:O+3_ioniz}).
A similar  effect is obtained with non-thermal electron distributions
\citep{dudik_etal:2014_o_4}.
 Non-equilibrium ionisation can also  significantly affect these
ratios, as discussed in Sect.~\ref{sec:non-eq}.

\subsubsection{$N_{\rm e}$ from C-like ions}

\begin{figure}[!htbp]
\centerline{\includegraphics[width=0.8\textwidth]{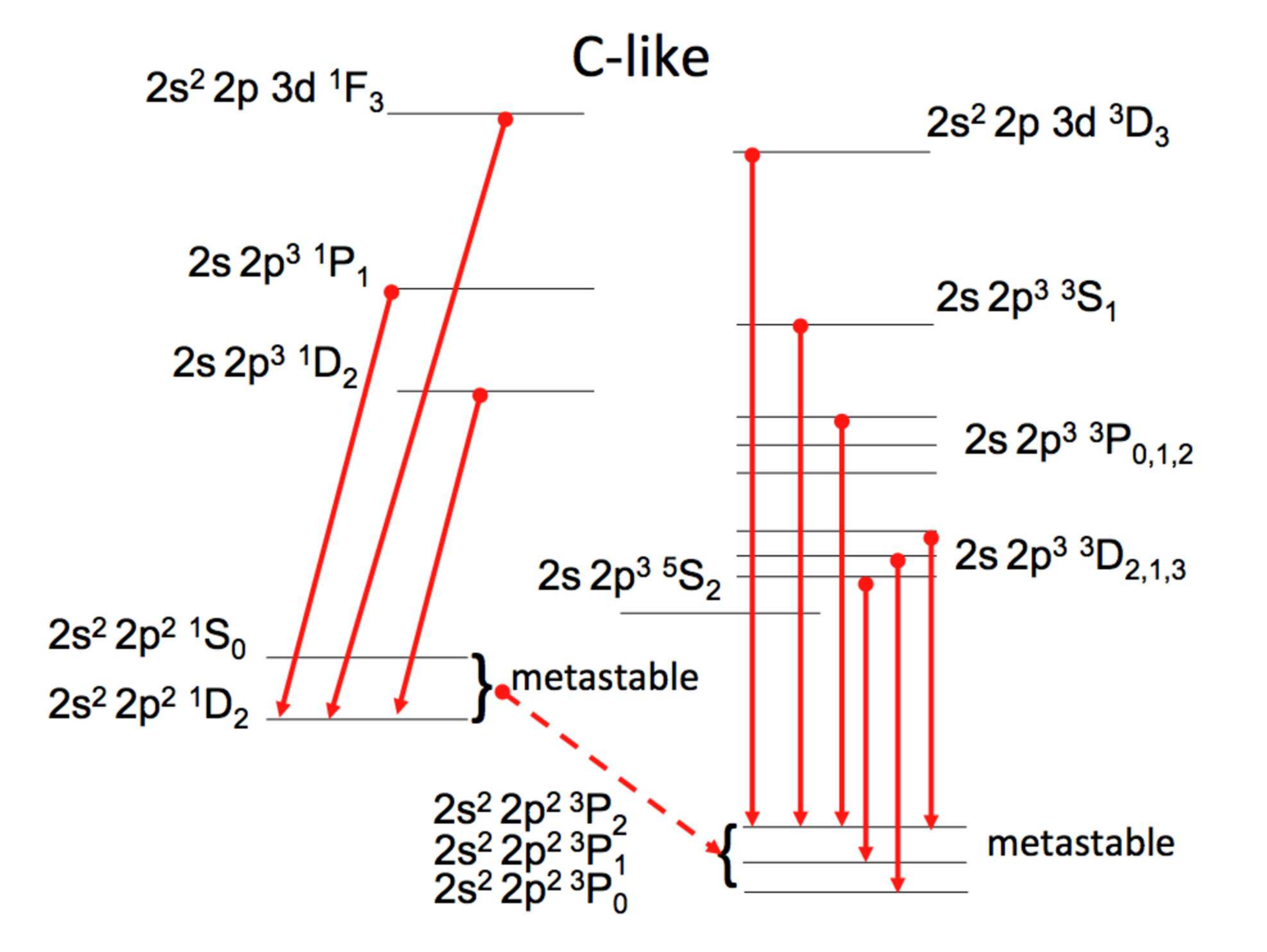}} 
\caption{A diagram (not to scale) of the main levels  in C-like ions.
The red downward arrows indicate the main spectral lines.
}
\label{fig:c-like}
\end{figure}

Transitions within the 2s$^2$ 2p$^2$ -- 2s 2p$^3$ configuration in 
the C-like ions fall in the EUV and are useful density diagnostics.
The density sensitivity arises from the fact that the four 
excited levels within the ground configuration are metastable 
(see  Fig.~\ref{fig:c-like}).

\cite{mason_bhatia:1978} discussed in detail the diagnostic ratios for 
Mg VII, Si IX, and S XI. 
A notable comprehensive study was based on Skylab observations with the 
NRL slitless spectrograph \citep{dere_etal:1979}.
Several other authors later improved the atomic data and presented comparisons
with observations. 
Table~\ref{tab:c-like} lists the main lines. 

Transitions within the 
2s$^2$ 2p$^2$ -- 2s$^2$ 2p 3d configurations fall in the X-rays 
or soft X-rays \citep{brown_etal:1986}
and have not been observed/used as much as the EUV lines.
Only lines from \ion{Ca}{xv} are shown in Table~\ref{tab:c-like},
although  similar line ratios from other ions 
are also potentially useful.

\begin{table}[!htbp]
\caption[C-like density diagnostics]{ C-like density diagnostics 
(layout as in Table~\ref{tab:ne_b-like}).}
\centering
\begin{tabular}{llllllll}
\toprule
Transitions & Ion & $\lambda_1$  (\AA)  & $\lambda_2$ (\AA) & 
log $T_{\rm e}$ &  log N$_{\rm e}$  \\
\midrule

 2s$^2$ 2p$^2$ $^3$P$_{2}$ - 2s 2p$^3$ $^3$D$_{3}$ / & &&&\\
2s$^2$ 2p$^2$ $^3$P$_{0}$ - 2s 2p$^3$ $^3$D$_{1}$ & 
  Si IX & 349.86  & 341.95 & 6.0 & 7--9 \\

& S XI & 291.58    & 281.41 & 6.3 & 8--12 \\

& Ar XIII & 248.70  & 236.28  & 6.45 & 9--11 \\ 

& Ca XV &  215.38 & 200.98  & 6.65 & 10--12 \\  

  & Fe XXI & 145.73 & 128.75 & 7.1 & 11--15 \\

\\

 2s$^2$ 2p$^2$ $^3$P$_{1}$ - 2s 2p$^3$ $^3$D$_{2}$ / & &&&\\ 
2s$^2$ 2p$^2$ $^3$P$_{0}$ - 2s 2p$^3$ $^3$D$_{1}$ & 
 Si IX & 345.12  & 341.95 & 6.0 & 7--9 \\

 & S XI & 285.83 (bl \ion{O}{iv})  & 281.41 & 6.3 & 8--12 \\

 & Ar XIII & 242.24 & 236.28  & 6.45 & 9--11 \\ 

 & Fe XXI & 142.14 (bl) & 128.73 & 7.1 & 11--15 \\
\\

2s$^2$ 2p$^2$ $^3$P$_{2}$ - 2s 2p$^3$ $^3$P$_{2}$ / & &&&\\ 
2s$^2$ 2p$^2$ $^3$P$_{0}$ - 2s 2p$^3$ $^3$D$_{1}$ & 
   Si IX & 296.11 (sbl) & 341.95 & 6.0 & 7--9 \\

 & S XI & 246.9  & 281.41 & 6.3 & 8--12 \\

 & Ar XIII & 210.47  & 236.28  & 6.45 & 9-11 \\ 
 
 & Ca XV &  181.90 & 200.98  & 6.65 & 10--12 \\  

 & Fe XXI & 121.21 (bl) & 128.73 & 7.1 & 11--15 \\

\\ 

2s$^2$ 2p$^2$ $^1$D$_{2}$ - 2s 2p$^3$ $^1$D$_{2}$ / & &&&\\ 
 2s$^2$ 2p$^2$ $^3$P$_{2}$ - 2s 2p$^3$ $^3$P$_{2}$ & 
  Mg VII & 319 (bl Ni XV) &  367.7 (bl) & 5.8  & 8--10 \\

 &   Si IX &  258.08 (bl Ar XIV) & 296.11 (sbl) &  6.0 & 9--12 \\

 &    S XI & 215.97 (bl Ni XV) & 246.89 & 6.3 & 8--12 \\

\\

2s$^2$ 2p$^2$ $^1$D$_{2}$ - 2s 2p$^3$ $^1$P$_{1}$ / & &&&\\ 
2s$^2$ 2p$^2$ $^3$P$_{2}$ - 2s 2p$^3$ $^3$S$_{1}$ & 
 Mg VII & 280.74 & 278.41 (bl Si VII) &   5.8  & 8--10 \\

 & Si IX & 227.36 (bl)  & 227.0 & 6.0 & 9--12 \\

 & S XI & 190.35 (bl Fe XI) & 191.26 (bl Fe IX) & 6.3 & 8--12 \\

2s$^2$ 2p$^2$ $^1$D$_{2}$ - 2s$^2$ 2p 3d $^1$F$_{3}$ / & &&&\\ 
2s$^2$ 2p$^2$ $^3$P$_{2}$ - 2s$^2$ 2p 3d $^3$D$_{3}$ & 
  Si IX & 56.03 & 55.40  & 6.1 & 7--11 \\

\\

2s$^2$ 2p$^2$ $^3$P$_{2}$ -- 2s$^2$ 2p 3d  $^3$D$_{3}$ / & &&&\\ 
2s$^2$ 2p$^2$ $^3$P$_{0}$ -- 2s$^2$ 2p 3d  $^3$D$_{1}$ & 
\ion{Ca}{xv} & 22.78 & 22.73 & 6.6 & 10--12 \\


 &  \ion{Fe}{xxi} & 12.32  & 12.282 & 7.1 & 11--15 \\



2s$^2$ 2p$^2$ $^3$P$_{1}$ -- 2s$^2$ 2p 4d  $^3$P$_{2}$ / & &&&\\ 
2s$^2$ 2p$^2$ $^3$P$_{1}$ -- 2s$^2$ 2p 4d  $^3$D$_{1}$ &
 \ion{Fe}{xxi} & 9.548  & 9.542 & 7.1 & 11--15 \\

\bottomrule
\end{tabular}
\label{tab:c-like}
\end{table}


A few diagnostics in the X-rays (around 10~\AA) are available, from 
 \ion{Fe}{xxi} 2s$^2$ 2p$^2$-- 2s$^2$ 2p 4d transitions 
 (cf.  Table~\ref{tab:c-like} and  \citealt{fawcett_etal:87,phillips_etal:1996}).

The  Mg VII 319.0 / 367.7~\AA\ ratio was observed by SOHO CDS,
however the 319.0~\AA\ line  is often blended with Ni XV and the 
367.7~\AA\ line is in the blue-wing of the Be-like Mg resonance line, so careful
deblending is needed.
There are other lines in the CDS spectral range, which could be used instead of the 
367.7~\AA\ line, but they are weaker and also blended.
The ratio of the  Mg VII 280.74~\AA\ line with several other lines 
(e.g. 434.70~\AA)
observed by SOHO/GIS is also a good diagnostic.
The  Mg VII 280.74/278.41~\AA\ ratio was observed by Hinode EIS,
but there are various problems in deblending one of the lines,
as discussed in  \cite{delzanna:12_atlas}.
S XI has some excellent diagnostic ratios
observed by Hinode EIS \citep{delzanna:12_atlas}.

The Si IX lines at 341.95, 345.10, 349.9~\AA\ (a self-blend of three transitions) 
were observed by SOHO/NIS and have been used extensively to measure 
densities.
The 227.36~\AA\ line is blended with several transitions, as discussed by 
\cite{dere_etal:1979}.

At high densities, typical of flares, \ion{Ca}{xv} lines 
are a good density diagnostic.
Transitions within the 2s$^2$ 2p$^2$ -- 2s 2p$^3$ fall in the EUV and are 
very good  diagnostics, as shown by  \cite{dere_etal:1979}.
The ratios of the 215.38 or 181.90~\AA\ lines vs. the resonance
EUV line at 200.97~\AA\ are  excellent density diagnostics around 4 MK
at high densities typical of flares. 
Skylab observations showed consistent density values \citep{keenan_etal:1988_ca_15}.
The \ion{Ca}{xv}  181.90/200.98~\AA\  ratio is also useful for Hinode EIS observations, as shown by
\cite{warren_etal:2008_flare}, who analysed the first EIS full-spectrum of a 
small flare, also first discussed in \cite{delzanna:08_bflare}.
In principle, there is another useful transition, at 208.72~\AA.
However, this  line is blended with \ion{Fe}{xiii}.
Transitions within the 
2s$^2$ 2p$^2$ -- 2s$^2$ 2p 3d configurations fall in the X-rays 
and are also useful density diagnostics, as shown by 
\cite{brown_etal:1986}. The lines are very close in wavelength, so 
high-resolution spectrometers are needed to resolve them.

A few \ion{Ar}{xiii} lines are also potentially useful, 
 although it is not clear if the problems 
reported by \cite{dere_etal:1979} are due to blending or 
the atomic data.

Fe XXI soft X-ray lines  offer excellent diagnostics for solar flares
\citep{mason_etal:79,mason_etal:84,delzanna_woods:2013}, 
although several of them are blended.
The 121.21 or the 145.73~\AA\ lines vs. the resonance line
at 128.75~\AA\ are the best ratios.

\subsubsection{$N_{\rm e}$ from N-like ions}

\begin{figure}[!htbp]
\centerline{\includegraphics[width=0.8\textwidth]{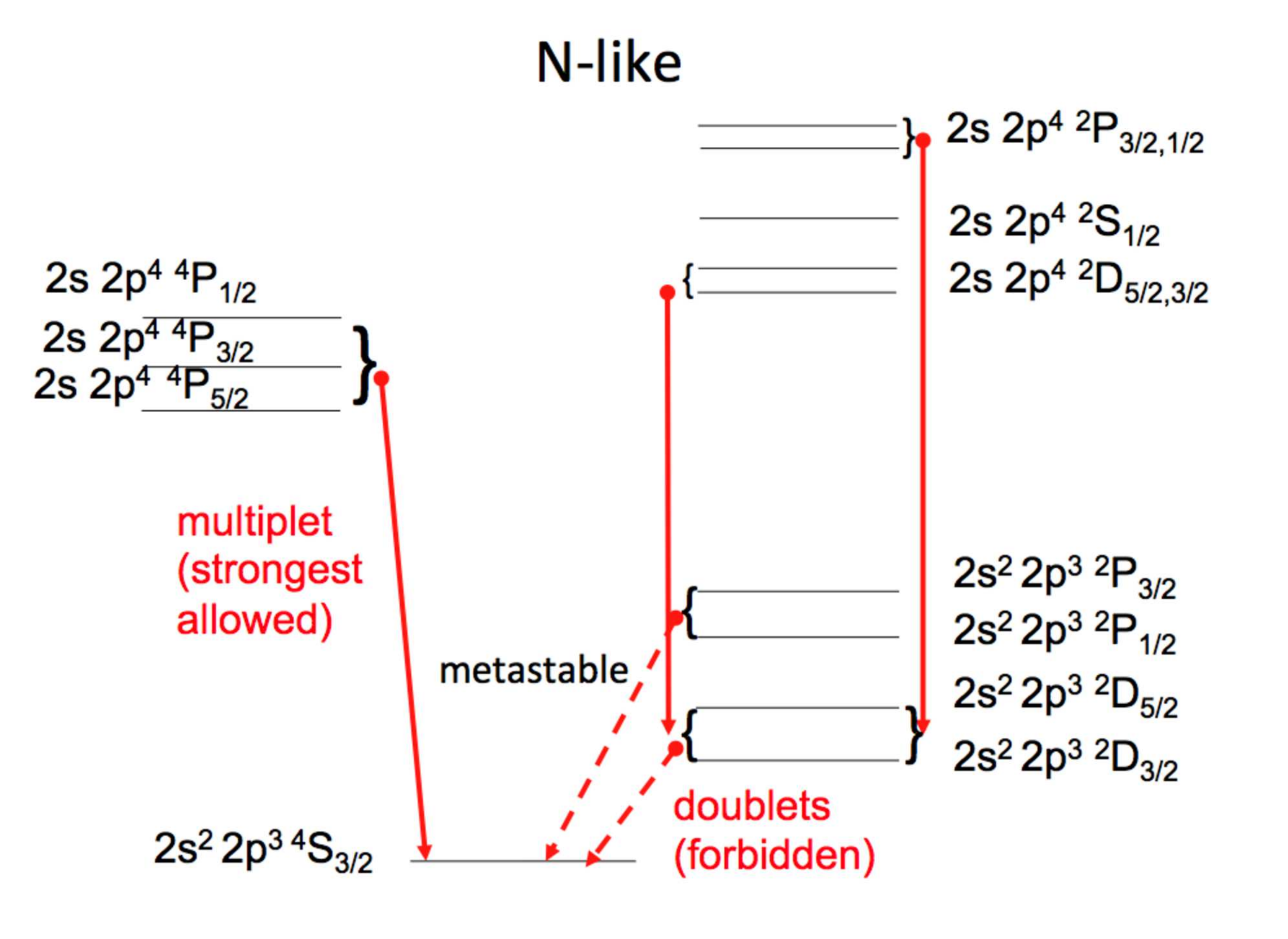}} 
\caption{A diagram (not to scale) of the main levels  in N-like ions.
The red downward arrows indicate the main spectral lines.
}
\label{fig:n-like}
\end{figure}

N-like ions offer several diagnostic ratios, as shown in 
Table~\ref{tab:ne_n-like}. The density sensitivity arises 
because the four excited levels, within the 
ground configuration 2s$^2$ 2p$^3$ are metastable (see Fig.~\ref{fig:n-like}).
Ratios of lines within the ground 
configuration are found in the UV. The advantage is that 
lines are close in wavelength and ratios are insensitive to temperature.
They have been observed mainly with Skylab and SOHO SUMER. 
The disadvantage is that these forbidden lines are intrinsically very weak and in 
on-disk observations are virtually unobservable because they are blended 
with cool emission. 
 The strongest lines providing density diagnostics are in the EUV for
 most ions in this sequence.

\cite{feldman_etal:1978} considered two line ratios within the ground 
configuration 
for Mg VI, Si~VIII, S X, and  Ar XII. These lines are intrinsically very weak and 
Skylab observations only allowed measurements of the 
 S X 1213 / 1196~\AA\ ratio.
The main ratio is the 2s$^2$ 2p$^3$ $^4$S$_{3/2}$ -- 2s$^2$ 2p$^3$ $^2$D$_{5/2}$  / 
 2s$^2$ 2p$^3$ $^4$S$_{3/2}$ -- 2s$^2$ 2p$^3$ $^2$D$_{3/2}$.
The Si VIII 1445 / 1440~\AA\ ratio is  also a  good diagnostic 
for quiet Sun densities.

Indeed, the S X and Si VIII ratios were used 
to obtain densities off-limb with SOHO SUMER \citep{doschek_etal:1997}.
\cite{mohan_etal:2003} extended the discussion of these  
forbidden lines within the
ground configuration to more ions observed by SOHO SUMER. 
These other line ratios have rarely been observed because lines are 
weak and formed at higher temperatures. One example 
is the off-limb flare observations of SOHO SUMER reported by 
\cite{landi_etal:2003}, where lines from P, Ar, K, and Ca were observed. 
For active regions, probably the best ratio is the \ion{Ca}{xiv}
one, since the lines are relatively bright. 
The main ratios are shown in Table~\ref{tab:ne_n-like}.
\cite{mohan_etal:2003} also listed a few weaker line ratios 
(2s$^2$ 2p$^3$ $^4$S$_{3/2}$ -- 2s$^2$ 2p$^3$ $^2$P$_{3/2}$  / 
 2s$^2$ 2p$^3$ $^4$S$_{3/2}$ -- 2s$^2$ 2p$^3$ $^2$P$_{1/2}$)
from Al, Si,  P, S, and  Ar, but resulting densities were inconsistent. 
We point out that some inconsistencies for ions of this sequence are 
to be expected because accurate atomic data are still not available.

Another diagnostic possibility is to use EUV lines, which are intrinsically
much stronger. 
The EUV lines of the  N-like ions Mg VI, Si VIII, S X, Ar XII, Ca XIV were discussed 
by \cite{bhatia_mason:1980} and later by several authors, including 
\cite{dwivedi_raju:1988}. The main ratios are shown in Table~\ref{tab:ne_n-like}.
\cite{dwivedi_mohan:1995}  used them to analyse  SERTS observations.

\begin{table}[!htbp]
\caption[N-like density diagnostics]{N-like density diagnostics 
(layout as in Table~\ref{tab:ne_b-like}; w: weak, bl:blended).}
\centering
\begin{tabular}{llllllll}
\toprule
Transitions  & Ion & $\lambda_1$  (\AA)  & $\lambda_2$ (\AA) & 
log $T$ &  log $N_{\rm e}$  \\
\midrule
 
 2s$^2$ 2p$^3$ $^4$S$_{3/2}$ -- 2s$^2$ 2p$^3$ $^2$D$_{5/2}$  / & &&&  \\  
 2s$^2$ 2p$^3$ $^4$S$_{3/2}$ -- 2s$^2$ 2p$^3$ $^2$D$_{3/2}$ & 
  \ion{Si}{viii} & 1440.51   & 1445.73  & 5.9 &  6--8.5 \\

  & \ion{P}{ix} & 1307.57 w  & 1317.65 w  &  6.0  &   \\

  & \ion{S}{x} & 1196.21   & 1212.93  &  6.2  &  7--9.5  \\

  & \ion{Ar}{xii} &  1018.89 (bl) & 1054.57  &  6.4  &  \\ 

  & \ion{K}{xiii} &  945.83 w & 994.52 w &  6.4    &   \\ 

  &   \ion{Ca}{xiv} &  880.35 & 943.70 & 6.5  &  \\ 

\\

 2s$^2$ 2p$^3$ $^2$D$_{5/2}$ - 2s 2p$^4$ $^2$P$_{3/2}$ / & &&& \\ 
2s$^2$ 2p$^3$ $^4$S$_{3/2}$ - 2s 2p$^4$ $^4$P$_{5/2}$ &  
 Si VIII & 216.92 (bl \ion{Fe}{ix}) & 319.84 & 5.9  & 7--12.5 \\ 

& S X     & 180.73 & 264.23 & 6.1 &  7--12.5 \\ 

 &  Ar XII  & 154.42 (?) & 224.25 & 6.4 & 8.5--13 \\

 &  Ca XIV & 134.27 (?) & 193.87  & 6.55 & 9.5--14 \\

\\

2s$^2$ 2p$^3$ $^2$D$_{5/2}$ - 2s 2p$^4$ $^2$D$_{5/2}$ / & &&&\\ 
2s$^2$ 2p$^3$ $^4$S$_{3/2}$ - 2s 2p$^4$ $^4$P$_{5/2}$ &  
 Si VIII & 277.06 (bl) &  319.84 &  & 7--12.5 \\ 

 & S X     & 228.69 & 264.23 &  6.1 & 7--12.5 \\ 

 &  Ar XII  & 193.70 (bl Fe X) & 224.25 & 6.4 & 8.5--13 \\

&  Ca XIV & 166.96 (?) & 193.87  & 6.55 & 9.5--14 \\

2s$^2$ 2p$^3$ $^2$D$_{5/2}$ --  2s$^2$ 2p$^2$ 3d $^2$F$_{7/2}$ / & &&&\\ 
2s$^2$ 2p$^3$ $^4$S$_{3/2}$ -- 2s$^2$ 2p$^2$ 3d $^4$P$_{3/2}$ & 
Fe XX  & 12.86 & 12.83 &  7.0 & 11.5--15 \\
\bottomrule
\end{tabular}
\label{tab:ne_n-like}
\end{table}

The Mg VI lines are weak and blended.
The Si VIII lines would be useful but are normally blended
in medium or low-resolution spectra.
The  S X  180.73/ 264.23~\AA\ is an excellent diagnostic ratio
observed by Hinode EIS \citep{delzanna:12_atlas}.
The Ar XII lines are normally weak and blended.
The Ca XIV 134.27 and 166.96~\AA\ lines  have not been observed
in solar spectra, but in principle could be used 
simultaneously with the resonance line 193.87~\AA\ 
to measure densities.

At flare temperatures, Fe XX lines offer  the possibility of measuring
electron densities above 10$^{11}$ cm$^{-3}$, with the 
110.63/121.85~\AA\ or 113.35/121.85~\AA\ ratios 
\citep{mason_etal:84,delzanna_woods:2013}.
In fact, the strongest transition, 
2s$^2$ 2p$^3$ $^4$S$_{3/2}$ - 2s 2p$^4$ $^4$P$_{5/2}$
falls at 132.85~\AA, where this line is blended with the 
Fe XXIII resonance line.

 \subsubsection{$N_{\rm e}$ from O-like ions}

\cite{raju_dwivedi:1978} discussed the diagnostics for Mg V, Si VII,
S IX, Ar XI, however these ions produce weak lines in solar spectra.


The N IV 923 vs. 765~\AA\ ratio would be a good density diagnostic
for the transition region, but these lines are blended, with an 
H I Lyman line and N III, respectively.


 



 \subsubsection{$N_{\rm e}$ from Mg-like ions}

\begin{figure}[!htb]
\centerline{\includegraphics[width=0.8\textwidth]{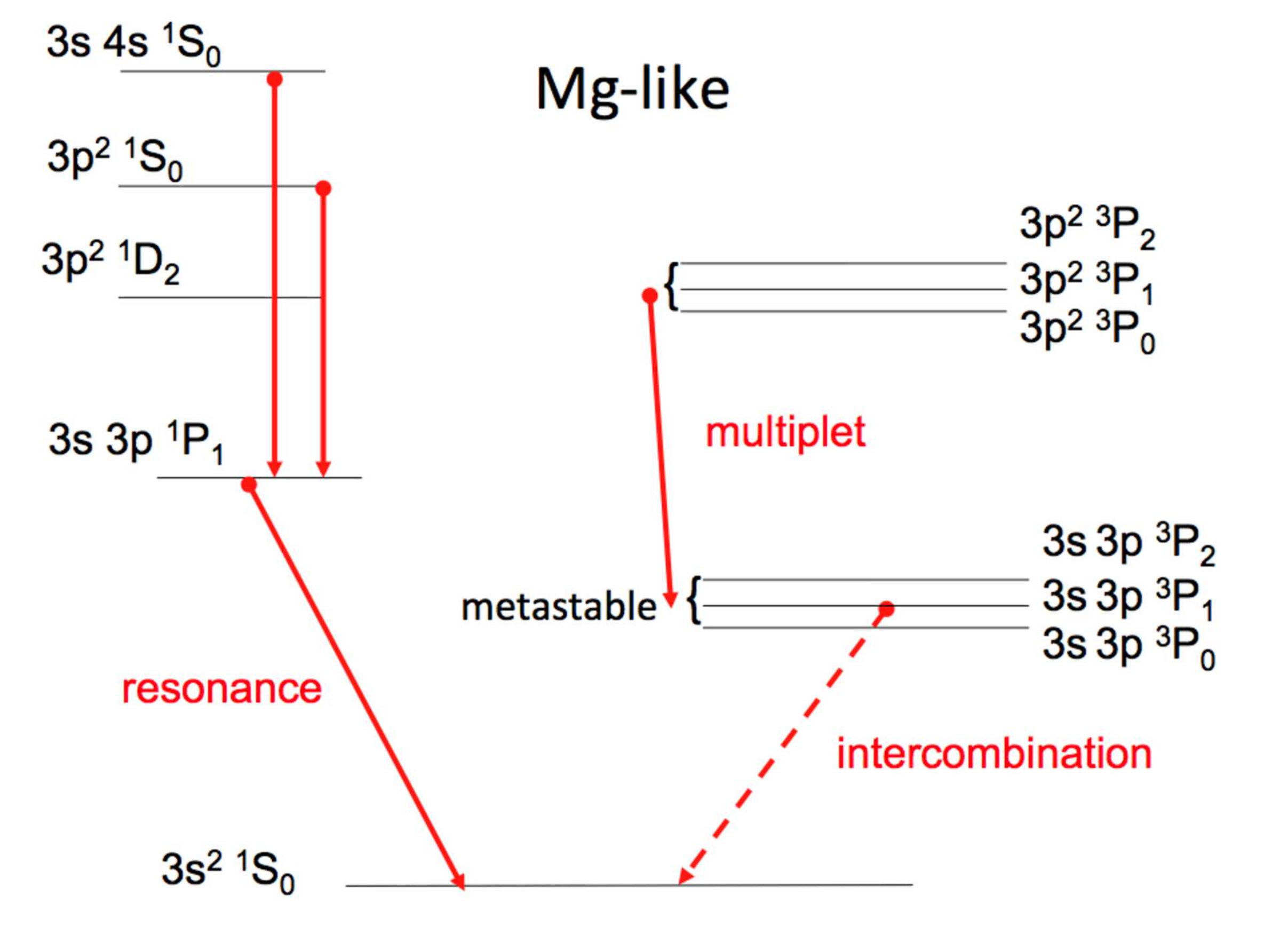}} 
\caption{ A diagram (not to scale) of the main levels  in Mg-like ions.
The red downward arrows indicate the main spectral lines.
}
\label{fig:mg-like}
\end{figure}

Ions in this sequence have a 3s$^2$ $^1$S$_{0}$ ground state, i.e. have a 
similar atomic structure as the Be-like (where the ground state 
is 2s$^2$ $^1$S$_{0}$), hence have similar density diagnostics
(see  Fig.~\ref{fig:mg-like}).
As in the Be-like case, the ratio of any lines within the 
3s 3p $^3$P - 3p$^2$ $^3$P multiplet with the 
resonance 3s$^2$ $^1$S$_{0}$ - 3s 3p $^1$P$_{1}$ line is density-sensitive, although 
there is also a strong temperature dependence. 
The most famous ion is  \ion{Si}{iii}
(see, e.g. \citealt{nicolas_etal:1977,kjeldseth-moe_nicolas:1977,dufton_etal:1983}),
where there is the additional problem, as in \ion{C}{iii}, that the 
resonance line at 1206.499~\AA\ is often  affected by opacity.
The ratio of the 1301.15~\AA\ with any of the other lines 
within the multiplet is a good density diagnostic in the 
N$_{\rm e}$ = 10$^{9}$ -- 10$^{12}$ cm$^{-3}$ range as it varies by a 
factor of 5 and is not temperature dependent. 
Indeed \cite{doschek:1997} used the \ion{Si}{iii} 
1296.73/1301.15~\AA\ ratio to get electron densities from  Skylab spectra.
A recent review of the \ion{Si}{iii} diagnostics can be found in 
\cite{delzanna_etal:2015_si_3}.

The equivalent ratios for \ion{S}{v} are observed in the 850--860~\AA\
range, but they are only sensitive to very high densities,
above 10$^{11}$ cm$^{-3}$.
\cite[][]{dufton_etal:1986} suggested another  density 
diagnostic, the ratio of the 1501.766 vs. the  1199.136~\AA\ lines for 
densities above 10$^{12}$ cm$^{-3}$, although we note  
that the lines are weak
and the ratio is strongly temperature sensitive.

Other ions in the sequence such as  Ar VII and Ca IX
produce weak lines, except \ion{Fe}{xv} and \ion{Ni}{xvii}
which are discussed below.


\begin{table}[!htbp]
\caption[Mg-like density diagnostics.]{Mg-like density diagnostics. 
The log $T_{\rm e}$ [K] values  
indicate the  temperatures of peak ion abundance in equilibrium.
Wavelengths are in \AA.
}
\centering
\begin{tabular}{llllllll}
\toprule
Transition & \ion{Si}{iii} & \ion{S}{v} & \ion{Fe}{xv} & \ion{Ni}{xvii} \\
           &     4.7       &      5.1  &    6.3  &   6.5 \\
\midrule

3s 3p $^3$P$_{2}$--3s 3d $^3$D$_{3}$ &  1113.232 (sbl)  &  & 233.866 & 207.520  \\ 
3s 3p $^1$P$_{1}$--3s 3d $^1$D$_{2}$ &  1206.557 (sbl)  &  & 243.794 & 215.910 (bl S XI) \\ 

 3s$^2$ $^1$S$_{0}$ - 3s 3p $^1$P$_{1}$ & 1206.502 (sbl) & & 284.15 &  249.185 \\  

 3s 3p $^3$P$_{1}$--3p$^2$ $^3$P$_{2}$ & 1294.548 & 849.240 & 292.275 & 251.952 \\ 

 3s 3p $^3$P$_{0}$--3p$^2$ $^3$P$_{1}$ & 1296.728 & 852.178 & 302.334 & 263.591 \\ 
 3s 3p $^3$P$_{2}$--3p$^2$ $^3$P$_{2}$ &1298.948 (sbl) & 854.770 (sbl) & 304.894 (bl \ion{Fe}{xvii}) & 266.064 \\
 3s 3p $^3$P$_{1}$--3p$^2$ $^3$P$_{1}$ &  1298.894 (sbl) & 854.870 (sbl) & 307.747 & 269.416 (bl \ion{Fe}{xvii})\\ 


3s 3p $^3$P$_{1}$--3p$^2$ $^3$P$_{0}$ & 1301.151 & 857.828 & 317.597 & 281.469  \\  
3s 3p $^3$P$_{2}$--3p$^2$ $^3$P$_{1}$ & 1303.325 & 860.473 & 321.769 & 285.616 \\ 
3s 3p $^3$P$_{2}$--3p$^2$ $^1$D$_{2}$ & 1447.191  &        & 327.033 & 289.743 \\ 
\\
3s$^2$ $^1$S$_{0}$ - 3s 3p $^3$P$_{1}$ &  &  1199.136 &  \\ 
3s 3p $^1$P$_{1}$ - 3p$^2$ $^1$D$_{2}$ &  &  1501.766 &  \\ 
\bottomrule 
\end{tabular}
\label{tab:mg-like}
\end{table}


\smallskip\smallskip\noindent
{\bf  \ion{Fe}{xv}}
\smallskip

There are several diagnostics in the EUV within  \ion{Fe}{xv} (formed around 3 MK). 
Two ratios that would in principle be excellent are, among others, the 
 233.87/243.79~\AA\ \citep[see, e.g.][]{cowan_widing:1973} and 321.78/327.03~\AA\ ratios.
The 233.87~\AA\ line is collisionally populated from the 3s 3p $^3$P
which is metastable, while the 243.79~\AA\ is not populated from a 
metastable, hence the density sensitivity of the ratio, for densities
between  10$^{9}$  and 10$^{11}$ cm$^{-3}$. The lines are close in wavelength. 
The  321.78/327.03~\AA\ ratio is particularly useful because it changes by a factor
of 3 in the same density interval, and the lines are close.

However, significant problems in these ratios 
have been noted by several authors 
(see, e.g. \citep{dere_etal:1979,dufton_etal:1990,bhatia_mason:1997}).
It was not clear if the problem was due to blending of the lines
or with the atomic data.
\cite{keenan_etal:2006}  suggested that the main problem is 
blending of virtually all the lines.
The 321.78/327.03~\AA\ ratio was however found to be reliable 
 in \cite{fernandez-menchero_etal:2015_unitary}, using the 
 most accurate atomic data for this ion 
(UK APAP data, see \citealt{fernandez-menchero_etal:2014_mg-like}).

The analogous of the \ion{Si}{iii} 1296.73/1301.15~\AA\ ratio is the 
 317.597 vs. 302.334~\AA. It is not very sensitive  as it varies by 
only  a factor of 5 within  10$^{8}$  and 10$^{10}$ cm$^{-3}$. 
The ratio of the 304.98~\AA\ line with e.g. the resonance line 
is also a good density diagnostic, but this line is blended with 
\ion{He}{ii}.
There are other possible diagnostics in the soft X-ray region
\citep[see, e.g.][]{bhatia_etal:1997},
but the lines are relatively weak. 


\smallskip\smallskip\noindent
{\bf  \ion{Ni}{xvii}}
\smallskip

The 207.520 vs. 215.910~\AA\ ratio is in principle a good density 
diagnostic in the  10$^{9}$  -- 10$^{12}$ cm$^{-3}$ range, but the 
215.91~\AA\ line is blended with  S XI. Instead of the 215.91~\AA\ line,
the 289.743~\AA\ could be used, although this line can become blended
with a \ion{Fe}{xix} 289.76~\AA\ if high temperatures are present. 
Another option is to use the ratio of the 266.064~\AA\  with 
the 289.743~\AA\ line.

 \subsubsection{$N_{\rm e}$ from Al-like ions}

\begin{figure}[!htbp]
\centerline{\includegraphics[width=0.8\textwidth]{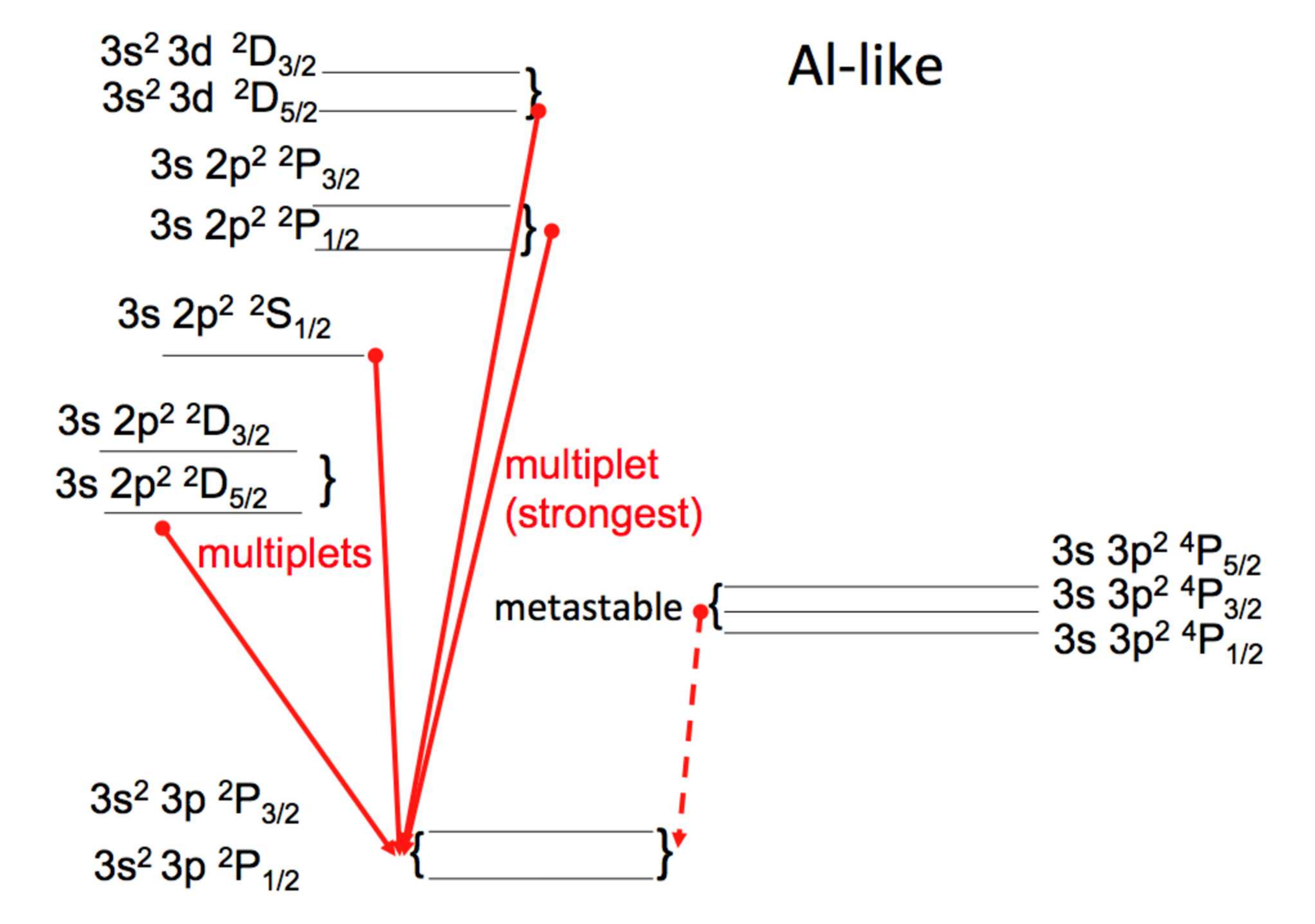}} 
\caption{ A diagram (not to scale) of the main levels  in Al-like ions.
The red downward arrows indicate the main spectral lines.
}
\label{fig:al-like}
\end{figure}


\begin{table}[!htbp]
\caption[Some of the common density diagnostics for \ion{Fe}{xiv}]{
Some of the common density diagnostics for \ion{Fe}{xiv}. 
(bl) indicates a blend, while (sbl) indicates a self-blend of more than one transition from the same ion.}
\label{tab:dens_fe_14}
\centering
\begin{tabular}{llllllll}
\toprule
 &  Transition & $\lambda$  (\AA) & Observed &  log N$_{\rm e}$  \\
\midrule
       & 3s$^2$ 3p $^2$P$_{3/2}$ -- 3s$^2$ 4d $^2$D$_{5/2}$ & 59.58 & & \\ 
       & 3s$^2$ 3p $^2$P$_{1/2}$ -- 3s$^2$ 4d $^2$D$_{3/2}$ & 58.96 & &  9--11 \\

       & 3s$^2$ 3p $^2$P$_{3/2}$ - 3s$^2$ 3d $^2$D$_{5/2}$ & 219.12 & Skylab  & \\ 
       & 3s$^2$ 3p $^2$P$_{3/2}$ - 3s$^2$ 3d $^2$D$_{3/2}$  & 220.08 & Skylab       &  9--11 \\

\\

       & 3s$^2$ 3p $^2$P$_{3/2}$ - 3s 3p$^2$ $^2$P$_{3/2}$ & 264.79 (bl) & EIS &   9--11 \\

        & 3s$^2$ 3p $^2$P$_{1/2}$ - 3s 3p$^2$ $^2$P$_{3/2}$   & 252.20 &   EIS &  9--11 \\
       & 3s$^2$ 3p $^2$P$_{1/2}$ - 3s$^2$ 3d $^2$D$_{3/2}$ & 211.32 &   &  \\


\\

       & 3s$^2$ 3p $^2$P$_{3/2}$ - 3s 3p$^2$ $^2$D$_{5/2}$ & 353.84 (bl) & & 9--11 \\
      &  3s$^2$ 3p $^2$P$_{3/2}$ - 3s$^2$ 3d $^2$D$_{5/2}$ & 334.18 &  CDS     & \\ 
\bottomrule
\end{tabular}
\end{table}

The most widely used ion in this sequence is \ion{Fe}{xiv}.
The main density sensitivity is associated with decays of levels that 
are populated from the  metastable  $^2$P$_{3/2}$ within the 
ground configuration 3s$^2$ 3p (see Fig.~\ref{fig:al-like}).

\ion{Fe}{xiv} (2 MK) offers many excellent  density diagnostics 
involving strong lines that have been observed by 
several instruments.
 We note that reliable densities have been obtained 
only with recent atomic calculations \citep{storey_etal:00}.

The best \ion{Fe}{xiv} diagnostic is the ratio of two strong lines close in wavelength,
at 219.12/211.3~\AA.
Another good ratio is the 219.12/220.08~\AA.
There are several other good ratios observed by e.g. Hinode EIS
(see \citealt{delzanna:12_atlas} for details), for example 
the 211.32 or 274.2~\AA\ lines (which is blended with Si VII)
vs. the 264.79~\AA\ (or 252.20~\AA).

Other good ratios are found at longer wavelengths observed by e.g.  
SOHO/CDS  \citep[see, e.g.][]{mason_etal:1997} 
or SERTS \citep[see, e.g.][]{young_etal:98}. 
The  best one is  the  353.84  / 334.18~\AA, although it 
 should be used with caution, since
the \ion{Fe}{xiv}   353.83 \AA\ becomes blended with an Ar XVI    353.920 \AA\ line
in active region spectra. 
There are other possible diagnostics in the soft X-ray region,
but lines are relatively weak. 

The accuracy of the atomic data in CHIANTI v.8 to measure densities 
in EBIT spectra of \ion{Fe}{xiv} has recently been confirmed by 
\cite{weller_etal:2018}.

After \ion{Fe}{xiv}, the most widely used ion in the sequence is  \ion{S}{iv}. 
As discussed by e.g. \cite{dufton_etal:1982}, the 
1416.93/1406.06 and  1423.88/1416.93~\AA\ ratios are in 
principle good density diagnostics. 
The \ion{S}{iv} 1398.08, 1416.93 \AA\ and 1423.88 \AA\ lines are  weak and are 
often not  detected. 
 Table~\ref{tab:ne_s_4} lists the main  \ion{S}{iv} transitions. 

We note that, as in the \ion{O}{iv} case, 
 there has been some confusion in the literature regarding the wavelengths of the
\ion{S}{iv} lines.  Table~\ref{tab:ne_s_4} lists the wavelengths
as recommended by \cite{delzanna_badnell:2016}, which are based on 
a reassessment of all the experimental data for this ion.
Of particular importance is the wavelength of the line at 
1404.85~\AA, estimated to be very close to the 
wavelength of the  \ion{O}{iv}, 1404.82~\AA. Indeed there is no indication
of a wider separation of the two lines from the IRIS spectra.

As we have mentioned when discussing \ion{O}{iv}, several inconsistencies
in the  \ion{S}{iv}  densities were reported
 \citep[see, e.g.][]{cook_etal:1995}. They were mostly due to the inaccuracy 
of the excitation data. The more recent calculations by 
\cite{tayal:2000_s_4} resolved the main discrepancies. 
However, given the inconsistencies in the 1404.8~\AA\ blend 
\citep{delzanna_etal:02_aumic}, a new $R$-matrix scattering calculation was carried
out by \cite{delzanna_badnell:2016}. The new data corrected 
a few problems with the previous calculation, but still 
did not fix the problem with the  1404.8~\AA\ blend, which
could  however be resolved by assuming that the plasma is 
isothermal  \citep[see details in][]{polito_etal:2016b}.

The  \ion{S}{iv} lines are particularly important because they 
allow measurements of  densities up to about 10$^{13}$ cm$^{-3}$,
i.e. higher than those measurable with the \ion{O}{iv}
intercombination lines. \cite{polito_etal:2016b}
presented IRIS measurements in kernels of chromospheric 
evaporation, which showed high densities, reaching the 
10$^{13}$ cm$^{-3}$ limit, during the impulsive phase of a flare.

\begin{table}[!htb]
\caption{The  \ion{S}{iv}  transitions  commonly used to measure densities 
(wavelengths from \cite{delzanna_badnell:2016}).}
\label{tab:ne_s_4}
\centering
\begin{tabular} [c]{lcll}
\toprule
Ion & Terms & Wavelength (\AA) & \\
\midrule
\ion{S}{iv} & $^2$P$_{1/2}$ -  $^4$P$_{3/2}$ & 1398.08 & \\
\ion{S}{iv} & $^2$P$_{1/2}$ -  $^4$P$_{1/2}$ & 1404.85 (bl \ion{O}{iv}) &  \\ 
\ion{S}{iv} & $^2$P$_{3/2}$ -  $^4$P$_{5/2}$ & 1406.059 & \\ 
\ion{S}{iv} & $^2$P$_{3/2}$ -  $^4$P$_{3/2}$ & 1416.928  & \\
\ion{S}{iv} & $^2$P$_{3/2}$ -  $^4$P$_{1/2}$ & 1423.885  & \\ 
\bottomrule
\end{tabular}
\end{table}


\subsubsection{$N_{\rm e}$ from other sequences - iron coronal ions,
  \ion{Fe}{ix} - \ion{Fe}{xiii}}

The coronal iron ions offer a very large number of diagnostics.
Iron lines dominate the XUV spectra, with a large number of strong lines.
Table~\ref{tab:dens_pairs}  presents  a list of 
EUV line ratios useful to measure densities.
Only the main ratios are listed, involving pairs of strong lines
close  in wavelength. There is 
a very large number of possible other combinations.
There is an extensive literature on the possible diagnostics of
iron ions. A fairly complete early review of the iron EUV lines 
is given in \cite{dere_etal:1987}. 

The density ranges where  the ratios are sensitive 
are also shown in Table~\ref{tab:dens_pairs}.
However, this does not mean that all the ratios listed 
in the Table are usable in all cases. 
 Ions like \ion{Fe}{xiii}  that emit above a million degrees, for example, are simply not seen 
in coronal holes or in very quiet regions of the Sun. 
Many lines are blended to some degree, and blending changes depending on 
the source region.

\begin{figure}[!htb]
 \centerline{\includegraphics[width=0.7\textwidth,angle=90]{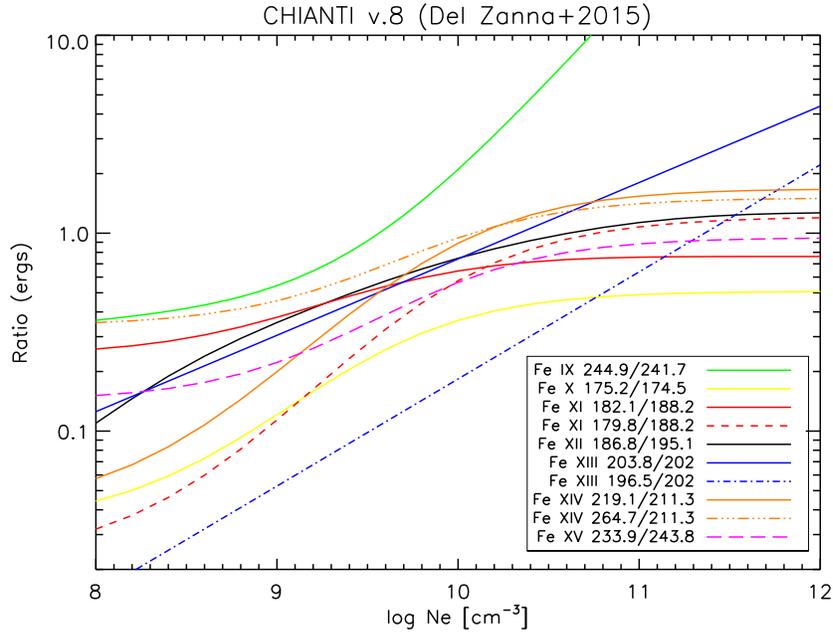}}
  \caption{The main diagnostic ratios in the EUV for the iron coronal ions.
Many of them are available to the Hinode EIS.
The atomic data are from CHIANTI v.8 \citep{delzanna_chianti_v8}.
}
 \label{fig:ne_ratios_eis}
\end{figure}

We emphasize \ion{Fe}{xii} and \ion{Fe}{xiii} below because the density diagnostics involve strong lines,
many of which are routinely observed by Hinode EIS.
Some of the main  diagnostic ratios (in the EUV), many of 
which are available to the Hinode EIS, are shown in Fig.~\ref{fig:ne_ratios_eis}.

\begin{table}[!htbp]
\caption[Some of the common density diagnostics of coronal iron ions.]
{Some of the common density diagnostics of coronal iron ions. 
(bl) indicates a blend, while (sbl) indicates a self-blend of more than one transition from the same ion.
The column R indicates which combination of lines is a good density diagnostic: a/b or b/a.}
\label{tab:dens_pairs}
\centering
\begin{tabular}{llllllll}
\toprule
Ion       & Transition & $\lambda$  (\AA) & R & Observed & log $T_{\rm e}$  & log N$_{\rm e}$  \\
\midrule

\ion{Fe}{ix} & 3s$^2$ 3p$^6$ $^1$S$_{0}$ -- 3s$^2$ 3p$^5$ 3d $^3$P$_{2}$  & 241.739 & a & Skylab &&\\
      & 3s$^2$ 3p$^6$ $^1$S$_{0}$ -- 3s$^2$ 3p$^5$ 3d $^3$P$_{1}$  & 244.909 & b &
 Skylab & 6.0  &  9--12 \\

\\

\ion{Fe}{x}       & 3s$^2$ 3p$^5$ $^2$P$_{1/2}$ - 3s$^2$ 3p$^4$  3d $^2$D$_{3/2}$ & 175.27  & a &  CDS, EIS &&  9.0 -- 11.0 \\ 
  & 3s$^2$ 3p$^5$ $^2$P$_{3/2}$ - 3s$^2$ 3p$^4$   3d $^2$D$_{5/2}$ &  174.534   & b & CDS, EIS  & 6.05 & \\


\\

\ion{Fe}{xi} &  3s$^2$ 3p$^4$ $^1$D$_{2}$ - 3s$^2$ 3p$^3$   3d $^1$F$_{3}$ & 179.764 & a & CDS, EIS &&  9.0 -- 11.0 \\
      &  3s$^2$ 3p$^4$ $^3$P$_{1}$ - 3s$^2$ 3p$^3$   3d $^3$D$_{2}$ & 182.17 & a & CDS, EIS & &  9.0 -- 11.0 \\
      &  3s$^2$ 3p$^4$ $^3$P$_{0}$ - 3s$^2$ 3p$^3$ 3d $^3$D$_{1}$ & 181.13  & a & CDS, EIS & &  9.0 -- 11.0 \\
      &  3s$^2$ 3p$^4$ $^1$D$_{2}$ - 3s$^2$ 3p$^3$   3d $^1$D$_{2}$  & 184.79 & a &  CDS, EIS & &  9.0 -- 11.0 \\

      &  3s$^2$ 3p$^4$ $^3$P$_{2}$ - 3s$^2$ 3p$^3$   3d $^3$P$_{2}$ & 188.216 & b & CDS, EIS &   6.1 &  \\


\\

\ion{Fe}{xii}  &   3s$^2$ 3p$^3$ $^2$D$_{5/2}$ - 3s$^2$ 3p$^2$   3d $^2$F$_{7/2,5/2}$  & 186.8 (sbl) & a & CDS, EIS && 8.5--12 \\

       &  3s$^2$ 3p$^3$ $^2$D$_{5/2}$ - 3s$^2$ 3p$^2$   3d $^2$D$_{5/2}$ & 196.65 (bl) & a & CDS, EIS  & & 8.5--12 \\
               & 3s$^2$ 3p$^3$ $^4$S$_{3/2}$ - 3s$^2$ 3p$^2$   3d $^4$P$_{5/2}$ &  195.12 (sbl) & b & CDS, EIS  & 6.16 & \\  %

\\
         & 3s$^2$ 3p$^3$ $^2$D$_{5/2}$ - 3s 3p$^4$ $^2$D$_{5/2}$ & 338.26 & a & CDS  & &  7.0 -- 12.0 \\
         & 3s$^2$ 3p$^3$ $^4$S$_{3/2}$ - 3s 3p$^4$ $^4$P$_{5/2}$ & 364.47 & b &  CDS  & 6.16 &  \\

\\

\ion{Fe}{xiii}    & 3s$^2$ 3p$^2$ $^1$D$_{2}$ - 3s$^2$ 3p 3d $^1$F$_{3}$ & 196.52 & a & CDS, EIS &&  9.0 -- 11.0 \\ 
          & 3s$^2$ 3p$^2$ $^3$P$_{1}$ - 3s$^2$ 3p 3d $^3$D$_{2}$  & 200.02 & a &  CDS, EIS & &  9.0 -- 11.0 \\
          & 3s$^2$ 3p$^2$ $^3$P$_{2}$ - 3s$^2$ 3p 3d $^3$D$_{3,2}$  & 203.8 (sbl) & a &  CDS, EIS & &  9.0 -- 11.0 \\

          & 3s$^2$ 3p$^2$ $^3$P$_{0}$ - 3s$^2$ 3p 3d $^3$P$_{1}$ & 202.04 & b &  CDS, EIS &  6.2 &  \\

\\


          & 3s$^2$ 3p$^2$ $^3$P$_{1}$ - 3s 3p$^3$ $^3$D$_{2}$ & 359.64 (sbl)  & a &  & & 8.5 -- 10.5 \\
          & 3s$^2$ 3p$^2$ $^3$P$_{0}$ - 3s 3p$^3$ $^3$D$_{1}$ &   348.18 & b & CDS  & 6.2 &   \\



\bottomrule
\end{tabular}
\end{table}

\smallskip\smallskip\noindent
{\bf \ion{Fe}{xiii} }
\smallskip

\noindent
\ion{Fe}{xiii}  (1.5 MK) also has many  excellent  density diagnostics
in the EUV, as recently  reviewed e.g.  by 
\cite{young_etal:2009,delzanna:11_fe_13,delzanna:12_atlas}.
At the Hinode EIS wavelengths, any ratio involving the main decay to the ground state (at 202.044~\AA)
and any permitted line decaying to an excited state is 
density-dependent. Depending  on which excited state is involved,
the ratio is sensitive to a different density range.

One of the strongest lines is the 203.8~\AA, however this line is actually 
a self-blend of two strong \ion{Fe}{xiii} lines, plus other weaker ones
tentatively identified by \cite{delzanna:11_fe_13},  and a relatively strong \ion{Fe}{xii} line.
Other good options are the 196.52 or the 200.02~\AA\ lines.
We recall that  Fig.~\ref{fig:fe_13_dens} shows the main populating 
processes for these transitions.

\begin{figure}[htbp]
\centerline{\includegraphics[width=0.9\textwidth]{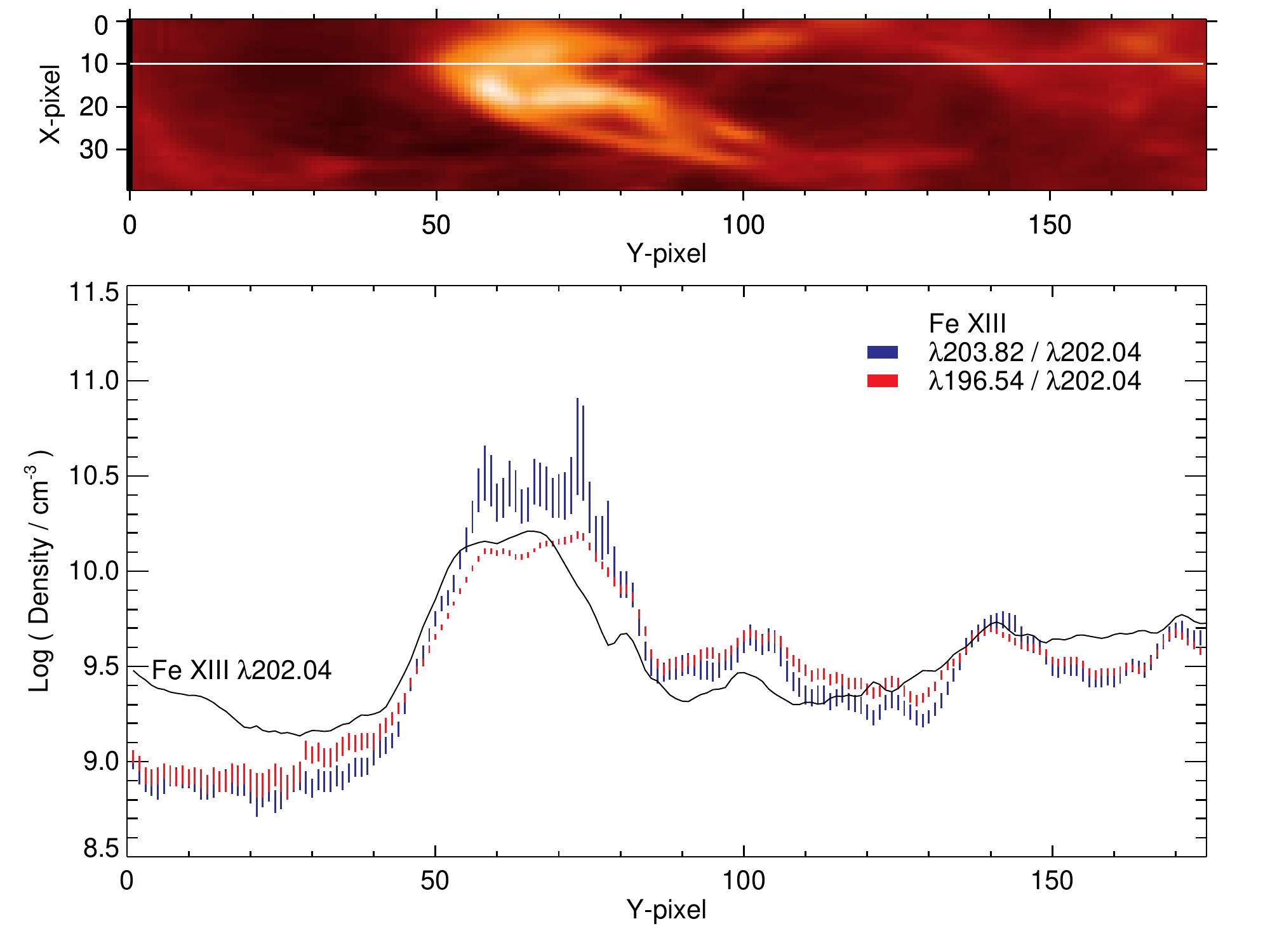}}
 \caption{The upper panel shows an image of the intensity 
of the  \ion{Fe}{xiii} 202.02~\AA\ line across some active region loops.
 In the lower panel the red and blue lines show densities derived from two  \ion{Fe}{xiii} 
ratios. The black line shows the variation in intensity of the  the  \ion{Fe}{xiii} 202.02~\AA\ line
(the plotted quantity is log (10$^5$ $\times$ I)). Figure  from  \cite{young_etal:2009}. }
  \label{fig:young_etal:2009}
\end{figure}

An example of the high accuracy achieved on both the observational (Hinode EIS)
and atomic physics (CHIANTI) sides is given in Figure~\ref{fig:young_etal:2009}.
Excellent agreement in the densities obtained from two different line ratios
from \ion{Fe}{xiii}  is obtained with a cut across some active region loops.

At longer wavelengths, observed by e.g. SOHO/CDS \citep[see, e.g.][]{mason_etal:1997} and 
the SERTS rocket flights \citep[see, e.g.][]{young_etal:98,landi:2002},  other good density diagnostics 
are the ratios involving the decay to the ground state at 
348.18~\AA\ and any decay to an excited level 
(e.g. the 359.64~\AA\ and possibly the 311.55, 312.87, 320.8~\AA\ lines,
although the last three can  be blended).

The accuracy of the atomic data in CHIANTI v.8 to measure densities 
in EBIT spectra of \ion{Fe}{xiii} has recently been confirmed by 
\cite{weller_etal:2018}.

\smallskip\smallskip\noindent
{\bf \ion{Fe}{xii}}
\smallskip

\begin{figure}[htbp]
\centerline{\includegraphics[width=0.9\textwidth]{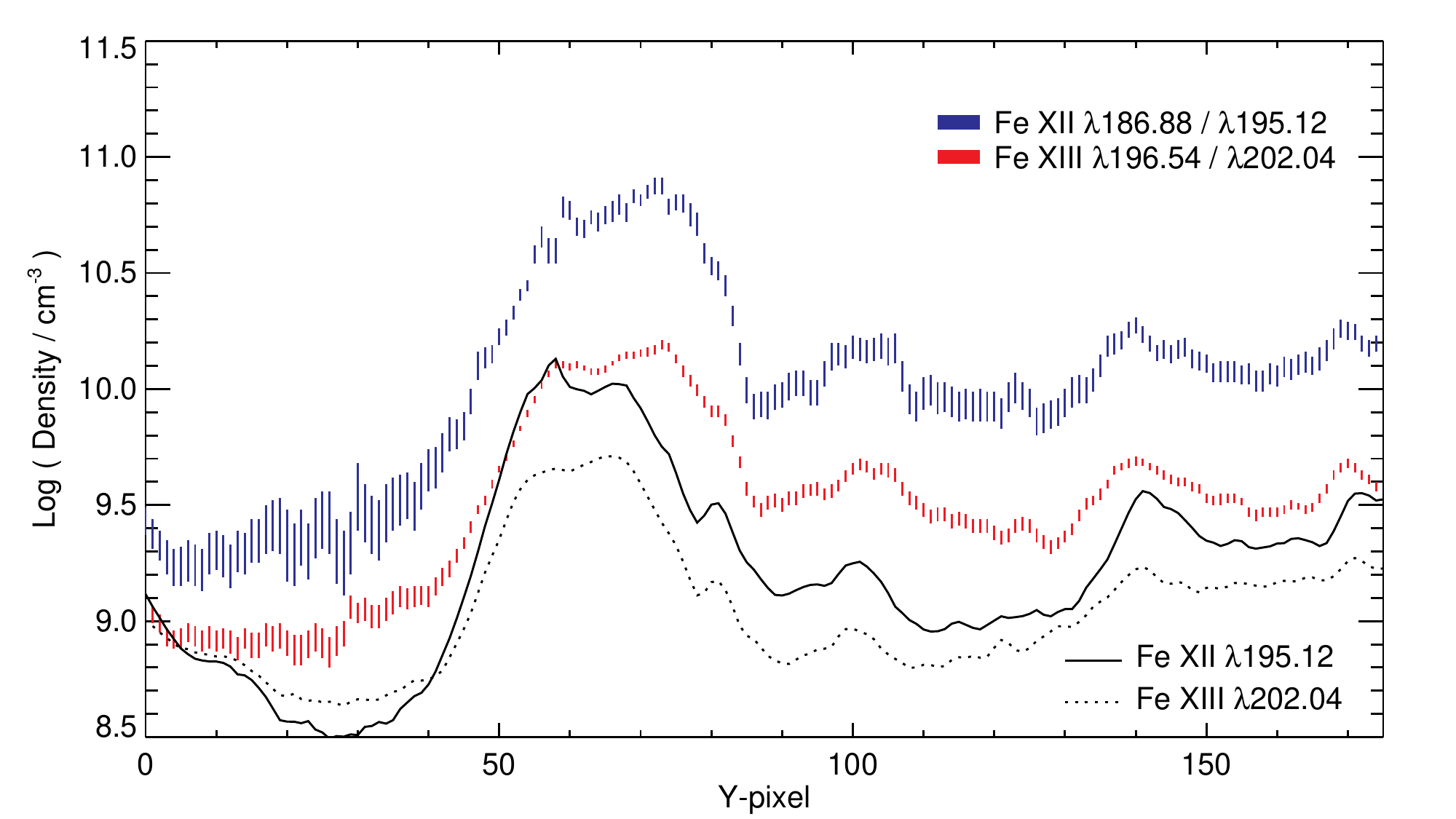}}
 \caption{Same as  Figure~\ref{fig:young_etal:2009},
but with the densities obtained from \ion{Fe}{xiii} and \ion{Fe}{xii} lines.}
  \label{fig:young_etal:2009b}
\end{figure}

\noindent
\ion{Fe}{xii} (1.5 MK) also has many  excellent  density diagnostics
in the EUV and UV \citep[see, e.g.][]{delzanna_mason:05_fe_12,delzanna:12_atlas}, 
although the complexity of this ion is such that 
only recent large-scale scattering calculations \citep{delzanna_etal:12_fe_12} 
seem to have resolved previous long-standing discrepancies.
As discussed by e.g.  \cite{young_etal:2009}, 
 some line ratios were providing too high densities, by about a factor of three,
compared to those obtained from \ion{Fe}{xiii}  lines, as shown in 
Figure~\ref{fig:young_etal:2009b}.

\begin{figure}[htb]
 \centerline{\includegraphics[width=0.5\textwidth,angle=90]{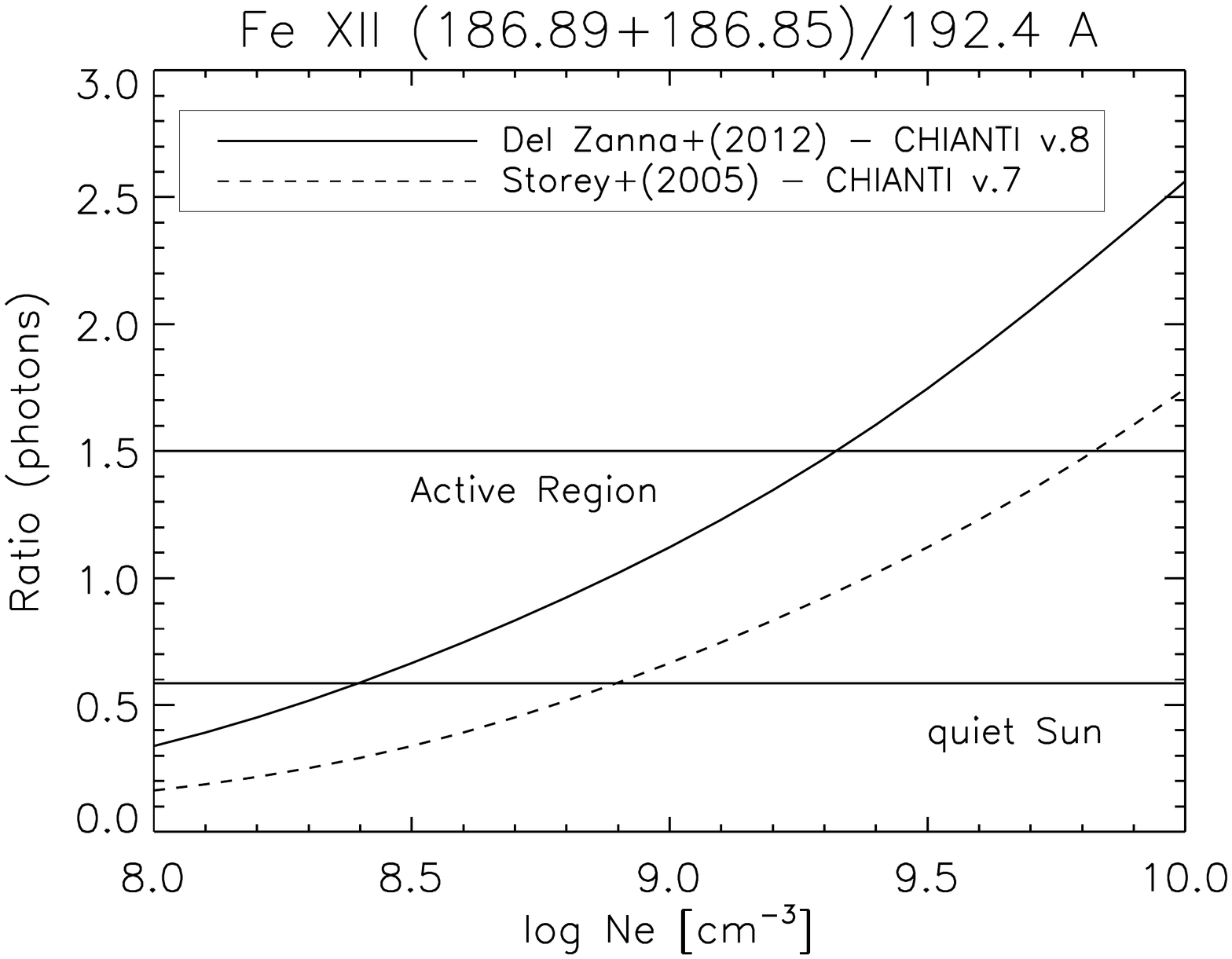}}
 \centerline{\includegraphics[width=0.6\textwidth,angle=0]{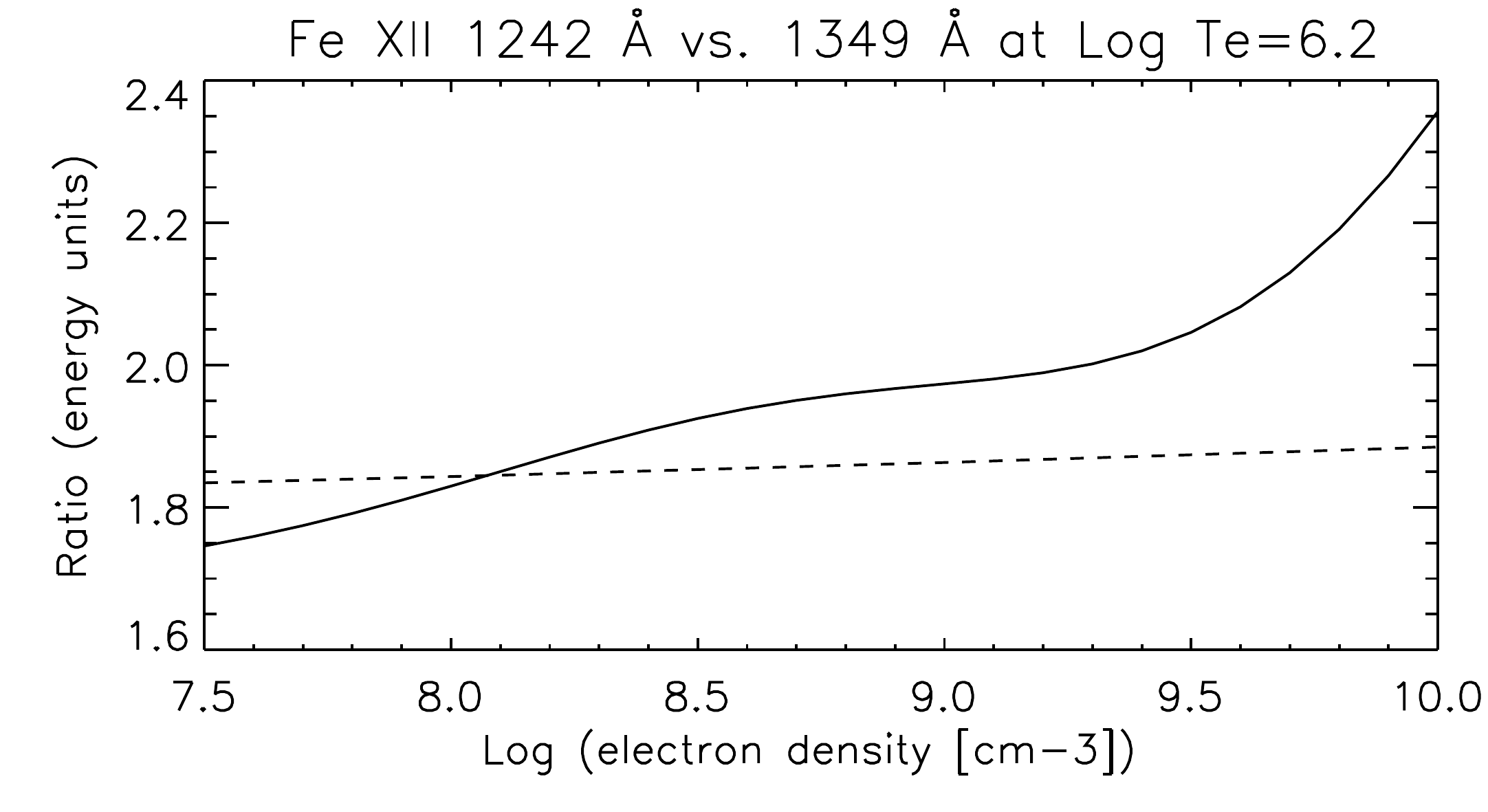}}
  \caption{Top: a comaprison of different calculations for
    \ion{Fe}{xii} - one of the main density diagnostic ratio in the EUV. Bottom:
the ratio of the UV forbidden lines is density-sensitive with the latest atomic data.
}
 \label{fig:fe_12_ratios}
\end{figure}

A large-scale calculation for  \ion{Fe}{xii} \citep{delzanna_etal:12_fe_12}
resulted in significantly lower densities obtained from this ion 
(about 0.5 dex), apparently resolving the discrepancy with the 
\ion{Fe}{xiii}  results. This occurred because the 
 population of the lower levels is mainly driven by cascading, so even 
small changes in the excitation rates to higher levels have a large 
cumulative effect.  \cite{delzanna_etal:12_fe_12} found increased populations
of the ground configuration of about 30\%. 
As a result, intensities of the forbidden lines are also affected, as shown in  Fig.~\ref{fig:fe_12_ratios}.

It should be noted  that even larger discrepancies between observed and predicted intensities 
were present until the scattering calculations of  \cite{storey_etal:04}.
\cite{delzanna_mason:05_fe_12} used these data to identify several new
lines and point out the best diagnostics for this ion, 
in particular the  ratio of the two self-blends at 186.88  to  195.12~\AA,
identified as such by \cite{binello_etal:2001}, and later confirmed 
using laboratory plates (where the lines were actually resolved) by 
\cite{delzanna_mason:05_fe_12}.
The 186.88  and  195.12~\AA\ lines  are routinely observed by Hinode EIS,
 as discussed e.g. in \cite{young_etal:2009}.

At high densities, the 196.65~\AA\ is to be preferred to the 186.88~\AA\ line, 
although it appears to be slightly  blended in on-disk observations.
Several other  ratios are useful, although many lines are blended.
At longer wavelengths, the ratio of the  338.3~\AA\ with e.g.
the 364.5~\AA\ line is also an excellent diagnostic and was observed by the 
SOHO CDS and SERTS, as discussed e.g. in \cite{young_etal:98,delzanna_thesis99}.
In the UV, a few forbidden lines can also be used to 
obtain densities.
Also, the ratio of the 1242 and 1349~\AA\ lines is seen to be 
density-sensitive with the latest atomic data.

\smallskip\smallskip\noindent
{\bf \ion{Fe}{xi}}
\smallskip

\noindent
This ion has been the most difficult of all coronal ions to provide
reliable atomic data for, with a history of several discrepancies and incorrect line identifications for 
some of the strongest EUV lines.
Now, with reliable atomic data and identifications, 
it is possible to use several line ratios from this ion,
as discussed in \cite{delzanna:10_fe_11,delzanna:12_atlas}.
The best ratios, with the strongest lines, fall at the 
 lower Hinode EIS wavelengths. They are obtained from one of the decays 
to the ground state at  180.4 (blended with Fe X), 188.2,188.3~\AA\ 
and from one of several decays to an excited level (179.76, 184.79,  181.13, 182.17~\AA).
Most of the other lines are blended or weaker.

\smallskip\smallskip\noindent
{\bf \ion{Fe}{x}}
\smallskip

\noindent
A review of the density  diagnostics for this ion was provided by \cite{delzanna_etal:04_fe_10}.
Within the EUV, only the ratio of the weak 175.27~\AA\ line with any of the strong
decays to the ground state (e.g. 174.53~\AA) is density-sensitive. 
Several forbidden lines, not listed here, are also useful density diagnostics in other wavelength regions.
The latest scattering calculations \citep{delzanna_etal:12_fe_10} for this ion 
changed the population of the lowest levels significantly, removing previous discrepancies 
 between observed and predicted line intensities for the forbidden lines \citep{delzanna_etal_2014_fe_9}.

\smallskip\smallskip\noindent
{\bf \ion{Fe}{ix}}
\smallskip

\noindent
One of the best EUV density diagnostic  for the 1 MK corona is the 
\ion{Fe}{ix} 241.739~\AA\   / 244.909~\AA\ ratio, as discussed  in  \cite{storey_etal:02}.
\cite{feldman_etal:1978_fe_9} used Skylab observations and older 
atomic data to point out that in principle the 
\ion{Fe}{ix} 241/244~\AA\ should be an excellent diagnostic.
However, they obtained values that were largely at odds with those obtained from other ions. 
This led \cite{feldman:1992_fe_9} to suggest that non-equilibrium effects were at play. 
However, it was shown by \cite{storey_etal:02} that the main problem was in the 
earlier scattering calculations, as  has often been the case for many of these complex coronal ions.
A weak density sensitivity is also present in other EUV lines observed by 
Hinode EIS, as discussed by \cite{young:09}.


%

\section{Measurements of Electron Densities } 
\label{sec:ne_obs}

In what follows, we provide several 
examples of  measurements of electron densities for different 
solar regions, to show which diagnostics have been applied.
Measurements of densities in the off-limb (outer) corona are fundamental 
for our understanding of coronal heating processes and as a boundary 
for solar wind modelling. 
Measurements of densities during flares are also extremely important as 
they provide information on the timescales of  ionisation/recombination 
processes.

\subsection{Densities in coronal holes}
\label{sec:ch_ne}

Following the first XUV  observations, coronal holes were defined 
 as those areas with much lower (by factors up to 10) intensities in  coronal lines.
They should not be confused, however, with \emph{filament channels},
where the decreased in the intensity is mostly due to absorption by 
neutral hydrogen and helium. 
Also, they should not be confused with \emph{dark halos} around active regions,
which have been described by \citep{andretta_delzanna:2014}.
Studies based on Skylab and HRTS data 
have shown that  transition region lines have almost the same 
intensity distribution inside and outside coronal holes,
and also show the same pattern of supergranular network.

Various features occur in coronal holes: bright points,
macrospicules, jets, etc. 
However, during solar minimum, (especially the 1995--1996 one)
the main feature was that of coronal hole plumes.
They appear as ray-like, extended (up to few \rsun) structures in white-light
eclipse coronagraph images of the polar regions.
Their bases are visible in 1 MK and lower temperature lines in the EUV.
It is important to study  plumes because they 
 are the only stable observable structures which 
trace the open field lines within coronal holes,
thought to be the source regions of the fast solar wind.

At large distances from the limb, electron densities can be
measured from the white-light images, following 
the \cite{vandehulst50}  method (see e.g. \cite{guhathakurta_etal:1993}).

Direct measurements from ratios of coronal lines were sparse 
until SOHO. 
There were some measurements at lower temperatures, but these
were somewhat unreliable, mostly because of the low spectral resolution
or low signal of early instruments 
(see, e.g. \citealt{munro_etal:1971,doschek_etal:1978,vernazza_mason:1978}).
Off-limb densities from S X were obtained from Skylab 
\cite{feldman_etal:1978}  but did not have any spatial 
resolution along the slit.

\subsubsection{SOHO}

There is quite an extended literature on the topic of coronal hole
densities from SOHO, so we just mention 
here some of the earlier studies, and point out that 
\cite{wilhelm_etal:2011} has recently provided a review 
of coronal hole plumes and lane measurements obtained from SOHO.

\begin{figure}[htbp]
\centerline{\includegraphics[width=0.8\textwidth,angle=0]{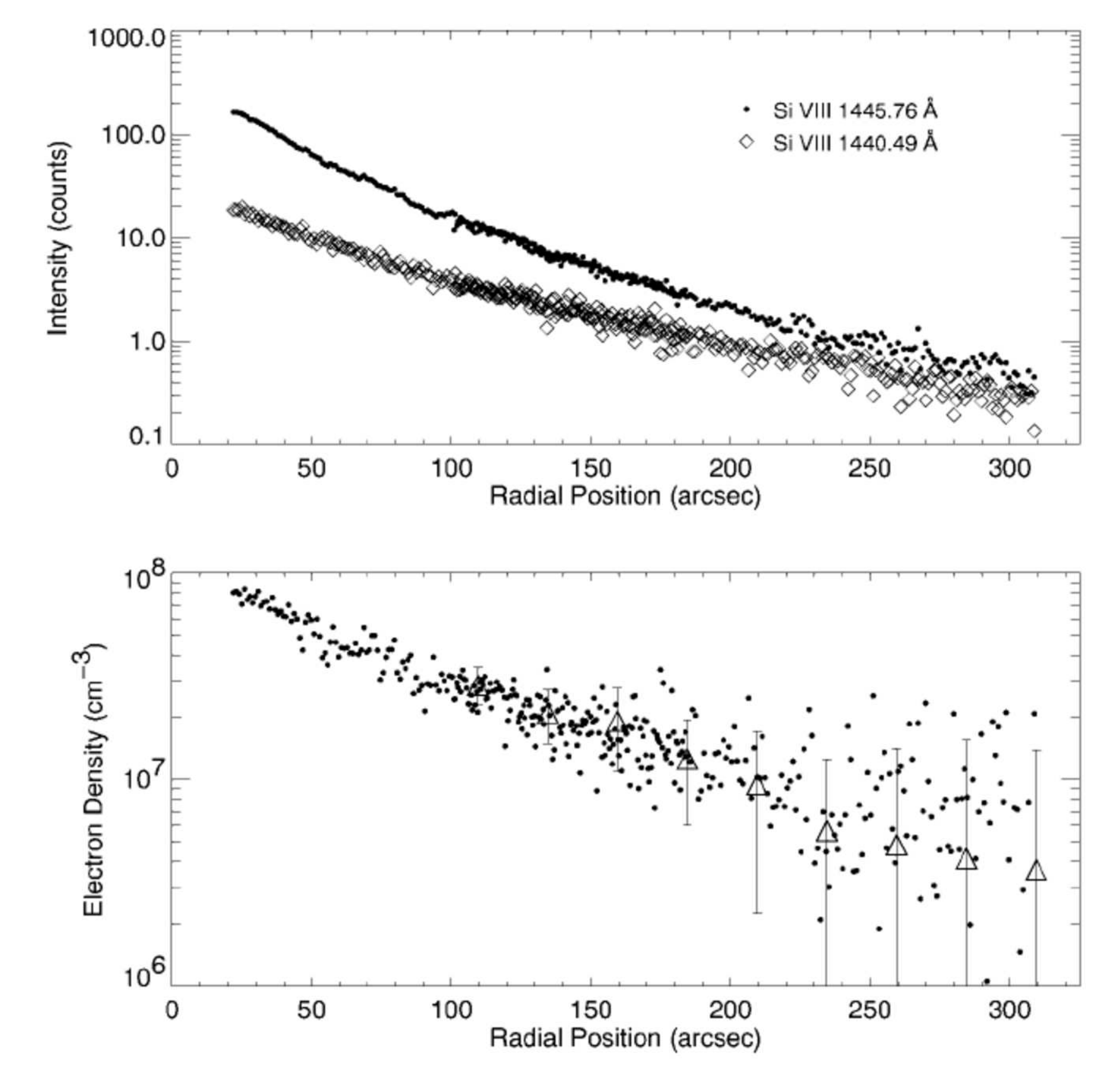}} 
  \caption{Top: intensities in two Si VIII lines observed off the limb with SoHO SUMER.
Bottom: electron densities obtained from the ratio of the two lines
 \citep{doschek_etal:1997}.
}
  \label{fig:doschek_etal:1997}
\end{figure}

\cite{doschek_etal:1997}   applied a ratio technique
to coronal lines (Si VIII and S X)  in SUMER off-limb observations (see Figure~\ref{fig:doschek_etal:1997}).
These are important observations as they are the first to provide
spatially-resolved densities off-limb at high resolution.
\cite{doschek_etal:1997} found the coronal hole densities to be systematically lower
 than those in the quiet sun,  by a factor of about 2. 
At the base of the corona, densities are about 10$\up{8}$ cm$\up{-3}$.
 Similar results were obtained by others also using SUMER, see e.g. 
\cite{laming_etal:1997, banerjee_etal:1998,warren:1999}. 
Densities from plumes and lanes were obtained from 
\cite{wilhelm_etal:98, wilhelm:2006} among others. 
Densities decreased from  10$\up{8}$ cm$\up{-3}$ to 
 10$\up{7}$ cm$\up{-3}$ around 300\arcsec\ above the limb.
Densities in lanes were normally found to be lower than those in plumes.
 \cite{doschek_etal:1998} used the  O V 759.441 and 761.128~\AA\  lines
observed at the limb by SUMER to derive  densities inside coronal
holes of $N\lo{e} =  2-3 \times 10\up{9}$ cm$\up{-3}$,
a factor of two lower than in the quiet Sun.

SoHO CDS also provided on-disk and off-limb measurements of densities
for coronal holes.
CDS was able to resolve the weak O IV 625~\AA\ from the nearby strong
Mg X line, and many density measurements have been obtained using this line,
as well as other lines formed at higher temperatures.
Several coronal ions were observed by CDS which allowed  on-disk and off-limb measurements of densities.
A  complete dataset of coronal hole 
observations was presented in \cite{delzanna_thesis99}.
The Whole Sun Month (WSM) international campaign was set up to obtain 
coordinated observations from all the SOHO instruments during August 1996.
Two workshops followed. 
As part of the Whole Sun Month  measurements of 
densities from the Si IX 349/341~\AA\  ratio in the polar regions
were obtained with  CDS.
Results are discussed in  \cite{delzanna_thesis99}; \cite{fludra99b}.
Further polar observations in 1997 and 1998 were analysed 
\citep{delzanna_thesis99}, and  similar values 
to those of  August 1996 were found (see also \citep{fludra_etal:2002}). 
Such observations were important as they showed that solar cycle 
variations did not significantly affect the density of the 1 MK coronal plasma.
The CDS coronal hole densities were found to be about a 
factor of two lower than in the quiet Sun regions, as 
in the case of the SUMER measurements.
Similar results have been obtained by others, see e.g. \cite{gallagher_etal:1999}.

During the WSM, a large equatorial coronal hole crossed the 
meridian. It was named the Elephant's Trunk  \citep{delzanna_jgr99a}.
Supergranular network cell centres and boundaries were selected,
and averaged densities were obtained from Si IX and O IV line ratios.
The  O IV ratio showed 
 a tendency to have slightly higher densities in the cell centres
($N_{\rm e} \simeq$  0.6--1.1 $\times 10^{10} $ cm$^{-3}$) than in 
the boundaries ($N_{\rm e} \simeq$  0.5--0.7 $\times 10^{10} $ cm$^{-3}$).
The Si IX ratio produced  $N_{\rm e} \simeq $ 1--2 $\times 10^{8} $ cm$^{-3}$.
 The fact that coronal holes are virtually indistinguishable from 
QS areas in TR lines suggests that it is the spectroscopic filling 
factor that changes, assuming the density measurements are correct.

A plume was identified in the Elephant's Trunk  \citep{delzanna_jgr99a}.
The Si IX ratio produced  $N_{\rm e} \simeq  2 \times 10^{8} $ cm$^{-3}$
at its base, i.e. a density comparable to those of the QS.
Further CDS observations of polar coronal holes showed similar values
\citep{delzanna_thesis99, delzanna_etal:03}.
Higher densities were found in a bright plume 
observed with CDS  \citep{young_etal:1999}.
\cite{antonucci_etal:2004}
obtained densities and outflowing velocities in a polar coronal hole
using the  O VI 1032, 1037~\AA\  doublet observed by SOHO UVCS.

\subsection{Densities in the quiet Sun}
\label{sec:qs_ne}

Direct measurements from line ratios in the quiet Sun were also sparse 
until SOHO, and large discrepancies between results 
obtained from different lines or instruments were  common. 
\cite{dupree_etal:1976}  analysed Skylab HCO data and found  significant variations
in the C III 1176/977 \AA\ ratio, for different solar regions,  with  
an average value for the quiet sun of 0.29, yielding a density  $N_e = 4.6 \times 10^9 $ cm$^{-3}$.
A  trend for  higher densities  in the cell centres  was found, although in the 
quiet Sun regions the uncertainty was large.
\cite{vernazza_reeves:1978}  derived densities from the same C III   1176/977 \AA\ lines observed by 
the Skylab HCO instrument, but  found  $N_e \simeq  10^{10}$ cm$^{-3}$ from the
  averaged spectra.
Moreover, 
differences were  found between the cell centres and network regions, but this time 
with the network having larger (by about a factor of two) densities than the cell centres, 
in contradiction to the previous results by \cite{dupree_etal:1976}.
As we mentioned previously, this ratio is not ideal, because it is
also temperature-dependent
and the strong resonance line at 977~\AA\ can have optical depth effects.

Various authors used the Skylab NRL S082-B to measure densities. However, only 
observations close to the limb could be used.
\cite{doschek_etal:1978,doschek:1997}  presented densities for coronal holes and the  quiet sun 
close to  the limb, using C III,  O III, Si III,  and Si IV lines, finding an averaged 
value $N_e \simeq  1 \times 10^{10} $ cm$^{-3}$.
Similar results were obtained by \cite{cook_nicholas:1979} using C III, Si III and Si IV lines.

The atomic data and Skylab, HRTS, and SOHO SUMER observations 
of the \ion{Si}{iii} lines have recently been reviewed in \cite{delzanna_etal:2015_si_3}.
Reliable measurements involve only the 1301~\AA\ and nearby lines.
Typical quiet Sun on disk or at the limb values are 1--3 10$^{10}$ cm$^{-3}$.

\cite{doschek:1997} obtained 
 densities of about 1--2 $\times$ 10$^{9}$ cm$^{-3}$ from the N III 1749/1753 ratio 
using  QS and CH Skylab S083B spectra 4'' above the limb. They adopted  
 the atomic data used by  \cite{brage_etal:1995}.
For on-disk spectra, the lines are blended with chromospheric lines and are not usable.

There are several observations of the O IV lines around 1400~\AA,
however as we mentioned the variations of the ratios are small, and
accurate observations and atomic data are required for reliable measurements.
Pre-SOHO measurements were reviewed by \cite{brage_etal:1996}.
A significant scatter in the results was present, however most 
QS observations yielded values around 10$^{10}$ cm$^{-3}$.


\cite{keenan_etal:1995_o_5}
obtained densities of  log $N_{\rm e}$ [cm$^{-3}$]=10.5 from Skylab S082B observations of 
the  O V 1371.29A/1218.35~\AA\  near the QS limb,
using new calculations for O V.
However, as \cite{doschek:1997} pointed out, this ratio 
has little density sensitivity for CH and QS densities, 
and is actually rather temperature-sensitive.
The uncertain Skylab calibration makes the results very uncertain.
\cite{keenan_etal:1995_o_5} also presented 
HRTS observations near disk centre,  indicating similar values.
\cite{keenan_etal:1995_o_5} found no differences 
between coronal holes and the quiet Sun.

\cite{keenan_etal:1994} presented 
densities obtained from the N IV 1718.6,1486.5~\AA\ line ratio measured near
the limb from the 
NRL S082B spectrograph on board Skylab.
They obtained large variations across the limb, suggesting that 
the 1718.6~\AA\ line is blended with a chromospheric line.

\cite{vernazza_mason:1978} obtained a QS density of 2 $\times$ 10$^{8}$ cm$^{-3}$ 
from the \ion{Si}{x} 347/356~\AA\  ratio.
\cite{feldman_etal:1978} measured rather high 
densities from the N-like  S X  forbidden lines at 1196.3 and 1213~\AA\
above the  solar limb with the NRL Skylab spectrograph,
about 10$^{9}$ cm$^{-3}$.
The problem with these lines is that they are intrinsically weak,
and blended with cool lines in on-disk spectra.

\subsubsection{SOHO}

SUMER provided many observations of the O IV multiplet at 1400~\AA.
The lines are intrinsically weak and vary little with density,
hence results are somewhat unreliable.
Moreover, the 1407.4~\AA\ is blended in SUMER spectra with the strong O III 
multiplet at 703.8~\AA\ in second order, and the 1408.8~\AA\ line
is blended with S IV. Most of the results have been obtained from the 
1399.8/1401.2~\AA\ ratio.

\cite{griffiths_etal:1999} and \cite{landi_etal:2000} obtained QS O IV densities 
of about 10$^{10}$ cm$^{-3}$, with a large uncertainty and scatter of values.  
No significant variation with the supergranular network was found.
In contrast, \cite{doschek_mariska:2001} found some evidence that
the average densities increase from 6 to 19 $\times$ 10$^{9}$ cm$^{-3}$ 
between quiet and brighter network regions (while the 1399.8/1401.2~\AA\ ratio
only changes on average  from 0.19 to 0.22).

SUMER was able to resolve the O V multiplet at 760~\AA.
\cite{doschek_etal:1998} obtained QS densities near disk centre
in the range 5--8 $\times$ 10$^{9}$ cm$^{-3}$.  
SUMER also observed the weak O V forbidden line  at 1213.9~\AA.
\cite{pinfield_etal:1998} used the ratio with the 
intercombination line at 1218.35~\AA\  to find
a density for the QS of 10$^{8.5}$ cm$^{-3}$, a value at odds
with most other measurements.

At the SUMER wavelengths, many diagnostic lines are blended with low-T lines 
 on-disk, but several off-limb observations have been run, up to 
about 1.3 solar radii.
Results from the Si VIII 1445.7/1440.5~\AA\ forbidden ratio
 have been published by 
several authors, e.g. \cite{laming_etal:1997,doschek_etal:1997,feldman_etal:99a}.
Typical QS densities near the limb are 2 $\times$ 10$^{8}$ cm$^{-3}$,
and decrease by about one order of magnitude at 1.3 \rsun.
\cite{doschek_etal:1997} also considered the ratio of the 
S X forbidden lines at 1213.0 and 1196.3~\AA, but concluded that 
typical QS densities are too close to the low-density limit 
for reliable measurements to be obtained.
\cite{laming_etal:1997} measured densities near the limb 
from several other line ratios, generally finding good agreement
with the Si VIII results.

With CDS it was possible to resolve the weak O IV 625~\AA\ from the nearby strong
Mg X line and density measurements have been obtained 
\citep{delzanna_bromage:1999}.
On-disk observations of the  O IV ratio showed 
 a tendency towards higher densities in the cell centres
($N_{\rm e} \simeq  1.2 \times 10^{10} $ cm$^{-3}$) than in 
the boundaries ($N_{\rm e} \simeq  0.5 \times 10^{10} $ cm$^{-3}$).
The Si IX ratios produced  $N_{\rm e} \simeq  4 \times 10^{8} $ cm$^{-3}$.
\cite{landi_landini:1998} found slightly higher values from the 
Si IX ratios in two QS areas, in the range 5--8 $\times 10^{8} $ cm$^{-3}$.
\cite{warren:2005} obtained somewhat lower values around 3 $\times 10^{8} $ cm$^{-3}$
from the Si IX and Si X ratios.
\cite{young:2005} also measured QS densities using the 
Si IX 349.9/345.1~\AA, finding a remarkable constant value around 
2.6 $\times 10^{8}$~cm$^{-3}$ during the 1996\,--\,1998 period.
A similar value was also obtained by \cite{brooks_warren:2006}.

\begin{figure}[htbp]
\centerline{\includegraphics[width=8cm,angle=90]{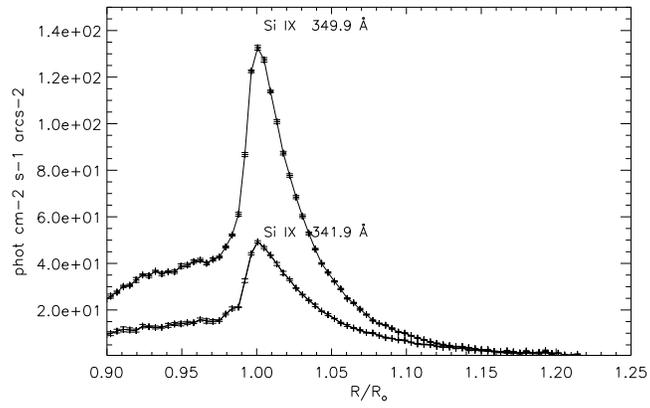}}
\centerline{\includegraphics[width=8cm,angle=90]{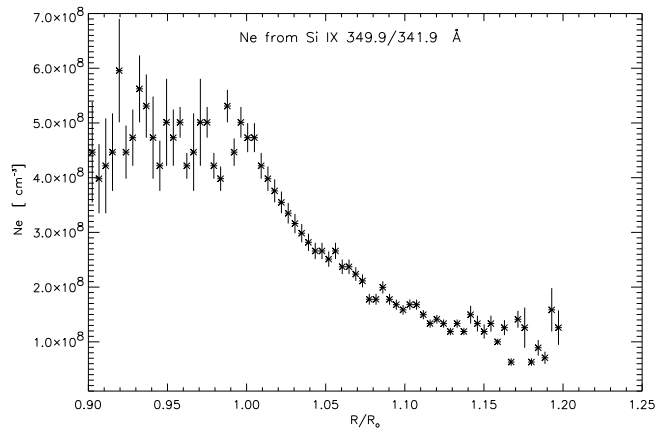}}
 \caption{Radiances in a quiet Sun off-limb area and relative 
electron densities  obtained from a 
SoHO CDS NIS  \ion{Si}{ix} ratio and CHIANTI v.2 \citep{delzanna_thesis99}.
}
  \label{fig:qs_ne_off_limb}
\end{figure}

As part of  Whole Sun Month campaign   CDS obtained measurements of 
densities from the Si~IX 349/341~\AA\ ratio in off-limb observations.
Results are described in \cite{delzanna_thesis99,fludra99b}.
The  density was found to decrease from 5 $\times$ 10$^8$ 
at the limb,  to 1 $\times$ 10$^8$ cm$^{-3}$ at 1.2 \rsun,
see Figure~\ref{fig:qs_ne_off_limb}.
 \cite{gibson99a} showed that  the densities determined from CDS 
and those determined from white-light observations and 
the \cite{vandehulst50}  inversion technique were very similar,
an important result.
Off-limb observations were regularly carried out  by CDS, but no 
significant differences to the earlier results were found 
\citep{delzanna_thesis99,fludra_etal:2002}.

\subsubsection{Densities from Hinode/EIS}

Hinode EIS has several density diagnostics for coronal 
temperatures. 
Measurements of the off-limb QS corona have produced 
results in broad agreement with the SOHO CDS ones
(see, e.g. \citealt{warren_brooks:2009}, \citealt{delzanna:12_atlas}).
On-disk measurements are also in good agreement with previous ones
(see, e.g. \citealt{brooks_warren:2009} using Fe XII).

One advantage of the Hinode EIS measurements is the fact that 
observations have been carried out continuously since 2006, while 
e.g. SoHO CDS measurements of the coronal diagnostic lines were 
significantly affected by degradation (in 1998) of the NIS. 
This has allowed us to study the variation 
of the coronal density over a solar cycle.
For example, 
\cite{kamio_mariska:12} used Hinode EIS synoptic observations 
of the quiet Sun and measured densities using the Si X 258.4/261.0~\AA\ line ratio. 
Values in the range 3--5  $\times 10^{8} $ cm$^{-3}$ were found.

\begin{figure}[htb]
\centerline{\includegraphics[width=7cm,angle=90]{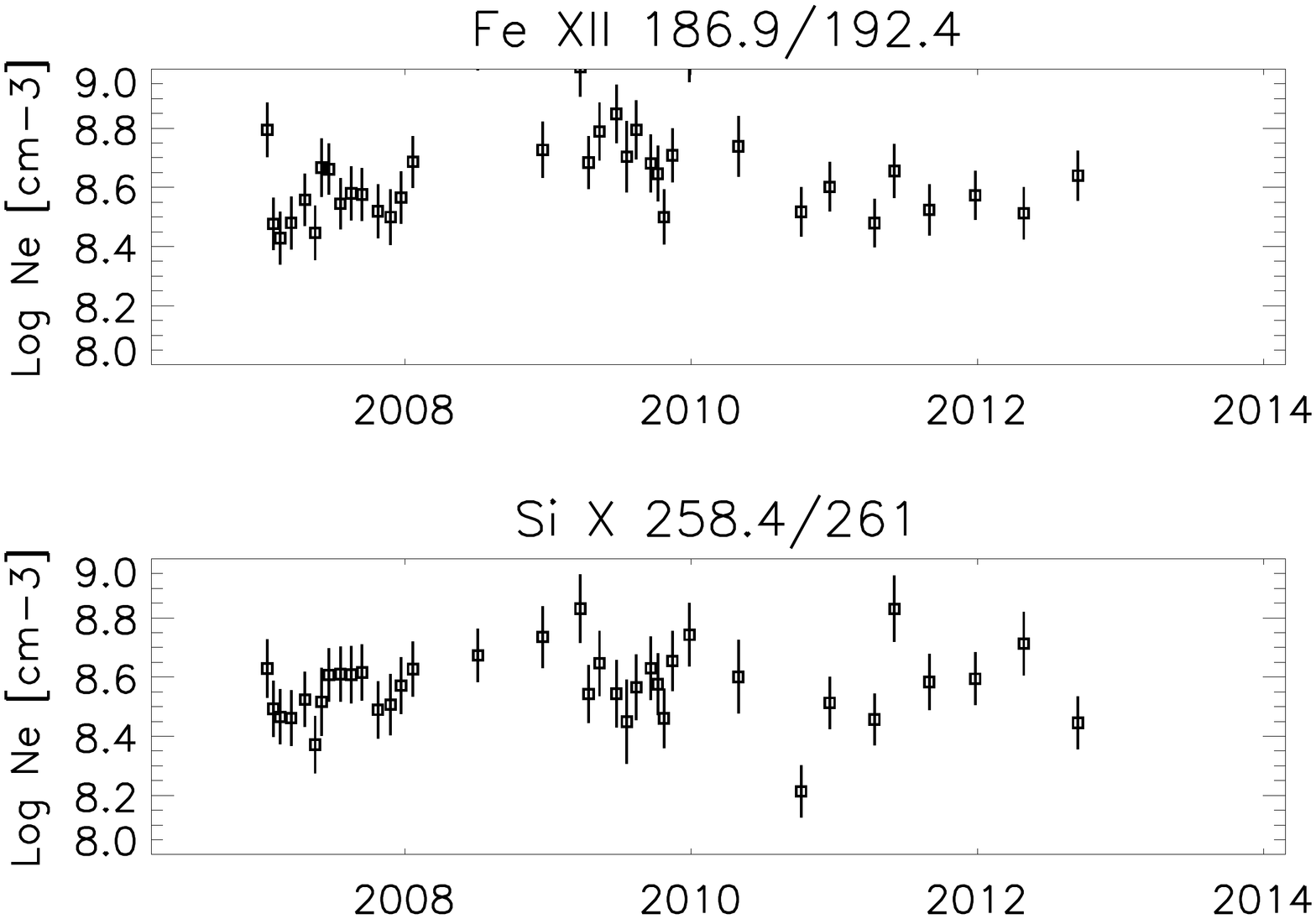}}
  \caption{Electron densities in quiet Sun areas obtained from 
Hinode EIS \ion{Si}{x}  and  \ion{Fe}{xii} lines overan eight year period
(adapted from \citealt{delzanna:13_eis_calib}).}
  \label{fig:qs_ne}
\end{figure}

We note, however, that for some line ratios the latest atomic data provide
 different values, especially for Fe XII. 
The above studies also used the EIS ground calibration which needed  revision. 
\cite{delzanna:13_eis_calib} presented a new EIS calibration 
and used \ion{Si}{x} and the new atomic data for \ion{Fe}{xii} \citep{delzanna_etal:12_fe_12}
to obtain  quiet Sun densities from carefully 
selected regions  between 2006 and 2013.
The densities were reasonably constant  around a value of  10$^{8.5}$ cm$^{-3}$ 
(see  Figure~\ref{fig:qs_ne}), 
i.e. around 3 $\times 10^{8} $ cm$^{-3}$, in close agreement with the 
Si IX results for the 1996--1998 period by \cite{young:2005}.

It is important to keep in mind, however, that the so-called 
quiet Sun areas as observed in lines formed above 1 MK become 
increasingly affected by the presence of the active regions 
during increased solar activity, as shown in 
\cite{delzanna_andretta:11,andretta_delzanna:2014}.

\subsection{Active regions and bright points}
\label{sec:ar_ne}

There is an extended literature on measurements of densities in 
active regions, although we would like to point out that the spatial 
resolution of previous and current spectrometers has not allowed 
a complete and detailed description of how densities vary along  fine structures
(e.g. warm loops) that are observed with the highest-resolution imagers.
A good review of the diagnostic possibilities in the EUV
offered by the Skylab NRL S082A slitless instrument for 
active regions is given by \cite{dere_mason:1981}.
For  general reviews on active regions 
see, e.g.  \cite{reale:2012_lr} and \cite{mason_tripathi:2008}.
 
\cite{kastner_etal:1974_limb} used OSO-7 off-limb observations 
of coronal lines 
with the Goddard spectroheliogram to measure  densities and temperatures.
They used the \ion{Fe}{xiv} lines and the atomic data 
available at the time to find densities ranging from 
1$\times$10$^{9}$ to 1$\times$10$^{8}$ cm$^{-3}$ above 
an active region.

Several results have been obtained from the Skylab instruments.
For example, \cite{feldman_doschek:1978}
used Skylab observations at the limb with the NRL slit spectrometer 
(1175--1940~\AA\ range) to obtain densities in an active region from 
Si III, C III, O IV, and O V that ranged from 2$\times$10$^{10}$
to 3$\times$10$^{11}$  cm$^{-3}$.

\begin{figure}[htb]
\centerline{\includegraphics[width=0.8\textwidth]{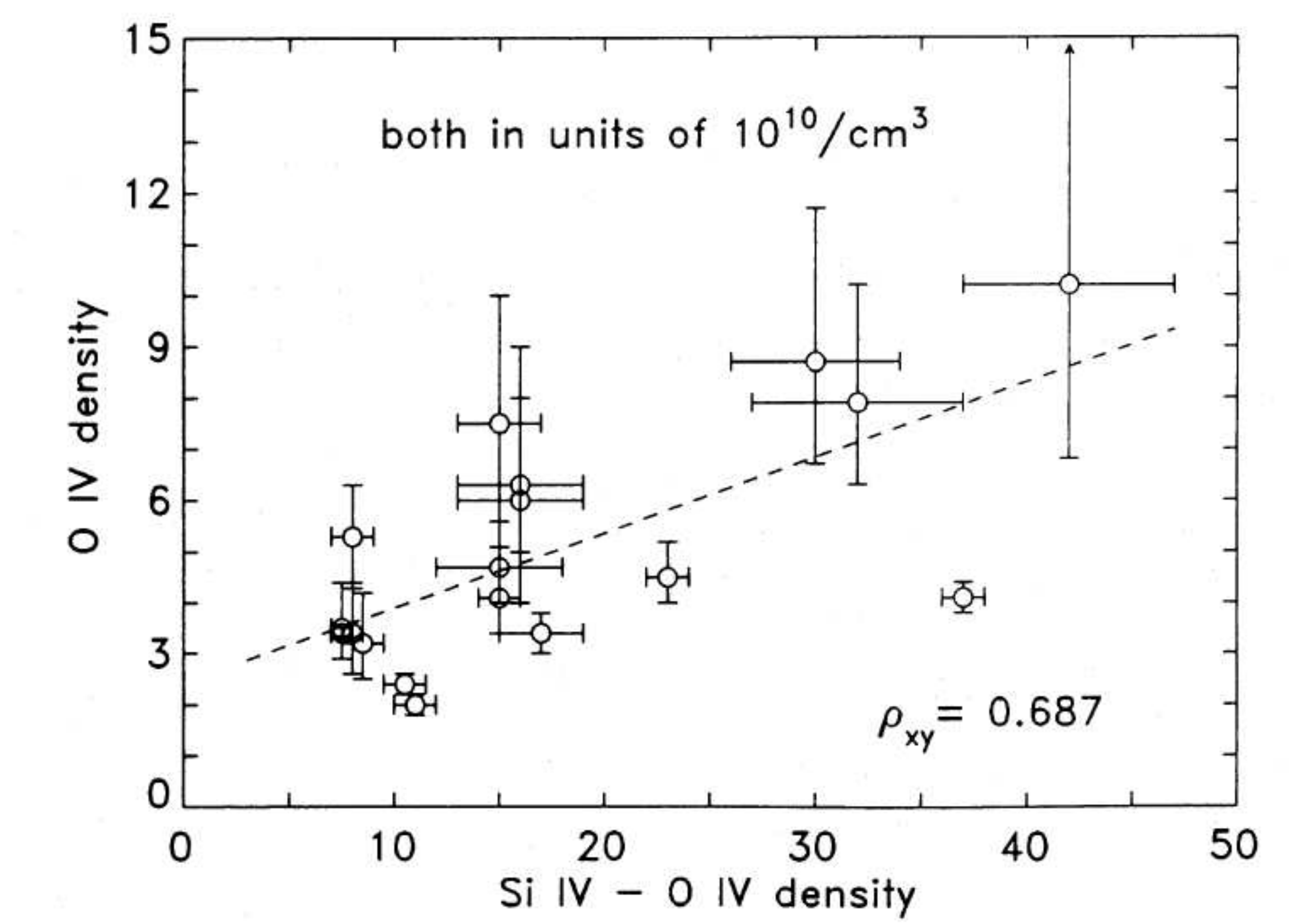}}
 \caption{Electron densities obtained from the SMM UVSP
observations of the O IV intercombination lines  
versus those obtained from the ratio of the O IV to 
the Si IV lines \citep{hayes_shine:1987}.
}
  \label{fig:hayes_shine:1987}
\end{figure}

There are several results from UV lines, measured from 
e.g. the HRTS and SMM UVSP instruments. 
For example, \cite{hayes_shine:1987} used the  SMM UVSP
observations of the O IV intercombination lines
to obtain densities in the  brightenings that occur frequently in 
active regions. Values around  5$\times$10$^{10}$  cm$^{-3}$ were obtained,
as shown in Figure~\ref{fig:hayes_shine:1987}.
There is in principle another way to estimate densities, that is 
from the ratio of the O IV intercombination lines with the 
Si IV allowed lines.
It is interesting to note, however,  that the densities obtained with this
method are found to be quite different than those obtained from the O IV ratios,
as shown in Figure~\ref{fig:hayes_shine:1987}.
This issue is still hotly debated in the literature regarding the interpretation
of similar active region  brightenings recently observed by IRIS 
\citep{peter_etal:2014,judge:2015}.

HRTS measurements of plage regions from the Si III lines 
provided values around 10$^{11}$--10$^{12}$  cm$^{-3}$
(see, e.g. \citealt{nicolas_etal:1979,delzanna_etal:2015_si_3}).

\begin{figure}[!htb]
 \centerline{\includegraphics[width=0.45\textwidth,angle=0]{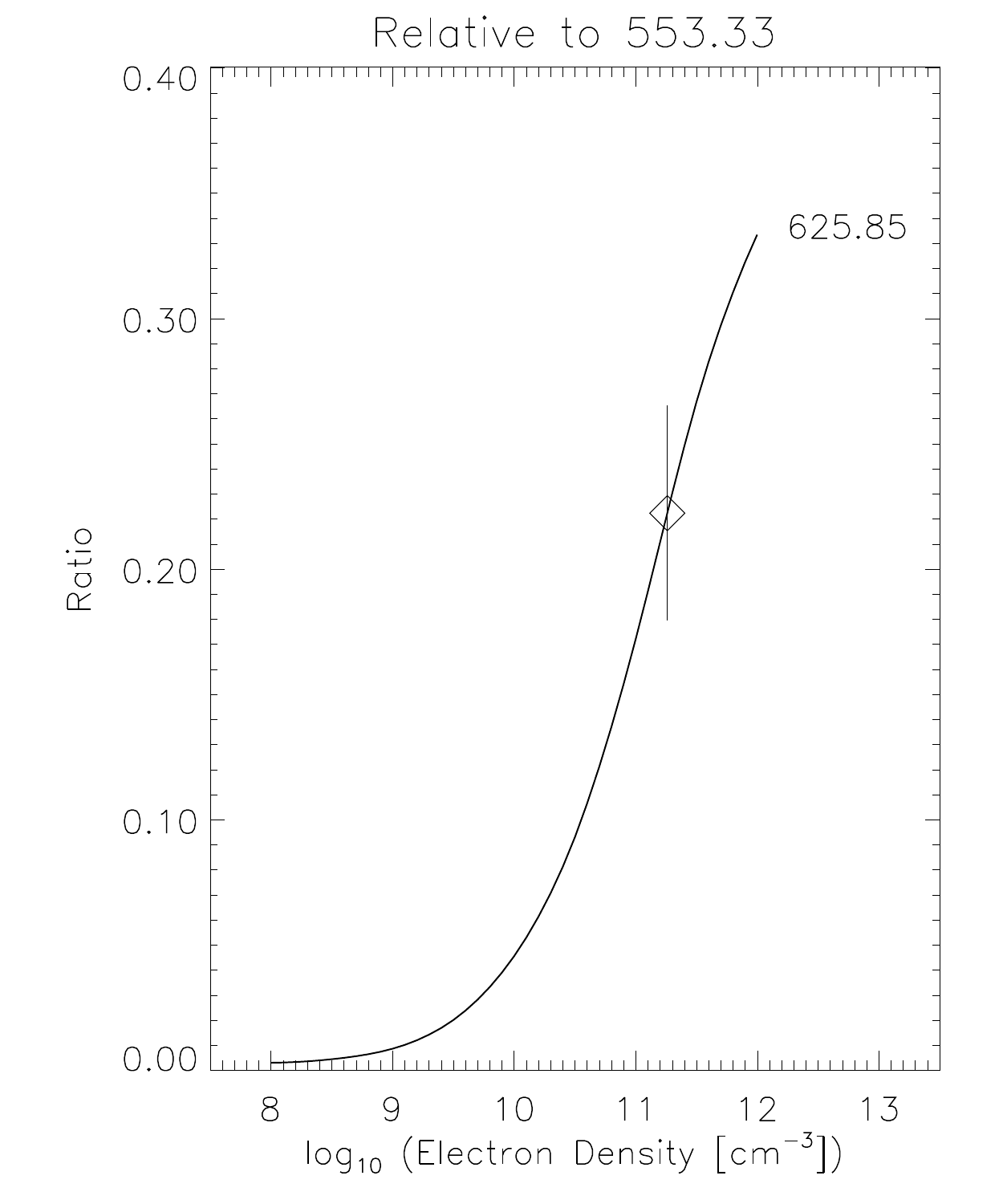}
\includegraphics[width=0.45\textwidth,angle=0]{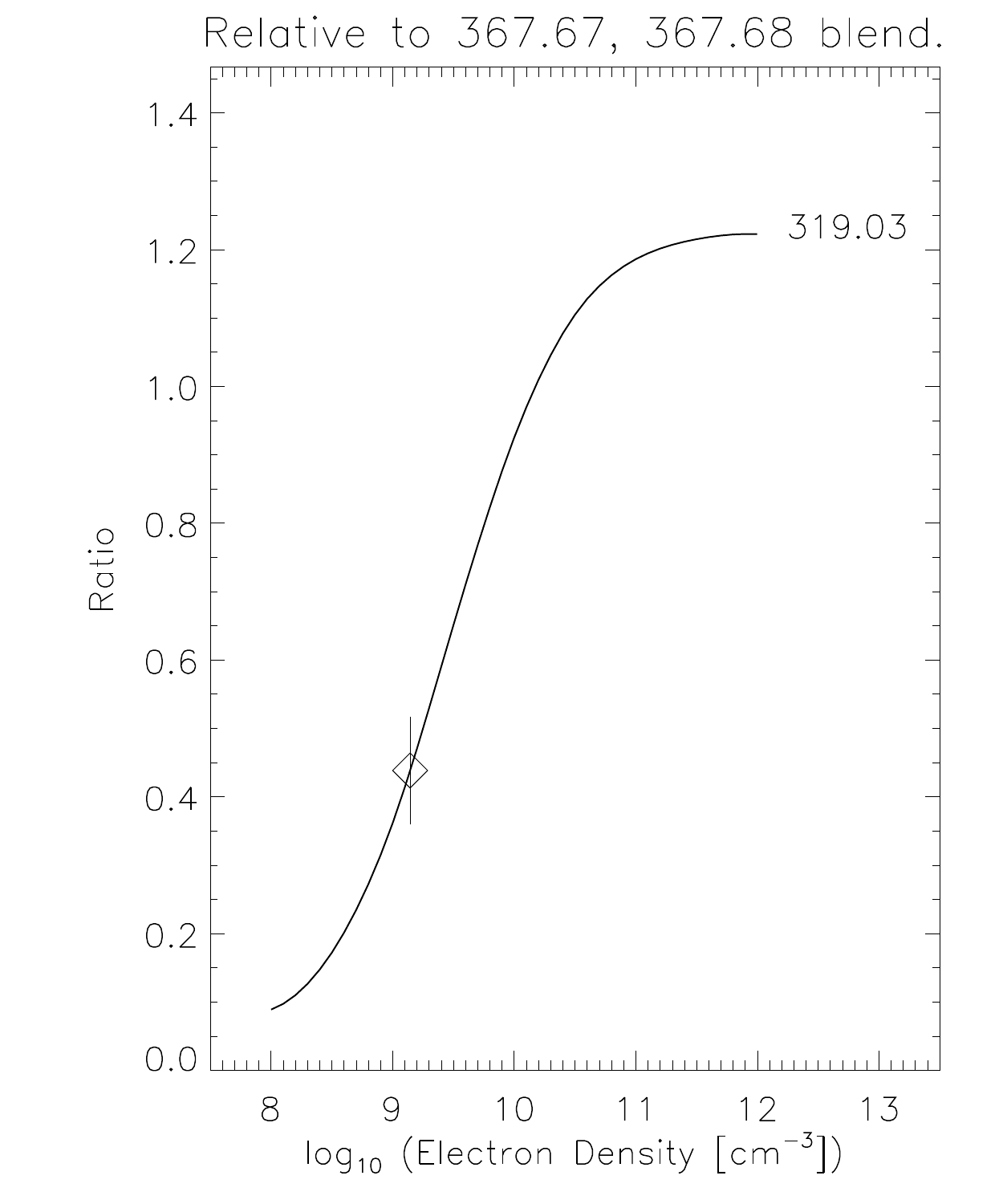}}
 \caption{Densities measured by \cite{young_mason:97} 
in TR brightenings occurring  in the cores of an active region
using SOHO NIS. The left plot is using  O IV lines, while the 
right plot using Mg VII lines.}
\label{fig:young_mason:97}
\end{figure}

The SOHO/SUMER observations of active regions were somewhat 
limited (to preserve the life of the detectors), so there are not
 many studies of active regions.
SOHO/CDS with its NIS instrument offered 
several density diagnostics, from the transition region
(O IV, Mg VII) to the 1~MK (Si IX, Si X) and higher temperatures
(Fe XIII, Fe XIV), however the degraded NIS spectral resolution 
after the SOHO loss in 1998 limited the results.
\citep{young_mason:97} measured densities of about 10$^{11}$ cm$^{-3}$
from the O IV and Mg VII lines in TR brightenings in the cores of an active region
using SOHO NIS, as shown in  Fig.~\ref{fig:young_mason:97}.
Density variations were found to be associated to  flux emergence.   

Averaged densities in the cores of active regions were found 
to be 2--3  $\times$ 10$^9$ cm$^{-3}$, with higher densities at lower heights
\citep[see, e.g.,][]{mason_etal:99}.
These results pertain to the 3 MK emission of the hot loops,
as is clearly visible off-limb.
The legs of the `warm' 1~MK loops could be easily 
resolvable with NIS, and densities of about 
2  $\times$ 10$^9$ cm$^{-3}$ were obtained  at their bases  
\citep[see, e.g.,][]{delzanna:03}.
Several density measurements have been obtained from EUV
lines measured by the various SERTS rocket flights,
with results in broad agreement with those from SOHO CDS.
\citep{young_etal:1994,brosius96,young_etal:98,brosius_etal:00}.

Hinode EIS has offered several excellent density 
diagnostics for active regions, from Mg VII, Fe XI, Fe XII, Fe XIII, Fe XIV,
Si X. Several studies have been carried out, see for example 
 \cite{doschek_etal:2007,watanabe_etal:2007,watanabe_etal:2009,young_etal:2009,odwyer_etal:11,young_etal:2012}.
The densities in active region cores were found to vary 
between 10$^9$ and 10$^{10}$ cm$^{-3}$.
A summary of the densities  obtained from  the various 
ions observed by Hinode EIS is provided in 
\cite{delzanna:12_atlas}.

\begin{figure}[htbp]
\centerline{\includegraphics[width=\textwidth]{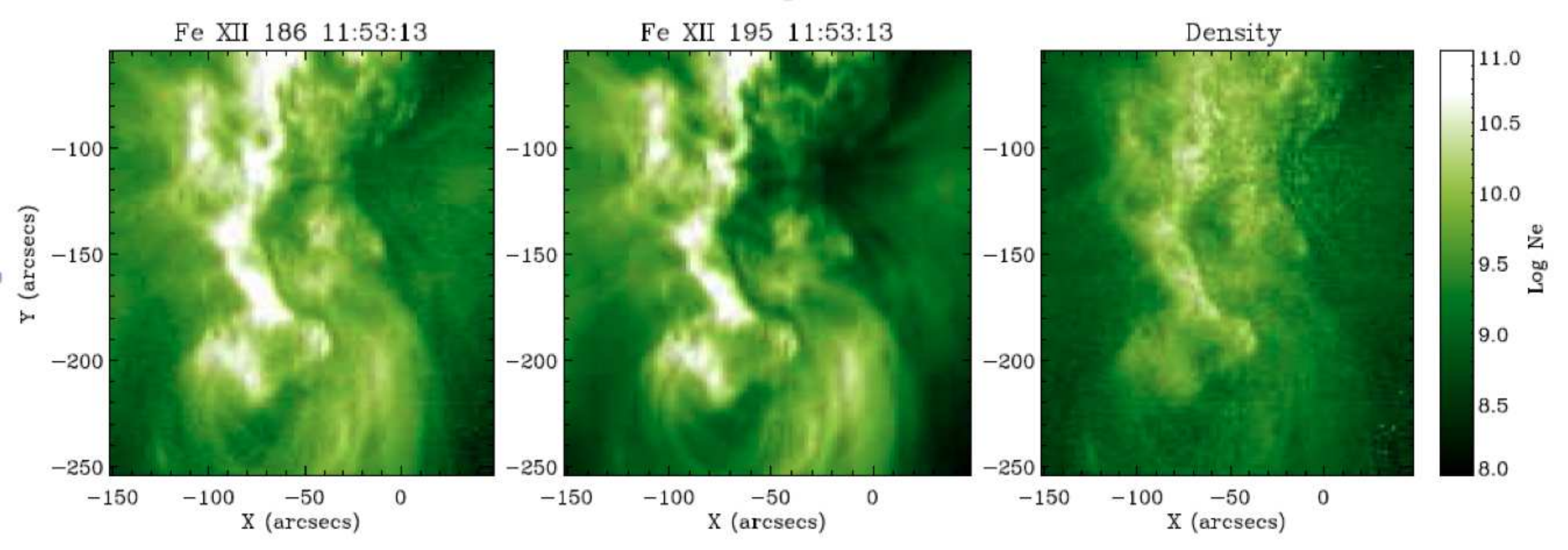}}
 \caption{A 2-D map of the average electron density in  active region moss 
as obtained from the ratio of two Fe XII lines observed by Hinode EIS.
Figure adapted from \cite{tripathi_etal:2008}.}
  \label{fig:tripathi_etal:2008}
\end{figure}

Off-limb observations of ARs with Hinode/EIS indicated similar 
densities as found from SOHO/CDS 
(see, e.g. \citealt{odwyer_etal:11}). 
%
%
On-disk observations of the hot cores of ARs 
produced averaged densities in agreement with 
 earlier results (see, e.g. \citealt{tripathi_etal:2008,tripathi_etal:2011, delzanna:2013_multithermal}).
An example is shown in Figure~\ref{fig:tripathi_etal:2008}. The highest densities are
in the `moss' regions, and correspong to the strongest magnetic fields.

\begin{figure}[htbp]
\centerline{\includegraphics[width=\textwidth]{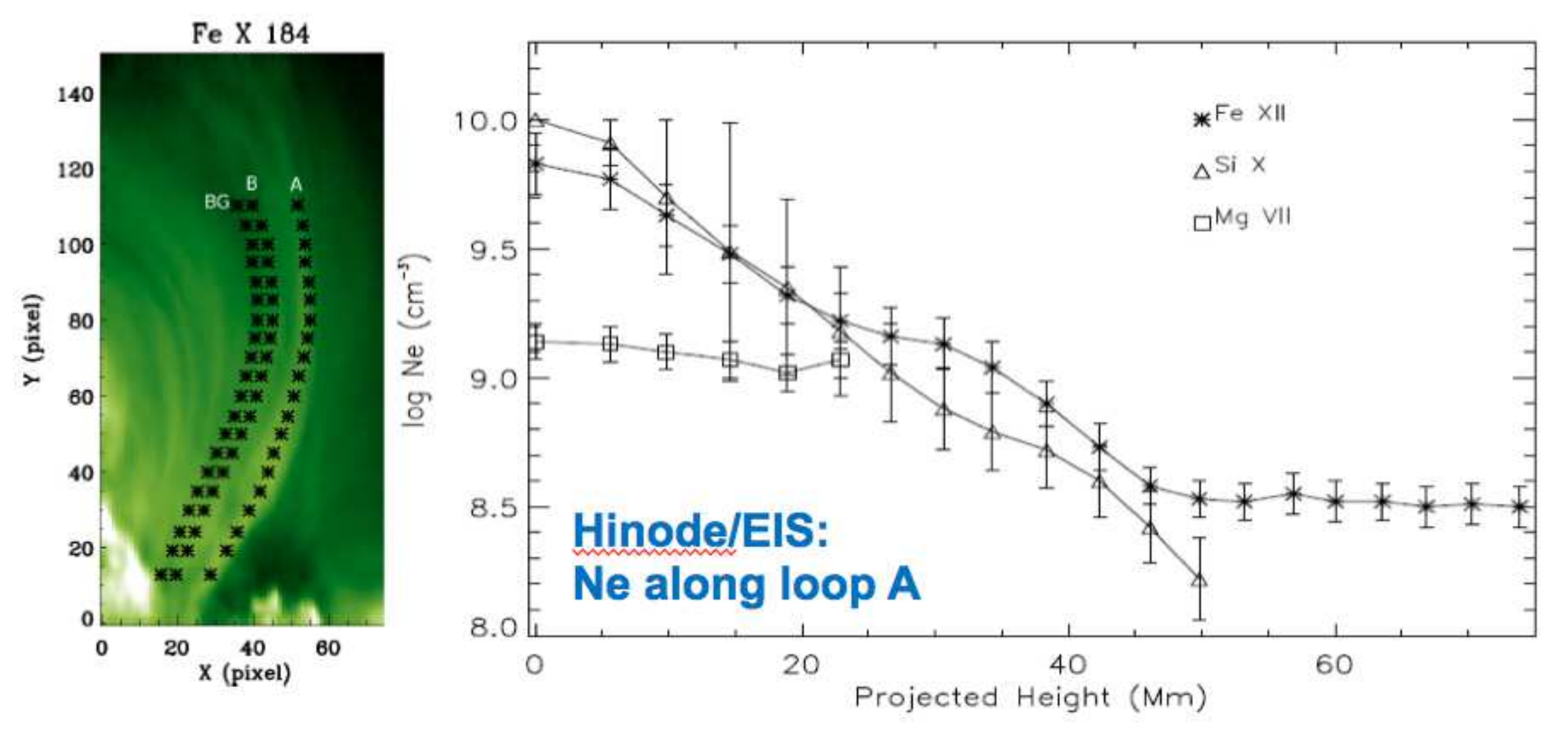}}
 \caption{Electron densities along the leg of a warm loop as measured from 
line ratios observed with Hinode EIS (adapted from \cite{tripathi_etal:2009}.)
}
 \label{fig:tripathi_etal:2009}
\end{figure}

At lower temperatures, Hinode/EIS has a  Mg VII diagnostic ratio
 \citep{young_etal:07a,young_etal:07b}, although some 
problems with this ratio  have been noted  \citep{delzanna:09_fe_7}. 
Hinode EIS measurements in the Mg VII lines at 
the legs of AR  warm loops have generally provided densities
of about  3 $\times$ 10$^9$ cm$^{-3}$ and filling factors
between 0.2 and 1 (see e.g. \citealt{tripathi_etal:2009,young_etal:2012}),
 in agreement with the earlier SOHO CDS results.

Hinode EIS measurements of the densities of coronal loops 
are  limited by the effective spatial resolution of the 
instrument (3--4\arcsec). Results have been published by various authors,
see e.g. \cite{warren_etal:2008,tripathi_etal:2009,gupta_etal:2014}.
An example is shown in Figure~\ref{fig:tripathi_etal:2009}.
\cite{tripathi_etal:2009} used Hinode EIS measurements of densities 
from Mg VII, Si X and Fe XII to estimate the spectroscopic filling factor
of a warm loop.
Values around unity were found, with smaller values from Si X and Fe XII
(0.02) near the base of the loop.  The values based on Fe XII  need to be revised in 
light of the revised atomic data for this ion \citep{delzanna_etal:12_fe_12}.

\cite{brooks_etal:2012} used a combination of Hinode EIS  Fe XIII density 
diagnostics, Fe XII intensities and SDO AIA images to infer that many of the 
warm loops that are visible at the EIS spatial scales are probably composed of 
just a few sub-resolution structures.

\subsubsection{Active region jets and bright points}

\begin{figure}[htbp]
\centerline{\includegraphics[width=0.9\textwidth]{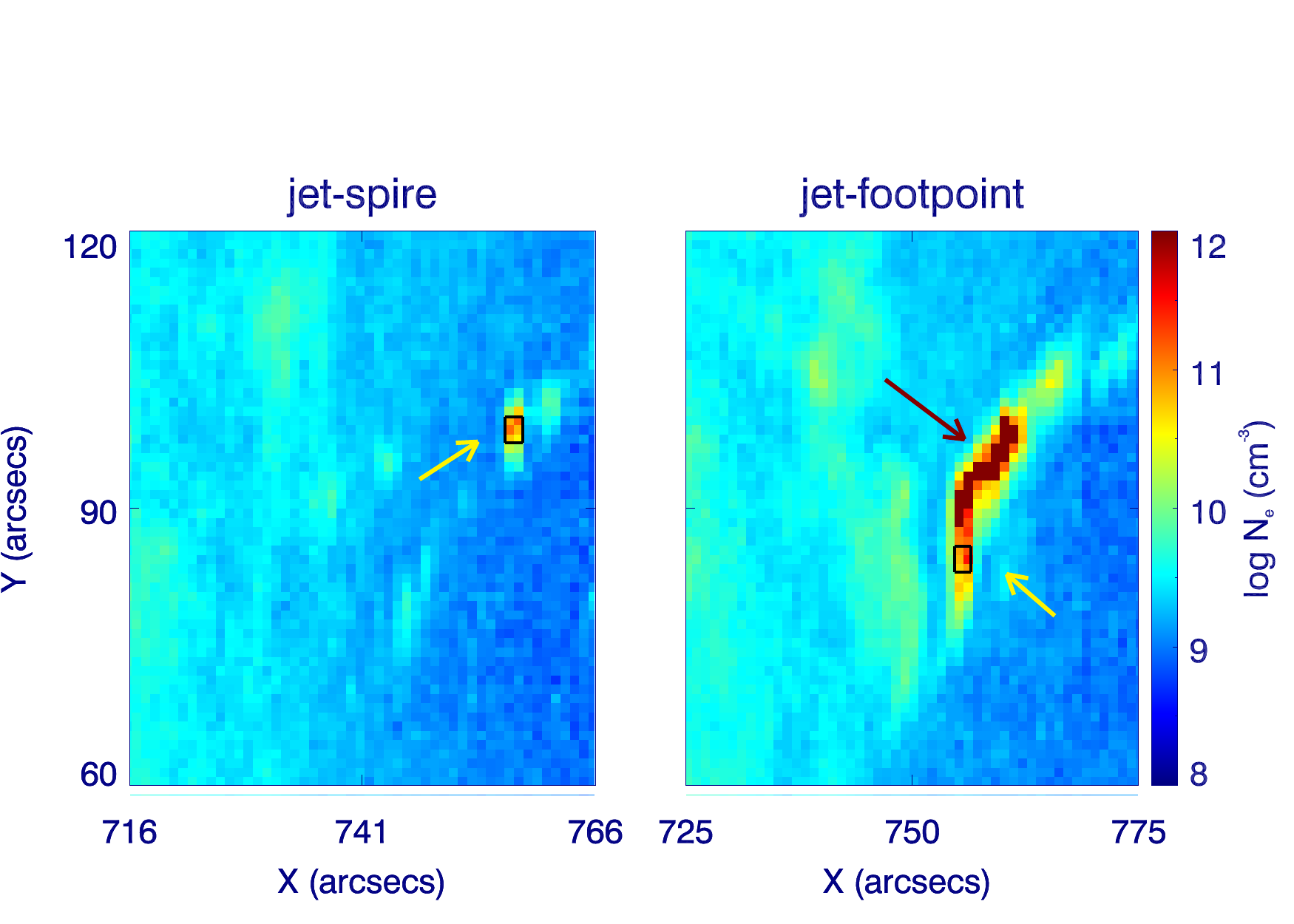}}
\caption{Density maps obtained from the \ion{Fe}{xii} lines observed by 
Hinode EIS for a jet spire and footpoint region (adapted from \citealt{mulay_etal:2017}.)
}
 \label{fig:mulay_fe_12}
\end{figure}

An interesting measurement of the density of an active region jet was obtained
by \cite{chifor_etal:2008_jet} with Hinode EIS. 
The density-sensitive \ion{Fe}{xii} 186.8~\AA\
self-blend showed strong blue-shifted components, so different 
densities for the various components were estimated. 
Densities reached values of about 10$^{11}$ cm$^{-3}$ in the blue-shifted component.
Densities from  Hinode EIS  \ion{Fe}{xii} observations of a jet footpoint
and spire  were also obtained by \cite{mulay_etal:2017}.
Also in this case,  values of about 10$^{11}$ cm$^{-3}$  or more were
obtained as shown in
Figure~\ref{fig:mulay_fe_12}.


Hinode EIS measurements of densities of bright points have also been carried out,
see for example  \cite{dere:2008,alexander_etal:2011}.
Densities  of about  5 $\times$ 10$^9$ cm$^{-3}$ and small filling factors
were found.
We note, however, that the densities obtained from Fe XII should be 
revisited.

\subsection{Flares}
\label{sec:flare_ne}

\begin{figure}[htb]
 \centerline{\includegraphics[width=0.8\textwidth,angle=0]{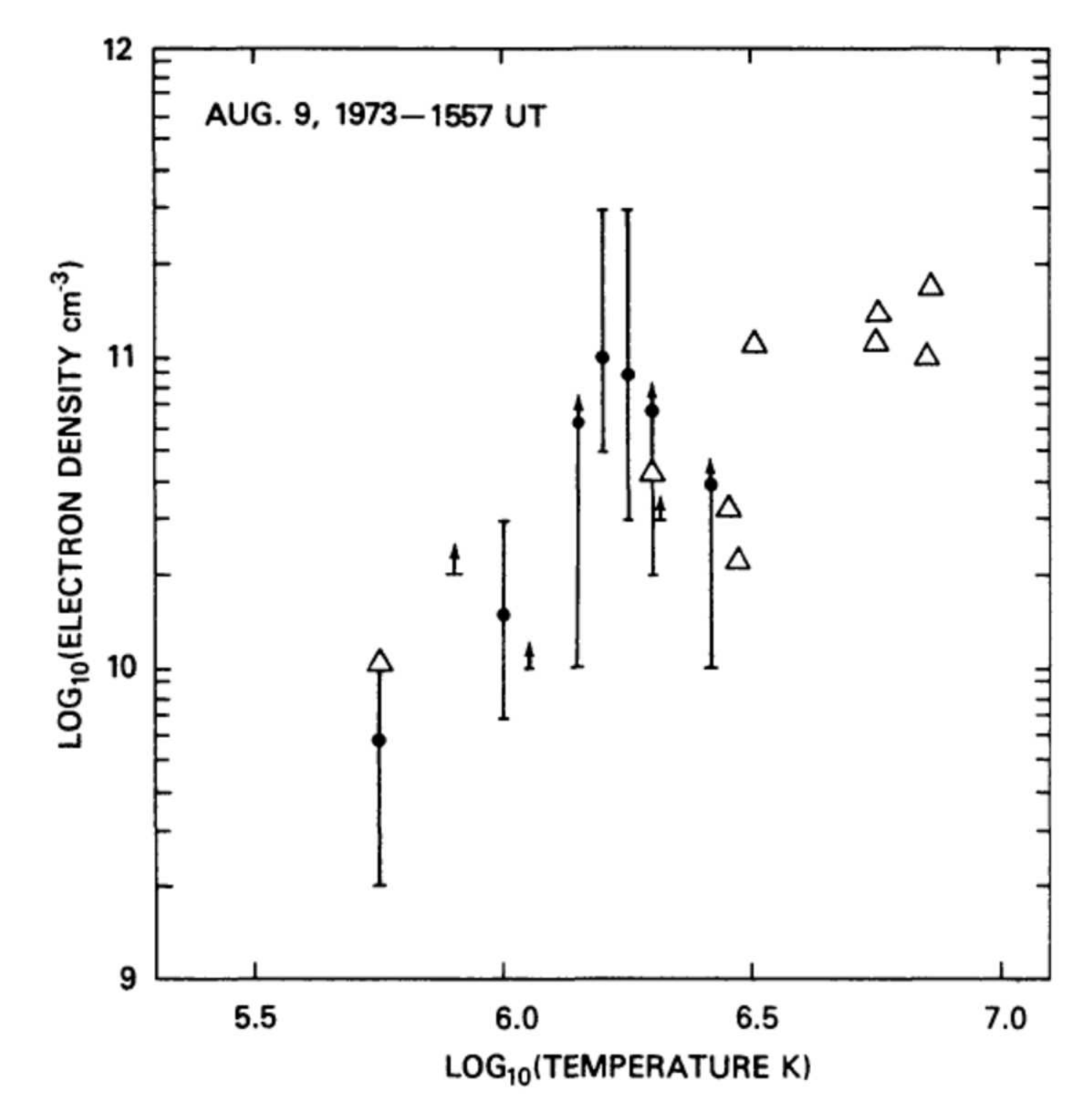}} 
 \caption{Electron densities obtained from Skylab during a small compact flare 
\citep{dere_etal:1979}. The points were obtained from line ratios,
while the triangles from the emission measures and the flare size.
}
 \label{fig:dere_etal:1979}
\end{figure}

A general review of solar flares and  diagnostics 
is given by \cite{doschek:1990_review}. 
One of the most comprehensive reviews of electron densities 
measured during solar flares in the EUV is given by  \cite{dere_etal:1979},
where observations with the Skylab NRL  slitless S082A instrument were used.
One example is shown in  Fig.~\ref{fig:dere_etal:1979}. That
instrument was ideal for compact, bright, isolated features.

The observations were excellent in that several line ratios 
of many ions could be used at once. Such a comprehensive diagnostic has not
been available since. 
Electron densities ranged between  0.1  and 1 \x 10$^{11}$  cm\up{-3},
from ions formed in the 1--6 MK range. The densities increase with the 
temperature of formation of the line.
Most of the diagnostics covered the 1-3 MK range. Only one diagnostic 
at higher temperatures was available, from Ca XVII. It  provided 
a density of about  10$^{11}$  cm$^{-3}$.
A similar result (5 $\times$ 10$^{11}$ cm$^{-3}$)
was obtained by \cite{doschek_etal:1977} also from Ca XVII for the 1973 
Aug 9 flare.

A good review of the diagnostic potential in the UV
offered by the Skylab NRL S082B normal incidence slit 
spectrometer is provided by 
\cite{widing_cook:1987},  who analysed the  1973 December 17 flare.

In what follows we give a short overview of the main results,
 organised by the formation temperature of the lines.

\subsubsection{Flare lines (about 10 MK)}

During flares, a significant amount of plasma is heated to 10 MK or more.
It is therefore  very important to measure the density of this high-temperature
plasma. However, results so far have been limited.
There are several diagnostics available in the X-rays, but they are 
mainly useful for  densities much higher than those typical of most solar flares,
 10$^{11}$ cm$^{-3}$ (see, e.g. \citealt{mckenzie_etal:85}).
For example, the \ion{Si}{xiii} and \ion{S}{xv} lines become sensitive to densities
well above 10$^{13}$ cm$^{-3}$, so are not  useful for solar flares.

A few diagnostics in the X-rays (around 10~\AA) are available, mostly from 
\ion{Fe}{xxi} and \ion{Fe}{xxii} 2p--4d transitions, which were identified by 
\cite{fawcett_etal:87} from  SMM FCS observations and discussed by \cite{phillips_etal:1996}.
The SMM FCS observations 
suggested that very high densities, about log $N_{\rm e}$ [cm$^{-3}$]  = 13
could be present. However, the lines were very weak, so accurate measurements 
were not possible. 
SMM BCS observations of satellite lines indicated 
lower densities, which was interpreted by  \cite{phillips_etal:1996}
as due to the much larger spatial area observed by this instrument, compared 
to FCS. 

\begin{figure}[!htbp]
\centerline{\includegraphics[width=0.88\textwidth]{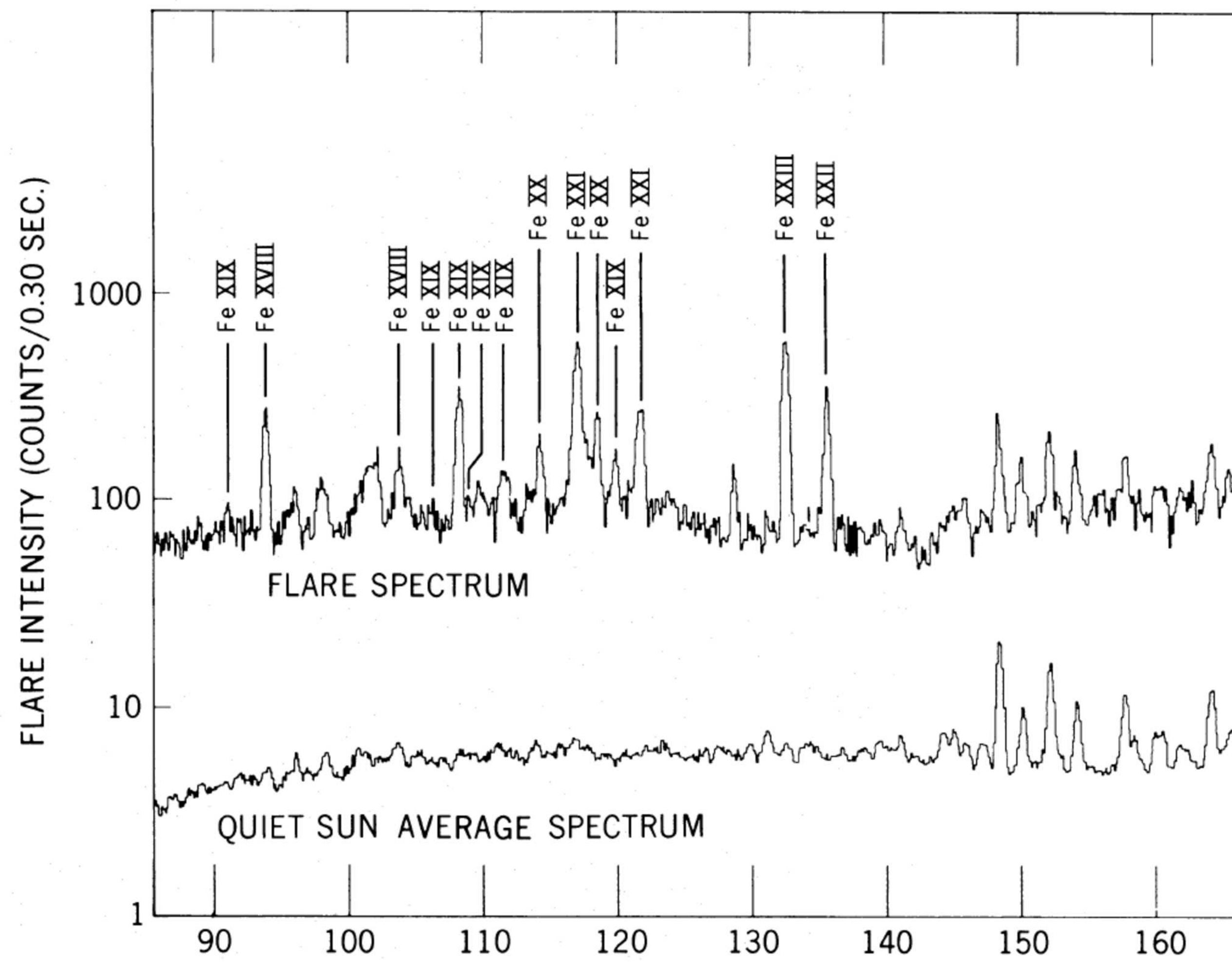}} 
\centerline{\hspace*{0.45cm}\includegraphics[width=0.885\textwidth]{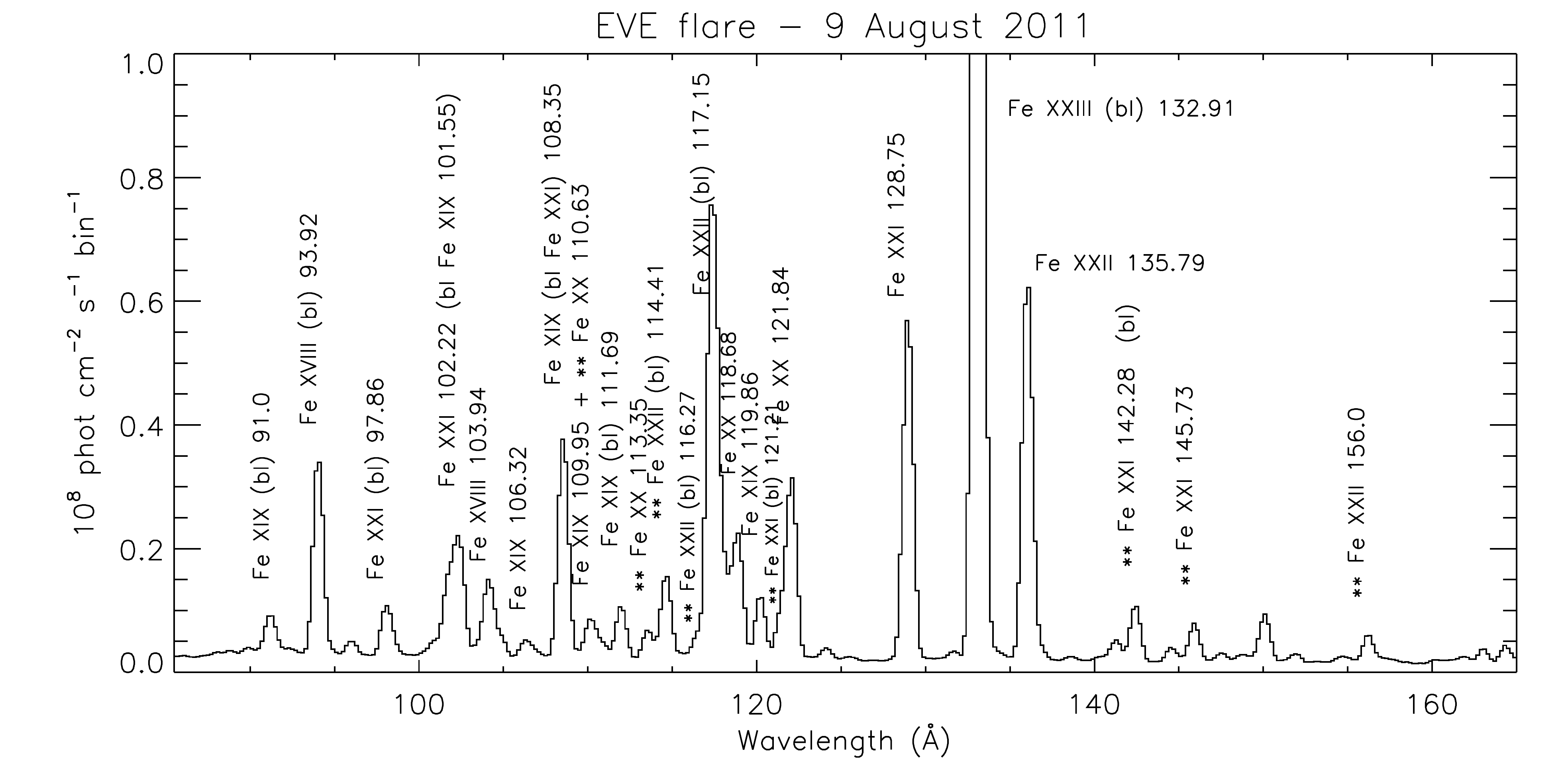}}
\caption{ Top: flare `E' spectrum from OSO-5 \citep{kastner_etal:1974_flare}.
Note that the quiet Sun spectrum is displayed on a different scale. 
Bottom: SDO EVE  spectrum of the X-class flare on 2011 Aug 9, 
with a  pre-flare spectrum subtracted \citep{delzanna_woods:2013} and 
in the same wavelength range as the OSO-5 spectrum. 
Lines that are density-sensitive (decays to excited levels) 
are highlighted with **.
}
\label{fig:sxr_flare}
\end{figure}

The strongest lines emitted by flare iron ions fall in the soft X-rays.
Table~\ref{tab:list_soft_xrays} provides a list of the main lines.
OSO-5 obtained the first solar flare spectra of these lines
\citep{kastner_etal:1974_flare}. An example spectrum (flare E) is shown in 
Fig.~\ref{fig:sxr_flare}, together with a more recent flare spectrum from 
SDO EVE  \citep{delzanna_woods:2013}.

\begin{table}[!htbp]
\caption[List of the main soft X-rays Fe  flare lines.]{List of the main soft X-rays Fe  flare lines.
The lines useful as density-diagnostics are highlighted with **.
$\lambda$ (\AA) is the  experimental  wavelength, 
T$_{\rm max}$ (in logarithm)  the approximate temperature of formation
of the ion in equilibrium;  bl means the line is blended, although we note that 
all the lines are blended to some degree  with lower-temperature transitions.
}
\begin{center} 
\begin{tabular}{@{}lllllllll@{}}
\toprule
 Ion &  $\lambda$  & Transition & T$_{\rm max}$ & Notes \\ 
\midrule

 \ion{Fe}{xix} &  91.013 & 2s$^2$ 2p$^4$ $^1$D$_{2}$ - 2s 2p$^5$ $^1$P$_{1}$ &  7.0 & \\ 

 \ion{Fe}{xxi} & 91.27 & 2s$^2$ 2p$^2$ $^3$P$_{0}$ - 2s 2p$^3$ $^3$S$_{1}$ &  7.1 &   \\ 

\ion{Fe}{xx} &  93.781 & 2s$^2$ 2p$^3$ $^2$D$_{5/2}$ - 2s 2p$^4$ $^2$P$_{3/2}$ &  7.0 & \\
\ion{Fe}{xviii} &  93.923 & 2s$^2$ 2p$^5$ $^2$P$_{3/2}$ - 2s 2p$^6$ $^2$S$_{1/2}$ &  6.9 &  \\ 

 \ion{Fe}{xxi} & 97.864 & 2s$^2$ 2p$^2$ $^3$P$_{1}$ - 2s 2p$^3$ $^3$S$_{1}$ &  7.1 & \\ 

 \ion{Fe}{xvii} &  98.25 & 2s$^2$2p$^5$3s $^3$P$_{1}$--2p$^6$3s $^1$S$_{0}$ &  6.9 & \\

\ion{Fe}{xxii} & 100.775 & 2s$^2$ 2p $^2$P$_{1/2}$ - 2s 2p$^2$ $^2$P$_{3/2}$ &  7.1 &   \\ %

\ion{Fe}{xix} &  101.55 & 2s$^2$ 2p$^4$ $^3$P$_{2}$ - 2s 2p$^5$ $^3$P$_{1}$ &  7.0 & \\

** \ion{Fe}{xxi} &  102.217 & 2s$^2$ 2p$^2$ $^3$P$_{2}$ - 2s 2p$^3$ $^3$S$_{1}$ &  7.1 & \\

\ion{Fe}{xviii} & 103.948 & 2s$^2$ 2p$^5$ $^2$P$_{1/2}$ - 2s 2p$^6$ $^2$S$_{1/2}$ &  6.9 & \\ 


\ion{Fe}{xix}  & 106.317 &  2s$^2$ 2p$^4$ $^3$P$_{1}$ - 2s 2p$^5$ $^3$P$_{0}$ &  7.0 & weak \\

\ion{Fe}{xxi} &  108.118 & 2s$^2$ 2p$^2$ $^3$P$_{0}$ - 2s 2p$^3$ $^3$P$_{1}$ &  7.1 & weak \\
\ion{Fe}{xix} &  108.355 & 2s$^2$ 2p$^4$ $^3$P$_{2}$ - 2s 2p$^5$ $^3$P$_{2}$ &  7.0 & strong \\ 

\ion{Fe}{xix} &  109.952 & 2s$^2$ 2p$^4$ $^3$P$_{0}$ - 2s 2p$^5$ $^3$P$_{1}$ &  7.0 & weak \\ 

\ion{Fe}{xx} &   110.627 & 2s$^2$ 2p$^3$ $^2$D$_{3/2}$ - 2s 2p$^4$ $^2$D$_{3/2}$ &  7.0 &  \\ 

\ion{Fe}{xix} &  111.695 & 2s$^2$ 2p$^4$ $^3$P$_{1}$ - 2s 2p$^5$ $^3$P$_{1}$ &  7.0 & weak \\ 

\ion{Fe}{xx} &   113.349 &  2s$^2$ 2p$^3$ $^2$D$_{5/2}$  2s 2p$^4$ $^2$D$_{5/2}$ &  7.0 &  \\ 

** \ion{Fe}{xxii} &  114.410 & 2s$^2$ 2p $^2$P$_{3/2}$ - 2s 2p$^2$ $^2$P$_{3/2}$ &  7.1 & \\

\ion{Fe}{xxii} &  116.268 & 2s$^2$ 2p $^2$P$_{3/2}$ - 2s 2p$^2$ $^2$S$_{1/2}$ &  7.1 & \\ 

\ion{Fe}{xxii} &  117.154 & 2s$^2$ 2p $^2$P$_{1/2}$ - 2s 2p$^2$ $^2$P$_{1/2}$ &  7.1 &  strong \\
** \ion{Fe}{xxi} &  117.50 & 2s$^2$ 2p$^2$ $^3$P$_{1}$ - 2s 2p$^3$ $^3$P$_{1}$ &  7.1 & \\ 

\ion{Fe}{xx} & 118.680  & 2s$^2$ 2p$^3$ $^4$S$_{3/2}$ - 2s 2p$^4$ $^4$P$_{1/2}$ &  7.0 & strong \\ 

** \ion{Fe}{xix} &  119.983 & 2s$^2$ 2p$^4$ $^3$P$_{1}$ - 2s 2p$^5$ $^3$P$_{2}$ &  7.0 &  \\ 

**  \ion{Fe}{xxi} &  121.213 & 2s$^2$ 2p$^2$ $^3$P$_{2}$ - 2s 2p$^3$ $^3$P$_{2}$ &  7.1 & weak \\ 

\ion{Fe}{xx} &  121.845 & 2s$^2$ 2p$^3$ $^4$S$_{3/2}$ - 2s 2p$^4$ $^4$P$_{3/2}$ &  7.0 & strong \\ 

** \ion{Fe}{xxi} &  123.831 & 2s$^2$ 2p$^2$ $^3$P$_{2}$ - 2s 2p$^3$ $^3$P$_{1}$ &  7.1 &  weak  \\ 

\ion{Fe}{xxi} &  128.753 & 2s$^2$ 2p$^2$ $^3$P$_{0}$ - 2s 2p$^3$ $^3$D$_{1}$ &  7.1 &  strong \\ 

\ion{Fe}{xx} & 132.840 & 2s$^2$ 2p$^3$ $^4$S$_{3/2}$ - 2s 2p$^4$ $^4$P$_{5/2}$ &  7.0 & strong\\ 
\ion{Fe}{xxiii} &  132.906 & 2s$^2$ $^1$S$_{0}$ - 2s 2p $^1$P$_{1}$ &  7.2 & strong \\ 

 \ion{Fe}{xxii} &  135.791 & 2s$^2$ 2p $^2$P$_{1/2}$ - 2s 2p$^2$ $^2$D$_{3/2}$ &  7.1 & strong \\ 

** \ion{Fe}{xxi} &  142.144 & 2s$^2$ 2p$^2$ $^3$P$_{1}$ - 2s 2p$^3$ $^3$D$_{2}$ &  7.1 & weak \\ 
** \ion{Fe}{xxi} &  142.281 & 2s$^2$ 2p$^2$ $^3$P$_{1}$ - 2s 2p$^3$ $^3$D$_{1}$ &  7.1 & weak   \\ 
** \ion{Fe}{xxi}  &  145.732  & 2s$^2$ 2p$^2$ $^3$P$_{2}$ - 2s 2p$^3$ $^3$D$_{3}$ &  7.1 &  weak \\

** \ion{Fe}{xxii}  &  156.019 &   2s$^2$ 2p $^2$P$_{3/2}$ - 2s 2p$^2$ $^2$D$_{5/2}$ &  7.1 & bl  \\ 
\bottomrule 
\end{tabular}
\end{center}
\label{tab:list_soft_xrays}
\end{table}

\begin{figure}[!htbp]
\centerline{\includegraphics[width=0.6\textwidth, angle=90]{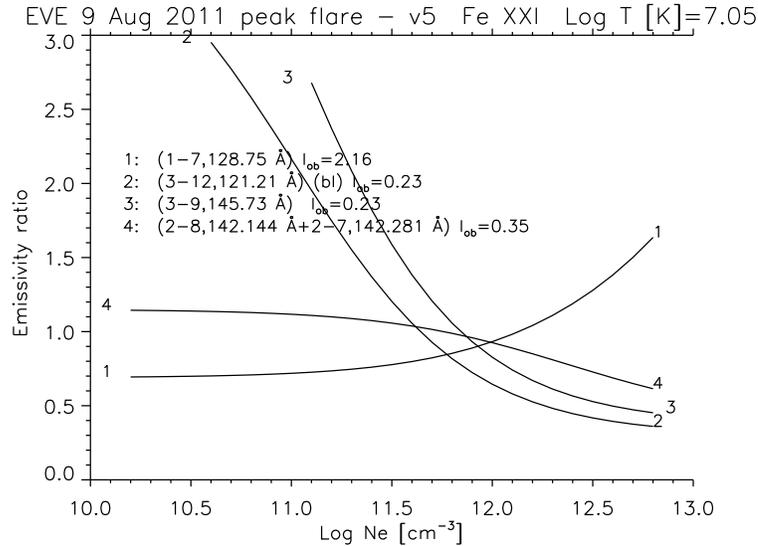}}
\caption{Emissivity ratio plots  for \ion{Fe}{xxi} lines 
observed by SDO EVE during  the X-class flare on 2011 Aug 9 
\citep[revised from ][]{delzanna_woods:2013}. 
}
\label{fig:sxr_flare_ne}
\end{figure}

Several density-sensitive line ratios are available in this wavelength region, 
from \ion{Fe}{xx}, \ion{Fe}{xxi}, \ion{Fe}{xxii}, as discussed in \cite{mason_etal:84}.
An analysis  of these spectra is also given in 
\cite{mason_etal:84}, where a density  of 4 $\times$ 10$^{11}$
cm$^{-3}$  was found from the \ion{Fe}{xxi} lines.

A recent review of the diagnostics available from flare iron lines 
in the soft X-rays, with a discussion of 
all the blends and recent atomic data is given by \cite{delzanna_woods:2013}, 
using SDO/EVE version 3  flare 
spectra. The low spectral resolution of the EVE instrument is a 
limiting factor, as many of the density-sensitive lines
are blended with stronger lines.
 \cite{delzanna_woods:2013} found 
 densities between  10$^{11}$ and 10$^{12}$ cm$^{-3}$ from one of the flares,
in agreement with the previous OSO-5 results. 
A density of about 10$^{11}$ cm$^{-3}$  was also found by 
\cite{warren_etal:2013} uing SDO EVE spectra. 
On the other hand, \cite{milligan_etal:2012} obtained 
from  the EVE  spectra slightly  higher densities, reaching 10$^{12}$ cm$^{-3}$.
We have re-analysed one of the large X-class flares discussed 
in  \cite{delzanna_woods:2013}, where the density-sensitive lines
are more visible. We used the more recent version 5  EVE data,
and also found densities of the order of  10$^{12}$ cm$^{-3}$,
as shown in Fig.~\ref{fig:sxr_flare_ne}.
In any case, it is important to keep in mind that such values are 
averaged values over the whole  flare, so it is perfectly reasonable to assume that
 higher densities could be present in localised places.

Densities obtained from the emission measures and estimates of the sizes
of the post-flare loops in smaller flares are often lower, of the order 
of 5$\times$10$^{10}$ cm$^{-3}$ (cf. \citealt{delzanna_etal:2011_flare,polito_etal:2015}),
if the plasma is assumed to be homogeneously distributed within the flare loops
(i.e. a spectroscopic filling factor of unity is assumed). 
In other words, such measurements from the emission measures are  
lower estimates of what the densities actually are.
The important issue of the actual densities of high-temperature flare plasma 
remains to be resolved with a next-generation spectrometer that 
could spatially resolve flares.
Measuring densities is important for flare modelling, as cooling times 
for very high densities  (10$^{13}$ cm$^{-3}$) 
and temperatures (12 MK) can reach very short timescales, of the order 
of seconds.

\subsubsection{Coronal lines (1--6 MK)}

\begin{figure}[htb]
\centerline{\includegraphics[width=0.8\textwidth]{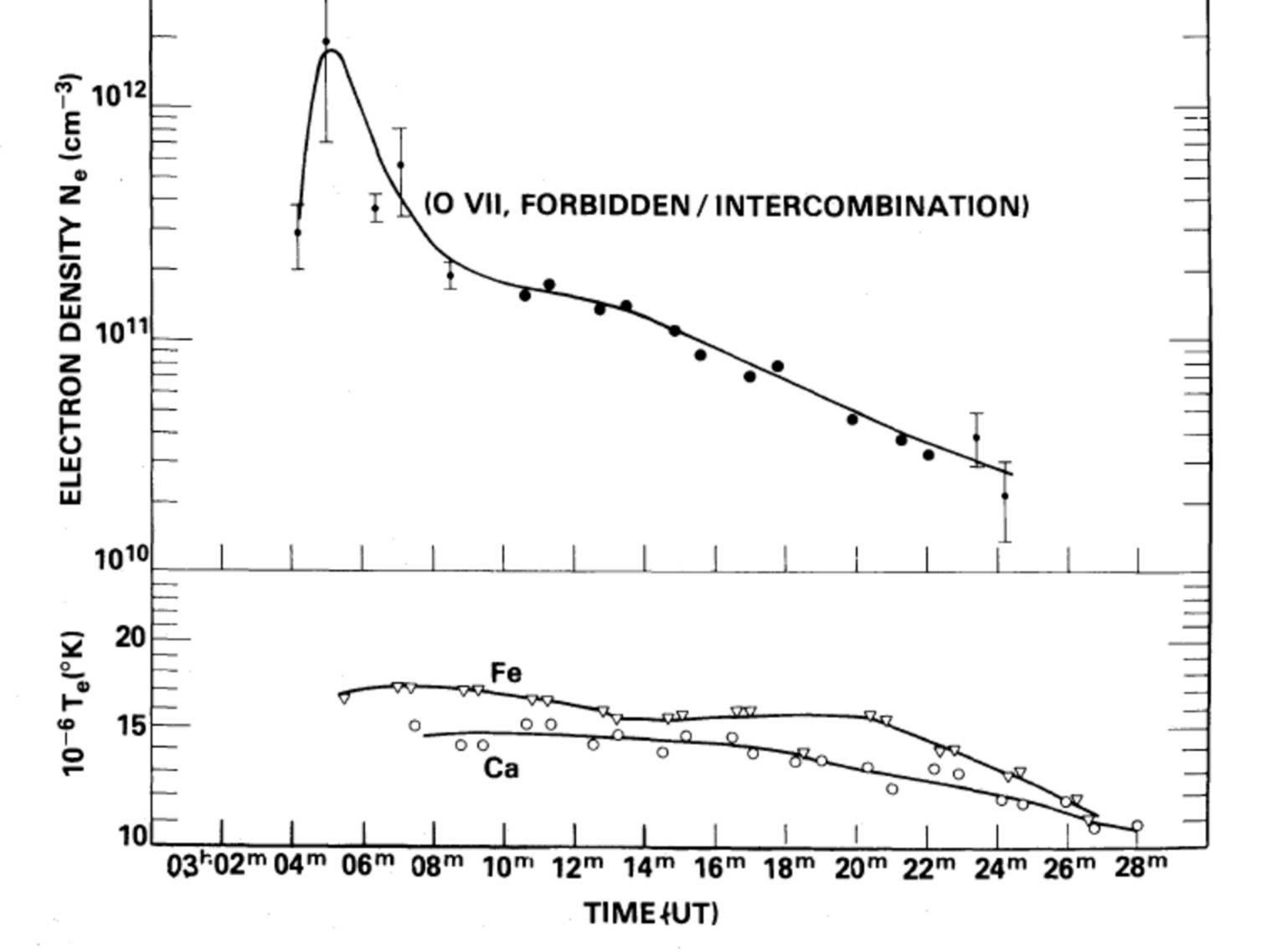}} 
\caption{Top: Electron densities obtained from \ion{O}{vii} SOLEX observations %
during a flare \citep{doschek_etal:1981}.
Bottom: temperatures obtained from the same observation.
}
\label{fig:ne_flare_o_7}
\end{figure}

As already mentioned,  He-like ions provide 
some diagnostics.
O~VII, formed around 2 MK,  has been used quite extensively in the past.
One example comes from the P78-1  SOLEX  observations of the O VII lines 
during a major solar flare, analysed by \cite{mckenzie_etal:80a}.
The density during the flare
 raised from about 0.5 to 2 $\times$  10$^{11}$ cm$^{-3}$.
A similar study of  SOLEX O VII observations
of two flares was published  by \cite{doschek_etal:1981}.
Similar densities were obtained, with a (single) peak value 
for both flares of about 10$^{12}$ cm$^{-3}$.
Fig.~\ref{fig:ne_flare_o_7} shows one example.

Among the other He-like ions,
the Ne IX lines would be useful but they are blended in flares
with Fe XIX lines, and accurate deblending is needed.
Mg XI lines are useful only for densities higher than 
10$^{12}$ cm$^{-3}$, and require very high resolution spectra.

Aside from the He-like ions, excellent diagnostics of 1--6 MK 
plasma are found in the soft X-rays and in the EUV. 
While there are plenty of measurements in the EUV, mostly 
from Skylab and Hinode/EIS, only one observation
 in the soft X-rays is available.
This was obtained from a huge (5 meters) grazing-incidence 
spectrograph on board a  sounding rocket on 1982 July 13,
about 2 minutes after the peak of a C9-class flare. 
The instrument and the observations are described in 
\cite{acton_etal:85}, while the diagnostics are discussed in 
\cite{brown_etal:1986}.
These latter authors obtained densities from \ion{Fe}{xiv} (59~\AA)
and \ion{Si}{ix} (56~\AA) line ratios of about 2--3 $\times$ 10$^{10}$ cm$^{-3}$,
and about  5$\times$10$^{10}$ cm$^{-3}$ from \ion{Ca}{xv} lines at 22.7~\AA.
Densities from the He-like C, N , O lines 
were similar, about  1--3$\times$10$^{10}$ cm$^{-3}$, while those from 
\ion{Ne}{ix} were significantly higher, perhaps because of the 
known blending issues.

\cite{widing_cook:1987}
 discussed  Skylab NRL S082A spectroheliograph observations in the EUV 
of the  1973 December 17 flare,
where lines from Ca XV and Ca XVI (3--5 MK)  showed densities slightly above 
log $N_{\rm e}$ [cm$^{-3}$]=11,
while lines from Fe XIV and Fe XV (2 MK) indicated much smaller values.

\cite{keenan_etal:1993_ar_13} used interpolated 
collision strengths for Ar XIII to measure densities from 
EUV flare observations with the Skylab NRL S082A instrument.
They consistently obtained values around  10$^{11}$ cm$^{-3}$,
 in very good  agreement with those
obtained from Ca XV \citep{keenan_etal:1988_ca_15}.

\cite{keenan_etal:1989_ar_15} used lines from the 
Be-like Ar XV and EUV observations of a flare with the Skylab 
NRL S082A slitless spectrograph to attempt a measurement 
of the density from the 221.12/266.23~\AA\ ratio.
However, the ratio was below  the theoretical density limit,
which was interpreted as being caused by blending of the 
266.23~\AA\ line with an unidentified line.

Lines from the B-like S XII were used by 
\cite{keenan_etal:2002_s_12} to obtain densities for active regions and flares
(log $N_{\rm e}$ [cm$^{-3}$] $\simeq$ 9.5, 10, respectively) 
from Skylab NRL S082A and SERTS observations.
The results are in agreement with those obtained from Fe XIV and Fe XV lines,
emitted at similar temperatures (2 MK).

With Hinode EIS, it has been possible to spatially resolve the various flare
emission during flares and measure densities from coronal lines.
\cite{watanabe_etal:2009} showed that the Fe XIII 203.8/202~\AA\ ratio
seemed to reach a high-density limit (of about 4.34) in a kernel of a flare,
indicating densities of about 10$^{11}$ cm$^{-3}$ or higher. 
Discrepancies with previous atomic data were noted by these authors.
However, we note that this observationally-based  high-density limit
is in very good agreement with the theoretical value of 4.4 obtained
from the latest calculations \citep{delzanna_storey:12_fe_13}.

\cite{delzanna_etal:2011_flare}
measured densities from Fe XIV line ratios observed by Hinode EIS 
during a small B-class flare. The  Fe XIV line profiles in the ribbons,
kernels of chromospheric evaporation during the impulsive phase, were 
a superposition of two components, a rest and a blue-shifted component.
The rest component, which was due to the overlying foreground plasma, had 
a density of about 6 $\times$ 10$^{9}$ cm$^{-3}$. The blue-shifted component
had densities in some places close to the high-density limit,
i.e. indicating densities of about 10$^{11}$ cm$^{-3}$ or higher.
From the emission measures, it was estimated that 
the  depth of the chromospherically-evaporated material was
only about  10 km.
\cite{young_etal:2013_flare} used the same Fe XIV diagnostics
but obtain  a lower density of  3 $\times$ 10$^{10}$ cm$^{-3}$
in the blue-shifted component, in a kernel of a large solar flare.

\subsubsection{Cooler TR lines (below 1 MK)}

\cite{widing_cook:1987} obtained from the Skylab NRL S082B normal incidence slit 
spectrometer observations of the  1973 December 17 flare
densities from low-temperature lines (O IV, O V, and S IV) and found 
values in the range log $N_{\rm e}$ [cm$^{-3}$]=11--12 (S V showed higher values).


\cite{feldman_etal:1977_flare} obtained  densities 
from the intersystem O IV 1401/1407~\AA\ ratio using a flare 
observation with the 
Skylab NRL  S082B  slit spectrometer. 
The spectral profiles showed a stationary component but a significant
non-thermal width.
Densities were not very high, a few times 10$^{11}$ cm$^{-3}$. 
Conflicting results concerning the O V 1371/1218~\AA\ ratio were found.
The authors  noted that the 1371 line becomes blended with a line at 1371.37~\AA.
\cite{keenan_etal:1995_o_5}
obtained   higher densities, 
log $N_{\rm e}$ [cm$^{-3}$] $\simeq$ 12,   from the same instrument and ratio,
during the 1973 August 9 flare.

\cite{dufton_etal:1982} used Skylab  NRL S082B AR and flare observations 
of the Al-like S IV lines, obtaining values in the range 
 log $N_{\rm e}$ [cm$^{-3}$] $\simeq$ 11--12.

\cite{keenan_etal:1991_o_5} re-analysed 
Skylab NRL S082A observations of two flares with updated atomic data for the 
2--3 transitions in  O V.
From the lines at 
192.80, 192.90, 215.10, 215.25, 220.35, and 248.46~\AA,
they obtained densities of approximately 10$^{12}$ cm$^{-3}$, significantly 
higher than those measured by \cite{widing_etal:1982}
using previous atomic data.
The 192.80, 192.90 (as self-blend) and 248.46~\AA\ lines
are observed by Hinode EIS, as discussed in detail 
by \cite{young_etal:07b} and \cite{delzanna:09_fe_7}.
However, one problem in flare observations is that the lines become 
broad and the 192.9~\AA\ normally becomes blended with 
the strong resonance line from \ion{Ca}{xvii}. The 
248.46~\AA\ line might also be blended. 
Other problems are the temperature sensitivity of the ratio, and  
 the degradation of the EIS instrument at the 
248.46~\AA\ wavelength.

\begin{figure}[!htbp]
\centerline{\includegraphics[width=0.6\textwidth]{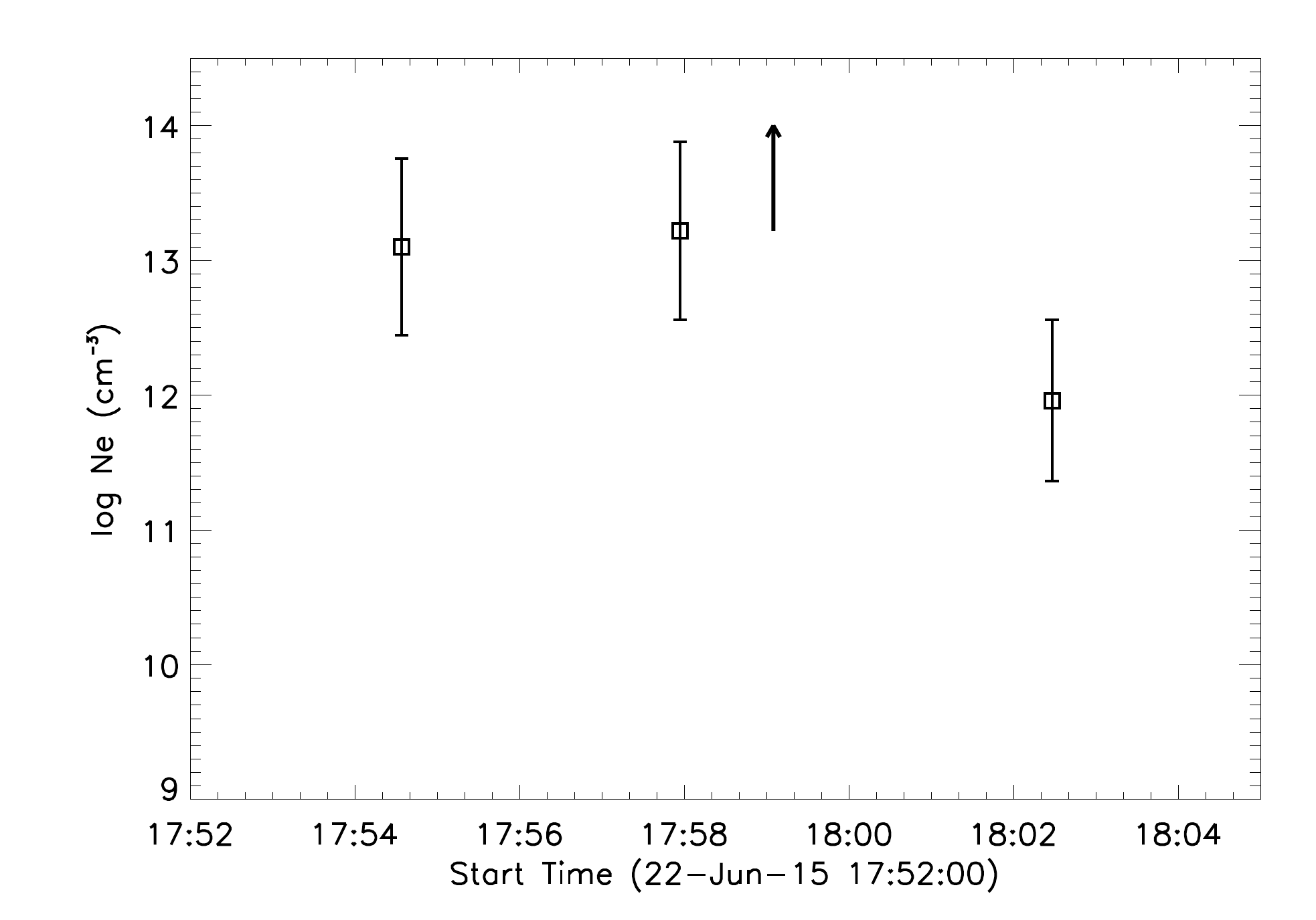}}
\caption{ Electron densities obtained from \ion{S}{iv} IRIS observations
during a flare (adapted from \citealt{polito_etal:2016a}).
}
\label{fig:ne_flare_s_4}
\end{figure}

With the launch of IRIS, it has become possible to measure 
densities using the  \ion{S}{iv} and \ion{O}{iv}
lines around 1400~\AA\ during flares. An up-to-date discussion 
of the atomic data and measurements during the impulsive phase
of flare ribbons can be found in  \cite{polito_etal:2016b}.
As already mentioned in the diagnostic section, the 
\ion{O}{iv} are not sensitive to densities higher than 
10$^{12}$ cm$^{-3}$, but  \ion{S}{iv} lines are.
Fig.~\ref{fig:ne_flare_s_4} shows the evolution over time 
of  \ion{S}{iv} densities, indicating very high values.
Indeed the high-density limit of  10$^{13}$ cm$^{-3}$
was reached in one instance.

\clearpage
\section{Diagnostics of Electron Temperatures} 
\label{sec:te_diagn}

There is ample literature where temperatures have been obtained 
from spectral lines  emitted using different ions. 
 These results are strongly dependent on the knowledge of the relative ion populations,
which even in equilibrium are sometimes quite uncertain.
There is also  ample literature where temperatures have been obtained 
from broad-band imaging, where additional uncertainties such as 
off-band contributions and elemental abundance variations are present. 
In this section, we mainly focus on results based on lines emitted from the 
same ion, which in principle are more reliable.

\subsection{Temperatures from line ratios within  the same ion}

A direct way to measure the electron temperature is to consider the intensity 
ratio of two allowed lines that have 
different excitation energies.
If e.g. the two lines are  excited from the ground level $g$:
\beq \frac{I_{g,j}}{I_{g,k}} = \frac{\Delta E_{g,j} \Upsilon_{g,j}}{\Delta 
E_{g,k} \Upsilon_{g,k}} \exp \left[ \frac{\Delta E_{g,k} - \Delta 
E_{g,j}}{k_B T} \right], 
\eeq 
where $j$ and $k$ denote the excited levels, 
$\Upsilon$ is the thermally averaged collision strength, and $k_B$
is the Boltzmann constant. The ratio is  temperature-dependent
 if the thermal energy of 
the electrons is much smaller than the difference between the excitation 
energies
\beq
 \frac{\Delta E_{g,k} - \Delta E_{g,j}}{k_B T} \gg 1
\eeq

This method of determining Te was first used in solar physics by 
\cite{heroux_etal:1972} using Li-like ions 
and \cite{flower_nussbaumer:1975_na-like} using Na-like ions.
One problem with this method is that normally 
such spectral lines are far apart in wavelength, which often means that 
they need to be observed by different instruments.

Alternatively, one of the lines which is observed might be collisionally excited
from the ground level, but radiatively decay to an excited level. 
The temperature sensitivity
will remain the same, but the two spectral lines might be closer in wavelength.

In general, such line ratio methods have been applied mostly to 
lines formed in the transition region, where non-equilibrium ionisation 
effects are likely to be important. 
Another major problem is that the temperature variation in the 
transition region is very steep, so different lines from the same ion
are naturally emitted in spatially different regions, as pointed out 
for the Si III case in \cite{delzanna_etal:2015_si_3}. Therefore,
it is not surprising to see that 
different line ratios  imply different temperatures, 
as is often found in the literature.

As with the density case, a useful way to measure temperatures when 
there are observations of more than two lines is to 
plot the emissivity ratios as a function of temperature, although 
they obviously also have the same limitations as the line ratios.


We now briefly review a few diagnostics that have been explored
in the literature. Compared to the density case, there are relatively 
few results on this topic.

\subsubsection{$T_{\rm e}$ from He-like ions}

\begin{figure}[htb]
 \centerline{\includegraphics[width=0.6\textwidth,angle=90]{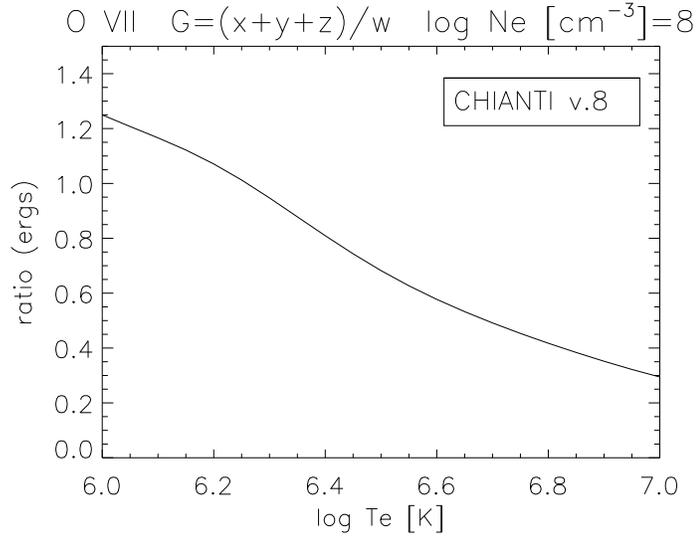}}
  \caption{ G-ratio for the He-like \ion{O}{vii}.}
  \label{fig:he-like_g-ratio}
\end{figure}

\cite{gabriel_jordan:1969a} discussed the use of the 
so called G ratio $(x+y+z)/w$  (recall  Fig.~\ref{fig:he-like} and Table~\ref{tab:he-like}) 
of the  intercombination and forbidden lines vs. the resonance line
$w$
 as a good temperature diagnostic  for the  helium-like ions,
as it does not have a strong dependence on electron density. 
There is an extended literature on the use of the G ratios;
see for example 
\cite[][]{keenan_etal:1990_si_13} for Si XIII,
\cite[][]{keenan_etal:1992_mg_11} for Mg XI. 
 Figure~\ref{fig:he-like_g-ratio} shows \ion{O}{vii} as an example.

\begin{figure}[!htbp]
 \centerline{\includegraphics[width=0.5\textwidth,angle=90]{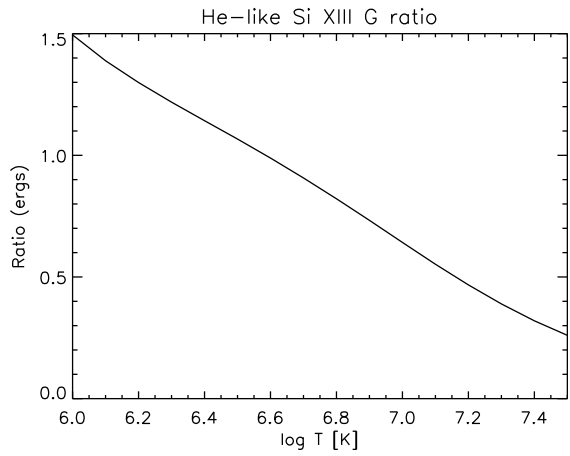}}
\centerline{\includegraphics[width=0.5\textwidth,angle=90]{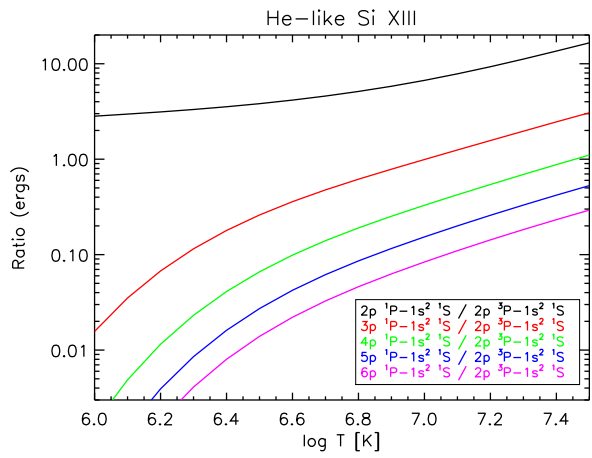}}
  \caption{ Top: G-ratio for \ion{Si}{xiii}. Bottom: other temperature-sensitive ratios 
from highly excited levels.}
  \label{fig:he-like_si_13_ratios}
\end{figure}

It has been known that 
ratios of the decays from the 3p, 4p, 5p, etc.  to the 
decay from the 2p to the ground state are also temperature sensitive
\citep[see, e.g. ][for a discussion on \ion{Si}{xiii}]{keenan_etal:1990_si_13}.
 Figure~\ref{fig:he-like_si_13_ratios} (right) shows these ratios for \ion{Si}{xiii},
as obtained from the atomic data discussed in  \cite{fernandez-menchero_etal:2016_h_he}.
They are in principle better than the  G-ratio (shown on the left of the figure)
for two main reasons. First, the ratios  vary more than the G-ratio, as  is clear from 
the figure. Second, the resonance and forbidden lines are more difficult to 
calculate accurately. For example, the contribution of the satellite lines 
 (not included in the plot), which is particularly important for the resonance 
line,  needs to be included in the modelling.
One drawback of the transitions from highly excited levels is that 
these lines are weaker.

\cite{kepa_etal:2006} however found significant discrepancies
between observed and predicted ratios in various He-like ions  during the impulsive phase
of solar flares observed with RESIK. Previously, discrepancies were also 
noted  \citep[see, e.g. ][]{keenan_etal:1990_si_13}.
Therefore, the results obtained from these ratios should be treated 
with caution until these discrepancies are understood.  
The atomic data for these highly excited states have been 
reviewed by \cite{fernandez-menchero_etal:2016_h_he}, where the possibility 
that non-Maxwellian electron distributions would increase the ratios
was investigated. The  non-Maxwellian distributions only increased the 
ratios at lower temperatures, so other processes are likely increasing
these ratios.

\subsubsection{$T_{\rm e}$ from Li-like ions}

 \cite{heroux_etal:1972} and other authors have proposed the use of 
2s--2p / 2s--3l (l=$s,p,d$) ratios of  Li-like ions to obtain information
about the temperature. 
However, the temperature sensitivity of these transitions is so different that 
the lines could easily be formed over different temperatures.
The $G(T)$ of these lines are so extended that it is possible that 
the lines are not formed within the same volume, hence a temperature
obtained from the ratio does not have much physical meaning.
Indeed \cite{heroux_etal:1972} did not obtain a temperature
from the observed ratios (obtained with an excellent rocket spectrograph
that observed the Sun as a star), but rather predicted the intensities of the 
lines using a $DEM$ obtained from other lines. With the ionization equilibrium 
tables available at the time, relative good agreement was found. 
\cite{heroux_etal:1972} discuss the various lines and possible blending issues.
Examples  are the O~VI 1032, 173~\AA\ lines, later 
 observed by SOHO SUMER and CDS/GIS \citep{david_etal:1998}, as discussed below.
Off-limb measurements of the quiet Sun are clearly an opportunity to obtain direct measurements
without depending too much on the $DEM$, since plasma is normally nearly isothermal.
\cite{delzanna:12_sxr1} showed that the Mg X soft X-ray lines have some 
temperature sensitivity.

\subsubsection{$T_{\rm e}$ from Be-like ions}

\begin{figure}[!htbp]
\centerline{\includegraphics[width=12cm,angle=0]{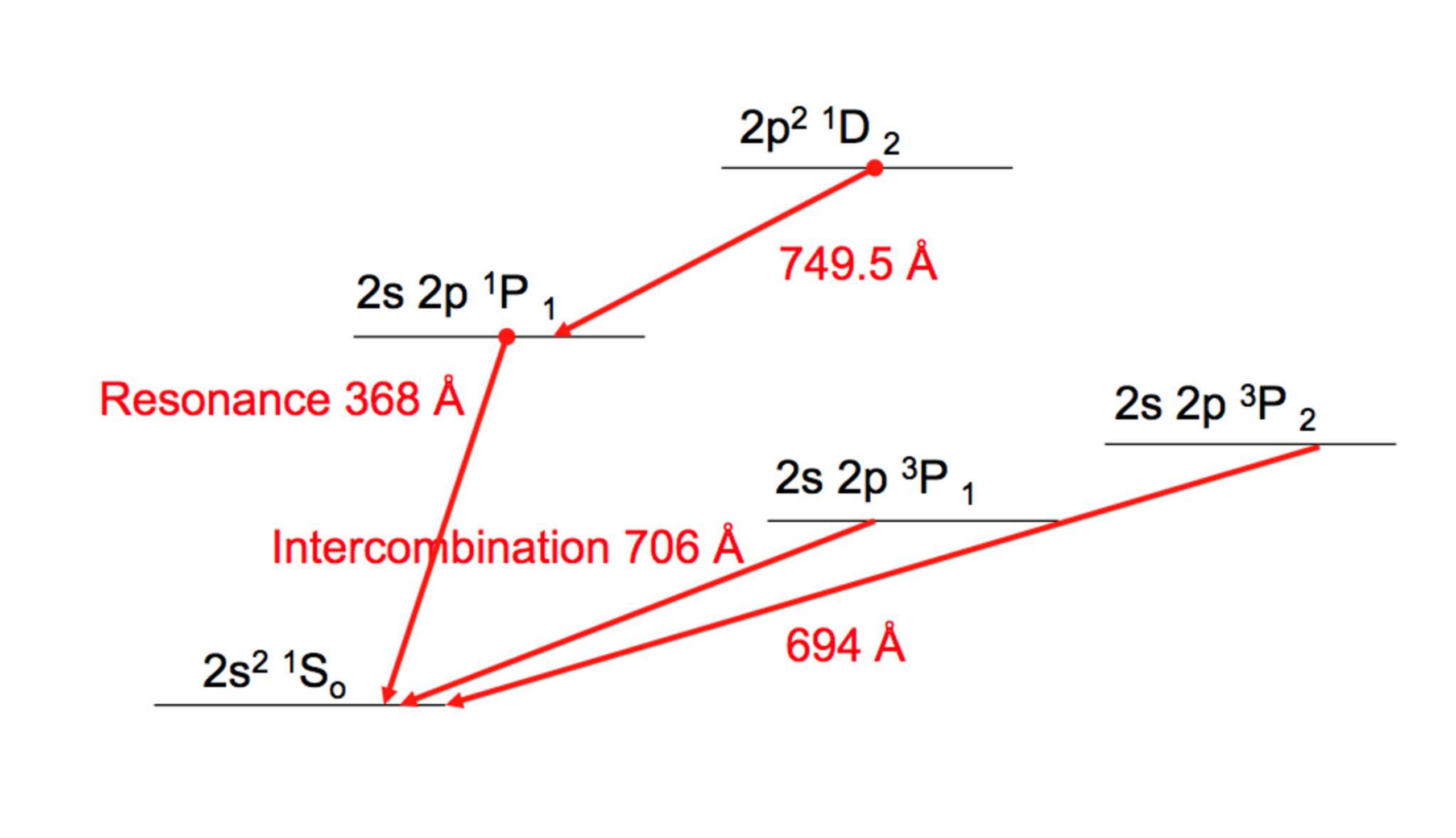}} 
  \caption{Diagram (not to scale) of the main diagnostic levels 
for \ion{Mg}{ix}.
}
\label{fig:mg_9_diagram}
\end{figure}

The Be-like ions offer excellent temperature  diagnostics, 
in particular with  the ratios of the 
strong resonance and intercombination lines, which have very little
density sensitivity and are very strong.
The atomic structure of Be-like 
ions is such that the intercombination line has a 
wavelength about twice the resonance line, so both lines 
appear close in wavelength, if the instrument is sensitive
to  second order diffraction.
A list of the main ions is provided in Table~\ref{tab:te_be-like},
while  Fig.~\ref{fig:mg_9_diagram} shows a sketch of an 
energy level diagram with the main diagnostic levels 
listed in the Table.

Of all the Be-like ions, \ion{Mg}{ix} is the one formed 
closest to the average coronal temperature (1~MK), so it is the best 
diagnostic for the average solar corona.
The lines at  368.07, 
706.06\AA\ have been observed with e.g. SOHO CDS GIS 
(see, e.g. \citealt{delzanna_thesis99,delzanna_etal:08_mg_9}).
The resonance line is somewhat blended though with several 
transitions, depending on the instrument resolution.
Excellent agreement between observations and theory has been found
only recently with new atomic data, as described below.

\begin{table}[!htbp]
\caption[Be-like temperature diagnostics.]{Be-like temperature  diagnostics.
The ratios of the resonance ($\lambda_2$) and intercombination ($\lambda_3$) lines
are temperature-sensitive. The ratios of the lines in the fifth
column ($\lambda_4$) with the intercombination ($\lambda_3$) lines
are also temperature-sensitive. Finally, another possibility is to use the 
ratios of the lines in the first column  ($\lambda_1$) with the resonance lines. 
The log $T$  values  [K] in the last column indicate the approximate range where 
the ions can be used. Lines denoted with `bl' are blended, those with `w'
are so weak that we are not aware of any  observed intensities.  
Wavelengths are in \AA.
}
\begin{center}
\begin{tabular}{llllllll}
\toprule
 Ion & 2s$^2$ $^1$S$_{0}$-- &  
2s$^2$ $^1$S$_{0}$-- & 
2s$^2$ $^1$S$_{0}$-- & 2s 2p $^1$P$_{1}$-- &  \\
   &   2s 3p $^1$P$_{1}$ & 2s 2p $^1$P$_{1}$ & 2s 2p $^3$P$_{1}$ & 2p$^2$ $^1$D$_{2}$ & \\

\toprule
         & $\lambda_1$   & $\lambda_2$  & $\lambda_3$ &  $\lambda_4$ &  log $T$   \\
\midrule

\ion{C}{iii} & 386.20 w &  977.02 & 1908.73 & 2297.58 w & 4--4.8 \\ 
\ion{N}{iv}  & 247.20 &  765.15 & 1486.50 & 1718.55 w & 4--5.2  \\ 
\ion{O}{v}   & 172.17 & 629.73 & 1218.34 & 1371.30 & 4--6 \\ 
\ion{Ne}{vii}& 97.49 w  & 465.22 &  895.17 & 973.33 (bl?) & 5--6 \\ 
\ion{Mg}{ix} & 62.75  & 368.07 (bl) &  706.06 & 749.55 & 5--7 \\ 
\ion{Al}{x}  & 72.99 w & 332.79 & 637.76 & 670.05 & 5--7  \\ 
\ion{Si}{xi}  & 43.75   & 303.32 (bl \ion{He}{ii}) & 580.92 & 604.15 & 5--7 \\ 
\ion{S}{xiii} & 32.24   & 256.68 & 491.46 & 500.33 w & 6--7 \\ 
\ion{Ar}{xv}  & 221.13 & 423.97 & ?  & ? & 6--7 \\  
\ion{Ca}{xvii}  & 19.56 w & 192.85 (bl) & 371.05 & 357.79 w & 6--8 \\ 
\ion{Fe}{xxiii} & 11.02 & 132.91 (bl) & 263.76 & 221.34 w & 6--8 \\ 
\ion{Ni}{xxv}   &  9.35 & 117.94   & 238.86 & 188.15 w & 6--8 \\  
\bottomrule
\end{tabular}
\end{center}
\label{tab:te_be-like}
\end{table}

A good temperature diagnostic for the TR is the ratio of 
the  O V 1218~\AA\ with the 
resonance line at 629~\AA\ \citep[see, e.g.][]{keenan_etal:1994}.
\cite[][]{dufton_etal:1978} used \ion{N}{iv} and 
\ion{Ne}{vii} Skylab observations of temperature-sensitive ratios 
to find values close to the peak abundance in ionization equilibrium.
\cite{keenan:1991_ne_7} used Skylab observations of various features 
and obtained from three  temperature-sensitive \ion{Ne}{vii} ratios 
values close to the peak abundance in ionization equilibrium (log T=5.7).
A similar study on \ion{N}{iv} was carried out by \cite{keenan:1990_n_4}.


The ratio of the resonance (192.8~\AA) with the 
intercombination line (371.1~\AA) in \ion{Ca}{xvii} can in principle be used to 
measure the temperature  \citep[see, e.g.][]{dufton_etal:1983_ca_17}.
The resonance line is however blended with strong \ion{Fe}{xi}
and \ion{O}{v} transitions (cf. Hinode EIS spectra, \citealt{delzanna_etal:2011_flare})
so accurate deblending needs to be applied to 
the observations. 
The resonance/intercombination ratio for 
Ar XV is discussed in \cite{keenan_etal:1989_ar_15}. Good agreement 
between theory and Skylab observations was found.
A similar study on \ion{S}{xiii}, with similar conclusions,
was presented in \cite{keenan_etal:1988_s_13}.
The ratio of the intercombination with the resonance line 
in Fe XXIII is a very good temperature diagnostic, although the 
resonance line is blended with a relatively strong 
\ion{Fe}{xx} line. 
 Examples ar provided e.g. in 
\cite{keenan_etal:1993_fe_23,delzanna_etal:2005_fe_23}.


Another diagnostic possibility is the ratio of the 
 2s 2p $^1$P$_{1}$--2p$^2$ $^1$D$_{2}$ transition with the 
intercombination. For a recent review of this
ratio see \cite{landi_etal:2001}.  Table~\ref{tab:te_be-like}
lists the main ions. 
The advantage of this ratio is that the lines are close in wavelength,
and as in the previous case the ratio is insensitive to density.
However, the $^1$P$_{1}$--$^1$D$_{2}$ transitions are extremely weak
and only those for a few ions have been observed. 
The \ion{Mg}{ix} ratio has been  applied to  SOHO SUMER
measurements, but will  also be available to the Solar Orbiter 
SPICE spectrometer, in off-limb observations, since on-disk the 
weak 749.55~\AA\ line is blended with cool emission.


A third diagnostic possibility is the ratio of the 
2s$^2$ $^1$S$_{0}$--2s 3p $^1$P$_{1}$ with the resonance line. 
This is similar to the Li-like one. The downside is that lines 
are very far in wavelength, so only some simultaneous observations exist.
\cite{malinovsky_etal:1973} used the O V [629/172~\AA] ratio to find 
an average temperature log $T$=5.4 from the full-Sun grazing-incidence
spectrum.
Finally, 
\cite{keenan_etal:2006_fe_15b} and \cite{delzanna:12_sxr1} showed that the 
Mg IX soft X-ray lines have some 
temperature and density sensitivity, but it is only with recent atomic data 
that good agreement with observations has been  found.

\subsubsection{$T_{\rm e}$ from  B-like ions}


\cite{dwivedi_gupta:1994} suggested using the ratio of the
\ion{O}{iv} 790/554~\AA\  multiplets to measure TR temperatures.
\cite{oshea_etal:1996} used Skylab S-055 observations of the 
ratio of the \ion{O}{iv} 790/554~\AA\  multiplets to find averaged temperatures
close to those of  peak abundance in ionization equilibrium.
\cite{peng_pradhan:1995} suggested the use of similar ratios 
in other ions along the sequence, but it turns out that 
some are sensitive to density and the variation with temperature
is very small.
\cite{doron_etal:2003} suggested the use of 
the \ion{N}{iii} lines around 990~\AA\ to infer temperatures
(see below).
\cite{keenan_etal:1995_ne_6} investigated the use of 
\ion{Ne}{vi} line ratios 
 using Skylab observations, but found significant discrepancies
between theory and observations, mostly due to blending and 
temporal variability.
\cite{foster_etal:1997} studied the diagnostics available 
within \ion{Mg}{viii} and found several useful ratios in the EUV.
However, all ratios depended on both density and temperature.
The approach suggested is to use ratio-ratio plots to infer
both temperatures and densities at the same time.

\subsubsection{$T_{\rm e}$ from  C-like ions}

\cite{keenan_aggarwal:1989}
studied the Te-sensitive 703/599.6 and 834/599.6~\AA\
 line ratios from the C-like O III.
They considered the quiet Sun Skylab observations of 
\cite{vernazza_reeves:1978} and obtained isothermal 
temperatures in the range log $T$[K]=4.7--5.2, i.e. close to the 
maximum ion abundance value in ionisation equilibrium (log $T$[K]=5.0).
However, the temperature sensitivity of the  703/599.6~\AA\ ratio is very 
small around those temperatures, so the ratios 
834/599.6~\AA\ or even better the 835/599.6~\AA\ are to be preferred.
This and similar ratios along the isoelectronic sequence 
are shown in Table~\ref{tab:te_c-mg-al-like}.
\cite{keenan_etal:1989_o_3} suggested the use of several 
\ion{O}{iii} lines in the 500-600~\AA\ range to measure 
temperatures, and obtained, using observations from the 
Skylab slitless spectrograph, temperatures close to 10$^5$ K,
i.e. around the peak abundance in equilibrium.
However, at such temperatures, recent atomic data 
show that the variations of the suggested ratios are very small,
hence these ratios are not very useful temperature diagnostics.
\cite{keenan_etal:1986_ne_5} suggested the use of 
either the \ion{Ne}{v} 569.8 or 572.3~\AA\ vs. the 416.2~\AA\
line to measure temperatures, since these ratios have a small
density sensitivity at the typical densities where this ion
is formed. These authors used Skylab  observations of a flare and 
a prominence obtained with the   NRL  slitless spectrometer and found values close to the peak 
temperature in equilibrium (log $T$[K]=5.5).
Between the two lines, the 572.3~\AA\ is a better choice
because the 569.8~\AA\ line is weaker and blended.
The 572.3~\AA\ vs. the 416.2~\AA\ ratio is indeed a very good 
one for the TR.
\cite{keenan_etal:1992_ne_5} suggested the use of further line
ratios from this ion, but their temperature sensitivity is 
not very good. 
We found that 
another good temperature diagnostic is the Mg VII ratio of 
the 278.404 vs. the 434.923~\AA\ lines, although the 
278.404~\AA\ line is blended with a Si VII line even at 
high resolution.  

\begin{table}[htb]
\caption[Temperature diagnostics: C-like, Mg-like, Al-like.]{
Temperature  diagnostics for C-like, Mg-like, and Al-like ions.
The log $T_{\rm e}$  values  [K] indicate the range of sensitivity.
Wavelengths are in \AA. bl indicates that the transition is blended.
sbl indicates that the line is a self-blend of transitions from the same ion.
}
\begin{center}
\begin{tabular}{llllllll}
\toprule
Transition 1  &  $\lambda_1$  (\AA)  & Transition 2  & $\lambda_2$ (\AA) & Ion &
log $T_{\rm e}$   \\
\midrule
2s$^2$ 2p$^2$ $^1$D$_{2}$ - 2s 2p$^3$ $^1$D$_{2}$ & 599.59 & 
2s$^2$ 2p$^2$ $^3$P$_{1}$ - 2s 2p$^3$ $^3$D$_{1,2}$ & 833.7 (sbl) &  O III & 4--5.5 \\

2s$^2$ 2p$^2$ $^1$D$_{2}$ - 2s 2p$^3$ $^1$D$_{2}$ & 599.59 & 
 2s$^2$ 2p$^2$ $^3$P$_{2}$ - 2s 2p$^3$ $^3$D$_{1,2,3}$ & 835 (sbl) & O III & 4--5.5 \\

2s$^2$ 2p$^2$ $^1$D$_{2}$ - 2s 2p$^3$ $^1$D$_{2}$ &  416.21 & 
2s$^2$ 2p$^2$ $^3$P$_{2}$ - 2s 2p$^3$ $^3$D$_{3}$ & 572.34  &  Ne V & 4.5--6. \\


2s$^2$ 2p$^2$ $^3$P$_{2}$ - 2s 2p$^3$ $^3$S$_{1}$ & 278.40 (bl) &  
2s$^2$ 2p$^2$ $^3$P$_{2}$ - 2s 2p$^3$ $^3$D$_{3}$ & 434.92 & Mg VII & 5--6.5 \\
\midrule
 3s$^2$ $^1$S$_{0}$ - 3s 3p $^1$P$_{1}$ &  1206.50 &  
3s$^2$ $^1$S$_{0}$ - 3s 3p $^3$P$_{1}$ &  1892.03   &  \ion{Si}{iii} & 4.3--5.5  \\

 3s$^2$ $^1$S$_{0}$ - 3s 3p $^1$P$_{1}$ &  786.47 &  
3s$^2$ $^1$S$_{0}$ - 3s 3p $^3$P$_{1}$ &  1199.13  &  \ion{S}{v} & 4.5--5.7  \\

 3s 3p $^1$P$_{1}$ - 3s 3d $^1$D$_{2}$ & 696.62 & 
 3s$^2$ $^1$S$_{0}$ - 3s 3p $^1$P$_{1}$ &  786.47 &  \ion{S}{v} & 4.5--5.7  \\
\midrule

 3s$^2$ 3p $^2$P$_{1/2}$ - 3s$^2$ 3d $^2$D$_{3/2}$ & 657.319 & 
 3s$^2$ 3p $^2$P$_{1/2}$ - 3s 3p$^2$ $^2$D$_{3/2}$ & 1062.664 &  \ion{S}{iv}  & 4.5--6. \\

\bottomrule
\end{tabular}
\end{center}
\label{tab:te_c-mg-al-like}
\end{table}

\subsubsection{$T_{\rm e}$ from Na-like ions}

\cite{flower_nussbaumer:1975_na-like}
suggested the use of ratios of lines from the 3s--3p and 3p--3d in 
  Na-like ions, because these lines fall relatively close in wavelength.
They applied the technique to Si IV (3p-3d)/(3s-3p)  [1128.3/1393.8] and 
S VI lines, but did not obtain 
agreement between theory and observations, because it turns out that lines 
are blended in  low-resolution spectra.

\cite{keenan_etal:1986} and \cite{doschek_feldman:1987}
studied  Al III, obtaining some results (see below).
\cite{keenan_etal:1986} also studied  \ion{Si}{iv},
but concluded that higher-resolution spectra for the \ion{Si}{iv}
 lines (1128.3, 1393.8, 815.0~\AA) were needed.
\cite{keenan_doyle:1988} later actually obtained good agreement 
between theory (using updated calculations)  and observations of the 
\ion{Si}{iv}  multiplets at 1120 and 815~\AA\
 for a solar active region at the limb obtained with the Harvard S-055 spectrometer on board Skylab,
but noted that the 1122.5~\AA\ must be severely blended with a cool emission line.
\cite{keenan_etal:1994_ca_10} investigated the use of 
Ca X line ratios using Skylab  observations of flares  with the   NRL  slitless spectrometer, but
found several discrepancies, which they ascribed to blending problems.
Indeed the 419.74~\AA\ line is blended with a strong C IV, 
and the 574.0~\AA\ line with O III.

\cite{keenan_etal:1994_fe_16} suggested the use of 
Fe XVI line ratios involving the 251, 263, and 335.4~\AA\ lines, 
and compared predicted with observed
values obtained with  the Skylab   NRL  slitless spectrometer. 
Some discrepancies were found, and ascribed to a 
problem in the calibration of the strong 335.4~\AA\ line (which is also blended).
However, aside from these discrepancies, the main 
limitation of such ratios is that they have very little
temperature sensitivity when  this ion is formed in 
ionization equilibrium.
\cite{keenan:1988_s_6} studied the \ion{S}{vi} 712.8 / 944.5~\AA\ ratio,
but concluded that both lines are blended at the resolution of the 
Skylab Harvard S055 spectrometer.

\subsubsection{$T_{\rm e}$ from  Mg-like ions}

The ratio between the intercombination and resonance line  in 
 \ion{Si}{iii} is in principle a good temperature diagnostic.
However, the ratio is also sensitive to densities above 10$^{10}$  cm$^{-3}$,
so the density has to be measured independently 
\citep[see][for a recent review on this ion]{delzanna_etal:2015_si_3}.

\cite{doyle_etal:1985_plumes, keenan_doyle:1990}
 studied  temperature-sensitive  
\ion{S}{v} line ratios, finding relatively good agreement with theory. 
The ratios involved the resonance line at 786~\AA\ and the 663.2, 852.2, and 854.8~\AA\ lines.
However, these ratios have a very small temperature sensitivity around 
10$^5$ K where  this ion is formed. 
We suggest that a  much better ratio is when the 696.62~\AA\ is considered.
Alternatively, the ratio with the UV line at 1199.13~\AA\ is also a good 
temperature diagnostic, unless very high densities above 10$^{12}$  cm$^{-3}$ are reached,
in which case the ratio also becomes density-sensitive. 
The main ratios we recommend are listed in Table~\ref{tab:te_c-mg-al-like}.


\subsubsection{$T_{\rm e}$ from Al-like ions}

\cite{doyle_etal:1985} pointed out that ratios of  
\ion{S}{iv} multiplets at 657, 750, 810~\AA\ with the 
1070~\AA\ multiplet are temperature-sensitive.
Several combinations are available, also involving the
strongest line, at 1406~\AA.
One of the best ratios is the 657.319 vs. 1062.664~\AA,
given that this has no density sensitivity below 10$^{12}$  cm$^{-3}$.
The 657.319~\AA\ line is the resonance (strongest) line for this ion.
The analogous ratio for \ion{Fe}{xiv} (211.32 vs. 334.18~\AA) does not vary much 
with temperature.

\subsubsection{$T_{\rm e}$ from S-like ions}

\cite{doschek_feldman:1987} suggested that ratios of lines from 
S III  3s$^2$ 3p 4p  levels (e.g. 1328.12, 1328.52~\AA) 
with those from the  3s 3p$^3$ levels 
(i.e. the strong line at 1200.96~\AA) would be a good temperature diagnostic,
because these lines are close in wavelength.

\subsubsection{$T_{\rm e}$ from iron ions}

Given that iron lines are very abundant in XUV spectra, it would clearly
be very useful to use diagnostics based on them.
There are many possible diagnostics, but most of them involve 
lines at very different wavelengths, so results have been scarce.
Table~\ref{tab:te_fe_ions} lists the main diagnostics which are available for 
\ion{Fe}{x}, \ion{Fe}{xi}, \ion{Fe}{xii}, and \ion{Fe}{xiii}. 
In principle, very good diagnostics are offered by the ratio of one 
of the forbidden lines within the ground configuration 
($\lambda_4$ in Table~\ref{tab:te_fe_ions}) and one of the
strongest dipole-allowed  EUV lines ($\lambda_1$  in Table~\ref{tab:te_fe_ions}). 
Such measurements are however not 
easy, as it is not simple to obtain well-calibrated simultaneous 
observations at very different wavelengths. 
Also, the accuracy of atomic data 
plays an important  role: the recent large-scale calculations for these iron ions
(discussed in the atomic data Section~\ref{sec:atomic}) have modified the predicted 
intensities of the forbidden lines of the iron coronal ions by 
large factors (two to three, relative to the dipole-allowed lines).

\begin{table}[!htb]
\caption[Temperature diagnostics from coronal Fe ions.]{
Temperature diagnostics from coronal Fe ions.
Three main wavelength ranges  (values in \AA) are shown. Any ratio among them 
is a good temperature diagnostic. Only the strongest lines are shown.
 The first group of lines ($\lambda_1, \lambda_2, \lambda_3$) 
are the strongest lines in the EUV, while the UV/visible 
lines ($\lambda_4$) are the forbidden lines within the ground configuration.
The log $T$  values  [K] indicate the approximate formation temperature of the ions.
}
\begin{center} 
\begin{tabular}{llllllll}
\toprule
Ion   &  $\lambda_1$  (\AA)  & $\lambda_2$ (\AA) & $\lambda_3$ (\AA)  & $\lambda_4$ (\AA)  & log $T$   \\
\midrule
\ion{Fe}{viii}  & 186.6, 185.21 & 130.94 & 253.96, 255.11  &  &  5.7  \\
                &               &        &  255.35,255.68  &  &        \\
\ion{Fe}{ix}    & 171.07 & 188.49,197.85    &           &   &  5.8  \\ 
\ion{Fe}{x}  & 174.53, 184.53 & 257.26(sbl) & 345.74     &  6376. & 6.0  \\
\ion{Fe}{xi}  & 180.40, 188.22 & 257.55(sbl) & 352.67 &  2649.50 & 6.1 \\ 
\ion{Fe}{xii}  & 193.51, 192.39 &             & 364.47 & 1242.0, 1349.4 & 6.2  \\
               &                &              &      &  2169.76, 2406.41 &  \\
\ion{Fe}{xiii}  & 202.04 &      &     & 10749 & 6.3  \\ 

\bottomrule
\end{tabular}
\end{center}
\label{tab:te_fe_ions}
\end{table}

On the other hand, there are several other ratios of iron lines which are 
closer in wavelength and which are temperature sensitive. 
  Some ratios for \ion{Fe}{x}, \ion{Fe}{xi}, \ion{Fe}{xii}
(see Table~\ref{tab:te_fe_ions}) involve lines around 180 and 350~\AA,
listed in the $\lambda_1$, $\lambda_3$ columns in  Table~\ref{tab:te_fe_ions}.
These diagnostics  have been known for some time, but 
lines have usually  been observed by different instruments. 
Note that the recent large-scale calculations for these iron ions
have modified the predicted intensities of the $\lambda_3$ lines  by  typically  30\%.

Several other ratios offer  temperature diagnostics, including ratios
where both lines were observed by the same instrument, 
 Hinode EIS. These  have recently been identified: 
\ion{Fe}{viii}  \citep{delzanna:09_fe_8}; 
\ion{Fe}{ix} \citep{young:09, delzanna:09_fe_7};
\ion{Fe}{xi} \citep{delzanna:10_fe_11}. Some of them are listed in 
the $\lambda_1$, $\lambda_2$ columns of Table~\ref{tab:te_fe_ions}.
Also in this case, we point out that it was only recently, with large-scale 
scattering calculations, that significant discrepancies 
between observed and predicted line intensities were resolved for 
\ion{Fe}{viii}  \citep{delzanna_badnell:2014_fe_8},
\ion{Fe}{ix} \citep{delzanna_etal_2014_fe_9}, 
\ion{Fe}{xi}  \citep{delzanna_etal:10_fe_11}.
This improvement was due to  the increases in intensities, by 30--50\%,  
for  the lines listed in  the $\lambda_2$ column. 
Temperature measurements from these ions therefore now appear  to be 
 feasible, although  results also depend on an accurate 
radiometric calibration of  Hinode EIS. This turns out to be 
particularly important as the sensitivity  of these ratios is not as large as that 
available when the forbidden lines are used.
Line ratios useful for measuring the electron temperatures from \ion{Fe}{xvi}, \ion{Fe}{xvii},
\ion{Fe}{xviii}, \ion{Fe}{xxiii} are briefly discussed below in Section~\ref{sec:ar_te}.


\subsection{$T$ from ratios of two lines  from different  ions }

Another way to estimate temperatures is to use the observed ratio
of two line intensities produced by two ionisation stages of the same element.
If ionisation equilibrium is assumed, the observed ratio can then be
directly compared to the theoretical ratio, and a temperature obtained.
This method has been widely used in the past, although its accuracy relies 
on the reliability of the ionisation/recombination rates and, most 
of all, on the ionisation equilibrium assumption. 
This last assumption is usually reasonable in the solar corona, but in the 
transition region it is clear that dynamical timescales are often shorter
than recombination times, hence departures from ionisation are to be expected.

\subsection{$T$ from several lines}

Whenever more than two lines from different ionisation stages of the same 
element are observed, one could obtain an estimate of the temperature
from the line ratios assuming ionisation equilibrium, and compare the results. 
If all the temperatures agree, the plasma is likely to be isothermal.
A more elegant way to analyse the data is to use the EM loci method,
which we discussed in Section~\ref{sec:em_loci}.
If the EM loci curves cross at one temperature, it means that the data 
are consistent with the plasma being isothermal (although that is not 
necessarily the only solution). 
If no crossing is obtained, it means that the emission is multi-thermal,
and a $DEM$ analysis should be carried out.

\subsection{$T$ from dielectronic satellite lines}

We have seen in  Section~\ref{sec:satellites} that the ratio of a satellite line formed by 
dielectronic capture with the resonance line is an excellent temperature diagnostic,
because it is independent of the population of the ions and the electron density.
The main problems with such measurements are that very high resolution 
spectra are needed, and the  contribution of the satellite lines to the resonance
line needs to be taken into account.

\subsection{$T$ from the X-ray continuum}

The spectral distribution of the  X-ray continuum depends quite 
strongly on the electron temperature, so in principle
this  continuum could be used to measure the electron temperature.
For example,  it is very common to obtain an isothermal 
temperature from the ratio of the GOES X-ray fluxes during a solar 
flare. The GOES fluxes are dominated by the continuum, 
which is mostly free-free emission for the largest flares.
However, some contribution from the  free-bound emission is normally also 
present (recall Section~\ref{sec:continuum}). 
While the free-free is mostly due to 
hydrogen, hence  is directly proportional to the hydrogen (proton)
density,  the  free-bound depends on the chemical abundances,
 so in principle  an independent 
estimate of the composition of the plasma (see Section~\ref{sec:abund})
should be carried out when measuring the temperature.

Direct measurements of the spectral 
distribution of the X-ray continuum  are difficult, and only a few observations 
are available. 
 \cite{parkinson_etal:1978} was able to obtain with excellent 
OSO-8 X-ray spectra both measurements of continuum and  line intensities
of a small flare and an active region after the flare.
An isothermal model was able to reproduce both the 
continuum and the ratio of the H-like to He-like Si 
(i.e. \ion{Si}{xiv} vs. \ion{Si}{xiii}).
The RESIK spectrometer observed the spectral range where
a significant temperature sensitivity of the continuum exist.

\begin{figure}[htb]
 \centerline{\includegraphics[width=0.6\textwidth,angle=90]{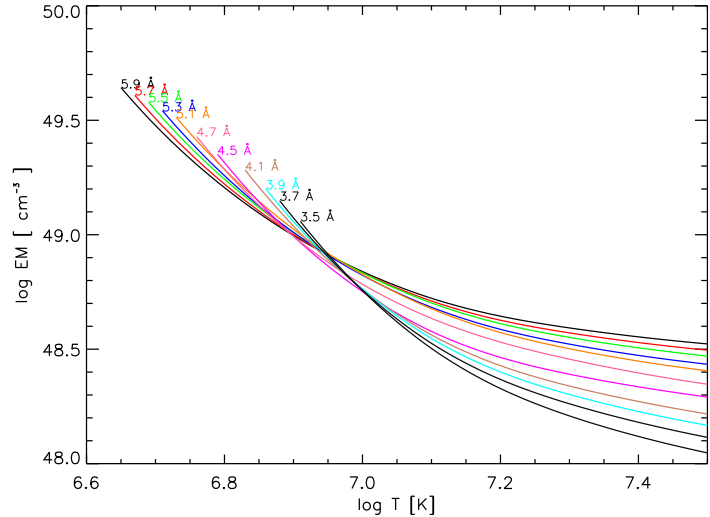}}
  \caption{EM loci curves for the X-ray continuum during the peak phase 
of a flare observed with the RESIK spectrometer 
(adapted from \citealt{chifor_etal:2007}). The crossing indicates an isothermal 
 temperature of almost 10 MK.
}
  \label{fig:resik_em_pott_cont}
\end{figure}

 A diagnostic similar to the EM loci curves, but applied to the 
X-ray continuum, was devised by one of us (GDZ) and 
applied to the X-ray continuum measured during a flare by the RESIK 
spectrometer. The emissivity of the continuum was calculated as a 
function of temperature for the observed RESIK wavelengths 
using the CHIANTI database, and compared to the observed continuum,
by plotting the ratio of the observed intensities with the  emissivity.
 Figure~\ref{fig:resik_em_pott_cont} shows the results for 
the peak phase of a flare.
As we discussed in the instrument's Section (\ref{sec:instruments}),
the RESIK instrumental fluorescence 
 created  a complex background emission
which needed to be subtracted to reliably use the continuum
for diagnostic purposes. This added some extra uncertainty on the 
continuum measurements. However, as described in  \cite{chifor_etal:2007}, 
the  continuum EM loci curves were consistent with a nearly isothermal plasma,
in good agreement with the results of the  
line EM loci curves. Furthermore, the isothermal temperature was decreasing during the flare.

\section{Measurements of Electron Temperatures} 
\label{sec:te_obs}

In what follows, we provide several 
examples of  measurements of electron temperatures for different 
solar regions, with emphasis on more recent results, to show
which diagnostics have been applied.
We stress that the types of measurements described in this 
Section are fundamental to constrain the wide-range
of coronal heating and solar wind theories \cite[see, e.g.][]{cranmer:09}

\subsection{$T$ in coronal holes}
\label{sec:ch_te}

Measurements of electron (and ion) temperatures in coronal holes are very 
 important, especially when defining the lower boundary of models
which attempt to explain how the fast solar wind becomes accelerated
inside coronal holes (see, e.g. the Living Reviews
 by \citealt{cranmer:09,poletto_lrsp}).

Estimates of the coronal temperatures have historically
been  obtained
from the ratios of lines emitted by different ions or  DEM analyses.
These were effectively 'ionisation' temperature, and not strictly
electron temperatures.
A good  review of the range of temperatures
that have been obtained from remote  measurements of coronal holes
before SoHO was given by \cite{habbal_etal:1993}.
Coronal temperatures were found to be about 1~MK, but with a large scatter and 
significant uncertainties. One of the issues was, and still is, how much of the 
emission observed off-limb in coronal holes originates from 
 the foreground/background quiet Sun, 
and whether the plasma emitted by different ions is co-spatial.

\cite{esser_etal:1995} analysed daily intensity measurements at 1.15\rsun\ of the green 
\ion{Fe}{xiv}  5303~\AA\ and red \ion{Fe}{x} 6374~\AA\ lines. 
Large temperature variations (more than 0.8~MK) were found over
the time interval of the observations which covered about four solar rotations.
  They also
discussed how the regions surrounding the coronal holes might 
have influenced the temperature measurements.

\cite{guhathakurta_etal:1992} derived temperatures from the 
green \ion{Fe}{xiv}  5303~\AA\ and red \ion{Fe}{x} 6374~\AA\
 forbidden lines, 
together with a broad-band image centred at 173~\AA\ and recorded by a sounding rocket.
They found an average coronal temperature of 1.34~MK for a south
polar coronal hole and 1.27~MK for a north polar coronal hole, at 1.15\rsun, during
the 1988 March 17-18 solar eclipse.

\subsubsection {$T$ from SOHO observations}

SOHO enabled  more direct measurements of electron temperatures. 
Also, coronal hole plumes could be differentiated from 
the inter-plume regions (lanes). There were  several studies on coronal hole and plume 
temperatures obtained from various methods. 
The most common ones were based on emission measure analyses,
see e.g.  the `Elephant's trunk' equatorial 
coronal hole measurements \citep{delzanna_jgr99a}.
The plume bases were shown to be nearly isothermal, at 
temperatures below 1~MK (cf. \citealt{delzanna_etal:03}).
Off-limb SUMER observations near the limb indicated nearly isothermal
plasma at a temperature just below 1 MK \citep{feldman_etal:98b}. 
For a recent review on coronal hole
measurements, see \cite{wilhelm_etal:2011}.

Measurements from line ratios, which are in principle more accurate,
were also obtained. 
The  Mg IX lines around 700~\AA\  are potentially an excellent 
temperature diagnostic, available for SOHO SUMER and CDS observations.
The Mg IX  resonance and intercombination lines 
are also an excellent temperature diagnostic, as shown in 
 \cite{delzanna_etal:08_mg_9} using SOHO/GIS observations, where
the resonance 368.07~\AA\  line is observed in second order 
near the intercombination line at 706.06~\AA.
\cite{wilhelm_etal:98} presented 
SOHO SUMER off-limb  (in the range 1.03 - 1.6 \rsun) 
observations of plume and inter-plume regions, and measured
temperatures  using the  Mg IX 749/706~\AA\  line ratio.
Plume temperatures  were found to be  approximately 
constant around $T=780\ 000$ K and then  decrease at higher 
radial distances,  while inter-plume 
temperatures were found  to increase over the  same height range.
These  results were limited by the fact that 
 plumes and  inter-plume regions were compared using data over a time span of six months, 
and  at different spatial locations. 
Another limiting factor was the use of atomic data that were 
interpolated, not calculated. 
The first $R$-matrix scattering calculation for Mg IX was 
carried out by  \cite{delzanna_etal:08_mg_9}. 
This  new atomic data produced significantly different temperatures.
For example,  a coronal hole inter-plume 
temperature of 850\ 000 K  found by \cite{wilhelm_etal:98}
was  revised  to 1160\ 000 K.

\cite{david_etal:1998} used the O VI 173/1032~\AA\ ratio
to measure temperatures from  off-limb spectra
in coronal holes and quiet Sun areas. 
Close to the limb, coronal holes were found to be about 0.8~MK,
rising to about 1~MK around 1.15 \rsun, and then decreasing down to 
about 0.4~MK at 1.3 \rsun. 
However, these results 
should be treated with caution for several reasons. 
First, SOHO was rotated and the slit were therefore tangential
to the limb of the Sun, i.e. plume and lanes could not easily be distinguished.
Second, the observations were obtained by 
two different instruments (CDS/GIS and SUMER), with an uncertain 
relative calibration. This was not in principle a major problem since the
measurements were made relative to the quiet Sun.
However,  the GIS observations used the 
long slit and not one of the usual pinhole slits. 
The spectra obtained with the pinhole slit have been calibrated
in-flight \citep{delzanna01_cdscal,kuin_delzanna:07}, but the 
long slit observations are impossible to calibrate, since the detectors were not
fully illuminated in this case.

Finally, it is worth mentioning that 
SUMER (e.g.,  \citealt{hassler97}) and  UVCS  
(e.g. \citealt{noci97})  off-limb observations  
have shown that the spectral line widths in the inter-plume lanes 
are  larger than in the plumes, indicating lower ion temperatures in 
plumes as a possible explanation.
Very large non-thermal widths in the 
O VI lines were  also discovered by UVCS.
They are  signatures of strong anisotropies, much larger than
those of the protons, already measured pre-SOHO.
These observations have prompted  new theories on the heating and
acceleration of the fast solar wind. For more details, see the reviews by 
\cite{kohl_etal:2006} and \cite{cranmer:09}.

\subsection{$T$ in the quiet Sun}
\label{sec:qs_te}

There is a vast literature on ratios of lines emitted by 
different ions and emission measure analyses of 
quiet Sun areas, most of which indicate that the quiet Sun 
has a near isothermal DEM  distribution around 1~MK 
(see, e.g. \citealt{feldman_etal:99a,landi_etal:2002_sumer,young:2005,brooks_warren:2009}).
These results are not reviewed here since we focus 
on diagnostics from individual ions.

\cite{keenan_etal:1986} used Skylab NRL S082B spectra of the QS and obtained 
a range of values, log $T$[K]= 4.4--4.7,
from the  Al III 3p--3d 1611.9~\AA,  3p--4s 1379.7~\AA, and 
3s--3p 1862.8 1854.7~\AA\ lines 
(in equilibrium the Al III abundance peaks at log $T$[K]= 4.6).
Similar values were obtained by \cite{doschek_feldman:1987_al_3},
i.e. temperatures slightly below those of the ion peak abundance in ionisation
equilibrium. However, slightly lower temperatures would be expected,
given the steep gradient in the emission measure distribution.

Lines from Si III have been used in the literature to measure 
temperatures, however as suggested by  \cite{doschek:1997} and 
 discussed in detail in  \cite{delzanna_etal:2015_si_3} 
the line ratios lose meaning because of the steep temperature gradients
where the Si III lines are formed. Lines sensitive to different temperatures
are likely to be formed in different layers of the atmosphere,
so the line ratio method cannot really be applied.

\cite{doron_etal:2003}
obtained SUMER measurements of the B-like N III lines around 990~\AA\
and obtained, assuming  isothermal emission,
temperatures in the range 
5.7$\times 10^4$--6.7$\times 10^4$ K,
considerably lower than the calculated temperature of maximum abundance 
of N III, which is about 7.6$\times 10^4$ K.
The Na-like  3p--3d and 3s--3p transitions in Si IV required a UV instrument like 
SOHO SUMER.
\cite{doschek_etal:1997_si_4} used SUMER measurements of the 
1128.32, 1128.34, 1402.77~\AA\ Si IV lines  in the quiet Sun near disk centre 
and found temperatures systematically
higher than the temperature of maximum ion abundance in equilibrium.
\cite{ahmed_etal:1999} on the other hand measured Si IV ratios involving the 
1394, 1403 and 818~\AA\ lines for the quiet Sun 
at disk centre  with SOHO  SUMER finding log $T$[K]=4.8, which is close to 
the temperature of maximum ion abundance.
 \cite{doschek:1997} pointed out that the uncertainty in the 
Skylab S082B calibration and the density sensitivity of the 
Be-like  O V 1371.29A/1218.35~\AA\ ratio makes the temperature results 
from this ion very uncertain. 
The uncertainty is even larger for the Be-like C III 
1247/1908~\AA\ ratio.

\cite{muglach_etal:2010}
recently attempted a measurement of the QS temperature from 
ratios of \ion{O}{iv} and \ion{O}{vi}  lines observed by Hinode EIS and 
SOHO SUMER in 2007; however, large discrepancies with theory were found,
with measured ratios being factors between 3 and 5 less than those expected in 
ionisation equilibrium.
The authors checked the  relative radiometric 
calibration and increased the SUMER radiances by a factor of 1.3.
 However, the relative  calibration of the instruments is probably a 
source of error. Indeed the Hinode EIS  ground calibration was found to 
be incorrect even in early observations  \citep{delzanna:13_eis_calib}.

In contrast, \cite{giunta_etal:2012} analysed similar Hinode EIS and SOHO SUMER 
observations in 2009 of the same \ion{O}{iv} 787.72 / 279.93~\AA\
ratio and found  reasonable temperatures in the range 
log $T$=5.2--5.4, close to those expected from ionisation equilibrium.
The main difference was the adoption of a larger atomic model,
and an increase of the SUMER radiances by a factor of 1.5.
However, the \cite{giunta_etal:2012} results are even more affected by the
degradation of the EIS instrument and should be revisited.

\begin{figure}[!htbp]
 \centerline{\includegraphics[width=0.5\textwidth,angle=90]{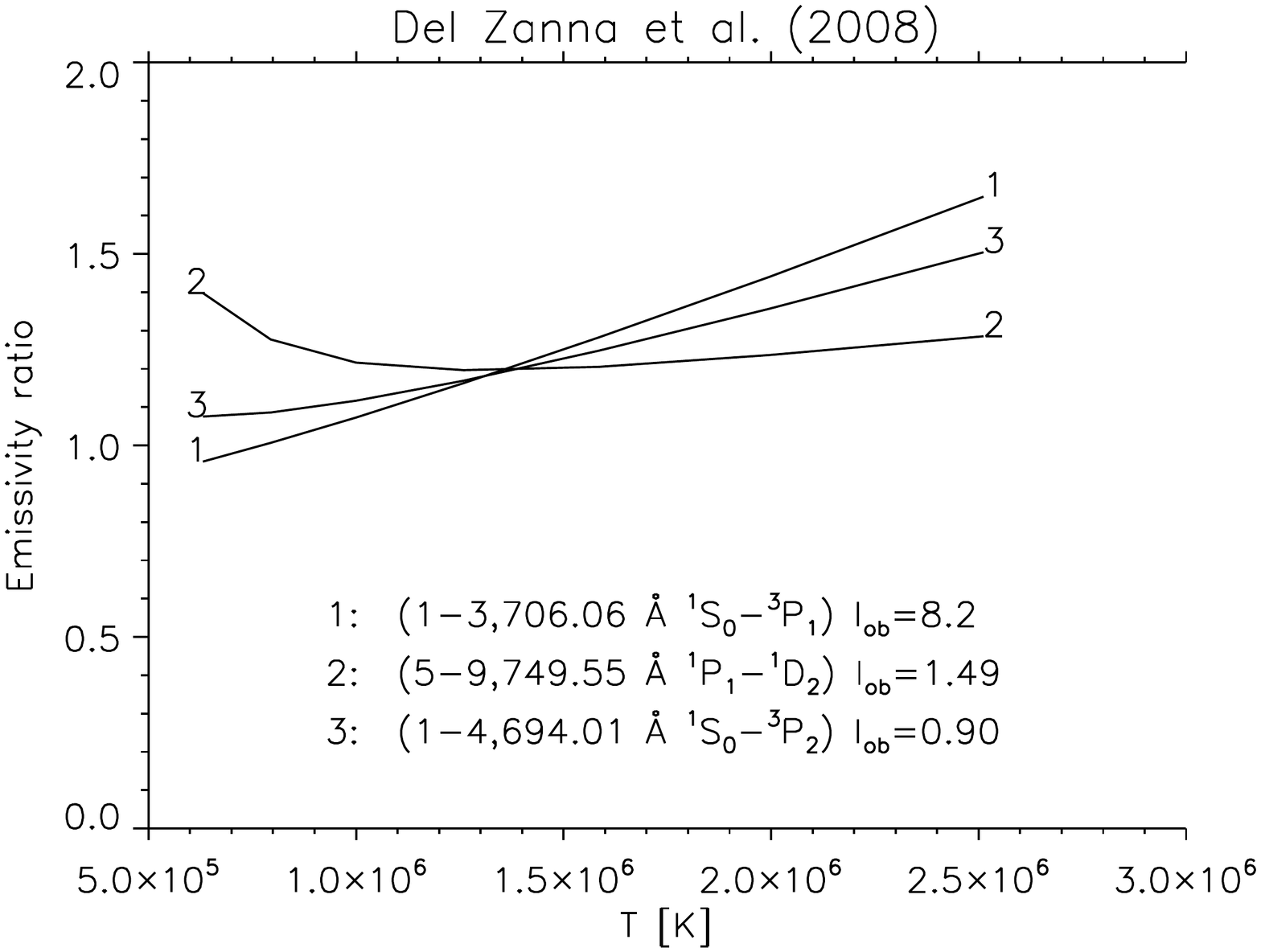}}
 \centerline{\includegraphics[width=0.7\textwidth]{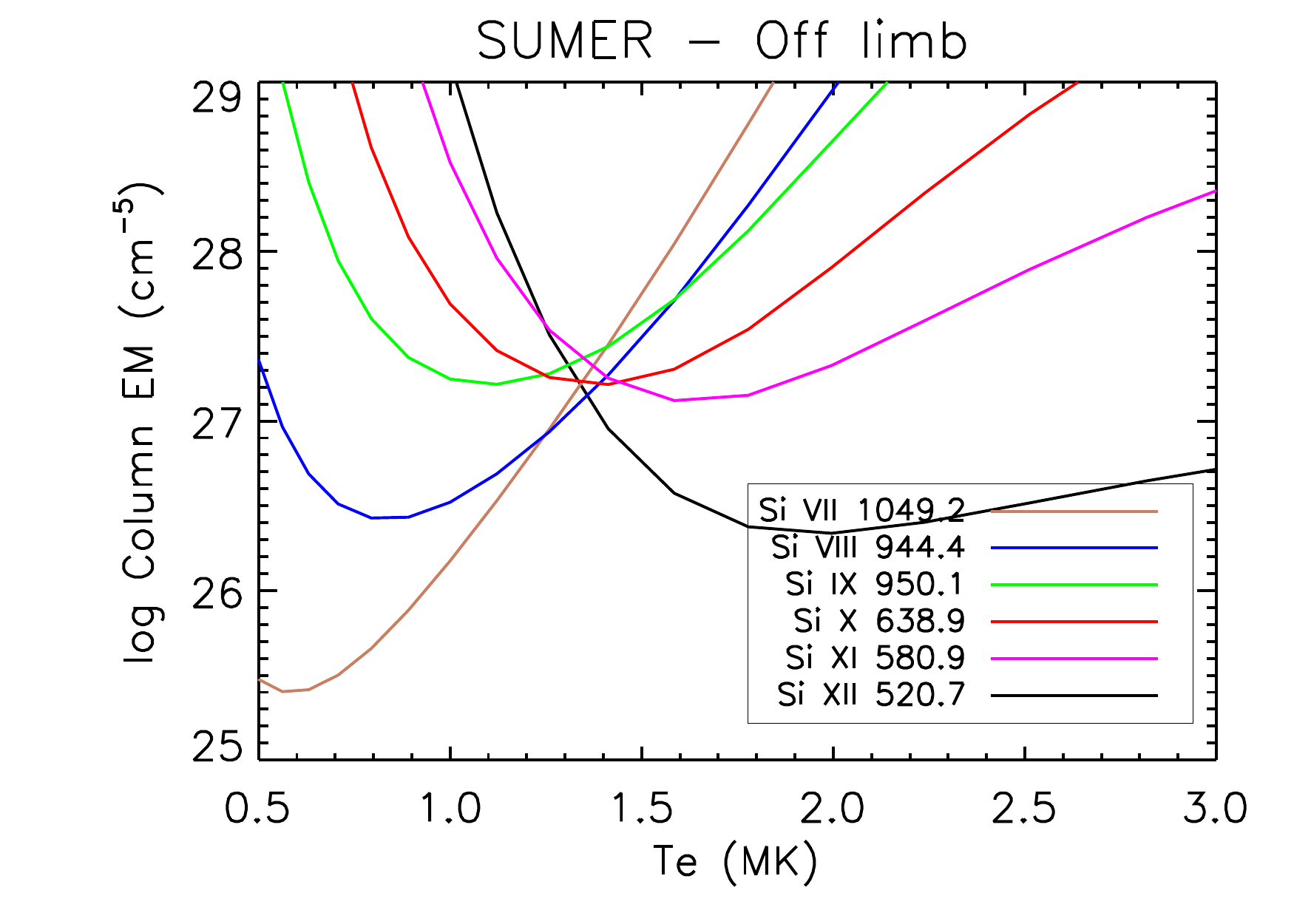}}
  \caption{Top: emissivity ratio curves  of  Mg IX lines observed by 
 SoHO SUMER in an off-limb quiet Sun area (figure adapted from \citealt{delzanna_etal:08_mg_9}).
Bottom: EM loci curves of silicon ions obtained from the same observation
and CHIANTI v.8 data. 
}
 \label{fig:mg_9_emratios}
\end{figure}
 
\cite{delzanna_etal:08_mg_9} presented new  $R$-matrix 
calculations for Mg IX and used them 
 to obtain an isothermal temperature of 1.35~MK in an off-limb quiet Sun area
measured with SoHO SUMER and discussed by  \cite{feldman_etal:99a,landi_etal:2002_sumer}. 
The emissivity ratio curves are shown in Fig.~\ref{fig:mg_9_emratios} (top).
This temperature is in excellent agreement with the one obtained 
from the EM loci curves, which indicates that the corona is approximately 
isothermal, as already noted by \cite{feldman_etal:99a}.
The EM loci curves shown in   Fig.~\ref{fig:mg_9_emratios} (bottom)
are  obtained with CHIANTI v.8 data. We note that previous atomic data
for Mg IX were interpolated and provided a lower temperature of 1~MK.
We also note that the EM loci plot in  Fig.~\ref{fig:mg_9_emratios} (bottom)
is slightly different than the one presented by \cite{feldman_etal:99a},
because of the use of more recent atomic data.

\subsection{$T$ in active regions and flares}
\label{sec:ar_te}

For  general reviews on studies of  active regions and the importance
of measuring temperatures 
see, e.g.  \cite{reale:2012_lr} and \cite{mason_tripathi:2008}.

\cite{keenan_doyle:1988} used Skylab 
 Harvard S-055 spectra  of an active region and a sunspot
to derive  log $T$[K]=4.6 from the Si IV 
ratio of the 1128.3 and 818.1~\AA\ lines.
This temperature is  only slightly lower than  the equilibrium 
value of log $T$[K]=4.8.
\cite{keenan_doyle:1990} 
measured  the temperature-sensitive  ratios R1 = 854.8~\AA\ / 786.9~\AA, 
R2 = 852.2~\AA\ / 786.9~\AA, R3 = 849.2~\AA\ / 786.9~\AA, 
and R4 = 1199.1~\AA\ / 786.9~\AA\ using sunspot data 
obtained with the Harvard S-055 spectrometer on board Skylab.
They found overall agreement between theory and observations, 
except in the case of R1,  probably due to blending of the 854.8~\AA\ line.

\begin{figure}[!htbp]
\centerline{
\includegraphics[width=7.5cm]{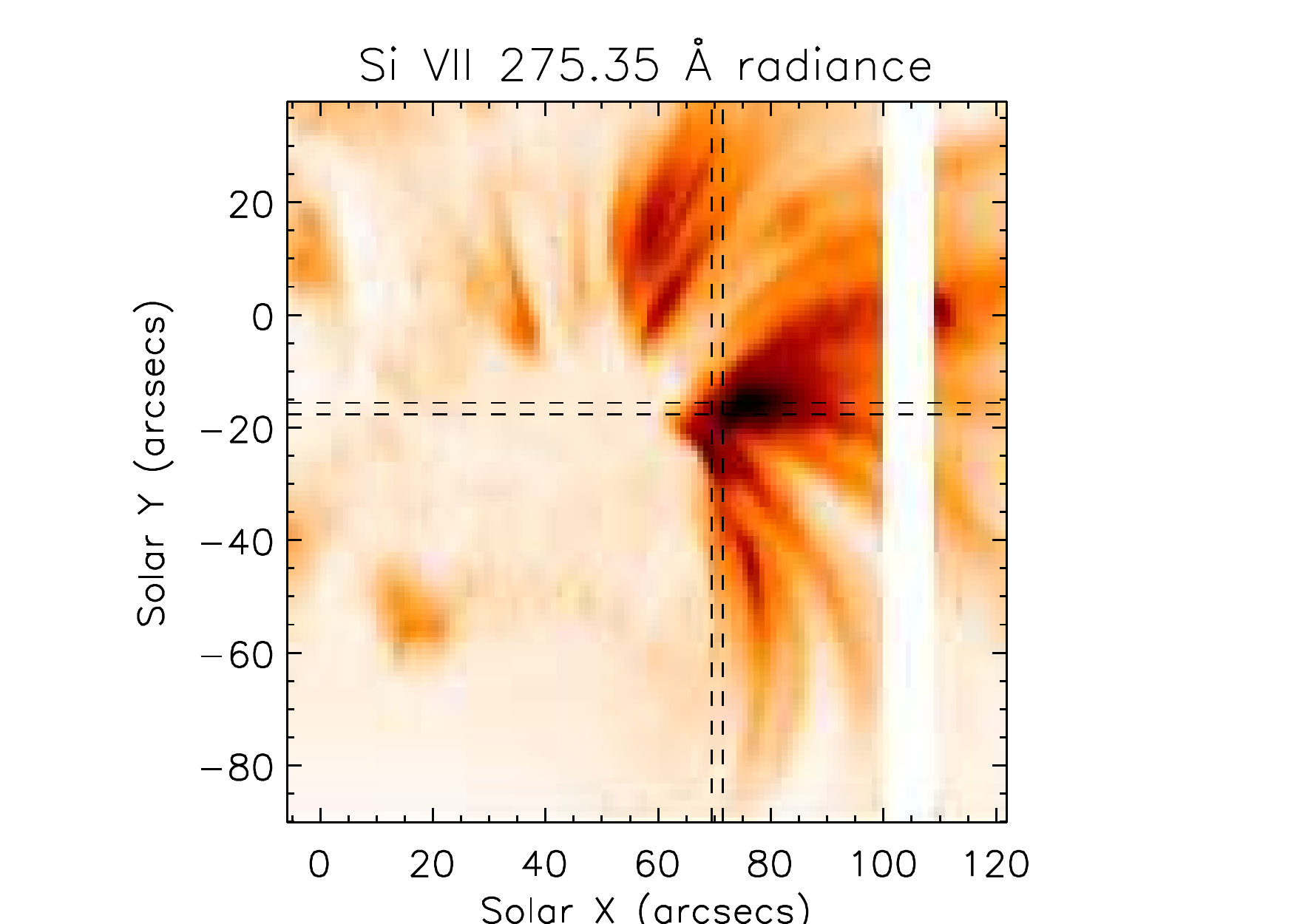}
\includegraphics[width=7.5cm]{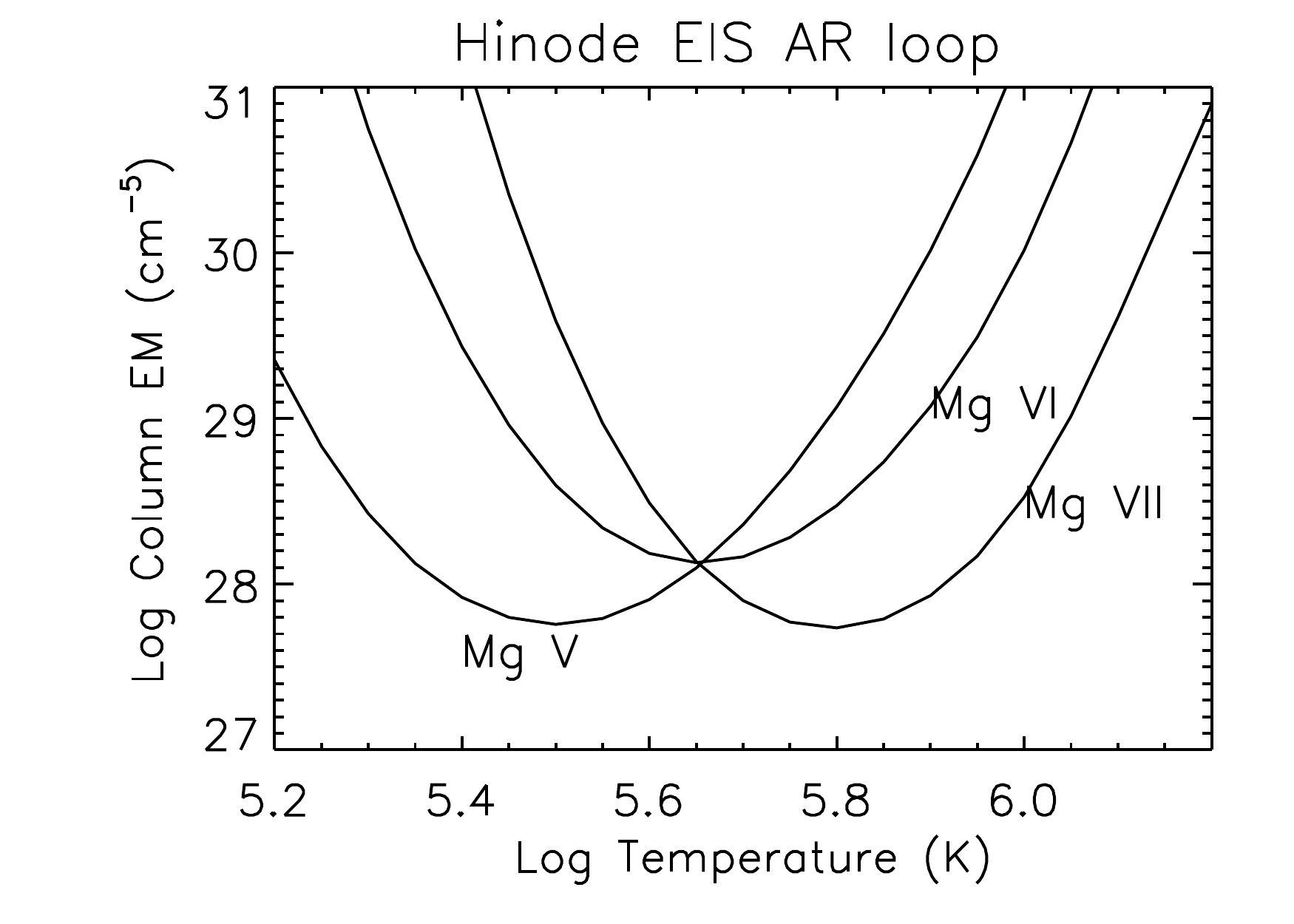} 
}
\centerline{\includegraphics[width=6cm,angle=90]{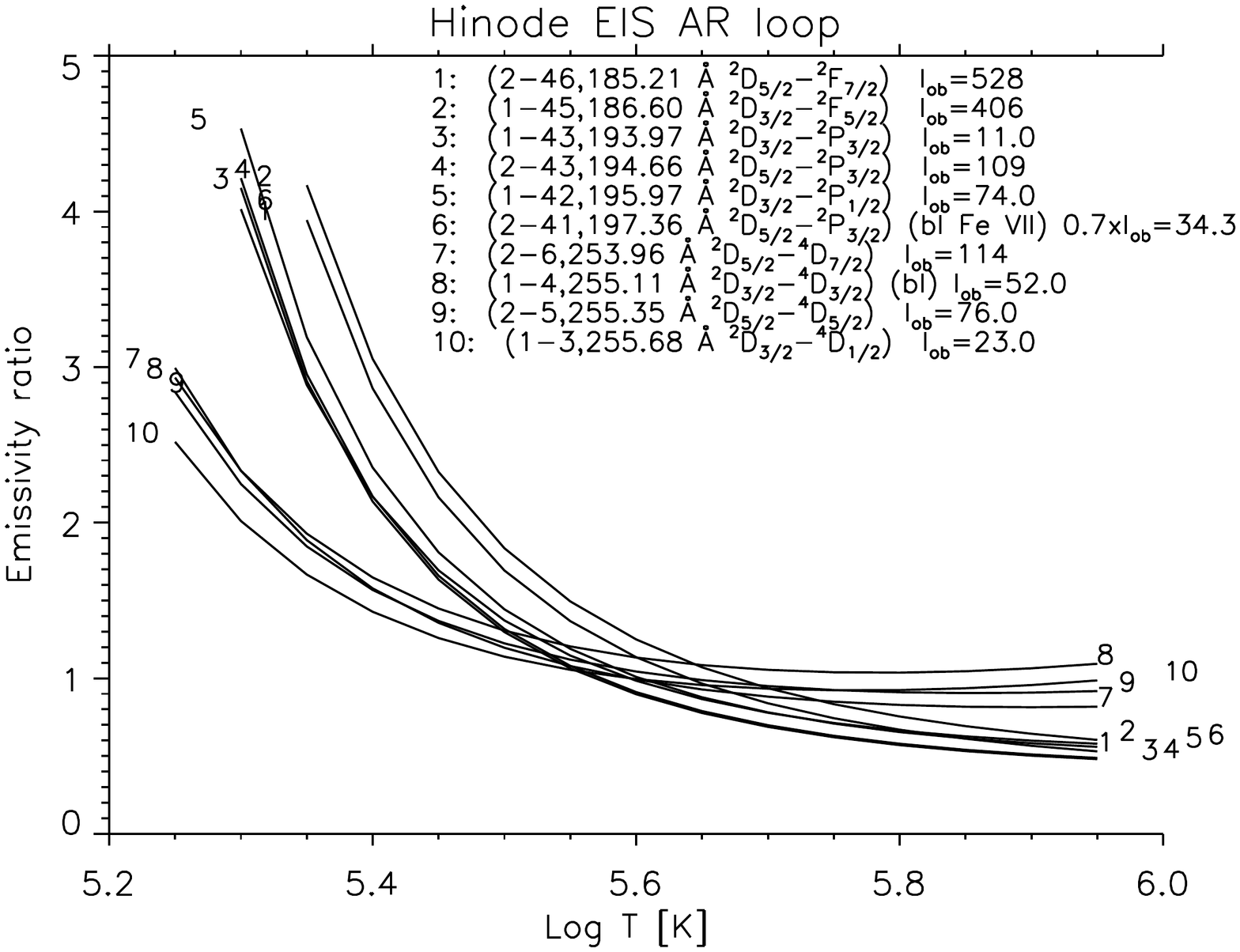}}
 \caption{Top left: a negative monochromatic image in \ion{Si}{vii} 
observed by Hinode EIS  showing a fan 
of AR loops rooted in a sunspot; top right:  the EM loci curves for Mg lines 
relative to a region near the base (indicated by a cross in the image). 
Bottom:  emissivity ratio curves for the \ion{Fe}{viii} lines, using the 
\citep{delzanna_badnell:2014_fe_8} atomic data and the \citep{delzanna:13_eis_calib}
 EIS calibration. 
Figures adapted from \cite{delzanna:09_fe_8,delzanna_badnell:2014_fe_8}.
}
 \label{fig:ar_loop_leg}
\end{figure}

Temperatures at the legs of AR loops obtained from 
\ion{Fe}{vii} \citep{delzanna:09_fe_7} and \ion{Fe}{viii} 
\citep{delzanna:09_fe_8,delzanna_badnell:2014_fe_8} line ratios 
observed by Hinode EIS
are in reasonable agreement with those obtained from the
 EM loci curves, suggesting that the plasma is nearly isothermal 
and close to ionization equilibrium.
One example is shown in  Fig.~\ref{fig:ar_loop_leg}.
However, the temperature-sensitive lines are relatively weak and 
results depend quite critically on the EIS radiometric calibration.

\cite{keenan_etal:1989_ar_15} used the 221.12 and 423.98~\AA\ lines from the 
Be-like Ar XV and EUV observations of two flares with the Skylab 
NRL S082A slitless spectrograph, finding log $T$[K]=6.46, 6.32, i.e.
close to the temperature of peak abundance in 
equilibrium,  log $T$[K]= 6.5.

\cite{keenan_etal:1994_fe_16} looked at the ratios of the Na-like 
Fe XVI 335.4 vs. the 251, 263, and 265~\AA\ lines observed in 
active regions and flares by the Skylab
NRL S082A instrument. Some of the  observations were consistent with 
isothermal temperatures of log $T$[K]=6.4, i.e. close 
to the peak ion temperature in equilibrium.
\cite{keenan_etal:2007_fe_16} compared Fe XVI observations with 
more recent atomic data, finding overall agreement. 
They used EUV  SERTS active region observations and the soft X-ray 
 XSST observation of a flare, but the temperature sensitivity was not discussed in 
detail.

SUMER measurements in the UV of flare temperatures were discussed in 
\cite{feldman_etal:00}.

\begin{figure}[!htbp]
 \centerline{
 \includegraphics[width=8cm]{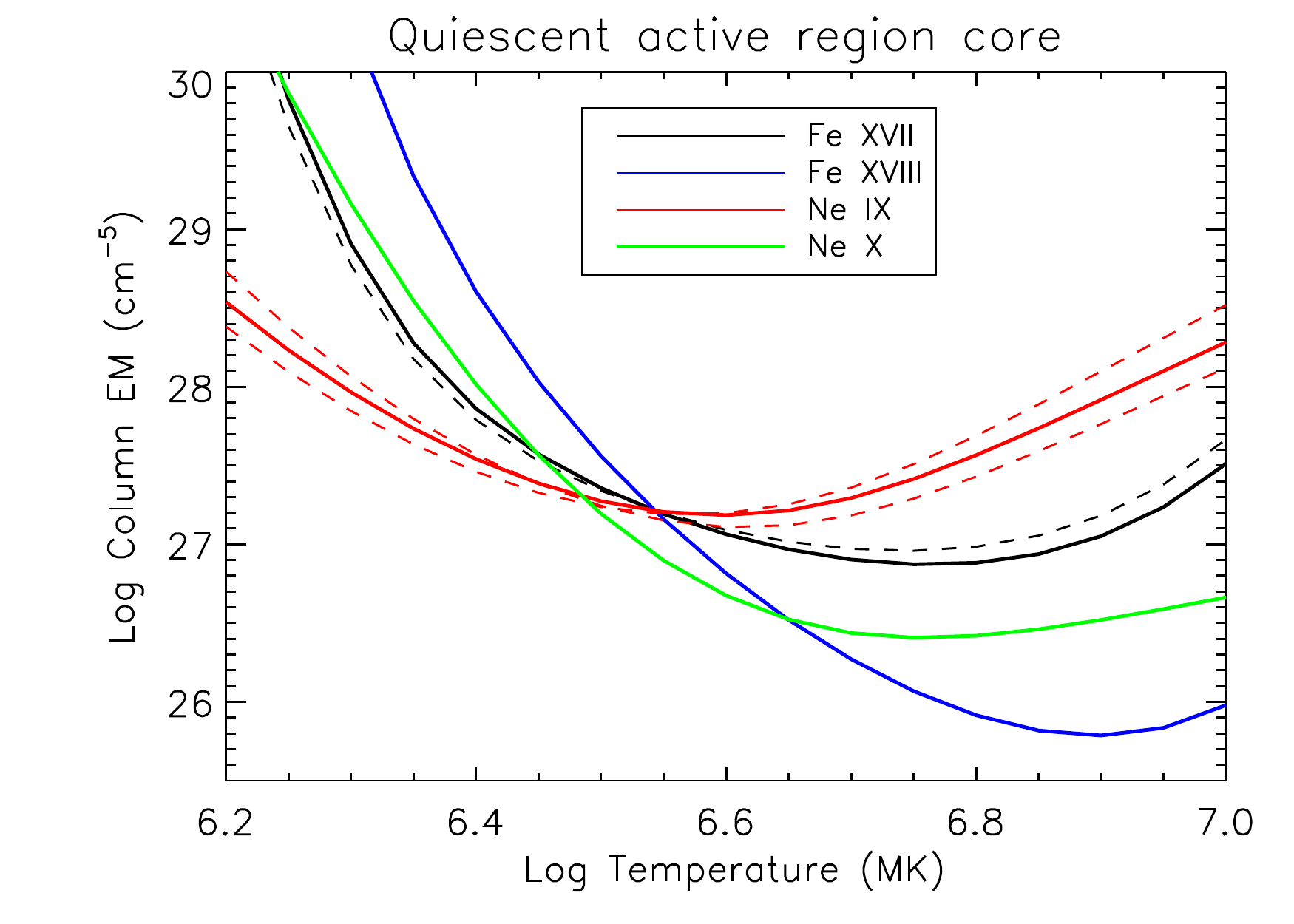}
 \includegraphics[width=6cm,angle=90]{fe_17_liang_parkinson_75b}
 }
 \caption{Left: EM loci curves of an active region core observed by 
\citep{parkinson:75}.  The EM loci curves have been recalculated with CHIANTI v.8.
Right: emissivity ratio curves for the \ion{Fe}{xvii} lines, using the 
\cite{liang_badnell:10_ne-like} atomic data.
Figures adapted from  \cite{delzanna:2011_fe_17}.
}
 \label{fig:delzanna:2011_fe_17}
\end{figure}

The soft X-ray lines have some temperature sensitivity, 
but results only improve with accurate scattering calculations,
as shown in \cite{delzanna:12_sxr1}, where the same XSST observation
was analysed with different atomic data.
Within the X-rays (5--50~\AA), the strongest lines are those from the 
Ne-like \ion{Fe}{xvii} around 15--17~\AA. 
It has been known for a long time that these strong X-ray lines 
from \ion{Fe}{xvii}  could be used to measure temperatures. However, large
discrepancies between theory and observations have also been present for a long time.
There is an extended literature on this important diagnostic, where various
attempts have (largely unsuccessfully)  been made to explain the discrepancies.
It was only  with  accurate $R$-matrix scattering calculations
\citep{loch_etal:06,liang_badnell:10_ne-like}
that the discrepancies with the astrophysical observations have been
resolved, as discussed in \cite{delzanna:2011_fe_17}.
Solar active region observations  from SMM/FCS and 
an excellent spectrum obtained from a Skylark sounding rocket 
\citep{parkinson:75} were used by 
\cite{delzanna:2011_fe_17} to obtain near-isothermal temperatures close to 
3~MK, the typical formation temperature of the hot core loops.
An example is shown in Fig.~\ref{fig:delzanna:2011_fe_17}.

We recall (see Section~\ref{sec:dem_obs}) that the fact that quiescent active region cores 
have near-isothermal distributions around 3~MK was already indicated by
Skylab X-ray imaging \citep{rosner_etal:78}, and then later 
confirmed with X-ray  (e.g. SMM FCS, Yohkoh BCS)
and EUV/UV (e.g. SoHO CDS, SUMER) spectroscopy.

Significant discrepancies between theory and observations for the 
\ion{Fe}{xviii} lines also existed for a long time. 
\cite{delzanna:2006_fe_18} showed  that the X-ray lines 
from \ion{Fe}{xviii} can reliably be used, in conjunction with accurate $R$-matrix
scattering calculations \citep{witthoeft_etal:06}, to measure temperatures in solar flares.
Values close to the peak ion temperature in equilibrium were found.

\cite{keenan_etal:1993_fe_23} noted that the OSO 5 flare observation 
 \citep{mason_etal:84} of the ratio of the 
 Fe XXIII resonance (132.8~\AA) and intercombination (263.8~\AA) lines
was consistent with a temperature close to the peak ion 
temperature in equilibrium,  log $T$[K]=7.1.
\cite{delzanna_etal:2005_fe_23} 
used updated atomic data for Fe XXIII \citep{chidichimo_etal:05}
to measure a temperature of about 10~MK from emissivity 
ratios involving 2--4  transitions and X-ray SMM BCS observations
of a flare. Higher temperatures were obtained by 
\cite{delzanna_etal:2005_fe_23} from the SOLEX 
observations.

As we have presented in  Section~\ref{sec:satellites},
the ratio of a satellite lines formed by 
dielectronic capture with the resonance line is an excellent temperature diagnostic,
independent of the population of the ions and the electron density.

\begin{figure}[!htb]
 \centerline{\includegraphics[width=1.0\textwidth,angle=0]{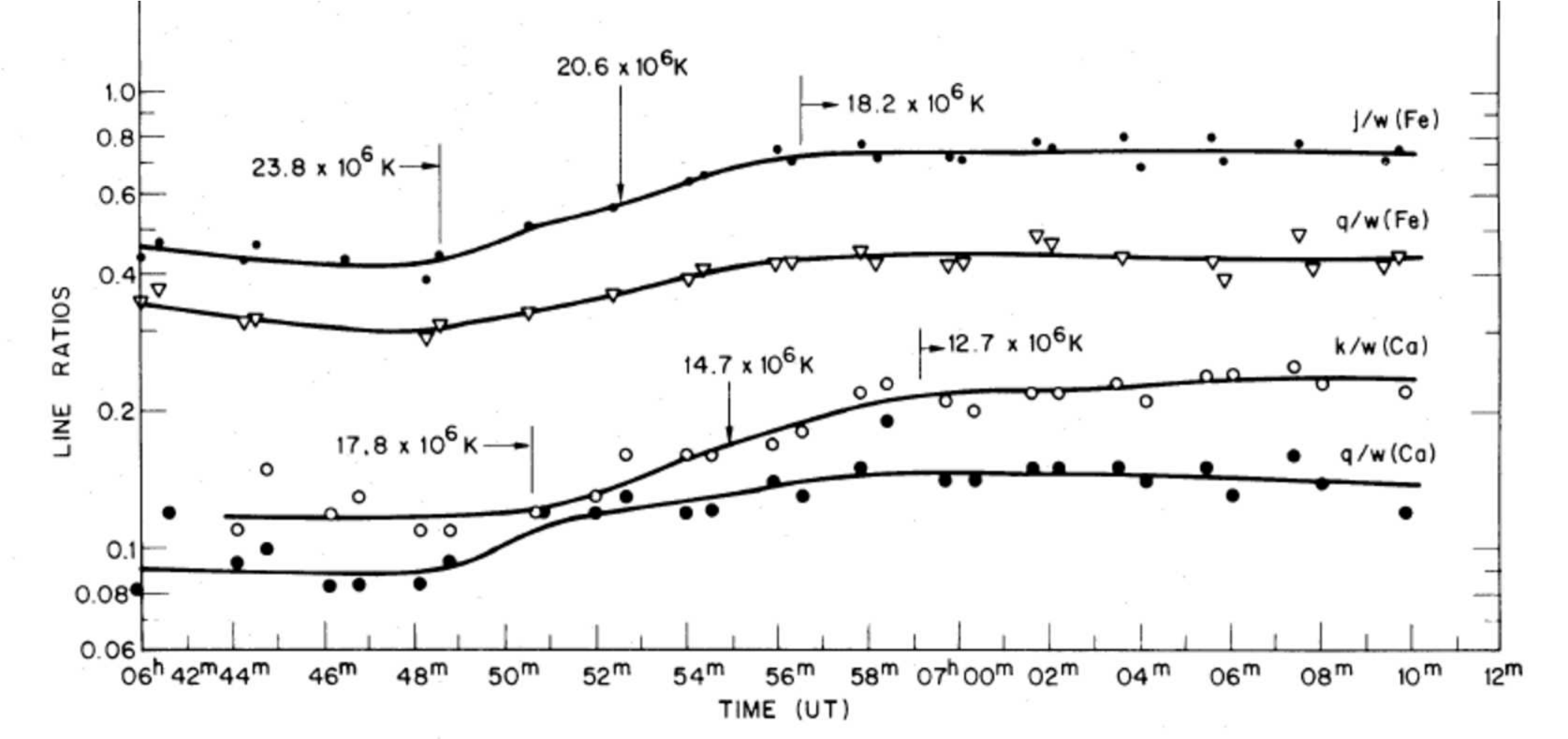}} 
\caption{A variation of ratios of lines within the He-like Fe and Ca 
complexes as observed during a large flare with SOLFLEX  \citep{doschek_etal:1980}.}
\label{fig:solflex_satellite}
\end{figure}

As an example we show the results obtained for one flare observed with 
 SOLFLEX \citep{doschek_etal:1980} in Fig.~\ref{fig:solflex_satellite}. 
The electron temperature as obtained from the ratio of the $j/w$ lines in the 
He-like Fe 
shows a marked increase from about 15 MK to at most 24 MK at the peak emission, then
it shows  a slow decay.
The temperatures obtained from the $k/w$ ratio in the He-like Ca show a similar
behaviour, but are consistently lower by about 5 MK. This has been interpreted as an indication
of the presence of multi-thermal plasma during the peak phase. 

Similar results have been obtained by the 
Hinotori, Yohkoh and SMM satellites
\citep[see, e.g.][]{doschek:1990_review}, although the 
Hinotori satellite also performed observations of the H-like 
Fe, discovering  that a super-hot
component was also present  in some of the largest flares.

In general, smaller flares reach lower peak temperatures. Indeed, there is a
good correlation between the class of a flare, as estimated e.g. by 
the total X-ray flux from GOES, and the isothermal temperatures 
at peak estimated from the He-like spectra, as shown by 
\cite{feldman_etal:1996}.

\clearpage

\section{Line widths and ion temperatures}
\label{sec:widths}

A significant amount of information on the solar plasma 
can be obtained from the shape of a  spectral line profile.
There are many effects that can  change this. Only a brief summary is given below,
without discussing detailed effects caused by any external
electromagnetic fields (e.g. Zeeman effect, Stark effect) or by opacity.

\subsection{Natural, collisional, and Doppler broadening}

The main effects which shape the profile of a 
spectral line are natural broadening, collisional broadening
and Doppler broadening.
The  natural broadening is  related to the uncertainty principle in 
Quantum Mechanics:
if we consider a transition with a  probability
of spontaneous decay $A_{ji}$, the average lifetime of the upper level 
is 1/$A_{ji}$, so the uncertainty principle is

\begin{equation}
\Delta E {1 \over A_{ji}} \simeq h
\end{equation}
where $\Delta E$ is the broadening in energy.

Collisional (or pressure) broadening is caused by the interaction between the 
ion  and other particles. It can be shown that the 
the line profile $\phi$, normalised in frequency, is a Lorentzian function:
\begin{equation}
 \phi(\nu - \nu_0) = {1 \over \pi} \, {\Gamma \over (\nu - \nu_0)^2 + \Gamma^2}  \;\;,
\end{equation}
where the damping constant $\Gamma$ is 
\begin{equation}
\Gamma = {A_{ji} \over 4 \pi} + {f \over 2 \pi} \;\;,
\end{equation}
and  $f$  is the number of collisions per unit time.

The Doppler broadening is due to the thermal motion of the ions.
The shift  caused by the Doppler effect in the frequency of the photons emitted
 is, for non-relativistic velocities, proportional to $v/c$:
\beq
{v \over c}= {\nu - \nu_0 \over \nu_0} \;\;,
\eeq
where $\nu_0$ is the frequency at rest.

If one assumes a thermal (Maxwellian) distribution for the velocities of the ions,
the  probability $f(v) \, {\rm d}v$ that the ion has a velocity 
(in the 3-D space) between $v$ and $v+{\rm d}v$ is: 
\beq
 f(v) = 4\, \pi \, v^2  \left( {M \over 2\, \pi \, kT_{\rm i}} \right)^{3/2} \, \eul^{-{M v^2 / 2kT_{\rm i}}}
  \;\; ,
\eeq
where $k$ indicates Boltzmann's constant, $M$ is the mass 
of the ion  (often written as the 
product of the atomic mass times the proton mass),  and $T_{\rm i}$ the ion temperature.
The normalised profile function is obtained by expressing the 
Maxwellian function in terms of frequency, and considering  ${\rm d}v = (c/\nu_0) {\rm d}\nu$:
\beq
\phi = {1 \over \pi^{1/2} \, \Delta\nu_D} \eul^{-{({\nu - \nu_0 \over \Delta\nu_D})^2}}  \;\;,
\eeq
where 
\beq
\Delta\nu_D = { \nu_0 \over c} \left(2 \, k \, T_{\rm i} \over M  \right)^{1/2}
\eeq
is the Doppler width of the line in frequency.
Note that the constant is such that the profile function is normalised to one:
$\int \phi(\nu)  {\rm d}\nu =1$.
%

The sum of the three broadening effects results in a 
Voigt profile. However, for the solar TR and corona both 
natural and collisional broadening are negligible, the 
Voigt profile can be approximated with a Gaussian, 
and the intensity of a spectral line can be written in terms of wavelength $\lambda$ as
\begin{equation}
I_\lambda ={I_{\rm tot}  \over \sqrt{2 \pi} \sigma}
e^{[- {(\lambda - \lambda_0)^2 \over  2 \sigma^2}]}  \;\;,
\label{gauss-prof}
\end{equation}
where $I_{\rm tot}=\int I_\lambda d\lambda$ is the integrated intensity and $\sigma$
is the Gaussian width given by:
\begin{equation}
\sigma^2 = {\lambda_0^2 \over 2 c^2} \left ({2 k T_{\rm i} \over M}  \right ) \;\; .
\end{equation}


Normally, the spectral resolution of XUV instruments
is  such that the instrumental width (assumed as a Gaussian 
with width $\sigma_I$)  is an important additional broadening.
In addition, local turbulent motions often provide further broadenings.
For an  optically thin line the profile is usually assumed Gaussian,
but with a Gaussian width given by:
\begin{equation}
\sigma^2 = {\lambda_0^2 \over 2 c^2} \left ({2 k T_{\rm i} \over M} + \xi^2 \right ) +
\sigma_I^2 \label{gauss-wid}
\end{equation}
where $\xi$ is the  non-thermal velocity, i.e. the most probable velocity
of the random bulk plasma motions, assuming they are Maxwellian.
For example, the thermal broadening (Gaussian width $\sqrt {2 k T_{\rm i} / M}$) of 
\ion{Si}{iv} is 6.88 km/s, assuming $T_{\rm i} = 80\, 000$ K, being $k= 1.38  \times 10^{-16}$
 and $M=28.086 \times 1.66  \times 10^{-24}$ (cgs units). 
For comparison, the instrumental width of the IRIS instrument is significantly lower,
3.9 km/s.  

Note that with the above definition of the Gaussian width
$\sigma$, the full-width-half-maximum FWHM=$2 \, \sqrt{(2 \, ln 2)} \sigma \simeq 2.35 \sigma$. 
Also note that in the literature the non-thermal velocity $\xi$ is sometimes 
obtained  from the measured FWHM:
\begin{equation}
FWHM^2 = w_{\rm I}^2 \, +  w_{\rm O}^2 =  w_{\rm I}^2 \, +  \, 4 \, \ln 2 \, \left ({\lambda_0 \over c} \right )^2 \, 
\left ({2 k T_{\rm i} \over M} + \xi^2 \right )  \;\;,
\end{equation}
where $w_{\rm I}$ is the instrumental FWHM and  $w_{\rm O}$ is the observed 
FWHM once the instrumental width is removed.

The above expression provides an important diagnostic for the solar corona,
i.e. the possibility of measuring the ion temperatures, if some additional 
constraints are adopted.
The approach suggested by \cite{tu_etal:1998} is to consider that 
the ion temperature $T_{\rm i}$ of a  line cannot be larger
than the value $T_{\rm max,i}$ obtained when $\xi=0$:
\begin{equation}
T_{\rm max,i}  =  {{M c^2}\over{8 k \ln 2}} 
 {\left({{ {w_{\rm O}} \over{\lambda_0}}}\right)}^2 \;\;.
\end{equation}

Similarly, the excess width cannot be larger than the value
 obtained when $T_{\rm i}=0$,  $\xi_{\rm max}$:
\begin{equation}
\xi_{\rm max}  = \sqrt{{{c^2}\over{4\ln 2}}{{\left({{{w_{\rm O}}\over{\lambda_0}}}\right)}^2}}
        = 1.80\times 10^{10} {\left({{{w_{\rm O}}\over{\lambda_0}}}\right)} \;\;.
\end{equation}

If we define with $\xi_{\rm m}$ as the smallest among a set of 
$\xi_{\rm max}$ values obtained from several lines,
 the ion temperature
of each ion must be larger than the value obtained when $\xi =\xi_{\rm m}$.
If we indicate with $m$ the line for which $xi=\xi_{\rm m}$,
the  ion temperatures should be higher than the values 
\begin{equation}
T_{\rm min,i}  =  {{M c^2}\over{8 k \ln 2}} 
{\left[{{\left({{{w_{\rm O}}\over{\lambda_0}}}\right)}_i^2-{\left({{{w_{\rm O}}\over{\lambda_0}}}\right)}_m^2}\right]} \;\; ,
\end{equation}
i.e. we obtain a lower limit for  $T_{\rm i}$ (except for the line $m$ where 
the lower limit is zero).

In addition, broadening due to opacity effects can also be present,
especially in low-temperature strong lines in the upper chromosphere / 
lower transition region.
If we assume that the source function $S_\lambda$ is constant 
along the direction of the observer, the emergent intensity $I$ is given by 
\begin{equation}
	I(\lambda) = \int\limits_{0}^{\tau(\lambda)} S_\lambda \exp{(-t_\lambda)} \mathrm{d}t_\lambda \\
	\nonumber = S_\lambda \left[1 - \exp(-\tau(\lambda))\right] \;\; .
	\label{Eq:tau_profile}
\end{equation}
For a Gaussian profile 
\beq
\phi(\lambda) \,=\, {I(\lambda) \over I_0} \,=\, \mathrm{exp}\left(-\frac{(\lambda-\lambda_0)^2}{2 \sigma^2} \right) \,,
\eeq
where the  total intensity is 
\begin{equation}
	I_\mathrm{tot} 	= \int\limits_{-\infty}^{+\infty} I(\lambda) \mathrm{d}\lambda
			= \sigma  I_0 (2\pi)^{1/2} \;\; ,
	\label{Eq:total_intensity_Gauss}
\end{equation}
we have 
\begin{equation}
	{\mathrm{FWHM}(\tau_0)}^2 = 8 \sigma^2 \left[\mathrm{ln}(\tau_0) - \mathrm{ln}\left(\mathrm{ln}(2) - \mathrm{ln}(1+ \mathrm{e}^{-\tau_0})\right)\right] \;\; ,
	\label{Eq:FWMH_Gauss_tau}
\end{equation}
from which one can estimate how much the line width increases for values of the optical depth
at line centre much greater than 1.

\subsection{An overview of line broadening measurements}

The main observational results are that 
 profiles of transition-region (TR) lines  always exhibit non-thermal
(also called excess) broadenings of the order of 10-40 km\,s$^{-1}$.
The profiles of coronal lines also often exhibit non-thermal
broadenings of the order 18 km\,s$^{-1}$  or larger
during flares. 
The interpretation of these non-thermal broadenings
 is  complex and has not yet been fully resolved.

This excess broadening has particular scientific relevance 
as it is most probably  related to the physical processes that 
heat or accelerate the plasma.  For example, it has long been thought that 
the excess broadening is a signature of waves that are propagating
in the solar atmosphere, which could contribute to the coronal heating
and acceleration of the solar wind. 
The observed non-thermal broadenings in such a case 
provide severe constraints on the types of waves 
\citep[see, e.g.][]{hollweg:1978,hollweg:1984,parker:1988,van_ballegooijen_etal:2011}.
For example, a decrease in the excess broadening with distance from the Sun 
has often been reported (see below) and interpreted as caused by the
 damping of Alfv\'en waves in the corona.
It is in fact thought that convective motions at the footpoints
of magnetic flux tubes  generate wave-like fluctuations that
propagate up into the extended corona \citep[see, e.g.][]{cranmer:2005}.
The  wave turbulence is therefore intimately related to the 
heating of the solar corona and the acceleration of the solar wind 
\citep[see, e.g.][]{ofman:2010,arregui:2015}.

Nanoflare heating could also cause spectral line broadening in the coronal lines
\citep[see, e.g.][]{cargill:1996,patsourakos_klimchuk:2006} through turbulent 
processes.

On the other hand,  there is always the possibility that the 
 excess broadenings are due to a superposition of velocity fields with 
scales smaller than the instrument resolution. In this case, 
one would expect to observe a decrease in the excess broadening with 
increased resolution, at least in the transition region
(in fact, in the corona the long path lengths and the superposition of
structures could limit such a decrease).
 
In principle, the possibility that the excess broadening in the TR lines 
is due to random mass flows is  not unreasonable.  In fact, if the spectroscopic
filling factor in these lines is assumed to be a real indication of sub-resolution structures, 
then HRTS results indicate the presence of scales down to a few km
\citep{dere_etal:1987}, i.e. much below the resolution of any previous and 
current instruments.
Recently, IRIS observations at 0.33\arcsec\ (slit width) resolution (about 250 km) have 
been obtained. \cite{depontieu_etal:2015} suggests that the 
 non-thermal broadening in  \ion{Si}{iv} in the quiet Sun is 
still of the same  order (about 20 km\,s$^{-1}$) as that  previously observed
with earlier instruments, which had a resolution of about 1\arcsec. 
Therefore, any velocity fields which could be responsible for the broadenings
must have  scales smaller than the IRIS resolution, 
unless they are consistent along the line of sight.
However, as we show below, where we present the IRIS results in the context
of earlier ones, our analysis shows that IRIS did actually observe lower excess widths.

Finally, other processes could influence the excess widths in the TR lines.
For example, \cite{depontieu_etal:2015}  found 
indications that  non-equilibrium ionisation 
could account for a significant fraction of the observed  non-thermal
broadening. That is that ions are not formed at the temperature
corresponding to peak abundance in the ionisation balance in equilibrium.

In the present section we provide a brief overview of the main observational 
constraints, organised in different sub-sections, depending on the solar feature observed.
 Before going into detail, however, it is worth pointing out 
 a few key general  issues which can affect the observations and their interpretation.
In fact, contradictory  results have often been reported in the literature. 
Variations in the excess widths  have often  been attributed
to a variety of reasons such as the choice of ion temperatures, non-equilibrium
effects, curvature of the magnetic field, instrumental effects and solar cycle variations.

On the observational side, we point out that for most instruments,
measurements of line widths are subject to large uncertainties, some 
being systematic. For several instruments, the instrumental line 
width is a significant fraction of the observed width. 
Characterising the  instrumental line width and its variations is 
notoriously difficult. Instrumental line widths
are normally assumed to be Gaussians but in reality they are not. 
The instrumental PSF often shows unexplained characteristics as well. 
Furthermore, several instruments (e.g. SoHO/CDS, Hinode/EIS) were 
not specifically designed to study line profiles, so the instrumental line
width is sampled with only a few pixels across it. 
Hence, accurate measurements of line profiles are very difficult.

On the issue of interpreting the observed widths, 
 when we say that the profiles exhibit  non-thermal broadening we mean that 
the line width is larger  than the one expected from thermal broadening in 
equilibrium conditions. For the equilibrium temperature of an ion 
it is often assumed  that the ion and electron temperatures 
are the same, although in reality what one measures is an ion 
temperature from the line width.
This assumption is justified by the fact that 
the equilibration time for the ion and electron to thermalise is very short.
For example, in the transition region, 
the equilibration time is about 0.04 s at a typical TR density of 10$^{10}$ cm$^{-3}$.  
The electron temperatures in turn are normally estimated from the intensity 
ratios of lines from different ions, assuming that 
the ion charge state distributions are in equilibrium, or occasionally 
by more direct measurements of line ratios from the same ion
 (cf. Section~\ref{sec:te_diagn}).
The ion temperatures as obtained from the line widths 
are typically a factor 4--5 higher than the ionisation temperatures,
corresponding to peak ion abundance, in the TR.

It would however be difficult to find a mechanism that 
keeps ion and electron temperatures so different.
\cite{bruner_mcwhirter:1979} estimate that maintaining such a difference
would also require a significant amount of energy. 
So,  a likely explanation
would be that the ionisation temperatures are not representing 
the real electron temperatures, which could actually  be much higher. 
This could be caused by several factors. We have seen in 
 Section~\ref{sec:charge_state} that various effects can change the ion charge 
state distribution even assuming equilibrium. Non-equilibrium 
ionization is likely to occur in the TR if one considers the 
observed dynamics and flows, temperature gradients and temporal variability in the 
line intensities. 
On the other hand, we have seen in  Section~\ref{sec:qs_te} that the few
measurements of direct electron temperatures in the TR do not suggest 
significant differences with the ionisation temperatures calculated assuming
equilibrium, in most cases. However, there are a few cases where
differences were reported. The situation is therefore unclear. 
More accurate observations and modelling of plasma processes, taking
into account of non-equilibrium and time dependent processes, are  needed
to resolve these issues.

Another approach often adopted in the literature is to assume that the 
excess widths of lines formed at similar temperatures should be the same,
and the  ion kinetic
temperature is independent of the mass and charge of the ions.
The first assumption is reasonable, as one would expect that lines
formed at similar temperatures should experience, within the same volume, 
the same processes. The second assumption is based on the idea that 
the ions are  in thermal equilibrium between themselves.
For example, \cite{seely_etal:1997} made these two assumptions  and determined
the turbulent velocity from the observed widths by assuming that 
the ion temperature was equal to the electron temperature.

More generally,  if ions of different masses are observed, one could relax the 
second assumption and obtain 
limits on the  ion temperatures 
 from the measured excess widths \citep[see, e.g.][ and below]{tu_etal:1998}. 
\cite{tu_etal:1998} applied this method to polar coronal hole observations
 by SUMER, finding ion temperatures significantly higher than electron temperatures. 
Other authors have applied the same method to observations of different regions.
For example, \cite{landi:2007} used SUMER observations and this 
method to find that the ion temperatures were close to the 
electron temperatures,  during solar minimum conditions. 
On the other hand, \cite{landi_cranmer:2005}
found in coronal holes  ion temperatures significantly higher than electron temperatures.

\subsection{TR lines in the quiet Sun}

The profiles of transition region lines are typically 
 much broader than their thermal width in the quiet Sun.
Observations have been obtained  with e.g. the 
Skylark rocket \citep{boland_etal:1973,boland_etal:1975}, the
Skylab NRL slit spectrometer 
\citep[see, e.g.][]{doschek_etal:1976,kjeldseth-moe_nicolas:1977,mariska_etal:1978,mariska_etal:1979},
OSO 8 \citep[see, e.g.][]{shine_etal:1976},
the  HRTS rocket \citep[see, e.g.][]{dere_etal:1987,dere_mason:1993}, 
 SoHO SUMER \citep[see, e.g.][]{chae_etal:1998,teriaca_etal:1999,peter:2001}
at about 1--2\arcsec\  resolution, and more recently with the IRIS satellite
at 0.3\arcsec\  resolution.

It should be noted that variations in the measured line widths are
present in the literature. They are partly due to real spatial variations,
but also to many other effects, such as the differences in the 
instrumental broadening, the spatial resolution, the exposure times, opacity effects, etc.
We note that considering its spatial resolution, sensitivity and  instrumental broadening,
the best instrument,  until the launch of IRIS,  was probably  HRTS.
The FWHM of the HRTS instrument was about 0.07~\AA\ \citep{dere_mason:1993},
i.e. similar to the value of the Skylab NRL slit spectrometer 
 \citep[about 0.06~\AA, cf.][]{doschek_etal:1976} which however did
 not have a stigmatic slit.
The FWHM of the SUMER instrument was about 2.3 pixels, equal to
about 0.1~\AA\ \citep[see, e.g.][]{chae_etal:1998},
while the spatial resolution was about 2''.
The second Skylark rocket had an excellent 
 FWHM resolution of 0.03~\AA\ \citep{boland_etal:1975}, but only measured
the line widths of a few lines, mostly at longer wavelengths (compared to the 
other instruments).
The excellent Skylab NRL slit spectrometer observations were performed near 
the limb, where some additional opacity effect were probably present.

The sensitivity of the HRTS instrument was such that exposures of only 1s
could be used to record the strongest lines. 
The \cite{dere_mason:1993} analysis used a combination of the 1, 2.8, 
8, and 20s exposures.
On the other hand, typical quiet Sun SUMER exposures were much longer.
For example, \cite{peter:2001} used the reference spectra obtained with 115s exposures,
while \cite{chae_etal:1998} used  data with 400--500s exposures.
Some additional  broadenings of the lines is clearly expected for longer exposures,
given the dynamical nature of the TR lines. Indeed  \cite{chae_etal:1998}
found averaged differences of about 4 km/s between the 20s and 300s exposures.

\begin{figure}[htb]
\centerline{\includegraphics[width=0.7\textwidth,angle=90]{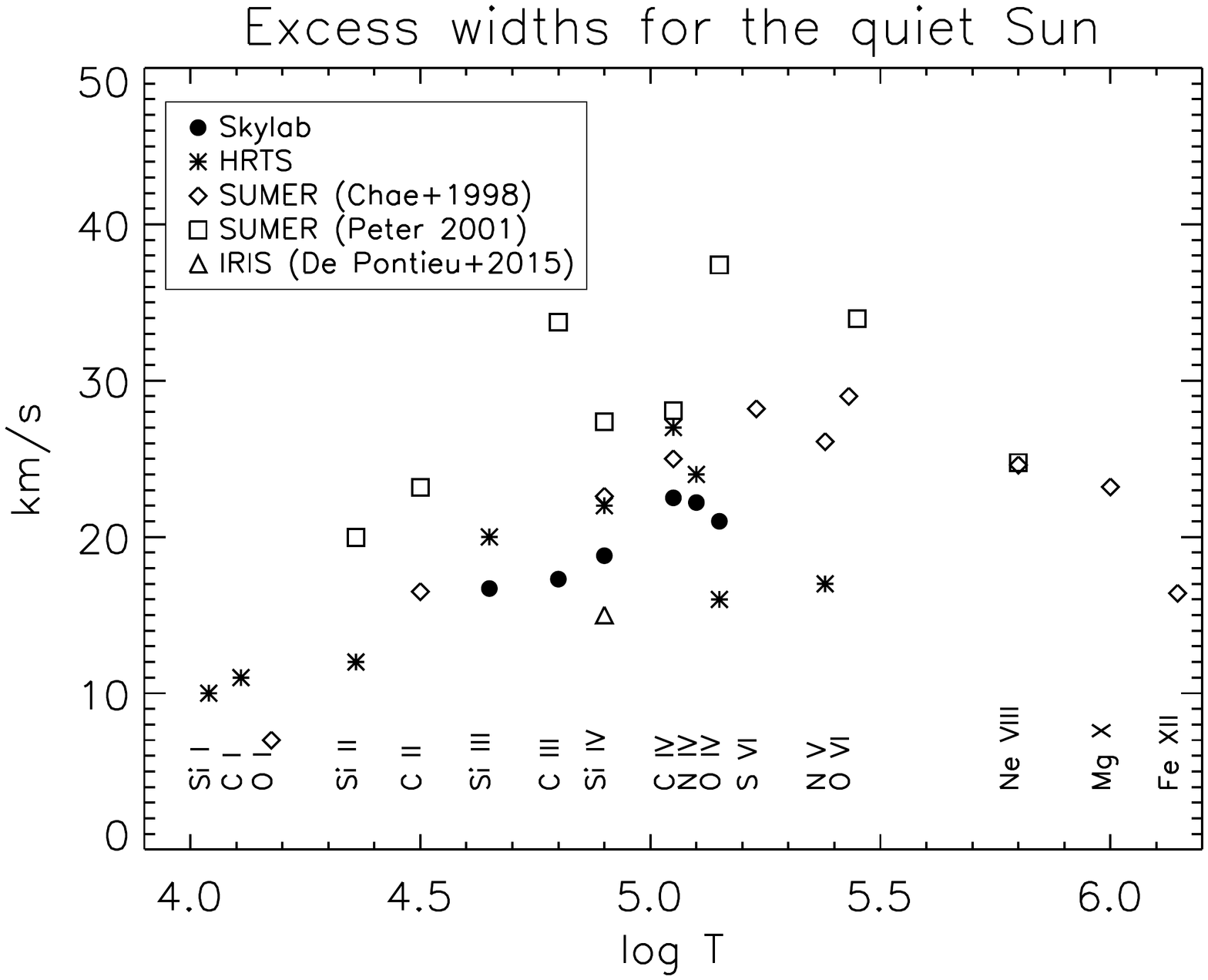}}
\caption{Nonthermal line widths as a function of temperature for the quiet Sun,
as obtained from the  Skylab NRL slit spectrometer \citep{feldman_etal:1988},  
from HRTS by \cite{dere_mason:1993},
 SoHO SUMER by \cite{chae_etal:1998,peter:2001}, and IRIS by \cite{depontieu_etal:2015}.}
\label{fig:ntw}
\end{figure}

The average excess broadening varies with the temperature of formation of the line,
with a peak in the middle of the transition region 
around  2 10$^5$ K of about 25 km\,s$^{-1}$ for the quiet Sun.
An example is shown in  Fig.~\ref{fig:ntw}, with  measurements from the 
 Skylab NRL slit spectrometer \citep[see the review by][]{feldman_etal:1988},  HRTS 
\citep{dere_mason:1993},  SoHO SUMER \citep{chae_etal:1998,peter:2001}, and the last
IRIS measurement from \citep{depontieu_etal:2015}.
We note that the Skylab NRL slit spectrometer 
results reported by  \citep{doschek_etal:1976,mariska_etal:1978}
 are consistent. 
We also note that there are several of such plots in the literature, however the 
actual formation temperature of a line is a non-trivial issue.
Due to the emission measure structure in the transition region, the formation temperature
of an ion can in fact  be quite different from the value at 
 peak ion abundance in ionization equilibrium, which would result in quite different plots.

The data points in Fig.~\ref{fig:ntw} are plotted at an approximate 
effective temperature of formation of the lines, and not at the temperature
of peak ion abundance in ionization equilibrium.
This was obtained by us from average quiet Sun radiances obtained from HRTS spectra. 
The actual formation temperatures within each observation could be 
slightly different, but the trends would remain. 
 The observed line widths 
reported by \cite{peter:2001} have been converted into excess widths 
adopting the temperature of peak ion abundance as reported  by 
\cite{peter:2001}, for consistency.

Another issue to keep in mind when interpreting
line width observations is that  lines formed at lower temperatures such as 
the dipole-allowed transitions of \ion{O}{i}, \ion{Si}{ii},
\ion{C}{ii}, \ion{Si}{iii}
are often optically thick, which tends to increase the width of a line. 
In fact, the widths of different lines from the same ion often differ.

\begin{figure}[!htb]
\centerline{\includegraphics[width=0.6\textwidth,angle=0]{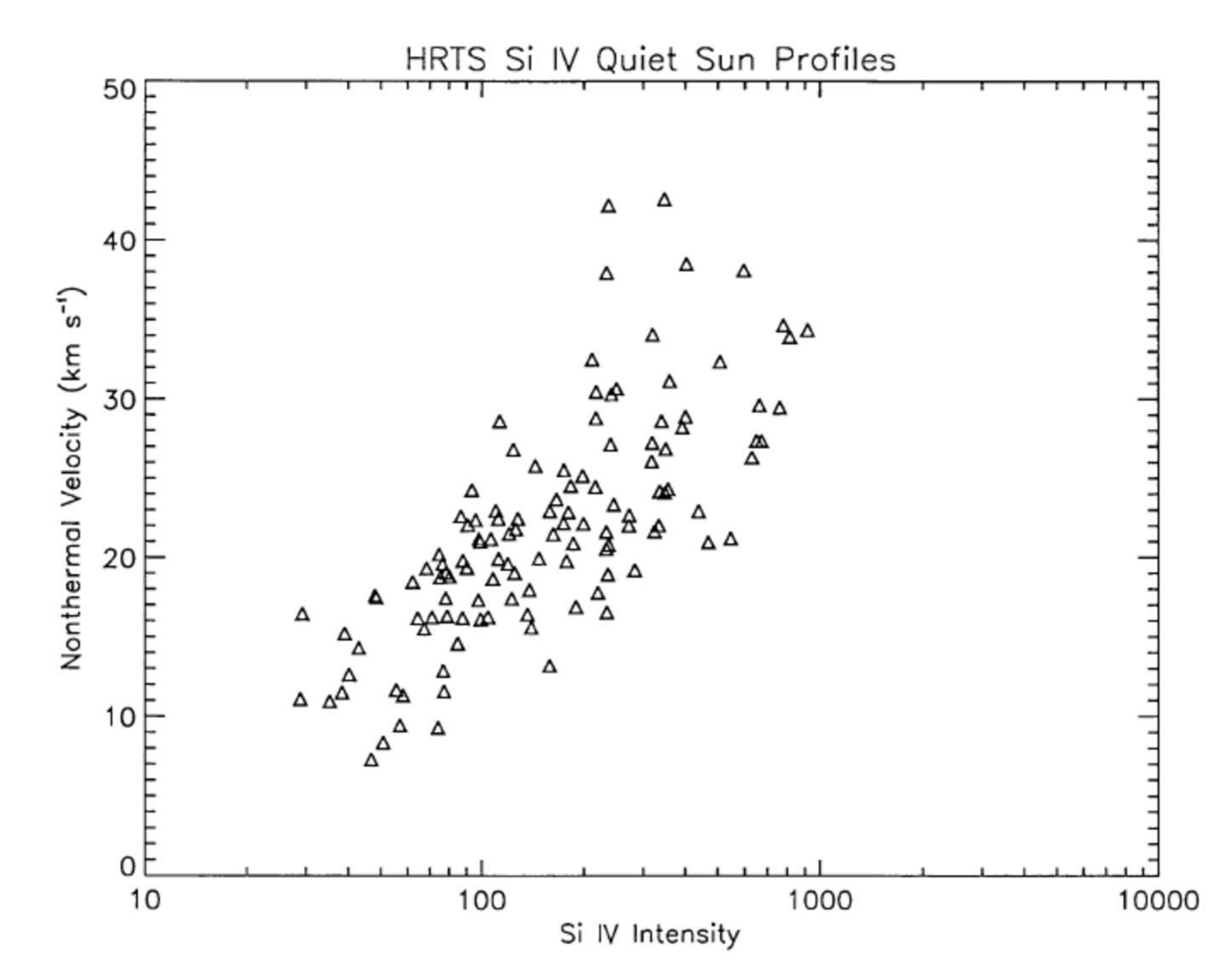} 
\includegraphics[width=0.4\textwidth,angle=0]{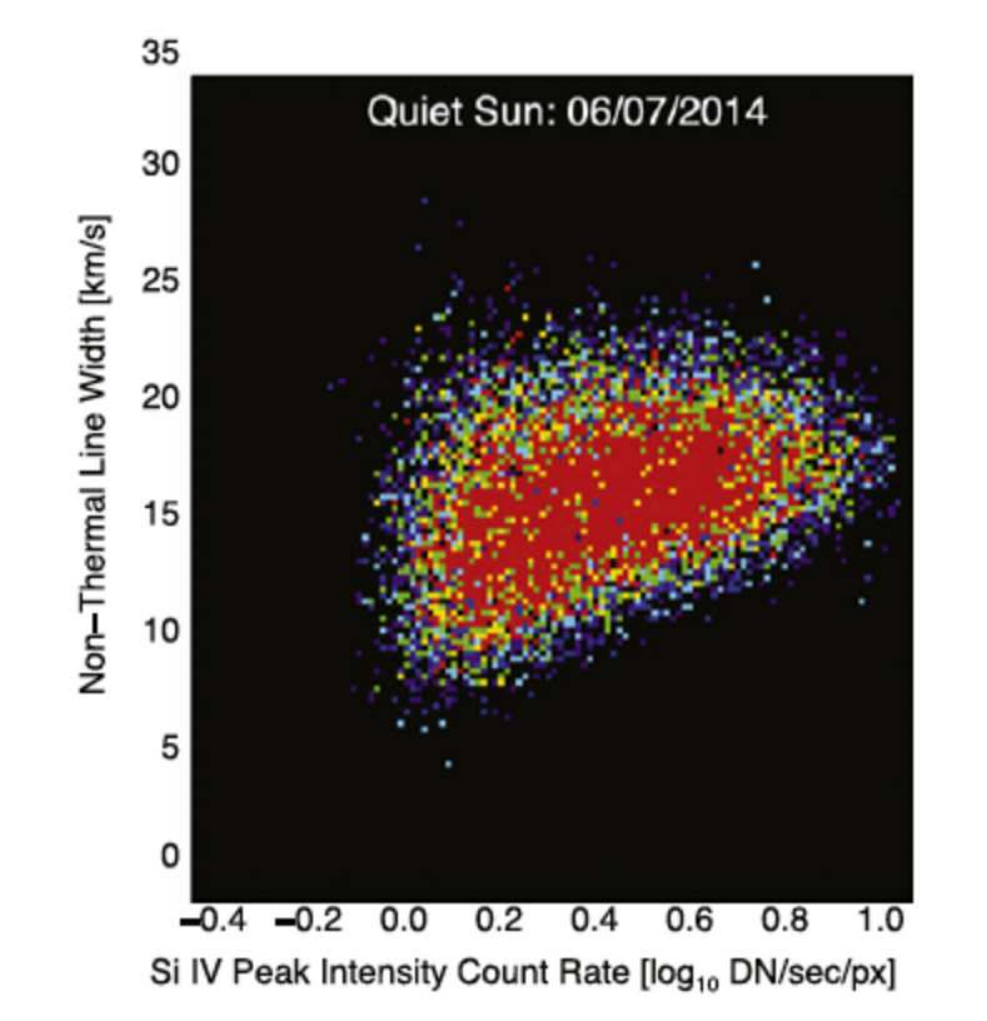}} 
\caption{Nonthermal line widths of the \ion{Si}{iv} 1403~\AA\ line as a function of its 
intensity for the quiet Sun, as obtained from HRTS by \cite{dere_mason:1993} on the left,
and IRIS  \citep{depontieu_etal:2015}, on the right. Note the different scales.}
\label{fig:ntw_si_4}
\end{figure}

Also, we stress that the values often found in the literature are average values. For example, 
the quiet Sun  excess broadening in \ion{Si}{iv} observed with HRTS
varied from about 10 to 40 km/s, when the radiance changed by over an 
order of magnitude (see \citealt{dere_mason:1993} and Fig.~\ref{fig:ntw_si_4}).

We recall that 
 IRIS observations at 0.33\arcsec (slit width) resolution and short
exposure times (seconds)  have shown \citep{depontieu_etal:2015}
that the non-thermal broadening in  \ion{Si}{iv} in the quiet Sun is 
still of the same  order, although we note that values are 
overall smaller (10--20 km\,s$^{-1}$) than those measured by HRTS,
as  Fig.~\ref{fig:ntw_si_4} shows. So, we find that the higher IRIS spatial 
resolution does indeed produce somewhat lower non-thermal broadenings, in 
contrast to the conclusions by  \cite{depontieu_etal:2015}.

It is in fact well established that the excess broadening is correlated with the intensity of the lines,
for the lines formed in the middle of the transition region such as  \ion{Si}{iv}.
There is very little excess broadening in the supergranular cell centres,
while the largest values are at the cell boundaries.
Interestingly, the network boundaries are typically the places where 
all sorts of explosive events occur (where widths are much larger). 
Lines formed at lower and higher temperature have a much weaker correlation
between intensities and excess widths.

Coronal holes on average have very similar characteristics. 
Active regions  have similar, or just a bit larger, 
excess broadening  than the quiet Sun.

Interestingly, observations of several quiescent prominences with the Skylab NRL slit 
spectrometer have shown  that in several (but not all) cases the excess
broadening was almost non-existent, of the order of just a few km/s
\cite{feldman_doschek:1977_prominences}.

\begin{figure}[!htb]
\centerline{\includegraphics[width=0.6\textwidth,angle=0]{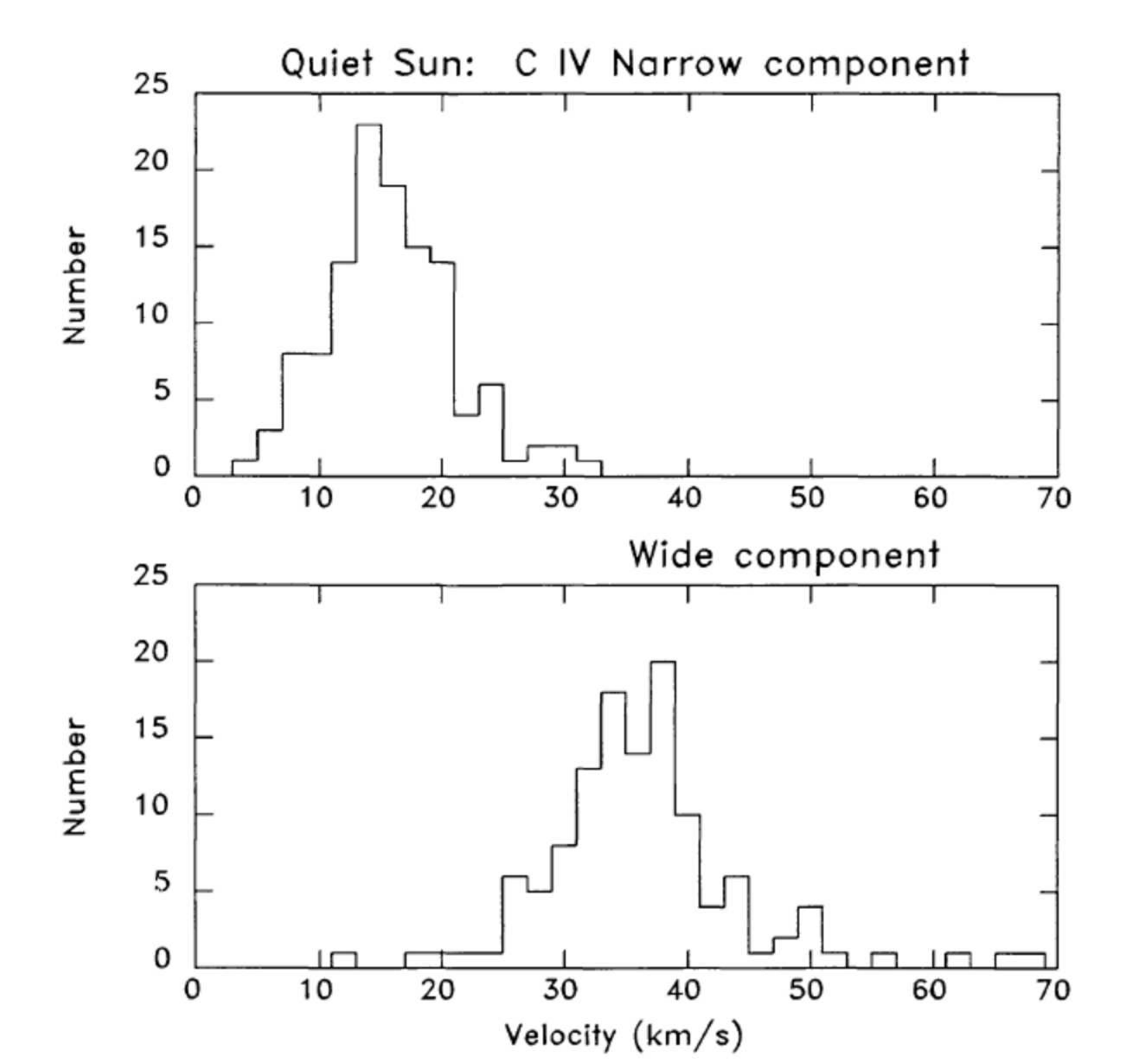}} 
\caption{Distribution of the nonthermal line widths of the core and broad 
component of the \ion{C}{iv} line 
 for the quiet Sun, as obtained from HRTS by \cite{dere_mason:1993}.}
\label{fig:ntw_c_4}
\end{figure}

Earlier observations   showed 
nearly Gaussian line profiles, although distortions of the 
profiles are very common, especially in TR lines, due to the intrinsic transient nature
of the emission and the ubiquity of explosive events, when very large 
red- and blue-wing asymmetries are observed.
In the more quiescent  observations,   it was however found, especially from HRTS, that 
the core of the lines was nearly Gaussian, but 
 a much broader and weaker  component, of the order of  50 km\,s$^{-1}$ was also present.
 Fig.~\ref{fig:ntw_c_4} shows the distributions of the nonthermal line widths of the core and broad 
component of one of the \ion{C}{iv} lines
 for the quiet Sun, as obtained from HRTS by \cite{dere_mason:1993}.

\begin{figure}[htb]
\centerline{\includegraphics[width=0.7\textwidth,angle=90]{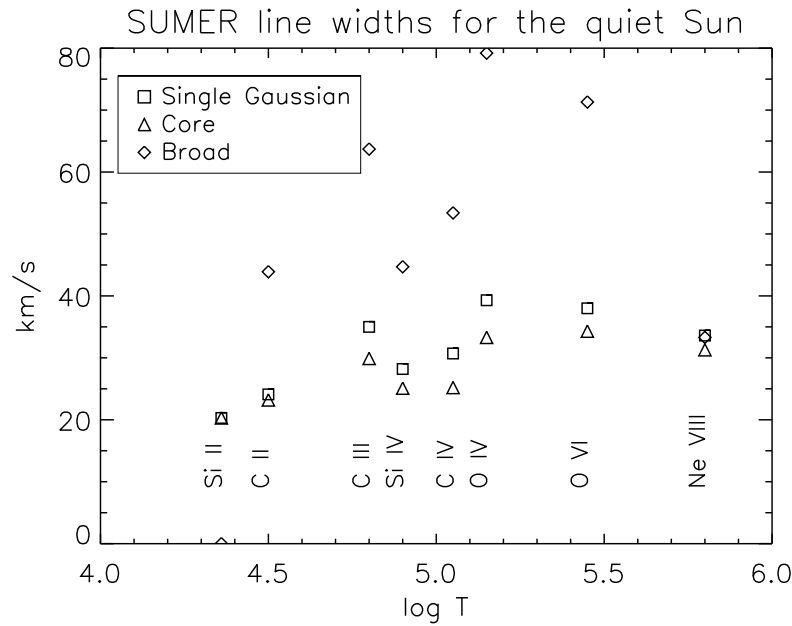}}
\caption{Observed  line widths of the core and broad components in the quiet Sun
from  SoHO SUMER \citep{peter:2001}.}
\label{fig:peter_2001_ntw}
\end{figure}

Traditionally, the profiles of TR lines have therefore been fitted with two Gaussian
profiles. Fig.~\ref{fig:peter_2001_ntw} shows the variation of the 
core and broad components in TR lines as measured with SoHO SUMER \citep[see, e.g.][]{peter:2001}.
It is interesting to note that the contribution of the broad component 
changes with the temperature of formation of the line. The different behaviour of
the narrow and broad components led to the interpretation that they 
originate  in two different  magnetic structures: 
the narrow one in closed magnetic loops and the broad one in open coronal funnels
\citep{peter:2001}.

An alternative possibility is that TR line profiles reflect a non-thermal distribution
of the velocities of the ions. \cite{dudik_etal:2017_ntw} showed that IRIS 
line profiles  in an active region are better fitted  with a $\kappa$-distribution 
rather  than a double Gaussian, although an extra non-thermal broadening
is still needed to explain the profiles.

Another interesting fact is the lack of a clear center-to-limb variation in the average
values of the excess widths in TR lines, which indicates that if 
the interpretation is in terms of Doppler motions, they are nearly isotropic.
Some center to limb changes in the broadening have been
observed, but were mostly attributed to optical depth effects.

Interestingly, broadening in cooler lines 
(e.g. \ion{C}{ii}, \ion{Si}{iii}, up to \ion{C}{iv}) 
is observed to increase with height when observations at the limb 
are performed \citep[see,e.g.][]{doschek_etal:77,nicolas_etal:1977,mariska_etal:1978,mariska_etal:1979}.
For example, the excess velocity in the resonance lines of 
\ion{C}{iv} 12\arcsec  above the visible limb was found to be about 33  km\,s$^{-1}$,
i.e. twice that which was observed on-disk \citep{dere_etal:1987}.

Systematic differences in the excess line widths of  
allowed and intersystem TR lines have been noted from Skylab and HRTS observations 
\citep{brueckner_etal:1977, kjeldseth-moe_nicolas:1977,dere_mason:1993}. 
The widths of the allowed oxygen lines appear to be about 50\% larger than those
of  the intersystem lines. 
This was also noted  by 
\cite{doschek_feldman:2004} using SoHO SUMER observations of the quiet Sun.
In principle, the larger widths of the allowed lines could be due to 
opacity effects. However, center-to-limb measurements of TR lines
normally indicate very small opacities in these lines \citep[see, e.g.][]{doyle_mcwhirter:80}.
Estimates of the opacities based on oscillator strengths confirm this
\citep{doschek_feldman:2004}.
\cite{doschek_feldman:2004} also discussed various other possible causes
for the differences in the line widths, 
without finding an explanation. 
The allowed lines are more sensitive to higher densities and temperatures, so the
 steep gradients would cause some effects. Standard DEM results have been used to 
show that this effect is too small.
If the plasma had  high-density regions, 
the intersystem lines would be depressed, compared to the allowed
ones. If the same plasma 
was more turbulent, the net effect would be to increase and broaden
the allowed lines, compared to the intersystem lines.
However, it is not very clear if such high densities components exist
\citep[see, e.g.][]{polito_etal:2016b}.

\cite{dere_mason:1993} suggested, however, that the broad wings in the 
 intersystem lines are not observed because the lines are weak in the 
spectra. Indeed  recent IRIS spectra of the same lines in active regions do 
indicate that allowed lines such as \ion{Si}{iv} and 
intersystem lines from \ion{O}{iv} have very  similar profiles and widths
\citep[see, e.g.][]{polito_etal:2016b,dudik_etal:2017_ntw}. 
We note that IRIS, with its very small instrumental width (one or two pixels)
and its good sampling of line profiles, is  an excellent instrument to 
measure line profiles. 
Recently, \cite{doschek_etal:2016} reviewed earlier Skylab observations,
and recalled the  high-density interpretation, but when comparing 
IRIS line widths also found (cf. their Fig.~8) similar line profiles for the 
\ion{O}{iv} and  \ion{Si}{iv} lines.

\subsection{Coronal lines}

Early observations of the green and red forbidden visible lines
showed  excess line widths (18--20 km\,s$^{-1}$) \citep[see, e.g.][]{billings_lehman:1962}.
Rocket observations of the full-Sun 
 EUV spectrum also showed significant excess broadening in coronal lines,
of the order of 30 km\,s$^{-1}$
\citep[][]{feldman_behring:1974}.
Skylab observations above the limb (around 10--20\arcsec) 
with the NRL slit spectrograph of the coronal forbidden lines 
also showed excess line widths of the order of 20 km\,s$^{-1}$
\citep[see, e.g.][]{doschek_feldman:1977_ntv,cheng_etal:1979}.
The  excess line widths above active regions were somewhat smaller.

The SMM - UVSP measurements indicated somewhat lower values 
of 15 km\,s$^{-1}$ or less in the weak 
forbidden coronal lines from Si VIII, Fe X, Fe XI, and Fe XII 
\citep{mason:1990}.

The excess widths of the coronal lines increase as a function of height above the limb. 
\cite{hassler_etal:1990} used a rocket spectrum to show this behaviour in the 
\ion{Mg}{x} lines, with broadenings up to 30 km\,s$^{-1}$.

SoHO CDS off-limb observations of the quiet corona 
 showed a decrease  in the line widths of \ion{Mg}{x}  at larger heights
\citep{harrison_etal:2002}. However, it turned out that there 
was an instrumental variation that was not taken into account
\citep{wilhelm_etal:2005}. With a correction obtained from on-disk observations,
the widths were found to be constant.

SoHO SUMER off-limb observations of quiet regions 
have shown little variations in the excess line widths up to 1.45~\rsun
\citep[see, e.g.][]{doyle_etal:1998,doschek_feldman:2000,wilhelm_etal:2004_widths}.
There is, however, a scatter of values, which ultimately depend on the assumed 
ion temperatures.
For example, \cite{seely_etal:1997} reported 10 km\,s$^{-1}$ at 100--200\arcsec above the limb,
i.e. a value  smaller than that reported by \cite{hassler_etal:1990}.
\cite{landi_feldman:2003} found instead relatively constant
excess widths of  25--30 km\,s$^{-1}$ up to 1.3~\rsun.

Interestingly, changes in the solar cycle appear to affect the  excess line widths:
\cite{landi:2007}, assuming that all the excess line widths for a range of ions
were the same, found that the ion temperatures were closer to the 
electron temperatures during solar minimum conditions, while 
differences increased with the solar activity.

SoHO LASCO C1 coronagraph observations showed that 
the width of the \ion{Fe}{xiv} green line during the 1996 solar minimum 
was roughly constant up to  1.3~\rsun\  and then decreased 
\citep{mierla_etal:2008}.

Hinode/EIS observations of the quiet Sun have 
indicated a small decrease with height \citep[see, e.g.][]{hahn_savin:2014}.


\subsubsection{Coronal lines in active regions}

Ground-based observations with the 
coronagraph at the Norikura Solar Observatory (Japan) over the years 
have  shown a range of excess broadenings  (14--26 km\,s$^{-1}$) and some 
variations along  loop structures \citep{hara_ichimoto:1999} or with height 
above the solar limb \citep[see, e.g.][and references therein]{singh_etal:2006}.
In their review, \cite{singh_etal:2006} report 
that the broadening of the red \ion{Fe}{x} line  tends to 
increase up to about 250\arcsec\ above the limb, but then remains unchanged. On the other hand,
the width of the green \ion{Fe}{xiv} line  decreases up to 
about 300\arcsec, and then remains unchanged. The widths of the 
\ion{Fe}{xi}, \ion{Fe}{xiii} lines show an intermediate behaviour.
The difference in the behaviour of the lines is puzzling.
We should point out, however, that such observations were carried out
above active regions and not in quiet Sun areas. Therefore, the 
various lines show emission that is not cospatial, and in fact 
it is likely to have quite different characteristics \citep[see, e.g.][]{delzanna_mason:03}.

Off-limb observations of the coronal forbidden lines with the Skylab NRL
instrument showed  excess line widths of about 10--25km\,s$^{-1}$
 in active regions that were
either similar or smaller than those of quiet Sun regions
\citep[see, e.g.][]{cheng_etal:1979}.

Regarding on-disk observations of active regions, 
SMM-XRP FCS observations of  X-ray lines (Mg XI, Ne IX, O VIII)
formed around 4 MK
indicated much higher values, of the order of 40--60  km\,s$^{-1}$
\citep{acton_etal:1981, saba_strong:1991a} in the cores of active regions.
We recall that the FCS instrument had an aperture of about 14'', i.e. 
had some spatial resolution. 
The broadest line profiles were found along the magnetic neutral line.
It is interesting to note that  
these excess widths did not significantly change with the location
of the active regions, i.e. no centre-to-limb effects were observed
 \citep{saba_strong:1991b}, which suggests that the widths are not due to a 
superposition of up- and down-flowing motions. 
The widths of these X-ray lines are significantly larger than most of the 
reported  widths of lines formed at similar temperatures observed in the EUV, UV or 
visible. This is puzzling and it is not easy to explain. 
Perhaps such large widths are associated with the increases 
(up to about  40--60  km\,s$^{-1}$) which have been 
recently observed by Hinode EIS 
in active regions before the occurrence of large flares
\citep{harra_etal:2009}, although in this case the widths were observed in 
\ion{Fe}{xii}, which is formed at a lower temperature, around 1.5 MK.

Further on-disk observations of widths of coronal 
lines formed around 1--3 MK in active region loops have been provided by Hinode EIS.
\cite{doschek_etal:2007_ntv} found that 
the excess width 
in \ion{Fe}{xii} is somewhat smaller (by 6 km\,s$^{-1}$) in coronal 
loops than in the surrounding regions.

\cite{imada_etal:2009} analysed Hinode EIS observations of 
an active region quiescent core observed off the limb.
By assuming that lines from different ions have the same excess widths, 
they obtained an ion temperature of 2.5 MK and an 
 excess width of about 13 km\,s$^{-1}$.

\cite{brooks_warren:2016} performed a statistical study of 
excess widths in coronal lines formed in the 1--4 MK range, using  observations with Hinode EIS
of quiescent AR cores. 
They found values around 18 km\,s$^{-1}$, in agreement with previous results,
and no significant trends with the temperature of formation of the line.
The authors conclude that these measurements are inconsistent  
with all the current  models. 

\subsubsection{Excess widths in active region coronal outflows}

Excess widths in lines formed around 1--3 MK
 have been discovered from Hinode EIS in the so called coronal 
outflow regions
\citep[see, e.g.][]{delzanna:07_dublin,delzanna:08_flows,doschek_etal:08,harra_etal:08}.
The non-thermal broadening and the blue-shifts are small (10--20 km\,s$^{-1}$), hence
were not previously clearly observed. \cite{delzanna:08_flows} showed that 
only lines formed at temperatures higher than 1 MK present this behaviour,
and that higher-temperature lines had stronger  blue-shifts. 
\cite{delzanna:08_flows} and  \cite{delzanna_etal:2011_outflows} indicated that 
the line profiles were mostly symmetric, as in the case of 
plage (AR moss) areas \citep[see, e.g.][]{tripathi_klimchuk:2013}.

\begin{figure}[htb]
\centerline{\includegraphics[width=0.6\textwidth]{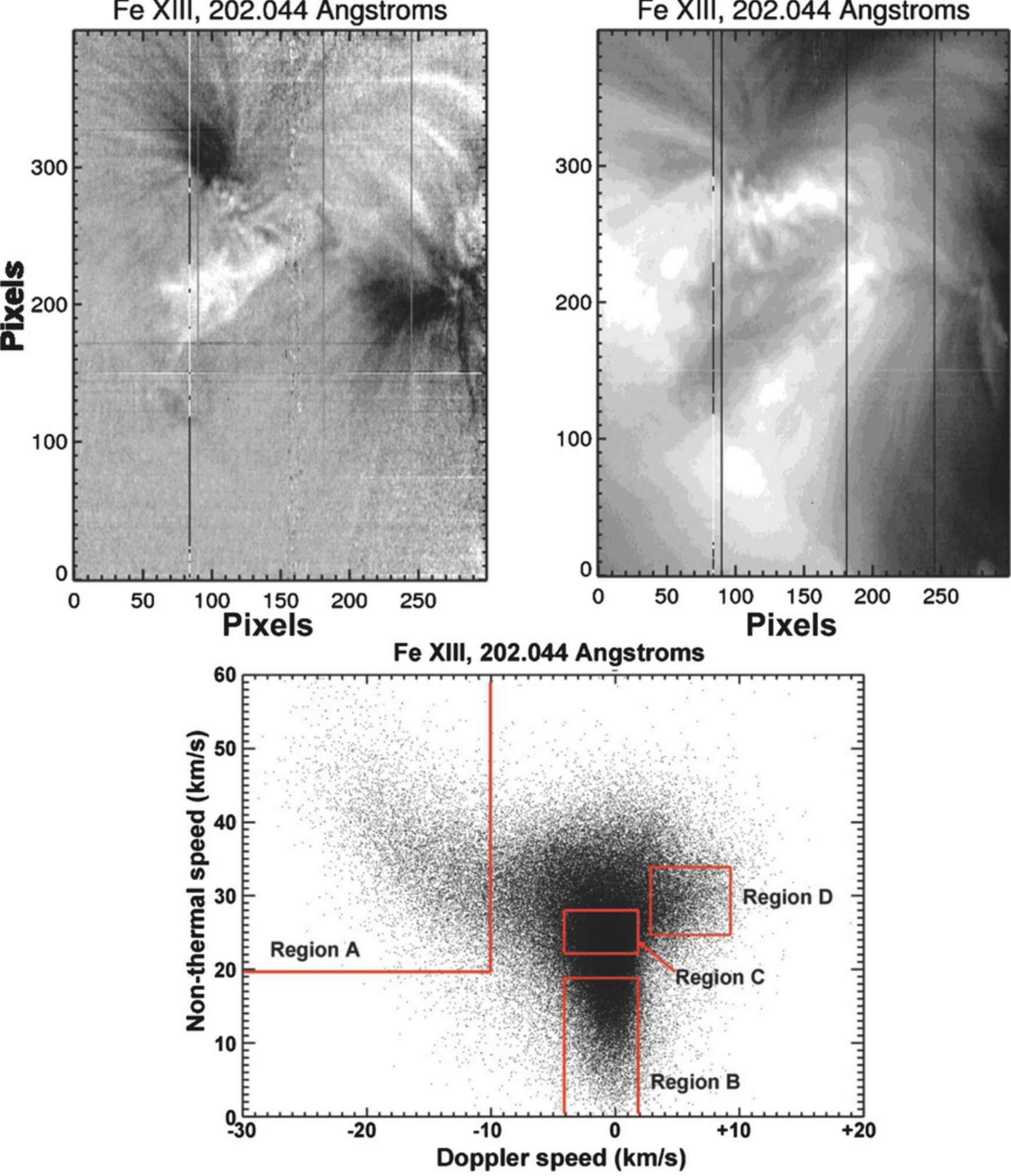}}
\caption{ Nonthermal line widths vs. blue-shifts in a Hinode EIS 
 \ion{Fe}{xiii} line (bottom panel) in an active region.
The values were obtained with single Gaussian fitting. 
The values of the Doppler flows  and the radiance in the line are shown
in the upper left and right plots, respectively 
\citep[figure from ][]{doschek:2012}. The largest non-thermal widths
are in region `A', which corresponds mainly with the dark (blue-shifted)
regions in the Doppler map.
}
\label{fig:ntw_co}
\end{figure}

On the other hand, asymmetric line profiles in some of the coronal lines
and some outflow regions have been reported by several authors
\citep[see, e.g.][]{hara_etal:08,de_pontieu_etal:09,peter:2010,bryans_etal:2010},
with large blue-shifts reaching values of the order of 100 km\,s$^{-1}$.
The  large blue-shifts were found to be correlated with high-speed
chromospheric jets or type II spicules by \cite{de_pontieu_etal:09},
and were interpreted as the signatures of coronal heating in the chromosphere.
However, such observations and subsequent  interpretations have 
been strongly debated in the literature (see, e.g. \citealt{reale:2012_lr}),
and more evidence was later provided to show that indeed 
 line profiles are  symmetric in most places
\citep{brooks_warren:2012_flows,doschek:2012}.
There is also a growing consensus that these excess widths are the 
result of a superposition, along the line of sight, of upflows with different 
velocities. 
Fig.~\ref{fig:ntw_co} (bottom panel, region A, from 
\citealt{doschek:2012}) shows the clear correlation that exist  between 
nonthermal line widths and blue-shifts, for one Hinode EIS 
coronal line from  \ion{Fe}{xiii} (other lines formed in the 2--3 MK range
show similar characteristics).

The coronal outflow regions are rooted in the centre of sunspots and in the middle
of plage regions but then have a significant expansion in the corona 
\citep{delzanna:07_dublin,delzanna:08_flows}. 
These basic observational results have later been confirmed by a number of 
authors.
A physical explanation for the excess widths and outflows was put forward
by \cite{delzanna_etal:2011_outflows}: interchange reconnection between 
active region core loops and the surrounding unipolar regions 
occurs in the outer corona and drives a pressure imbalance 
(a rarefaction wave) which  later produces effects observed in the lower corona.
Hydrodynamical modelling of the rarefaction wave gave 
outflow velocities close to the observed ones, increasing 
with the temperature of formation of the line \citep{bradshaw_etal:11}.
The process continues for a long time, but is intrinsically intermittent. 
The temporal and spatial (along the line of sight) superposition 
of different rarefaction waves would explain the broadening.


\subsubsection{Coronal lines in coronal holes}

Coronal holes are the places where one would expect 
 broadening of coronal lines caused by  the Alfv\'en waves which are 
expected to be present, both on a theoretical and observational basis.
In fact, counter-propagating  Alfv\'en waves  are expected to be present in 
the outer corona and with non-linear processes 
(turbulent dissipation) could provide energy to the 
solar wind  \citep[see, e.g.][but note that there is an extensive literature on this topic]{parker:1965,hollweg:1986,velli:1993,verdini_etal:2010,vanballegooijen_Asgari-Targhi:2016}.
The fast solar wind, streaming from coronal holes, is strongly 
 Alfv\'enic  as observed in-situ \citep[see, e.g.][]{cranmer_etal:2007}.
Pure Alfv\'en waves are transverse, i.e. the oscillations are perpendicular
to the direction of the magnetic field, which is mostly radial 
in the polar coronal holes. Therefore, outward propagating  Alfv\'en waves
would cause a broadening of the spectral lines above polar coronal holes,
as indeed observed. 
The non-thermal widths of the lines in the coronal holes are larger than in equatorial regions.
If the excess broadening decreases with height, then it would be a signature
of damping of the Alfv\'en waves, which would resolve a long-standing puzzle
about the heating of the fast solar wind.

SoHO SUMER off-limb observations of coronal hole regions 
have shown that excess line widths sometimes increase, with a tendency
to flatten off
\citep[see, e.g.][]{tu_etal:1998,banerjee_etal:1998,doyle_etal:1999,doschek_etal:2001,wilhelm_etal:2004_widths}.
Large uncertainties are often present in the data.

Hinode/EIS observations of coronal holes
have shown an increase in the excess line widths, then some decrease after 1.15-1.2~\rsun above the limb
\citep[see, e.g.][]{banerjee_etal:2009,hahn_etal:2012,bemporad_abbo:2012,hahn_savin:2013}. 

Similar results have been obtained from 
ground-based observations. Those  taken at the National Solar Observatory (USA)
of  coronal hole regions off-limb
 show large non-thermal broadenings 
increasing with height and ranging from 40 - 60 km\,s$^{-1}$ up to 1.16~\rsun
in the red line from \ion{Fe}{x} \citep{hassler_moran:1994}.
Ground-based observations with the 
coronagraph at the Norikura Solar Observatory (Japan) above coronal holes
show the same behaviour in the iron forbidden lines: the 
width of the red \ion{Fe}{x} line increases with height, while that of the 
green \ion{Fe}{xiv} line  decreases \citep[see, e.g.][]{prasad_etal:2013}.

\subsection{Flare lines}

It is well established that flare lines, i.e. lines formed at 10 MK or more,
show significant excess widths during the impulsive phase of flares,
in particular a few minutes before the onset of the HXR bursts
and also show blue-shifted components, which are signatures of the 
chromospheric evaporation. 
These excess widths were observed with e.g. the NRL SOLFLEX instrument
\citep[see, e.g.][]{doschek_etal:1979} and  the SMM  BCS 
\citep[see, e.g.][]{antonucci_etal:1982,antonucci_etal:1995}. 
One of the limitations of such measurements was the limited
spatial information about the location of the non-thermal broadenings.
With SMM/UVSP, which had much better spatial resolution than the X-ray
instruments, excess widths and asymmetric line profiles (with blue-shifts of the order of 
200 km/s) were observed during the early stages of flares at the
footpoint regions in the \ion{Fe}{xxi} flare line \citep{mason_etal:86}.

\begin{figure}[htb]
\centerline{\includegraphics[width=0.6\textwidth]{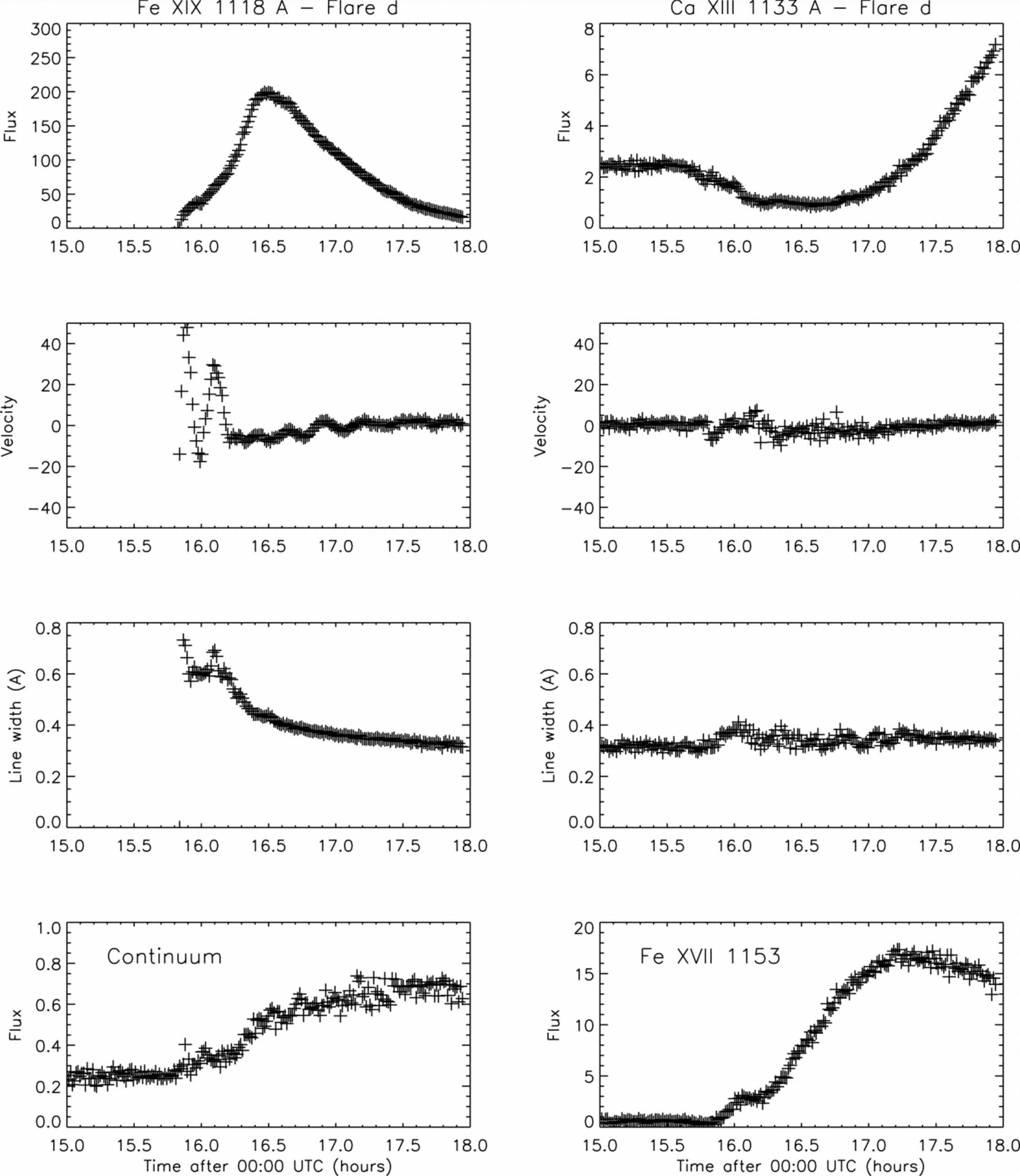}}
\caption{Nonthermal line widths in the \ion{Fe}{xix} SUMER flare line
(third plot form the top, left column) \citep{feldman_etal:2004}.
}
\label{fig:ntw_sumer_fe_19}
\end{figure}

Several  spatially-resolved observations of flares were obtained with 
SoHO SUMER in flare lines while the slit was positioned off the limb,
normally in a sit-and-stare mode. 
Most studies focused on extremely interesting Doppler flows and 
damped oscillations observed
in the flare lines. However, a few published studies also 
presented  measurements of the excess widths as a function of time.
For example, \cite{kliem_etal:2002} reported a significant 
excess width of almost 100 km\,s$^{-1}$ in the \ion{Fe}{xxi} flare line
during the peak phase of a flare. 
Similar results were obtained by 
 \cite{landi_etal:2003} from SUMER observations of an M7 class flare.
The widths during the initial phases were difficult to 
estimate because the line profiles were broadened by large 
(600 km\,s$^{-1}$) systematic bulk motions, as described by \cite{innes_etal:2001}.
  \cite{feldman_etal:2004} measured excess widths during the 
initial phases of several flares, in the \ion{Fe}{xix} SUMER flare line
at 1118~\AA. 
At flare onset, nonthermal mass motions were found to be between 50 and
100 km\,s$^{-1}$, as shown in  Fig.~\ref{fig:ntw_sumer_fe_19}.

SoHO/CDS allowed spatially-resolved measurements of  non-thermal broadenings on-disk
in the \ion{Fe}{xix} flare line.
 It was a challenge to find an 
observation where the  \ion{Fe}{xix} flare line was  visible during the impulsive phase 
of a flare, when the CDS slit was at the right time and right place. 
One such observation is the M1-class flare described in  
\cite{delzanna_etal:2006_m1_flare}. During the impulsive phase of the flare,
 large excess widths (and totally blueshifted line profile) were 
observed  in the \ion{Fe}{xix} line, at the  kernels of chromospheric evaporation.
The line intensity was very weak, at the limit of detection.
 The authors also showed that the 
line width, as well as the blue-shift, decreased over time and 
quickly disappeared during the peak phase. 
Similar (but progressively smaller) blueshifted line profiles were found in lines formed 
at lower temperatures (1--3 MK). 
On the other hand, the bright post-flare loops
that were formed during the impulsive phase did not show any appreciable
non-thermal broadenings. 
Large non-thermal widths and a blue-shifted \ion{Fe}{xix} line profile were also found
in a sit-and-stare CDS observation by  \cite{brosius:2003} during the impulsive phase of
a flare.
In contrast, most other CDS studies found  that the \ion{Fe}{xix} line profile was asymmetric,
with an enhancement in the blue wing \cite{teriaca_etal:2003, milligan_etal:2006}.

The hydrodynamic response of the atmosphere during chromospheric 
evaporation has been studied for a long time with 1-D modelling, which 
usually predicts that line profiles should be totally blueshifted with 
increasing velocities in lines formed at higher temperatures,
as seen in the CDS observations reported by \cite{delzanna_etal:2006_m1_flare}.

As we explain below, there is now convincing evidence that the flare lines 
are broad but blueshifted. However at the time the vast majority of 
the community thought that the asymmetric line widths were real.  
A possible explanation for this was that   
the asymmetric excess widths were composed of a 
 superposition of multi-thread loops  activated at different times 
\citep[see, e.g.][]{doschek_warren:2005}.

As in the CDS case, most literature based on Hinode EIS observations of the flare 
 \ion{Fe}{xxiii} and  \ion{Fe}{xxiv}  lines reported asymmetric line profiles, with strong 
blue-shifted components \citep[see, e.g.][]{milligan_dennis:2009,young_etal:2013_flare}.
As in the CDS case, the observations are challenging because the EIS slit had to be 
at the right place at the right time and the flare lines are very weak during the impulsive phase.
Furthermore, these EIS flare lines are all blended to some degree, as discussed
e.g. in \cite{delzanna:08_bflare}.
\cite{delzanna_etal:2011_flare} however reported, as in the CDS case, weak but totally 
blueshifted and broad  \ion{Fe}{xxiii} emission in kernels during the 
impulsive phase of a small B-class  flare, which behaved like a
textbook standard model case.  
Similarly, Hinode EIS sit-and-stare \ion{Fe}{xxiii} observations reported by 
\cite{brosius:2013a}  of a C1-class flare also showed large widths and blue-shifts,
although at some times two components were observed.

As an aside, \cite{delzanna_etal:2011_flare} clearly showed that lower-temperature lines
formed around 3 MK appear to be broader, but in reality their line profiles are 
composed of a superposition of a stationary component which is produced by
foreground emission (which is always present in active region at such temperatures),
and a blue-shifted component which only appears during the impulsive phase 
in the kernels of chromospheric evaporation. 
This is an example of the complexity and richness of information which is 
present in any line profile. 

\begin{figure}[htb]
\centerline{\includegraphics[width=0.6\textwidth]{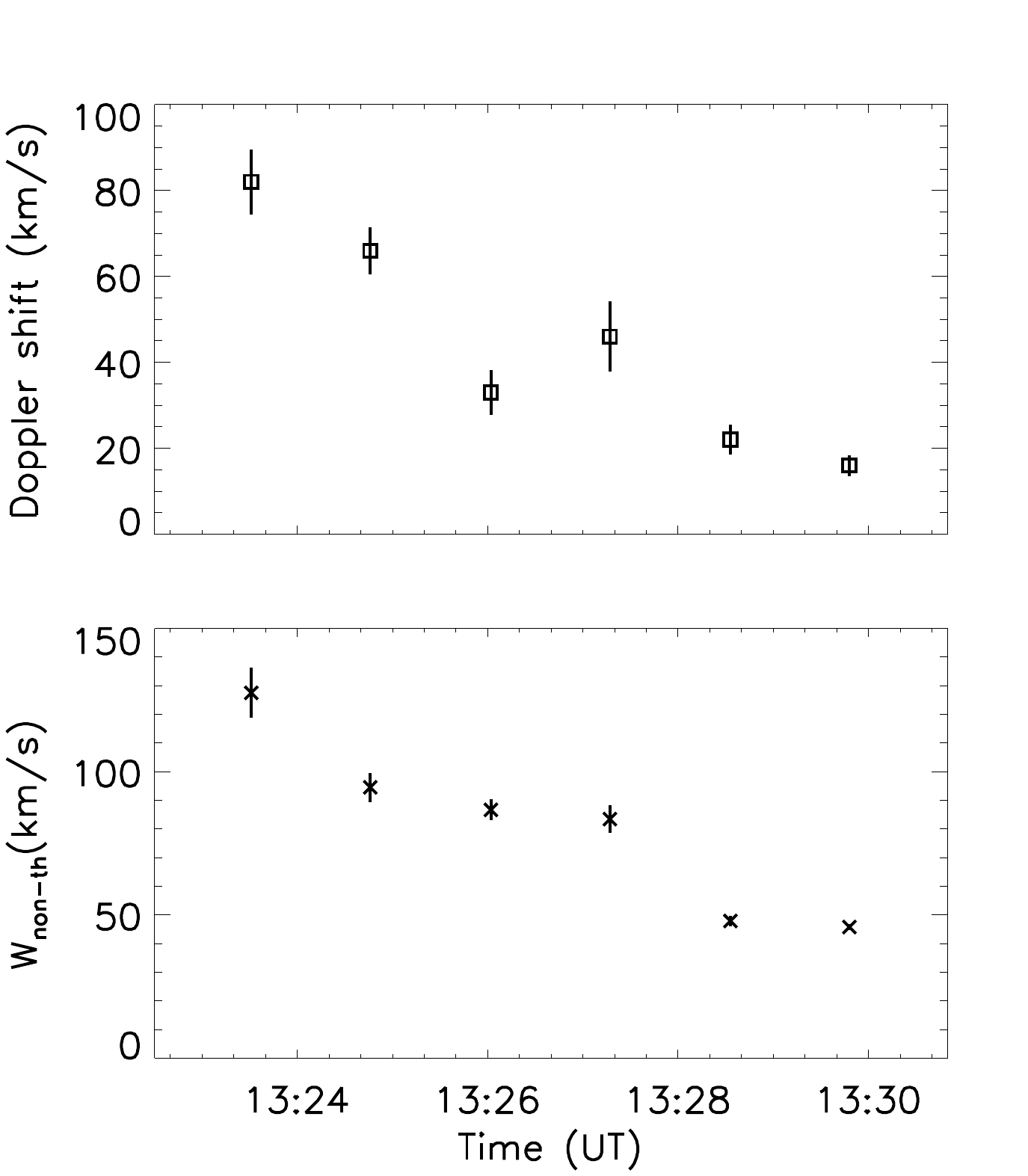}}
\caption{Nonthermal line widths and blue-shifts in the IRIS 
 \ion{Fe}{xxi} line  as a function of time during the impulsive phase of a 
flare, at one kernel  location in a ribbon (adapted from \citealt{polito_etal:2015}).
}
\label{fig:ntw_fe_21}
\end{figure}

With the launch of IRIS, new observations of the \ion{Fe}{xxi} flare 
line at high spatial resolution and high cadence have been possible.
In all cases the line was found to have  strong non-thermal broadening and
a totally blueshifted line profile  during the impulsive phase 
at the kernels of chromospheric evaporation  
\citep[see e.g.][]{polito_etal:2015, young_etal:2015,tian_etal:2015}, as in the
\ion{Fe}{xix} CDS observations reported by \cite{delzanna_etal:2006_m1_flare}. 
One example from \cite{polito_etal:2015} is shown in  Fig.~\ref{fig:ntw_fe_21}. 
There is a remarkable agreement between the decrease of the excess width and the 
blue-shift in the line profile. This confirms earlier work with
SoHO/CDS and the 
\ion{Fe}{xix} line  by  \cite{delzanna_etal:2006_m1_flare}, although
CDS had a much lower
temporal, spectral and spatial resolution.

The above differences are puzzling. It is possible that different sub-resolution
strands are activated in different ways, as suggested by the observations 
reported by \cite{brosius:2013a}. 
On the other hand, the large differences in spatial resolution between EIS and IRIS
could cause such differences, as shown by \cite{polito_etal:2016a}.
\cite{polito_etal:2016a} found a rare occurrence, a flare simultaneously
observed  at the same place by Hinode EIS and IRIS. The EIS flare lines 
showed asymmetric line profiles while the IRIS \ion{Fe}{xxi} was totally blueshifted. 
However, the IRIS data showed that within the EIS effective spatial resolution (3--4''), 
both a stationary and a blueshifted component were present. 
This provides a simple explanation as to
 why the profiles of EIS flare lines are often asymmetric.

\clearpage
 \section{Measuring Chemical  Abundances from XUV Spectroscopy} 
\label{sec:abund}

\subsection{Introduction}

\cite{pottasch:63} used full-Sun EUV observations and developed
an approximate emission measure method (described below) to
investigate the abundances of elements. He found that 
the abundances of elements such as magnesium and iron were greatly 
enhanced compared to their photospheric values.
These early results had large uncertainties (about a factor of 2),
 mainly because of the use of approximate methods to 
estimate the cross-sections for electron impact excitation.

During the 1980's it became well established (cf. \citealt{meyer:85}) that 
the coronal abundances were variable and different from  the 
 photospheric values.
 Patterns in the variations also emerged, the most famous one being the 
correlation between the  abundance of an element   and its 
first ionization potential (FIP).
The low-FIP ($\leq$ 10 eV) elements are more abundant 
than the high-FIP ones, relative to the photospheric values (the FIP bias).

Support for  the presence of a FIP bias 
later came from the in-situ measurements
of the chemical composition of the  solar wind,
where it was clear that the fast wind streaming from the 
coronal holes has a near-photospheric composition, while 
the slow wind has a variable composition, on average 
showing a FIP bias (see, e.g. \citealt{vonsteiger00}).

There is an extended literature on 
measurements of solar chemical abundances, and only 
a few  key issues are discussed here.
We point out that 
over the years  several review articles have been written, see e.g.
\cite{meyer:85, feldman:92, mason:95, bochsler:07,feldman_laming:2000, feldman_widing:2002,feldman_widing:2003,feldman_widing:2007,asplund_etal:09,lodders_etal:10,schmelz_etal:12_abundances,delzanna_mason:2013_book,delzanna_mason:2014}.
Various  articles can also be found in 
\cite{soho_ace_workshop:01}, the proceedings of a 
joint SOHO/ACE workshop on solar and galactic composition in 2001.
There is also a recent \textit{Living Review} by \cite{laming:2015},
which also discusses stellar observations and the theoretical aspects
of abundance variations in the solar and stellar atmospheres.
 
We focus here on remote-sensing measurements of abundances in 
the low corona from XUV spectroscopy.
We note that accurate abundance measurements are important in many respects.
As described in  \cite{laming:2015}, the chemical fractionation 
is intimately related to the processes that heat the solar corona, so 
FIP bias variations provide important  diagnostics. 

As briefly mentioned in this chapter, there are clear correlations between  
remote-sensing and the in-situ abundance measurements. 
Accurate measurements of remote-sensing abundances will be 
particularly important
for Solar Orbiter, because they can be used to help
identifying the source regions of the solar wind that 
will be measured close to the Sun (at 0.3 AU). 
There is now a growing consensus that element fractionation is 
perhaps the only plasma characteristic that does not change, along open 
field lines from source regions into the heliosphere. 
 Once the FIP fractionation has occurred in the chromosphere,
it is not going to change in the low corona, unlike other characteristics such as the 
temperature, ionization state, particle distributions, etc. 
In the outer corona, other effects such as gravitational settling 
can occur and modify the abundances until the plasma becomes collisionless.

The abundances of several elements such as Helium, Neon and Argon cannot be 
measured in the photosphere, since they do not produce any absorption lines.
Therefore, remote-sensing measurements are one direct way to 
measure the solar abundances of these elements.

Knowing the chemical abundances is also important for modelling
the energy budget in coronal structures.
In fact, the radiative losses in the chromosphere-corona
are a major sink of energy, and depend directly on the 
absolute abundances.

\subsection{Diagnostic methods to measure elemental abundances using spectral lines}

Several diagnostic methods have been developed to measure chemical abundances in 
the low corona. The most common ones rely on some sort of EM/DEM analysis.
We recall that for optically thin lines the observed radiance is proportional
to the chemical abundance of the element ${Ab(Z)}$:

\begin{equation}
{I(\lambda_{ij})}= {Ab(Z)} {\int\limits_T ~{C(T,\lambda_{ij},N_e)} ~DEM (T) ~ dT},
\end{equation}
\noindent
so for example once a $DEM(T)$ is obtained from e.g. lines of the same element,
the relative abundances of the other elements are directly obtained.

However, what is really needed is the `absolute' abundance of an element, 
i.e. its abundance relative to hydrogen. 
One way to establish this is when hydrogen lines are also observed. 
Hydrogen lines are normally formed in the chromosphere, so radiative 
transfer effects should be taken into account. 
However, in off-limb observations the observed hydrogen emission is 
formed at coronal temperatures, and lines are optically thin, so direct
abundance measurements  can be obtained
\citep[see, e.g.][]{laming_feldman:2001}. The lines are very weak and
can be affected by instrumental stray light and resonance scattering 
higher up in the corona. Gravitational settling also affects measurements
higher up.

Several EM/DEM methods have been developed and used to measure 
relative elemental abundances.
Whenever the plasma is nearly isothermal, the EM loci method (see Section~\ref{sec:em_methods})
is  often used instead,
adjusting the relative abundances so all the EM loci curves cross at one point. 

Various approximate methods have also been developed.
One method for example  is to  define an 
average emission measure $<EM>$ for each observed line, while another is to 
define an averaged $<DEM>$ for each observed line.
These approximations and their limitations are described below.

\subsubsection{The Pottasch approximation}

Following \cite{pottasch:63}, many authors
have approximated the integral by  assuming an averaged value of 
the $G(T)$ (i.e. $A_b(Z) ~ C(T)$): 
\beq
I\lo{th} =   A_b(Z) ~ <C(T)> ~ {\int\limits_h  ~N_e N_H dh }
\eeq
A \textit{line emission measure} $EM_L$ 
can therefore  be defined, for each observed line of intensity I\lo{ob}:
\beq
 EM_L \equiv  {I\lo{ob} \over  A_b(Z) ~ <C(T)>}  \quad ~{\rm [cm^{-5}]}
\label{eq:em_l}
\eeq

The approximation  applied by  \cite{pottasch:63}
was to take $C(T)=0$ when $C(T)$ is less than one-third of its maximum
value, and equal to a constant value otherwise.
For $T_1, T_2$ such that $ C(T_1)=C(T_2)={1\over3} C(T\lo{max})$, define:

\beq
C_{P}(T)= \left\{ \begin{array}{lll}
                            C_{0} &\quad T_1 \leq T \leq T_2 & \\
                            \noalign{\smallskip}
                            0 &\quad T < T_1 , \quad T > T_2  & \\
                               \end{array}
                         \right. 
\eeq
and requiring that 

\begin{equation}
<C(T)> = { {\int C(T) dT} \over {|T_2-T_1|}}
\end{equation}
which gives $<C(T)> = 0.7  ~C(T_{\max})$.

\begin{figure}[htb]
\centerline{\includegraphics[width=10cm]{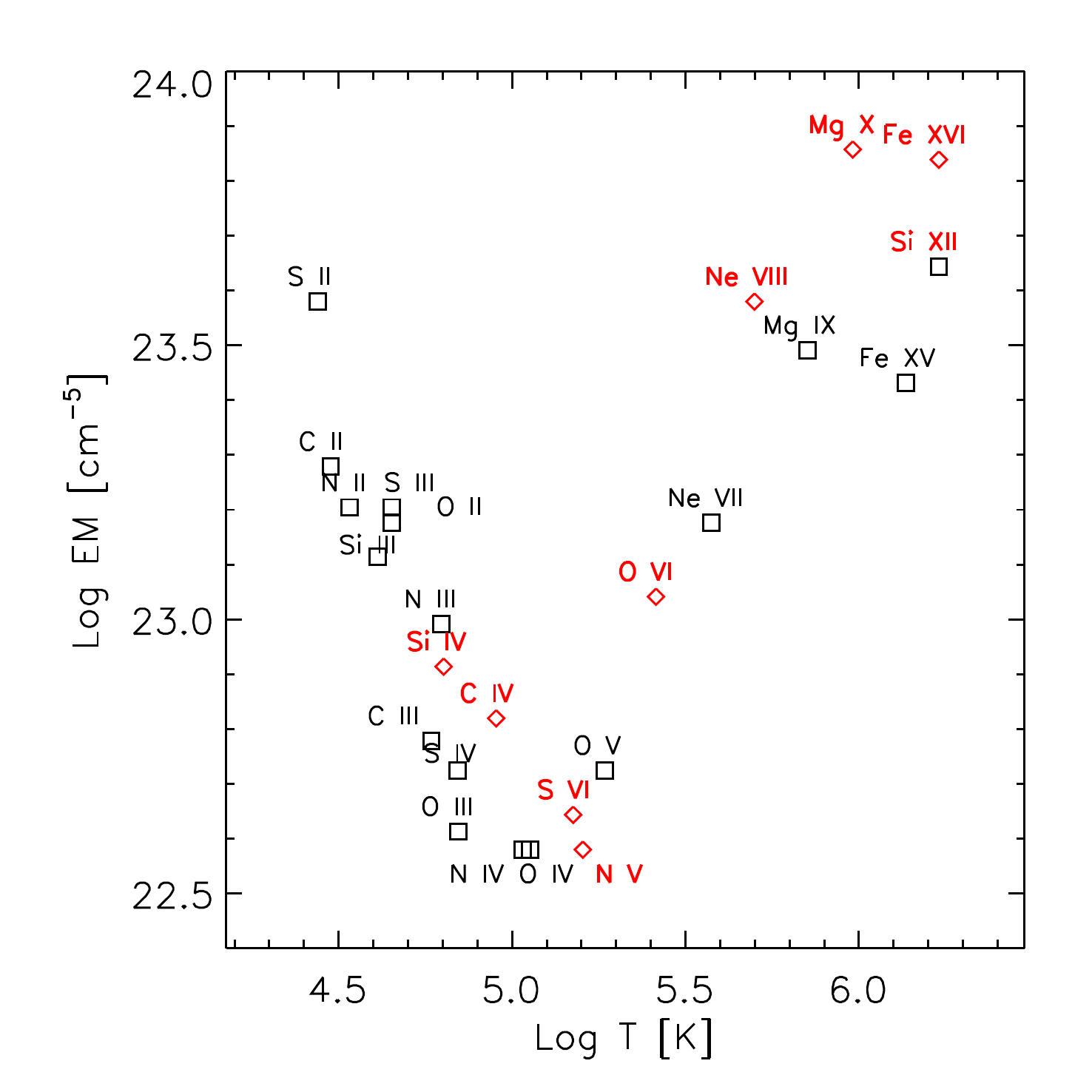}}
  \caption{The line emission measures as obtained by 
\cite{pottasch:63} with older atomic data.
Lines from ions of the  Li- and Na-like sequences, 
such as \ion{C}{iv} and \ion{Si}{iv}, are shown in red. These 
are clearly at odds with the other ones.
}
\label{fig:pott}
\end{figure}

\cite{pottasch:63} and following authors have produced  figures of the 
\textit{line emission measures}, multiplied by the corresponding abundance value:
\beq
A_b(Z) ~EM_L = {I\lo{ob} \over 0.7 ~ C(T_{\max})}
\eeq
plotted at the  temperature T\lo{max} corresponding to the  peak value 
of the contribution function of the spectral line.
An example is given in Figure~\ref{fig:pott}, obtained from the 
original data points published by  \cite{pottasch:63}.
The relative abundances of the elements are derived in order to have all
the \textit{line emission measures} of the various ions lie along a 
common smooth curve.

\subsubsection{The  Jordan and Wilson (1971) approximation}

\cite{jordan_wilson:71}  adopted the  \cite{pottasch:63}  approach but 
 proposed a different approximation, assuming that $C(T)$ has a 
constant value over a narrow temperature interval $\Delta log T=0.3$
 around $C(T_{\max})$:
\beq
C_{J}(T)= \left\{ \begin{array}{ll}
                            C_{0} &\quad |\log\,T - \log\,T_{\max}| < 0.15 \\
                            \noalign{\smallskip}
                            0 &\quad |\log\,T - \log\,T_{\max}|  > 0.15\\
                               \end{array}
                         \right. 
\eeq
and requiring that 
\beq
\int C(T) dT = \int C_{J}(T) dT
\eeq
so that
\begin{equation}
\label{C-lamb-def}
C_{0}={ \int C(T) ~dT \over T_{\max}(10^{+0.15} - 10^{-0.15})} = { \int C(T) ~dT \over  0.705 ~T_{\max}}
\end{equation}
and thus deducing the relative \textit{line emission measure} $EM_L$, sometimes 
indicated by EM(0.3), to make it clear that the contributions are calculated over a 
temperature interval $\Delta$ log $T=0.3$.
This approximation has been used by various authors to  derive  element abundances.
As an example, values of EM(0.1) (at intervals of $\Delta$ log $T=0.1$)
 relative to a Hinode EIS spectrum of an active region core
 are  displayed  in  Figure~\ref{fig:em_pott} as triangles.
The EM loci are  upper limits and can be quite different
(as Figure~\ref{fig:em_pott} shows) from the 
actual EM values, which should be obtained once the full DEM has been calculated.

\subsubsection{The Widing and Feldman (1989) approximation}
\label{sec:em_methods_wf}

A different approach was proposed by \cite{widing_feldman:89}.
The idea is to extract  from the integral  
an averaged value of the DEM of the line:
\beq
DEM_L \equiv \left< N_e N_H {dh \over dT}  \right>  \quad [cm^{-5} K^{-1}]  
\label{eq:obs_dem_l}
\eeq
such that  for each line of observed intensity I\lo{ob}:
\beq
DEM_L \equiv  { I\lo{ob} \over  A_b(Z)  ~  {\int\limits_T  ~{C(T) dT }}}
\eeq
A plot of the $A_b(Z) ~ DEM_L= I\lo{ob} / {\int\limits_T  ~{C(T) dT }}$ values 
displayed at the temperatures T\lo{max}  gives the $DEM_L$ values of the observed lines.

This method has been widely used in the literature 
to obtain relative element abundances, 
 adjusting the abundances in order to have a  continuous sequence of the  $A_b(Z) ~ DEM_L$ values.
In reality, the approximation only works when 
a smooth continuous distribution of the plasma temperature is present.
In  fact, 
given two elements $X_1$ and $X_2$, the ratio of the observed intensities
can be written:
\begin{equation}
{I_1 \over I_2} = { {A_b(X_1) ~\int\limits_T~ {C_1(T, N\lo{_e})}  ~DEM (T) ~dT}
\over {A_b(X_2) ~\int\limits_T~ {C_2(T, N\lo{_e})}  ~DEM (T) ~dT} }
\label{eqn:ab1}
\end{equation}
from which the relative element abundance $A_b(X_1)/A_b(X_2)$ can be deduced, from the 
observed intensity ratio $I_1 / I_2$, once the DEM distribution is known.
Only when the two lines have similar $C(T)$ and the DEM
distribution is relatively flat would one  expect that the DEM factors out 
from the integrals:
\begin{equation}
{\int\limits_T~ {C_1(T, N\lo{_e})}  ~DEM (T) ~dT \over \int\limits_T~ {C_2(T, N\lo{_e})}  ~DEM (T) ~dT}
 = 
{\int\limits_T~ {C_1(T, N\lo{_e})} ~dT \over \int\limits_T~ {C_2(T, N\lo{_e})}  ~dT}
\label{eqn:ab12}
\end{equation}
If the above equality holds, then 
it is possible to  deduce the relative abundances directly from the observed
intensities and the contribution functions, because:
\begin{equation}
{A_b(X_1) \over A_b(X_2)} = { {I_1 \cdot  ~\int\limits_T~ {C_2(T, N\lo{_e})} ~dT }  \over
 {I_2 \cdot ~\int\limits_T~ {C_1(T, N\lo{_e})} ~dT} } = {DEM_L(X_2) \over DEM_L(X_1)}
\label{eqn:ab2}
 \end{equation}
i.e. the $DEM$ method and the $DEM_L$ method are equivalent.
Of course, if two lines have exactly the same contribution function, then the 
$DEM_L$ method should produce the same results as the $DEM$ method.
 However, the contribution functions of lines from two ions are 
never exactly the same, and any small difference can be amplified if the 
$DEM$ is a steep function where the two lines differ. 
In such a case,  the $DEM$ and  $DEM_L$  methods can be produce very 
different results. 

\begin{figure}[htbp]
\centerline{\includegraphics[width=8cm,angle=90]{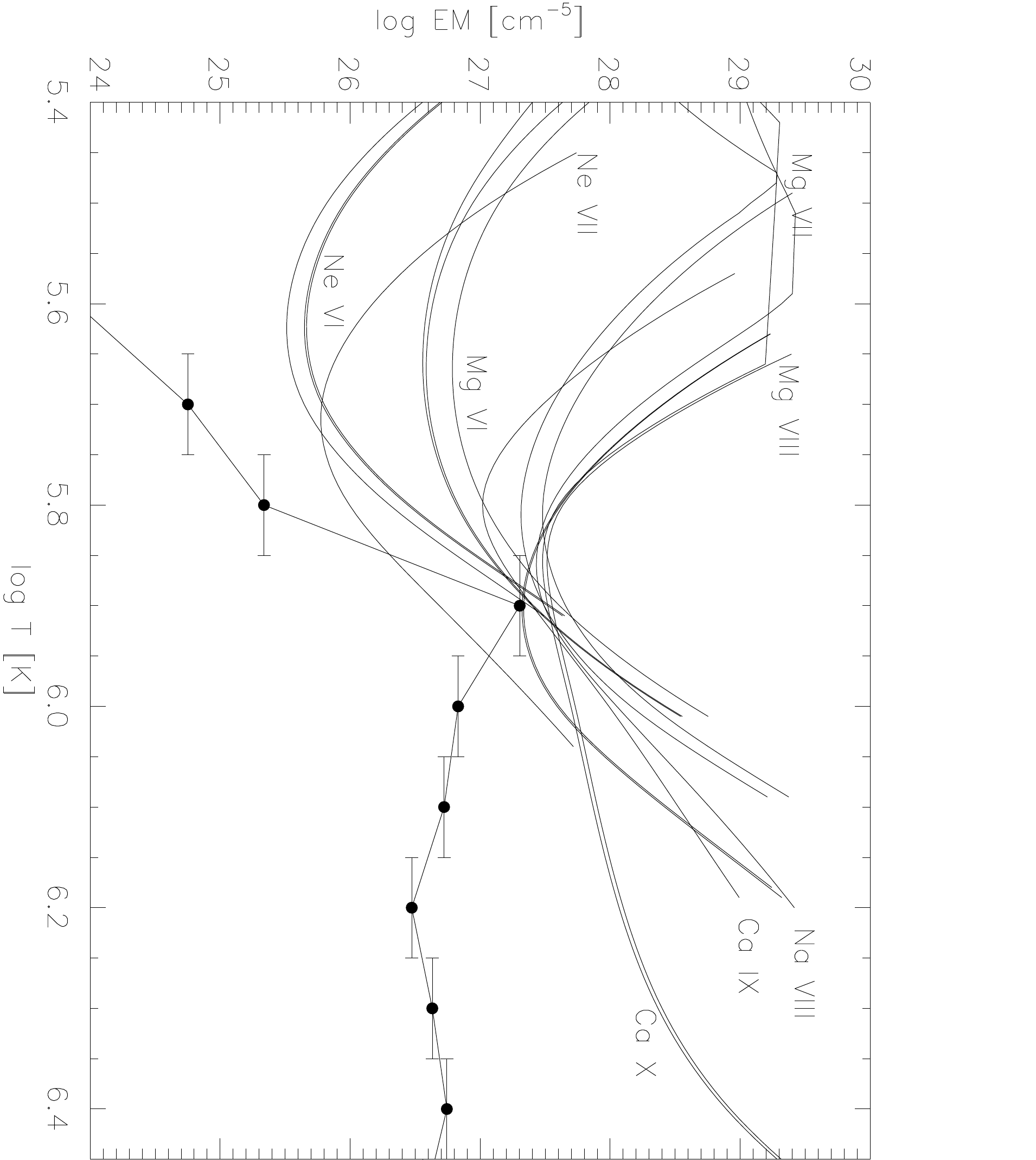}}
  \caption{EM loci $I\lo{ob} / (G(T))$ curves obtained 
from the coronal hole plume observed by Skylab
\citep{widing_feldman:89}, assuming
  photospheric abundances. The data indicate an isothermal 
distribution at log $T$ [K]=5.9 and are consistent with no FIP effect being present. 
The emission measure $EM(0.1)$ values  are also shown (filled circles). 
Figure from \cite{delzanna_etal:03}.
}
  \label{fig:wf_pl_deml}
\end{figure}

There are several features  on the Sun
 where the plasma distribution appears to be  nearly isothermal,
and the $DEM_L$  approximation can produce significant errors in the 
derived abundances. 
For example, several abundance measurements were obtained from the Skylab 
slit-less spectrometer 
on active region loop legs or coronal hole plumes.
In these cases, it has been shown by e.g. \cite{delzanna01_ace,delzanna:03} that 
the method can seriously overestimate  the relative 
chemical abundances, whenever the DEM distributions are close to isothermal.
As an example, Fig.~\ref{fig:wf_pl_deml} shows the EM loci curves obtained 
from the coronal hole plume observed by Skylab
\citep{widing_feldman:89}, which are consistent with photospheric abundances,
and not an increased  Mg/Ne ratio by a factor of 10 as obtained with the \cite{widing_feldman:89} approximation.

\subsection{Absolute elemental abundances using the continuum}

One way to get absolute abundances is to measure both line radiances and the continuum. 
As already pointed out by \cite{woolley_allen:1948},
the visible lines allow a direct measurement of the 
absolute (i.e. relative to hydrogen) values of the chemical abundances, 
by comparing line radiances with the visible continuum, whenever observed.
This is because the continuum is due to  the 
Thomson scattering of the photospheric radiation
by the coronal electrons.
Given that this Living Review is focused on XUV diagnostics, 
we do not discuss this diagnostic further here. 

Another possibility which has been exploited in the X-rays is to 
use the X-ray line to continuum measurements
\citep[see, e.g.][]{veck_parkinson:81}. 
The X-ray continuum emissivity at some temperatures and wavelengths is dominated
by  free-free radiation, that is mostly by hydrogen, 
hence the continuum is directly proportional to the hydrogen density.
These types of measurements are normally limited to the X-rays when 
free-free emission becomes significant during solar flares (at temperatures
of the order of 10 MK). However, some free-bound emission, which depends on 
the chemical abundances of several elements, is also present 
(see  Figure~\ref{fig:continuum})  and 
needs to be taken into account in the analysis.
We provide several examples below on the use of this diagnostic.

Finally, we point out that the same measurements 
can be carried out in the UV using the free-free emission during a 
flare. Results from  SUMER have been obtained by  \cite{feldman_etal:2003}.

\subsection{Overview of abundance measurements}

It has become commonly accepted  in the literature
that the  solar corona has a FIP bias of about 4,
while coronal holes have near photospheric abundances. 
Results on active regions and flares have been  
controversial, and several issues are still very much debated
in the literature.

We start by noting that 
several issues have complicated  the interpretation of the 
measurements.
First, most of the remote-sensing and in-situ measurements
are about \emph{relative abundances}, while only occasionally 
have results been obtained relative to hydrogen.
Indeed it is still debated  whether the low-FIP elements are \emph{enhanced}
in the corona, or high-FIP ones are \emph{ depleted}, 
or if a `hybrid' solution (cf. \citealt{fludra_schmelz:99}) applies.
Second, the FIP effect is normally measured \emph{relative to 
a standard set of photospheric abundances}. 
Third, it has not been possible to measure directly the solar 
photospheric abundances for some important elements (such as the 
high-FIP He, Ar, Ne). 
Fourth, results for the solar photospheric abundances have been changing
(and debated) considerably in the literature, especially in the past few years, when 
3D time-dependent hydrodynamic simulations of the solar convection zone
have been introduced. Measurements of the FIP bias have therefore been changing
as a consequence.
Fifth, several measurements had little or no 
spatial resolution, many were full Sun. 
Finally, several approximate or inaccurate methods have been used.

A major problem of full-Sun measurements is related to the 
fact that while most of the contributions to the 
irradiance of lines formed up to 1~MK comes from QS areas,
a significant contribution for the higher-temperature lines
comes from the hot loops in active regions, 
so relative abundances in lines formed at different temperatures refer
to different solar regions.
One example of an often-cited result based on irradiance 
measurement is the \cite{laming_etal:1995} study.
They  used the excellent 
grazing-incidence spectrum published by \cite{malinovsky_heroux:73}
and an emission measure analysis. They found  
near photospheric abundances when considering 
lines formed in the transition region, up to 1~MK.
However, they found  a FIP bias of 3\,--\,4 at temperatures above 1~MK.
\cite{malinovsky_heroux:73}  presented  a calibrated  spectrum 
covering the 50-300 \AA\ range with a medium resolution (0.25 \AA),
taken with  a grazing-incidence spectrometer  flown on a
rocket on 1969 April 4, 
when the Sun was `active', and significant contributions from 
active regions were present.
The \cite{laming_etal:1995} results  therefore seem to be consistent
with the overall picture that we discuss below, i.e. that 
the quiet Sun has nearly photospheric abundances, while 
the hot loops in active regions show a FIP bias of 3\,--\,4.

On the other hand, there is a significant body of literature
where  observations have been interpreted  in a different way: 
that the lower TR and the corona above have different FIP 
biases \citep[see also][]{feldman_laming:1994,feldman:1998}.
This is the interpretation that \cite{laming_etal:1995} made, 
in terms of unresolved fine structures, ufs, small TR loops.

Due to the nature of the Skylab NRL slitless 
spectrometer, only small-size bright features could be 
readily observed with this instrument.
Many of the abundance results were obtained from nearby lines of 
Mg VI (low-FIP) and Ne VI (high-FIP).

SOHO CDS  could resolve lines from a sequence of Magnesium 
 (Mg V, Mg VI, Mg VII,Mg VIII) and Neon (Ne IV, Ne V, Ne VI, and Ne
 VII) ions
before degradation due to the temporary SOHO loss,  so the 
FIP effect could be studied in terms of Mg/Ne relative abundances
at transition region temperatures.
SOHO SUMER also offered several abundance diagnostic ratios,
but again mostly in transition-region lines. 

On the other hand, some diagnostics in coronal lines 
have been available with Hinode EIS.
As described in \cite{delzanna:12_atlas}, Hinode EIS observes several 
lines formed around 3~MK, from the low-FIP Fe, Si, Mg, Ni, Ca and 
the high-FIP Ar. 
Strong lines from S are also present. They are very useful 
because S abundances normally show the same 
FIP effects as those of the  high-FIP elements.

Finally, we note that there are many studies of solar 
abundances that are not strictly measurements of the FIP effect.
For example, see \cite{young:2005b} on the O/Ne ratio in the quiet Sun
and  \cite{schmelz_etal:2005} on the O/Ne ratio in active regions.
In what follows, we focus mainly on results relating to the FIP effect.

\subsection{Abundances in coronal holes and plumes}

\cite{feldman_widing:1993} obtained near photospheric 
Mg/Ne abundances (FIP bias about 1\,--\,2) from the rims of the limb brightening 
in coronal holes 
imaged by the Skylab NRL slitless spectrometer in  Mg VI and Ne VI lines.

No significant FIP effect was observed using SOHO CDS observations
in coronal holes
\citep{delzanna01_ace, delzanna_etal:03}.
 \cite{delzanna_jgr99a} found indications of a small FIP bias between 
cell centres and supergranular network areas, with the cell
centres showing enhanced Mg/Ne.

\cite{feldman_etal:98b} used SOHO SUMER observations 
above the north polar coronal hole  at a radial distance  $R \le
1.03$\,\rsun
and found no significant FIP effect. 

\cite{doschek_etal:1998_si_ne} used SOHO SUMER observations of the off-limb corona
to measure the Si/Ne abundance using primarily lines from Si VII, Si VIII,
Ne VII and Ne VIII. They found 
 that the Si/Ne abundance ratio in inter-plume polar coronal hole regions 
was about a factor of 2 greater than the photospheric value.

Finally, \cite{laming_feldman:2001,laming_feldman:2003} measured the He abundance using 
ratios of He/H lines observed with SoHO SUMER off limb. 
The authors obtained a value similar to the He abundance measured 
in-situ in the fast solar wind.

The main features within coronal holes are plumes. It is important to assess
if plumes have non-photospheric abundances, as in that case this would be a 
strong indication that plumes are probably not the sources of the fast 
solar wind, since in-situ measurements indicate nearly photospheric abundances
for this type of wind (see the Living Review by \cite{poletto_lrsp}).

The legs of coronal hole plumes are very bright in the
transition-region low-FIP Mg VI, Mg VII and high-FIP Ne VI, Ne VII
lines, which have been observed with e.g. Skylab, SoHO CDS and SUMER.
\cite{widing_feldman:92} analysed one Skylab observation of a plume and
found an FIP effect of 10, using the \cite{widing_feldman:89} approximation.
As discussed above (cf. Fig.~\ref{fig:wf_pl_deml}), the isothermal 
nature of the plumes led to an overestimation of the FIP effect.
Indeed, using  the EM loci method and SoHO CDS observations,  plumes 
were found by \cite{delzanna01_ace, delzanna_etal:03} to have 
near-photospheric abundances.

\cite{young_etal:1999} found a small FIP bias of only 1.5
at the base of a coronal hole plume, from the Mg and Ne lines observed by CDS. 
\cite{curdt_etal:2008} pointed out that the ratio of the 
\ion{Ne}{viii} vs. \ion{Mg}{viii} as observed by SUMER varies 
across plume structures, but could not perform a temperature analysis
(using lines from other ionisation stages) to be able to measure the 
FIP bias. 
Recently, \cite{guennou_etal:2015}  monitored 
coronal hole plumes over  a few hours using Hinode EIS, and found no significant
variation in the plume abundances with time. 
The values obtained were 
consistent with photospheric abundances, confirming the CDS results.
 However, we point out that only a weak an blended line 
from a  mid-FIP element (\ion{S}{viii} 198.5~\AA) was used to measure the 
FIP bias relative to Fe and Si. 

\subsection{Abundances in the quiet Sun}

\cite{feldman_widing:1993} obtained near photospheric 
Mg/Ne abundances (FIP bias = 1\,--\,2) from the rims of the QS limb brightening 
in Skylab NRL slitless spectrometer, i.e. basically no 
difference with the coronal hole case.

Skylab observations of the Mg and Ne lines in small 
quiet Sun brightenings have generally shown photospheric 
abundances (cf. \citealt{sheeley:1996}).

\begin{figure}[htbp]
\centerline{\includegraphics[width=8cm,angle=0]{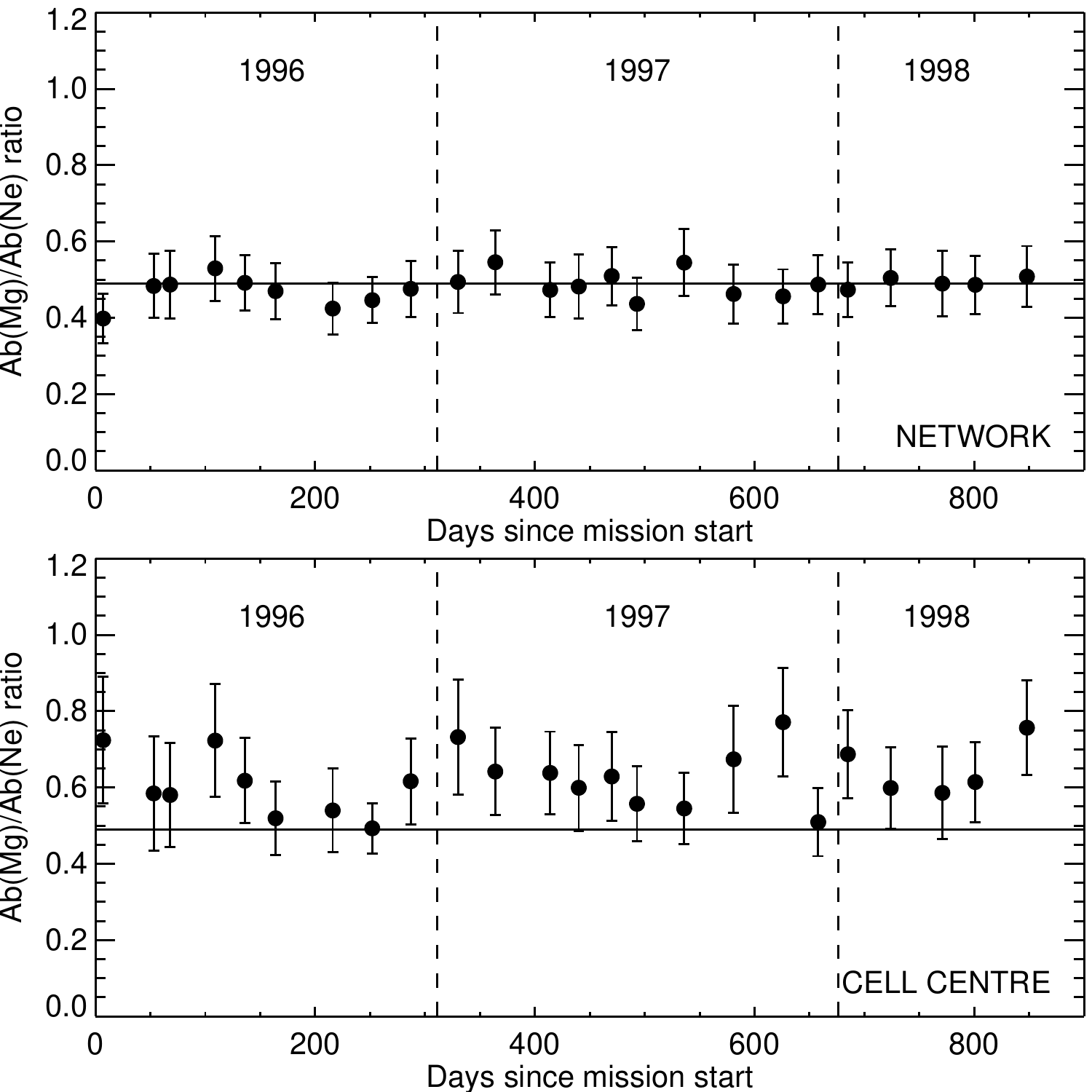}}
  \caption{The  Mg/Ne abundance ratios for the network (upper panel) and cell centre (lower panel) 
quiet Sun regions as a function of time, as obtained from SoHO CDS by P.R. Young
(a revision of the analysis presented in \citealt{young:2005}). 
The black horizontal line denotes the photospheric Mg/Ne abundance ratio. 
}
  \label{fig:young:2005}
\end{figure}

SOHO CDS results in terms of Mg/Ne abundance in the transition region
 have generally shown 
that the quiet Sun has  nearly photospheric abundances 
\citep{delzanna_jgr99a}.
These authors found indications of a small FIP bias between 
cell centres and supergranular network areas, 
with the cell-centre regions showing a Mg/Ne enhancement 
of about a factor of 2.5 compared to the coronal hole network.
Similar results were obtained by \cite{young:2005}, where it 
was further shown that no significant variations with the solar cycle 
were present (see  Fig.~\ref{fig:young:2005}).

\cite{feldman_etal:98b} used SOHO SUMER observations in 1996
above a quiet Sun region at a radial distance  $R$ $\le$ 1.03 \rsun\
and found a FIP effect of about 4. 
A similar FIP bias was found using SUMER by other authors
(e.g. \citealt{laming_etal:1999}), although in several 
other cases the FIP bias was found to be much smaller, about a factor of 2
(see, e.g. \citealt{warren:1999,widing_etal:2005}).
The differences  were ascribed to local variations or 
changes  due to the solar cycle. 

\cite{doschek_etal:1998_si_ne} used SOHO SUMER observations of the off-limb corona
to measure the Si/Ne abundance using primarily lines from Si VII, Si VIII,
Ne VII and Ne VIII. They found  photospheric abundances.

The discrepancy among the SUMER results is puzzling  and appears to be
related to the choice of spectral lines/ions and potential issues with the atomic data.
We note that \cite{feldman_etal:98b} obtained measurements relative to 
the Li-like O VI 1032 \AA\ line, and  noted that 
`{\em the Ne VIII, Na IX, Mg X lines, all of which 
belong to the Li-like isoelectronic sequence, appear to indicate a systematic 
lower effective FIP bias value than the rest of the lines}'.
As the emission measures of lines from Li-like and Na-like ions 
are often different than those from other sequences (see Section~\ref{sec:anomalous_ions}),
it is therefore likely that, depending on the choice of lines, different 
results could be obtained. 
The quiet Sun off-limb measurement are ideal to measure abundances 
because the temperature is nearly isothermal  
 (see Section~\ref{sec:qs_te}).
It was noted by 
\cite{landi_etal:2002_sumer} that the SUMER lines of different isoelectronic sequences in such 
off-limb observations have different emission measures. The authors suggested a FIP bias of 4.
On the other hand, a re-analysis of the same observation 
by \cite{delzanna_deluca:2017} using more recent
atomic data (CHIANTI v.8) and a different selection of ions indicates 
 abundances close to the  photospheric ones recommended by
 \cite{asplund_etal:09}. Due to discrepancies found, it was suggested
 by \cite{landi_etal:2002_sumer} and 
\cite{widing_etal:2005}
that some of the atomic data available at the time were not accurate.
This affected e.g. the complex coronal iron ions, which were significantly
improved in CHIANTI v.8, and the forbidden lines in various ions,
which still do not have accurate enough atomic data, 
as discussed in \cite{delzanna_deluca:2017}.

Another issue to be considered is the photospheric  abundance of oxygen,
which was used as a reference by  \cite{feldman_etal:98b}, and which
has been modified in  recent times. 
Interestingly, \cite{widing_etal:2005} analysed another SUMER observation
in 1999, and found a smaller FIP bias of about 2 between Mg,Si and 
O, from the ratios of  \ion{Mg}{ix} , \ion{Si}{viii} and \ion{O}{vi}
lines. However, a comparison against the \ion{H}{i} Ly~$\beta$
indicated nearly photospheric abundances for Mg and Si, with a depletion of O
by almost a factor of 2. 
This differs from the result of \cite{feldman_etal:98b}, where the 
relative O/H ratio was found to be close to photospheric.

Finally, we note that \cite{laming_feldman:2001,laming_feldman:2003} 
found  the He abundance to be close 
to its `photospheric' value (He/H about 5\% by number), using 
ratios of He/H lines observed with SOHO SUMER off the limb.

\subsubsection{Abundances in the quiet Sun outer corona}

\cite{raymond_etal:97}  used SOHO UVCS measurements (off limb) and
found that high-FIP elements were 
depleted by an order of magnitude
(compared with photospheric abundances) in the core of a quiescent
equatorial streamer, during solar minimum. On the other hand, 
low-FIP ions were also depleted by about a factor of 3. 
 The abundances along the edges of  streamer legs were similar 
to those measured in the slow solar wind, which is a possible indication 
 that they are the source regions for the slow wind.
Several abundance studies based on SOHO UVCS observations have followed,
but have found similar results. 

It has been noted from the SUMER and UVCS observations by e.g.
\cite{feldman_etal:98b,raymond_etal:97} 
 that  gravitational settling
is an important effect in the outer corona, which complicates
any interpretation of the observations.
Another complication in the analysis is the process of photo-excitation,
which becomes important in lower density plasma and affects different lines in different ways. 

Since the present Living Review does not discuss  
  photo-excitation in detail or the outer corona in general, further 
analyses of  UVCS observations that are available in 
the literature are not reviewed  here.

Finally, it is worth mentioning that 
CHASE observations in the outer corona 
of the  ratio of He~II at 304~\AA\ to the H Lyman-$\alpha$ 
at 1218~\AA\ suggested a helium abundance 
 of 0.079$\pm$0.011 for the quiet Sun  \citep{gabriel_etal:95}.

\subsection{Abundances in active regions}

 \subsubsection{Abundances in hot (3~MK) core loops - the $\pi$ factor}

The abundance measurements of the hot core loops in quiescent 
active regions have recently been reviewed in 
\cite{delzanna_mason:2014}. Here we give a summary of the main results.
The hot core loops are nearly isothermal around 3~MK so 
FIP measurements require that lines formed at these 
temperatures are observed.

There are several results on the FIP bias in terms of the 
Fe vs. the O, Ne abundances from X-ray spectra obtained by  e.g.
the P78-1 SOLEX B spectrometer \citep{mckenzie_feldman:92} and the
SMM/FCS \citep{saba_strong:1993,schmelz_etal:1996}. 
Significant variability in the FIP bias was reported.
However, a re-analysis of the same SMM/FCS observations by \cite{delzanna_mason:2014}, 
selecting only quiescent active region cores and 
using the most recent atomic data and the DEM analysis, has shown  
that in most cases the apparent variability can be explained by 
slightly different temperature distributions.
The  FIP effect for almost all the active region cores,
independently of their age and size, was 
remarkably  consistent at around a value of  $\pi$.

\cite{delzanna:2013_multithermal}
analysed several Hinode EIS observations of quiescent active region core loops,
finding a remarkable constancy of the FIP effect,  around 
a factor of  $\pi$, in excellent agreement with the FIP 
bias obtained from  SMM/FCS  X-ray observations.

Table~\ref{tab:abund} shows a summary of some of the abundance measurements
in quiescent  active region cores.

\begin{table}[htbp]
\caption[Abundance measurements relative to iron in quiescent  active region cores.]{Abundance measurements relative to iron in quiescent  active region cores.
The abundances are listed with decreasing FIP of the element (eV,
the FIP of Fe is 7.9).
Previous values  from quiescent active region observations:
E68: \cite{evans_pounds:1968}; D75: \cite{davis_etal:1975}; 
E: \cite{mason:1975} for Ca and  \cite{young_etal:1997} for Ar; P77: \cite{parkinson:77}; 
W94: \cite{waljeski_etal:1994}; M94: \cite{monsignori_etal:1994};
D13: \cite{delzanna:2013_multithermal},obtained from Hinode EIS in the EUV;
DM14: \cite{delzanna_mason:2014}, obtained from a sample of 
quiescent active region cores observed by SMM/FCS.
For reference,   we show the \cite{asplund_etal:09} [A09] photospheric  abundances
($^\dagger$ the Ne and Ar values were obtained with  
indirect measurements) and the  `coronal' abundances of \cite{feldman_etal:92a}
[F92], which show an average FIP bias of a factor of 4.
}
\centering
\begin{tabular}{lrlllllllllll}
\toprule
 ~        & ~      & ~    & ~    & ~    & ~    & ~    & ~   & ~    & ~    & Phot.          & Coronal \\
 Fe / El. & (FIP)  & E68  & D75  & E    & P77  & W94  & M94 & D13  & DM14 & A09            & F92     \\
\midrule
Fe/Ne     & (21.6) & 2.5  & 1.7  & -    & 1.0  & 1.1  & -   & -    & 1.2  & 0.37$^\dagger$  & 1.05    \\
Fe/Ar     & (15.8) & -    & -    & 13.2 & -    & -    & 25  & 50   & -    & 12.6$^\dagger$  & 33      \\
Fe/O      & (13.6) & 0.28 & 0.26 & -    & 0.16 & 0.16 & -   & -    & 0.2  & 0.065          & 0.16    \\
Fe/S      & (10.4) & -    & -    & -    & -    & -    & 4.0 & 6.8  & -    & 2.4            & 6.8     \\
Fe/Si     & (8.1)  & -    & -    & -    & -    & -    & -   & 1.0  & -    & 1.0            & 1.0     \\
Fe/Mg     & (7.6)  & -    & -    & -    & 1.0  & 1.05 & -   & 0.8  & -    & 0.8            & 0.9     \\
Fe/Ni     & (7.6)  & 6.0  & -    & -    & 17.8 & -    & 8.3 & 29.5 & -    & 19.1           & 18.2    \\
Fe/Ca     & (6.1)  & -    & -    & 11.2 & -    & -    & -   & 13.5 & -    & 14.5           & 14.8    \\
\bottomrule 
\end{tabular}
\label{tab:abund}
\end{table}

Using the Hinode EIS absolute in-flight calibration 
\citep{wang_etal:2011, delzanna:13_eis_calib} and density measurements from 
Fe XIV, \cite{delzanna:2013_multithermal} measured the 
resulting path lengths and  showed that it is the low-FIP elements 
that must be \emph{enhanced} at least by a factor of 3,
otherwise the path lengths would be unreasonably larger than the 
size of the observed structures.

If one assumes a filling factor of one, the EUV and X-ray 
measurements of the AR cores are therefore consistent 
with the abundances of the high-FIP elements being 
photospheric, and those of the low-FIP elements increased 
by about a factor of 3.

This FIP effect is consistent with the Ca/H abundance obtained
by  \cite{mason:1975} from the  \ion{Ca}{xv} intensity above an active region
(a coronal condensation) observed during the total solar eclipse in  1952 in Kharthoum. 

\subsubsection{Abundances in plage and sunspots}

\cite{feldman_etal:1990} studied HRTS UV observations of low charge states of C and Si 
and found an increased Si/C abundance in plage areas, compared to a sunspot.
The increase was  about a factor of 3. A comparison with the \ion{H}{i} L$\alpha$ 
indicated that the sunspot had photospheric abundances.
Abundance variations from the HRTS data were also studied by 
\cite{doschek_etal:1991}, where increased Si/C abundances in active regions
were confirmed.

\subsubsection{TR brightenings}

\cite{young_mason:97} studied the relative Mg/Ne abundance
using SOHO CDS observations of active region brightenings,
some of which were associated with emerging flux regions.
They found variable abundances.
The brightenings associated with emerging flux regions in 
the core of an active region showed near photospheric abundances,
while those associated with  long-lasting fans of loop
showed an increased  Mg/Ne abundance. 

The fact that abundance variations in transition-region lines 
seemed to be associated with the emergence of the magnetic field
was previously pointed out by \cite{sheeley:1995} using Skylab 
observations.

\subsubsection{Abundances in warm (1~MK) loops}

The legs of 
AR `warm' 1~MK loops were clearly observed by the  Skylab NRL slitless 
spectrometer in lines of the low-FIP Mg and high-FIP Ne 
formed at low temperatures (Mg VI, Mg VII,Mg VIII and Ne VI, Ne VII), 
hence there is ample literature on the FIP effect in terms of Mg/Ne relative abundances.
A particularly interesting result is the 
fact that the FIP seems  to increase with the age of 
an active region \citep{widing_feldman:01}, 
from being non-existent  at birth to a typical increase
of a factor of 4\,--\,5 in 2\,--\,3~days. 
One issue, however, is that the above results were obtained 
using the \cite{widing_feldman:89} approximation, which assumes 
that a smooth emission measure distribution exists.

However, AR warm loops are normally consistent with the plasma
being nearly isothermal (across their cross-section, not along the
length), 
so the FIP bias obtained using the 
\cite{widing_feldman:89} method may be overestimated  \citep{delzanna:03}.
In one case,  the EM loci method applied by \cite{delzanna:03} 
to Skylab observations indicated  an FIP bias 4 times smaller, but the
FIP bias was still about a factor of 4.

\begin{figure}[htbp]
 \centerline{
   \includegraphics[width=7.0cm]{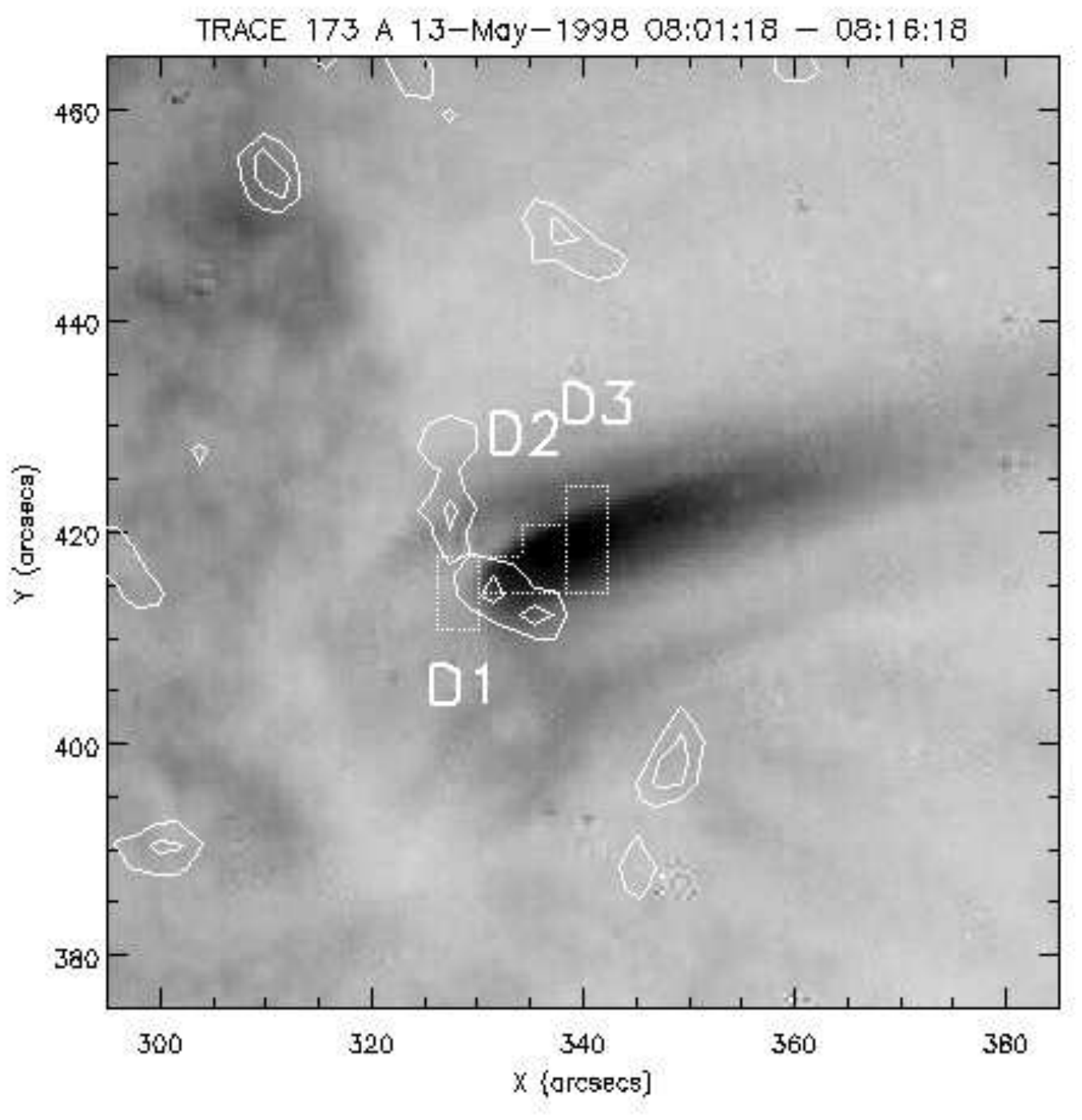}
   \includegraphics[width=6.5cm,angle=90]{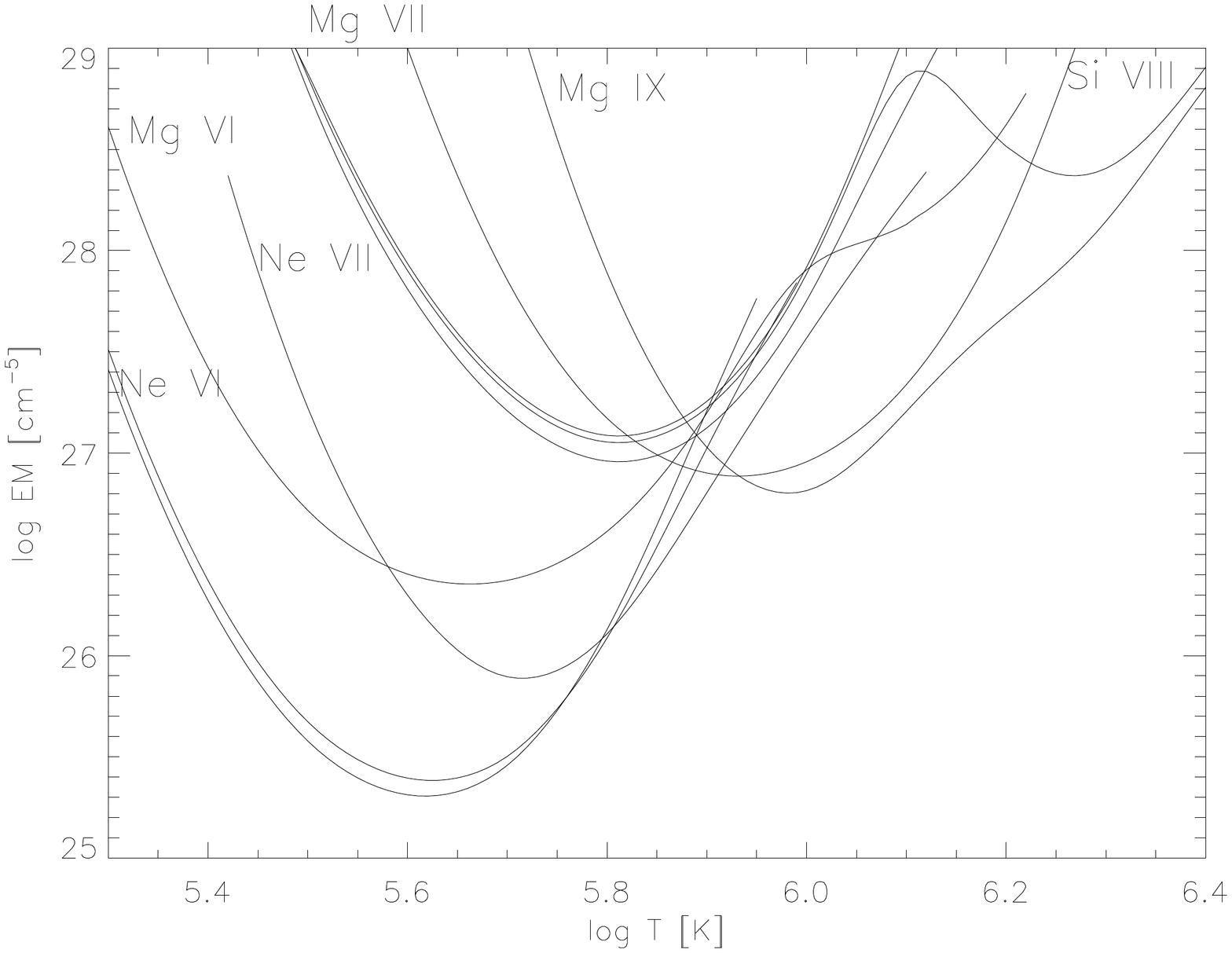}
 }
 \caption{Left: TRACE 171~\AA\ image of the leg of a warm loop, with contours
of the photospheric magnetic fields and three areas selected.
Right: EM loci curves for region D2, using photospheric abundances
(adapted from \citealt{delzanna:03}).
}
  \label{fig:delzanna:2003}
\end{figure}

Results based on SOHO CDS spectra 
indicated  that the Mg/Ne abundances in the legs of 
quiescent 1~MK  loops have either an FIP effect 
of about a factor of 4 \citep[cf.][]{delzanna_mason:03}, or 
have near photospheric abundances \citep{delzanna:03}.
 Figure~\ref{fig:delzanna:2003} shows one of the many 
warm loops that showed photospheric abundances.

We note that \cite{delzanna_etal:2011_outflows}
proposed that a significant fraction of the warm loops
are formed by interchange reconnection of the hot core loops,
which show a FIP bias of a factor of 3\,--\,4.

\subsubsection{Abundances in the unresolved 1--2~MK AR emission}

A FIP bias  of about 2 was found using Hinode EIS off-limb 
observations of the unresolved diffuse 2~MK emission in 
active regions \citep{delzanna:12_atlas}.
Similar results were found in several active regions 
\citep{delzanna:2013_multithermal}.

\cite{baker_etal:2013} have reported 
similar relatively small FIP biases (2--3)  in the core of a
young active region, obtained from lines formed around 1.5~MK,
\ion{S}{x} and \ion{Si}{x}. It is however unclear if this bias 
is associated with the unresolved 1--2~MK emission or 
to the warm loops.

\subsubsection{Abundances in coronal outflows}

The so-called 
coronal outflows in active regions have been discovered with Hinode EIS.
The coronal lines observed by 
Hinode EIS show persistent blue-shifts that are 
 progressively stronger  in lines formed at T $>$ 1~MK and are 
 located exactly above regions of strong magnetic field (sunspots
 umbrae and plage) (\cite{delzanna:08_flows}).
It was also shown that 
they are typically persistent over several days and have a large spatial expansion.
Line profiles are generally symmetric but broadened. 

Several other studies  
\citep{doschek_etal:08,harra_etal:08,hara_etal:08,baker_etal:09}
have been carried out, giving a consistent picture, although 
some  details have been debated in the literature 
(e.g. the asymmetry of the line profiles).
\cite{delzanna_etal:2011_outflows,bradshaw_etal:11}
developed a physical model which is able to explain the general characteristics
of these outflows, triggered by interchange reconnection between the magnetically closed
 hot (3 MK) core loops and  the surrounding open magnetic field.
The model predicts that part of the outflows would contribute  mass and 
momentum  into the slow solar wind, and the reconnection would create 
some of the warm (1 MK) loops.

\cite{brooks_warren:2011}
found a rather large (3--5) FIP bias (in terms of 
the \ion{S}{x} vs. \ion{Si}{x} abundance) in the coronal outflows 
associated with an active region.
Such FIP biases are close to those observed in the slow
solar wind by in-situ instruments, which is suggestive that 
indeed coronal outflows are directly contributing to the solar wind 
(see also \citealt{brooks_warren:2012b,brooks_etal:2015}).

Further measurements of this kind will  be very useful in conjunction 
with the in-situ abundance measurements of Solar Orbiter.

\subsection{Abundances in flares}

There is quite an extensive literature on abundances in solar flares,
although some  results are contradictory. 
It remains to be seen if appropriate methods and updated atomic data 
will  lead to a revision  of the previous results,
as  in the case of quiescent active regions cores.

The FIP bias of 3\,--\,5~MK plasma can be measured from the 
Fe, Mg vs.\ the H-like and He-like O, Ne X-ray lines. 
One of the first studies was that of \cite{veck_parkinson:81},
where X-ray spectra from OSO-8 of solar flares were analysed.
Both  H-like, He-like lines and  the continuum were measured.

\cite{mckenzie_feldman:92} analysed many flares observed with the
 P78-1 SOLEX B spectrometer  and found a variable FIP
effect, up to a factor of 4.
\cite{doschek_etal:1985_flare_ab} used X-ray spectra from the 
P78-1 Bragg crystal spectrometers to obtain the following relative abundances:
Ar/Ca=0.65, K/Ca=0.1, and Ca/Fe=0.1.
\cite{sterling_etal:1993} analysed 25 flares observed with the 
NRL SOLFLEX Bragg crystal spectrometer and obtained an average
Ca abundance of 5$\times$10$^{-6}$.

Several results have been published based on the Skylab spectroheliograms. For example, 
\cite{feldman_widing:1990} used TR lines from O, Ne, Ar, and Mg
during the peak phase of an impulsive flare to find that the relative O/Mg 
was close to its photospheric value. 
The Ar/Mg and Ne/Mg abundances were close to galactic values and have 
often been used to estimate the photospheric abundances of Ar and Ne.

\cite{widing_feldman:2008} re-analysed a Skylab observation of an 
arcade of post-flare loops. EUV lines formed around  3\,--\,5~MK
from Ca, Ni, and Ar indicated a FIP bias between 1.7 and 4.6,
i.e. unexpectedly large for a flare. The authors suggested that 
perhaps  some  filament material contributed to the arcade.

Many results  also exist from SMM. 
\cite{schmelz:1993} analysed two flares observed by SMM/FCS.
Whilst one showed coronal abundances, the second one 
had anomalous Ne and S abundances.
\cite{schmelz_fludra:1993} found a FIP bias of about 4 from two flares
observed with  FCS and BCS, with the 
abundances of high-FIP elements lower than the photospheric 
values (but anomalous Ne and S abundances were found).
\cite{fludra_schmelz:1995} analysed two solar flares observed 
with the SMM FCS and BCS instruments, finding generally that the high-FIP
elements were depleted, compared to their photospheric values,
while low-FIP elements such as Ca were slightly enhanced. 
Anomalous Ne and Ar abundances were also reported.

\begin{figure}[!htbp]
\centerline{\includegraphics[width=10cm]{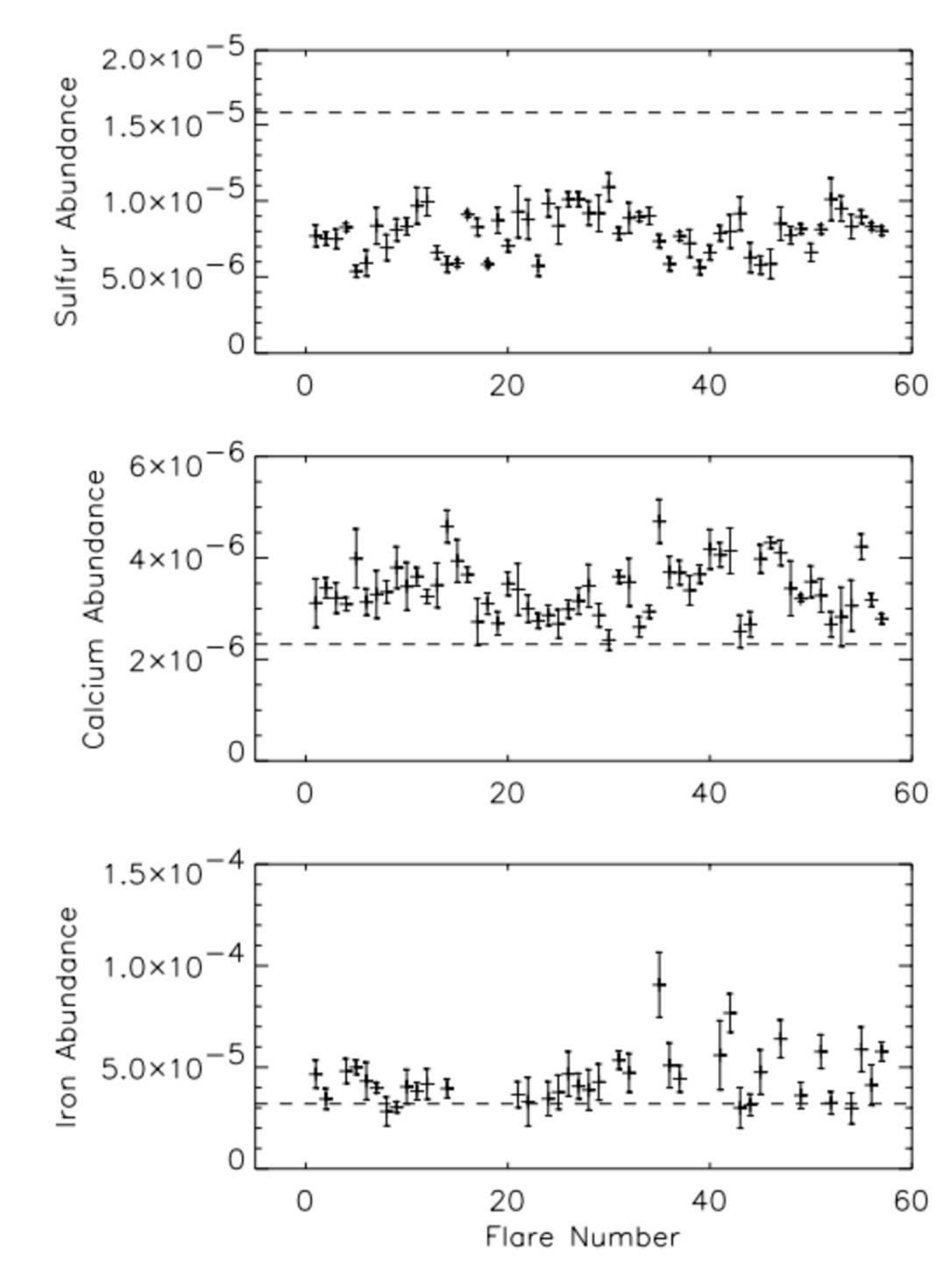}} 
\caption{The absolute abundances of Ca, S, and Fe from several 
flares observed with Yohkoh BCS by \cite{fludra_schmelz:99}.
The dashed lines were the photospheric values known at the time.
Note that the photospheric abundance of S has been significantly 
revised to 1.3$\times$10$^{-5}$ by \cite{asplund_etal:09}.
}
\label{fig:fludra_schmelz_99}
\end{figure}

Line-to-continuum measurements of the He-like
Fe, S, and Ca during flares in the X-rays were possible with Yohkoh BCS. 
\cite{fludra_schmelz:99} obtained absolute abundances for 
many flares, which showed a remarkable constancy,
as shown in Figure~\ref{fig:fludra_schmelz_99}.
There are several similar studies, but often limited to one element.
For example, \cite{bentley_etal:1997} measured an average Ca abundance
for 177 flares to be 3.6$\times$10$^{-6}$.
Overall, the Yohkoh BCS results indicate a small increase in the Ca abundance,
and a S abundance still lower than the revised 
photospheric abundance of 1.3$\times$10$^{-5}$ \citep{asplund_etal:09}.
The Fe abundance is pretty much photospheric.

\cite{landi_etal:2007} analysed SOHO SUMER observations of a flare and  
obtained a Ne  abundance of 1.3$\times$10$^{-4}$  using line vs. continuum 
measurements. This  Ne abundance is in agreement with 
older estimates but significantly higher than the value recommended 
by \cite{asplund_etal:09} in their controversial revision of the
photospheric abundances.

\cite{feldman_etal:2005} also used SOHO SUMER observations
of H, He lines together with the continuum to measure the He
abundance, which on average was found to be 12.2\%.
On the other hand, 
\cite{andretta_etal:2008} used SOHO CDS and ground-based 
observations of another (C-class) flare and obtained 
a He chromospheric abundance of 7.5\%.
This value is closer to the He abundance obtained from 
helioseismology, 8.5\% $\pm$0.2 \citep{asplund_etal:09}.

\begin{figure}[!htbp]
\centerline{\includegraphics[width=7cm]{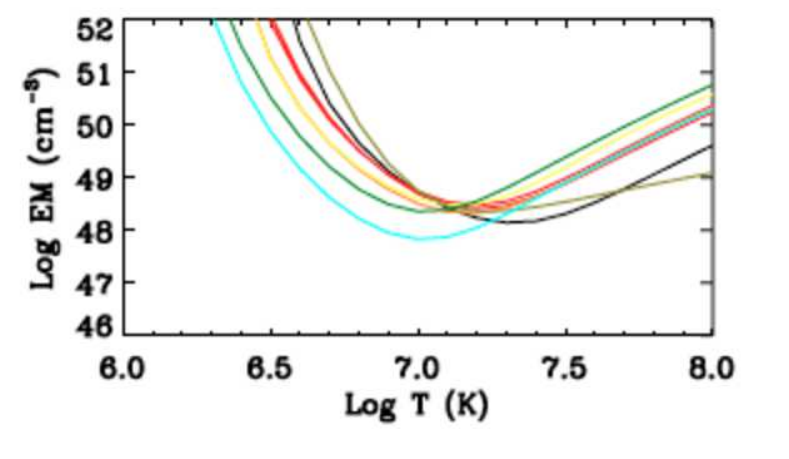}} 
  \caption{EM loci curves during the peak phase of a flare observed with 
RESIK  \citep{chifor_etal:2007}.
}
\label{fig:chifor_etal_resik}
\end{figure}

The RESIK crystal spectrometer 
on-board the CORONAS-F spacecraft has  allowed line-to-continuum 
analyses of solar flares and measurements of the FIP effect in 
terms of e.g. Si vs. Ar and S abundances.
A study by \cite{chifor_etal:2007} found abundances
close to the  photospheric ones of \cite{asplund_etal:09}
and Ar and S abundances consistent with earlier measurements 
from SMM.
Several similar studies have since been carried out, in a 
series of papers on K, Cl, Ar, S by J. Sylwester and 
collaborators, with a variety of results (K enhanced by a large factor, Si 
by a factor of 2). However, in the latest revised analysis
\citep{sylwester_etal:2014}
the results broadly agree with the  \cite{chifor_etal:2007} ones,
i.e. with  abundances close to the \cite{asplund_etal:09}
recommended values, with the exception of K and Ar,
as can be seen from Table~\ref{tab:abund_flares}, where a few 
measurements are listed.

It is interesting to see a relatively good consistency in the results
listed in Table~\ref{tab:abund_flares}, obtained from OSO-8, SMM,  Yohkoh BCS, and RESIK.
The relative abundances of Ar/Ca and Ca/Fe measured by 
\cite{doschek_etal:1985_flare_ab} with P78-1 are in good agreement, while 
the K abundance is half-way between the \cite{chifor_etal:2007} and
\citep{sylwester_etal:2014} RESIK results.
The NRL SOLFLEX Ca abundance obtained by \cite{sterling_etal:1993} is also in 
good agreement with the measurements reported in the Table.

\begin{table}[htbp]
\caption[Abundance measurements relative to hydrogen during flares from X-ray He-like lines.]
{Abundance measurements relative to hydrogen during flares from X-ray He-like lines.
VP1981: OSO-8 measurements by \cite{veck_parkinson:81};
FS1995: SMM results by \cite{fludra_schmelz:1995};  FS1999: results 
from Yohkoh BCS by \cite{fludra_schmelz:99};  C+2007: results from RESIK by \cite{chifor_etal:2007};
S+2014: results from RESIK by \cite{sylwester_etal:2014}.  Photospheric abundances are the
 \cite{asplund_etal:09} recommended values (those in brackets are derived, not measured).
}
\centering
\begin{tabular}{llllllllllll}
\toprule
Element & VP1981 &  FS1995 &  FS1999   & C+2007 &   S+2014 & Photospheric \\
\midrule

O  & -                   &       2.5$\times$10$^{-4}$ & -  & - & - & 4.9$\times$10$^{-4}$ \\
Ne & -                   & 3.8--8.6$\times$10$^{-5}$ & -  & - & - & (8.5$\times$10$^{-5}$) \\
K  & -                   &       -                   & - &  1.2$\times$10$^{-7}$& 7.2$\times$10$^{-7}$  & 1.1$\times$10$^{-7}$ \\
Ar & 2.4$\times$10$^{-6}$ &       -                   &      -  &  1.5$\times$10$^{-6}$ & 3.8$\times$10$^{-6}$ & (2.5$\times$10$^{-6}$) \\
S  & 8.1$\times$10$^{-6}$ &  9.7--14$\times$10$^{-6}$  &  7.8--8.2$\times$10$^{-6}$ &  7.9$\times$10$^{-6}$ & 8.7$\times$10$^{-6}$ & 1.3$\times$10$^{-5}$ \\

Mg  & -                   & 4.0--4.9$\times$10$^{-5}$ & -  & - & - & 4.0$\times$10$^{-5}$  \\

Si & 4.5$\times$10$^{-5}$ &  3.3--4.2$\times$10$^{-5}$ & - & 3.2$\times$10$^{-5}$  & 3.6$\times$10$^{-5}$  &  3.2$\times$10$^{-5}$ \\
Ca & 3.2$\times$10$^{-6}$ &  4.2--4.4$\times$10$^{-6}$ &  2.9--3.1$\times$10$^{-6}$ & -                     &  -                &  2.2$\times$10$^{-6}$\\
Fe & -                    & 3.8--4.8$\times$10$^{-5}$ &  3.9--4.1$\times$10$^{-5}$ & -                     &  -                &  3.2$\times$10$^{-5}$\\

\bottomrule 
\end{tabular}
\label{tab:abund_flares}
\end{table}

Direct measurements of inner-shell Fe lines formed during large solar flares
observed by the crystal spectrometers 
aboard  Yohkoh, SMM and P78-1 were thought to be consistent with the
coronal Fe abundance being equal to the 
photospheric value, within a factor of two, according to
 \citep{phillips_etal:95}. 
\cite{phillips:2012} revised 
observations of the Fe K$\alpha$ lines observed by
 P78-1 to obtain an  iron abundance about  1.6 times photospheric.

RHESSI observations of flares offer in principle the 
opportunity to obtain the iron abundance from the line-to-continuum
measurements in the X-rays. 
The 6.65 keV iron lines are observed as a small `bump' in 
the RHESSI spectra.
\cite{phillips_dennis:2012} revised a previous analysis to 
obtain an iron abundance from 20 flares of  8.1$\times$10$^{-5}$,
or 2.6 times the  photospheric abundance.

As shown by \cite{delzanna_woods:2013}, it is possible to use 
SDO/EVE spectra to obtain information on the FIP effect during 
large flares.  The relative abundance of high-temperature high-FIP
argon  lines vs. low-FIP elements (iron, calcium) was measured.
Several flares were studied and in all cases near 
photospheric abundances were found. 

\cite{warren:2014} analysed the SDO/EVE  line-to-continuum emission 
of 21 flares to obtain nearly photospheric iron 
abundances in all cases. 

Finally,  measurements of an inverse FIP effect 
have recently been reported in solar observations. 
They have been obtained with 
Hinode EIS, using lines from Ar and  Ca, during a solar flare
but close to a sunspot \citep{doschek_etal:2015,doschek_warren:2016}.
These measurements are very interesting because it is the first time 
that an inverse FIP effect has been reported on the Sun. 
On the other hand, inverse FIP effects are common in 
active stars \citep[see, e.g. the review by][]{laming:2015}.

\subsection{Conclusions on the abundance measurements}

In conclusion, there is little evidence that at transition region 
temperatures the coronal holes and the quiet Sun 
present any significant FIP bias.

Some of the SUMER results on the measurements of the 
quiet Sun above 1 MK appear to be affected by the choice of spectral lines and 
the atomic data, which still need further improvement. 

The 3~MK hot loops in the cores of the active regions generally show 
a FIP bias of about a factor of three, consistently using 
EUV and X-Ray measurements. 

The situation with regard to  the warm 1~MK  loops 
is not clear, in the sense that the FIP bias is not as large 
as previously thought on the basis of the Skylab observations and 
the approximate methods previously used.

We end by noting that results from  in-situ 
measurements of the solar wind have similar contradictions and 
large uncertainties as the remote-sensing ones. 
However, the overall picture that is emerging is that 
the fast solar wind has nearly photospheric abundances,
while the slow solar wind has abundances that vary with the 
solar cycle. They are intrinsically very variable, but still 
normally between the photospheric and the abundances of the 
hot core loops (FIP bias of 3).

\clearpage

\section{Conclusions and Future}

There has been tremendous progress in the plasma diagnostics
for XUV spectroscopy since the early rocket observations, in particular
in the determination of the physical state of the emitting plasma.
Great strides have been made with high cadence imaging instruments,
such as SDO/AIA. A better knowledge of the contribution of spectral
lines to the Hinode/XRT and SDO/AIA channels, for example using Hinode/EIS
spectra has been very fruitful. A very powerful tool is provided by
combining imaging and spectroscopic observations, in particular with
the combination of SDO, Stereo, Hinode and IRIS instruments.

XUV spectrometers have produced a large amount of data in the
past couple of decades, however improvements in terms of spectral
resolution, spatial resolution and radiometric calibration have
been modest, when compared to earlier observations,
for example Skylab and HRTS. More recent spectrometers, such as
 IRIS, have provided much improved spatial and spectral
resolution, with high cadence, but only over a limited wavelength range and FOV.

One of the most significant advances has been with regard to
the atomic data calculations, e.g. from the APAP team, 
reaching an accuracy of 10\,--\,20\%
for the strongest lines, especially when lines from the same ion are considered.
However, a significant amount of work is still needed before
accurate atomic data are available for all astrophysically-important ions.
Together with these new atomic data, another major advance has been
the identification of all the strongest lines in the XUV.
However, a significant fraction of the weaker EUV and UV
lines is remain unidentified. These new atomic data have been
consolidated in the recent release of CHIANTI v8. 

Some important spectral regions for solar diagnostics have not
received enough attention in the past few decades. In particular, the
X-ray wavelength range. In this regard, we look forward to seeing
data for the first X-ray spectra (after SMM) in the 6\,--\,20~\AA\
region with the forthcoming MAGIXS sounding rocket.

Our review of the status of atomic data and observations
has shown that several outstanding issues still remain.

High-cadence observations of the transition region with IRIS
and of the corona with, e.g., SDO/AIA have clearly shown variability and
dynamics on  short  timescales, likely leading to the
consideration of  non-equilibrium  atomic processes, for example,
time-dependent ionisation/recombination. 
More studies in this regard will will be needed for future analyses.

However, emission measure analyses of `quiescent' features
normally suggest that equilibrium ionization holds, at least for the
majority of ions. One outstanding issue relates to the frequently
enhanced intensities of lines from the Li-like and Na-like ions,
which produce the strongest UV solar lines. The EUV Helium lines are
also often enhanced. These enhancements have not to-date been adequately explained.

Other non-equilibrium effects such as departures from
Maxwellian electron distributions might be common in the corona,
but are still hard to diagnose with the current instrumentation.
Much work in this regard has been carried out to model them,
but further refinements are needed.

Regarding electron densities, there is now better agreement in the various
measurements of different solar regions, although simultaneous measurements
along single structures such as coronal loops are still beyond the current
instrumental capabilities. Electron densities in the transition-region flare 
plasma  can reach high values, but measurements based on high-temperature
lines are still uncertain.  Further measurements of diagnostic line
ratios from flare ions are needed.

Regarding electron temperatures, relatively few direct
measurements based on ratios of lines from the same ion exist.
Measurements of the electron temperatures in the outer corona
are particularly important for solar wind modelling.
We therefore look forward to off-limb Mg IX observations from
the SPICE spectrometer on-board Solar Orbiter.

Regarding elemental abundances and the FIP effect, we have presented
an overview of the many  results avaiable  in the literature,
highlighting that in several cases significantly different FIP biases
have been obtained. We have pointed out that such differences can 
partly be explained by the use of different diagnostic techniques,
a different selection of the spectral lines and the limitations of the 
atomic data which was used at the time of the analyses.
Most of the discordant results concern the quiet Sun, while a 
consensus on a FIP bias of about 3--4 in the hot (3 MK) 
active region cores is emerging. Coronal holes and coronal 
hole plumes appear to have close to photospheric abundances. 
However,  puzzling results 
have  also been obtained on the `warm' 1~MK loops, and further 
analyses would be useful.
The connection between elemental abundances as measured remotely
with those measured in-situ is still extremely hard to study, but will feature
prominently in the future with Solar Orbiter observations closer to the Sun.
Further advances will be obtained by linking more realistic models 
with time-dependent plasma processes.

\section{Acknowledgements}
\label{sec:acknowledgements}
We acknowledge funding from STFC (UK).

We thank the anonymous referee for the careful reading and useful comments
which have enhanced the quality of this review. 
We also thank  various colleagues for their comments on this
review which have helped to improve the manuscript.
In particular, we thank George Doschek, Peter Young, Martin Laming, 
Nigel Badnell, Ian Grant, Elmar  Tr{\"a}bert.


\newcommand\aj{{AJ}}%
\newcommand\araa{{ARA\&A}}%
\newcommand\apj{{ApJ}}%
\newcommand\apjl{{ApJ}}%
\newcommand\apjs{{ApJS}}%
\newcommand\ao{{Appl.~Opt.}}%
\newcommand\apss{{Ap\&SS}}%
\newcommand\aap{{A\&A}}%
\newcommand\aapr{{A\&A~Rev.}}%
\newcommand\aaps{{A\&AS}}%
\newcommand\azh{{AZh}}%
\newcommand\baas{{BAAS}}%
\newcommand\jrasc{{JRASC}}%
\newcommand\memras{{MmRAS}}%
\newcommand\mnras{{MNRAS}}%
\newcommand\pra{{Phys.~Rev.~A}}%
\newcommand\prb{{Phys.~Rev.~B}}%
\newcommand\prc{{Phys.~Rev.~C}}%
\newcommand\prd{{Phys.~Rev.~D}}%
\newcommand\pre{{Phys.~Rev.~E}}%
\newcommand\prl{{Phys.~Rev.~Lett.}}%
\newcommand\pasp{{PASP}}%
\newcommand\pasj{{PASJ}}%
\newcommand\qjras{{QJRAS}}%
\newcommand\skytel{{S\&T}}%
\newcommand\solphys{{Sol.~Phys.}}%
\newcommand\sovast{{Soviet~Ast.}}%
\newcommand\ssr{{Space~Sci.~Rev.}}%
\newcommand\zap{{ZAp}}%
\newcommand\nat{{Nature}}%
\newcommand\iaucirc{{IAU~Circ.}}%
\newcommand\aplett{{Astrophys.~Lett.}}%
\newcommand\apspr{{Astrophys.~Space~Phys.~Res.}}%
\newcommand\bain{{Bull.~Astron.~Inst.~Netherlands}}%
\newcommand\fcp{{Fund.~Cosmic~Phys.}}%
\newcommand\gca{{Geochim.~Cosmochim.~Acta}}%
\newcommand\grl{{Geophys.~Res.~Lett.}}%
\newcommand\jcp{{J.~Chem.~Phys.}}%
\newcommand\jgr{{J.~Geophys.~Res.}}%
\newcommand\jqsrt{{J.~Quant.~Spec.~Radiat.~Transf.}}%
\newcommand\memsai{{Mem.~Soc.~Astron.~Italiana}}%
\newcommand\nphysa{{Nucl.~Phys.~A}}%
\newcommand\physrep{{Phys.~Rep.}}%
\newcommand\physscr{{Phys.~Scr}}%
\newcommand\planss{{Planet.~Space~Sci.}}%
\newcommand\procspie{{Proc.~SPIE}}%


\newpage
\bibliography{refs}

\end{document}